\pdfoutput=1
\documentclass[11pt]{article}
\usepackage[utf8]{inputenc}
\usepackage[T1]{fontenc}
\usepackage{lmodern}
\usepackage[letterpaper,margin=1in]{geometry}%
\usepackage{amsmath,amssymb,amsthm,mathtools}
\usepackage{booktabs}
\usepackage{array}
\usepackage{tabularx}
\usepackage{enumitem}
\usepackage{xcolor}
\usepackage[hidelinks]{hyperref}
\usepackage[capitalize,noabbrev]{cleveref}
\newcommand{\companionurl}{https://serggasp.github.io/sample-MLS-site/}
\usepackage[disable]{todonotes}%

\makeatletter
\renewenvironment{abstract}{%
  \begin{center}\bfseries\abstractname
\end{center}\list{}{\leftmargin=\leftmargini\rightmargin=\leftmargin\listparindent=1.5em\parsep=0pt}\item\relax}{\endlist}
\makeatother

\usepackage{aliascnt}
\newtheorem{theorem}{Theorem}[section]
\newcommand{\sharedtheorem}[2]{%
\newaliascnt{#1}{theorem}\newtheorem{#1}[#1]{#2}\aliascntresetthe{#1}}
\sharedtheorem{lemma}{Lemma}
\sharedtheorem{proposition}{Proposition}
\sharedtheorem{corollary}{Corollary}
\theoremstyle{definition}
\sharedtheorem{definition}{Definition}
\theoremstyle{remark}
\sharedtheorem{remark}{Remark}
\sharedtheorem{example}{Example}
\crefname{algorithm}{Algorithm}{Algorithms}
\usepackage{thmtools}

\usepackage[most]{tcolorbox}
\newaliascnt{algorithm}{theorem}\aliascntresetthe{algorithm}
\definecolor{algframe}{RGB}{40,70,120}
\newenvironment{algorithm}[1][]{%
  \refstepcounter{algorithm}%
  \begin{tcolorbox}[enhanced,breakable,colback=white,colframe=algframe,
      boxrule=0.6pt,arc=1.5pt,left=6pt,right=6pt,top=4pt,bottom=4pt,
      before skip=\medskipamount,after skip=\medskipamount,
      fonttitle=\sffamily\bfseries,coltitle=white,
      title={Algorithm~\thealgorithm\if\relax\detokenize{#1}\relax\else\ \textmd{(#1)}\fi},
      before upper={\parindent=0pt\parskip=3pt}]}%
  {\end{tcolorbox}}
\makeatletter
\newcommand{\statementswithin}[1]{\counterwithin{theorem}{#1}%
  \aliascntresetthe{lemma}\aliascntresetthe{proposition}\aliascntresetthe{corollary}%
  \aliascntresetthe{definition}\aliascntresetthe{remark}\aliascntresetthe{example}%
  \aliascntresetthe{algorithm}%
  \@for\st@name:=theorem,lemma,proposition,corollary,definition,remark,example,algorithm\do{%
    \expandafter\edef\csname theH\st@name\endcsname{\expandafter\noexpand\csname theH#1\endcsname.\noexpand\arabic{\st@name}}}}
\makeatother
\newcommand{\Input}{\par\textsf{\textbf{Input:}}\ }
\newcommand{\Parameters}{\par\textsf{\textbf{Parameters:}}\ }
\newcommand{\Body}{\par\vspace{2pt}{\color{algframe!50}\hrule}\vspace{4pt}}

\newcommand{\Ostar}{O^{*}}
\newcommand{\E}{\mathbb{E}}
\newcommand{\N}{\mathbb{N}}
\newcommand{\R}{\mathbb{R}}
\newcommand{\cA}{\mathcal{A}}
\newcommand{\cC}{\mathcal{C}}
\newcommand{\cD}{\mathcal{D}}
\newcommand{\cE}{\mathcal{E}}
\newcommand{\cF}{\mathcal{F}}
\newcommand{\cN}{\mathcal{N}}%
\newcommand{\cT}{\mathcal{T}}
\newcommand{\poly}{\operatorname{poly}}
\newcommand{\YES}{\textup{\textsc{yes}}}
\newcommand{\NO}{\textup{\textsc{no}}}
\newcommand{\MLS}{\textup{\textsf{MLS-$\theta$}}}

\newcommand{\comp}{\operatorname{comp}}
\newcommand{\gc}{\gamma_{\mathrm{c}}}
\newcommand{\OPT}{\operatorname{OPT}}
\newcommand{\Tm}{\mathrm{T}}

\newcommand{\pname}[1]{\textup{\textsc{#1}}}

\newcommand{\ancfile}[1]{\texttt{#1}}
{}

\title{Monotone Local Search with Sample-Dependent Running Times}
\author{Serge Gaspers\\ UNSW Sydney, Australia\\ \href{mailto:serge.gaspers@unsw.edu.au}{\texttt{serge.gaspers@unsw.edu.au}}}
\hypersetup{pdftitle={Monotone Local Search with Sample-Dependent Running Times},pdfauthor={Serge Gaspers}}
\date{}

\begin{document}
\pagenumbering{alph}%
\begin{titlepage}
\maketitle
\thispagestyle{empty}
\begin{abstract}
  Monotone Local Search turns an algorithm that extends a given partial solution into an algorithm
  that finds a solution from scratch: sample a few elements, hope that they lie in an unknown solution,
  and complete them. Its known analyses bound every call of the extension algorithm by a worst-case time
  that depends only on the number of sampled elements, and Esmer, Kulik, Marx, Neuen and Sharma (SODA 2024)
  showed that no algorithm that uses the extension algorithm only as a black box with this worst-case running
  time beats the original form by more than a polynomial factor.
  We let the running time of the extension algorithm depend on the actual sample. If it decides in time
  $\Ostar(2^{g(I,X,k)})$, for a polynomial-time computable $g$, whether a sample $X$ can be completed to a solution with at most $k$ elements, our main
  theorem gives a randomized algorithm with one-sided error in time $\Ostar(2^{\psi n})$ that accepts with
  probability at least $1/2$ whenever, for some solution $S$ of size $k$ in a universe of $n$ elements and some
  $t$, the \emph{exponential moment} $\sum_{X\subseteq S,|X|=t}2^{-g(I,X,k)}$ is at least $2^{-\psi n}\binom nt$.
  A corollary, multivariate Monotone Local Search, generalizes the theorems of Fomin, Gaspers, Lokshtanov and
  Saurabh (STOC 2016 \& J.~ACM 2019) and of Gaspers and Lee (ICALP 2017).

  For \pname{Max Leaf Spanning Tree} and, equivalently, \pname{Connected Dominating Set}, the previous fastest
  known algorithms run in time $\Ostar(1.8619^n)$ in polynomial space and $\Ostar(1.8139^n)$ in exponential
  space. Completing a sample is a Steiner tree problem, solved in time exponential in the number of terminals,
  which varies strongly with the sample. We obtain a randomized polynomial-space algorithm with one-sided error
  in time $O(1.6991^n)$ and a deterministic polynomial-space algorithm in time
  $(\sqrt5-\frac12)^{n}2^{o(n)}=O(1.7361^n)$. For connectivity problems in which the running time of the
  extension algorithm doubles with every component of the sample, an analysis through a partition function
  gives randomized polynomial-space algorithms in time $\Ostar(1.8^n)$ for \pname{Connected Feedback Vertex Set}
  and $O(1.6444^n)$ for \pname{Connected Vertex Cover}, both also with vertex weights.
\end{abstract}
{}
\begin{center}
  Companion site (reading version, ancillary programs and certificates, Lean formalization):\\
  \url{\companionurl}
\end{center}
\end{titlepage}

\pagenumbering{roman}
\setcounter{tocdepth}{1}%
\tableofcontents
\clearpage
\pagenumbering{arabic}

\section{Introduction}\label{sec:intro}

Many exact exponential-time algorithms~\cite{FominK10} solve a \emph{subset problem}: given an instance whose
solutions are subsets of a universe of $n$ elements, find a solution of a given size $k$. We write
$\Ostar(f(n))$ for $f(n)$ times a factor polynomial in the input length. When the decimal base of our running
times is rounded up from a strictly smaller constant, the big-$O$ notation absorbs the polynomial factor, so we
write $O(1.7204^n)$ rather than $\Ostar(1.7204^n)$.

A standard way to solve a subset problem faster than $2^n$ combines a parameterized algorithm with
enumeration. Suppose that solutions of size $k$ can be found in time $\Ostar(c^k)$ with
$c<4$. For each $k$, run this algorithm or enumerate all $\binom nk$ candidate sets, whichever is
faster; the running time is $\Ostar\left(\max_k\min\left(c^k,\binom nk\right)\right)$.
At $k=n/2$ the two terms are $c^{n/2}$ and about $2^n$,
and if $c<4$ then for every fraction $k/n$ one of the two terms is exponentially smaller than $2^n$.

Monotone Local Search~\cite{FominGLS19} beats
$2^n$ whenever the \emph{extension} version of the subset problem is solved in time $\Ostar(c^{k-t})$, for \emph{any} constant $c$. In the extension version, a size-$t$ part
of the solution is given and the question is whether the part can be completed to a full solution of size $k$. Monotone Local Search draws a random set $X$ of $t$ elements, hopes that it lies inside an unknown
solution $S$, and asks an extension algorithm whether $X$ can be completed by at most $k-t$ further
elements. If the extension algorithm runs in time $\Ostar(c^{k-t})$, the resulting algorithm runs in time
$\Ostar((2-1/c)^n)$. Gaspers and Lee~\cite{GaspersL17} allowed a further factor $b^{n-t}$ in the
running time of the extension algorithm, giving running time $\Ostar((b+1-1/c)^n)$. We show that an extension algorithm with running time
$\Ostar(b^{n-t}c^{k-t}d^{t})$, for constants $b,c,d\ge1$ with $d\le bc$, gives the multivariate
bound $\Ostar((b+d-d/c)^n)$ (\cref{cor:co-multivariate}). In all these versions the running-time
bound of the extension algorithm depends only on $n$, $t$ and $k$: every sample of a given size is
charged the same, worst-case, running time.

As long as the extension algorithm is used only through such a worst-case running time, the bound
$(2-1/c)^n$ cannot be improved,
and by a remark in~\cite{EsmerKMNS24} neither can
$(b+1-1/c)^n$. \cite{EsmerKMNS24} consider algorithms whose only problem-specific access is a
polynomial-time feasibility test and an extension algorithm with running time $\Ostar(c^{k-|X|})$,
and they show that every randomized algorithm of this kind takes time at least $(2-1/c)^n$
divided by a polynomial in $n$; their concluding section remarks that the same technique shows the bound of
Gaspers and Lee to be optimal. %
To go below these
bounds, an algorithm must therefore use more information about
the extension algorithm than its worst-case running time (\cref{sec:fw-related}).

This paper lets the running time of the extension algorithm depend on the sample. Our main theorem
(\cref{thm:fw-main}) applies to every extension algorithm that, given $(I,X,k)$,
decides whether some set of at most $k-|X|$ further elements completes $X$ to a solution,
in time $\Ostar(2^{g(I,X,k)})$, for some \emph{running-time exponent} $g$ that is computable in polynomial
time before the extension algorithm is called (\cref{def:fw-cost}).
For a solution $S$ of
size $k$ and a sample size $t$, the \emph{exponential moment} of $S$ is
\[
  \mu_t(S)=\sum_{X\in\binom{S}{t}}2^{-g(I,X,k)} .
\]
The theorem requires one solution $S$ whose exponential moment, for some $t$, is large compared with
the number $\binom nt$ of possible samples.

\begin{theorem}[informal version of \cref{thm:fw-main}]\label{thm:intro-main}
  Suppose that the extension algorithm runs in time $\Ostar(2^{g(I,X,k)})$ for a running-time exponent
  $g$ that is computable in polynomial time, and let $\psi\ge0$ be a rational constant. There is a randomized
  algorithm that, given $I$ and $k$, runs in time $\Ostar(2^{\psi n})$, and in polynomial space if the
  extension algorithm uses polynomial space; it rejects if $I$ has no solution of size at most $k$,
  and it accepts with probability at least $1/2$ if some solution $S$ of size $k$ satisfies
  $\mu_t(S)\ge\binom nt2^{-\psi n}$ for some $t$.
\end{theorem}

\Cref{thm:intro-main} follows from \cref{thm:fw-main} with $\Gamma=\kappa=1$.
The algorithm is simple: it tries every sample size and every integer threshold
$\theta$, and for each pair it draws about $2^{\psi n-\theta}$ samples and calls the extension
algorithm only on those with $g(I,X,k)\le\theta$. It never needs to know which sample size or threshold is the right one. For one sample size, the
probabilistic core of the theorem is the counterpart of Scaman's analysis of restart strategies~\cite{Scaman23},
and the mean-exponent form (\cref{thm:co-mean}) is the counterpart of Zamir's Markov cutoff~\cite{Zamir23jensen}
(\cref{sec:fw-mechanisms}).

\paragraph{What is new compared with restarts and cutoffs.} One might summarize the algorithm as ``run the
extension algorithm only on samples whose running time is below a threshold, and optimize the restart
schedule''. Five features separate our setting from restart and cutoff strategies for Las Vegas algorithms,
and the analyses in this paper use all five.
\begin{enumerate}[label=(\roman*)]
  \item The running-time exponent $g(I,X,k)$ is computed in polynomial time \emph{before} the extension
    algorithm is called, so a sample above the threshold is skipped at polynomial cost instead of being run
    and cut off.
  \item A call can succeed only if $X\subseteq S$ for an unknown solution $S$, so the moment $\mu_t(S)$ sums
    $2^{-g(I,X,k)}$ over the subsets of $S$ only; a sample outside every solution plays the role of a run that
    never succeeds. The algorithm thus balances the number of samples inside $S$ with a small running-time
    exponent against the running times of their extension calls.
  \item The running-time exponent depends on the structure of the sample $X$, not only on its size.
  \item The generating function $W_S$ defined below collects the moments of all sample sizes, so no sample
    size has to be guessed.
  \item $W_S$ is bounded from below by combinatorial recursions over the structure of the sampled
    subsets of $S$, such as dynamic programming along a spanning tree of $S$; all our applications are
    analyzed in this way.
\end{enumerate}
Using the exponential moment instead of the mean of $g$ can gain an exponential factor. If half of the
$t$-subsets of $S$ have running-time exponent $n/4$ and the other half have exponent $3n/4$, then the worst
case charges running time $2^{3n/4}$ for a sample in $S$ and the mean charges $2^{n/2}$, whereas the
exponential moment charges at most $2^{n/4+1}$, so the faster half determines it.
\Cref{sec:ov-moments} gives the extreme case of exponents $0$ and $n$.
In general Jensen's inequality shows that the exponential moment never certifies a
larger running time than the mean.
Within the class of algorithms that sample uniformly, threshold on $g$ and are charged $2^\theta$ for a call at
threshold $\theta$, exponential moments measure, up to polynomial factors, exactly what thresholded sampling
can use (\cref{thm:fw-optimal}), and in a black-box model the theorem cannot in general be derandomized
without exponential loss (\cref{sec:dr-barrier}). These optimality statements are limited to the models in
which they are proved; they are not lower bounds against other algorithms for the problems themselves.
For concrete extension algorithms, we bound exponential moments from below via generating functions along a spanning tree of the solution. Adapting an
entropy argument of Csikv\'ari and Lin~\cite{CsikvariL14} to weighted counts, we show that paths are the worst
connected solutions for the tree partition function that bounds these generating functions from below, up to a
constant factor.

\subsection{Why sample-dependent running times}

The motivating example is \pname{Max Leaf Spanning Tree} (MaxLeaf): given a connected graph $G$, find
a spanning tree with as many leaves as possible. For $n\ge3$, the maximum number of leaves is
$n-\gc(G)$, where $\gc(G)$ is the minimum size of a connected dominating set, and optimal solutions of
MaxLeaf and of \pname{Connected Dominating Set}, the problem of finding a connected dominating set of
minimum size, translate into each other in polynomial time (\cref{lem:cds-maxleaf}). So a running-time
bound for one of them holds for the other up to a polynomial factor, and we state our results for both.
\Cref{tab:intro-bounds} lists the
previous bounds for the two problems, in polynomial and in exponential space, together with ours; the
fastest previously published polynomial-space bound, $1.8619^n$, dates from 2011, setting aside a
claimed $k$-leaf algorithm that is deemed incorrect (\cref{sec:fw-related}).

Completing a sampled set $X$ to a connected dominating set is a Steiner tree problem, as in the
algorithm of Abu-Khzam, Mouawad and Liedloff~\cite{Abu-KhzamML11}: the terminals
are the vertices that $X$ does not dominate and the components of $G[X]$. Inclusion--exclusion over branching walks~\cite{Nederlof13} solves it in time $2^{|\Tm(J_X)|}n^{O(1)}$ and polynomial space through a
directed Steiner encoding (\cref{lem:cv-W,lem:cds-dir-steiner}), where $J_X$
is the residual instance left by $X$ (\cref{sec:cds-states}). Denoting the closed neighborhood of
$X$ by $N[X]$ and the number of connected components of $G[X]$ by $\comp(G[X])$, the number of terminals is
\[
  |\Tm(J_X)|=n-|N[X]|+\comp(G[X])
\]
by \cref{cor:cds-steiner}. This running time depends on the sample, and a single
sampled vertex can be worthless: if its closed neighborhood is already dominated and it touches exactly one
component of $G[X]$, adding it to $X$ leaves the number of terminals unchanged. Accounting that charges
every sample of a given size its worst case therefore credits sampling with nothing. Sampling leaves
of the spanning tree instead meets another obstacle: the extension problem of MaxLeaf parameterized by the
number $\ell'$ of additional leaves is W[1]-hard (\cref{thm:co-w1}), so it has no algorithm with running time
$\Ostar(c^{\ell'})$ for a constant $c$ unless $\mathrm{FPT}=\mathrm{W}[1]$, and for running times
$b^{n-|X|}d^{|X|}$ (the product form with $c=1$) the main theorem certifies, for the worst solution size, nothing below a single call of the extension
algorithm on the empty sample (\cref{sec:co-limits}).

An average sample, by contrast, is valuable. If $S$ is a connected dominating set of size $s=\sigma n$
and $X$ is a uniform $t$-subset of $S$ with $t\ge1$, then
\[
  \E\,|\Tm(J_X)|\le n-1-\frac{t(n-2)}{s}
\]
by \cref{lem:cds-C}: a sampled vertex removes on average about $1/\sigma$ terminals, which divides the
running time of the extension algorithm by about $2^{1/\sigma}$, while the requirement that it lie in $S$
lowers the success probability only by a factor of about $1/\sigma$. As $2^{1/\sigma}>1/\sigma$, an average sampled
vertex pays for itself, and sample-dependent running times turn this average gain into a faster algorithm.

\subsection{Our results}

We make three contributions. First, sample-dependent Monotone Local Search: the main theorem with exponential
moments and its generating-function form (\cref{thm:fw-main,lem:fw-gf}). Second, Algorithm RSC with graded
closure sampling, which solves \pname{Max Leaf Spanning Tree} and \pname{Connected Dominating Set} in time
$O(1.6991^n)$, together with a deterministic algorithm (\cref{thm:intro-cds}). Third, the component-count
principle: for extension algorithms whose running time doubles with every component of the sample, paths are
the worst connected solutions up to a constant factor (\cref{lem:cc-gibbs}), which gives our algorithms for
\pname{Connected Feedback Vertex Set} and \pname{Connected Vertex Cover} (\cref{thm:intro-cc}). Further
consequences of these techniques are collected in \cref{tab:intro-applications}.
The companion site~\cite{Gaspers26companion} has a reading version of this paper together with the
ancillary programs, the certificates and the Lean development.

\paragraph{Monotone Local Search with exponential moments.}
Classical Monotone Local Search with
sample size $t$ draws about $\binom nt/\binom kt$ samples before one of them lies inside $S$, and it
budgets each call of the extension algorithm its worst-case running time $2^{\gamma}$, so its running
time is about $\binom nt\binom kt^{-1}2^{\gamma}$. In $\mu_t(S)$, each $t$-subset $X$ of $S$ contributes
$2^{-g(I,X,k)}$, the reciprocal of the running time of the extension algorithm on $X$. If $g$ takes the
same value $\gamma$ on all these subsets, then $\mu_t(S)=\binom kt2^{-\gamma}$ and the running time
$\binom nt/\mu_t(S)$ of the main theorem is the classical one. In general, let $\gamma$ be the largest value of $g$ on these subsets. A subset $X$ with
$g(I,X,k)=\gamma-j$, whose running-time bound is $2^j$ times smaller, contributes $2^j$ times as
much to $\mu_t(S)$ as a subset with $g(I,X,k)=\gamma$. So the subsets on which the extension algorithm
is fast can make $\mu_t(S)$ exponentially larger than $\binom kt2^{-\gamma}$, and the running time
$\binom nt/\mu_t(S)$ exponentially smaller than the classical one.

The main theorem (\cref{thm:fw-main}) is stated for arbitrary implicit set systems, real-valued
running-time exponents, and a polynomial loss factor. Its condition~\eqref{eq:fw-star} asks that
$\mu_t(S)\ge\binom nt2^{-\psi n}$ for some $t$, up to that factor. To avoid a separate analysis for
each $t$, \cref{lem:fw-gf} collects all sample sizes in one polynomial in a variable $y$, the
\emph{generating function} $W_S(y)=\sum_t\mu_t(S)\,y^t=\sum_{X\subseteq S}y^{|X|}2^{-g(I,X,k)}$: up to a
polynomial factor, the smallest running time $2^{\psi n}$ for which the theorem applies to a solution
$S$ is $\inf_{y\ge0}(1+y)^n/W_S(y)$. Our analyses of the applications bound $W_S(y)$ from below.
Exponential moments are the right quantity for this class of algorithms.
Condition~\eqref{eq:fw-star} is necessary, up to a polynomial factor, for the algorithm to call the
extension algorithm on any subset of $S$ at all (\cref{prop:fw-necessary}), and no algorithm that draws uniform samples, chooses
sample sizes and thresholds $\theta$ on $g$, even adaptively, and calls the extension algorithm only on samples
with $g\le\theta$, is faster on any instance by more than a polynomial factor, when each trial is
counted with the running time $2^\theta$ that its threshold allows (\cref{thm:fw-optimal}). The theorem
extends to randomized trials, such as samples from any polynomial-time samplable distribution and
extension algorithms with one-sided error; it needs only that a trial at threshold $\theta$ runs in time at
most $2^\theta$ and succeeds whenever a random variable attached to it is at most $\theta$
(\cref{thm:fw-budgeted,cor:fw-onesided,thm:fw-samplable}).

\paragraph{Corollaries.} If only a bound $\lambda$ on the \emph{mean} of $g$ over the $t$-subsets of
every solution is known, Jensen's inequality turns it into a bound on the exponential moment, and the
running time is $\Ostar\left(\binom nt\binom kt^{-1}2^{\lambda}\right)$ (\cref{thm:co-mean}). This
bound also has a simple, direct proof, by a cutoff at $\lambda+1$ and Markov's inequality (\cref{rem:co-markov}). If the mean decreases by
$\rho$ bits per sampled element, the exponent has a closed form, and sampling pays exactly when
$\rho>\log(1/\sigma)$ at solution fraction $\sigma=k/n$ (\cref{cor:co-linear}). For running times of product form
$b^{n-|X|}c^{k-|X|}d^{|X|}$ the generating function has a closed form, which gives the multivariate bound
$\bigl(b+\min(d,bc)(1-1/c)\bigr)^n$ for all $b,c,d\ge1$, and hence the theorems
of~\cite{FominGLS19} ($b=d=1$) and~\cite{GaspersL17} ($d=1$); conversely, the main theorem certifies
nothing better for such running times (\cref{prop:co-product,cor:co-multivariate}). Every gain in this paper therefore comes from the
dependence of the running time on the samples.

\paragraph{Applications.} \Cref{tab:intro-bounds,tab:intro-applications} compare our running-time bounds with the previous ones. The
rest of this subsection describes the algorithms and the ideas behind them. All our algorithms use
polynomial space, and the randomized ones have one-sided error.

\paragraph{Max Leaf Spanning Tree and Connected Dominating Set.} All four algorithms complete a sample
with the Steiner tree algorithm described above, whose running time is exponential in the number of
terminals. The randomized ones never output a value below $\gc(G)$, and they output $\gc(G)$ with
constant probability.

\begin{theorem}[\cref{thm:rsc,thm:det-cds}]\label{thm:intro-cds}
  \pname{Max Leaf Spanning Tree} and \pname{Connected Dominating Set} can be solved in polynomial space
  by a randomized algorithm with one-sided error in time $O(1.6991^n)$, and by a deterministic algorithm
  in time $(\sqrt5-\frac12)^n2^{o(n)}=O(1.7361^n)$.
\end{theorem}

\begin{itemize}
  \item Algorithm RS uses only the mean number of terminals of a sample (\cref{lem:cds-C}), through the
    linear-savings corollary, and runs in time $O(1.7449^n)$ (\cref{thm:rs}).
  \item Algorithm RS-$\theta$ is the same algorithm run with a smaller target running time $2^{\psi n}$, that is, a smaller
    \emph{rate}~$\psi$; the improvement lies in the
    analysis. Exponential moments are not additive, so the mean bound no longer suffices. Instead we bound the number of terminals by a potential that
    is local along a spanning tree of the solution, evaluate the generating function of this potential by
    a three-state recursion, and bound the recursion from below by a convex region that the merge rule maps into itself,
    following Rote~\cite{Rote19} and Rosenfeld~\cite{Rosenfeld21} (\cref{sec:ov-cds}). This gives
    $O(1.7204^n)$ (\cref{thm:rstheta}). On the uniform $(2,2)$-lobsters, and on graphs whose unique
    maximum-leaf spanning tree is such a lobster, the algorithm needs time
    $\Omega(1.7192^n)$, so the analysis is nearly tight for this algorithm.
  \item Algorithm RSC adds \emph{closure sampling}: every vertex with at least two sampled neighbors is
    added to the sample independently, with probability $0.07$ if it has exactly two and $0.077$ if it has more. In the tight case of the mean bound the
    solution induces a star, and its center is the only vertex that can have two or more sampled neighbors. The budgeted form of the main theorem
    applies to these random trials and gives $O(1.6991^n)$ (\cref{thm:rsc}). %
  \item The deterministic algorithm uses more than the values of the running-time exponent. In a black-box
    model, every deterministic algorithm is, on some instances, slower than the randomized one by a factor exponential in $n$, or has cost $\binom nk$ up to a polynomial factor, even for the mean form of the main theorem (if the mean bound, at
    least~$1$, is below $\log\binom kt$ by a linear amount; \cref{sec:derandomisation,sec:app-barrier}). But the number of Steiner terminals is at most an affine function of the sample
    with nonnegative integer weights, an \emph{integer affine certificate}. Balanced set-inclusion families,
    which we list deterministically in polynomial space, derandomize the exponential moments of such
    certificates (\cref{thm:dr-det-moments}). This gives a deterministic polynomial-space algorithm in time
    $O(1.7361^n)$ (\cref{thm:det-cds}).
\end{itemize}

\begin{table}[tb]
  \centering
  \renewcommand{\arraystretch}{1.12}
  \begin{tabularx}{\textwidth}{@{}lllll>{\raggedright\arraybackslash}X@{}}
    \toprule
    Bound & Type & Space & Year & Source & Method \\
    \midrule
    $1.9407^n$ & deterministic & polynomial & 2008 & \cite{FominGK08} & measure and conquer \\
    $1.8966^n$ & deterministic & polynomial & 2011 & \cite{FernauKKLLRR11} & measure and conquer \\
    $1.8962^n$ & deterministic & polynomial & 2014 & \cite{Binkele-RaibleF14} & measure and conquer \\
    $1.8619^n$ & deterministic & polynomial & 2011 & \cite{Abu-KhzamML11} & branching, Steiner tree \\
    $1.8139^n$ & deterministic & exponential & 2012 & \cite{Binkele-RaibleF12dir} & algorithm for digraphs \\
    \midrule
    $1.7449^n$ & randomized & polynomial & & \cref{thm:rs} & mean number of terminals \\
    $\mathbf{1.7361^n}$ & \textbf{deterministic} & polynomial & & \cref{thm:det-cds} & affine certificate \\
    $1.7204^n$ & randomized & polynomial & & \cref{thm:rstheta} & exponential moments \\
    $\mathbf{1.6991^n}$ & \textbf{randomized} & polynomial & & \cref{thm:rsc} & closure sampling \\
    \bottomrule
  \end{tabularx}
  \caption{Bounds for \pname{Connected Dominating Set} and \pname{Max Leaf Spanning Tree} on
    $n$-vertex graphs.
    The $1.8139^n$ algorithm is
    for digraphs and applies to $G$ with each edge replaced by two opposite arcs. Randomized
    algorithms have one-sided error. Our four bases rest on computations checked in exact or ball arithmetic
  (\cref{sec:ancillary}); \cref{tab:rsc-audit} maps the proof of the base $1.6991$.}
  \label{tab:intro-bounds}
\end{table}

\begin{table}[tb]
  \centering
  \renewcommand{\arraystretch}{1.1}
  \setlength{\tabcolsep}{3pt}
  \begin{tabular*}{\textwidth}{@{\extracolsep{\fill}}lllll@{}}
    \toprule
    Problem & Previous & New & Type & Source \\
    \midrule
    \multicolumn{5}{@{}l}{\emph{Main applications of the component-count principle}}\\
    \pname{Connected Feedback Vertex Set} & $1.9526^n$~\cite{CyganNPPRW22}$^*$ & $1.8^n$ & rand. & Thm.~\ref{thm:cc-cfvs} \\
    \pname{Connected Vertex Cover} & $1.7088^n$~\cite{GolovachHK18} & $1.6444^n$ & rand. & Thm.~\ref{thm:cc-cvc} \\
    \quad weighted & $1.7088^n$~\cite{Cygan12}$^*$ & $1.6444^n$ & rand. & Thm.~\ref{thm:cc-cvc} \\
    \midrule
    \multicolumn{5}{@{}l}{\emph{Further consequences}}\\
    \pname{Graph Motif} & $1.7088^n$~\cite{BjorklundKK15motif}$^*$ & $1.6444^n$ & rand. & Thm.~\ref{thm:cc-motif} \\
    \pname{Connected $r$-Dominating Set}, $r\ge2$ & $1.8619^n$~\cite{Abu-KhzamML11}$^\dagger$ & $1.7361^n$ & det. & Thm.~\ref{thm:rdom} \\
    & & $1.7204^n$ & rand. & Thm.~\ref{thm:rdom} \\
    \pname{Connected Dominating Set}, weighted & $1.9896^n$~\cite{Abu-KhzamFGLM25}$^\dagger$ & $1.7361^n$ & det. & Thm.~\ref{thm:cv-weighted} \\
    & & $1.7204^n$ & rand. & Thm.~\ref{thm:cv-weighted} \\
    \pname{Weakly Connected Dominating Set} & $1.8619^n$~\cite{Abu-KhzamML11}$^\dagger$ & $1.7361^n$ & det. & Thm.~\ref{thm:cv-wcds} \\
    \pname{Directed Max Leaf Spanning Tree} & $1.9044^n$~\cite{Binkele-RaibleF12dir} & $1.7361^n$ & det. & Thm.~\ref{thm:cv-dir} \\
    \pname{Directed Steiner Tree} & $2^{n/2}$ (see \S\ref{sec:fa-dst}) & $1.3727^n$ & rand. & Thm.~\ref{thm:fa-dst} \\
    \pname{Paired Domination} & $1.7159^n$~\cite{AndreevBKSTV24} & $1.6302^n$ & rand. & Thm.~\ref{thm:fa-pd} \\
    \bottomrule
  \end{tabular*}
  \caption{Component-count applications and further consequences, on $n$-vertex graphs.
    All bounds in the table use polynomial space, and the randomized ones have
    one-sided error. Every new base except $1.8$ rests on a computation checked in exact or ball arithmetic
    (\cref{sec:ancillary}). Weighted problems have polynomially bounded integer weights. $^*$:
    baseline derived by balancing the cited algorithm against enumeration.
    $^\dagger$: by a reduction to the
    cited algorithm (\cref{sec:rdom,sec:cv-weighted,sec:cv-wcds}).
    Other previous bounds:
    $1.3533^n$ for undirected
    \pname{Steiner Tree}~\cite[Theorem~5]{FominGKLS13}, and
    $1.8139^n$ in exponential space for
    \pname{Directed Max Leaf Spanning Tree}~\cite{Binkele-RaibleF12dir}.
  }
  \label{tab:intro-applications}
\end{table}

\paragraph{Connected Feedback Vertex Set.} Our extension algorithm is a Cut \& Count algorithm, built on the $\Ostar(3^k)$ algorithm of \cite{CyganNPPRW22}, that, given a sample $X$, runs in time $\Ostar(3^{k-|X|}2^{\comp(G[X])})$: its running time
doubles with every component of the sample, so its product form $(b,c,d)=(1,3,2)$ gives nothing. A
random subset of a connected solution, however, has far fewer components than vertices once the sample
is a sizeable part of the solution. We bound the generating function $W_S$ from below by a partition function on a spanning tree of $S$
(\cref{cor:cc-T}), and a weighted form of a lower bound of Csikv\'ari and
Lin~\cite{CsikvariL14} on the number of homomorphisms of a tree into a graph shows that, up to a constant
factor, this partition function is smallest on paths (\cref{lem:cc-gibbs}).
The main theorem
then gives time $\Ostar(1.8^n)$ (\cref{thm:cc-cfvs}), against the baseline $O(1.9526^n)$ obtained by
balancing $\Ostar(3^k)$ against enumeration. The analysis is tight for this algorithm: on a family of graphs, every base below $1.8$ leads to success probability $o(1)$
(\cref{prop:cc-cfvs-tight}).

\paragraph{Other problems with component-count running times.} Let $c>1$ be a constant, and consider a problem on a graph $G$ whose solutions induce connected subgraphs of $G$. For every extension algorithm that runs in time $\Ostar(c^{k-|X|}2^{\comp(G[X])})$, the Gibbs-type
inequality and the main theorem give an algorithm with running time $\Ostar(((4c-3)/(2c-1))^n)$
(\cref{cor:cc-rate}). %
The case $c=2$, where
a sharper analysis gives $1.6444$ (\cref{lem:cc-cert}), covers two
problems:%
\begin{itemize}
  \item \pname{Connected Vertex Cover}. With Cygan's weighted $\Ostar(2^k)$
    algorithm~\cite{Cygan12} as the extension step, we find a minimum-weight connected vertex cover, for polynomially
    bounded integer weights, in time $O(1.6444^n)$ (\cref{thm:cc-cvc}).
  \item \pname{Graph Motif}. Our extension algorithm contracts each component of the sample to one
    vertex with a new color and then runs the $\Ostar(2^k)$ algorithm of Björklund, Kaski and
    Kowalik~\cite{BjorklundKK15motif}, in time $\Ostar(2^{k-|X|}2^{\comp(G[X])})$. The exponential
    moments give $O(1.6444^n)$ (\cref{thm:cc-motif}).
\end{itemize}
For polynomially bounded integer weights, the same analysis gives $O(1.6444^n)$
for \pname{Maximum-Weight Connected $k$-Subgraph} and $\Ostar(1.8^n)$ for vertex-weighted
\pname{Connected Feedback Vertex Set} (\cref{thm:cc-mwcs,thm:cc-wcfvs}). For the two problems that we
consider the main applications of the component-count principle, we obtain the following.

\begin{theorem}[\cref{thm:cc-cfvs,thm:cc-wcfvs,thm:cc-cvc}]\label{thm:intro-cc}
  \pname{(Weighted) Connected Feedback Vertex Set} and \pname{(Weighted) Connected Vertex Cover} can be
  solved in polynomial space by randomized algorithms with one-sided error in time $\Ostar(1.8^n)$ and
  $O(1.6444^n)$, respectively, for vertex weights in $\{1,\dots,w_{\max}\}$ with $w_{\max}\le n^{O(1)}$.
\end{theorem}

\paragraph{Further applications.} The analyses of Algorithm RS-$\theta$ and of the deterministic
algorithm use only a spanning tree of the solution and an assignment of each vertex outside the
solution to a solution vertex that dominates it, so they extend to \pname{Connected $r$-Dominating Set}
and to vertex-weighted \pname{Connected Dominating Set} and weighted MaxLeaf, with randomized
$O(1.7204^n)$ and deterministic $O(1.7361^n)$ running times, and to \pname{Weakly Connected Dominating Set} and
\pname{Directed Max Leaf Spanning Tree}, with a deterministic $O(1.7361^n)$ running time
(\cref{thm:rdom,thm:cv-weighted,thm:cv-wcds,thm:cv-dir}); closure sampling does not extend in this way
(\cref{rem:cv-limits}). The inclusion--exclusion formula for vertex-weighted directed Steiner trees
behind these extensions (\cref{lem:cv-W}) is also the extension algorithm for \pname{Directed Steiner
Tree}, which gives randomized $O(1.3727^n)$ (\cref{thm:fa-dst}). For \pname{Paired Domination}, a
Tutte-matrix finishing step gives randomized $O(1.6302^n)$ (\cref{thm:fa-pd}); both bounds also hold with
polynomially bounded integer vertex weights.

\subsection{Techniques and related work}\label{sec:fw-related}

\paragraph{Monotone Local Search.}
\Cref{sec:co-product} recalls classical and multivariate Monotone Local
Search~\cite{FominGLS19,GaspersL17}. \cite{EsmerKMNS22,EsmerKMNS26} developed approximate Monotone Local Search,
after Lee's thesis~\cite{Lee21} had applied Monotone Local Search with exact and approximate extension
algorithms to approximation and computed the running times numerically, and they
showed it to be optimal in the extension-oracle model described earlier%
~\cite{EsmerKMNS24}.
Their oracle lower bounds use the same planting idea as our black-box barrier (\cref{sec:dr-barrier}).
Esmer and Kulik~\cite{EsmerK25}
sample with a black box in which the number of correct samples is random, but the running time of
each black-box call is fixed by the algorithm. \Cref{thm:fw-main} reads a sample-dependent running-time
exponent before the extension call, which is information outside the oracle model
of~\cite{EsmerKMNS24}, so their lower bounds do not apply; for product-form running times it gives the classical bound up to a polynomial factor (\cref{prop:co-product}), so it does not contradict them.

\paragraph{Restarts, cutoffs and exponential moments.}
The probabilistic core of \cref{thm:fw-main} has counterparts outside Monotone Local Search: restart
strategies with cutoffs for Las Vegas algorithms~\cite{LubySZ93,AltGMKW96}, the Markov cutoff of
Zamir~\cite{Zamir23jensen}, which corresponds to the mean-exponent theorem (\cref{thm:co-mean}), the bounds
of Scaman~\cite{Scaman23} on restart strategies through the mean of the reciprocal running time, and the
exponential moment of the number of guessed variables in the analysis of~\cite{PaturiPSZ05}. \cite{FominGLS19,GaspersL17} already used generating functions in
Monotone Local Search, for product-form running times. \Cref{sec:fw-mechanisms} compares these with
our setting.

\paragraph{Certificates.} The running times of RS-$\theta$ and RSC follow from lower bounds on the
generating function $W_S$ of a solution $S$, which obeys a recursion along a spanning tree of $S$. We
prove these lower bounds by antinorm certificates in a cumulative cone. This adapts the method that
Rote~\cite{Rote19} introduced to count minimal dominating sets in trees, generalized by
Rosenfeld~\cite{Rosenfeld21}, from upper to lower bounds; for lower bounds, antinorms in the sense of
Guglielmi and Protasov~\cite{GuglielmiP13} take the place of norms. Measure-and-conquer analyses of
branching algorithms~\cite{FominGK09} bound the number of leaves of a search tree from above; our
certificates bound generating functions from below.

\paragraph{A caveat.} Koutis and Williams~\cite{KoutisW09} claimed a randomized $\Ostar(2^k)$ algorithm that
decides whether a graph has a spanning tree with at least $k$ leaves; balanced against enumeration, %
it would give randomized time $O(1.7088^n)$.
This algorithm is deemed incorrect \cite[footnote~(i)]{Binkele-RaibleF14}\cite[footnote~1]{Zehavi15arxiv}, and the $k$-leaf application does not appear in the
journal version~\cite{KoutisW16}. We know of no
repair; our $1.6991^n$ is below $1.7088^n$ in any case.

\paragraph{Derandomization.} \cite{FominGLS19} derandomize product-form Monotone Local
Search with set-inclusion families, which multiplies the running time by $2^{o(n)}$. For sample-dependent
running-time exponents this route fails in general. Consider the class of instances whose running-time exponents satisfy
the mean bounds that Algorithm RS for \pname{Connected Dominating Set} uses. On this class,
every deterministic algorithm that may use the running-time exponent and the extension algorithm only as oracles
must spend running time at least $1.7547^n/\poly(n)$ in that model, where an extension call on a sample
$X$ takes time $2^{g(I,X,k)}$ and every other oracle call takes unit time, whereas the randomized
algorithm of the linear-savings corollary needs time $\Ostar(\sqrt3^{\,n})$ when the solution size is $n/2$ (\cref{cor:dr-bb-rs}; the proofs are in \cref{sec:app-barrier}). Our deterministic algorithms use the affine structure of the running-time exponent instead, so the
bound $1.7547^n$, which concerns the class defined by the mean bounds alone, does not apply to them.
\cite{EsmerKMNS22,EsmerKMNS26} generalize set-inclusion families to set-intersection families and to known weights;
in their extension-oracle model a query costs $c^{\ell}$ for an extension budget $\ell$, whatever the
sample, and deterministic and randomized running times match up to a factor
$2^{o(n)}$~\cite{EsmerKMNS24,EsmerKMNS26}. The lopsided universal families of \cite{FominLPS16} also give set-inclusion families, but in general exponentially larger ones.

{}
\section{Overview of the key ideas}\label{sec:overview}

This section explains informally why the main
theorem has the form it has, and which ideas give the bounds for \pname{Connected Dominating Set} and
for the problems with component-count running times. Every statement in the first four subsections is proved
later, at the place cited. \Cref{sec:ov-eth} reports recent claims about the Exponential Time Hypothesis
together with our assessment of what they imply for the problems of this paper, supported by informal arguments
that are not developed into proofs here.

\subsection{Why exponential moments}\label{sec:ov-moments}

Fix an instance with a solution $S$ of size $k$ in a universe of $n$ elements, and a sample size $t$.
A uniformly random $t$-subset $X$ of the universe lies in $S$ with probability $\binom kt/\binom nt$.
In classical Monotone Local Search every extension call has the same running time $\Ostar(c^{k-t})$, so it finds $S$ in
about $\binom nt/\binom kt$ trials of running time $\Ostar(c^{k-t})$ each, and optimizing over $t$
gives the running time $\Ostar((2-1/c)^n)$. When the running time $2^{g(X)}$ of an extension call depends on $X$, where we write
$g(X)$ for the running-time exponent of the sample $X$ when the target size $k$ is fixed, the natural first
attempt runs the extension algorithm on every sample. Its expected running time per trial is the mean
$\E[2^{g(X)}]$, which the slowest samples dominate.

The algorithm of \cref{thm:fw-main} instead runs the extension algorithm only on samples whose running-time
exponent, computed in polynomial time beforehand, is at most a threshold $\theta$, and discards the
other samples. A trial of the pair $(t,\theta)$ then runs in time $\Ostar(2^\theta)$, and it succeeds with probability $p_\theta=\Pr[X\subseteq S\text{ and }g(X)\le\theta]$, so this
pair finds $S$ in expected time about $2^\theta/p_\theta$. The best threshold is characterized by the
following two-sided estimate (\cref{lem:fw-threshold-prob}). Write $\xi=g(X)$ if $X\subseteq S$ and
$\xi=\infty$ otherwise. Since $2^{-\theta}\mathbf 1[\xi\le\theta]\le2^{-\xi}$ pointwise, every threshold
satisfies $2^{-\theta}p_\theta\le\E[2^{-\xi}]$. Conversely, splitting $\E[2^{-\xi}]$ according to the
integer part of $\xi$ shows that one of the thresholds up to about $\log(1/\E[2^{-\xi}])+1$
attains $2^{-\theta}p_\theta\ge\E[2^{-\xi}]/\poly(n)$. Moreover,
\[
  \E[2^{-\xi}]=\frac{1}{\binom nt}\sum_{X\in\binom St}2^{-g(X)}=\frac{\mu_t(S)}{\binom nt},
\]
the normalized exponential moment of the introduction. So the best
threshold for sample size $t$ finds $S$ in time $\Ostar\left(\binom nt/\mu_t(S)\right)$. The algorithm knows neither the
right $t$ nor the right $\theta$, but there are only polynomially many pairs, and it gives each the
same running-time budget $2^{\psi n}$; it succeeds as soon as
$\binom nt/\mu_t(S)=\Ostar(2^{\psi n})$ for some $t$, which is condition~\eqref{eq:fw-star}.

Thresholding can be exponentially better than paying the mean. Suppose that half of the $t$-subsets of
$S$ have running-time exponent $0$ and the other half $n$. Running the extension algorithm on every sample in
$S$ takes time about $2^{n-1}$ on average, the mean-exponent theorem (\cref{thm:co-mean}) charges
$2^{\E[g]}=2^{n/2}$ per sample, and the exponential moment charges $1/\E[2^{-g}]\le2$. In general,
Jensen's inequality gives $\E[2^{-g}]\ge2^{-\E[g]}$, so the exponential moment never certifies a worse
running time than the mean, and it certifies a better one whenever the running-time exponent varies.

The same estimate shows that nothing in this class of algorithms does better. Any algorithm that
draws uniform samples, picks the sample size and the threshold of each trial, possibly adaptively,
and is charged $2^\theta$ for a trial with threshold $\theta$, succeeds with probability at most its
total charge times $\max_{t,\theta}2^{-\theta}\Pr[X\text{ extends to a solution and }g(X)\le\theta]$,
and by the estimate above this maximum is at most $\max_t\mu_t/\binom nt$, where $\mu_t$ sums
$2^{-g(X)}$ over all $t$-sets $X$ that extend to a solution (\cref{thm:fw-optimal}). The exponential
moments are therefore not an artifact of the analysis: they measure exactly what thresholded sampling
can exploit.

\paragraph{One generating function instead of $k+1$ moments.}
To verify condition~\eqref{eq:fw-star} we use the generating function
$W_S(y)=\sum_{X\subseteq S}y^{|X|}2^{-g(X)}=\sum_t\mu_t(S)y^t$. Comparing it termwise with
$(1+y)^n=\sum_t\binom nty^t$ shows that the running time certified for $S$ is
$\inf_{y\ge0}(1+y)^n/W_S(y)$, up to a factor $n+1$ (\cref{lem:fw-gf}). Upper bounds on running times
thus become lower bounds on a sum over the subsets of $S$, for a single well-chosen value
of~$y$; this is the form in which all our applications are analyzed. For a product-form running time,
$2^{-g(X)}=b^{-(n-|X|)}c^{-(k-|X|)}d^{-|X|}$, the sum is
$W_S(y)=b^{-n}c^{-k}(1+ybc/d)^k$, and at $y=(c-1)d/(bc)$ the second factor becomes $c^k$; this gives
$(1+y)^n/W_S(y)=(b+d-d/c)^n$. This is the multivariate bound when $d\le bc$; for $d>bc$ the
choice $y=0$ gives at most $(bc)^n$, which is smaller, hence the $\min(d,bc)$ in \cref{cor:co-multivariate}.
This bound subsumes the bounds of \cite{FominGLS19} ($b=d=1$) and of Gaspers and Lee~\cite{GaspersL17} ($d=1$)
(\cref{prop:co-product,cor:co-multivariate}). For such running times the main theorem certifies nothing
better, so the gains below come only from the dependence of the running time on the sample.

\subsection{Connected Dominating Set: from averages to the shape of the solution}\label{sec:ov-cds}

\paragraph{The extension problem and its running time.} Suppose we have a sample $X$ that is a subset of an unknown minimum connected
dominating set $S$. The extension algorithm must add a set $Y$ of vertices that dominates the vertices outside
$N[X]$ and such that $G[X\cup Y]$ is connected: the added vertices must join the components of $G[X]$
into one connected subgraph.
This is a Steiner tree problem with one terminal for each
undominated vertex and one for each component of $G[X]$, which inclusion--exclusion over branching
walks~\cite{Nederlof13} solves in polynomial space and in time $\Ostar(2^{|\Tm(J_X)|})$, where $|\Tm(J_X)|=n-|N[X]|+\comp(G[X])$ is the number of terminals
(\cref{lem:cds-dir-steiner,cor:cds-steiner}).

\paragraph{An affine bound and the mean.} Take a spanning tree $F$ of $G$ in which all vertices outside
$S$ are leaves, and give each $u\in S$ the weight $w(u)=d_F(u)-1$. The weights add up to $n-2$, and
counting the vertices and edges of $F$ around a nonempty sample gives the pointwise bound
$|\Tm(J_X)|\le n-1-w(X)$ (\cref{lem:cds-pointwise}). Each vertex of $S$ lies in a uniform $t$-subset
of $S$ with probability $t/|S|$, so on average a sampled vertex removes $(n-2)/|S|\approx1/\sigma$
terminals, where $\sigma=|S|/n$ (\cref{lem:cds-C}). Each further sampled vertex multiplies the
probability that the sample lies in $S$ by about $\sigma$, so it multiplies the expected number of
trials by about $1/\sigma=2^{\log(1/\sigma)}$; the terminals it removes divide the running time of
the extension algorithm by about $2^{1/\sigma}$. Since $1/\sigma>\log(1/\sigma)$ for every
$\sigma\in(0,1]$, the second factor wins at every solution size; the linear-savings corollary
(\cref{cor:co-linear}) turns this into Algorithm RS and time $O(1.7449^n)$ (\cref{thm:rs}). The
bound on the mean is tight, for instance, when the solution induces a star and every outside vertex
has exactly one neighbor in it (\cref{prop:cds-tight}).

\paragraph{Exponential moments along a tree.} The mean cannot take advantage of the variation of the number of
terminals between samples, but the exponential moments reward this variation. To use it, we bound the number of
terminals from above by a potential $\Pi_{F,\nu}(X)$ along a spanning tree $F$ of $G[S]$
(\cref{lem:rst-E}). Each vertex outside $S$ is assigned a neighbor in $S$, its \emph{owner}, and
$\nu:S\to\mathbb Z_{\ge0}$ is the map that gives, for each solution vertex $v$, the number $\nu_v$ of
outside vertices that $v$ owns. The potential adds up three
counts: the outside vertices whose owner is not sampled, the solution vertices that are neither
sampled nor adjacent in $F$ to a sampled vertex, and the components of $F[X]$.
The potential is \emph{local} along $F$: since $\comp(F[X])=|X|-e(F[X])$ for the forest $F[X]$, where
$e(F[X])$ is the number of edges of $F$ with both ends in $X$, the potential is a sum over the vertices
$v\in S$ of terms that each depend only on $\nu_v$ and on which vertices of the closed neighborhood
$N_F[v]$ are sampled. Because of this, its generating function $\sum_{X\subseteq S}y^{|X|}2^{-\Pi_{F,\nu}(X)}$ can be computed by
dynamic programming from the leaves of $F$ to its root, with a vector of three numbers per vertex
and a fixed bilinear rule that merges a child into its parent (\cref{lem:rst-treerec}). What we need is
a lower bound on the result that holds for every tree at once. Following the method of
Rote~\cite{Rote19} and Rosenfeld~\cite{Rosenfeld21} for counting dominating sets in trees, which we
turn from upper into lower bounds, we look for a convex region $\mathcal P$ in $\R^3$ that is mapped into itself by the merge rule
and contains the vector of every single vertex $v$ up to a factor $e_1e_2^{\,\nu_v}$, where $\nu_v$ is the
number of outside vertices that $v$ owns and $e_1,e_2>0$ are two constants chosen together
with $\mathcal P$: a scale for each solution vertex and a scale for each outside vertex.
Then the vector of every tree on the $s$ solution
vertices lies in $e_1^{\,s}e_2^{\,n-s}\mathcal P$, and a lower bound of order $e_1^{\,s}e_2^{\,n-s}$ on
the generating function follows (\cref{lem:rst-F}). The region is the convex hull of finitely many
points, enlarged by a cone that the merge rule preserves; for lower bounds this plays the role that
a norm plays in the method for upper bounds, and is an antinorm in the sense of Guglielmi and
Protasov~\cite{GuglielmiP13}. Two such regions for the middle solution sizes, with the closed form of RS elsewhere, give Algorithm
RS-$\theta$, which is RS with a smaller rate, and time $O(1.7204^n)$ (\cref{thm:rstheta}). On a family of
lobsters it needs time at least $1.7192^n$, so little remains to be
gained for this algorithm (\cref{prop:rst-caterpillar}).

\paragraph{Changing the samples.} The tight case of the mean bound, a star, suggests a change of the
algorithm. A sample of leaves of the star dominates the hub but leaves every sampled leaf in a
component of its own; adding the hub
would merge
these components into one. Algorithm RSC therefore adds
every vertex with at least two sampled neighbors to the sample independently, with probability
$0.07$, or $0.077$ if it has three or more sampled neighbors. The resulting samples are no longer uniform sets of a given size, but the budgeted form of the
main theorem (\cref{thm:fw-budgeted}) only needs trials that are fast, sound, and accept with probability at least
$\Pr[\xi\le\theta]$ for a random variable $\xi$ whose distribution does not depend on $\theta$; then $\E[2^{-\xi}]$ averages over the random closure. An eight-state version of the tree recursion, which bounds the generating function from below, and a region of the same kind give time $O(1.6991^n)$ (\cref{thm:rsc}).

\paragraph{Derandomization.} Classical Monotone Local Search is derandomized by replacing random samples
with a family of $t$-sets that has a member inside every $k$-set~\cite{FominGLS19}. With
sample-dependent running times this is not enough, since the member inside the solution may be one of the
slowest samples. In a black-box model, where an algorithm sees the running-time exponent and the extension
algorithm only through queries, an adversary can make slow exactly the samples that a
deterministic algorithm queries. For target size $k=n/2$ and the instances that satisfy the mean bounds of
\pname{Connected Dominating Set}, this forces every deterministic algorithm to incur cost at least $1.7547^n/\poly(n)$, when an extension call on $X$ takes time $2^{g(X)}$ and every other query
unit time, while the randomized algorithm has cost only $\Ostar(\sqrt3^{\,n})$
(\cref{cor:dr-bb-rs}). The affine bound $|\Tm(J_X)|\le n-1-w(X)$ above has more structure than
the mean: the weights are unknown, but their total is known. A \emph{balanced} family contains,
inside every $k$-set and for every nonnegative weighting of it, a sample whose weight is at least the
average; such a family plays the role of the random sample in the mean-exponent analysis. We construct
balanced families deterministically and in polynomial space (\cref{thm:dr-balanced}). For integer
weights, the exponential moments of the affine bound,
$2^{-(n-1)}\prod_{u\in S}\bigl(1+y2^{w(u)}\bigr)$ in generating-function form, can also be achieved
deterministically, with two-class balanced families, and this gives a deterministic polynomial-space
algorithm in time $O(1.7361^n)$ (\cref{thm:det-cds}).

\paragraph{Other domination variants.} These arguments use the graph only through a spanning tree of
the solution and the assignment of every outside vertex to a solution vertex that dominates it. They
therefore carry over to distance-$r$ domination, vertex weights, weakly connected domination and
directed graphs, and similar finishing steps give the bounds for \pname{Directed Steiner Tree} and
\pname{Paired Domination} (\cref{sec:further}).

\subsection{Component-count running times: paths are the worst case}\label{sec:ov-cc}

Consider \pname{Connected Feedback Vertex Set}, \pname{Graph Motif} and \pname{Connected Vertex Cover}.
When given a sample $X$ of a connected solution $S$ of size $k$, an extension algorithm runs in time
$\Ostar(c^{k-|X|}2^{\comp(G[X])})$, with $c=2$ or $c=3$. A sample of size $t$ can
have up to $t$ components, so the product-form bound is $c^{k-t}2^t$, which gives nothing better than
enumeration. Let $F$ be a spanning tree of $G[S]$. Every edge of $F$ inside $X$ merges two components,
so $\comp(G[X])\le|X|-e(F[X])$, where $e(F[X])$ is the number of edges of $F$ with both endpoints in $X$,
and therefore
\[
  W_S(y)\ \ge\ c^{-k}\sum_{X\subseteq V(F)}\zeta^{|X|}\,2^{e(F[X])}=c^{-k}\mathcal Z_F(\zeta),
  \qquad\zeta=cy/2,
\]
a partition function on the tree in which every sampled vertex has weight $\zeta$ and every edge
between two sampled vertices has weight $2$ (\cref{cor:cc-T}). On a path with $k$ vertices, $\mathcal Z_F(\zeta)$ is computed by the $2\times2$
transfer matrix $M_\zeta=\left(
  \begin{smallmatrix}1&\zeta\\1&2\zeta
\end{smallmatrix}\right)$, whose rows and
columns are indexed by the states ``not sampled'' and ``sampled'' of a vertex, and it grows like
$\lambda_{\max}(M_\zeta)^k$, where $\lambda_{\max}(M_\zeta)$ is the largest eigenvalue of $M_\zeta$.
The key lemma (\cref{lem:cc-gibbs}) shows that no tree
does worse: $\mathcal Z_F(\zeta)\ge\min(1,\zeta)\lambda_{\max}(M_\zeta)^{k-1}$ for every tree with $k$
vertices. The proof, a weighted form of an argument of Csikv\'ari and Lin~\cite{CsikvariL14} for
homomorphisms of trees, bounds $\ln\mathcal Z_F$ from below by the Gibbs variational inequality, using as
test distribution the distribution of the labels given by a reversible two-state Markov chain run from the root down the tree,
started in its stationary distribution. Then every vertex and every edge has the same label distribution,
whatever the shape of the tree, and the bound is the same as for a path. Concretely, the Gibbs inequality
bounds $\ln\mathcal Z_F(\zeta)$ from below by the entropy of the test distribution plus the expected logarithm of
the weight $\zeta^{|X|}2^{e(F[X])}$. For the Markov chain both terms split into a term for the root and one term
for each of the $k-1$ edges, and since every edge has the same joint label distribution, whichever endpoint is
the parent, all edge terms are equal. The chain is built from the Perron eigenvector of a symmetric matrix
with the same eigenvalues as $M_\zeta$, and this choice makes each edge term exactly $\ln\lambda_{\max}(M_\zeta)$.
The lemma is a statement about the partition function $\mathcal Z_F$, which bounds $W_S$ from below; it does not
say that paths are the hardest instances of the problems. With paths identified as the worst case of this lower
bound, the choice of $y$ is a one-dimensional optimization. It gives
$\Ostar(((4c-3)/(2c-1))^n)$ in closed form for every $c>1$ (\cref{cor:cc-rate}), which is
$\Ostar(1.8^n)$ for \pname{Connected Feedback Vertex Set}, and no smaller base works for this
algorithm (\cref{prop:cc-cfvs-tight}). For $c=2$, further optimization gives $O(1.6444^n)$ for \pname{Graph Motif} and
\pname{Connected Vertex Cover} (\cref{thm:cc-motif,thm:cc-cvc}).

\subsection{Computer certificates}\label{sec:ov-certificates}

The following bounds rest on computations: the bases $1.7449$, $1.7204$, $1.6991$ and $1.7361$ of RS,
RS-$\theta$, RSC and the deterministic algorithm (\cref{thm:rs,thm:rstheta,thm:rsc,thm:det-cds}), the
bases $1.6444$ for $c=2$, $1.3727$ for \pname{Directed Steiner Tree} and $1.6302$ for \pname{Paired
Domination} (\cref{thm:cc-motif,thm:fa-dst,thm:fa-pd}), and the lower bound $1.7192^n$ for RS-$\theta$ on the
uniform lobsters. For RS-$\theta$ and RSC, the points that span
the regions $\mathcal P$ of \cref{sec:ov-cds} are found by computer: the certificate of RSC has $1024$
rational points. Each certificate lists its points and is checked by programs that verify the finitely
many inclusions in exact rational arithmetic. The two certificates of RS-$\theta$ are checked by three independent
programs and the certificate of RSC by two; the second program for RSC was written from the definitions
without reusing code of the first, re-derives the recursion and rejects perturbed copies of the certificate
(\cref{sec:anc-rstheta,sec:anc-rsc}). Both RS-$\theta$ certificates, and the reduction of the bound for RSC to the
finite conditions of its certificate, are also checked in Lean~4 (\cref{sec:anc-lean}); there the conditions of the RSC certificate
are checked as well, by compiled evaluation, which also trusts the Lean compiler. The lobster bound and the three certificates for
\pname{Directed Steiner Tree} are also checked in Lean, by the kernel alone, without compiled evaluation. The rates of these certificates, the constant
of \cref{thm:rs}, the optimization over $y$ behind the base $1.6444$ for $c=2$, and the finitely many cases behind the base $1.6302$ are checked in exact
rational arithmetic, or in ball arithmetic with outward rounding where logarithms, exponentials
and square roots occur. \Cref{sec:ancillary} says which statement rests on which program or certificate.
The programs, the certificates and the Lean development are on the companion site~\cite{Gaspers26companion},
together with a reading version of the paper that links each statement to them.
The base $1.8$ for \pname{Connected Feedback Vertex Set} has a proof without computation (\cref{cor:cc-T,cor:cc-rate}), and the proof of the base $\sqrt5-\frac12$ needs one algebraic identity and ten rational inequalities checked in exact arithmetic (\cref{lem:det-cert}).

\subsection{The Exponential Time Hypothesis}\label{sec:ov-eth}

Recent preprints by OpenAI~\cite{OpenAI26eth} and Claude (Anthropic)~\cite{Claude26eth} claim to refute the Exponential Time Hypothesis. This would give subexponential time algorithms for 3-SAT and imply subexponential time algorithms for every problem with a SERF reduction to 3-SAT~\cite{ImpagliazzoPZ01}. This includes all problems in SNP and size-constrained SNP; for example, \pname{Vertex Cover}, \pname{Independent Set} and \pname{$k$-Set
Cover}.
We are not aware of
SERF reductions to \pname{3-SAT} for any of the problems in
\cref{tab:intro-bounds,tab:intro-applications}, nor of subexponential-time algorithms for them under either
claim. The obstacle that we see is a global constraint on the solution. All these
problems except \pname{Paired Domination} require the solution to be connected, and so they are not evidently in SNP:
universal first-order constraints on a witness are preserved under induced substructures, whereas connectivity is not, since deleting the middle vertex of a path on three
vertices disconnects it. Moreover, a natural CNF encoding of connectivity, by parent pointers of a
spanning tree, uses $\Theta(n\log n)$ variables, and with $\Theta(n\log n)$ variables the claimed bounds do not give subexponential time. For \pname{Paired Domination}, the subgraph induced by the solution
must have a perfect matching, and a natural encoding by a partner pointer for each vertex also uses
$\Theta(n\log n)$ variables.

\subsection{Organization}\label{sec:ov-organization}

\Cref{sec:aprime} sets up implicit set systems, running-time exponents and exponential moments, and proves the
main theorem and its general forms. It also shows that the algorithm is optimal in its class. \Cref{sec:corollaries} derives the mean-exponent theorem,
the linear-savings corollary and multivariate Monotone Local Search. The applications are
grouped by problem family. \Cref{sec:cds} prepares the applications to \pname{Connected Dominating Set}
and MaxLeaf and states the vertex-weighted directed Steiner lemma that the later sections use;
\cref{sec:rs,sec:rstheta,sec:rsc} present Algorithms RS, RS-$\theta$ and RSC; \cref{sec:derandomisation}
states the black-box barrier and gives the derandomization for affine certificates; and \cref{sec:det}
gives the deterministic algorithm. \Cref{sec:cc} treats component-count running times: it proves that paths are the
worst case, up to a constant factor, of the partition function that bounds the exponential moments from below, and then treats \pname{Connected Feedback Vertex Set}, \pname{Graph Motif},
\pname{Connected Vertex Cover} and their variants, one problem at a time. The section on further
applications (\cref{sec:further}) transfers the algorithms for \pname{Connected Dominating Set} to connected
$r$-domination, vertex weights, weak connectivity and directed graphs, and treats
\pname{Directed Steiner Tree} and \pname{Paired Domination} (\cref{sec:fa-dst,sec:fa-pd}). \Cref{sec:conclusion}
lists open problems. The appendices treat samplable distributions and other levers
(\cref{sec:app-levers}), related mechanisms outside Monotone Local Search (\cref{sec:fw-mechanisms}), linear
savings, product-form running times and the hardness of MaxLeaf-Extension (\cref{sec:app-co-hardness}), and the computer certificates
(\cref{sec:ancillary}). \Cref{sec:app-proofs} contains the proofs that the main text defers: optimality in
its class (\cref{sec:app-optimal}), linear savings and product-form running times (\cref{sec:app-co-proofs}),
the Steiner computations
(\cref{sec:app-steiner}), the analysis of RS-$\theta$ and its lower bound (\cref{sec:app-rst}), the closure
bound of RSC (\cref{sec:app-rsc}), the black-box barrier (\cref{sec:app-barrier}), the balanced families and
the count of trees across block partitions (\cref{sec:app-balanced}), the deterministic algorithms
(\cref{sec:app-det}), the extension algorithm and the tightness results for \pname{Connected Feedback Vertex
Set} (\cref{sec:app-fvs}), the weighted finishing steps (\cref{sec:app-weighted}), and the further
applications, among them \pname{Directed Steiner Tree} and \pname{Paired Domination} with their
vertex-weighted versions (\cref{sec:app-variants}); \cref{sec:app-audit} lists, for every application, the
imported results, the randomness, the weights and the output.
{}
\section{Monotone Local Search with sample-dependent running times}\label{sec:aprime}

This section proves the main theorem, on which the rest of the paper is built. It concerns an
extension algorithm whose running time is exponential in a quantity that depends on the sample, and it
says that sampling and completing finds a solution quickly as soon as the extension algorithm is
fast on enough subsets of a single solution. ``Enough'' is measured by the sum of $2^{-x}$, for running time $2^x$,
over the subsets of the solution of a given size, which we call an exponential moment, compared with
the number of subsets of the universe of that size.

\Cref{sec:fw-setting} fixes the model. \Cref{sec:fw-main} presents Algorithm~\MLS{}, the main
theorem and its proof. \Cref{sec:fw-gf,sec:fw-necessary,sec:fw-optimal} show
that the condition of the theorem can be phrased through a generating function, that
this condition is necessary for the algorithm, and that the algorithm is optimal, up to polynomial factors,
among all algorithms that draw uniform samples, run the extension algorithm only on samples whose
running-time exponent is at most a chosen threshold, and are charged $2^\theta$ for a trial with threshold $\theta$. \Cref{sec:fw-general} gives the general form that
later sections use: budgeted trials, which include one-sided-error extension algorithms.
\Cref{sec:fw-mechanisms} compares the main theorem with restart strategies, the Markov
cutoff and earlier uses of exponential moments.
\Cref{sec:app-levers} treats arbitrary samplable distributions and records which further changes of
the trials can and cannot lower the running time by more than a polynomial factor.
The mean-exponent theorem,
linear mean savings and multivariate Monotone Local Search are derived from \cref{thm:fw-main} in
\cref{sec:corollaries}.

\subsection{Setting}\label{sec:fw-setting}

We write $\log$ for the logarithm to base~$2$ and $\ln$ for the natural logarithm.
For a finite set $S$ and an
integer $t$, $\binom St$ is the family of $t$-element subsets of~$S$. For a statement or event $A$,
$\mathbf 1[A]$ is $1$ if $A$ holds and $0$ otherwise. We write $0^0=1$ and
$0\log0=0$. For $x\in[0,1]$, $H(x)=-x\log x-(1-x)\log(1-x)$ is the binary entropy, so that
$2^{H(x)}=x^{-x}(1-x)^{-(1-x)}$. We use the standard entropy estimates for binomial coefficients, valid for integers $0\le j\le m$:
\begin{equation}\label{eq:fw-binom}
  \frac{2^{mH(j/m)}}{m+1}\;\le\;\binom mj\;\le\;2^{mH(j/m)} .
\end{equation}
Throughout
the paper, $\Ostar(f)$ denotes $f\cdot N^{O(1)}$, where $N$ is the length of the input. Every decimal base of a running time of ours
is rounded up from a constant that the analysis proves to be strictly smaller, so the polynomial
factors are absorbed and we write $O(c^n)$. Exact and parametric bases of ours, such as $1.8^n$, $2^{n/2}$
or $c^k$, are stated with the $\Ostar$-notation. So are the two
bases of ours that are only computed numerically, the mean-exponent bound for \pname{Graph Motif} after
\cref{thm:cc-motif} and the multivariate bound for \pname{Max Leaf Spanning Tree} in \cref{sec:co-limits}; no
proof uses them.
A \emph{randomized algorithm with one-sided
error} never answers \YES{} on a \NO{}-instance; its error probability is the probability of
answering \NO{} on a \YES{}-instance. We assume that a randomized algorithm can draw a uniform integer from
$\{1,\dots,M\}$ in time polynomial in $\log M$; so it draws a uniform $t$-element subset of an $n$-element set,
or a Bernoulli variable with a rational probability, in time polynomial in $n$ or in the bit length of the
probability. (With fair coins alone, such draws take this time in expectation but not in the worst case.)

The framework is that of \cite{FominGLS19}: a problem is
given by a polynomial-time membership test for its solutions.

\begin{definition}[Implicit set systems]\label{def:fw-iss}
  An \emph{implicit set system} $\Phi$ assigns to every instance $I\in\{0,1\}^*$ a universe $U_I$
  with $n:=|U_I|\le N:=|I|$ and a family $\cF_I\subseteq2^{U_I}$, such that $U_I$ can be listed and
  membership in $\cF_I$ decided in time $N^{O(1)}$. The members of $\cF_I$ are the \emph{solutions}
  of~$I$.
  \begin{itemize}
    \item \emph{$\Phi$-Subset}: given $I$, decide whether $\cF_I\neq\emptyset$.
    \item \emph{$\Phi$-Extension}: given $I$, a set $X\subseteq U_I$ and an integer $k\ge0$, decide
      whether some $Y\subseteq U_I\setminus X$ with $|Y|\le k-|X|$ satisfies $X\cup Y\in\cF_I$, that
      is, whether $X$ lies in a solution with at most $k$ elements. For $k<|X|$ the answer is \NO{};
      the extension algorithms of this paper reject such inputs in polynomial time, and their
      running-time bounds are stated for $k\ge|X|$.
  \end{itemize}
\end{definition}

All theorems below are stated for the decision versions. If the extension algorithm returns a
witness $Y$ when it accepts, the algorithms below return the solution $X\cup Y$.

\paragraph{Sample-dependent running-time exponents.}
In classical and multivariate Monotone Local Search (\cref{cor:co-multivariate}) the running-time
bound of an extension call depends on the sample $X$ only through $|X|$; \cref{sec:intro} explains
why such a bound gains nothing from sampling for \pname{Connected Dominating Set}. The model below
bounds the running time of the extension algorithm on each input separately, by a function that the
subset algorithm can evaluate in advance.

\begin{definition}[Running-time exponent]\label{def:fw-cost}
  Let $\cA$ be an algorithm for $\Phi$-Extension. A \emph{running-time exponent} for $\cA$ is a function
  $g$ from inputs $(I,X,k)$ with $k\ge|X|$ to $[0,\infty)$ such that $g(I,X,k)$ can be computed in time
  $N^{O(1)}$ and $\cA$ runs in time $2^{g(I,X,k)}N^{O(1)}$ on every
  such input $(I,X,k)$.
\end{definition}

The running-time bound must hold on every input, including samples $X$ that are contained in no
solution. A \emph{product-form}
running-time exponent is $g(I,X,k)=(n-|X|)\log b+(k-|X|)\log c+|X|\log d$ with constants $b,c,d\ge1$; it
depends on $X$ only through $|X|$, and it expresses the running-time bound
$b^{n-|X|}c^{k-|X|}d^{|X|}N^{O(1)}$ of classical and multivariate Monotone Local Search. An
example of a sample-dependent running-time exponent is one computed from the number of Steiner
terminals that a sample of vertices leaves (\cref{sec:cds}).
Once $I$ and a target size $k$ are fixed, we write $g(X)=g(I,X,k)$ for $|X|\le k$.

\paragraph{Comparisons with thresholds.}
The algorithms below compare $g(I,X,k)$ with thresholds that are integers or rationals of polynomial
bit length, and \cref{def:fw-cost} guarantees that such a comparison
is decided in time $N^{O(1)}$. This holds for all running-time exponents of this paper.

The main theorem needs, for one solution, the following sums.

\begin{definition}[Exponential moments and generating function]\label{def:fw-moments}
  Fix an instance $I$, a running-time exponent $g$, and a solution $S\in\cF_I$ with $|S|=k$. For
  $0\le t\le k$ and real $y\ge0$ let
  \[
    \mu_t(S)=\sum_{X\in\binom St}2^{-g(I,X,k)},\qquad
    W_S(y)=\sum_{t=0}^k\mu_t(S)\,y^t=\sum_{X\subseteq S}y^{|X|}\,2^{-g(I,X,k)} .
  \]
\end{definition}

If $X$ is drawn uniformly from $\binom{U_I}t$, then
$\mu_t(S)/\binom nt=\E\bigl[\mathbf 1[X\subseteq S]\,2^{-g(I,X,k)}\bigr]$: the normalized moment
is the expected value of $2^{-g}$ on the event that the sample lies inside~$S$. Every $\mu_t(S)$ is
positive, because $g$ is finite, and $\mu_t(S)\le\binom kt$, because $g\ge0$.

\subsection{The algorithm and the main theorem}\label{sec:fw-main}

\paragraph{How the algorithm works.}
The algorithm monitors one quantity: the running-time exponent $g(I,X,k)$ of each sample $X$, which it
computes in polynomial time before deciding whether to call the extension algorithm. It works
through all pairs $(t,\theta)$ of a sample size $t\in\{0,\dots,k\}$ and an integer threshold
$\theta\ge0$. A \emph{trial} of the pair $(t,\theta)$ draws a uniformly random $t$-set $X$. If
$g(I,X,k)\le\theta$, the trial runs $\cA$ on $(I,X,k)$, in time $\Ostar(2^\theta)$; otherwise it
discards $X$ in polynomial time. Low thresholds make trials fast but rarely successful, and high
thresholds do the opposite. To balance the two, the algorithm gives every pair the same running-time
budget $\Ostar(2^{\psi n})$, where $2^{\psi n}$ is the target running time: the pair
$(t,\theta)$ receives about $2^{\psi n-\theta}$ trials. Thresholds above a cap
$\Theta\approx\psi n$ are not used, since a single such trial could exceed the budget. As there
are only polynomially many pairs, the total time is $\Ostar(2^{\psi n})$.

The algorithm has three parameters. The \emph{rate} $\psi\ge0$ is the exponent of the target running
time $2^{\psi n}$. The \emph{loss factor} $\Gamma\ge1$ weakens the success condition~\eqref{eq:fw-star}
of \cref{thm:fw-main} to $\mu_t(S)\ge\binom nt2^{-\psi n}/\Gamma$, so that an application may prove
a lower bound on $\mu_t(S)$ that is smaller by a factor $\Gamma$, typically polynomial in $n$, without
changing $\psi$; the cap and the trial counts grow with $\Gamma$, and the running time stays
$2^{\psi n}N^{O(1)}$ as long as $\Gamma\le N^{O(1)}$. The \emph{confidence} $\kappa>0$ sets the bound
$e^{-\kappa}$ on the error probability in item~(iii) of \cref{thm:fw-main}.

\begin{algorithm}[\MLS$(\psi,\Gamma,\kappa)$]\label{alg:fw-mls}
  \Input an instance $I$ of $\Phi$ and an integer $k\le n$.
  \Parameters $\psi$, $\Gamma$, $\kappa$, an integer cap $\Theta\ge0$ and integer trial counts $m_{t,\theta}$ such that
  \begin{equation}\label{eq:fw-params}
    \begin{gathered}
      x+1\le\Theta\le x+3,\qquad
      M_{t,\theta}\le m_{t,\theta}\le 2M_{t,\theta}+1,\\
      \text{where }x=\psi n+\log\Gamma\text{ and } M_{t,\theta}=4\kappa\Gamma(\Theta+1)\,2^{\psi n-\theta} .
    \end{gathered}
  \end{equation}
  \Body
  For every $t\in\{0,\dots,k\}$ and every integer
  $\theta\in\{0,\dots,\Theta\}$, repeat $m_{t,\theta}$ times:
  \begin{enumerate}[label=\arabic*.]
    \item draw $X$ uniformly from $\binom{U_I}t$ and compute $g(I,X,k)$;
    \item if $g(I,X,k)\le\theta$, run $\cA$ on $(I,X,k)$, and if it accepts, answer \YES{}.
  \end{enumerate}
  After all trials, answer \NO{}.
\end{algorithm}

When $x$ and $M_{t,\theta}$ are rational, $\Theta=\lceil x\rceil+1$ and
$m_{t,\theta}=\lceil M_{t,\theta}\rceil$ qualify. In general, integers in these ranges are computable in
time $N^{O(1)}$ from rational approximations: $\Theta=\lceil\tilde x+\frac12\rceil+1$ for a rational
$\tilde x$ with $|\tilde x-x|<\frac12$, and $m_{t,\theta}=\lceil\hat m\rceil$ for a rational $\hat m$
with $M_{t,\theta}\le\hat m\le2M_{t,\theta}$. The ranges in~\eqref{eq:fw-params} are part of the
statement of the theorem, so no separate rounding argument is needed in the applications.

\begin{theorem}[Exponential-moment Monotone Local Search]\label{thm:fw-main}
  Let $\Phi$ be an implicit set system and let $\cA$ solve $\Phi$-Extension with running-time exponent $g$.
  Let $\psi\ge0$, $\Gamma\ge1$ and $\kappa>0$ be numbers that can be computed from $(I,k)$ in time
  $N^{O(1)}$ and whose values are at most $N^{O(1)}$.
  With any cap $\Theta$ and trial counts $m_{t,\theta}$ satisfying~\eqref{eq:fw-params},
  Algorithm~\MLS$(\psi,\Gamma,\kappa)$, given $I$ and $k\le n$,
  \begin{enumerate}[label=\textup{(\roman*)}]
    \item runs in time $2^{\psi n}N^{O(1)}$ and in space $N^{O(1)}$ plus the space of $\cA$;
    \item answers \NO{} if $\cF_I$ contains no set of size at most $k$;
    \item answers \YES{} with probability at least $1-e^{-\kappa}$ if $\cF_I$ contains a set $S$ with
      $|S|=k$ that satisfies
      \begin{equation*}\label{eq:fw-star}
        \max_{0\le t\le k}\frac{\mu_t(S)}{\binom nt}\ \ge\ \frac{2^{-\psi n}}{\Gamma}.
        \tag{$\star$}
      \end{equation*}
  \end{enumerate}
  Condition~\eqref{eq:fw-star} holds in particular if
  $W_S(y)\ge(1+y)^n\,2^{-\psi n}/\Gamma$ for some $y\ge0$.
\end{theorem}

By items~(ii) and~(iii), Algorithm~\MLS{} is a randomized algorithm with one-sided error. Two
remarks complete the statement.

\paragraph{Optimization problems.} %
For an optimization problem, we say that the optimum can be computed (or that the problem can be
solved) by a randomized algorithm with one-sided error, with success probability $p$, if the
following holds on every input. On every outcome of its coins, the algorithm stops within the stated
time and space bounds. Its output is never better than the optimum, which means never below it for a
minimization problem and never above it for a maximization problem. A report that the input has no
solution counts as the value $+\infty$ for a minimization problem and $-\infty$ for a maximization
problem. With probability at least $p$, the output is the optimum, or a correct report that there is no
solution. Where the algorithm also outputs a solution, that solution is feasible on every outcome of
the coins and its value is the output value. Unless a statement names another probability, $p=1-e^{-\kappa}$
for an arbitrary constant $\kappa>0$ fixed in advance, and the time bound holds for every such $\kappa$.
The wrappers that provide this form are \cref{lem:cds-first-accepted}, which outputs the first accepted
size, \cref{lem:cv-budget-loop}, which outputs the least accepted budget, and equivalent loops over the
value and the size, such as the one in the proof of \cref{thm:cc-cvc}.

\begin{remark}[Decision problem]\label{rem:fw-decision}
  Running the algorithm for $k=0,1,\dots,n$, with rates
  $\psi=\psi(I,k)$, decides $\Phi$-Subset with one-sided error at most $e^{-\kappa}$ in time
  $N^{O(1)}\max_k2^{\psi(I,k)n}$, provided every instance with $\cF_I\neq\emptyset$ has a solution
  $S$ that satisfies~\eqref{eq:fw-star} with $k=|S|$.
\end{remark}

\begin{remark}[Real-valued running-time exponents and parameters]\label{rem:fw-real}
  If a running-time exponent $g$ can only be approximated, as the product-form exponent with irrational $\log
  b$, $\log c$ or $\log d$, replace it by a rational $\hat g(I,X,k)$ with $g\le\hat g\le g+1$,
  computable in time $N^{O(1)}$ from an approximation of $g$ to within $\frac12$. Then $\hat g$ is again
  a running-time exponent for $\cA$, and $2^{-\hat g}\ge2^{-g}/2$, so every exponential moment $\mu_t(S)$ and
  every value $W_S(y)$ of the generating function (\cref{def:fw-moments}) is at least half of its value
  for $g$, and the mean of $\hat g$ over any set of samples exceeds that of $g$ by at most $1$. All
  statements of the paper therefore hold for $\hat g$ once their success conditions are weakened
  accordingly: the loss factor $\Gamma$ of \cref{thm:fw-main} is replaced by $2\Gamma$, and an upper
  bound on the mean running-time exponent of a random sample, as assumed in the mean-exponent theorem
  (\cref{thm:co-mean}), is increased by $1$. Parameters of the algorithms that are real numbers, such
  as a bound on the mean or the constants of the affine certificates of
  \cref{sec:derandomisation}, are likewise replaced by rational approximations from above with slack at
  most $1$; this multiplies the running-time bound only by a constant factor.
\end{remark}

\paragraph{Proof overview.}
The proof has three steps. \emph{Running time:} the trials of every pair $(t,\theta)$ take time
$2^{\psi n}N^{O(1)}$, because there are about $2^{\psi n-\theta}$ of them and each takes time at
most $2^\theta N^{O(1)}$.
\emph{A good threshold
exists below the cap:} if $\mu_t(S)\ge\binom nt2^{-\psi n}/\Gamma$, some threshold $\theta\le\Theta$
is passed by at least $2^\theta\mu_t(S)/(4(\Theta+1))$ sets $X\in\binom St$. The sets $X\in\binom St$
with running-time exponent above the cap contribute at most $\mu_t(S)/2$ to $\mu_t(S)$.
\emph{Success:} a trial of that pair succeeds with probability at least
$2^{\theta-\psi n}/(4\Gamma(\Theta+1))$, and the number $m_{t,\theta}$ of trials of the pair is at
least $\kappa$ divided by this probability. The second step is the following lemma, stated for a
general random variable so that \cref{sec:fw-general} can use it as well. In this section the random variable is the
running-time exponent of a uniformly random sample of size $t$, set to $\infty$ when the sample is not inside the
solution under consideration. We write $2^{-\infty}=0$.

\begin{lemma}[Threshold selection]\label{lem:fw-threshold-prob}
  Let $\xi$ be a random variable with values in $[0,\infty]$ and $\Pr[\xi<\infty]>0$, so that $\E[2^{-\xi}]>0$.
  \begin{enumerate}[label=\textup{(\alph*)}]
    \item If $\Theta\ge0$ is an integer with $2^{-\Theta}\le\E[2^{-\xi}]/2$,
      then some
      $\theta\in\{0,\dots,\Theta\}$ satisfies $\Pr[\xi\le\theta]\,2^{-\theta}\ge\E[2^{-\xi}]/(4(\Theta+1))$.
    \item Conversely, $\Pr[\xi\le\theta]\,2^{-\theta}\le\E[2^{-\xi}]$ for every integer $\theta$.
  \end{enumerate}
\end{lemma}

\begin{proof}
  (a) On the event $\xi>\Theta$ we have $2^{-\xi}<2^{-\Theta}$, so
  $\E[2^{-\xi}\mathbf 1[\xi>\Theta]]\le2^{-\Theta}\le\E[2^{-\xi}]/2$ and hence
  $\E[2^{-\xi}\mathbf 1[\xi\le\Theta]]\ge\E[2^{-\xi}]/2$. On the event $\xi\le\Theta$ put
  $\theta_\xi=\lceil\xi\rceil$; then $\theta_\xi\in\{0,\dots,\Theta\}$, because $0\le\xi\le\Theta$ and
  $\Theta$ is an integer. Moreover $2^{-\xi}<2^{1-\theta_\xi}$ and
  $\{\theta_\xi=\theta\}\subseteq\{\xi\le\theta\}$. Hence
  \[
    \frac{\E[2^{-\xi}]}2\le\sum_{\theta=0}^{\Theta}2^{1-\theta}\Pr[\xi\le\Theta,\ \theta_\xi=\theta]
    \le2\sum_{\theta=0}^{\Theta}2^{-\theta}\Pr[\xi\le\theta]
    \le2(\Theta+1)\max_{0\le\theta\le\Theta}2^{-\theta}\Pr[\xi\le\theta].
  \]

  (b) $\E[2^{-\xi}]\ge\E[2^{-\xi}\mathbf 1[\xi\le\theta]]\ge2^{-\theta}\Pr[\xi\le\theta]$.
\end{proof}

\begin{proof}%
  Throughout, a \emph{trial} of the pair $(t,\theta)$ is one execution of steps~1 and~2 of
  Algorithm~\MLS{} with these values.

  \emph{(ii) Soundness.} If $\cA$ accepts $(I,X,k)$, then $X\cup Y\in\cF_I$ for some
  $Y\subseteq U_I\setminus X$ with $|Y|\le k-t$, and this set has size at most $t+(k-t)=k$.

  \emph{(i) Time and space.} Drawing $X$, computing $g(I,X,k)$ and comparing it with
  $\theta$ take time $N^{O(1)}$.
  If
  $\cA$ is run, then $g(I,X,k)\le\theta$, so the trial takes time $2^{g(I,X,k)}N^{O(1)}\le2^\theta
  N^{O(1)}$. The
  trials of a pair $(t,\theta)$ therefore take time at most
  \[
    m_{t,\theta}2^{\theta+1}N^{O(1)}\le\bigl(2M_{t,\theta}+1\bigr)2^{\theta+1}N^{O(1)}
    \le\bigl(16\kappa\Gamma(\Theta+1)2^{\psi n}+2^{\Theta+1}\bigr)N^{O(1)},
  \]
  and $\Theta\le\psi n+\log\Gamma+3$ gives $2^\Theta\le8\Gamma2^{\psi n}$. There are $(k+1)(\Theta+1)$ pairs. Since $\kappa$, $\Gamma$
  and $\psi$ are at most $N^{O(1)}$ and $n\le N$, also $\Theta+1\le N^{O(1)}$, and the total time
  is $2^{\psi n}N^{O(1)}$. Besides the space of $\cA$, the algorithm stores one sample, the current
  pair and a trial counter of $O(\Theta+\log(\kappa\Gamma))$ bits.

  \emph{(iii) Success probability.} Let $S\in\cF_I$ with $|S|=k$ satisfy~\eqref{eq:fw-star}, and fix
  $t$ with $\mu_t(S)\ge\binom nt2^{-\psi n}/\Gamma$. Let $X$ be uniform in $\binom{U_I}t$, and put
  $\xi=g(I,X,k)$ if $X\subseteq S$ and $\xi=\infty$ otherwise. Then
  $\E[2^{-\xi}]=\mu_t(S)/\binom nt>0$, and for an integer $\theta$, $\Pr[\xi\le\theta]=q_\theta/\binom nt$,
  where $q_\theta$ is the number of $X\in\binom St$ with $g(I,X,k)\le\theta$. The hypothesis of
  \cref{lem:fw-threshold-prob}(a) holds, because $\Theta\ge\psi n+\log\Gamma+1$ gives
  \[
    2^{-\Theta}\le\frac{2^{-\psi n}}{2\Gamma}\le\frac{\mu_t(S)}{2\binom nt} .
  \]
  So some $\theta\in\{0,\dots,\Theta\}$ satisfies
  \[
    p:=\frac{q_\theta}{\binom nt}=\Pr[\xi\le\theta]\ \ge\ \frac{2^\theta\mu_t(S)}{4(\Theta+1)\binom nt}
    \ \ge\ \frac{2^{\theta-\psi n}}{4\Gamma(\Theta+1)} .
  \]
  A trial of the pair $(t,\theta)$ draws an $X\in\binom St$ with $g(I,X,k)\le\theta$ with probability
  $p$. Then $\cA$ is run on $(I,X,k)$, and it accepts, because $Y=S\setminus X$ satisfies $|Y|=k-t$ and
  $X\cup Y=S\in\cF_I$. The $m_{t,\theta}\ge M_{t,\theta}=4\kappa\Gamma(\Theta+1)2^{\psi n-\theta}\ge\kappa/p$
  trials of this pair are independent, so all of them fail with probability at most
  $(1-p)^{m_{t,\theta}}\le e^{-pm_{t,\theta}}\le e^{-\kappa}$.

  \emph{The generating-function condition.} For $y\ge0$,
  \[
    W_S(y)=\sum_{t=0}^k\binom nty^t\,\frac{\mu_t(S)}{\binom nt}
    \le\max_{t}\frac{\mu_t(S)}{\binom nt}\sum_{t=0}^n\binom nty^t
    =\max_{t}\frac{\mu_t(S)}{\binom nt}\,(1+y)^n,
  \]
  so $W_S(y)\ge(1+y)^n2^{-\psi n}/\Gamma$ implies~\eqref{eq:fw-star}.
\end{proof}

\paragraph{Mean versus exponential moments.}
By Jensen's inequality, $\E[2^{-g(X)}]\ge2^{-\E[g(X)]}$ for $X$ uniform in $\binom St$, so
condition~\eqref{eq:fw-star} is never more demanding than a hypothesis on the mean of $g$
(\cref{thm:co-mean}), and it can certify a smaller running time when $g$ varies between samples: for \pname{Connected
Dominating Set} it lowers the base from $1.7449$ to $1.7204$ (\cref{thm:rs,thm:rstheta}). The same inequality,
$\log(1/\E[2^{-\xi}])\le\E[\xi]$, shows that the condition on $\Theta$ in \cref{lem:fw-threshold-prob}(a)
holds whenever $\Theta\ge\E[\xi]+1$. The converse fails, and the gap can be large: if $\xi=0$ and
$\xi=2m$ with probability $\frac12$ each, the condition holds for $\Theta\ge2$, whereas $\E[\xi]+1=m+1$.

\subsection{The generating function}\label{sec:fw-gf}

Condition~\eqref{eq:fw-star} involves a maximum over sample sizes. The generating function $W_S$
replaces it by an optimization over one real variable, which is how the later sections bound the
exponential moments (\cref{sec:rstheta,sec:rsc}). The next lemma shows that the two forms agree up
to a factor $n+1$.

\begin{lemma}[Generating function versus exponential moments]\label{lem:fw-gf}
  For every instance $I$, running-time exponent $g$ and solution $S$ with $|S|=k$, the moments $\mu_t(S)$ with
  $0\le t\le k$ are positive, since $g$ is finite, and
  \[
    \min_{0\le t\le k}\frac{\binom nt}{\mu_t(S)}\ \le\ \inf_{y\ge0}\frac{(1+y)^n}{W_S(y)}\ \le\
    (n+1)\min_{0\le t\le k}\frac{\binom nt}{\mu_t(S)} .
  \]
\end{lemma}

\begin{proof}
  The first inequality is the last display of the proof of \cref{thm:fw-main}, read as
  $(1+y)^n/W_S(y)\ge1/\max_t(\mu_t(S)/\binom nt)$; note that $W_S(y)\ge\mu_0(S)>0$.

  For the second, fix $t$; we exhibit $y\ge0$ with $(1+y)^n/W_S(y)\le(n+1)\binom nt/\mu_t(S)$, and
  the minimum over $t$ gives the claim. If $t=0$, then $y=0$ and
  $(1+y)^n/W_S(y)=1/\mu_0(S)=\binom n0/\mu_0(S)$.
  If $0<t<n$, put $y=t/(n-t)$ and $\tau=t/n$, so that $y/(1+y)=\tau$ and $1/(1+y)=1-\tau$. Since
  $W_S(y)\ge\mu_t(S)\,y^t$,
  \[
    \frac{(1+y)^n}{W_S(y)}\le\frac{(1+y)^n}{\mu_t(S)\,y^t}=\frac1{\mu_t(S)\,\tau^t(1-\tau)^{n-t}}
    \le\frac{(n+1)\binom nt}{\mu_t(S)} .
  \]
  The last step uses $\binom nt\tau^t(1-\tau)^{n-t}\ge1/(n+1)$, which holds because
  $\tau^t(1-\tau)^{n-t}=2^{-nH(\tau)}$ and $\binom nt\ge2^{nH(\tau)}/(n+1)$ by~\eqref{eq:fw-binom}.

  If $t=n\ge1$, then $k=n$ and $y=n$. Now $W_S(n)\ge\mu_n(S)\,n^n$, and
  $(1+n)^n/n^n=(1+1/n)^n\le\min(e,2^n)\le n+1$, so
  $(1+y)^n/W_S(y)\le(n+1)/\mu_n(S)=(n+1)\binom nn/\mu_n(S)$.
\end{proof}

The infimum in \cref{lem:fw-gf} need not be attained: for $k=n$ the ratio may decrease toward
$1/\mu_n(S)$ as $y\to\infty$; \cref{thm:fw-main} uses the generating-function condition only at a
specific~$y$.

\subsection{The criterion is necessary for the algorithm}\label{sec:fw-necessary}

Exponential moments are the right quantity for Algorithm~\MLS{}: if the left-hand side
of~\eqref{eq:fw-star} is smaller than $2^{-\psi n}$ by more than a polynomial factor, namely
$4P_{\mathrm{nec}}$ with the polynomial factor $P_{\mathrm{nec}}$ defined in the next proposition, then the algorithm runs $\cA$ on a subset of $S$ with
probability at most $1/4$. It might still succeed through another solution, or through a sample
that is not inside $S$; the class-optimality result of \cref{sec:fw-optimal} accounts for those as
well. \Cref{prop:fw-necessary} is used in \cref{sec:rstheta} to show that no analysis of
Algorithm RS-$\theta$ can lower its base below $1.7192$.

The proposition also covers runs of the algorithm on $(I,k)$ with $k$ larger than the size of $S$,
which the tightness results of \cref{sec:cc} need. In such a run, a trial of sample size $t$ runs
$\cA$ on $(I,X,k)$, with the size bound $k$ rather than $|S|$. For $S\in\cF_I$, an integer $k\ge|S|$
and $0\le t\le|S|$ we therefore put
\begin{equation}\label{eq:fw-moment-budget}
  \mu^{(k)}_t(S)=\sum_{X\in\binom St}2^{-g(I,X,k)} .
\end{equation}
For $|S|=k$ this is the exponential moment $\mu_t(S)$ of \cref{def:fw-moments}.

\begin{proposition}[Necessity of~\eqref{eq:fw-star}]\label{prop:fw-necessary}
  Let $k\le n$ be an integer, fix $S\in\cF_I$ with $|S|\le k$ and run Algorithm~\MLS$(\psi,\Gamma,\kappa)$
  on $(I,k)$. The expected number of trials that run $\cA$ on a subset of $S$ is at most
  \[
    P_{\mathrm{nec}}\cdot2^{\psi n}\max_{0\le t\le|S|}\frac{\mu^{(k)}_t(S)}{\binom nt},
  \]
  where $P_{\mathrm{nec}}=8\Gamma(k+1)(\Theta+1)\bigl(\kappa(\Theta+1)+1\bigr)\le N^{O(1)}$ and $\Theta$ is
  the cap used. In particular, if
  $\max_{t\le|S|}\mu^{(k)}_t(S)/\binom nt\le2^{-\psi n}/(4P_{\mathrm{nec}})$, the probability that some trial runs
  $\cA$ on a subset of $S$ is at most $1/4$.
\end{proposition}

\begin{proof}
  A trial of $(t,\theta)$ draws a $t$-set $X$ and runs $\cA$ on $(I,X,k)$ exactly when
  $g(I,X,k)\le\theta$. So it runs $\cA$ on a subset of $S$ exactly when $t\le|S|$ and it draws
  $X\in\binom St$ with $g(I,X,k)\le\theta$; for $t>|S|$ this never happens. Let $t\le|S|$. With $X$ uniform in $\binom{U_I}t$ and $\xi=g(I,X,k)$ if $X\subseteq S$,
  $\xi=\infty$ otherwise, we have $\E[2^{-\xi}]=\mu^{(k)}_t(S)/\binom nt$, so by
  \cref{lem:fw-threshold-prob}(b) the event has probability
  $\Pr[\xi\le\theta]\le2^\theta\mu^{(k)}_t(S)/\binom nt$. Here
  $m_{t,\theta}\le2M_{t,\theta}+1=8\kappa\Gamma(\Theta+1)2^{\psi n-\theta}+1$ and
  $2^\theta\le2^\Theta\le8\Gamma2^{\psi n}$ by~\eqref{eq:fw-params}, so
  $m_{t,\theta}2^\theta\le8\Gamma(\kappa(\Theta+1)+1)2^{\psi n}$. By linearity of expectation, the
  expected number of trials that run $\cA$ on a subset of $S$ is at most the sum of
  $m_{t,\theta}2^\theta\mu^{(k)}_t(S)/\binom nt$ over the pairs $(t,\theta)$ with $t\le|S|$. There are at
  most $(k+1)(\Theta+1)$ such pairs, and each term is at most
  $8\Gamma(\kappa(\Theta+1)+1)2^{\psi n}\max_{t'\le|S|}\mu^{(k)}_{t'}(S)/\binom n{t'}$; this gives the
  bound on the expectation. By Markov's inequality, the probability that some trial runs $\cA$ on a
  subset of $S$ is at most this expectation, which gives the bound $1/4$.
\end{proof}

The moments with a larger size bound are easy to bound for the running-time exponents of this paper, which
grow with the size bound. If $g(I,X,k)$ is nondecreasing in $k$, then for every $S\in\cF_I$,
every $k\ge|S|$ and every $t\le|S|$, each term of~\eqref{eq:fw-moment-budget} satisfies
$2^{-g(I,X,k)}\le2^{-g(I,X,|S|)}$; hence
\begin{equation}\label{eq:fw-moment-monotone}
  \mu^{(k)}_t(S)\le\mu^{(|S|)}_t(S)=\mu_t(S),
\end{equation}
where $\mu_t(S)$ is the exponential moment of \cref{def:fw-moments} for $S$ and its own size $|S|$.
For such running-time exponents, \cref{prop:fw-necessary} therefore bounds the runs on $(I,k)$ for all
$k\ge|S|$ by the moments of $S$ for $k=|S|$.

\subsection{Optimality in its class}\label{sec:fw-optimal}

\Cref{prop:fw-necessary} concerns one algorithm and one solution. We now show that Algorithm~\MLS{}
is optimal, up to polynomial factors and on every instance, among all algorithms that sample
uniformly, run $\cA$ only on samples whose running-time exponent is at most a chosen threshold $\theta$, and
are charged $2^\theta$ per trial (the class $\cC_{\mathrm{thr}}$ defined below). Consequently a
running time below the one certified by \cref{thm:fw-main}, by more than a polynomial factor, needs
trials outside this class (\cref{sec:app-levers}); a sharper analysis of Algorithm~\MLS{} cannot
give it.

Fix $I$, $k\le n$ and a correct extension algorithm $\cA$ with running-time exponent $g$, and write
$U=U_I$ and, as declared after \cref{def:fw-cost}, $g(X)=g(I,X,k)$.

\begin{definition}[Extendable sets and the class optimum]\label{def:fw-extendable}
  The family $\cE$ of \emph{extendable sets} consists of the sets $X\subseteq U$ with $|X|\le k$
  such that $X\cup Y\in\cF_I$ for some $Y\subseteq U\setminus X$ with $|Y|\le k-|X|$; put
  $\cE_t=\cE\cap\binom Ut$. Since $\cA$ is correct, it accepts $(I,X,k)$ exactly when
  $X\in\cE$, and $\binom St\subseteq\cE_t$ for every $S\in\cF_I$ with $|S|=k$. For an integer
  $\theta\ge0$ let $q_\theta(t)$ be the number of $X\in\cE_t$ with $g(X)\le\theta$, let
  $\mu_t(\cE)=\sum_{X\in\cE_t}2^{-g(X)}$, and define the \emph{class optimum}
  \[
    \pi^*=\max_{0\le t\le k,\ \theta\in\mathbb Z_{\ge0}}\frac{q_\theta(t)\,2^{-\theta}}{\binom nt}.
  \]
\end{definition}

The maximum exists, because for each $t$ the count $q_\theta(t)$ is constant for $\theta\ge\max g$,
so only finitely many values of $\theta$ are candidates. Note that $\mu_t(\cE)\ge\mu_t(S)$ for
every $S\in\cF_I$ with $|S|=k$.

Let $\cC_{\mathrm{thr}}$ be the class of algorithms that perform a sequence of trials of the
following kind. Before trial $i$ the algorithm chooses $t_i\in\{0,\dots,k\}$ and an integer
$\theta_i\ge0$, as any, possibly randomized, function of everything observed so far; negative
thresholds would pass nothing, as $g\ge0$. Trial $i$ draws $X_i$ uniformly from $\binom U{t_i}$,
independently of the past. If $g(X_i)\le\theta_i$ it runs $\cA$ on $(I,X_i,k)$, and it answers
\YES{} if $\cA$ accepts; these are the only \YES{} answers of the algorithm. Trial $i$ is \emph{charged} $2^{\theta_i}$, whether or not $\cA$ is run.
Algorithm~\MLS{} belongs to $\cC_{\mathrm{thr}}$, and its running time is its total charge times
$N^{O(1)}$. Under this charge the threshold filter is the filter with the largest success
probability: a filter must reject every $X$ with $g(X)>\theta$ to keep the running-time bound of the
trial within $2^\theta N^{O(1)}$, and rejecting more samples only lowers the success probability.

\begin{restatable}[Optimality in its class]{theorem}{RestThmFwOptimal}\label{thm:fw-optimal}
  Let $I$, $k$, $\cA$ and $g$ be as fixed above, and $\cE_t$, $q_\theta(t)$, $\mu_t(\cE)$ and $\pi^*$
  as in \cref{def:fw-extendable}. Then:
  \begin{enumerate}[label=\textup{(\alph*)}]
    \item every algorithm in $\cC_{\mathrm{thr}}$ whose total charge is at most $\mathsf B$ on every run
      answers \YES{} with probability at most $\mathsf B\pi^*$;
    \item if $\pi^*\ge2^{-\psi n}/\Gamma$, then Algorithm~\MLS$(\psi,\Gamma,\kappa)$ answers \YES{} with probability at least $1-e^{-\kappa}$, and it
      runs in time $2^{\psi n}N^{O(1)}$ by \cref{thm:fw-main}(i);
    \item for every $t$ and every integer $\theta$, $q_\theta(t)2^{-\theta}\le\mu_t(\cE)$. If, for an integer $g_{\max}\ge0$, $g$ takes
      integer values in $\{0,\dots,g_{\max}\}$ on $\cE_t$, then
      $\max_{0\le\theta\le g_{\max}}q_\theta(t)2^{-\theta}\ge\mu_t(\cE)/(g_{\max}+1)$.
  \end{enumerate}
\end{restatable}

The proof is in \cref{sec:app-optimal}. For~(a) it bounds the success probability of each trial by $\pi^*$
times its charge and sums over the trials; for~(b) it uses the trials of a pair $(t,\theta)$ that attains
$\pi^*$; and~(c) compares each threshold with the running-time exponents below it.

By~(a) and~(b), Algorithm~\MLS{} with $\Gamma=1$ and $2^{\psi n}=1/\pi^*$ is optimal in
$\cC_{\mathrm{thr}}$ up to polynomial factors; in a worst-case analysis, $2^{\psi n}$ is an upper
bound on $1/\pi^*$ proven for a whole family of instances. By~(c), $1/\pi^*$ equals
$\min_t\binom nt/\mu_t(\cE)$ up to a factor $g_{\max}+1$ for integer running-time exponents bounded by
$g_{\max}$. For real-valued $g$, \cref{lem:fw-threshold-prob}(a) takes the place of the integer case
of~(c): with $\xi=g(X)$ if $X\in\cE_t$ and $\xi=\infty$ otherwise, for $X$ uniform in $\binom Ut$,
the definitions of $\mu_t(\cE)$ and $q_\theta(t)$ give $\E[2^{-\xi}]=\mu_t(\cE)/\binom nt$ and
$\Pr[\xi\le\theta]=q_\theta(t)/\binom nt$. Since Algorithm~\MLS{} succeeds on
every extendable sample, (b) and~(c) also show that \cref{thm:fw-main}(iii) remains true when $\mu_t(S)$
in~\eqref{eq:fw-star} is replaced by the sum $\mu_t(\cE)$ over the $t$-subsets of all solutions of size at
most $k$: for Algorithm~\MLS$(\psi,(g_{\max}+1)\Gamma,\kappa)$ when $g$ takes integer values in
$\{0,\dots,g_{\max}\}$ on $\cE$ for an integer $g_{\max}\le N^{O(1)}$, and for Algorithm~\MLS$(\psi,\Gamma,\kappa)$
itself, for any $g$, when the whole left-hand side of~\eqref{eq:fw-star} is replaced by $\pi^*$.

\subsection{General forms}\label{sec:fw-general}

The proof of \cref{thm:fw-main} uses only two properties of a trial: it runs in time at most
$2^\theta N^{O(1)}$, and it succeeds whenever a random quantity attached to it is at most $\theta$. This subsection
isolates that structure in a version for arbitrary budgeted trials, which covers randomized extension
algorithms and the closure sampling of \cref{sec:rsc}.

The general form rests on \cref{lem:fw-threshold-prob}. The next theorem replaces the trials of Algorithm~\MLS{} by arbitrary randomized procedures. A
trial indexed by $\iota$ with threshold $\theta$ must run in time $2^\theta N^{O(1)}$ and must
succeed at least when a random variable $\xi_{\iota}$, which the analysis attaches to it, is at most $\theta$.
The variable $\xi_{\iota}$ is only used in the analysis; it may depend on an unknown solution.

\begin{algorithm}[Budgeted trials]\label{alg:fw-budgeted}
  \Input an instance $I$ and an integer $k\le n$.
  \Parameters a randomized algorithm $\mathrm{Tr}$ with inputs $(I,k,\iota,\theta)$ and, for each
  $(I,k)$, a set $\cT=\cT_{I,k}$ of indices; for $\iota\in\cT$ and an integer $\theta\ge0$, the run of
  $\mathrm{Tr}$ on $(I,k,\iota,\theta)$ is the \emph{trial} $\mathrm{Tr}_{\iota,\theta}$.
  Further $\psi$, $\Gamma$ and $\kappa$, with a cap $\Theta$ and trial counts
  $m_{\iota,\theta}$ as in~\eqref{eq:fw-params} with $\iota$ in place of $t$.
  \Body
  For every $\iota\in\cT$ and $\theta\in\{0,\dots,\Theta\}$, run
  $m_{\iota,\theta}$ independent trials $\mathrm{Tr}_{\iota,\theta}$, and answer \YES{} as soon as one
  accepts. Answer \NO{} after all trials.
\end{algorithm}

\begin{theorem}[Budgeted trials]\label{thm:fw-budgeted}
  Let $\Phi$ be an implicit set system, and let $\mathrm{Tr}$ and the index sets $\cT=\cT_{I,k}$ be as
  in \cref{alg:fw-budgeted}, such that $\cT_{I,k}$ can be listed in time $N^{O(1)}$. Suppose that
  there is a polynomial $p$ such that for every instance $I$, every integer $k\le n$, every
  $\iota\in\cT_{I,k}$ and every integer $\theta\ge0$
  \begin{enumerate}[label=\textup{(T\arabic*)}]
    \item\label{it:fw-T1} $\mathrm{Tr}_{\iota,\theta}$ runs in time $2^\theta p(N)$ and in space $p(N)$
      plus $\mathrm{sp}$, where $\mathrm{sp}\ge0$ depends only on $I$ (for example, the space of an
      extension algorithm on $I$), and
    \item\label{it:fw-T2} $\mathrm{Tr}_{\iota,\theta}$ rejects if $\cF_I$ contains no set of size at most
      $k$.
  \end{enumerate}
  Fix $I$ and $k$, let $S\in\cF_I$ with $|S|=k$, and suppose that for each $\iota\in\cT$ there is a random variable
  $\xi_{\iota}$ with values in $[0,\infty]$, whose distribution may depend on $S$ but not on $\theta$, such that
  \begin{enumerate}[label=\textup{(T\arabic*)},resume]
    \item\label{it:fw-T3} $\Pr[\mathrm{Tr}_{\iota,\theta}\text{ accepts}]\ge\Pr[\xi_{\iota}\le\theta]$ for every
      integer $\theta\ge0$.
  \end{enumerate}
  Let $\psi$, $\Gamma$, $\kappa$ be as in \cref{thm:fw-main}. Then \cref{alg:fw-budgeted}, run with
  these trials and parameters, runs in time $2^{\psi n}N^{O(1)}$ and space $N^{O(1)}+\mathrm{sp}$, answers \NO{} if $\cF_I$ contains no
  set of size at most $k$, and answers \YES{} with probability at least $1-e^{-\kappa}$ if
  \[
    \max_{\iota\in\cT}\E[2^{-\xi_{\iota}}]\ \ge\ \frac{2^{-\psi n}}\Gamma .
  \]
\end{theorem}

\begin{proof}
  Time, space and soundness follow from~\ref{it:fw-T1} and~\ref{it:fw-T2} exactly as items~(i)
  and~(ii) of \cref{thm:fw-main}; the number of pairs is now $|\cT|(\Theta+1)\le N^{O(1)}$, as $\cT$ is listed in time $N^{O(1)}$. Fix $\iota$ with $\E[2^{-\xi_{\iota}}]\ge2^{-\psi n}/\Gamma$.
  Then $2^{-\Theta}\le2^{-\psi n}/(2\Gamma)\le\E[2^{-\xi_{\iota}}]/2$, so
  \cref{lem:fw-threshold-prob}(a) gives $\theta\le\Theta$ with
  $\Pr[\xi_{\iota}\le\theta]\ge2^\theta\E[2^{-\xi_{\iota}}]/(4(\Theta+1))\ge2^{\theta-\psi n}/(4\Gamma(\Theta+1))$.
  By~\ref{it:fw-T3} each trial of $(\iota,\theta)$ accepts with probability
  $\pi_{\iota,\theta}\ge\Pr[\xi_{\iota}\le\theta]$, the trial count satisfies $m_{\iota,\theta}\ge M_{\iota,\theta}\ge\kappa/\pi_{\iota,\theta}$, and by
  independence all these trials fail with probability at most $(1-\pi_{\iota,\theta})^{m_{\iota,\theta}}\le e^{-\kappa}$.
\end{proof}

In each application, \ref{it:fw-T1} and~\ref{it:fw-T2} follow directly from the description of the
trial, and only~\ref{it:fw-T3} needs an argument. \Cref{thm:fw-main} is the instance
$\cT=\{0,\dots,k\}$, in which a trial of $(t,\theta)$ is steps~1--2 of Algorithm~\MLS{}, and
$\xi_t=g(I,X,k)$ if $X\subseteq S$ and $\xi_t=\infty$ otherwise; \ref{it:fw-T3} holds because
$\cA$ accepts when $X\subseteq S$, and $\E[2^{-\xi_t}]=\mu_t(S)/\binom nt$. The closure algorithm RSC of
\cref{sec:rsc} is another instance: its trials add a random closure to a uniform sample before
the Steiner computation, so the uniform-sample form of \cref{thm:fw-main} does not apply
literally, but \cref{thm:fw-budgeted} does. The next instance allows extension algorithms with
one-sided error, as used for \pname{Connected Feedback Vertex Set} in \cref{sec:cfvs}.

\begin{corollary}[One-sided-error extension algorithms]\label{cor:fw-onesided}
  Let $\cA$ be a randomized algorithm for $\Phi$-Extension that never accepts a \NO{}-instance,
  accepts every \YES{}-instance with probability at least $1/2$ using its own independent coins, and
  runs in time $2^{g(I,X,k)}N^{O(1)}$ for every input and every outcome of its coins, where $g$ is
  polynomial-time computable. Then \cref{thm:fw-main} holds for $\cA$ with $\Gamma$ replaced by
  $2\Gamma$ in the cap $\Theta$ and in the trial counts $m_{t,\theta}$.
\end{corollary}

\begin{proof}
  Apply \cref{thm:fw-budgeted} with $\cT=\{0,\dots,k\}$ and the trials of Algorithm~\MLS{}, now with
  the randomized $\cA$. Property~\ref{it:fw-T1} holds because the time bound holds for every outcome
  of the coins, and~\ref{it:fw-T2} because $\cA$ never accepts a \NO{}-instance. Let $\xi_t$ be as
  above, and let $\xi'_t$ equal $\xi_t$ if an independent fair coin shows heads and $\infty$
  otherwise. If $X\subseteq S$ and $g(I,X,k)\le\theta$, then $\cA$ is run on a \YES{}-instance and
  accepts with probability at least $1/2$, independently of $X$. Hence
  \[
    \Pr[\mathrm{Tr}_{t,\theta}\text{ accepts}]\ge\tfrac12\Pr[\xi_t\le\theta]=\Pr[\xi'_t\le\theta],
    \qquad \E[2^{-\xi'_t}]=\tfrac12\E[2^{-\xi_t}]=\frac{\mu_t(S)}{2\binom nt}.
  \]
  If $S$
  satisfies~\eqref{eq:fw-star} with $\Gamma$, then $\max_t\E[2^{-\xi'_t}]\ge2^{-\psi n}/(2\Gamma)$,
  and \cref{thm:fw-budgeted} with $2\Gamma$ in place of $\Gamma$ gives the claim.
\end{proof}

\paragraph{Other levers.}
Changes of the trials outside the class $\cC_{\mathrm{thr}}$ of \cref{thm:fw-optimal} are examined in \cref{sec:app-levers}.
\Cref{thm:fw-budgeted} covers every sampling distribution that can be sampled in polynomial time
(\cref{thm:fw-samplable}); non-uniform sampling can lower the running time by an exponential factor
(\cref{prop:fw-twoclass}), whereas sampling in several adaptive stages cannot (\cref{prop:fw-multistage}).
{}
\section{Corollaries: mean exponents, linear savings and product form}\label{sec:corollaries}

This section derives earlier and intermediate forms of Monotone Local Search from
\cref{thm:fw-main}: the mean-exponent theorem (\cref{sec:co-mean}), which is the reference point of the black-box
barrier (\cref{sec:derandomisation}) and of the comparisons in \cref{sec:cc}; a closed form for linear
mean savings (\cref{sec:co-linear}), used for Algorithm
RS (\cref{sec:rs}); and multivariate Monotone Local Search, for whose product-form running-time exponents
the main theorem certifies no smaller bound (\cref{sec:co-product}). \Cref{sec:app-co-hardness}
explains why \pname{Max Leaf Spanning Tree} needs sample-dependent running times.

Recall the notation of \cref{sec:aprime}: $\Phi$ is an implicit set system, $I$ an instance with
$N=|I|$, universe $U_I$ with $n=|U_I|\le N$ and solution family $\cF_I$; the algorithm $\cA$ solves
$\Phi$-Extension with running-time exponent $g$ (\cref{def:fw-cost}); and for $S\in\cF_I$ with $|S|=k$ the
exponential moments and the generating function are
\[
  \mu_t(S)=\sum_{X\in\binom{S}{t}}2^{-g(I,X,k)}\quad(0\le t\le k),\qquad
  W_S(y)=\sum_{t=0}^{k}\mu_t(S)\,y^t\quad(y\ge0),
\]
as in \cref{def:fw-moments}. By the first inequality of \cref{lem:fw-gf}, the
\emph{generating-function condition} $W_S(y)\ge(1+y)^n2^{-\psi n}/\Gamma$, for a single $y\ge0$,
implies condition~\eqref{eq:fw-star} of \cref{thm:fw-main}, and by the second inequality the
converse holds up to a factor $n+1$: with
\[
  M(S)=\min_{0\le t\le k}\frac{\binom nt}{\mu_t(S)}\qquad\text{and}\qquad
  G(S)=\inf_{y\ge0}\frac{(1+y)^n}{W_S(y)},
\]
we have $M(S)\le G(S)\le(n+1)M(S)$. Note that \eqref{eq:fw-star} holds exactly when
$2^{\psi n}\Gamma\ge M(S)$; thus, with $\Gamma=1$, $M(S)$ is the smallest running-time bound
$2^{\psi n}$ (up to the factor $N^{O(1)}$) that \cref{thm:fw-main} certifies for the solution $S$,
and $G(S)$ is the smallest one that the generating-function condition certifies. The binary entropy
$H$ and the entropy estimates~\eqref{eq:fw-binom} are as in \cref{sec:fw-setting}.

\subsection{Mean exponents: the Markov cutoff}\label{sec:co-mean}

The simplest hypothesis on a sample-dependent running-time exponent is a bound $\lambda$ on its mean over the
$t$-subsets of a solution, for one sample size $t$. We derive the resulting bound from the main theorem by
Jensen's inequality; \cref{rem:co-markov} describes the direct algorithm, the Markov-cutoff mechanism of
Lemma~2 of Zamir~\cite{Zamir23jensen}, which discards samples whose running-time exponent exceeds $\lambda$
by more than one.

In \cref{thm:co-mean}, $t(I,k)$ is the sample size and $\lambda(I,k)$ the bound on the mean running-time
exponent of samples of this size inside a solution of size $k$; the running time is the expected
number $\binom nt/\binom kt$ of uniform $t$-sets drawn until one lies inside a fixed solution, times the
running-time bound $2^{\lambda(I,k)}$ of an extension call at the mean running-time exponent.

\begin{theorem}[Mean-exponent Monotone Local Search]\label{thm:co-mean}
Let $t(I,k)\in\{0,\dots,k\}$ and $\lambda(I,k)\ge0$ be computable in time $N^{O(1)}$, with values
$\lambda(I,k)\le N^{O(1)}$, such that for every instance $I$, every $k\le n$ and every $S\in\cF_I$
with $|S|=k$, writing $t=t(I,k)$,
\begin{equation}\label{eq:co-mean-hyp}
  \E_X\bigl[g(I,X,k)\bigr]\le\lambda(I,k),\qquad X\text{ uniform in }\binom{S}{t}.
\end{equation}
Then there is a randomized algorithm that, given $I$ and $k\le n$, runs in time
\[
  \frac{\binom nt}{\binom kt}\,2^{\lambda(I,k)}\,N^{O(1)}
\]
and in space $N^{O(1)}$ plus the space of $\cA$. It answers \NO{} if $\cF_I$ contains no set of
size at most $k$, and it answers \YES{} with probability at least $1-1/e$ if $\cF_I$ contains a set
of size exactly $k$. Running it for $k=0,1,\dots,n$ decides $\Phi$-Subset as in
\cref{rem:fw-decision}.
\end{theorem}

\begin{proof}
Fix $I$ and $k$, and write $t=t(I,k)$ and $\lambda=\lambda(I,k)$. If $n=0$, then $\cF_I\subseteq\{\emptyset\}$
and $k=0$, and the membership test for $\cF_I$ decides the question in time $N^{O(1)}$. So let $n\ge1$.
Define $\psi$ by
\[
  2^{\psi n}=\frac{\binom nt}{\binom kt}\,2^{\lambda},
\]
and take $\Gamma=1$ and $\kappa=1$. Since $\binom kt\le\binom nt\le2^n$ and $\binom kt\ge1$, we
have $0\le\psi\le1+\lambda/n\le N^{O(1)}$, and $\psi$ is computable from $(I,k)$ in time
$N^{O(1)}$ because $t$ and $\lambda$ are. Let $S\in\cF_I$ with $|S|=k$. The function $x\mapsto2^{-x}$
is convex, so Jensen's inequality and \eqref{eq:co-mean-hyp} give
\[
  \mu_t(S)=\binom kt\,\E_X\bigl[2^{-g(I,X,k)}\bigr]\ \ge\ \binom kt\,2^{-\E_X[g(I,X,k)]}
  \ \ge\ \binom kt\,2^{-\lambda}=\binom nt\,2^{-\psi n},
\]
with $X$ uniform in $\binom St$. Hence \eqref{eq:fw-star} holds with $\Gamma=1$, and
\cref{thm:fw-main} gives the time and space bounds, soundness, and success probability
$1-e^{-1}$.
\end{proof}

\begin{remark}[The Markov cutoff]\label{rem:co-markov}
The algorithm of this proof is Algorithm \MLS{}, which tries every sample size and every threshold.
A direct algorithm uses only the sample size $t=t(I,k)$ and the single threshold $\lambda+1$, where
$\lambda=\lambda(I,k)$: it repeats the following trial $m=\lceil(\lambda+1)\binom nt\binom kt^{-1}\rceil$
times and answers \NO{} if no trial succeeds. A trial draws $X$ uniformly from $\binom{U_I}{t}$,
computes $g(I,X,k)$, and, if $g(I,X,k)\le\lambda+1$ (with a rational threshold in
$[\lambda+1,\lambda+2]$ if needed, \cref{rem:fw-real}), runs $\cA$ on $(I,X,k)$ and answers \YES{}
if $\cA$ accepts. Every \YES{} answer is correct,
each trial takes time $2^{\lambda+2}N^{O(1)}$, so the total time is the bound of the theorem, and
the space is that of $\cA$ plus $N^{O(1)}$. For completeness, let $S\in\cF_I$ with $|S|=k$. A trial
has $X\subseteq S$ with probability $\binom kt/\binom nt$; conditioned on this event $X$ is uniform in
$\binom St$, and by Markov's inequality and \eqref{eq:co-mean-hyp} the running-time exponent exceeds
$\lambda+1$ with probability at most $\lambda/(\lambda+1)$. Otherwise the trial runs $\cA$, which
accepts because $Y=S\setminus X$ extends $X$ to $S$. So a trial succeeds with probability at least
$p_{\mathrm{trial}}=\binom kt/((\lambda+1)\binom nt)\ge1/m$, and all $m$ trials fail with probability
at most $(1-p_{\mathrm{trial}})^m\le e^{-1}$. This is the Markov-cutoff mechanism of Lemma~2 of
Zamir~\cite{Zamir23jensen}.
\end{remark}

\subsection{Linear mean savings}\label{sec:co-linear}

In many applications the bound on the mean running-time exponent of a sample decreases roughly linearly in
the number of sampled elements: for samples $X$ of size $t$ inside a solution of size $k$, the mean of
$g(I,X,k)$ is at most $\beta n-\rho t+C$ for reals $\beta,\rho,C\ge0$, so that each sampled element of
the solution lowers this bound by $\rho$. The next corollary gives the resulting exponent of
the running time in closed form. At solution fraction $\sigma=k/n$, a uniformly random element lies in
a fixed solution with probability $\sigma$, so each further sampled element divides the probability
that the sample lies inside the solution by about $1/\sigma$. This multiplies the number of trials by
about $1/\sigma$; in exchange it divides the running-time bound $2^{\beta n-\rho t+C}$ of an extension
call by $2^{\rho}$. We show (\cref{cor:co-linear}(2)) that sampling lowers the exponent exactly when
$\rho>\log(1/\sigma)$.

For $\sigma\in(0,1)$, reals $\beta,\rho\ge0$ and $p=2^{-\rho}$ define
\begin{equation}\label{eq:co-Fstar}
  F^*(\sigma)=
  \begin{cases}
    \beta & \text{if }\sigma\le p,\\[2pt]
    \beta+\sigma\log\dfrac p\sigma+(1-\sigma)\log\dfrac{1-p}{1-\sigma}=\beta-\sigma\rho+H(\sigma)+(1-\sigma)\log(1-p) & \text{if }\sigma>p .
  \end{cases}
\end{equation}
For $\sigma>p$ the corresponding base is
\[
  2^{F^*(\sigma)}=2^{\beta}\left(\frac p\sigma\right)^{\sigma}\left(\frac{1-p}{1-\sigma}\right)^{1-\sigma}
  \le2^{\beta}\left(\sigma\cdot\frac p\sigma+(1-\sigma)\cdot\frac{1-p}{1-\sigma}\right)=2^{\beta}
\]
by the weighted arithmetic--geometric mean inequality, with equality only if $p/\sigma=(1-p)/(1-\sigma)$,
that is $\sigma=p$. So $F^*(\sigma)\le\beta$, with equality exactly when $\sigma\le p$.

In the corollary, $\beta n+C$ bounds the mean running-time exponent of the unsampled call ($t=0$),
$\rho$ is the decrease of this bound per sampled element, and $C$ is an additive slack.

\begin{restatable}[Linear mean savings]{corollary}{RestCorCoLinear}\label{cor:co-linear}
Let $\beta=\beta(I,k)$, $\rho=\rho(I,k)$ and $C=C(I,k)$ be nonnegative reals computable from $(I,k)$
in time $N^{O(1)}$, with $\beta n+C\le N^{O(1)}$. Let $I$ be an instance and $k$ an integer with
$\sigma=k/n\in(0,1)$ such that for every $t\in\{0,\dots,k\}$ and every $S\in\cF_I$ with $|S|=k$,
\begin{equation}\label{eq:co-linear-hyp}
  \E_X\bigl[g(I,X,k)\bigr]\le\beta n-\rho t+C,\qquad X\text{ uniform in }\binom St .
\end{equation}
Let $F^*(\sigma)$ be as in \eqref{eq:co-Fstar}, and let $\kappa>0$ be as in \cref{thm:fw-main}. Then:
\begin{enumerate}[label=(\arabic*)]
\item Every such $S$ satisfies $G(S)\le2^{C+nF^*(\sigma)}$. Algorithm \MLS{} with $\Gamma=1$,
confidence $\kappa$ and a rate computable from $(I,k)$ in time $N^{O(1)}$ gives a randomized algorithm
that, given $I$ and $k$, runs in time $2^{\max(0,\,C+nF^*(\sigma))}N^{O(1)}$, answers \NO{} if
$\cF_I$ contains no set of size at most $k$, and answers \YES{} with probability at least
$1-e^{-\kappa}$ if $\cF_I$ contains a set of size $k$.
\item $F^*(\sigma)<\beta$ if and only if $\rho>\log(1/\sigma)$.
\end{enumerate}
\end{restatable}

The corollary uses the bound \eqref{eq:co-linear-hyp} for every $t$, and $\rho$ only
has to be the decrease of a bound on the \emph{mean} running-time exponent over the samples inside a
solution, not a decrease of the running-time exponent guaranteed for every sampled element.
\Cref{rem:co-fsigma} compares this with the algorithm of \cref{thm:co-mean} at a single sample
size.

The proof is in \cref{sec:app-co-linear}. It bounds the moments $\mu_t(S)$ from below by Jensen's
inequality, as in the proof of \cref{thm:co-mean}, sums them by the binomial theorem, and minimizes the
resulting bound on $(1+y)^n/W_S(y)$ over $y$; the minimum lies at $y=0$ if $\sigma\le p$ and at
$y_0=(\sigma-p)/(1-\sigma)$ otherwise.

\begin{remark}[One sample size]\label{rem:co-fsigma}
The algorithm of \cref{thm:co-mean} at a single sample size $t$ with the mean bound $\beta n-\rho t+C$
of \eqref{eq:co-linear-hyp} runs in time $\binom nt\binom kt^{-1}2^{\beta n-\rho t+C}N^{O(1)}$, and
\begin{equation}\label{eq:co-product}
  \binom nt\binom kt^{-1}2^{\beta n-\rho t}=2^{\beta n}\prod_{i=0}^{t-1}\frac{n-i}{2^{\rho}(k-i)} .
\end{equation}
The factor with index $i$ weighs the two effects of sampling one more element: the probability that the
sample lies inside a fixed solution of size $k$ is divided by $(n-i)/(k-i)$, and the running-time bound
$2^{\beta n-\rho t+C}$ of the extension call is divided by $2^{\rho}$. As $n\ge k$, the factors are
nondecreasing in $i$, and with $p=2^{-\rho}$ the factor with index $i$ is below $1$ exactly when
$(1-p)i<k-pn$. So the best sample size is the number of factors below~$1$, and sampling at all beats
$t=0$ exactly when the first factor $n/(2^{\rho}k)=p/\sigma$ is below~$1$, that is $\sigma>p$, the
condition of \cref{cor:co-linear}(2).

By the entropy estimates~\eqref{eq:fw-binom}, the running time above is at most
$(k+1)2^{C}2^{nf_\sigma(\tau)}N^{O(1)}$ with $\tau=t/n$ and
\begin{equation}\label{eq:co-fsigma}
  f_\sigma(\tau)=H(\tau)-\sigma H(\tau/\sigma)+\beta-\rho\tau\qquad(0\le\tau\le\sigma),\qquad p=2^{-\rho}.
\end{equation}
Expanding the entropies gives
$H(\tau)-\sigma H(\tau/\sigma)=-(1-\tau)\log(1-\tau)+(\sigma-\tau)\log(\sigma-\tau)-\sigma\log\sigma$,
so on $[0,\sigma)$ the derivative is
$f_\sigma'(\tau)=\log\frac{1-\tau}{\sigma-\tau}-\rho$, the logarithm of the factor of~\eqref{eq:co-product}
with index $i=\tau n$. It is strictly increasing in $\tau$, so $f_\sigma$ is
convex on $[0,\sigma]$. If $\sigma\le p$ (which includes $\rho=0$), then
$f_\sigma'(\tau)\ge f_\sigma'(0)=\log\frac1\sigma-\rho\ge0$, and the minimum of $f_\sigma$ over $[0,\sigma]$ is
$f_\sigma(0)=\beta=F^*(\sigma)$. If $\sigma>p$ (so $p<1$), then $f_\sigma'$ vanishes exactly at
$\tau^*=\frac{\sigma-p}{1-p}\in(0,\sigma)$, the unique minimizer. With $A=\frac{1-\sigma}{1-p}$ we have
$1-\tau^*=A$, $\sigma-\tau^*=pA$ and $-\rho=\log p$, so
\[
  f_\sigma(\tau^*)=-(1-p)A\log A+pA\log p+\tau^*\log p-\sigma\log\sigma+\beta
  =\beta+\sigma\log\frac p\sigma+(1-\sigma)\log\frac{1-p}{1-\sigma}=F^*(\sigma),
\]
using $(1-p)A=1-\sigma$ and $pA+\tau^*=\sigma$. In both cases the minimum of $f_\sigma$ over $[0,\sigma]$
is $F^*(\sigma)$. Since $H(\tau)-\sigma H(\tau/\sigma)$ is nondecreasing on $[0,\sigma]$ (its derivative is
$\log\frac{1-\tau}{\sigma-\tau}\ge0$), the single-size algorithm run at $t=\lfloor\tau^*n\rfloor$
(with $\tau^*=0$ if $\sigma\le p$) attains the exponent $C+nF^*(\sigma)$ of \cref{cor:co-linear}(1) up to the factors $2^{\rho}$ and $k+1$.
Algorithm RS (\cref{sec:rs}), the black-box barrier (\cref{sec:app-barrier}) and the two-class family
of \cref{sec:app-levers} use $f_\sigma$ in this form.
\end{remark}

\subsection{Product-form running times and multivariate Monotone Local Search}\label{sec:co-product}

In the classical forms of Monotone Local Search the running-time bound of the extension algorithm
depends on the sample $X$ only through $|X|$. \cite{FominGLS19}
showed that an extension algorithm running in time $c^{k-|X|}N^{O(1)}$ on input $(I,X,k)$ gives a randomized algorithm for
$\Phi$-Subset running in time $(2-1/c)^nN^{O(1)}$; Gaspers and Lee~\cite{GaspersL17} allowed a further
factor $b^{\,n-|X|}$ in the running time of the extension algorithm and obtained the running time
$(1+b-1/c)^nN^{O(1)}$; and the multivariate version, with a further factor $d^{|X|}$, gives
$(b+d-d/c)^nN^{O(1)}$ when $d\le bc$ (\cref{cor:co-multivariate}). We now derive all three from the main theorem and show that for such running-time
exponents the main theorem certifies no smaller running-time bound.

For constants $b,c,d\ge1$, the \emph{product-form} running-time exponent is
\begin{equation}\label{eq:co-product-g}
  g(I,X,k)=(n-|X|)\log b+(k-|X|)\log c+|X|\log d .
\end{equation}
It is nonnegative on inputs with $k\ge|X|$, the domain of \cref{def:fw-cost}, because $b,c,d\ge1$. Put
\[
  \mathrm{base}(b,c,d)=b+\min(d,bc)\Bigl(1-\frac1c\Bigr)=
  \begin{cases}b+d-d/c&\text{if }d\le bc,\\ bc&\text{if }d\ge bc.\end{cases}
\]
Since $b+d-d/c-bc=(c-1)(d-bc)/c$, also $\mathrm{base}(b,c,d)=\min(b+d-d/c,\,bc)$. If $S\in\cF_I$ with $|S|=k$ and
$g$ has the form \eqref{eq:co-product-g} on all inputs $(I,X,k)$ with $X\subseteq S$, then every
$X\in\binom St$ has $2^{g(I,X,k)}=b^{\,n-t}c^{\,k-t}d^{\,t}$, so
\begin{equation}\label{eq:co-product-mu}
  \mu_t(S)=\binom kt\,b^{-(n-t)}c^{-(k-t)}d^{-t}\qquad(0\le t\le k).
\end{equation}
Thus $M(S)$ and $G(S)$ depend only on $n$, $k$, $b$, $c$ and $d$; we write $M_k$ and $G_k$ for them
(with $n,b,c,d$ fixed), and define them by \eqref{eq:co-product-mu} for every integer $0\le k\le n$.

The evaluation of the product form through a generating function goes back to Gaspers and
Lee~\cite{GaspersL17} and \cite{FominGLS19} (see \cref{sec:fw-related}). The
generating function $W_S$ used here is a different series, but for product-form running times it has the
following closed form.

\begin{restatable}[Product form]{proposition}{RestPropCoProduct}\label{prop:co-product}
Let $b,c,d\ge1$, let $0\le k\le n$, and let $\mu_t$, $M_k$ and $G_k$ be given by
\eqref{eq:co-product-mu}. Then:
\begin{enumerate}[label=(\alph*)]
\item The generating function is $W(y)=\sum_{t=0}^{k}\mu_ty^t=b^{-n}c^{-k}(1+ybc/d)^k$, and
\[
  M_k=\min_{0\le t\le k}\frac{\binom nt}{\binom kt}\,b^{\,n-t}c^{\,k-t}d^{\,t} .
\]
\item $M_k\le G_k\le \mathrm{base}(b,c,d)^n$. More precisely, $y^*=(c-1)d/(bc)$ gives
$(1+y^*)^n/W(y^*)=(b+d-d/c)^n$, and $y=0$ gives $(1+0)^n/W(0)=b^nc^k\le(bc)^n$.
\item Conversely, the following hold:
  \begin{itemize}
  \item if $c=1$, then $G_0=M_0=b^n=\mathrm{base}(b,c,d)^n$;
  \item if $c>1$ and $d<bc$, let $\sigma^*=cd/(bc+cd-d)$; then $\sigma^*\in(0,1)$; if $\sigma^*n$ is an
  integer, then $G_{\sigma^*n}=(b+d-d/c)^n$; and for every $n\ge1$,
  \[
    \max_{0\le k\le n}G_k\ \ge\ (b+d-d/c)^n\Bigl(\frac{b}{b+d-d/c}\Bigr)^{1/\sigma^*};
  \]
  \item if $c>1$ and $d\ge bc$, then $G_n=M_n=(bc)^n=\mathrm{base}(b,c,d)^n$.
  \end{itemize}
  Consequently, with $\mathsf K=\bigl((b+d-d/c)/b\bigr)^{1/\sigma^*}$ in the second case and $\mathsf K=1$ otherwise,
  \[
    \frac{\mathrm{base}(b,c,d)^n}{(n+1)\mathsf K}\ \le\ \max_{0\le k\le n}M_k\ \le\ \mathrm{base}(b,c,d)^n .
  \]
\end{enumerate}
\end{restatable}

The expression for $M_k$ in~(a) is the classical Monotone Local Search expression. So for
product-form running-time exponents, $\max_{0\le k\le n}M_k$, the smallest running-time bound
$2^{\psi n}$ that \cref{thm:fw-main} with $\Gamma=1$ certifies for every solution size, is
$\mathrm{base}(b,c,d)^n$ up to the factor $(n+1)\mathsf K$, where $\mathsf K$ depends only on $b,c,d$.

\begin{proof}
Parts (a) and (b) are direct computations. For (c) we treat the solution size as a real parameter
$x$: the minimization over $y$ in (a) then defines a concave function $\mathsf f$ of $x$,
whose value at $x=\sigma^*n$ is exactly $n\log(b+d-d/c)$, and concavity transfers this value
to the integer $\lfloor\sigma^*n\rfloor$, losing in $G_{\lfloor\sigma^*n\rfloor}$ only the factor $\mathsf K$,
which does not depend on $n$.

(a) Grouping by $t$ and using the binomial theorem,
\[
  W(y)=\sum_{t=0}^{k}\binom kt y^t\,b^{-(n-t)}c^{-(k-t)}d^{-t}
  =b^{-n}c^{-k}\sum_{t=0}^{k}\binom kt\Bigl(\frac{ybc}{d}\Bigr)^t=b^{-n}c^{-k}\Bigl(1+\frac{ybc}{d}\Bigr)^k .
\]
The formula for $M_k=\min_t\binom nt/\mu_t$ is immediate from \eqref{eq:co-product-mu}.

(b) The inequality $M_k\le G_k$ is the first inequality of \cref{lem:fw-gf}: for every $y\ge0$,
$W(y)\le(1+y)^n/M_k$. At $y^*\ge0$ we have $1+y^*bc/d=c$, so $W(y^*)=b^{-n}$ and
\[
  \frac{(1+y^*)^n}{W(y^*)}=\bigl(b(1+y^*)\bigr)^n=\Bigl(b+\frac{(c-1)d}{c}\Bigr)^n=\Bigl(b+d-\frac dc\Bigr)^n .
\]
At $y=0$ the ratio is $1/W(0)=b^nc^k\le(bc)^n$, because $c\ge1$ and $k\le n$. Hence
$G_k\le\min(b+d-d/c,\,bc)^n=\mathrm{base}(b,c,d)^n$.

Part~(c) and the final inequalities are proved in \cref{sec:app-co-product}.
\end{proof}

With the product form evaluated, the multivariate theorem follows; its cases $d=1$ and $b=d=1$ are
the two classical statements. In the running-time bound $b^{\,n-|X|}c^{k-|X|}d^{|X|}N^{O(1)}$ of the
extension algorithm, $b$ is the factor per element outside the sample, $c$ the factor per element
that may still be added, and $d$ the factor per sampled element.

\begin{corollary}[Multivariate Monotone Local Search]\label{cor:co-multivariate}
Let $b,c,d\ge1$ be constants such that $\log b$, $\log c$ and $\log d$ are computable to polynomially
many bits in polynomial time. If $\Phi$-Extension can be solved in time
$b^{\,n-|X|}c^{k-|X|}d^{|X|}N^{O(1)}$ on every input $(I,X,k)$, then $\Phi$-Subset can be decided by a
randomized algorithm with one-sided error at most $1/e$ in time
\[
  \Bigl(b+\min(d,bc)\Bigl(1-\frac1c\Bigr)\Bigr)^nN^{O(1)}=
  \begin{cases}(b+d-d/c)^nN^{O(1)}&\text{if }d\le bc,\\ (bc)^nN^{O(1)}&\text{if }d\ge bc,\end{cases}
\]
and in space $N^{O(1)}$ plus the space of the extension algorithm.
\end{corollary}

If the extension algorithm returns a completion when it accepts, the algorithm returns a solution. As
$bc\ge1$, the case $d=1$ gives $(1+b-1/c)^nN^{O(1)}$, the theorem of Gaspers and Lee~\cite{GaspersL17},
and the case $b=d=1$ gives $(2-1/c)^nN^{O(1)}$, the randomized theorem of \cite{FominGLS19}. The case $d\ge bc$ needs no sampling: a single call of the extension
algorithm with $X=\emptyset$ takes time $b^nc^kN^{O(1)}\le(bc)^nN^{O(1)}$.

\begin{proof}
The product-form exponent \eqref{eq:co-product-g} is nonnegative, computable in time $N^{O(1)}$, and
by hypothesis a running-time exponent for the given extension algorithm. Fix $k\le n$, and put
$2^{\psi}=\mathrm{base}(b,c,d)$, $\Gamma=1$ and $\kappa=1$; then $0\le\psi\le\log(bc)$ is a constant
(if $\log b$, $\log c$ or $\log d$ is irrational, $g$ is rounded up as in \cref{rem:fw-real}, which replaces
$\Gamma$ by $2$; if $\psi$ is irrational, a rational rate in $[\psi,\psi+1/n]$ is used; both change the running
time by a constant factor). Let
$S\in\cF_I$ with $|S|=k$. On the inputs $(I,X,k)$ with $X\subseteq S$ the running-time exponent has the
form \eqref{eq:co-product-g}, so $W_S$ is the function $W$ of \cref{prop:co-product}(a), and by
\cref{prop:co-product}(b) the generating-function condition holds with $2^{\psi n}=\mathrm{base}(b,c,d)^n$: at
$y=y^*$ if $d\le bc$, and at $y=0$ if $d\ge bc$. By the first inequality of \cref{lem:fw-gf}, $S$
satisfies \eqref{eq:fw-star}. So \cref{thm:fw-main} gives an algorithm that runs in time
$\mathrm{base}(b,c,d)^nN^{O(1)}$, answers \NO{} if $\cF_I$ has no set of size at most $k$, and answers \YES{}
with probability at least $1-1/e$ if $\cF_I$ has a set of size $k$. Running it for $k=0,\dots,n$
decides $\Phi$-Subset as claimed (\cref{rem:fw-decision}). A witness $Y$ returned by
the extension algorithm gives the solution $X\cup Y$. For $d=1$ we have $\min(1,bc)=1$, since
$bc\ge1$, so $\mathrm{base}(b,c,1)=1+b-1/c$; for $b=d=1$ this is $2-1/c$.
\end{proof}

\paragraph{Max Leaf Spanning Tree.}
For \pname{Max Leaf Spanning Tree}, the extension problem that the original Monotone Local Search
needs is W[1]-hard. Product-form running-time exponents certify nothing below a single unsampled call
when $k/n\le d/(bc)$, nor, when $c=1$, for the worst solution size. Both statements and their proofs are in \cref{sec:app-co-hardness}.
{}
\section{Connected Dominating Set and Max Leaf Spanning Tree}\label{sec:cds}

This section prepares the applications of \cref{sec:rs,sec:rstheta,sec:rsc,sec:det}: the
equivalence of \pname{Max Leaf Spanning Tree} and \pname{Connected Dominating Set}
(\cref{lem:cds-maxleaf}), a vertex-weighted directed Steiner tree computation by inclusion--exclusion
(\cref{lem:cv-W}), the Steiner extension algorithm built on it, whose running-time exponent is the number of
terminals that the sample leaves (\cref{cor:cds-steiner}), and the mean bound on this running-time exponent
(\cref{lem:cds-C}).

All graphs are finite, simple and undirected. Throughout, $G$ is a graph with vertex set $V=V(G)$
and $n=|V|$ vertices; when $G$ is fixed we write $N(v)$, $N[v]$, $N[X]=\bigcup_{x\in X}N[x]$ and
$N(X)=N[X]\setminus X$ for $N_G(v)$, $N_G[v]$, $N_G[X]$ and $N_G(X)$. For a subgraph $F\subseteq G$
we keep the subscript: $N_F(v)$, $N_F[X]$ and $d_F(v)$ are neighborhoods and degrees in~$F$. For
a graph, $\comp(\cdot)$ denotes the number of its connected components; in particular
$\comp(G[\emptyset])=0$. A \emph{connected dominating set} of $G$ is a nonempty set $S\subseteq V$
with $N[S]=V$ such that $G[S]$ is connected; a graph that has one is connected, and for a connected
graph $G$ we write $\gc(G)$ for the minimum size of a connected dominating set.
\pname{Connected Dominating Set} asks for $\gc(G)$, and \pname{Max Leaf Spanning Tree} (MaxLeaf)
asks for the maximum number of leaves of a spanning tree of a connected graph~$G$.

The following equivalence of the two problems is folklore; it is stated, for example, by Caro, West
and Yuster~\cite{CaroWY00}, by Fomin, Grandoni and Kratsch~\cite{FominGK08}, by \cite[Section~1]{FernauKKLLRR11} and by Abu-Khzam, Mouawad and Liedloff~\cite[Section~1]{Abu-KhzamML11}. We include its short proof because the construction in the
second half of the proof is reused for the mean bound later in this section.

\begin{lemma}[Max leaf and connected domination]\label{lem:cds-maxleaf}
Let $G$ be a connected graph with $n$ vertices. If $n\ge3$, the maximum number of leaves of a spanning
tree of $G$ equals $n-\gc(G)$, and a spanning tree with at least $n-|S|$ leaves can be computed in
polynomial time from any connected dominating set $S$. If $n\le2$, the only spanning tree of $G$ is $G$
itself, with no leaf if $n=1$ and two leaves if $n=2$.
\end{lemma}
\begin{proof}
For $n\le2$ the claim is immediate, as a connected graph on at most two vertices is a tree. Let $n\ge3$,
let $F$ be a spanning tree of $G$ and let $\operatorname{leaves}(F)$ be its set of leaves. Since $n\ge3$, two leaves of $F$
are never adjacent in $F$ (they would form a component $K_2$ of $F$), so the set $S=V\setminus\operatorname{leaves}(F)$
of non-leaves is nonempty and contains the neighbor in $F$ of every leaf; hence $N[S]=V$.
Deleting leaves from a tree leaves a tree, so $F-\operatorname{leaves}(F)$ is a spanning tree of $G[S]$ and $G[S]$ is
connected. Thus $S$ is a connected dominating set and $|\operatorname{leaves}(F)|=n-|S|\le n-\gc(G)$.

Conversely, let $S$ be a connected dominating set. Take a spanning tree of $G[S]$ and attach every
vertex of $V\setminus S$ by one edge to a neighbor in $S$, which exists because $N[S]=V$. The
result is a spanning tree of $G$ in which every vertex of $V\setminus S$ is a leaf, so it has at
least $n-|S|$ leaves, and it is computed in polynomial time. With $|S|=\gc(G)$ this gives the
reverse inequality.
\end{proof}

All algorithms of \cref{sec:rs,sec:rstheta,sec:rsc,sec:det} and of
\cref{sec:further} run, for $s=1,\dots,n$ in turn, a procedure that rejects $s$ if no solution of
size at most $s$ exists and accepts the optimum size (with probability bounded away from $0$, for
the randomized ones), and output the first accepted $s$. The following lemma records what this
wrapper needs from the inner procedure; it is stated for an
arbitrary implicit set system (\cref{def:fw-iss}).

\begin{lemma}[First accepted size]\label{lem:cds-first-accepted}
Let $\Phi$ be an implicit set system and let $I$ be an instance of $\Phi$ with a universe of $n$
elements whose family $\cF_I$ of solutions is nonempty and does not contain the empty set; let
$\operatorname{opt}(I)\in\{1,\dots,n\}$ be the least size of a solution. Let $\mathrm{Dec}$ be a procedure which,
on input $(I,s)$ with $1\le s\le n$, runs in time $T$ and space $\mathrm{Sp}$ and rejects if
$\cF_I$ contains no set of size at most $s$. Then running $\mathrm{Dec}$ on
$(I,1),(I,2),\dots,(I,n)$ and outputting the first accepted $s$, or $n$ if none is accepted, runs
in time $nT$ and space $\mathrm{Sp}$ and never outputs a value below $\operatorname{opt}(I)$; if $\mathrm{Dec}$
accepts $(I,\operatorname{opt}(I))$ with probability at least $p$, the output is $\operatorname{opt}(I)$ with
probability at least $p$ (deterministic form: $p=1$). For $\Phi_{\mathrm{CDS}}$ (\cref{def:cds-ast}), a
connected graph $G$ and $\operatorname{opt}(G)=\gc(G)$, the same procedure solves \pname{Max Leaf Spanning
Tree} within the same bounds, by \cref{lem:cds-maxleaf}, where graphs with $n\le2$ vertices are handled
directly in constant time.
\end{lemma}
\begin{proof}
The time and space bounds are clear. An accepted $s$ satisfies $s\ge\operatorname{opt}(I)$, and the default
output $n$ satisfies $n\ge\operatorname{opt}(I)$ because every solution is a subset of the universe; so the output
is never below $\operatorname{opt}(I)$. If the run on $(I,\operatorname{opt}(I))$ accepts, which happens with probability at
least $p$, then, as no smaller $s$ is ever accepted, the output is $\operatorname{opt}(I)$. For \pname{Max
Leaf Spanning Tree}, output $n-\gc(G)$ leaves for $n\ge3$ by \cref{lem:cds-maxleaf}; for $n\le2$, output $0$ if
$n=1$ and $2$ if $n=2$ without running $\mathrm{Dec}$, which is correct by the last sentence of \cref{lem:cds-maxleaf}.
\end{proof}

\subsection{Vertex-weighted directed Steiner trees}\label{sec:cv-steiner}

The extension algorithm of \cref{sec:cds-states}, the variants of \cref{sec:further} and \pname{Directed
Steiner Tree} in \cref{sec:fa-dst} all use one Steiner computation, on a digraph with vertex weights.
This subsection states it, \cref{lem:cv-W}; its proof is in \cref{sec:app-steiner}.

Let $\vec G$ be a digraph (finite, without loops or parallel arcs) with $n$ vertices, let $r\in V(\vec G)$
be a \emph{root}, let $K\subseteq V(\vec G)\setminus\{r\}$ be a set of \emph{terminals}, and let
$\varphi\colon V(\vec G)\to\{0,1,\dots,\varphi_{\max}\}$ be vertex weights; zero weights are allowed. A set
$D\subseteq V(\vec G)\setminus(\{r\}\cup K)$ is \emph{feasible} if every terminal is reachable from
$r$ in $\vec G[\{r\}\cup K\cup D]$, and we put
\[
  \OPT_\varphi(\vec G,r,K)=\min\{\varphi(D):D\text{ feasible}\},
\]
with $\min\emptyset=\infty$. Equivalently, $\OPT_\varphi(\vec G,r,K)$ is the least weight of the non-root,
non-terminal vertices of an out-tree of $\vec G$ rooted at $r$ that contains $K$. The weights of the
root and of the terminals play no role.

\begin{restatable}[Vertex-weighted directed Steiner tree]{lemma}{RestLemCvW}\label{lem:cv-W}
For $\vec G$, $n$, $r$, $K$ and $\varphi$ as above, $\OPT_\varphi(\vec G,r,K)$ can be computed
deterministically in time $2^{|K|}(n+\varphi_{\max})^{O(1)}$ and space $(n+\varphi_{\max})^{O(1)}$.
\end{restatable}

The method is Nederlof's~\cite{Nederlof13}: count, by inclusion--exclusion over the
terminals, the branching walks from the root that visit every terminal, and read the optimum off the
least length at which this count is nonzero. Nederlof states it for undirected graphs with edge
weights; Lokshtanov and Nederlof obtain the same bound, $2^{|K|}$ times a factor polynomial in $n$ and in
the largest edge weight, from the Dreyfus--Wagner recurrence~\cite[Theorem~6.3]{LokshtanovN10}, and
\cite{FominKLPS19} remove the dependence on the weights at the price of a
larger base, in time $O(7.97^{|K|}n^4\log W)$ and polynomial space for edge weights at most
$W$. Two changes give the lemma: arcs are used only in their own direction, and the weight of a
vertex, multiplied by $n$, is added to the length of every arc that enters it, so that one unit of
weight contributes more to a length than all arcs of an out-tree together.
The directed case is not new: \cite[Lemma~3.5]{JonesLRSS17} state it for a
minimum-size set of non-terminals and attribute it to Nederlof and to~\cite{MisraPRSS12}, who prove it for
\pname{Directed Steiner Out-Tree}, with the number of vertices of the out-tree as the objective; the length encoding of vertex weights is a routine
device. We regard the lemma as folklore and include a proof
(\cref{sec:app-steiner}) because we need vertex weights, including zero weights, on a digraph.

\subsection{Red--blue states and the Steiner encoding}\label{sec:cds-states}

Abu-Khzam, Mouawad and Liedloff~\cite{Abu-KhzamML11} solve \pname{Connected
Dominating Set} by reducing it to a connected red--blue domination problem, which they finish with
a Steiner tree computation. We use their formulation, extended by \emph{dominating components}, the
connected pieces of a partial solution, which the encoding contracts to single vertices.

\begin{definition}[State]\label{def:cds-state}
A \emph{state} of a graph $G$ is a triple $J=(\cD,B,R)$ in which $\cD$ is a family of pairwise
disjoint nonempty vertex sets, each inducing a connected subgraph of $G$, with no edge of $G$ between
two distinct members, and, writing $V(\cD)=\bigcup_{C\in\cD}C$, the sets $B$ and $R$ satisfy
$B\subseteq V\setminus V(\cD)$ and $R\subseteq V\setminus N[V(\cD)]$. The vertices in $B$ are
\emph{blue} (still available), those in $R$ are \emph{red} (still to be dominated), and the members
of $\cD$ are the \emph{dominating components}. The \emph{solutions} of $J$ are the sets
\[
  Y\subseteq B\quad\text{with}\quad R\subseteq N[Y],\quad G[V(\cD)\cup Y]\text{ connected and}\quad
  V(\cD)\cup Y\neq\emptyset,
\]
and $\OPT(J)$ is the minimum size of a solution ($\OPT(J)=\infty$ if there is none). The
\emph{terminals} of $J$ are its red vertices and its dominating components; their set is
$\Tm(J)=R\cup\cD$, so $|\Tm(J)|=|R|+|\cD|$.
\end{definition}

The initial state $J_0=(\emptyset,V,V)$ has the connected dominating sets of $G$ as its solutions,
so $\OPT(J_0)=\gc(G)$ for connected~$G$. For a nonempty set $X\subseteq V$ let $\cD_X$ be the family
of vertex sets of the components of $G[X]$, and put
\begin{equation}\label{eq:cds-JX}
  J_X=\bigl(\cD_X,\;V\setminus X,\;V\setminus N[X]\bigr),
\end{equation}
and $J_\emptyset=J_0$. This is a state: the components of $G[X]$ are disjoint, connected and
pairwise non-adjacent, $V(\cD_X)=X$, and $V\setminus N[X]$ avoids $N[X]$. Unwinding the definition
shows that the solutions of $J_X$ are exactly the sets $Y\subseteq V\setminus X$ for which $X\cup Y$
is a connected dominating set (for $X\neq\emptyset$ use $N[X\cup Y]=N[X]\cup N[Y]$; for $X=\emptyset$
this is the statement about $J_0$). The number of terminals is
\begin{equation}\label{eq:cds-TX}
  |\Tm(J_X)|=n-|N[X]|+\comp(G[X])\qquad\text{for every }X\subseteq V,
\end{equation}
including $X=\emptyset$, where both sides equal $n$. Since $\comp(G[X])\le|X|\le|N[X]|$, this is an
integer in $\{0,\dots,n\}$; if $X$ is a connected dominating set, then $|\Tm(J_X)|=1$.

The Steiner encoding turns a state into an instance of the directed Steiner problem of
\cref{sec:cv-steiner}. Adjacencies among blue vertices and dominating components become pairs of opposite
arcs, and each red vertex becomes a sink that can be entered only from a blue vertex dominating it, so
that red terminals are leaves of every out-tree. Abu-Khzam, Mouawad and Liedloff~\cite[Lemma~2]{Abu-KhzamML11}
use an undirected encoding instead, in which heavy edge weights force the red terminals to be leaves, and
solve it with Nederlof's algorithm~\cite{Nederlof13}; this is the classical route, and the
directed encoding needs no heavy weights. We allow weights on the blue vertices, which
\cref{sec:cv-weighted} uses; for the unweighted problem all weights are~$1$.

\begin{definition}[Directed Steiner encoding]\label{def:cds-dir-encoding}
For a state $J=(\cD,B,R)$ of $G$ let $\vec G_J$ be the digraph with a vertex $\hat C$ for every $C\in\cD$, a
vertex $\hat b$ for every $b\in B$ and a vertex $\check r$ for every $r\in R$ (a red vertex $r$ may lie in $B$,
and then $\hat r$ and $\check r$ are different vertices), with the following arcs:
\begin{itemize}
\item $\hat b\hat b'$ and $\hat b'\hat b$ for every edge $bb'$ of $G$ with $b,b'\in B$;
\item $\hat C\hat b$ and $\hat b\hat C$ for every $C\in\cD$ and every $b\in N(C)\cap B$;
\item $\hat b\check r$ for every $r\in R$ and every $b\in N[r]\cap B$.
\end{itemize}
For weights $\varphi\colon B\to\{0,1,\dots,\varphi_{\max}\}$ let $\OPT_\varphi(J)$ be the minimum of $\varphi(Y)$
over the solutions $Y$ of $J$ ($\infty$ if there is none), and give $\hat b$ the weight $\varphi(b)$ and every
$\hat C$ and $\check r$ the weight~$0$. With $\varphi\equiv1$, $\OPT_\varphi(J)=\OPT(J)$.
\end{definition}

The members of $\cD$ are disjoint nonempty subsets of $V\setminus B$, so $|\cD|+|B|\le n$ and $\vec G_J$
has at most $2n$ vertices. The vertices of $\check R=\{\check r:r\in R\}$ have no outgoing arcs. The proofs of
the next lemma and of \cref{cor:cds-steiner} are in \cref{sec:app-steiner}; the first turns solutions of a
state into feasible sets of its encoding, and back, and the second is the first with unit weights.

\begin{restatable}[Directed Steiner encoding]{lemma}{RestLemCdsDirSteiner}\label{lem:cds-dir-steiner}
Let $J=(\cD,B,R)$ be a state of an $n$-vertex graph $G$ and $\varphi\colon B\to\{0,1,\dots,\varphi_{\max}\}$.
If $C_1\in\cD$, then $\OPT_\varphi(J)=\OPT_\varphi(\vec G_J,\hat C_1,K_1)$ for
$K_1=\{\hat C:C\in\cD\setminus\{C_1\}\}\cup\check R$. Consequently, for every state $J$ of $G$,
$\OPT_\varphi(J)$ can be computed deterministically in time $2^{|\Tm(J)|}(n+\varphi_{\max})^{O(1)}$ and
space $(n+\varphi_{\max})^{O(1)}$.
\end{restatable}

The extension algorithm used in all the applications is this computation with unit weights.

\begin{definition}[Set system and extension algorithm]\label{def:cds-ast}
Let $\Phi_{\mathrm{CDS}}$ be the implicit set system whose instances are graphs $G$, with universe
$V(G)$ and with the connected dominating sets of $G$ as solutions. Let $\cA_{\mathrm{St}}$ be the
algorithm that on input $(G,X,k)$ rejects at once if $k<|X|$, and otherwise computes $\OPT(J_X)$ and accepts if and only if
$\OPT(J_X)\le k-|X|$.
\end{definition}

\begin{restatable}[Steiner extension]{corollary}{RestCorCdsSteiner}\label{cor:cds-steiner}
For every state $J$ of an $n$-vertex graph, $\OPT(J)$ can be computed deterministically in time
$2^{|\Tm(J)|}n^{O(1)}$ and polynomial space. Consequently, the algorithm $\cA_{\mathrm{St}}$ of
\cref{def:cds-ast} solves $\Phi_{\mathrm{CDS}}$-Extension in polynomial space, and
\[
  g(G,X,k)=|\Tm(J_X)|=n-|N[X]|+\comp(G[X])
\]
is a running-time exponent for it in the sense of \cref{def:fw-cost}, with values in $\{0,\dots,n\}$.
\end{restatable}

This running-time exponent does not depend on the size bound $k$. For a fixed graph $G$ and target size $k$,
the shorthand of \cref{def:fw-cost} therefore reads $g(X)=g(G,X,k)=|\Tm(J_X)|$, and the later
sections use it in this form.

The running-time exponent depends on the sample itself, not only on its size, and adding a vertex to a
sample need not reduce it: if $x\notin X$ has $N[x]\subseteq N[X]$ and is adjacent to exactly one
component of $G[X]$, then $|\Tm(J_{X\cup\{x\}})|=|\Tm(J_X)|$. So product-form running times cannot credit a
sampled vertex with a guaranteed reduction (\cref{sec:corollaries}); the mean number of terminals,
however, does decrease with the sample size, as we show next.

\subsection{Sampling inside a connected dominating set}\label{sec:cds-sampling}

The main result of this subsection is the mean bound of \cref{lem:cds-C}. It follows from a
pointwise inequality for trees, applied to a spanning tree of $G$ in which the vertices outside the
solution are leaves.

For a tree $F$ and a set $X\subseteq V(F)$, let $F\{X\}$ denote the subgraph of $F$ with vertex set
$N_F[X]$ and the edges of $F$ with at least one end in $X$. It records
how the components of $F[X]$ are joined to the neighbors of $X$; the inequality below is an
equality exactly when it is connected.

\begin{lemma}[Tree inequality]\label{lem:cds-tree}
Let $F$ be a tree and let $X\subseteq V(F)$ be nonempty. Then
\[
  |N_F[X]|-\comp(F[X])\ \ge\ 1+\sum_{x\in X}\bigl(d_F(x)-1\bigr),
\]
with equality if and only if $F\{X\}$ is connected.
\end{lemma}
\begin{proof}
Let $e_X$ be the number of edges of $F[X]$ and $e_{XY}$ the number of edges of $F$ with exactly one
end in $X$; the other end of such an edge lies in $Y=N_F(X)$. Then $\sum_{x\in X}d_F(x)=2e_X+e_{XY}$,
and $\comp(F[X])=|X|-e_X$ because $F[X]$ is a forest. The graph $F\{X\}$ is a forest with
$|X|+|Y|$ vertices and $e_X+e_{XY}$ edges, so $e_X+e_{XY}\le|X|+|Y|-1$, with equality if and only if
$F\{X\}$ is connected. Therefore
\[
  |N_F[X]|-\comp(F[X])=|Y|+e_X\ \ge\ 2e_X+e_{XY}+1-|X|=1+\sum_{x\in X}\bigl(d_F(x)-1\bigr),
\]
with equality exactly when $e_X+e_{XY}=|X|+|Y|-1$.
\end{proof}

\begin{lemma}[Pointwise form]\label{lem:cds-pointwise}
Let $G$ be a connected graph with $n\ge2$ vertices and let $S$ be a connected dominating set of $G$.
Let $F$ be a spanning tree of $G$ in which every vertex of $V\setminus S$ is a leaf; one exists, by
the construction in the proof of \cref{lem:cds-maxleaf}, which needs only $n\ge2$. Put $w(u)=d_F(u)-1$ for $u\in S$. Then
$\sum_{u\in S}w(u)=n-2$, and for every nonempty $X\subseteq S$,
\[
  |N[X]|-\comp(G[X])\ \ge\ 1+\sum_{x\in X}w(x).
\]
\end{lemma}
\begin{proof}
The tree $F$ has $n-1$ edges and the $n-|S|$ vertices outside $S$ have degree $1$ in $F$, so
$\sum_{u\in S}d_F(u)=2(n-1)-(n-|S|)$ and $\sum_{u\in S}w(u)=n-2$. For nonempty $X\subseteq S$ we have
$N[X]\supseteq N_F[X]$ because $F\subseteq G$, and $\comp(G[X])\le\comp(F[X])$ because $F[X]$ is a
spanning subgraph of $G[X]$. Now \cref{lem:cds-tree}, applied to $F$ and $X$, gives the inequality.
\end{proof}

Averaging the pointwise form over a uniform $t$-subset of the solution gives the mean bound.

\begin{lemma}[Mean number of terminals]\label{lem:cds-C}
Let $G$ be a connected graph with $n\ge2$ vertices, let $S$ be a connected dominating set of $G$ with
$|S|=s$, let $1\le t\le s$ and let $X$ be uniform in $\binom St$. Then
\[
  \E\bigl[|N[X]|-\comp(G[X])\bigr]\ \ge\ 1+\frac{t(n-2)}s,
  \qquad\text{equivalently}\qquad \E\,|\Tm(J_X)|\ \le\ n-1-\frac{t(n-2)}s .
\]
\end{lemma}
\begin{proof}
Take $F$ and $w$ as in \cref{lem:cds-pointwise}. As $t\ge1$, the set $X$ is nonempty, so
\cref{lem:cds-pointwise} applies to it. Each $u\in S$ lies in $X$ with probability $t/s$, so by
linearity of expectation
\[
  \E\bigl[|N[X]|-\comp(G[X])\bigr]\ \ge\ 1+\frac ts\sum_{u\in S}w(u)=1+\frac{t(n-2)}s .
\]
The second form follows from~\eqref{eq:cds-TX}.
\end{proof}

The assumption $t\ge1$ of \cref{lem:cds-C} is needed: for $X=\emptyset$ the expectation
$\E[|N[X]|-\comp(G[X])]$ is $0$. The bound of \cref{lem:cds-C} cannot be improved in general:

\begin{proposition}[Tightness for stars]\label{prop:cds-tight}
Let $S$ be a connected dominating set of a connected graph $G$ with $|S|=s\ge2$. Suppose that $G[S]$
is a star $K_{1,s-1}$ and that every vertex of $V\setminus S$ has exactly one neighbor in $S$ (edges
between vertices of $V\setminus S$ are allowed). Then equality holds in \cref{lem:cds-pointwise} for
every nonempty $X\subseteq S$, and in \cref{lem:cds-C} for every $1\le t\le s$.
\end{proposition}

An example for \cref{prop:cds-tight} is the \emph{star of stars}: a star on $S$ in which every
vertex of $S$ carries $h$ pendant vertices, so that $n=s(h+1)$.
\begin{proof}
First, the tree $F$ of \cref{lem:cds-pointwise} is unique. Deleting the leaves of $F$ that lie outside
$S$ leaves a spanning tree of $G[S]$, which must be the star $G[S]$ itself. If $V\setminus S$ is
nonempty, then every $v\notin S$ is a leaf of $F$, and its neighbor in $F$ lies in $S$: otherwise that
neighbor is a leaf as well and $F$ would be a single edge, whereas $|V(F)|\ge s+1\ge3$. So $v$ is
attached to its unique neighbor in~$S$.

For $X\subseteq S$ we have $G[X]=F[X]$, and $N[X]=N_F[X]$ because the only edges of $G$ between $S$
and $V\setminus S$ are edges of $F$. So equality in \cref{lem:cds-pointwise} is equality in
\cref{lem:cds-tree}, and it remains to show that $F\{X\}$ is connected for nonempty
$X\subseteq S$. Let $z$ be the center of the star. If $z\in X$, then $F[X]$ is connected, as every
other vertex of $X$ is adjacent to $z$, and every vertex of $N_F(X)$ is joined to $X$ by an edge of
$F\{X\}$. If $z\notin X$, then the vertices of $X$ are all adjacent to $z\in N_F(X)$, and every
other vertex of $N_F(X)$ lies outside $S$ and is adjacent to one of them. In both cases $F\{X\}$ is
connected. Equality in \cref{lem:cds-C} follows,
because its proof uses only \cref{lem:cds-pointwise} and linearity of expectation.
\end{proof}

{}
\section{Warm-up: sampling a connected dominating set with the mean running-time exponent}\label{sec:rs}

This section combines the linear mean savings of \cref{cor:co-linear} with the Steiner extension
(\cref{cor:cds-steiner}) and the mean bound of \cref{lem:cds-C}. The resulting Algorithm RS is
Algorithm \MLS{} for connected dominating sets, run with a constant rate: it samples vertices of an
unknown connected dominating set and completes the sample by a Steiner tree computation. We prove
that it runs in randomized time $O(1.7449^n)$ and polynomial space. At each solution fraction, the
certified rate has a closed form (\cref{prop:rs-rate}), and at solution fraction
$\frac12$ the corresponding base is exactly $\sqrt3$. RS shows the mechanism in its simplest form:
its analysis uses, through Jensen's inequality, only the mean number of terminals of the samples
inside a solution. \Cref{sec:rstheta} runs the same algorithm with a smaller rate and bounds the
exponential moments directly on part of the range of solution sizes.

Throughout, $\sigma=s/n$ denotes the solution fraction of a candidate size $s$. The
\emph{certified rate} at $\sigma$ is the infimum of the rates $\psi$ (\cref{sec:aprime}) for which our
analysis gives condition~\eqref{eq:fw-star} for every connected dominating set of size $s=\sigma n$.
Algorithm RS uses one rate, $\psi_{\mathrm{RS}}$, above the maximum of the certified rate over
$\sigma$. The certified rate is governed by the function
\begin{equation}\label{eq:rs-F}
  F^*(\sigma)=H(\sigma)+(1-\sigma)\log\bigl(1-2^{-1/\sigma}\bigr)\qquad(0<\sigma<1),
\end{equation}
which is the function $F^*$ of~\eqref{eq:co-Fstar} with $\beta=1$ and $\rho=1/\sigma$: since $2^x>x$
for every real $x$, we have $2^{1/\sigma}>1/\sigma$, that is $p=2^{-1/\sigma}<\sigma$, so the second
case of~\eqref{eq:co-Fstar} applies, and $\sigma\rho=1$ turns its last expression into~\eqref{eq:rs-F}.
The corresponding base is
\[
  2^{F^*(\sigma)}=\frac{\bigl(1-2^{-1/\sigma}\bigr)^{1-\sigma}}{\sigma^{\sigma}(1-\sigma)^{1-\sigma}} ,
\]
whose denominator is $2^{-H(\sigma)}$, where $2^{nH(\sigma)}$ is, up to a polynomial factor, the number of
candidate sets of size $\sigma n$, and whose numerator is the saving. We put $F^*(1)=0$, the value of the right-hand side of~\eqref{eq:rs-F} at $\sigma=1$ and its limit as $\sigma\to1$. By \cref{rem:co-fsigma}, for $\sigma<1$, $F^*(\sigma)$ is the minimum over the
sample fraction $\tau\in[0,\sigma]$ of
\[
  f_\sigma(\tau)=H(\tau)-\sigma H(\tau/\sigma)+1-\frac\tau\sigma ,
\]
whose first part is, up to an additive $O(\log n)/n$ by~\eqref{eq:fw-binom}, the loss
$\frac1n\log\left(\binom nt/\binom st\right)$ of requiring a uniform $t$-subset of $V$ to lie inside a
fixed connected dominating set of size $s$, and whose second part, $1-\tau/\sigma$, is the mean bound
of \cref{lem:cds-C} on the number of terminals, divided by $n$. A single sample size and the Markov
cutoff of \cref{rem:co-markov} attain this exponent up to the factors $2^{1/\sigma}$ and $s+1$ (\cref{rem:co-fsigma}); Algorithm \MLS{}
attains it without rounding.

\paragraph{How RS decides.} For each candidate size $s$, RS runs Algorithm \MLS{} with a fixed rate.
For each sample $X$ it monitors the number of terminals $|\Tm(J_X)|=n-|N[X]|+\comp(G[X])$, which is
computed in polynomial time, and compares it with the current threshold $\theta$ (the thresholds of Algorithm \MLS{},
\cref{sec:fw-main}). A sample at or
below the threshold is completed by the Steiner tree algorithm, in time at most $2^{\theta}n^{O(1)}$;
a sample above it is discarded in polynomial time.

The next proposition evaluates the certified rate at a given solution fraction: it gives the value at
$\sigma=\frac12$ and an upper bound on the maximum over~$\sigma$, together with two bounds
on ranges of $\sigma$ that \cref{sec:rstheta} uses.

\begin{proposition}[Certified rate of RS]\label{prop:rs-rate}
  Let $F^*$ be the function of~\eqref{eq:rs-F}.
  \begin{enumerate}[label=(\alph*)]
    \item For every $\sigma\in(0,1)$, $F^*(\sigma)<H(\sigma)\le1$, and the minimum of $f_\sigma$ over
      $[0,\sigma]$ is attained exactly at $\tau^*(\sigma)=\dfrac{\sigma-2^{-1/\sigma}}{1-2^{-1/\sigma}}\in(0,\sigma)$.
    \item At $\sigma=\frac12$: $\tau^*=\frac13$ and $F^*(\frac12)=\frac12\log3$, so the base
      $2^{F^*(1/2)}$ is exactly $\sqrt3$.
    \item $F^*(\sigma)\le\log1.744835$ for every $\sigma\in(0,1]$. Moreover
      $F^*(\sigma)\le0.7807$ for $\sigma\in(0,0.31]$ and $F^*(\sigma)\le0.7802$ for $\sigma\in[0.54,1]$.
  \end{enumerate}
\end{proposition}
\begin{proof}
  (a) The first claim holds because $\log(1-2^{-1/\sigma})<0$ and $H\le1$. The second is \cref{rem:co-fsigma} with
  $p=2^{-1/\sigma}<\sigma$.

  (b) Here $2^{-1/\sigma}=\frac14$ and $\tau^*=\frac{1/2-1/4}{3/4}=\frac13$, and
  $2^{F^*(1/2)}=(\frac34)^{1/2}/\frac12=\sqrt3$.

  (c) The certificate of \cref{sec:anc-rs} gives $F^*(\sigma)\le0.8030906143<\log1.744835$ for every
  $\sigma\in[0.2,0.8]$, $F^*\le0.7807$ on $[0.15,0.31]$ and $F^*\le0.7802$ on $[0.54,0.8]$. For
  $\sigma\in(0,1)\setminus[0.2,0.8]$, (a) gives
  $F^*(\sigma)\le H(\sigma)\le H(0.2)=\log5-1.6<0.73$, because $H$ is increasing on $[0,\frac12]$ and
  symmetric about $\frac12$, and $5^{100}<2^{233}$; and $F^*(1)=0$. As $0.73<0.7802$, this bound covers
  the rest of the two ranges.
\end{proof}

The closed form bounds
the exponential moments of \cref{def:fw-moments} for every connected dominating set; the corollary
is used again in \cref{sec:rstheta}.

\begin{corollary}[Closed-form moment bound]\label{cor:rs-moments}
  Let $G$ be a connected graph with $n$ vertices and let $S$ be a connected dominating set of $G$ with
  $|S|=s$ and $\sigma=s/n$. Then, for the running-time exponent $g(X)=|\Tm(J_X)|$ of \cref{cor:cds-steiner},
  \[
    \max_{0\le t\le s}\frac{\mu_t(S)}{\binom nt}\ \ge\ \frac{2^{-nF^*(\sigma)}}2 .
  \]
\end{corollary}
\begin{proof}
  If $s=n$, then $S=V$, $\mu_n(S)=2^{-|\Tm(J_V)|}=\frac12$ by~\eqref{eq:cds-TX}, and $F^*(1)=0$.
  Otherwise $n\ge2$ and $\sigma\in(0,1)$. For $1\le t\le s$, \cref{lem:cds-C} gives
  \[
    \E_{X\in\binom St}|\Tm(J_X)|\le n-1-\frac{t(n-2)}s=n-\frac t\sigma+\frac{2t}s-1\le n-\frac t\sigma+1,
  \]
  as $t\le s$, and for $t=0$ we have $|\Tm(J_\emptyset)|=n\le n+1$. So \eqref{eq:co-linear-hyp} holds
  with $\beta=1$, $\rho=1/\sigma$ and $C=1$ for every connected dominating set of size $s$, and
  \cref{cor:co-linear}(1) gives $G(S)\le2^{1+nF^*(\sigma)}$ with $F^*$ as in~\eqref{eq:rs-F}. The first
  inequality of \cref{lem:fw-gf} gives $\min_{0\le t\le s}\binom nt/\mu_t(S)=M(S)\le G(S)$.
\end{proof}

We can now state the algorithm.

\begin{algorithm}[RS]\label{alg:rs-algorithm}
  \Input a connected graph $G$ with $n$ vertices.
  \Parameters $\psi_{\mathrm{RS}}=0.80309062$ and a constant $\kappa>0$.
  \Body
  For $s=1,2,\dots,n$ in turn, run Algorithm \MLS$(\psi_{\mathrm{RS}},2,\kappa)$
  (\cref{alg:fw-mls}) on $(G,s)$ for the implicit set system $\Phi_{\mathrm{CDS}}$, the extension
  algorithm $\cA_{\mathrm{St}}$ and the running-time exponent $g$ of \cref{cor:cds-steiner}, written
  $g(X)=g(G,X,s)=|\Tm(J_X)|$ as declared after that corollary. Output the first $s$ for which the run
  answers \YES{}, and output $n$ if there is none.
\end{algorithm}

\begin{theorem}[Algorithm RS]\label{thm:rs}
  For every constant $\kappa>0$, Algorithm RS runs in time $O(1.7449^n)$ and polynomial space. It
  never outputs a value below $\gc(G)$, and it outputs $\gc(G)$ with probability at least
  $1-e^{-\kappa}$. Consequently \pname{Connected Dominating Set} and
  \pname{Max Leaf Spanning Tree} (by \cref{lem:cds-maxleaf}, and directly when $n\le2$) can be solved by a randomized algorithm with one-sided error
  (one-sided Monte Carlo) in time $O(1.7449^n)$ and polynomial space.
\end{theorem}

\begin{proof}
  We verify the hypotheses of \cref{lem:cds-first-accepted} for the run on $(G,s)$. By
  \cref{cor:cds-steiner}, $\cA_{\mathrm{St}}$ solves $\Phi_{\mathrm{CDS}}$-Extension in polynomial space
  with the running-time exponent $g$, and the parameters $\psi_{\mathrm{RS}}$, $\Gamma=2$ and $\kappa$ are
  constants. So \cref{thm:fw-main} applies: the run takes time $2^{\psi_{\mathrm{RS}}n}n^{O(1)}$ and
  polynomial space, and it answers \NO{} if $\gc(G)>s$.

  \emph{Completeness.} Let $s=\gc(G)$ and let $S$ be a connected dominating set with $|S|=s$. By
  \cref{cor:rs-moments} and \cref{prop:rs-rate}(c),
  \[
    \max_{0\le t\le s}\frac{\mu_t(S)}{\binom nt}\ \ge\ \frac{2^{-n\log1.744835}}2\ \ge\ \frac{2^{-\psi_{\mathrm{RS}}n}}2,
  \]
  as $\psi_{\mathrm{RS}}>\log1.744835$; that is, $S$ satisfies~\eqref{eq:fw-star} with $\Gamma=2$. By
  \cref{thm:fw-main}(iii), the run on $(G,\gc(G))$ answers \YES{} with probability at least
  $1-e^{-\kappa}$.

  \Cref{lem:cds-first-accepted} now gives the statements about the output, and the total time is
  $n\cdot2^{\psi_{\mathrm{RS}}n}n^{O(1)}=O(1.7449^n)$, because
  $2^{\psi_{\mathrm{RS}}}=1.744835\ldots<1.7449$, where the constant $1.744835$ comes from
  \cref{prop:rs-rate}(c).
\end{proof}
{}
\section{Exponential moments of the Steiner extension: Algorithm RS-\texorpdfstring{$\theta$}{theta}}\label{sec:rstheta}

The analysis of Algorithm RS in \cref{sec:rs} uses, through Jensen's inequality, only the mean
number of terminals of the samples inside a solution. By \cref{prop:cds-tight} the bound of
\cref{lem:cds-C} on this mean cannot be improved, but how it is used can: when the number of
terminals fluctuates, Jensen's inequality $\E[2^{-|\Tm(J_X)|}]\ge2^{-\E|\Tm(J_X)|}$ is strict, and
the samples with few terminals account for the difference. Algorithm RS-$\theta$ is Algorithm RS run
with a smaller rate: it is Algorithm \MLS{} (\cref{alg:fw-mls}) for connected dominating sets with
the Steiner extension of \cref{cor:cds-steiner}, and its success condition in \cref{thm:fw-main} is
condition~\eqref{eq:fw-star} on the exponential moments $\mu_t(S)$ of the number of terminals, which
we now bound directly. We prove that it runs in randomized time $O(1.7204^n)$ and polynomial space
(\cref{thm:rstheta}).

The analysis needs a lower bound on these moments for \emph{every} connected dominating set. Since
$\E[2^{-|\Tm(J_X)|}]$ is not linear in $|\Tm(J_X)|$, the linearity-of-expectation argument behind
\cref{lem:cds-C} does not apply. We bound the number of terminals by a potential that is local along
a spanning tree of the solution (\cref{lem:rst-E}), evaluate its generating function by a
three-state recursion (\cref{lem:rst-treerec}), and bound the recursion from below, uniformly over
all trees, by an antinorm certificate (\cref{lem:rst-F}), following Rote~\cite{Rote19} and Rosenfeld~\cite{Rosenfeld21}.
The proofs of these three steps are in \cref{sec:app-rst}.
Condition~\eqref{eq:fw-star} is then verified at every solution fraction by two certificates or by
the closed form of \cref{sec:rs} (\cref{tab:rst-coverage}). On the uniform
$(2,2)$-lobsters the analysis is nearly tight for this algorithm (\cref{prop:rst-caterpillar}).

For a connected graph $G$ and the running-time exponent $g$ of \cref{cor:cds-steiner}, written
$g(X)=g(G,X,s)=|\Tm(J_X)|$ in the shorthand of \cref{def:fw-cost} with target size $k=s$, the
quantities of \cref{def:fw-moments} for a connected dominating set $S$ with $|S|=s$ are
\begin{equation}\label{eq:rst-mu}
  \mu_t(S)=\sum_{X\in\binom St}2^{-|\Tm(J_X)|}\quad(0\le t\le s),\qquad
  W_S(y)=\sum_{X\subseteq S}y^{|X|}\,2^{-|\Tm(J_X)|}\quad(y\ge0);
\end{equation}
the running-time exponent does not depend on the size bound $k$. Recall from~\eqref{eq:cds-TX} that
$|\Tm(J_X)|$ is an integer in $\{0,\dots,n\}$, that $|\Tm(J_\emptyset)|=n$ and that $|\Tm(J_S)|=1$.

\subsection{The algorithm and the result}\label{sec:rst-algorithm}

\begin{algorithm}[RS-$\theta$]\label{alg:rst-algorithm}
\Input a connected graph $G$ with $n$ vertices.
\Parameters $\psi^*=0.7827016$ and a constant $\kappa>0$.
\Body
For $s=1,2,\dots,n$ in turn, run Algorithm \MLS$(\psi^*,2,\kappa)$ on $(G,s)$ for the
implicit set system $\Phi_{\mathrm{CDS}}$, the extension algorithm $\cA_{\mathrm{St}}$ and the running-time
exponent $g(X)=|\Tm(J_X)|$ of \cref{cor:cds-steiner}, as in Algorithm RS. Output the first accepted
$s$, or $n$ if no $s$ is accepted.
\end{algorithm}

So RS-$\theta$ differs from RS only in its rate, and the improvement lies in the analysis. The analysis
of RS bounds the moments $\mu_t(S)$ through the mean of \cref{lem:cds-C}, whereas the analysis below
chooses, for each minimum connected dominating set $S$, a pair $(t,\theta)$ at which the number of
samples $X\subseteq S$ of size $t$ with $|\Tm(J_X)|\le\theta$ is, up to a polynomial factor, at
least $2^\theta\mu_t(S)$ (\cref{lem:fw-threshold-prob}). For instance, the pair $t=s$, $\theta=1$
replaces enumeration of the $s$-sets: each of its trials draws a uniform $s$-set $X$, and the
threshold test passes exactly when $X$ is a connected dominating set. The main result of this section
is the following.

\begin{theorem}[Algorithm RS-$\theta$]\label{thm:rstheta}
For every constant $\kappa>0$, Algorithm RS-$\theta$ runs in time $\Ostar(2^{\psi^*n})$, where
$2^{\psi^*}=1.7203\ldots<1.7204$, and in polynomial space. It never outputs a value below
$\gc(G)$, and it outputs $\gc(G)$ with probability at least $1-e^{-\kappa}$. Consequently
\pname{Connected Dominating Set}, and \pname{Max Leaf Spanning Tree} (by \cref{lem:cds-maxleaf}, and directly when $n\le2$), can be
solved by a randomized algorithm with one-sided error (one-sided Monte Carlo) in time
$O(1.7204^n)$ and polynomial space.
\end{theorem}

The proof is given at the end of \cref{sec:rst-coverage}; it rests on \cref{prop:rst-rate}, which
verifies condition~\eqref{eq:fw-star} for every connected dominating set. The constant $1.7204$
rests on \cref{prop:rst-certs} and \cref{prop:rs-rate}(c).

\subsection{Reduction to a tree with an owner map}\label{sec:rst-tree}

The first step replaces the graph by a tree. Let $S$ be a connected dominating set of $G$. An
\emph{owner map} is a map $o:V\setminus S\to S$ with $o(u)\in N(u)$ for every $u\notin S$; it exists
because $S$ is dominating. For $v\in S$ let $\nu_v=|o^{-1}(v)|$ be the number of vertices that $v$
owns, so that $\sum_{v\in S}\nu_v=n-|S|$. For a tree $F$ with $V(F)=S$, a map
$\nu:S\to\mathbb Z_{\ge0}$ and $X\subseteq S$ put
\begin{equation}\label{eq:rst-Phi}
\begin{gathered}
  \Pi_{F,\nu}(X)=\sum_{v\in S\setminus X}\nu_v\;+\;\bigl|\{v\in S:N_F[v]\cap X=\emptyset\}\bigr|
  \;+\;\comp(F[X]),\\
  W_{F,\nu}(y)=\sum_{X\subseteq S}y^{|X|}\,2^{-\Pi_{F,\nu}(X)} .
\end{gathered}
\end{equation}
The three terms count, in order: the owned vertices whose owner is not sampled, the solution vertices
not dominated by a sample within the tree, and the components of the sample within the tree. We call
$\Pi_{F,\nu}$ the \emph{potential}; the next lemma shows that it is at least the running-time exponent
$|\Tm(J_X)|$ of every sample $X\subseteq S$.

\begin{restatable}[Tree reduction]{lemma}{RestLemRstE}\label{lem:rst-E}
Let $G$ be a connected graph, let $S$ be a connected dominating set of $G$, let $F$ be a spanning tree
of $G[S]$ and let $o$ be an owner map with counts $\nu$. Then for every $X\subseteq S$,
\[
  |\Tm(J_X)|\ \le\ \Pi_{F,\nu}(X).
\]
Equality holds for every $X\subseteq S$ if $G[S]=F$ and every vertex of $V\setminus S$ has exactly one
neighbor in $S$. Consequently $W_S(y)\ge W_{F,\nu}(y)$ for every $y\ge0$.
\end{restatable}

\subsection{A three-state recursion}\label{sec:rst-recursion}

The second step computes $W_{F,\nu}(y)$ by dynamic programming over the tree. Root $F$ at a vertex
$r$; for $v\in V(F)$ let $F_v$ be the subtree formed by $v$ and its descendants. A \emph{message} is a
vector $P=(P_1,P_2,P_3)\in\R^3$ whose coordinates are sums of weights $y^{|X|}2^{-\Pi^{\mathrm{in}}(X)}$, where
$\Pi^{\mathrm{in}}(X)$ is the part of the potential of $X$ that is determined inside the subtree (made precise
in the proof of \cref{lem:rst-treerec}); the sums run over the subsets $X$ of that subtree in each of three
\emph{states} of its top vertex $v$: in state~$1$, $v\in X$; in state~$2$,
$v\notin X$ and some child of $v$ lies in $X$; in state~$3$, $v\notin X$ and no child of $v$ lies in
$X$. For $P,Q\in\R^3$ let
\begin{equation}\label{eq:rst-merge}
  P\circ Q=\Bigl(P_1(2Q_1+Q_2+Q_3),\ \ P_2\bigl(Q_1+Q_2+\tfrac{Q_3}2\bigr)+P_3Q_1,\ \
  P_3\bigl(Q_2+\tfrac{Q_3}2\bigr)\Bigr),
\end{equation}
and let $\mathrm{fin}(P)=P_1+P_2+\tfrac12P_3$ and $\mathrm{Init}_y(j)=\bigl(\tfrac y2,0,2^{-j}\bigr)$
for integers $j\ge0$; the argument $j$ of $\mathrm{Init}_y$ will be the number $\nu_v$ of vertices
owned by $v$. The operation $\circ$ is bilinear, and it adds a completed child with message $Q$ to a
partially built message $P$. For $y\ge0$, order the children of every vertex arbitrarily, and for
$v\in V(F)$ with children $w_1,\dots,w_d$ in this order define the messages from the leaves upwards by
\begin{equation}\label{eq:rst-msg}
  \mathrm{msg}_v=\Bigl(\cdots\bigl(\mathrm{Init}_y(\nu_v)\circ\mathrm{msg}_{w_1}\bigr)\circ\cdots\Bigr)\circ\mathrm{msg}_{w_d}.
\end{equation}
The output of this recursion is as follows.

\begin{restatable}[Tree recursion]{lemma}{RestLemRstTreerec}\label{lem:rst-treerec}
Let $y\ge0$, and let the messages $\mathrm{msg}_v$ be defined by~\eqref{eq:rst-msg} for any order of the
children. Then every $\mathrm{msg}_v$ has nonnegative entries and $W_{F,\nu}(y)=\mathrm{fin}(\mathrm{msg}_r)$.
\end{restatable}

\subsection{Antinorm certificates in the cumulative cone}\label{sec:rst-antinorm}

The third step bounds $\mathrm{fin}(\mathrm{msg}_r)$ from below for all trees at once. The idea is to find a
convex region that the merge operation maps into itself up to a known scale; the scale then
multiplies along the tree. The region must be closed upwards in an order that the recursion respects.
Let
\[
  \mathcal K_{\mathrm{cum}}=\bigl\{P\in\R^3:\ P_1\ge0,\ P_1+P_2\ge0,\ P_1+P_2+P_3\ge0\bigr\},
\]
the \emph{cumulative cone}. It is the preimage of $\R^3_{\ge0}$ under the cumulative map
$P\mapsto(P_1,P_1+P_2,P_1+P_2+P_3)$, so it is a convex cone generated by $(1,-1,0)$, $(0,1,-1)$ and
$(0,0,1)$, and it contains $\R^3_{\ge0}$. For a finite nonempty set $Z\subseteq\R^3_{\ge0}$ let
$\mathcal P_Z=\mathrm{conv}(Z)+\mathcal K_{\mathrm{cum}}$ and $c_Z=\min_{z\in Z}\mathrm{fin}(z)$. The facts about
$\mathcal K_{\mathrm{cum}}$ that the certificate argument uses (\cref{lem:rst-closure}) and the abstract
mechanism behind it (\cref{lem:rst-abstract}), which \cref{sec:rsc} reuses with another cone, are proved in
\cref{sec:app-rst}.

The next lemma is the core of the third step. If a finite set $Z$ of nonnegative messages is such that
$\mathcal P_Z$ absorbs the merge of any two of its generators and contains every initial message up to
a geometric scale, then that scale multiplies along any tree. Data $(y,e_1,e_2,Z)$ with these two
properties are the \emph{certificates} referred to above.

\begin{restatable}[Antinorm certificate]{lemma}{RestLemRstF}\label{lem:rst-F}
Let $y\ge0$ and $e_1,e_2>0$, and let $Z\subseteq\R^3_{\ge0}$ be finite and nonempty with
\begin{enumerate}[label=(\alph*)]
\item $z\circ z'\in\mathcal P_Z$ for all $z,z'\in Z$, and
\item $\mathrm{Init}_y(j)\in e_1e_2^{\,j}\,\mathcal P_Z$ for every integer $j\ge0$.
\end{enumerate}
Then for every tree $F$ with $s$ vertices and every $\nu:V(F)\to\mathbb Z_{\ge0}$, writing
$\nu(F)=\sum_{v\in V(F)}\nu_v$, we have $W_{F,\nu}(y)\ge c_Z\,e_1^{\,s}e_2^{\,\nu(F)}$.
If $e_2\le1$, then (b) holds for all $j$ as soon as it holds for $j\le j_0$ and
$(y/2,0,0)\in e_1e_2^{\,j_0+1}\mathcal P_Z$.
\end{restatable}

Together with \cref{lem:rst-E}, a certificate turns into the exponential-moment condition of the main
theorem.

\begin{restatable}[From certificates to exponential moments]{corollary}{RestCorRstF}\label{cor:rst-F}
Suppose that $(y,e_1,e_2,Z)$ satisfies the hypotheses of \cref{lem:rst-F}. Then for every connected
graph $G$ and every connected dominating set $S$ of $G$ with $|S|=s$, writing $\sigma=s/n$,
\[
  W_S(y)\ge c_Z\,e_1^{\,s}e_2^{\,n-s}
  \qquad\text{and}\qquad
  \max_{0\le t\le s}\frac{\mu_t(S)}{\binom nt}\ \ge\ c_Z\,2^{-n\,\psi_{\mathrm{cert}}(\sigma)},
\]
where $\psi_{\mathrm{cert}}(\sigma)=\log(1+y)-\sigma\log e_1-(1-\sigma)\log e_2$.
\end{restatable}

So a certificate gives condition~\eqref{eq:fw-star} of \cref{thm:fw-main} for $S$ with the rate
$\psi=\psi_{\mathrm{cert}}(\sigma)$ and the loss factor $\Gamma=\max(1,1/c_Z)$; we call $\psi_{\mathrm{cert}}$ the \emph{rate function}
of the certificate.

\subsection{The certificates and the coverage of the solution fraction}\label{sec:rst-coverage}

Two certificates for \cref{lem:rst-F} cover the middle range of the solution fraction; the closed
form of \cref{sec:rs} covers the rest. We record the certificates first and then combine the two
arguments.

\begin{proposition}[Antinorm certificates]\label{prop:rst-certs}
For $i\in\{0,1\}$, the data $(y_i,e_{1,i},e_{2,i},Z_i)$ of certificate~$i$ (\cref{sec:anc-rstheta})
are rational and satisfy the hypotheses of \cref{lem:rst-F}, with
$j_0=30$ and $e_{2,i}<1$:
\begin{center}
\setlength{\tabcolsep}{4pt}
\begin{tabular}{@{}lllllll@{}}
\toprule
$i$ & $y_i$ & $e_{1,i}$ & $e_{2,i}$ & $|Z_i|$ & $c_{Z_i}\ge$ & $\psi_i(\sigma)$ \\
\midrule
$0$ & $\frac{69}{200}$ & $\frac{747593601}{960952792}$ & $\frac{512099210}{652701491}$ & $27$ & $0.6212$ & $0.7776061742+0.0122113367\,\sigma$ \\[3pt]
$1$ & $\frac{17}{50}$ & $\frac{248231700}{312721579}$ & $\frac{568097827}{739289481}$ & $29$ & $0.6293$ & $0.8022330021-0.0468085913\,\sigma$ \\
\bottomrule
\end{tabular}
\end{center}
Here $\psi_i$ is the rate function $\psi_{\mathrm{cert}}$ of \cref{cor:rst-F}, with coefficients rounded to ten decimals. The
two lines cross at $\sigma_\times=0.4172629270\ldots$, and
\[
  \max\Bigl(\max_{\sigma\in[0.31,\,0.4172629]}\psi_0(\sigma),\ \max_{\sigma\in[0.4172629,\,0.54]}\psi_1(\sigma)\Bigr)
  =0.78270151357\ldots<\psi^* .
\]
\end{proposition}

\begin{proof}
The hypotheses of \cref{lem:rst-F} are finitely many conditions on the rational data; they are checked
for both certificates in \cref{sec:anc-rstheta}. Since $\psi_i$ is linear in $\sigma$, its maximum on an
interval is attained at an endpoint, so the displayed maximum is the largest of the values of $\psi_0$ at
$0.31$ and $0.4172629$ and of $\psi_1$ at $0.4172629$ and $0.54$; these four values are bounded from above in
\cref{sec:anc-rstheta}, and each is below $0.78270152<\psi^*$.
\end{proof}

\begin{table}[t]
\centering
\begin{tabular}{@{}lll@{}}
\toprule
range of $\sigma=s/n$ & argument & certified exponent \\
\midrule
$(0,0.31]$ & closed form (\cref{cor:rs-moments}, \cref{prop:rs-rate}(c)) & $F^*(\sigma)\le0.7807$ \\
$[0.31,0.4172629]$ & certificate $0$ & $\psi_0(\sigma)\le0.78270152$ \\
$[0.4172629,0.54]$ & certificate $1$ & $\psi_1(\sigma)\le0.78270152$ \\
$[0.54,1]$ & closed form (\cref{cor:rs-moments}, \cref{prop:rs-rate}(c)) & $F^*(\sigma)\le0.7802$ \\
\bottomrule
\end{tabular}
\caption{Coverage of the solution fraction $\sigma$ in \cref{prop:rst-rate}. The maximum,
$0.7827015136$ (base $2^{0.7827015136}=1.7203\ldots$), is attained at $\sigma=0.4172629$.}
\label{tab:rst-coverage}
\end{table}

The cases of \cref{tab:rst-coverage} combine into condition~\eqref{eq:fw-star} for every
connected dominating set.

\begin{proposition}[Rate of RS-$\theta$]\label{prop:rst-rate}
For every connected graph $G$ and every connected dominating set $S$ of $G$,
\[
  \max_{0\le t\le|S|}\frac{\mu_t(S)}{\binom nt}\ \ge\ \frac{2^{-\psi^*n}}{2} .
\]
\end{proposition}
\begin{proof}
Let $s=|S|$ and $\sigma=s/n$; we follow \cref{tab:rst-coverage}.

\emph{Closed form: $\sigma\notin[0.31,0.54]$.} By \cref{cor:rs-moments} and \cref{prop:rs-rate}(c),
$\max_t\mu_t(S)/\binom nt\ge2^{-nF^*(\sigma)}/2\ge2^{-0.7807n}/2\ge2^{-\psi^*n}/2$, as
$0.7807<\psi^*$.

\emph{Certificates: $\sigma\in[0.31,0.54]$.} Take $i=0$ if $\sigma\le0.4172629$ and $i=1$ otherwise.
\Cref{cor:rst-F,prop:rst-certs} give
$\max_t\mu_t(S)/\binom nt\ge c_{Z_i}2^{-n\psi_i(\sigma)}\ge0.62\cdot2^{-\psi^*n}\ge2^{-\psi^*n}/2$.
\end{proof}

\begin{proof}[Proof of \cref{thm:rstheta}]
We verify the hypotheses of \cref{lem:cds-first-accepted}. For each $s$, RS-$\theta$ is Algorithm
\MLS{} for $\Phi_{\mathrm{CDS}}$, $\cA_{\mathrm{St}}$ and $g(X)=|\Tm(J_X)|$, with $k=s$,
$\psi=\psi^*$, $\Gamma=2$ and confidence $\kappa$. By \cref{cor:cds-steiner}, $\cA_{\mathrm{St}}$
solves $\Phi_{\mathrm{CDS}}$-Extension in polynomial space and $g$ is a running-time exponent for it, and
the parameters are constants. So \cref{thm:fw-main} applies: the run for $s$ takes time
$2^{\psi^*n}n^{O(1)}$ and polynomial space, and it rejects if $\gc(G)>s$. For $s=\gc(G)$ let
$S$ be a minimum connected dominating set; \cref{prop:rst-rate} gives condition~\eqref{eq:fw-star}
for $S$ with $\Gamma=2$, so $s$ is accepted with probability at least $1-e^{-\kappa}$. Finally,
$2^{\psi^*}=2^{0.7827016}=1.7203\ldots<1.7204$.
\end{proof}

\subsection{A lower bound for RS-\texorpdfstring{$\theta$}{theta}: uniform lobsters}\label{sec:rst-caterpillar}

The analysis above is close to tight for Algorithm RS-$\theta$ itself. Recall that a
\emph{caterpillar} is a tree that becomes a path when its leaves are deleted, and a \emph{lobster} is a
tree that becomes a caterpillar when its leaves are deleted.
The \emph{uniform $(2,2)$-lobster} $\Lambda_L$ has a spine path $x_1x_2\cdots x_L$; every spine vertex
has exactly two neighbors off the spine, its \emph{legs}, and every leg has exactly two further
neighbors, both of degree $1$, its \emph{claws}. So the legs are the vertices at distance $1$ from the
spine, the claws are the vertices at distance $2$ and the leaves of $\Lambda_L$, and $n=7L$. The proposition identifies the minimum solution of $\Lambda_L$, bounds the
acceptance probability of RS-$\theta$ on it in terms of the normalized moments, and bounds these
moments by a certificate.

\begin{restatable}[Uniform $(2,2)$-lobsters]{proposition}{RestPropRstCaterpillar}\label{prop:rst-caterpillar}
Let $L\ge1$, let $n=7L$ be the number of vertices of $\Lambda_L$, and let $S^*$ be the set of its
spine vertices and legs. Then:
\begin{enumerate}[label=(\alph*)]
\item $S^*$ is the unique minimum connected dominating set of $\Lambda_L$; so $\gc(\Lambda_L)=3L$
and $\sigma=3/7$.
\item Run Algorithm RS-$\theta$ with any rate $\psi\ge0$ in place of $\psi^*$ (and the
corresponding cap and trial counts). For $s=\gc(\Lambda_L)$, the probability that $s$ is accepted
is at most $P_{\mathrm{nec}}\,2^{\psi n}\max_t\mu_t(S^*)/\binom nt$, where $P_{\mathrm{nec}}\le n^{O(1)}$ is
the polynomial factor of \cref{prop:fw-necessary} (with $\Gamma=2$).
\item For all $L\ge1$ and $0\le t\le n$, $\mu_t(S^*)/\binom nt\le(n+1)\,1.71927^{-n}$. Conversely,
$\min_t\binom nt/\mu_t(S^*)\le C\cdot1.71928^{n}$ for a constant $C$ and all $L\ge1$. So
$\bigl(\min_t\binom nt/\mu_t(S^*)\bigr)^{1/n}$ lies in $[1.71927,1.71928]$ up to factors $(n+1)^{1/n}$
and $C^{1/n}$; numerically its limit is $1.7192725\ldots$, attained near $y^*=0.3453$ in the
generating-function form.
\end{enumerate}
Consequently, RS-$\theta$ run with a rate $\psi$ in place of
$\psi^*$ outputs $\gc(\Lambda_L)$ with probability bounded away from $0$ on all $\Lambda_L$ only if
$2^{\psi}\ge1.71927$; so the certified base $1.7204$ exceeds the least base with which this
algorithm succeeds on all $\Lambda_L$ by at most $0.0012$.
\end{restatable}

The proposition holds for every graph whose unique maximum-leaf spanning tree is $\Lambda_L$ with
$L\ge2$ (\cref{sec:app-lobster}; parts~(b) and~(c) with their proofs verbatim, part~(a) by a separate
argument), so the lower bound does not rest on
instances on which \pname{Max Leaf Spanning Tree} is easy.

{}
\section{Closure sampling under exponential moments: Algorithm RSC}\label{sec:rsc}

The analysis of \cref{sec:rstheta} is nearly tight for its algorithm, and the instances on which it
is nearly attained have a common feature: some vertices of the solution have many neighbors inside
the solution. A random sample of the solution then contains several neighbors of such a vertex, and
the vertex stands out among the vertices with at least two sampled neighbors. Algorithm RSC adds each
such \emph{candidate} to the sample independently with a small probability, before it computes the
number of terminals and decides whether to call the Steiner tree algorithm. The probability is
\emph{graded}: a candidate with exactly two sampled neighbors is added with probability $p_2$, a
candidate with three or more with probability $p_3\ge p_2$. The main theorem in its budgeted-trial
form (\cref{thm:fw-budgeted}) applies to these trials directly, and the exponential moment in its
success condition becomes an average over the random closure as well as over the sample. For
$(p_2,p_3)=(0.07,0.077)$ we prove the following bound, the best one in this paper.

\begin{theorem}[Algorithm RSC]\label{thm:rsc}
  For every constant $\kappa>0$, Algorithm RSC (\cref{alg:rsc-algorithm}) runs in time
  $\Ostar\bigl(\varrho^{\,n}\bigr)$, where $\varrho=1.6990\ldots<1.6991$, and in polynomial space, on
  every connected graph $G$ with $n$ vertices. It never outputs a value below $\gc(G)$, and it outputs
  $\gc(G)$ with probability at least $1-e^{-\kappa}$. Consequently \pname{Connected Dominating Set},
  and \pname{Max Leaf Spanning Tree} (by \cref{lem:cds-maxleaf}, and directly when $n\le2$), can be solved by a randomized algorithm
  with one-sided error (one-sided Monte Carlo) in time $O(1.6991^n)$ and polynomial space.
\end{theorem}

The proof of the value of $\varrho$ rests on \cref{prop:rsc-cert}. The base is $\varrho=(1+y)/\hat e$ for the rationals $y=31/100$ and $\hat e=145729958/189014575$ of the
certificate, so the running time is $2^{\psi n}n^{O(1)}$ with $\psi=\log\varrho=0.76476681\ldots$.

\paragraph{Route of the proof.} \Cref{sec:rsc-alg} describes the algorithm, and
\cref{sec:rsc-budget} shows that \cref{thm:fw-budgeted} reduces \cref{thm:rsc} to a lower bound
$W^{\mathrm{cl}}_S(y)\ge0.6060\,\hat e^n$ on a generating function
$W^{\mathrm{cl}}_S$ of the closure trials (\cref{def:rsc-gf}), for every \emph{minimum} connected
dominating set $S$. We prove this lower bound (\cref{prop:rsc-bound}) by extending the analysis of
\cref{sec:rstheta} in three steps.
\begin{enumerate}[label=(\arabic*)]
  \item \emph{Tree model} (\cref{sec:rsc-tree}). \Cref{lem:rsc-model} replaces the graph by a
    spanning tree $F$ of $G[S]$, keeping for every vertex outside $S$ a factor $h_u$ that records whether
    it is dominated and whether it was a candidate, and keeping of the closure only what $F$ certifies.
    Every outside vertex is assigned to an owner in $S$ (\cref{def:rsc-owner}), and
    \cref{lem:rsc-pointwise} compares $h_u$ with a factor of the owner; the comparison depends on the
    \emph{type} of $u$, that is, on whether $u$ has two or more neighbors in $S$. Most factors $h_u$ are
    then bounded below by owner factors times an \emph{exposure}, a factor $q_\tau=1-p_\tau$ whenever
    the owner is in the closed sample, together with a \emph{credit} $\varpi_\tau\ge1$ on the factor of an
    owner outside the closed sample, which records that an exposed vertex is usually dominated by one of
    its other neighbors (\cref{lem:rsc-credit}). For the other shared vertices, Jensen's inequality gives
    a correction term instead (\cref{lem:rsc-jensen}), which the pointwise comparison bounds from below.
  \item \emph{Light owners} (\cref{sec:rsc-rec,sec:rsc-light}). The tree model is evaluated by a
    ten-state recursion along $F$ (\cref{lem:rsc-recursion}). Inequalities between the
    coordinates of its messages (\cref{lem:rsc-messages}) show that the Jensen correction of an outside
    vertex is nonnegative when its owner has few owned vertices and small degree in $F$
    (\cref{lem:rsc-jlemma}), and they bound the odds that a vertex of $S$ is sampled from below
    (\cref{lem:rsc-odds}), which is what the credits rest on. The other owners are exposed; minimality of
    $S$ limits the exposures at the leaves of $F$ (\cref{lem:rsc-leaf}).
  \item \emph{Certificate} (\cref{sec:rsc-cert}). An antinorm in an eight-dimensional poset cone,
    in the style of \cref{lem:rst-F}, bounds a relaxation of the recursion from below for every tree that
    can arise (\cref{lem:rsc-cert}); the relaxation (\cref{lem:rsc-relaxed}) forgets whether a
    vertex has more than two sampled children except where the eight states can tell, and it never
    overestimates the tree model. The certificate consists of $1024$ rational points and staged
    conditions for exposed vertices, and \cref{prop:rsc-cert} states the conditions it satisfies.
\end{enumerate}
The three steps are carried out in \cref{sec:app-rsc}, and \cref{sec:rsc-limits} records how
tight the analysis is and how the parameters were chosen.

\paragraph{Proof map.} \Cref{tab:rsc-audit} lists the ingredients of \cref{thm:rsc}, with the kind of
argument behind each and the machine checks that exist for it.
\begin{table}[!htb]
  \centering
  \renewcommand{\arraystretch}{1.12}
  \small
  \begin{tabularx}{\textwidth}{@{}>{\raggedright\arraybackslash}p{3.6cm}>{\raggedright\arraybackslash}p{3.3cm}>{\raggedright\arraybackslash}X>{\raggedright\arraybackslash}p{3.4cm}@{}}
    \toprule
    Ingredient & Statements & Argument & Machine checks \\
    \midrule
    Steiner extension in time $2^{|\Tm(J_X)|}n^{O(1)}$ and polynomial space
      & \cref{lem:cv-W,lem:cds-dir-steiner,cor:cds-steiner}
      & inclusion--exclusion over branching walks~\cite{Nederlof13}; proofs in \cref{sec:app-steiner}
      & Lean: the computed value, not running time or space \\
    Budgeted trials: running time, rejection, success probability
      & \cref{thm:fw-budgeted,lem:rsc-trials,cor:cds-budgeted}
      & mathematical proof
      & Lean: rejection and success probability, not running time \\
    Graph to tree model, owners, exposures, credits, ten-state recursion, light owners
      & \cref{lem:rsc-model} to \cref{cor:rsc-reduction}
      & mathematical proof
      & Lean \\
    Finitely many conditions imply the bound for every tree
      & \cref{lem:rsc-relaxed,lem:rsc-cone,lem:rsc-cert}
      & invariant region and induction from the leaves
      & Lean \\
    Light pairs (condition \ref{it:rsc-C3}) and credit inequalities
      & \cref{lem:rsc-lightpairs,lem:rsc-rsccredits}
      & rational bounds on logarithms
      & exact program; Lean \\
    Certificate data: \ref{it:rsc-C1}, \ref{it:rsc-C2}, \ref{it:rsc-C4}, \ref{it:rsc-C5}, \ref{it:rsc-C6} and $c_Z>0.6060$
      & \cref{prop:rsc-cert}
      & $1024$ rational points and about $3.7$ million relations, each checked exactly
      & two independently written programs, exact rational arithmetic; Lean with compiled evaluation \\
    Base $\varrho<1.6991$
      & proof of \cref{thm:rsc}
      & comparison of rationals
      & Lean \\
    \pname{Max Leaf Spanning Tree} from \pname{Connected Dominating Set}
      & \cref{lem:cds-maxleaf}
      & mathematical proof
      & Lean \\
    \bottomrule
  \end{tabularx}
  \caption{Proof map of \cref{thm:rsc}. ``Lean'' refers to the statements listed in \cref{sec:anc-lean};
    the programs and their running times are described in \cref{sec:anc-rsc}.}
  \label{tab:rsc-audit}
\end{table}
Apart from the Lean kernel, the bound therefore rests on the mathematical proofs of the running time and space
bounds, including those of the Steiner extension, which are not formalized, and on the two programs that check the data
of the certificate; the Lean check of these data (\cref{sec:anc-lean}) is a third check, which also trusts the
Lean compiler. The two programs are independent: the second was written from the definitions of this
section without reusing code of the first, it re-derives the recursion, and it rejects perturbed copies of the
certificate (\cref{sec:anc-rsc}).

Throughout this section $0<p_2\le p_3<\frac12$, $q_2:=1-p_2$ and $q_3:=1-p_3$, so $q_2\ge q_3>\frac12$; for an
integer $k\ge2$ we write $p_k:=p_2$ if $k=2$ and $p_k:=p_3$ if $k\ge3$, and $q_k:=1-p_k$. For a
connected graph $G$ with vertex set $V$ and $X'\subseteq V$ the \emph{candidate set} is
\[
  C(X')=\bigl\{v\in V\setminus X':\ |N(v)\cap X'|\ge2\bigr\}.
\]
With $p_2=p_3$ the closure is the Bernoulli closure of a single probability; the bound of the
theorem rests on the graded one (see \cref{rem:rsc-tight}).

Esmer and Kulik~\cite{EsmerK25} also use problem-specific random steps before a black-box call, for
parameterized approximation of vertex deletion problems: each of their
sampling steps returns a vertex whose removal lowers the optimum with at least a fixed probability, and the
running time of the black-box call is fixed by the algorithm. Here the closure is added to a uniform
sample inside Monotone Local Search, its probabilities depend on the sample, and it changes the
running-time exponent of the call, which the exponential moments account for.

\subsection{The algorithm}\label{sec:rsc-alg}

\paragraph{How RSC decides.} For each candidate size $s$, RSC runs the trials of RS-$\theta$ (\cref{alg:rst-algorithm}), indexed by
a sample size $t\in\{0,\dots,s\}$ and an integer threshold $\theta$, with one change, which takes
polynomial time per trial: after drawing a uniform $t$-set $X'$, it puts each candidate $v\in C(X')$
into a set $\Upsilon$ independently with probability $p_{|N(v)\cap X'|}$, that is, with probability $p_2=0.07$ if $v$
has exactly two sampled neighbors and $p_3=0.077$ if it has more, and it then treats the \emph{closed
sample} $X=X'\cup\Upsilon$ as RS-$\theta$ treats its sample, also discarding $X$ if $|X|>s$. Fix a
solution $S$. The closure never increases the number of terminals (\cref{lem:rsc-mono}), and an added
candidate with a neighbor outside $N[X']$ lowers it; but the closed sample leaves $S$ when the
closure adds a vertex outside $S$, which happens with probability $1-\prod_{v\in C(X')\setminus S}q_{|N(v)\cap X'|}$. The
small values of $p_2$ and $p_3$ keep this probability small; the analysis needs small
probabilities, and $p_3$ only slightly above $p_2$.

\begin{algorithm}[RSC]\label{alg:rsc-algorithm}
  \Input a connected graph $G$ with $n\ge1$ vertices.
  \Parameters $p_2=0.07$, $p_3=0.077$, $y=31/100$ and $\hat e=145729958/189014575$; $\psi=\log\varrho$ with
  $\varrho=(1+y)/\hat e$; a constant $\kappa>0$ and $\Gamma=2$; a cap $\Theta$ and
  trial counts $m_{t,\theta}$ within the ranges~\eqref{eq:fw-params}.
  \Body
  For $s=1,2,\dots,n$, decide whether to accept
  $s$ as follows, and output the first accepted $s$ (and $n$ if there is none). For every
  $t\in\{0,\dots,s\}$ and every integer $\theta\in\{0,\dots,\Theta\}$, repeat $m_{t,\theta}$ times:
  \begin{enumerate}[label=(\alph*)]
    \item draw $X'$ uniformly from $\binom Vt$ and compute $C(X')$;
    \item put each $v\in C(X')$ into $\Upsilon$ independently with probability $p_{|N(v)\cap X'|}$, and let $X=X'\cup\Upsilon$;
    \item if $|X|\le s$ and $|\Tm(J_X)|=n-|N[X]|+\comp(G[X])\le\theta$, compute $\OPT(J_X)$ by
      \cref{cor:cds-steiner}, and accept $s$ if $\OPT(J_X)\le s-|X|$.
  \end{enumerate}
\end{algorithm}

The trial counts $m_{t,\theta}$ should not be confused with the numbers
$\nu_v$ of owned vertices used in the analysis.

\subsection{Applying the budgeted-trials theorem}\label{sec:rsc-budget}

The trials of RSC are not those of Algorithm~\MLS{}: the extension algorithm is called on the closed
sample $X'\cup\Upsilon$, whose distribution is not uniform on sets of a given size. The budgeted form of
the main theorem (\cref{thm:fw-budgeted}) needs only that a trial of threshold $\theta$ runs in time
$2^\theta n^{O(1)}$, rejects if no solution of size at most $k$ exists, and accepts whenever a
random variable attached to it is at most $\theta$. The exponential moment in its success condition
is then a generating function that averages over the closure.

\begin{definition}[Closure generating function]\label{def:rsc-gf}
  Let $G$ be a connected graph, $S$ a connected dominating set of $G$, $0\le p_2\le p_3\le1$ and $y\ge0$. For
  $X'\subseteq V$ let $\Upsilon$ be the random subset of $C(X')$ that contains each candidate $v$ independently
  with probability $p_{|N(v)\cap X'|}$. Put
  \[
    W^{\mathrm{cl}}_S(y)=\sum_{X'\subseteq S}y^{|X'|}\,
    \E_\Upsilon\Bigl[\mathbf 1[\Upsilon\subseteq S]\,2^{-|\Tm(J_{X'\cup\Upsilon})|}\Bigr].
  \]
\end{definition}

For $p_2=p_3=0$ the set $\Upsilon$ is empty and $W^{\mathrm{cl}}_S$ is the generating function $W_S$
of~\eqref{eq:rst-mu}.

\begin{lemma}[The trials of RSC]\label{lem:rsc-trials}
  Let $G$ be a connected graph with $n$ vertices and let $1\le s\le n$. For $t\in\{0,\dots,s\}$ and
  an integer $\theta\ge0$, let $\mathrm{Tr}_{t,\theta}$ be one execution of steps (a)--(c) of
  \cref{alg:rsc-algorithm} for $s$, with any rational $0\le p_2\le p_3\le1$.
  \begin{enumerate}[label=(\alph*)]
    \item $\mathrm{Tr}_{t,\theta}$ runs in time $2^\theta r_0(n)$ and space $r_0(n)$ for a polynomial $r_0$ that
      depends only on $p_2$ and $p_3$, not on $G$, $s$, $t$ or $\theta$, and it rejects if $\gc(G)>s$.
    \item Let $S$ be a connected dominating set of $G$ with $|S|=s$. Let $\xi_t=|\Tm(J_X)|$ if $X\subseteq
      S$ and $\xi_t=\infty$ otherwise, where $X=X'\cup\Upsilon$ is the closed sample of the trial. Then
      the distribution of $\xi_t$ does not depend on $\theta$,
      $\Pr[\mathrm{Tr}_{t,\theta}\text{ accepts}]\ge\Pr[\xi_t\le\theta]$ for every $\theta$, and
      $W^{\mathrm{cl}}_S(y)\le(1+y)^n\max_{0\le t\le s}\E\bigl[2^{-\xi_t}\bigr]$ for every $y\ge0$.
  \end{enumerate}
\end{lemma}

\begin{proof}
  (a) Steps (a) and (b) and the test in (c) take polynomial time, by the convention on random draws in
  \cref{sec:fw-setting}, since $p_2$ and $p_3$ are rational. The Steiner extension is called only
  if $|\Tm(J_X)|\le\theta$, and then it takes time $2^{|\Tm(J_X)|}n^{O(1)}\le2^\theta n^{O(1)}$ and
  polynomial space by \cref{cor:cds-steiner}, where the exponent of $n$ is a fixed constant. Together these
  give one polynomial $r_0$ as stated. If the trial accepts, then some $Y\subseteq V\setminus X$
  with $|Y|\le s-|X|$ makes $X\cup Y$ a connected dominating set, by the description of the solutions
  of $J_X$ in \cref{sec:cds}; this set has at most $s$ vertices.

  (b) The trial's random choices do not involve $\theta$. Suppose that $\xi_t\le\theta$. Then $X\subseteq S$, so $|X|\le s$
  and $|\Tm(J_X)|\le\theta$; the trial computes $\OPT(J_X)$, and since $X\cup(S\setminus X)=S$ is a
  connected dominating set, $S\setminus X$ is a solution of $J_X$ and $\OPT(J_X)\le|S\setminus
  X|=s-|X|$. So the trial accepts. For the last inequality, $X\subseteq S$ holds if and only if $X'\subseteq S$
  and $\Upsilon\subseteq S$, and $X'$ is uniform in $\binom Vt$, so
  \[
    \E\bigl[2^{-\xi_t}\bigr]=\binom nt^{-1}\sum_{X'\in\binom Vt}\mathbf 1[X'\subseteq S]\,
    \E_\Upsilon\Bigl[\mathbf 1[\Upsilon\subseteq S]\,2^{-|\Tm(J_{X'\cup\Upsilon})|}\Bigr].
  \]
  Grouping the terms of \cref{def:rsc-gf} by $t=|X'|$ (only $t\le s$ occur) and using
  $\sum_{t=0}^n\binom nty^t=(1+y)^n$ gives
  $W^{\mathrm{cl}}_S(y)=\sum_{t=0}^s\binom nt\E[2^{-\xi_t}]\,y^t\le(1+y)^n\max_t\E[2^{-\xi_t}]$.
\end{proof}

The following corollary of \cref{thm:fw-budgeted} and \cref{lem:cds-first-accepted} packages the
wrapper around budgeted trials for connected dominating sets.

\begin{corollary}[Budgeted trials for connected domination]\label{cor:cds-budgeted}
  Let $\psi\ge0$ be a real constant that can be approximated to within $2^{-b}$ in time $b^{O(1)}$
  for every integer $b\ge1$, let $\Gamma_n\ge1$ be computable in time $n^{O(1)}$ with
  $\Gamma_n\le n^{O(1)}$, and let $\kappa>0$ be a constant. Suppose that there are a randomized
  algorithm $\mathrm{Tr}$ with inputs $(G,s,t,\theta)$, written $\mathrm{Tr}_{t,\theta}$ when $G$ and $s$
  are fixed, and a polynomial $r_0$ such that for every connected graph $G$ with $n$ vertices, every
  $s\in\{1,\dots,n\}$, every $t\in\{0,\dots,s\}$ and every integer $\theta\ge0$:
  \begin{enumerate}[label=\textup{(T\arabic*)}]
    \item $\mathrm{Tr}_{t,\theta}$ runs in time $2^\theta r_0(n)$ and space $r_0(n)$,
    \item $\mathrm{Tr}_{t,\theta}$ rejects if $\gc(G)>s$, and
    \item if $s=\gc(G)$, then there are random variables $\xi_t$ with values in $[0,\infty]$, whose distributions
      do not depend on $\theta$, such that $\Pr[\mathrm{Tr}_{t,\theta}\text{ accepts}]\ge\Pr[\xi_t\le\theta]$
      for all $t$ and $\theta$, and $\max_{0\le t\le s}\E[2^{-\xi_t}]\ge2^{-\psi n}/\Gamma_n$.
  \end{enumerate}
  Then the algorithm that, for $s=1,\dots,n$ in turn, runs \cref{alg:fw-budgeted} on $(G,s)$ with the
  index set $\{0,\dots,s\}$, these trials, rate $\psi$, loss factor $\Gamma_n$, confidence $\kappa$
  and a cap and trial counts within the ranges~\eqref{eq:fw-params}, and outputs the first accepted
  $s$, or $n$ if none, runs in time $2^{\psi n}n^{O(1)}$ and polynomial space, never outputs a value
  below $\gc(G)$, outputs $\gc(G)$ with probability at least $1-e^{-\kappa}$, and, when $n\ge3$, solves \pname{Max Leaf
  Spanning Tree} within the same bounds.
\end{corollary}
\begin{proof}
  Fix $s$. The cap and the trial counts are computed in polynomial time: the number
  $x=\psi n+\log\Gamma_n$ of~\eqref{eq:fw-params} is approximated to within $\frac12$, and
  $M_{t,\theta}=4\kappa\Gamma_n(\Theta+1)2^{\psi n-\theta}$ to within a factor $2$, by rationals obtained
  from an approximation of $\psi$ and from $\Gamma_n$, which suffices, as noted after~\eqref{eq:fw-params};
  the values of $\psi$, $\Gamma_n$ and $\kappa$ are at most $n^{O(1)}$. Extend $\mathrm{Tr}$ to all other
  inputs, that is, to instances that do not encode a connected graph and to values of $s$ or $t$ outside the
  ranges above, by rejecting at once after a test that takes time $r_1(N)$ for a polynomial $r_1$. We may
  take $r_0$ nondecreasing; as $n\le N$, conditions~\ref{it:fw-T1} and~\ref{it:fw-T2} of \cref{thm:fw-budgeted} then hold for every
  instance and every $k\le n$, with the polynomial $r_0+r_1$ and space bound $\mathrm{sp}=0$. By \cref{thm:fw-budgeted} for
  the implicit set system $\Phi_{\mathrm{CDS}}$ of \cref{def:cds-ast}, with $k=s$, the index sets
  $\cT_{I,k}=\{0,\dots,k\}$ (empty for $k<0$), which are listed in time $O(N)$,
  the run for $s$ takes time $2^{\psi n}n^{O(1)}$ and polynomial space by~(T1), rejects if
  $\gc(G)>s$ by~(T2), and, when $s=\gc(G)$, accepts with probability at least $1-e^{-\kappa}$
  by~(T3), with a minimum connected dominating set as the set $S$ of size $k$ in
  \cref{thm:fw-budgeted}. \Cref{lem:cds-first-accepted}, with success probability $1-e^{-\kappa}$, gives the rest, the
  total time being $n\cdot2^{\psi n}n^{O(1)}$.
\end{proof}

The analysis uses the solution $S$ only through $W^{\mathrm{cl}}_S$, and it needs $S$ to be a
minimum connected dominating set. The proof of the following proposition, given at the end of
\cref{sec:rsc-cert} in \cref{sec:app-rsc}, uses the certificate of \cref{prop:rsc-cert}.

\begin{restatable}[Closure generating function of a minimum solution]{proposition}{RestPropRscBound}\label{prop:rsc-bound}
  Let $p_2=0.07$, $p_3=0.077$, $y=31/100$ and $\hat e=145729958/189014575$. For every connected graph $G$ with $n$ vertices and every connected dominating set $S$ of $G$
  with $|S|=\gc(G)$,
  \[
    W^{\mathrm{cl}}_S(y)\ \ge\ 0.6060\,\hat e^n .
  \]
\end{restatable}

\begin{proof}[Proof of \cref{thm:rsc}, given \cref{prop:rsc-bound}]
  The steps of \cref{alg:rsc-algorithm} for $s$ are \cref{alg:fw-budgeted} with the trials
  $\mathrm{Tr}_{t,\theta}$ of \cref{lem:rsc-trials}, rate $\psi=\log\varrho$, loss factor
  $\Gamma_n=2$ and confidence $\kappa$, so \cref{thm:rsc} follows from \cref{cor:cds-budgeted}, with $n\le2$ handled directly; the
  rate $\psi$, the logarithm of a rational number, can be approximated to any precision in polynomial
  time. Properties (T1) and (T2) are \cref{lem:rsc-trials}(a). For (T3), let $s=\gc(G)$, let $S$ be a
  minimum connected dominating set, and take the variables $\xi_t$ of \cref{lem:rsc-trials}(b); their
  distributions do not depend on $\theta$ and $\Pr[\mathrm{Tr}_{t,\theta}\text{ accepts}]\ge\Pr[\xi_t\le\theta]$,
  and by \cref{lem:rsc-trials}(b) and \cref{prop:rsc-bound},
  \[
    \max_{0\le t\le s}\E\bigl[2^{-\xi_t}\bigr]\ \ge\ \frac{W^{\mathrm{cl}}_S(y)}{(1+y)^n}
    \ \ge\ 0.6060\Bigl(\frac{\hat e}{1+y}\Bigr)^n=0.6060\cdot2^{-\psi n}\ \ge\ \frac{2^{-\psi n}}2 .
  \]
  So RSC runs in time $2^{\psi n}n^{O(1)}=\Ostar(\varrho^n)$ and polynomial space, never outputs a value
  below $\gc(G)$, outputs $\gc(G)$ with probability at least $1-e^{-\kappa}$, and, for $n\ge3$ by \cref{cor:cds-budgeted}, solves \pname{Max Leaf
  Spanning Tree} within the same bounds; for $n\le2$ the maximum leaf spanning tree is $G$ itself (\cref{lem:cds-maxleaf}). Finally $\varrho=(1+y)/\hat e=1.6990\ldots<1.6991$, by exact
  evaluation of the rational number $\frac{131}{100}\cdot\frac{189014575}{145729958}=\frac{990436373}{582919832}$, so
  $\Ostar(\varrho^n)\subseteq O(1.6991^n)$.
\end{proof}

\subsection{Tightness of the analysis}\label{sec:rsc-limits}

The following remark is a numerical observation; it is not used in the proof of \cref{thm:rsc}.

\begin{remark}[How tight the analysis is]\label{rem:rsc-tight}
  On the uniform $(2,2)$-lobster $\Lambda_L$ of \cref{prop:rst-caterpillar}, the generating function
  $W^{\mathrm{cl}}_S$ of the unique minimum connected dominating set can be evaluated by a transfer
  matrix. With $(p_2,p_3)=(0.07,0.077)$, the running-time bound $\inf_{y>0}(1+y)^n/W^{\mathrm{cl}}_S(y)$ that
  \cref{lem:rsc-trials} and \cref{thm:fw-budgeted} give is about $1.6976^n$ (numerical, floating point), so the certified
  running time $O(1.6991^n)$ is within a factor $2^{0.0013n}$ of the best that this route can give for this algorithm.
  The grading is worth about $0.0008$ in the base: with $p_2=p_3=0.07$ an antinorm certificate of the same kind certifies about $1.6999^n$.
\end{remark}

The values of $p_2$, $p_3$ and $y$ come from a numerical exploration, which the following remark records; as
with \cref{rem:rsc-tight}, nothing in it is used in a proof.

\begin{remark}[Choice of the closure probabilities]\label{rem:rsc-params}
  Without closure ($p_2=p_3=0$), Algorithm RSC is RS-$\theta$, which needs time $\Omega(1.7192^n)$ on the
  uniform lobsters (\cref{prop:rst-caterpillar}). A candidate with $k$ sampled neighbors in different components
  of the sample can merge up to $k$ components into one and lower the number of terminals by up to $k-1$
  if it lies in $S$, whereas the trial is lost if it lies outside $S$, whatever $k$ is; this is why a larger
  probability for candidates with three or more sampled neighbors can pay. The analysis, however, charges a factor
  $q_\tau=1-p_\tau$ for every exposed outside vertex, and the light-owner tests of \cref{def:rsc-light} hold only
  for small probabilities. Numerically, the smallest base that the exact tree model of \cref{sec:app-rsc} can
  certify, over $(p_2,p_3,y)$, is attained near $(0.07,0.077,0.31)$, where it is about $1.6983$; the
  eight-state relaxation cannot go below about $1.6985$ there, and the certificate gives $1.69910$. The window for $p_3$ is
  narrow: at $(p_2,p_3)=(0.07,0.15)$, owners of two or more outside vertices fail the test for type-$3$
  vertices and the limit of the tree model rises to about $1.749$.

  The decimal $1.6991$ needs no numerical slack: $\varrho=990436373/582919832=1.699095\ldots$ is an exact rational,
  and every condition of the certificate is checked exactly, so the rounding up to four decimals is the only
  step between $\varrho$ and the stated base.
\end{remark}
{}
\section{Derandomization}\label{sec:derandomisation}

The algorithms of \cref{thm:fw-main,thm:co-mean} are randomized. Classical Monotone Local Search
is derandomized by \cite{FominGLS19} at the price of a factor
$2^{o(n)}$ in the running time. For sample-dependent running-time exponents the answer has two parts. If the
algorithm may use the running-time exponent and the extension algorithm only as oracles, no deterministic
algorithm matches the mean-exponent theorem in general: when the mean bound, at least $1$, is below $\log\binom kt$
by a linear amount, where $t$ is the sample size, its cost exceeds that of the randomized algorithm by an
exponential factor unless it reaches the cost $\binom nk$ of testing every $k$-set
(\cref{thm:dr-bb-one,thm:dr-bb-all,cor:dr-bb-gap}), and for the exponential-moment theorem the lower bound
can come close to $\binom nk$ (\cref{thm:dr-bb-moments}); for the mean bounds of \pname{Connected
Dominating Set} the gap is $1.7547^n/\poly(n)$ against $\sqrt3^{\,n}\poly(n)$ (\cref{cor:dr-bb-rs}).
If, however, the mean bound comes from a pointwise affine majorant $g(X)\le\alpha-w(X)$, where
$g(X)=g(I,X,k)$ is the shorthand declared after \cref{def:fw-cost} and $w$ are unknown nonnegative
weights of known total, then balanced set-inclusion families, which we list deterministically in
polynomial space (\cref{thm:dr-balanced}), derandomize, for integer weights and with a factor
$2^{O(n(\log\log n)^{2}/\log n)}$, the exponential-moment bound of the affine \emph{surrogate} exponent $\alpha-w(X)$
(\cref{thm:dr-det-moments}). This is not a derandomization of \cref{thm:fw-main} for the true running-time
exponent $g$. \Cref{sec:det} derives the consequences for \pname{Connected Dominating Set}.

Throughout we use the setting of \cref{def:fw-iss,def:fw-cost,def:fw-moments}. An implicit set
system $\Phi$ assigns to an instance $I$ with $N=|I|$ a universe $U_I$ with $n=|U_I|\le N$ and a
family $\cF_I$ of solutions; an extension algorithm $\cA$ decides, given $(I,X,k)$, whether some
$Y\subseteq U_I\setminus X$ with $|Y|\le k-|X|$ has $X\cup Y\in\cF_I$, in time
$2^{g(I,X,k)}N^{O(1)}$, where the running-time exponent $g\ge0$ is computable in polynomial time. For a
solution $S$ with $|S|=k$ we write $\mu_t(S)=\sum_{X\in\binom St}2^{-g(I,X,k)}$, and $H$ is the
binary entropy, with the estimates~\eqref{eq:fw-binom}.

\subsection{Set-inclusion families and what changes with sample-dependent running times}\label{sec:dr-fgls}

Classical Monotone Local Search samples a uniform $t$-subset $X$ of the universe and calls the
extension algorithm on it. The running-time bound $c^{k-t}N^{O(1)}$ of the extension call depends only
on $|X|$, so a sample $X$ of size $t$ can fail only by not being contained in the solution. \cite{FominGLS19} replace the random samples by a
deterministic family that contains a $t$-subset of every $k$-set.

\begin{definition}[Set-inclusion family~\cite{FominGLS19}]\label{def:dr-inclusion}
  Let $|U|=n$ and $0\le t\le k\le n$. A family $\mathcal J\subseteq\binom Ut$ is an
  \emph{$(n,k,t)$-set-inclusion family} if every $S\in\binom Uk$ contains some member of $\mathcal J$.
\end{definition}

Every member of $\binom Ut$ lies in $\binom{n-t}{k-t}$ sets of $\binom Uk$, so double counting gives
$|\mathcal J|\ge\binom nk/\binom{n-t}{k-t}=\binom nt/\binom kt$. \cite[Section~3]{FominGLS19} construct families of size
$\binom nt\binom kt^{-1}2^{o(n)}$ in time of the same order: they hash
the universe into about $\log n$ buckets with a pairwise independent family of functions, build a
near-optimal family on each bucket by greedy \pname{Set Cover}, and combine the bucket families.
The bucket families have size $2^{O(n/\log n)}$ and are stored, and the paper notes that the
deterministic algorithm needs an additional $2^{o(n)}$ space. Looping over such a family instead
of sampling gives a deterministic $(2-\frac1c)^{n+o(n)}N^{O(1)}$ algorithm from a $c^{k}N^{O(1)}$
extension algorithm.

A sample-dependent running-time exponent changes the picture. The family must now contain, inside every
solution, a sample of small running-time exponent, and the running-time exponents of the subsets of the solution are
visible only through evaluations of $g$. An adversary can give large running-time exponents to exactly the
evaluated subsets; this planting argument gives the lower bounds of \cref{sec:dr-barrier}. For running-time
exponents bounded by an affine function of unknown nonnegative weights, a family with a stronger
property suffices (\cref{sec:dr-balanced}): it contains, inside every solution, a sample of at least
average weight. This \emph{balance} is strictly stronger
than inclusion: $\{\{1\},\{2\},\{3\}\}$ is a $(4,2,1)$-set-inclusion family, but for $S=\{1,4\}$ and
a weight concentrated on the element $4$, no member inside $S$ has positive weight.

\subsection{The black-box barrier}\label{sec:dr-barrier}

The randomized algorithms of \cref{thm:fw-main,thm:co-mean} use the instance only through
evaluations of $g$ and calls of $\cA$. The barrier concerns deterministic algorithms with the same
access, and with membership tests added.

\begin{definition}[Black-box model]\label{def:dr-blackbox}
  An \emph{instance} is a pair $(\cF,g)$ with $\cF\subseteq2^{U}$, $|U|=n$, and
  $g\colon2^{U}\times\N\to\R_{\ge0}$. The algorithm knows $n$, $k$ and the parameters of the
  hypothesis ($t$ and $\lambda$, $\lambda_0,\dots,\lambda_k$, or $\psi$ and $\Gamma$, below). It may call
  \begin{itemize}
    \item $\mathrm{COST}(X,\ell)$, for a set $X\subseteq U$ and an integer $\ell\ge0$, which returns $g(X,\ell)$, at cost $1$;
    \item $\mathrm{EXT}(X,\ell)$, which returns whether some $Y\subseteq U\setminus X$ with $|Y|\le\ell-|X|$ has
      $X\cup Y\in\cF$, at cost $2^{g(X,\ell)}$;
    \item $\mathrm{MEM}(X)$, which returns whether $X\in\cF$, at cost $1$.
  \end{itemize}
  Computation is free. A deterministic algorithm is \emph{correct for a class} $\cC$ of instances if it
  answers \NO{} if $\cF$ contains no set of size at most $k$, and it answers \YES{} on every
  instance of $\cC$ in which $\cF$ contains a set of size exactly $k$. Its \emph{cost} on an instance
  is the total cost of its calls. With $\E_{X\in\binom St}$ the mean over $X\in\binom St$, the classes are:
  \begin{itemize}
    \item $\cC_{\mathrm{mean}}(t,\lambda)$: every $S\in\cF$ with $|S|=k$ satisfies
      $\E_{X\in\binom St}g(X,k)\le\lambda$;
    \item $\cC_{\mathrm{mean}}^{\mathrm{all}}(\lambda_0,\dots,\lambda_k)$: every $S\in\cF$ with $|S|=k$
      satisfies $\E_{X\in\binom Sj}g(X,k)\le\lambda_j$ for every $j\in\{0,\dots,k\}$;
    \item $\cC_{\mathrm{exp}}(\psi,\Gamma)$: some $S\in\cF$ with $|S|=k$ has
      $\max_j\mu_j(S)/\binom nj\ge2^{-\psi n}/\Gamma$, where $\mu_j(S)=\sum_{X\in\binom Sj}2^{-g(X,k)}$.
  \end{itemize}
\end{definition}

The model is generous to the algorithm: computation is free, every evaluation of $g$ costs one
step, and $\cF$ need not be monotone. The randomized algorithms of \cref{thm:co-mean} (on
$\cC_{\mathrm{mean}}(t,\lambda)$) and \cref{thm:fw-main} (on $\cC_{\mathrm{exp}}(\psi,\Gamma)$) run
in this model within their stated bounds, since they only evaluate $g$ and call the extension
algorithm. Write
\[
  \mathsf R_t=\binom nt\binom kt^{-1}2^{\lambda}
\]
for the running-time bound of the mean-exponent theorem at sample size $t$, without its polynomial factor;
\cref{thm:co-mean} decides the class $\cC_{\mathrm{mean}}(t,\lambda)$ with one-sided error in time $\mathsf R_t$ up to a factor polynomial in $N$
and $\lambda$ (the Markov-cutoff algorithm of \cref{rem:co-markov}, with the exact threshold $\lambda+1$, runs
  $\lceil(\lambda+1)\binom nt\binom kt^{-1}\rceil$ trials of cost at most $1+2^{\lambda+1}$ each, a
total cost of at most $6(\lambda+1)\mathsf R_t$).

\paragraph{The idea.} All lower bounds use one adversary, a form of the planting argument of the
oracle lower bounds of \cite{EsmerKMNS22,EsmerKMNS24}, where a solution is hidden among the
sets that a run leaves unqueried. It runs a deterministic algorithm on a
no-instance, $\cF=\emptyset$, whose running-time exponents depend only on the size of the sample. Call a set
\emph{touched} if the run evaluates its running-time exponent or calls the extension algorithm on it, and call
a $k$-set \emph{clean} if the run neither tests its membership nor calls the extension algorithm on one
of its subsets. A clean $k$-set can be planted as the only solution, with running-time exponent $0$ on its
untouched subsets, without changing any answer of the run, so the algorithm still answers \NO{}. A
cheap run leaves a random $k$-set clean with few touched subsets, and then the
planted instance satisfies the hypothesis of the class, which contradicts correctness. The proofs are
in \cref{sec:app-barrier}.

\paragraph{Mean exponents.} The planting argument gives lower bounds for the hypothesis of
\cref{thm:co-mean} at one sample size and for mean bounds $\lambda_0,\dots,\lambda_k$ at all sample
sizes; in both cases a deterministic black-box algorithm has cost about $\binom nt$ instead of
$\mathsf R_t$, up to the cap $\binom nk$ of testing every $k$-set, so the gap to the randomized algorithm is exponential
whenever $\log\binom kt$ exceeds the mean bound by a linear amount. The statements are in
\cref{sec:app-bb-mean} (\cref{thm:dr-bb-one,cor:dr-bb-gap,thm:dr-bb-all}). We evaluate \cref{thm:dr-bb-all} on the mean bounds $\lambda_0,\dots,\lambda_k$ that
\pname{Connected Dominating Set} provides. By the mean number of terminals (\cref{lem:cds-C}),
a uniform $j$-subset of a connected dominating set of size $k$ leaves on average at most $n-1-j(n-2)/k$ terminals; with the empty
sample, which leaves all $n$ vertices as terminals, this gives the bounds
$\lambda_0=n$ and $\lambda_j=n-1-j(n-2)/k$ for $1\le j\le k$.

\begin{restatable}[The mean bounds of Connected Dominating Set]{corollary}{RestCorDrBbRs}\label{cor:dr-bb-rs}
  Let $n$ be even, $k=n/2$, $\lambda_0=n$ and $\lambda_j=n-1-j(n-2)/k$ for $1\le j\le k$. Every
  deterministic black-box algorithm that is correct for
  $\cC_{\mathrm{mean}}^{\mathrm{all}}(\lambda_0,\dots,\lambda_k)$ has cost at least
  \[
    \frac{2^{H(1/4)n}}{6(n+1)(n+2)}=\frac{(4\cdot3^{-3/4})^{n}}{6(n+1)(n+2)}\ge\frac{1.7547^{n}}{6(n+1)(n+2)}
  \]
  on some no-instance, while Algorithm \MLS{} with the parameters of \cref{cor:co-linear}(1), which calls only
  $\mathrm{COST}$ and $\mathrm{EXT}$ and needs only the mean bounds $\lambda_0,\dots,\lambda_k$ as
  parameters, decides the same class with one-sided error at cost $\sqrt3^{\,n}n^{O(1)}$.
\end{restatable}

\Cref{cor:dr-bb-rs} is a statement about the abstract class, not about \pname{Connected Dominating Set}:
over all solution sizes, the randomized algorithm of \cref{thm:rs} runs in time $O(1.7449^n)$, and \cref{sec:det}
solves \pname{Connected Dominating Set} deterministically in time $O(1.7361^n)$
(\cref{thm:det-cds}), by using the affine structure of the running-time exponent and not only its values.

\paragraph{Exponential moments.} For the exponential-moment theorem the lower bound is much larger:
the cost of a deterministic black-box algorithm can come close to the cost $\binom nk$ of testing
every $k$-set with $\mathrm{MEM}$.

\begin{restatable}[Black-box lower bound for exponential moments]{theorem}{RestThmDrBbMoments}\label{thm:dr-bb-moments}
  Let $\psi\ge0$ and $\Gamma\ge1$, and let $\mathcal H$ be a deterministic black-box algorithm that is correct
  for $\cC_{\mathrm{exp}}(\psi,\Gamma)$. Call a level $t\in\{0,\dots,k\}$ \emph{feasible} if
  $J_t:=\lceil\binom nt2^{-\psi n}/\Gamma\rceil\le\binom kt/2$. Then for every feasible $t$, the cost of
  $\mathcal H$ on the no-instance $\cF=\emptyset$, $g\equiv n$, is at least $\frac13\min\{\binom nk,\binom nt\}$.
\end{restatable}

The strongest bound comes from the largest feasible $t$; for Algorithm RS-$\theta$ at the solution fraction
$k/n\approx0.4173$, where its running-time bound is attained (\cref{tab:rst-coverage}),
it is numerically about $1.9440^{n}$ (\cref{sec:ancillary}), against $O(1.7204^{n})$ for the randomized
algorithm; this value is not used later.

We conjecture that the same planting argument, combined with Yao's principle, shows that \cref{thm:co-mean} is
optimal up to polynomial factors also for \emph{randomized} black-box algorithms, in analogy with the
oracle lower bounds of \cite{EsmerKMNS24,EsmerKMNS26}.

\subsection{Balanced set-inclusion families}\label{sec:dr-balanced}

The structured derandomization replaces random samples by a family that contains, inside every
solution, a sample of at least average weight, for every nonnegative weighting at once. We allow
several classes of solution elements, because \cref{thm:dr-det-moments} needs two.

\begin{definition}[Balanced family]\label{def:dr-balanced}
  Let $|U|=n$ and $\mathsf r\ge1$, let $k_1,\dots,k_{\mathsf r}\ge0$ be integers with $\sum_jk_j\le n$, and let $t_j$ be
  integers with $0\le t_j\le k_j$; put $t=\sum_jt_j$. A family $\mathcal B\subseteq\binom Ut$ is
  \emph{$(n;k;t)$-balanced} if for all pairwise disjoint $S_1,\dots,S_{\mathsf r}\subseteq U$ with $|S_j|=k_j$ and
  every $w\colon\bigcup_jS_j\to\R_{\ge0}$, some $X\in\mathcal B$ satisfies $X\subseteq\bigcup_jS_j$,
  $|X\cap S_j|=t_j$ for all $j$, and
  \[
    w(X\cap S_j)\ \ge\ \frac{t_j}{k_j}\,w(S_j)\qquad\text{for every $j$ with $k_j\ge1$.}
  \]
  For $\mathsf r=1$ we say \emph{$(n,k,t)$-balanced}.
\end{definition}

An $(n,k,t)$-balanced family is in particular an $(n,k,t)$-set-inclusion family
(\cref{def:dr-inclusion}).

The construction partitions the universe into blocks of size about $\log n/\log\log n$, guesses how
the unknown solution classes meet the blocks, distributes the targets proportionally, and covers each
block greedily by subsets that capture their proportional share of every weighting; the details and
the proof of the following theorem are in \cref{sec:app-balanced}.

\begin{restatable}[Balanced families in polynomial space]{theorem}{RestThmDrBalanced}\label{thm:dr-balanced}
  Fix $\mathsf r\ge1$. Let $n\ge16$ and $\iota=\max(2,\lfloor\log n/\log\log n\rfloor)$. A deterministic algorithm,
  given $k_1,\dots,k_{\mathsf r}$ and $t_1,\dots,t_{\mathsf r}$ as in \cref{def:dr-balanced}, lists (possibly with
  repetitions) the members of an $(n;k;t)$-balanced family $\mathcal B$ with
  \[
    |\mathcal B|\ \le\ \iota^{12\mathsf r(n/\iota+1)}\,\frac{\binom nt}{\prod_j\binom{k_j}{t_j}}
    \ =\ \frac{\binom nt}{\prod_j\binom{k_j}{t_j}}\,2^{O(\mathsf rn(\log\log n)^{2}/\log n)} .
  \]
  It runs in time $n^{O(1)}$ times this bound, plus $n^{O(1)}$, and in space $n^{O(1)}$.
\end{restatable}

For $n<16$ the family $\binom Ut$ is balanced (for each class take the $t_j$ elements of $S_j$ of
largest weight) and has at most $2^{15}$ members, so the theorem holds for all $n$ after a constant
change in the bound. For $\mathsf r=1$ the size is optimal up to the factor $2^{O(n(\log\log n)^{2}/\log n)}$,
since every $(n,k,t)$-set-inclusion family has at least $\binom nt/\binom kt$ members
(\cref{sec:dr-fgls}).

\subsection{Exponential moments of integer affine certificates}\label{sec:dr-moments}

The balanced families apply to running-time exponents with an \emph{affine certificate}: a solution
$S\in\cF_I$ with $|S|=k$ for which some weights $w\colon S\to\R_{\ge0}$, which may depend on $S$ and
need not be computable, and a constant $\alpha$ satisfy $g(I,X,k)\le\alpha-w(X)$ for the samples
$X\subseteq S$ in question, where only $\alpha$ and a bound $\Xi$ on the total weight $w(S)$ are known
to the algorithm. The exponential-moment theorem can exploit such a certificate more strongly than
the mean-exponent theorem, because $\mu_t(S)$ sees the whole distribution of $w(X)$ and not only its
mean. We first show that if the weights may be arbitrary nonnegative reals, then in the worst case this certifies no smaller running-time bound than the
mean-exponent theorem, and then that for integer weights it certifies a smaller one, and that this smaller
bound can be achieved deterministically.

\paragraph{Forms of the certificate.} \Cref{thm:dr-det-moments} below tries every sample size
$t\ge1$ and needs $g(I,X,k)\le\alpha-w(X)$ for every nonempty $X\subseteq S$. The empty sample is
excluded because the algorithm never lists it and because the inequality can fail there: for
\pname{Connected Dominating Set}, $\alpha=n-1$ while the empty sample leaves all $n$ vertices as
terminals (\cref{lem:cds-pointwise} and~\eqref{eq:cds-TX}). It needs $w(S)=\Xi$ rather than
$w(S)\ge\Xi$. This is no restriction, since lowering integer weights until the total is $\Xi$ keeps it a certificate. And it needs integer weights, because only for them does the
exponential-moment bound of the surrogate gain over its mean-exponent bound (\cref{prop:dr-real}).

Let $S$ with $|S|=k$ have an affine certificate $g(I,X,k)\le\alpha-w(X)$ for all nonempty
$X\subseteq S$, with $w(S)=\Xi$. The \emph{surrogate exponent} $\alpha-w(X)$ is defined for every
$X\subseteq S$, including $X=\emptyset$, and the generating function of \cref{def:fw-moments} for
the surrogate exponent is
\[
  W_S(y)=\sum_{X\subseteq S}y^{|X|}2^{-\alpha+w(X)}=2^{-\alpha}\prod_{u\in S}\bigl(1+y2^{w(u)}\bigr).
\]
\Cref{lem:dr-chord} finds, for every $y$ at once, the weight profile that minimizes this product,
and hence $W_S(y)$.

\begin{restatable}[Chord lemma]{lemma}{RestLemDrChord}\label{lem:dr-chord}
  Let $k\ge1$, let $\Xi\ge0$ be an integer, $a=\lfloor\Xi/k\rfloor$, $k_{a+1}=\Xi-ak$ and
  $k_a=k-k_{a+1}$. Let $v\in\R^{k}$ have $\sum_uv_u=\Xi$ and no entry in the open interval $(a,a+1)$. Then
  for every $y>0$,
  \[
    \prod_u\bigl(1+y2^{v_u}\bigr)\ \ge\ \mathsf P^*(y):=\bigl(1+y2^{a}\bigr)^{k_a}\bigl(1+y2^{a+1}\bigr)^{k_{a+1}} .
  \]
  Without the restriction on the entries, $\prod_u(1+y2^{v_u})\ge(1+y2^{\Xi/k})^{k}$.
\end{restatable}

With the chord lemma we can compare the two theorems on the surrogate exponent.

\begin{restatable}[Real weights give only the mean-exponent bound]{proposition}{RestPropDrReal}\label{prop:dr-real}
  Let $S$ be a $k$-subset of a universe of size $n$, with $k\ge1$, and let $\alpha$ be real and
  $\Xi\ge0$. For $w\colon S\to\R_{\ge0}$ with $w(S)=\Xi$ let $\mu^w_t(S)=\sum_{X\in\binom St}2^{-\alpha+w(X)}$
  for $0\le t\le k$, the exponential moments of the surrogate exponent $\alpha-w(X)$, and put
  $\mathsf R(w)=\min_{0\le t\le k}\binom nt/\mu^w_t(S)$. Let $\bar w\equiv\Xi/k$ be the uniform profile. Then
  $\mathsf R(\bar w)=\min_{0\le t\le k}\binom nt\binom kt^{-1}2^{\alpha-t\Xi/k}$, and $\mathsf R(w)\le(n+1)\,\mathsf R(\bar w)$
  for every such $w$. If $\Xi$ is an integer and $w^*$ is the profile of \cref{lem:dr-chord}, then
  $\mathsf R(w)\le(n+1)\,\mathsf R(w^*)$ for every such $w$ with integer values.
\end{restatable}

Here $\mathsf R(w)$ is the smallest running-time bound $2^{\psi n}$ that \cref{thm:fw-main} with $\Gamma=1$
certifies for $S$ through the surrogate exponent, and $\mathsf R(\bar w)$ is the running-time bound of
\cref{thm:co-mean} with mean bound $\alpha-t\Xi/k$, without its polynomial factor.

So in the worst case over real weights the affine surrogate certifies, through \cref{thm:fw-main}, no smaller
running-time bound than the mean-exponent theorem. For \pname{Connected Dominating Set} the weights
of \cref{lem:cds-pointwise}, the degrees minus one of the vertices of the solution in a spanning tree,
are integers, and for the worst integer profile the running-time bound
certified through \cref{thm:fw-main} is exponentially smaller than that of the mean-exponent theorem
(\cref{sec:det}). The following algorithm derandomizes the
exponential-moment bound for the worst integer profile. It splits the unknown solution into a
\emph{light} class (weights at most $a$) and a \emph{heavy} class (weights at least $a+1$), guesses the
sizes and total weights of the two classes, and runs a two-class balanced family for each guess.

\begin{algorithm}[Integer affine certificates]\label{alg:dr-integer}
  \Input $(I,k)$ together with a rational $\alpha$ and an integer $\Xi\ge0$.
  \Body
  Put $a=\lfloor\Xi/k\rfloor$. If $\alpha<\Xi$, answer \NO{}. Otherwise run the following for every
  \emph{guess} $(k^-,\Xi^-)$ of integers with $1\le k^-\le k$ and $0\le\Xi^-\le ak^-$, where
  $k^+:=k-k^-$ and $\Xi^+:=\Xi-\Xi^-$, subject to
  \[
    \Xi^+\ge(a+1)k^+\ \text{ if }k^+\ge1,\qquad\Xi^+=0\ \text{ if }k^+=0 .
  \]
  \begin{enumerate}[label=(\arabic*)]
    \item Put $\bar w^-=\Xi^-/k^-$ and, if $k^+\ge1$, $\bar w^+=\Xi^+/k^+$. (If $k^+=0$ the heavy class is omitted,
      $t^+=0$, and $\bar w^+$ is not used.)
    \item Choose integers $0\le t^-\le k^-$ and $0\le t^+\le k^+$ with $t:=t^-+t^+\ge1$ that minimize
      $\binom nt\big/\left(\binom{k^-}{t^-}\binom{k^+}{t^+}\right)\cdot2^{\theta}$, where
      $\theta:=\alpha-t^-\bar w^--t^+\bar w^+$, up to a factor $2$, by taking a minimizer of approximations of the
      logarithms of the candidates to within $\frac14$.
    \item List the two-class $(n;k^-,k^+;t^-,t^+)$-balanced family of \cref{thm:dr-balanced} (a one-class family
      if $k^+=0$).
    \item For each listed $X$, compute $g(I,X,k)$; if $g(I,X,k)\le\theta$, run $\cA$ on $(I,X,k)$ and
      answer \YES{} if it accepts.
  \end{enumerate}
  After all guesses, answer \NO{}.
\end{algorithm}

The condition $\Xi^+=0$ for $k^+=0$ matters: without it, a guess with $k^+=0$ and $\Xi^-<\Xi$ would
run a one-class family with mean $\bar w^-<\Xi/k$, and hence a larger threshold $\theta$. Since every
guess is run, this guess, which takes time about $2^{n}$ for the parameters of \cref{sec:det}, would
dominate the total. \Cref{alg:dr-integer} has the following guarantee.

\begin{restatable}[Deterministic exponential moments of integer affine certificates]{theorem}{RestThmDrDetMoments}\label{thm:dr-det-moments}
  Let $k\ge1$, a rational $\alpha=\alpha(I,k)$ and an integer $\Xi=\Xi(I,k)$ with $0\le\Xi\le N^{O(1)}$
  be computable in time $N^{O(1)}$. Say that $S\in\cF_I$ with $|S|=k$ has an \emph{integer certificate}
  if some $w\colon S\to\mathbb Z_{\ge0}$ with $w(S)=\Xi$ satisfies
  \[
    g(I,X,k)\le\alpha-w(X)\qquad\text{for every nonempty }X\subseteq S .
  \]
  Put $a=\lfloor\Xi/k\rfloor$, $k_{a+1}=\Xi-ak$, $k_a=k-k_{a+1}$ and
  $\mathsf P^*(y)=(1+y2^{a})^{k_a}(1+y2^{a+1})^{k_{a+1}}$. \Cref{alg:dr-integer} is deterministic, answers \NO{} if
  $\cF_I$ has no set of size at most $k$, and answers \YES{} if some $S\in\cF_I$ with $|S|=k$ has an integer
  certificate. It runs in time
  \[
    N^{O(1)}\,2^{O(n(\log\log n)^{2}/\log n)}\,
    \max\Bigl(1,\ \inf_{y>0}\frac{\bigl((1+y)^{n}-1\bigr)\,2^{\alpha}}{\mathsf P^*(y)-1}\Bigr)
  \]
  and in space $N^{O(1)}$ plus the space of $\cA$.
\end{restatable}

The maximum with $1$ in the time bound covers the early answer \NO{} and the polynomial bookkeeping;
the infimum alone can be smaller than $1$ when $\alpha<\Xi$. The hypothesis that $\alpha$ is rational
makes the test $\alpha<\Xi$ and the comparisons with $\theta$ exact; a real $\alpha$ is replaced by a
rational $\hat\alpha$ with $\alpha\le\hat\alpha\le\alpha+1$, which preserves every integer certificate
and at most doubles the time.

The proof of \cref{thm:dr-det-moments} is in \cref{sec:app-det-moments}, with those of \cref{lem:dr-chord}
and \cref{prop:dr-real}. It shows that the guess formed by the light and heavy classes of an integer
certificate is tried, that the balanced family for this guess contains a nonempty sample whose weight is at
least its proportional share, and it bounds the running time of each guess by expanding the product over a
two-valued weight profile and applying \cref{lem:dr-chord}.
{}
\section{Deterministic algorithms in polynomial space}\label{sec:det}

The algorithms of \cref{sec:rs,sec:rstheta,sec:rsc} are randomized. With the balanced set-inclusion
families of \cref{sec:derandomisation}, we derandomize the exponential moments of an integer affine surrogate
of the number of terminals (\cref{thm:det-cds}). The resulting algorithm runs in polynomial space, below the
$\Ostar(1.8619^n)$ of Abu-Khzam, Mouawad and Liedloff~\cite{Abu-KhzamML11}. It escapes
the black-box barrier of \cref{cor:dr-bb-rs} because it uses an affine upper bound on the running-time
exponent, the pointwise tree inequality of \cref{lem:cds-pointwise}, and not only the values of the exponent. It does not
reach the randomized bounds of \cref{sec:rstheta,sec:rsc}: their analyses bound the running-time
exponent through tests along an unknown spanning tree of the solution, which are not separable over the
fixed block partitions of the balanced families.

\paragraph{Setting.} All graphs are connected, with $n$ vertices. We use the implicit set system
$\Phi_{\mathrm{CDS}}$ of connected dominating sets, the Steiner extension algorithm $\cA_{\mathrm{St}}$
and its running-time exponent (\cref{def:fw-cost}, with the graph $G$ as the instance $I$)
\[
  g(G,X,k)=|\Tm(J_X)|=n-|N[X]|+\comp(G[X])
\]
from \cref{cor:cds-steiner}: $\cA_{\mathrm{St}}$ decides whether $X$ extends to a connected dominating set
with at most $k$ vertices, deterministically, in time $2^{|\Tm(J_X)|}n^{O(1)}$ and polynomial
space. The running-time exponent does not
depend on $k$. The certificate is the pointwise form of the tree inequality: if
$S$ is a connected dominating set and $F$ a
spanning tree of $G$ in which every vertex outside $S$ is a leaf, the integer weights $w(v)=d_F(v)-1\ge0$
on $S$ satisfy
\begin{equation}\label{eq:det-cert}
  w(S)=n-2\qquad\text{and}\qquad|\Tm(J_X)|\le n-1-w(X)\ \text{ for every nonempty }X\subseteq S,
\end{equation}
by \cref{lem:cds-pointwise} and~\eqref{eq:cds-TX} (for $n\ge2$). In the language of
\cref{sec:derandomisation}, \eqref{eq:det-cert} is an integer certificate (\cref{thm:dr-det-moments}) with
$\alpha=n-1$ and total weight $\Xi=n-2$.

The algorithm decides, for $s=1,\dots,n$, whether $G$ has a connected dominating set of size $s$, by
enumeration for small and large $s$ and by \cref{alg:dr-integer} in between (\cref{lem:det-decide}),
and outputs the least $s$ at which it answers \YES{}. By \cref{lem:cds-maxleaf}, every bound below
holds for \pname{Max Leaf Spanning Tree} as well.

\subsection{Two weight classes}\label{sec:det-two}

For the integer weights of~\eqref{eq:det-cert}, we express the running-time bound of
\cref{thm:dr-det-moments} through the following function.
For $u\in(0,1]$ put $a=\lfloor1/u\rfloor$ and, for $y>0$,
\[
  \chi_y(u)=\log(1+y)+1-\bigl((a+1)u-1\bigr)\log\bigl(1+y2^{a}\bigr)-(1-au)\log\bigl(1+y2^{a+1}\bigr).
\]
Here $u$ will be $s/\Xi$, the ratio of the solution size to the total weight $\Xi$ ($\Xi=n-2$ here,
$\Xi=n-1$ in \cref{sec:cv-wcds}), and the two coefficients are the fractions (relative to $\Xi$) of solution vertices of weight $a$ and $a+1$ in the balanced
integer profile of \cref{lem:dr-chord}. The next lemma bounds $\inf_{y>0}\chi_y(u)$ on the range of $u$ that the algorithm meets, with one
explicit $y$ for each range of $u$.

\begin{restatable}[Rate certificate]{lemma}{RestLemDetCert}\label{lem:det-cert}
For every $u\in[0.15,0.85]$ there is
$y\in\{\frac{1+\sqrt5}8,\frac12,\frac5{16},\frac14,\frac{27}{128},\frac{29}{160}\}$ with
$\chi_y(u)\le\log(\sqrt5-\frac12)$.
\end{restatable}

Further optimization over $y$ does not appear to lower this bound (\cref{sec:ancillary}); the proof does not use this.

With the rate certificate in hand, the integer-affine algorithm gives the deterministic bound.

The running time of the integer-affine algorithm with a certificate of this shape depends only on
$n$, $s$, $\alpha$ and $\Xi$. We record it once, for the two parameter pairs that occur in this paper
($\Xi=n-2$ here, $\Xi=n-1$ for weak connectivity in \cref{sec:cv-wcds}).

\begin{restatable}[Running time of the integer-affine algorithm for tree-type certificates]{lemma}{RestLemDetTime}\label{lem:det-time}
Let $n\ge34$, $0.15n<s<0.8n$, $\Xi\in\{n-2,n-1\}$ and $\alpha=\Xi+1$. For every implicit set
system, extension algorithm and running-time exponent with values in $\{0,\dots,n\}$, \cref{alg:dr-integer} on
$(I,k=s,\alpha,\Xi)$ runs in time
\[
  \bigl(\sqrt5-\tfrac12\bigr)^{n}\,2^{O(n(\log\log n)^{2}/\log n)}\,N^{O(1)} .
\]
\end{restatable}

The next lemma packages one run of the integer-affine algorithm, together with enumeration for the
solution sizes that \cref{lem:det-time} excludes, as a deterministic decision procedure for one
solution size; \cref{sec:further} uses it for the variants.

\begin{restatable}[Deterministic decision for one size]{lemma}{RestLemDetDecide}\label{lem:det-decide}
Let $\Phi$ be an implicit set system, let $\cA$ be a deterministic algorithm that solves
$\Phi$-Extension with a running-time exponent $g$ with values in $\{0,\dots,n\}$, and let
$(\alpha,\Xi)\in\{(n-1,n-2),(n,n-1)\}$. For an instance $I$ with a universe of $n$ elements and
$1\le s\le n$, let $\mathrm{Det}(I,s)$ be the following procedure: if $n<34$, or $s\le0.15n$, or
$s\ge0.8n$, test every $s$-subset of $U_I$ for membership in $\cF_I$ and answer \YES{} if and only if
one of them is a solution; otherwise run \cref{alg:dr-integer} on $(I,k=s,\alpha,\Xi)$ with $\cA$ and
$g$. Then $\mathrm{Det}$ is deterministic, runs in time
$(\sqrt5-\frac12)^{n}2^{O(n(\log\log n)^{2}/\log n)}N^{O(1)}$ and in space $N^{O(1)}$ plus the space of
$\cA$, answers \NO{} if $\cF_I$ has no solution of size at most $s$, and answers \YES{} whenever
$\cF_I$ has a solution $S$ of size exactly $s$ which, in the case $n\ge34$ and $0.15n<s<0.8n$, has an
integer certificate for $(\alpha,\Xi)$ in the sense of \cref{thm:dr-det-moments}.
\end{restatable}

The proofs of \cref{lem:det-cert,lem:det-time,lem:det-decide} and of the following theorem are in
\cref{sec:app-det-cds}. The theorem runs $\mathrm{Det}(G,s)$ of \cref{lem:det-decide} for $s=1,\dots,n$ with
$(\alpha,\Xi)=(n-1,n-2)$, for which the weights of~\eqref{eq:det-cert} are an integer certificate of every
minimum connected dominating set.

\begin{restatable}[Deterministic \pname{Connected Dominating Set} and MaxLeaf]{theorem}{RestThmDetCds}\label{thm:det-cds}
\pname{Connected Dominating Set}, and hence \pname{Max Leaf Spanning Tree} (by \cref{lem:cds-maxleaf}, and directly when $n\le2$), can be solved by a
deterministic algorithm in time
\[
  O\Bigl(\bigl(\sqrt5-\tfrac12\bigr)^{n}\,2^{O(n(\log\log n)^{2}/\log n)}\Bigr)=O(1.7361^{n})
\]
and polynomial space, where $\sqrt5-\frac12=1.7360\ldots$.
\end{restatable}

\paragraph{What \cref{thm:det-cds} derandomizes.} By \cref{prop:dr-real}, the randomized
exponential-moment theorem (\cref{thm:fw-main}) applied to the surrogate exponent $\alpha-w(X)$ certifies,
in the worst case over integer certificates of total $\Xi$, the running-time bound
$\inf_y(1+y)^{n}2^{\alpha}/\mathsf P^*(y)$ up to polynomial factors; whenever the infimum may be taken over
$y$ with $\mathsf P^*(y)\ge2$, this agrees with the bound of \cref{thm:dr-det-moments} up to a factor $2$. So
\cref{thm:dr-det-moments,thm:det-cds} derandomize the exponential-moment bound of the \emph{integer affine
surrogate}, here $n-1-w(X)$, of the number of terminals, up to a factor $2^{o(n)}$ in the running time; they
do \emph{not} derandomize \cref{thm:fw-main} for the true running-time exponent, whose exponential moments can
be much larger. The surrogate bound $1.7361^n$ already improves on the mean-exponent analysis of
Algorithm RS ($1.7449^n$), but not on Algorithm RS-$\theta$ ($1.7204^n$, \cref{thm:rstheta}), whose tree
potential captures correlations that the surrogate discards, nor on RSC ($1.6991^n$, \cref{thm:rsc}). The
factor $2^{o(n)}$ that the balanced families contribute is large for practical $n$, about $2^{12718}$ at
$n=1000$ (\cref{sec:app-balanced}).
{}
\section{Component-count running times}\label{sec:cc}

This section treats problems whose solutions are vertex sets $S$ of a graph $G$ with $G[S]$
connected, and whose extension algorithms run in time exponential in the number of components of
the sample. Given a sample $X$ of size $t$ from a solution of size $k$, with $\comp(G[X])$ the
number of components of $G[X]$, the extension algorithm runs in time
\[
  \Ostar\bigl(c^{\,k-t}\,2^{\comp(G[X])}\bigr),
\]
with $c=3$ for \pname{Connected Feedback Vertex Set} (a Cut~\&~Count algorithm) and $c=2$ for
\pname{Graph Motif} and \pname{Connected Vertex Cover} (the algorithms of Björklund, Kaski and
Kowalik and of Cygan). A sample of size $t$ can have $t$ components, so a running-time bound that
depends on $X$ only through $t$ is the product-form bound $c^{k-t}2^t$, for which multivariate
Monotone Local Search (\cref{cor:co-multivariate} with $(b,c,d)=(1,c,2)$) gives the base
$3-2/c\ge2$ when $c\ge2$, no better than enumeration. Our improvements come from the fact that a
random part of a connected solution has few components once it is a sizeable fraction of the
solution.

We use this fact through a lower bound on $W_S(y)$ by a partition function of a spanning
tree of the solution (\cref{def:cc-Z}), which paths minimize up to a factor independent of $k$
(\cref{sec:cc-gibbs}). \Cref{sec:cc-cost} turns this into the running time
$\Ostar(((4c-3)/(2c-1))^n)$ for Algorithm~\MLS{}; the applications follow in
\cref{sec:cfvs,sec:motif,sec:cc-apps}.

Throughout, $H$ is the binary entropy of \cref{sec:fw-setting}, with the logarithm to base~$2$, and for a
probability distribution $\mathbb P$ on a finite set, $H_e(\mathbb P)=-\sum_z\mathbb P(z)\ln\mathbb P(z)$ is its
entropy with the natural logarithm.
For a graph $F$ and $X\subseteq V(F)$ we write
$e(F[X])=|E(F[X])|$.

\subsection{The partition function of a tree}\label{sec:cc-gibbs}

\begin{definition}[Tree partition function]\label{def:cc-Z}
  For a tree $F$ and a real $\zeta\ge0$ let
  \[
    \mathcal Z_F(\zeta)=\sum_{X\subseteq V(F)}\zeta^{|X|}\,2^{e(F[X])},
  \]
  let $M_\zeta=\bigl(
    \begin{smallmatrix}1&\zeta\\1&2\zeta
  \end{smallmatrix}\bigr)$, and let $\lambda_{\max}(M_\zeta)$ be its largest eigenvalue, that is,
  \[
    \lambda_{\max}(M_\zeta)=\tfrac12\Bigl(1+2\zeta+\sqrt{(1+2\zeta)^2-4\zeta}\Bigr)=\tfrac12\Bigl(1+2\zeta+\sqrt{1+4\zeta^2}\Bigr).
  \]
  Both eigenvalues of $M_\zeta$ are real and nonnegative, since the discriminant $1+4\zeta^2$ is positive and
  their sum $1+2\zeta$ and product $\zeta$ are nonnegative. So $\lambda_{\max}(M_\zeta)$ is also the largest
  absolute value of an eigenvalue of $M_\zeta$, its spectral radius.
\end{definition}

The function $\zeta\mapsto\lambda_{\max}(M_\zeta)$ is continuous and increasing on $[0,\infty)$, with $\lambda_{\max}(M_0)=1$, and
$\lambda_{\max}(M_\zeta)>1$ for $\zeta>0$. Since $\lambda_{\max}(M_\zeta)$ is the larger root of $x^2-(1+2\zeta)x+\zeta$, it is at least
half the trace: $\lambda_{\max}(M_\zeta)\ge(1+2\zeta)/2$. On a path, $M_\zeta$ is the transfer matrix of $\mathcal Z_F(\zeta)$: the
state of a vertex is $1$ if it lies in $X$ and $0$ otherwise, a vertex in state $1$ has weight
$\zeta$, and an edge between two vertices in state $1$ has weight~$2$. The first lemma says that no
tree on $k$ vertices has a partition function smaller than $\lambda_{\max}(M_\zeta)^{k-1}$, up to a factor
depending only on~$\zeta$.

The quantity $\mathcal Z_F(\zeta)$ is a weighted count of homomorphisms from $F$ to a target with two vertices,
$0$ and $1$, with vertex weights $w_0=1$ and $w_1=\zeta$ and the symmetric edge weights
$A=\bigl(\begin{smallmatrix}1&1\\1&2\end{smallmatrix}\bigr)$ (a loop of weight~$1$ at vertex~$0$, an edge of weight~$1$
between the two vertices, and a loop of weight~$2$ at vertex~$1$), as in the proof below. For graphs
without loops or weights, Csikv\'ari and Lin~\cite[Theorem~4.3]{CsikvariL14} proved that every tree $T$ with
$m$ vertices has $\hom(T,G)\ge e^{\eta}\lambda^{m-1}$ homomorphisms into a connected graph $G$, where $\lambda$ is the largest
adjacency eigenvalue of $G$ and $\eta\ge0$ is the natural-logarithm entropy of the probability vector formed by the squared entries of a unit Perron eigenvector. They
proved it by running a stationary Markov chain on $G$ down the tree, and the proof below follows theirs, with the
vertex weights absorbed into a symmetric matrix. Whether the path minimizes such counts exactly depends on the
target. Sidorenko~\cite{Sidorenko94} proved that among the trees with a given number of vertices the star
has the most homomorphisms into every graph, whereas for some graphs some trees have fewer homomorphisms than
the path~\cite{GalvinMNR25}. Galvin, Marmorino, McMillon, Nir and Redlich~\cite{GalvinMMNR25} give criteria for
paths to be the minimizers, for targets without edge weights (with weights on the vertices only); our target
carries the weight $2$ on a loop, and we know of no exact result for it. Our analysis needs only the bound up
to the factor $\min(1,\zeta)$.

\begin{lemma}[Gibbs bound, after Csikv\'ari and Lin]\label{lem:cc-gibbs}
  For every tree $F$ with $k\ge1$ vertices and every $\zeta>0$,
  \[
    \mathcal Z_F(\zeta)\ \ge\ \min(1,\zeta)\,\lambda_{\max}(M_\zeta)^{k-1}.
  \]
\end{lemma}

\begin{proof}
  We write $\mathcal Z_F(\zeta)$ as a sum of positive weights $f(z)$ over labelings $z$ of $V(F)$,
  construct a probability distribution $\mathbb P$ on the labelings from a reversible two-state Markov chain run
  down the tree, and apply the Gibbs inequality $\ln \mathcal Z_F(\zeta)\ge\E_{\mathbb P}\ln f+H_e(\mathbb P)$; for this $\mathbb P$
  the right-hand side is $(k-1)\ln\lambda_{\max}(M_\zeta)$ up to a term depending only on~$\zeta$.

  Write $w_0=1$, $w_1=\zeta$, $A_{ij}=2$ if $i=j=1$ and $A_{ij}=1$ otherwise, for $i,j\in\{0,1\}$. A
  subset $X$ of $V(F)$ is a labeling $z\colon V(F)\to\{0,1\}$, and
  $\mathcal Z_F(\zeta)=\sum_z f(z)$ with
  \[
    f(z)=\prod_{v\in V(F)}w_{z_v}\prod_{uv\in E(F)}A_{z_uz_v}>0 .
  \]

  \emph{A reversible chain.} Let $\mathbf K_{ij}=\sqrt{w_i}\,A_{ij}\sqrt{w_j}$, that is,
  $\mathbf K=\bigl(
    \begin{smallmatrix}1&\sqrt \zeta\\\sqrt \zeta&2\zeta
  \end{smallmatrix}\bigr)$. This is a symmetric
  matrix with the same characteristic polynomial $x^2-(1+2\zeta)x+\zeta$ as $M_\zeta$, so its largest
  eigenvalue is $\lambda_{\max}(M_\zeta)$. The vector $(\sqrt \zeta,\lambda_{\max}(M_\zeta)-1)$ is an eigenvector for
  $\lambda_{\max}(M_\zeta)$, by the first row of $\mathbf K-\lambda_{\max}(M_\zeta)\,\mathrm{Id}$, with $\mathrm{Id}$ the identity matrix, and both entries are positive, since $\lambda_{\max}(M_\zeta)>1$.
  Let $\mathbf v$ be this vector normalized to $\mathbf v_0^2+\mathbf v_1^2=1$, and put
  \[
    \pi_i=\mathbf v_i^2,\qquad p_{ij}=\frac{\mathbf K_{ij}\mathbf v_j}{\lambda_{\max}(M_\zeta)\mathbf v_i}\qquad(i,j\in\{0,1\}).
  \]
  Then $p_{ij}>0$ and $\sum_jp_{ij}=(\mathbf K\mathbf v)_i/(\lambda_{\max}(M_\zeta)\mathbf v_i)=1$, so $(p_{ij})$ is a stochastic matrix.
  Moreover $\pi_{ij}:=\pi_ip_{ij}=\mathbf v_i\mathbf K_{ij}\mathbf v_j/\lambda_{\max}(M_\zeta)$ is symmetric in $i$ and $j$, so
  $\sum_i\pi_ip_{ij}=\sum_i\pi_jp_{ji}=\pi_j$: the probability vector $\pi$ is stationary for $(p_{ij})$.

  \emph{A labeling of the tree.} Root $F$ at a vertex $r$. Let $\mathbb P$ be the distribution of the random
  labeling that gives $r$ a label drawn from $\pi$ and then, going down the tree, gives every child
  $v$ of a vertex $u$ a label drawn from the row $p_{z_u,\cdot}$, independently of everything
  else:
  \[
    \mathbb P(z)=\pi_{z_r}\prod_{u\to v}p_{z_uz_v}>0,
  \]
  the product over the $k-1$ pairs of a parent $u$ and a child $v$. By stationarity and induction on
  the depth, every vertex has label distribution $\pi$. Hence every edge, with parent $u$ and child $v$, has
  joint label distribution $\Pr[z_u=i,z_v=j]=\pi_{ij}$, which is symmetric and has both marginals
  equal to~$\pi$.

  \emph{The Gibbs inequality.} For any probability distribution $\mathbb P$ on the labelings, Jensen's inequality
  for the concave function $\ln$ gives
  \[
    \E_{\mathbb P}\ln f+H_e(\mathbb P)=\sum_{z:\,\mathbb P(z)>0}\mathbb P(z)\ln\frac{f(z)}{\mathbb P(z)}
    \le\ln\sum_{z}f(z)=\ln \mathcal Z_F(\zeta),
  \]
  where $H_e(\mathbb P)=-\sum_z\mathbb P(z)\ln\mathbb P(z)$. We evaluate the left-hand side for our
  $\mathbb P$. By the product form of $\mathbb P$ and the marginals above,
  \[
    H_e(\mathbb P)=H_e(\pi)-(k-1)\sum_{i,j}\pi_{ij}\ln p_{ij},\qquad
    \E_{\mathbb P}\ln f=k\sum_i\pi_i\ln w_i+(k-1)\sum_{i,j}\pi_{ij}\ln A_{ij}.
  \]
  Since $\ln A_{ij}=\ln \mathbf K_{ij}-\frac12\ln w_i-\frac12\ln w_j$ and both marginals of $(\pi_{ij})$ are $\pi$,
  we have $\sum\pi_{ij}\ln A_{ij}=\sum\pi_{ij}\ln \mathbf K_{ij}-\sum_i\pi_i\ln w_i$. Since
  $\ln p_{ij}=\ln \mathbf K_{ij}+\ln\mathbf v_j-\ln\mathbf v_i-\ln\lambda_{\max}(M_\zeta)$ and $(\pi_{ij})$ is symmetric, the terms in $\mathbf v$
  cancel, and $-\sum\pi_{ij}\ln p_{ij}=\ln\lambda_{\max}(M_\zeta)-\sum\pi_{ij}\ln \mathbf K_{ij}$. Adding up, the terms in
  $\mathbf K$ cancel as well, and
  \[
    \ln \mathcal Z_F(\zeta)\ \ge\ \sum_i\pi_i\ln w_i+H_e(\pi)+(k-1)\ln\lambda_{\max}(M_\zeta)
    =\pi_1\ln \zeta+H_e(\pi)+(k-1)\ln\lambda_{\max}(M_\zeta) .
  \]
  Finally $H_e(\pi)\ge0$ and $\pi_1\in[0,1]$, so $\pi_1\ln \zeta\ge\ln\min(1,\zeta)$.
\end{proof}

Paths attain the bound of \cref{lem:cc-gibbs} up to a factor depending only on~$\zeta$, by the next
lemma. This is why the tightness results below (\cref{prop:cc-tight}) use solutions whose tree is
a path.

\begin{lemma}[Paths]\label{lem:cc-path}
  For a path $F$ with $k\ge1$ vertices and every $\zeta\ge0$,
  $\mathcal Z_F(\zeta)\le(1+\zeta)\,\lambda_{\max}(M_\zeta)^{k-1}\le2\lambda_{\max}(M_\zeta)^k$.
\end{lemma}

\begin{proof}
  Let $v_1,\dots,v_k$ be the vertices of the path in order, and $\mathbf d=(1,\sqrt \zeta)$. With $\mathbf K$ as in the
  proof of \cref{lem:cc-gibbs} (also for $\zeta=0$), each labeling $z$ contributes
  $\prod_vw_{z_v}\prod_{i<k}A_{z_{v_i}z_{v_{i+1}}}=\mathbf d_{z_{v_1}}
  \prod_{i<k}\mathbf K_{z_{v_i}z_{v_{i+1}}}\,\mathbf d_{z_{v_k}}$, because every inner vertex lies on
  two edges and each end on one. Hence $\mathcal Z_F(\zeta)=\mathbf d^{\mathsf T}\mathbf K^{k-1}\mathbf d$. As $\mathbf K$ is symmetric with
  eigenvalues in $[0,\lambda_{\max}(M_\zeta)]$ (their product is $\zeta\ge0$ and their sum is positive), this is at
  most $\|\mathbf d\|^2\lambda_{\max}(M_\zeta)^{k-1}=(1+\zeta)\lambda_{\max}(M_\zeta)^{k-1}$. Finally $1+\zeta\le1+2\zeta\le2\lambda_{\max}(M_\zeta)$.
\end{proof}

\subsection{Component-count running times and exponential-moment Monotone Local Search}\label{sec:cc-cost}

We now state the property of a running-time exponent that the applications share, and the running time
that \cref{thm:fw-main} gives for it. Fix an implicit set system $\Phi$ and an algorithm $\cA$ for
$\Phi$-Extension with running-time exponent $g$ (\cref{def:fw-cost}); recall
$W_S(y)=\sum_{X\subseteq S}y^{|X|}2^{-g(I,X,k)}$ for a solution $S$ of size $k$
(\cref{def:fw-moments}).

\begin{definition}[Component-count bound]\label{def:cc-bound}
  Let $c\ge1$ and $\gamma\ge0$ be constants. The running-time exponent $g$ has a \emph{component-count bound
  with constants $(c,\gamma)$} if for every instance $I$ and every solution $S\in\cF_I$ with
  $|S|=k\ge1$ there is a tree $F$ with vertex set $S$ such that
  \begin{equation}\label{eq:cc-bound}
    g(I,X,k)\ \le\ (k-|X|)\log c+|X|-e(F[X])+\gamma\qquad\text{for every }X\subseteq S .
  \end{equation}
\end{definition}

The tree $F$ is used only in the analysis. Typically the solutions induce connected subgraphs of
a graph $G$, the running-time exponent is $(k-|X|)\log c+\comp(G[X])$, and $F$ is a spanning tree of $G[S]$: then
$F[X]$ is a spanning forest of a subgraph of $G[X]$, so
$\comp(G[X])\le\comp(F[X])=|X|-e(F[X])$, by the edge count for forests. For $\sigma\in[0,1]$ and $y\ge0$
put
\begin{equation}\label{eq:cc-phi}
  \phi_{c,\sigma}(y)=\log(1+y)-\sigma\log\frac{\lambda_{\max}(M_{cy/2})}{c}.
\end{equation}
For fixed $y$ this is an affine function of $\sigma$, so a bound on it over an interval of values of
$\sigma$ follows from its values at the two ends; the certificate of \cref{sec:cc-motif} uses this.

\begin{corollary}[Generating function of a component-count running time]\label{cor:cc-T}
  Let $g$ have a component-count bound with constants $(c,\gamma)$, let $S\in\cF_I$ with $|S|=k\ge1$,
  let $F$ be a tree for $S$ as in \cref{def:cc-bound}, and put $\sigma=k/n$.
  \begin{enumerate}[label=\textup{(\alph*)}]
    \item For every $y>0$, with $\zeta=cy/2$,
      \[
        W_S(y)\ \ge\ 2^{-\gamma}c^{-k}\mathcal Z_F(\zeta)\ \ge\ 2^{-\gamma}\,\frac{\min(1,\zeta)}{\lambda_{\max}(M_\zeta)}
        \Bigl(\frac{\lambda_{\max}(M_\zeta)}{c}\Bigr)^{k},
      \]
      and therefore
      \[
        \frac{(1+y)^n}{W_S(y)}\le2^\gamma\,\frac{\lambda_{\max}(M_\zeta)}{\min(1,\zeta)}\,2^{n\phi_{c,\sigma}(y)} .
      \]
    \item Put $y^*=\frac{2(c-1)}{2c-1}$ and $\zeta^*=cy^*/2=\frac{c(c-1)}{2c-1}$. Then $\lambda_{\max}(M_{\zeta^*})=c$,
      and if $c>1$,
      \[
        W_S(y^*)\ \ge\ 2^{-\gamma}\,\frac{\min(1,\zeta^*)}{c},\qquad
        \frac{(1+y^*)^n}{W_S(y^*)}\le\frac{2^\gamma c}{\min(1,\zeta^*)}\Bigl(\frac{4c-3}{2c-1}\Bigr)^n .
      \]
    \item $W_S(0)=\mu_0(S)\ge2^{-\gamma}c^{-k}$ and $\mu_k(S)\ge2^{-1-\gamma}$.
  \end{enumerate}
\end{corollary}

\begin{proof}
  (a) By~\eqref{eq:cc-bound}, each term of $W_S(y)$ satisfies
  $y^{|X|}2^{-g(I,X,k)}\ge2^{-\gamma}c^{-k}(cy/2)^{|X|}2^{e(F[X])}$. Summing over $X\subseteq S$
  gives the first inequality, and \cref{lem:cc-gibbs} the second. The last form follows by
  dividing $(1+y)^n$ by this bound and writing $(c/\lambda_{\max}(M_\zeta))^k=2^{-n\sigma\log(\lambda_{\max}(M_\zeta)/c)}$.

  (b) The number $c$ is a root of $x^2-(1+2\zeta^*)x+\zeta^*$, because
  $c^2-c-\zeta^*(2c-1)=c^2-c-c(c-1)=0$. The other root is $\zeta^*/c=(c-1)/(2c-1)<c$, so
  $\lambda_{\max}(M_{\zeta^*})=c$. Now (a) gives
  $W_S(y^*)\ge2^{-\gamma}(\min(1,\zeta^*)/c)\cdot1^k$, and $1+y^*=(4c-3)/(2c-1)$.

  (c) For $X=\emptyset$ the right-hand side of~\eqref{eq:cc-bound} is $k\log c+\gamma$; for $X=S$ it
  is $k-(k-1)+\gamma=1+\gamma$, because $F$ has $k-1$ edges.
\end{proof}

Part~(b) uses the same $y^*$ for every solution size and, through \cref{cor:cc-rate} below, gives
the closed-form base $(4c-3)/(2c-1)$, which is $1.8$ for $c=3$ and $5/3$ for $c=2$. Part~(a)
allows $y$ to depend on $\sigma$, and part~(c)
covers the two ends: no sampling ($t=0$) and enumeration ($t=k$). Together they give the following
consequence of \cref{thm:fw-main}.

\begin{corollary}[Algorithm \MLS{} for component-count running times]\label{cor:cc-rate}
  Let $\cA$ solve $\Phi$-Extension, either deterministically with running-time exponent $g$, or with
  one-sided error as in \cref{cor:fw-onesided}, and let $g$ have a component-count bound with
  constants $(c,\gamma)$. Let $\psi>0$ and $\Gamma_0\ge1$ be constants such that every $\sigma\in(0,1]$
  satisfies at least one of
  \begin{enumerate}[label=\textup{(R\arabic*)}]
    \item\label{it:cc-R1} $\sigma\log c\le\psi$;
    \item\label{it:cc-R2} $H(\sigma)\le\psi$;
    \item\label{it:cc-R3} $\phi_{c,\sigma}(y)\le\psi$ for some $y>0$ with
      $\lambda_{\max}(M_{cy/2})/\min(1,cy/2)\le\Gamma_0$.
  \end{enumerate}
  Then for every $k\ge1$, every instance $I$ and every $\kappa>0$ as in \cref{thm:fw-main},
  Algorithm~\MLS$(\psi,\Gamma,\kappa)$ with
  $\Gamma=2^{1+\gamma}\Gamma_0$, given $(I,k)$, runs in time $2^{\psi n}N^{O(1)}$ and in
  space $N^{O(1)}$ plus the space of $\cA$, answers \NO{} if $\cF_I$ contains no set of size at
  most $k$, and answers \YES{} with probability at least $1-e^{-\kappa}$ if $\cF_I$ contains a set of
  size exactly~$k$. In particular this holds with $\psi=\log((4c-3)/(2c-1))$ for every $c>1$.
\end{corollary}

\begin{proof}
  Items (i) and (ii) of \cref{thm:fw-main} give the time, space and soundness; for a randomized
  $\cA$ use \cref{cor:fw-onesided}, which doubles $\Gamma$ in the cap and in the trial counts. The
  rate $\psi$ is irrational in general, for instance $\psi=\log1.8$; the cap and the trial counts
  are then any integers in the ranges of~\eqref{eq:fw-params}, which are found in time $N^{O(1)}$ from
  an approximation of $\psi$ to within $1/(4n)$. For item~(iii), let
  $S\in\cF_I$ with $|S|=k$ and $\sigma=k/n$. If~\ref{it:cc-R1} holds, then by \cref{cor:cc-T}(c)
  $W_S(0)\ge2^{-\gamma}2^{-\sigma n\log c}\ge(1+0)^n2^{-\psi n}/\Gamma$. If~\ref{it:cc-R2} holds,
  then $\binom nk\le2^{nH(\sigma)}\le2^{\psi n}$, and \cref{cor:cc-T}(c) gives
  $\mu_k(S)/\binom nk\ge2^{-1-\gamma}2^{-\psi n}\ge2^{-\psi n}/\Gamma$, which
  is~\eqref{eq:fw-star}. If~\ref{it:cc-R3} holds with $y$, then \cref{cor:cc-T}(a) gives
  $(1+y)^n/W_S(y)\le2^\gamma\Gamma_02^{\psi n}\le\Gamma2^{\psi n}$. In the first and third case
  the condition $W_S(y)\ge(1+y)^n2^{-\psi n}/\Gamma$ of \cref{thm:fw-main} holds, with $y=0$ in the
  first case, and it implies~\eqref{eq:fw-star}. For the last claim,
  \ref{it:cc-R3} holds for every $\sigma$ with $y=y^*$ and $\Gamma_0=c/\min(1,\zeta^*)$, by
  \cref{cor:cc-T}(b).
\end{proof}

The parameters of \cref{cor:cc-rate} have the following roles. The constant $\psi$ is the rate:
the target running time is $2^{\psi n}$. Condition~\ref{it:cc-R1} covers the
solution fractions $\sigma$ for which running $\cA$ without a sample ($t=0$) is fast enough,
\ref{it:cc-R2} those for which enumerating $k$-subsets ($t=k$) is, and \ref{it:cc-R3} those for
which the generating-function bound of \cref{cor:cc-T}(a) is. The constant $\Gamma_0$ bounds the
factor $\lambda_{\max}(M_\zeta)/\min(1,\zeta)$ of that bound; together with the factor $2^{1+\gamma}$ it enters only
the loss factor $\Gamma$ of condition~\eqref{eq:fw-star}, not the rate. Real-valued running-time exponents are
handled as in \cref{rem:fw-real}. The decision problem is solved by
running the algorithm for every $k$ (\cref{rem:fw-decision}); membership of $\emptyset$ in
$\cF_I$ is tested directly.

\paragraph{The best rate for a given $c$.}
The best rate $\psi$ that \cref{cor:cc-rate} can certify for a given $c$, when $y$
in~\ref{it:cc-R3} may be chosen freely, is
\[
  \psi_c=\max_{0\le\sigma\le1}\ \min\Bigl(\sigma\log c,\ H(\sigma),\ \inf_{y>0}\phi_{c,\sigma}(y)\Bigr).
\]
By \cref{lem:cc-path}, solutions whose tree $F$ is a path with equality in~\eqref{eq:cc-bound} attain
this rate up to polynomial factors by the argument of \cref{prop:cc-tight}(a), which makes this precise for $c=3$; and
\cref{lem:cc-gibbs} shows that no other tree has a smaller partition function $\mathcal Z_F(\zeta)$, up to a
factor depending only on~$\zeta$. For $c=3$ the closed form is optimal: $\psi_3\le\log1.8$ by \cref{cor:cc-T}(b), and $\psi_3\ge\log1.8$,
since at $\sigma=26/45$ the infimum equals $\log1.8$ (\cref{lem:cc-inf}) and both $\sigma\log3$ (by the
proof of \cref{lem:cc-inf}) and $H(\sigma)$ are larger: as $1/2\le26/45<3/5$ and $H$ decreases on $[1/2,1]$,
$H(\sigma)\ge H(\tfrac35)=\log5-\tfrac35\log3-\tfrac25>2\log3-\log5=\log1.8$, which is equivalent to
$5^{10}=9765625>6377292=4\cdot3^{13}$. For $c=2$ it is
not: the base $5/3$ is larger than $2^{\psi_2}\approx1.6443$, because at the solution fraction $\sigma$
that attains the maximum in $\psi_2$ the infimum over $y$ is not attained at~$y^*$. The following
table compares, for a few $c$, three bases: the base $\max_\sigma\min(c^\sigma,2^{H(\sigma)})$ obtained by
running $\cA$ without a sample in time $\Ostar(c^k)$ or enumerating all $k$-subsets, whichever is
faster; the base given by the mean-exponent theorem (\cref{thm:co-mean}), for which the expected number
of components of a uniformly random $t$-subset $X$ of a connected $k$-set $S$ is at most
$t-t(t-1)/k$ (each of the $k-1$ edges of a spanning tree of $G[S]$ lies in $G[X]$ with probability
$t(t-1)/(k(k-1))$); and $2^{\psi_c}$.
All entries are floating-point evaluations and are not used in any proof, except $1.8$, which is
exact; for $c=2$, \cref{lem:cc-cert} certifies $2^{\psi_2}\le1.6444$, and for $c=4$ and $c=6$ the
entries in the column $2^{\psi_c}$ agree to five decimals with the closed-form upper bounds
$13/7=1.857142\ldots$ and $21/11=1.909090\ldots$ of \cref{cor:cc-T}(b).

\begin{center}
  \begin{tabular}{lccc}
    \toprule
    $c$ & $\max_\sigma\min(c^\sigma,2^{H(\sigma)})$ & mean exponent & $2^{\psi_c}$\\
    \midrule
    $2$ & $1.70871$ & $1.64877$ & $1.64429$\\
    $3$ & $1.95258$ & $1.80802$ & $1.8$\\
    $4$ & $2$ & $1.86182$ & $1.85714$\\
    $6$ & $2$ & $1.91121$ & $1.90909$\\
    \bottomrule
  \end{tabular}
\end{center}

\subsection{Connected Feedback Vertex Set}\label{sec:cfvs}

A \emph{connected
feedback vertex set} of a graph $G$ is a set $S\subseteq V(G)$ such that $G-S$ is a forest and
$G[S]$ is connected. We write $\OPT(G)$ for the minimum size of a connected feedback vertex set of
$G$, if one exists; \pname{Connected Feedback Vertex Set} asks, given $G$, for one of size
$\OPT(G)$. We are not aware of a published exact algorithm for \pname{Connected Feedback Vertex Set} that
runs in time $O((2-\varepsilon)^n)$ for a constant $\varepsilon>0$. Without the connectivity constraint,
\cite{FominGPR08} find a minimum feedback vertex set in time $O(1.7548^n)$ and show
that a graph with $n$ vertices has at most $1.8638^n$ minimal feedback vertex sets.
\cite{MisraPRSS12} gave the first fixed-parameter algorithm for
\pname{Connected Feedback Vertex Set}, in time $c^kn^{O(1)}$ for a constant $c$. \cite{CyganNPPRW22} solve the
problem by a Monte Carlo algorithm in time $3^kn^{O(1)}$ and polynomial space, where $k$ is the
solution size. For each $k$, running this algorithm or enumerating
all $k$-subsets, whichever is faster, gives time $\max_{0\le\sigma\le1}\min(3^{\sigma},2^{H(\sigma)})^n
n^{O(1)}$; a floating-point evaluation of the maximum gives the base $1.9526$, at $\sigma\approx0.6091$. This
combination is folklore; we do not know of it in print, and it is the bound that we improve.

We sample $X$, force it into the solution and complete it by a Cut~\&~Count algorithm that, after
contracting each component of $G[X]$ to a forced vertex, runs in time $3^{k-t}2^{\comp(G[X])}n^{O(1)}$
(\cref{sec:fvs-graphs,sec:fvs-cc,sec:fvs-ext}), and \cref{sec:cc-cfvs} analyzes the sampling with the
exponential moments.

\subsubsection{Problem and the role of the sample}\label{sec:fvs-problem}

Graphs are finite and simple. We use the convention that the graph with no vertices is connected,
so that $\emptyset$ is a connected feedback vertex set of every forest.
It suffices to treat connected graphs that are not forests. If $G$ is a forest, then $\OPT(G)=0$.
If $G$ is disconnected, a connected feedback vertex set lies inside one component of $G$, so all
other components must be forests; hence $G$ has a connected feedback vertex set if and only if at
most one component $G_1$ of $G$ contains a cycle, and then the connected feedback vertex sets of
$G$ that are nonempty are exactly those of $G_1$. Both checks take polynomial time. For a connected
graph $G$, the set $V(G)$ itself is a connected feedback vertex set, so $\OPT(G)$ exists and is at
most~$n$.

For a connected graph $G$ that is not a forest, the connected feedback vertex sets form an implicit
set system in the sense of \cref{def:fw-iss}: the instance is $G$, the universe is $V(G)$ with
$n=|V(G)|$, and membership is decided by testing that $G-S$ is acyclic and $G[S]$ is connected.
Every solution is nonempty. $\Phi$-Extension for this system (\cref{def:fw-iss}), which we call
\pname{Connected Feedback Vertex Set Extension}, asks, given a connected graph $G$, a set
$X\subseteq V(G)$ and an integer $k\ge0$, whether some $Y\subseteq V(G)\setminus X$ with $|Y|\le k-|X|$
makes $X\cup Y$ a connected feedback vertex set of $G$.

\paragraph{Why the improvement needs a sample-dependent running-time exponent.}
The following three facts show that the improvement does not come from classical or multivariate
Monotone Local Search. First, the extension problem is not fixed-parameter tractable in the number $k-|X|$ of
further vertices under standard assumptions (\cref{lem:fvs-w2}), so the theorem of \cite{FominGLS19}, which needs an extension algorithm with running time $c^{k-|X|}N^{O(1)}$,
does not apply. Second, the product-form running-time bound that our extension algorithm satisfies
in the worst case $\comp(G[X])=|X|$ is $3^{k-|X|}2^{|X|}$, that is, $(b,c,d)=(1,3,2)$, and \cref{cor:co-multivariate} turns it into
$(1+2-2/3)^n=(7/3)^n$, worse than enumeration. Third, under the Set Cover Conjecture, no extension
algorithm with a running-time bound $c^{k-|X|}d^{|X|}N^{O(1)}$ gives a base below $2$ through
\cref{cor:co-multivariate} (\cref{prop:fvs-sc}). The running-time exponent we use instead is
\begin{equation}\label{eq:fvs-g}
  g(G,X,k)=(k-|X|)\log3+\comp(G[X]),
\end{equation}
which depends on $X$ through the number of components of $G[X]$. The proofs of the
first and the third fact are in \cref{sec:app-fvs} (\cref{lem:fvs-w2,prop:fvs-sc}).

\subsubsection{The extension algorithm}\label{sec:fvs-ext}

The extension algorithm contracts each component of the sample to a forced vertex and then follows the
iterative compression of \cite{CyganNPPRW11arxiv,CyganNPPRW22} along a sequence of edge
contractions, with a Cut \& Count test that keeps the forced vertices in the solution; its running time is
exponential in the number of forced vertices, which is $\comp(G[X])$. The construction and the proof of
the following lemma are in \cref{sec:app-fvs}.

\begin{restatable}[The extension algorithm]{lemma}{RestLemFvsExt}\label{lem:fvs-ext}
  There is a randomized algorithm $\cA$ for \pname{Connected Feedback Vertex Set Extension} that,
  given a connected graph $G$ with $n$ vertices, a set $X\subseteq V(G)$ and an integer $k\ge|X|$,
  runs in polynomial space and, for every outcome of its coins, in time
  $3^{k-|X|}2^{\comp(G[X])}n^{O(1)}$. If it accepts, it outputs a set $Y\subseteq V(G)\setminus X$
  with $|Y|\le k-|X|$ such that $X\cup Y$ is a connected feedback vertex set of $G$. On a
  \YES{}-instance it accepts with probability at least $3/4$. In particular, the running-time
  exponent~\eqref{eq:fvs-g} bounds its running time.
\end{restatable}

\subsubsection{Exponential moments: the base \texorpdfstring{$1.8$}{1.8}}\label{sec:cc-cfvs}

We now run Algorithm~\MLS{} with the extension algorithm of \cref{sec:fvs-ext} and analyze it
through the exponential moments of the running-time exponent~\eqref{eq:fvs-g}.

\begin{theorem}[Connected Feedback Vertex Set]\label{thm:cc-cfvs}
  There is a randomized algorithm that, given a connected graph $G$ with $n$ vertices, outputs a
  connected feedback vertex set of $G$ in time $\Ostar(1.8^n)$ and polynomial space. The output
  always has size at least $\OPT(G)$, and it has size exactly $\OPT(G)$ with probability at least
  $1-1/e$. For the running-time exponent~\eqref{eq:fvs-g}, the rate $\log1.8$ of Algorithm~\MLS{} cannot
  be lowered by a positive constant (\cref{prop:cc-cfvs-tight}).
\end{theorem}

\begin{proof}
  If $G$ is a forest, output $\emptyset$. Otherwise the connected feedback vertex sets form an
  implicit set system whose solutions are nonempty (\cref{sec:fvs-problem}). The extension
  algorithm of \cref{lem:fvs-ext} has one-sided error, accepts a \YES{}-instance with probability
  at least $3/4$, and runs in time $2^{g(G,X,k)}n^{O(1)}$ for every outcome of its coins, with $g(G,X,k)=(k-|X|)\log3+\comp(G[X])$ as in~\eqref{eq:fvs-g}; comparisons of $g$ with the
  thresholds of Algorithm~\MLS{} reduce to integer comparisons (\cref{sec:fw-setting}). Every
  solution $S$ induces a connected subgraph, so a spanning tree $F$ of $G[S]$ gives the
  component-count bound~\eqref{eq:cc-bound} with constants $(3,0)$. By \cref{cor:cc-rate} with
  $c=3$, $y^*=4/5$, $\zeta^*=6/5$ and $\psi=\log1.8$, Algorithm~\MLS{} with $\kappa=1$, run for $k=1,\dots,n$,
  returns a solution of size at most $k$ only if one exists, and returns one with probability at
  least $1-1/e$ for $k=\OPT(G)$; each run takes time $1.8^n\,n^{O(1)}$ and polynomial space. Output
  the smallest set found, and $V(G)$, which is a connected feedback vertex set because $G$ is
  connected, if none is found. Every output is a connected feedback vertex set, so its size is at
  least $\OPT(G)$, and with probability at least $1-1/e$ the run for $k=\OPT(G)$ returns a set of
  size at most $\OPT(G)$, which is then the output.
\end{proof}

By \cref{cor:cc-T}(b) with $\gamma=0$, every solution satisfies $W_S(4/5)\ge1/3$. The next proposition
turns the path bound of \cref{lem:cc-path} into an upper bound on the probability that a run of
Algorithm~\MLS{} runs the extension algorithm on a subset of a solution $S$ with $|S|=26n/45$ whose
running-time exponent is bounded below as for a path, for every run on $(I,k)$ with $k\ge|S|$. Its proof,
in \cref{sec:app-fvs-tight}, uses that at the critical fraction $\sigma^*=26/45$ the function
$\phi_{3,\sigma^*}$ of~\eqref{eq:cc-phi} has $\inf_{y\ge0}\phi_{3,\sigma^*}(y)=\log1.8$, attained at
$y=4/5$ (\cref{lem:cc-inf}), and combines this with the necessity
of~\eqref{eq:fw-star} (\cref{prop:fw-necessary}), in the form for budgets $k\ge|S|$, and with the
bound~\eqref{eq:fw-moment-monotone} on the moments $\mu^{(k)}_t(S)$ of~\eqref{eq:fw-moment-budget}.

\begin{restatable}[Tightness of $1.8$]{proposition}{RestPropCcTight}\label{prop:cc-tight}
  Let $I$ be an instance and $S\in\cF_I$ with $|S|/n=26/45$. Suppose that $g(I,X,k)$ is
  nondecreasing in $k$ for every $X\subseteq S$, and that, for a path $F$ with vertex set $S$ and a
  constant $\gamma'\ge0$,
  \[
    g(I,X,|S|)\ \ge\ (|S|-|X|)\log3+|X|-e(F[X])-\gamma'\qquad\text{for every }X\subseteq S .
  \]
  \begin{enumerate}[label=\textup{(\alph*)}]
    \item The exponential moments of $S$ (\cref{def:fw-moments}, with $k=|S|$) satisfy
      \[
        \max_{0\le t\le|S|}\frac{\mu_t(S)}{\binom nt}\le2^{\gamma'+1}(n+1)(5/9)^n .
      \]
    \item For every integer $k$ with $|S|\le k\le n$, a run of Algorithm~\MLS$(\psi,\Gamma,\kappa)$
      on $(I,k)$ runs $\cA$ on a subset of $S$ with probability at most
      $P_{\mathrm{nec}}\cdot2^{\gamma'+1}(n+1)\bigl(2^{\psi}\cdot5/9\bigr)^n$, with $P_{\mathrm{nec}}\le N^{O(1)}$ as in
      \cref{prop:fw-necessary}.
    \item Suppose that $\cA$ never accepts a \NO{}-instance and that $S$ is the only set in $\cF_I$
      of size at most $|S|$. Then a run on $(I,k)$ with $k\le|S|$ answers \YES{}, and a run on
      $(I,k)$ with any $k\le n$ returns a solution of size at most $|S|$, only if it runs $\cA$ on a
      subset of~$S$.
    \item Under the hypotheses of (c), let $\varepsilon>0$ and $\psi\le\log1.8-\varepsilon$, and run
      Algorithm~\MLS$(\psi,\Gamma,\kappa)$ on $(I,k)$ for $k=1,\dots,n$, with any $\Gamma$ and
      $\kappa$ as in \cref{thm:fw-main} (for a one-sided-error $\cA$, with $2\Gamma$ as in
      \cref{cor:fw-onesided}). The probability that some run answers \YES{} at a $k\le|S|$, or that
      some run returns a set of size at most $|S|$, is at most
      $nP_{\mathrm{nec}}2^{\gamma'+1}(n+1)2^{-\varepsilon n}$, where $P_{\mathrm{nec}}\le N^{O(1)}$ is the
      largest of the factors of \cref{prop:fw-necessary} over these runs.
  \end{enumerate}
\end{restatable}

Families of instances that satisfy all hypotheses of \cref{prop:cc-tight} exist for
\pname{Connected Feedback Vertex Set}.

\begin{restatable}[A tight family for Connected Feedback Vertex Set]{proposition}{RestPropCcCfvsTight}\label{prop:cc-cfvs-tight}
  Let $\varepsilon>0$ be a constant. There is an infinite family of graphs on which the algorithm of
  the proof of \cref{thm:cc-cfvs}, run with rates $\psi\le\log1.8-\varepsilon$ in place of
  $\log1.8$ and with any $\Gamma$ and $\kappa$ as in \cref{thm:fw-main}, outputs a
  minimum connected feedback vertex set with probability $o(1)$ as the number $n$ of vertices grows.
\end{restatable}

\subsection{Graph Motif}\label{sec:motif}

The method of \cref{sec:cfvs} needs a finishing step whose running time is exponential in the
number of components of the sample, and a solution that is connected. \pname{Graph Motif} has both,
with a simpler finishing step: the randomized $\Ostar(2^k)$ algorithm of Björklund, Kaski and
Kowalik~\cite{BjorklundKK15motif}, used as a black box, after the components of the sample have been
contracted and given fresh colors.

An instance of \pname{Graph Motif} is a graph $G$ with $n$ vertices, a coloring
$\mathrm{col}:V(G)\to C$ and a multiset $M$ over $C$ with $|M|=k$. A \emph{solution} is a set
$S\subseteq V(G)$ such that $G[S]$ is connected and $\mathrm{col}(S)=M$ as multisets; every solution
has size exactly~$k$. Björklund, Kaski and Kowalik~\cite{BjorklundKK15motif} solve the problem in
randomized time $\Ostar(2^k)$ and polynomial space. They also show~\cite[Theorem~6]{BjorklundKK15motif} that if, for some constant $\varepsilon>0$, the problem can be solved
in time $O((2-\varepsilon)^k)$, even only on instances in which every color occurs at most once in
$M$, then there is a constant $\varepsilon'>0$ such that \pname{Set Cover} on a universe of $u$
elements can be solved in time $O((2-\varepsilon')^u)$. (They state this for \pname{Maximum Graph
Motif}, which asks for a connected set $S$ with $|S|=k$ and $\mathrm{col}(S)\subseteq M$; when
$|M|=k$, as in their reduction, this is the problem above.)
Running their algorithm or enumerating all $k$-subsets, whichever is faster, gives
$\max_{0\le\sigma\le1}\min(2^{\sigma},2^{H(\sigma)})^n=(1.7087\ldots)^n\le1.7088^n$, at
$\sigma\approx0.7729$.

\subsubsection{Finishing with forced vertices}

The finishing step contracts each component of the sample to one vertex and gives that vertex a fresh
color which the motif demands exactly once, so that the black-box algorithm must use every
contracted component.

\begin{lemma}[Extension]\label{lem:mot-ext}
  Given an instance $(G,\mathrm{col},M)$ with $|M|=k$ and a set $X\subseteq V(G)$ with $|X|=t\le k$, a
  randomized algorithm can decide whether some solution $S$ contains $X$. It runs in time
  $2^{(k-t)+\comp(G[X])}n^{O(1)}$ for every outcome of its coins and in polynomial space, never accepts
  if there is no such solution, and accepts with probability at least $1/2$ if there is one.
\end{lemma}
\begin{proof}
  If $\mathrm{col}(X)\not\subseteq M$ as multisets, reject. Otherwise let $X_1,\dots,X_q$ be the vertex
  sets of the components of $G[X]$, $q=\comp(G[X])$. Form $G'$ from $G$ by contracting each $X_i$ into
  one vertex $d_i$ (adjacent to every vertex outside $X$ with a neighbor in $X_i$; the vertices $d_i$ are pairwise
  non-adjacent, since the $X_i$ are components of $G[X]$), give $d_i$ a fresh color $c_i$, and let
  $M'=(M\setminus\mathrm{col}(X))+\{c_1,\dots,c_q\}$, a multiset of size $k-t+q$.

  We claim that the solutions $S\supseteq X$ of the original instance correspond to the connected sets
  $S'$ of $G'$ with $\mathrm{col}(S')=M'$, by $S\mapsto(S\setminus X)\cup\{d_1,\dots,d_q\}$. Such an $S'$
  contains every $d_i$, because $c_i$ occurs once in $M'$ and only $d_i$ has that color. Let $S'$ be such a
  set. Then $S=(S'\setminus\{d_1,\dots,d_q\})\cup X$ induces a connected subgraph of $G$, since each vertex of
  the connected graph $G'[S']$ expands to a connected set, and an edge of $G'$ at a vertex $d_i$ comes from
  an edge of $G$ at $X_i$. Its colors are $(M'-\{c_1,\dots,c_q\})+\mathrm{col}(X)=M$. Conversely, if
  $S\supseteq X$ is a solution, the image of $S$ in $G'$ is connected and has the colors $M'$.

  If $q\ge1$ and the vertices $d_i$ do not all lie in one component of $G'$, reject. Otherwise restrict
  $G'$ to the component that contains the vertices $d_i$ (to each component in turn if $q=0$), so that the host
  graph is connected. On this graph run the algorithm of Björklund, Kaski and
  Kowalik~\cite[Theorem~4 and Section~3.7]{BjorklundKK15motif} with motif $M'$. It decides the existence
  of a connected induced subgraph $S'$ with $|S'|=|M'|$ and $\mathrm{col}(S')\subseteq M'$ as multisets,
  that is, with $\mathrm{col}(S')=M'$, in time $2^{|M'|}n^{O(1)}$ and polynomial space; it never answers
  \YES{} on a \NO{}-instance and answers \YES{} on a \YES{}-instance with probability at least $1/2$.
  As $|M'|=(k-t)+\comp(G[X])$, the lemma follows.
\end{proof}

\subsubsection{Exponential moments: the base \texorpdfstring{$1.6444$}{1.6444}}\label{sec:cc-motif}

For $c=2$ the closed form of \cref{cor:cc-T}(b) gives the base $5/3$, which is not optimal, so we optimize $y$ for each
solution fraction and certify the result. Since $\phi_{2,\sigma}(y)$ is affine in $\sigma$ for fixed
$y$, a finite list of intervals in $\sigma$ suffices, and no rounding of a sample size has to be
absorbed: $y$ is a real parameter of the analysis only.

\begin{lemma}[Rate certificate for $c=2$]\label{lem:cc-cert}
  Every $\sigma\in(0,1]$ satisfies~\ref{it:cc-R1}, \ref{it:cc-R2}
  or~\ref{it:cc-R3} of \cref{cor:cc-rate} with $c=2$, $\psi=0.71755$ and $\Gamma_0=16$. Moreover
  $0.71755<\log1.6444$, so $2^{0.71755n}n^{O(1)}=O(1.6444^n)$.
\end{lemma}

\begin{proof}
  Let $\psi=0.71755$, $\sigma_0=0.7175$ and $\sigma_1=0.81$. The certificate of \cref{sec:anc-cc} gives
  (o)~$\psi<\log1.6444$;
  (i)~$\sigma_0\le\psi$; (ii)~$H(\sigma_1)<\psi$
  and $\sigma_1\ge1/2$; (iii)~for $925$ consecutive intervals $[\sigma^-_j,\sigma^+_j]$ of width $10^{-4}$ covering
  $[\sigma_0,\sigma_1]$, with explicit decimals $y_j>0$, we have
  $\max\bigl(\phi_{2,\sigma^-_j}(y_j),\phi_{2,\sigma^+_j}(y_j)\bigr)<\psi$ and $\lambda_{\max}(M_{y_j})/\min(1,y_j)\le16$.
  If $\sigma\le\sigma_0$, then~\ref{it:cc-R1} holds, as $\log c=1$. If $\sigma\ge\sigma_1$,
  then~\ref{it:cc-R2} holds, as $H$ is decreasing on $[1/2,1]$. Otherwise $\sigma\in[\sigma^-_j,\sigma^+_j]$ for
  some $j$, and since $\phi_{2,\sigma}(y_j)$ is affine in $\sigma$, it is at most the maximum of its
  values at $\sigma^-_j$ and $\sigma^+_j$, so~\ref{it:cc-R3} holds with $y=y_j$.
\end{proof}

\begin{theorem}[Graph Motif]\label{thm:cc-motif}
  \pname{Graph Motif} can be solved by a randomized algorithm with one-sided error in time
  $O(1.6444^n)$ and polynomial space.
\end{theorem}

\noindent The constant $1.6444$ rests on \cref{lem:cc-cert}; the algorithm of Björklund, Kaski and Kowalik~\cite{BjorklundKK15motif} is used as a
black box. The mean-exponent theorem (\cref{thm:co-mean}) with the same finishing step, run with the sample size
that minimizes its bound, gives, by a numerical evaluation, a running time of
$\Ostar(1.6488^n)$ (the entry $1.64877$ in the table of \cref{sec:cc-cost}).

\begin{proof}
  The solutions of an instance $(G,\mathrm{col},M)$ with $|M|=k$ are the sets $S$ with $G[S]$
  connected and $\mathrm{col}(S)=M$; they have size $k\ge1$ (for $k=0$ the answer is trivial). By
  \cref{lem:mot-ext} the extension problem is solved with one-sided error, success probability at
  least $1/2$ and polynomial space, in time $2^{g(X)}n^{O(1)}$ for every outcome of the coins, with
  the integer running-time exponent $g(X)=g(I,X,k)=(k-|X|)+\comp(G[X])$ (shorthand of
  \cref{def:fw-cost}, for the instance $I=(G,\mathrm{col},M)$). A spanning tree of $G[S]$ gives the
  component-count bound with constants $(2,0)$. \Cref{cor:cc-rate} with \cref{lem:cc-cert}, that is,
  with the rate $\psi=0.71755$, gives running time $2^{0.71755n}n^{O(1)}=O(1.6444^n)$; only the
  value $k=|M|$ is run.
\end{proof}

Numerically, the base $2^{\psi_2}\approx1.64429$ is
attained at $\sigma\approx0.71746$, where the running-time bound $2^{\sigma n}$ of the plain algorithm
equals the bound $2^{n\phi_{2,\sigma}(y)}$ of~\ref{it:cc-R3} with $y\approx1.294$; the best strategy jumps there from no
sampling to large samples. \Cref{lem:cc-cert} proves the upper bound $1.6444$; it is not sharp enough for the bound $1.6443$. The analysis needs paths to minimize $\mathcal Z_F$ up to a constant factor (\cref{lem:cc-gibbs}). We do
not know whether the exact inequality $\mathcal Z_F\ge\mathcal Z_{\text{path}}$ holds for all trees on the same
number of vertices.

\paragraph{Related work.} Graph motif problems were introduced by Lacroix, Fernandes and
Sagot~\cite{LacroixFS06} for the analysis of metabolic networks. \cite{FellowsFHV11} showed that \pname{Graph Motif} is fixed-parameter tractable for the
parameter~$k$, and a series of improvements~\cite{BetzlerBFKN11,GuillemotS13,Koutis12} led to the
$\Ostar(2^k)$ algorithm of Bj\"orklund, Kaski and Kowalik~\cite{BjorklundKK15motif}.
When $M$ is a set
(\pname{Colorful Graph Motif}), the problem can be solved in time $O(2^{n-k}|E(G)|)$~\cite[Theorem~3]{BetzlerBFKN11}; together with the $\Ostar(2^k)$ algorithm this gives $\Ostar(2^{n/2})$, below the
bound of \cref{thm:cc-motif}. For multisets this route is closed: \pname{Graph Motif} is W[1]-hard for the
parameter $n-k$, even when $M$ has only two colors~\cite[Theorem~4]{BetzlerBFKN11} (see also~\cite{FertinK20}).

\subsection{Connected Vertex Cover and further applications}\label{sec:cc-apps}

Each application below needs only a finishing step whose running-time exponent has a component-count
bound; the sampling analysis is \cref{cor:cc-rate}, with \cref{lem:cc-cert} for $c=2$ and the
closed form of \cref{cor:cc-T}(b) for $c=3$. For optimization problems with weights we use the
following standard reduction. If the weights are integers in $\{-w_{\max},\dots,w_{\max}\}$ with $w_{\max}\le n^{O(1)}$,
we guess the optimal value $\theta_w$ among $O(nw_{\max})$ candidates and the size $k$ of one optimal
solution $S^*$, and decide $\Phi$-Subset for the implicit set system $\Phi$ whose solutions are the
feasible sets of value at least (or at most) $\theta_w$; its membership test is polynomial. The sampling analysis concerns only $S^*$,
and the weights never enter the running-time exponent.

\subsubsection{Connected Vertex Cover, weighted and unweighted}

A \emph{connected vertex cover} of a graph $G$ is a set $S\subseteq V(G)$ such that $G-S$ has no
edges and $G[S]$ is connected. In \pname{Weighted Connected Vertex Cover} the vertices carry
weights $w\colon V(G)\to\{1,\dots,w_{\max}\}$, $w_{\max}\le n^{O(1)}$, and we seek a connected vertex cover of
minimum weight; unit weights give \pname{Connected Vertex Cover}. We may assume that $G$ is
connected and has an edge: otherwise a connected vertex cover exists only if at most one component
has an edge, and the other components can be discarded. The finishing step uses Cygan's algorithm
as a black box~\cite{Cygan12}: given $G$, positive integer weights bounded by a polynomial and
an integer $b$, it computes a minimum-weight connected vertex cover of size at most $b$ (or reports
that none exists), deterministically, in time $2^bn^{O(1)}$ and polynomial space (his Theorem~3).

\begin{lemma}[Finishing step]\label{lem:cc-cvc-ext}
  Given a connected graph $G$ with at least one edge, weights as above, a set $X\subseteq V(G)$ and
  integers $k\ge|X|$ and $\theta_w$, one can decide deterministically, in polynomial space and in time
  $2^{k-|X|+\comp(G[X])}n^{O(1)}$, whether some $Y\subseteq V(G)\setminus X$ with $|Y|\le k-|X|$ makes
  $X\cup Y$ a connected vertex cover of weight at most $\theta_w$, and output such a $Y$ if so.
\end{lemma}

\begin{proof}
  Put $k'=k-|X|$.
  If $X=\emptyset$, run Cygan's algorithm with $b=k'$; accept if it returns a cover of weight at most
  $\theta_w$, and output that cover as~$Y$. Otherwise let $G_X$ be the contracted
  graph of \cref{sec:fvs-graphs}, with forced set $\hat X=\{x_1,\dots,x_q\}$, $q=\comp(G[X])$. For
  $Y\subseteq V(G)\setminus X$, the set $X\cup Y$ is a connected vertex cover of $G$ if and only if
  $\hat X\cup Y$ is one of $G_X$: the proof of \cref{lem:fvs-contract}(2) applies verbatim, because $G-(X\cup
  Y)=G_X-(\hat X\cup Y)$, and a set is a vertex cover exactly when the graph without it has no edges. If $G_X$
  has one vertex, $q=1$, $Y=\emptyset$ is the only candidate, and the algorithm accepts if and only if
  $w(X)\le\theta_w$, with output $Y=\emptyset$. Otherwise $G_X$ is connected with at least two
  vertices. Let $G_X^+$ be $G_X$ with a new pendant vertex $r'$ attached to each $r\in \hat X$, and give
  the vertices of $\hat X$ and the new vertices weight~$1$.

  Every connected vertex cover $C$ of $G_X^+$ contains $\hat X$. Indeed, if $r\in \hat X\setminus C$, then
  $r'\in C$, to cover $rr'$, and the other end $u$ of an edge $ru$ of $G_X$ lies in $C$; but the
  only neighbor of $r'$ is $r\notin C$, so $G_X^+[C]$ is disconnected. Hence each $r'\in C$ is a
  leaf of $G_X^+[C]$, and removing all of them leaves a connected vertex cover of $G_X$ that contains $\hat X$,
  of no larger size and weight. Conversely, a connected vertex cover of $G_X$ that contains $\hat X$ is one of
  $G_X^+$. So the minimum weight of a connected vertex cover of $G_X^+$ of size at most $k'+q$ equals the
  minimum of $w(Y)+q$ over the sets $Y$ with $|Y|\le k'$ such that $\hat X\cup Y$ is a connected vertex
  cover of $G_X$. Compute it with Cygan's algorithm, $b=k'+q$, and accept if it is at most
  $\theta_w-w(X)+q$. In that case the cover $C$ returned by Cygan's algorithm gives the output
  $Y=C\setminus(\hat X\cup\{r':r\in\hat X\})$: by the above, $\hat X\cup Y$ is a connected vertex cover of
  $G_X$ with $|Y|\le|C|-q\le k'$ and $w(Y)\le w(C)-q\le\theta_w-w(X)$.
\end{proof}

\begin{theorem}[Connected Vertex Cover]\label{thm:cc-cvc}
  Let $w_{\max}\le n^{O(1)}$ and let $\kappa>0$ be a constant. There is a randomized algorithm that, given a
  connected graph $G$ with $n\ge2$ vertices and weights $w\colon V(G)\to\{1,\dots,w_{\max}\}$, outputs a
  connected vertex cover of $G$ in time $O(1.6444^n)$ and polynomial space. On every outcome of its coins
  the output is a connected vertex cover of $G$, so its weight is at least the minimum; with probability at
  least $1-e^{-\kappa}$ its weight is the minimum. In particular, this holds for \pname{Connected Vertex
  Cover}, and by the reduction above it extends to graphs that are not connected or have no edge.
\end{theorem}

\noindent The constant rests on \cref{lem:cc-cert}; Cygan's algorithm is used as a black box.

\begin{proof}
  For each guess of $\theta_w\in\{1,\dots,nw_{\max}\}$ and each $k\in\{1,\dots,n\}$, run Algorithm~\MLS{} for the
  connected vertex covers of weight at most $\theta_w$ and the size $k$, and output the witness $X\cup Y$ of an
  accepting run with the smallest $\theta_w$. If no run accepts, output $V(G)$, which is a connected vertex cover
  because $G$ is connected. An accepting run returns a connected vertex cover of weight at most its
  $\theta_w$ (\cref{lem:cc-cvc-ext}), so on every outcome of the coins the output is a connected vertex cover. The extension
  problem for these implicit set systems is solved by \cref{lem:cc-cvc-ext}, with running-time exponent
  $g(G,X,k)=k-|X|+\comp(G[X])$, deterministically. Every solution induces a connected subgraph, so a
  spanning tree of $G[S]$ gives the component-count bound with constants $(2,0)$, and
  \cref{cor:cc-rate} with \cref{lem:cc-cert} applies for every $k$, with the rate $\psi=0.71755$; there
  are polynomially many runs, each in time $2^{0.71755n}n^{O(1)}$, which is $O(1.6444^n)$ in total by
  \cref{lem:cc-cert}. A run for the optimal $\theta_w$
  and $k=|S^*|$ succeeds with probability at least $1-e^{-\kappa}$, and no run accepts a value below
  the optimum.
\end{proof}

\paragraph{Comparison with published bounds.}
For \pname{Connected Vertex Cover} the best published bound in terms of $n$ is
$O(1.7088^n)$~\cite[Corollary~1]{GolovachHK18}, from Cygan's $2^k$ algorithm and enumeration. For the
weighted problem, enumerating the minimal connected vertex covers gives
$\Ostar(1.8668^n)$~\cite[Theorem~3]{GolovachHK18} (with positive weights a minimum-weight connected vertex cover is
minimal), and balancing Cygan's Theorem~3~\cite{Cygan12} against enumeration gives $O(1.7088^n)$, a
combination we have not seen stated. Our algorithm is randomized through the sampling only; the
finishing step is deterministic.

\subsubsection{Maximum-Weight Connected \texorpdfstring{$k$}{k}-Subgraph}

Given a graph $G$, integer vertex weights $w\colon V(G)\to\{-w_{\max},\dots,w_{\max}\}$ with $w_{\max}\ge1$ and an
integer $1\le k\le n$, \pname{Maximum-Weight Connected $k$-Subgraph} asks for a set $S$ with $|S|=k$ and $G[S]$
connected that maximizes $w(S)$; we call the sets $S$ with $|S|=k$ and $G[S]$ connected, of any weight,
\emph{connected $k$-sets}. We know of no exact bound in terms of $n$ besides enumeration of all
$k$-sets; for the version without the size constraint, Buchanan, Wang and Butenko~\cite{BuchananWB18} give
$O(1.5875^n)$. The finishing step below with $X=\emptyset$ runs in time $2^k(nw_{\max})^{O(1)}$; balanced
against enumeration it gives $O(1.7088^n)$ for $w_{\max}\le n^{O(1)}$, the same base as for
\pname{Graph Motif}, and this folklore combination is the bound we compare with.

\paragraph{The finishing step.}
As for \pname{Graph Motif} (\cref{lem:mot-ext}), the components $X_1,\dots,X_q$ of the sample are
contracted to vertices $d_1,\dots,d_q$ of weight~$0$, and each new vertex receives a fresh color that must occur
exactly once, while all other vertices share one color of multiplicity $k-|X|$. The connected
$k$-sets $S\supseteq X$ then correspond to the connected sets of size $k-|X|+q$ of the contracted graph
that contain all vertices $d_i$, with $w(S)=w(X)$ plus the weight of the corresponding set. We detect these
sets with the algebraic sieve of Björklund, Kaski and Kowalik~\cite{BjorklundKK15motif}, to which we
add one indeterminate whose exponent records the weight of the set, and we recover the polynomial in
this indeterminate by interpolation. Their cost-constrained lemma lets the coloring vary, so we
derive the variant with a fixed coloring and a cost on each vertex from their proofs. A witness is
found by deleting vertices one at a time. The finishing step takes time $2^{(k-|X|)+\comp(G[X])}
(nw_{\max})^{O(1)}$ and polynomial space and has one-sided error.

\begin{restatable}[Maximum-Weight Connected $k$-Subgraph]{theorem}{RestThmCcMwcs}\label{thm:cc-mwcs}
  Let $1\le w_{\max}\le n^{O(1)}$. There is a randomized algorithm that, given a graph $G$ with $n$
  vertices, weights $w\colon V(G)\to\{-w_{\max},\dots,w_{\max}\}$ and an integer $1\le k\le n$, either
  reports correctly that $G$ has no connected $k$-set or outputs a connected $k$-set, in time
  $O(1.6444^n)$ and polynomial space. The output has maximum weight with probability at least
  $1-1/e$. In particular, the decision version (``is there a connected $k$-set of weight at least
  $\theta$?'') is solved with one-sided error.
\end{restatable}

\noindent The constant rests on \cref{lem:cc-cert}. The proof is in \cref{sec:app-weighted}. The variant without the size constraint is not improved by
this method.

\subsubsection{Vertex-weighted Connected Feedback Vertex Set}

In the vertex-weighted version of \pname{Connected Feedback Vertex Set} the vertices carry weights
$w\colon V(G)\to\{1,\dots,w_{\max}\}$, and we seek a connected feedback vertex set of minimum weight.
We know of no published algorithm in time $O((2-\varepsilon)^n)$ for a constant $\varepsilon>0$; the
Cut~\&~Count algorithm of \cite{CyganNPPRW22} is stated for the cardinality version.

\paragraph{The extension algorithm.}
After contracting the sample (\cref{lem:fvs-contract}(2)) and giving the forced vertices weight~$0$, the
weight of a completion is $w(X)$ plus the weight of its non-forced part. The weighted extension
algorithm first runs the unweighted extension algorithm of \cref{lem:fvs-ext}, only to obtain a
connected feedback vertex set of small excess, which serves as a structured modulator; so the
intermediate graphs of the iterative compression never need weights. It then counts the objects
of~\ref{it:fvs-CC1}, through the proved \cref{lem:wfvs-parity,lem:wfvs-core} rather than the cited dynamic
program, with a weight function that adds to each isolating weight
of a forest vertex a large multiple of its vertex weight, so that the weight of a counted candidate
determines both its isolating weight and its vertex weight, and it finds a solution by
self-reduction. It has the running-time exponent~\eqref{eq:fvs-g} of the unweighted algorithm. The
two properties of the counting that it needs are proved in \cref{sec:app-weighted-wcfvs}, so they are
not imported. They are a parity identity for the counted cuts (\cref{lem:wfvs-parity}) and a dynamic program
over the forest outside the modulator (\cref{lem:wfvs-core}). That program has four states per vertex for
every weight function, and the weights change only the range of its weight index.

\begin{restatable}[Vertex-weighted Connected Feedback Vertex Set]{theorem}{RestThmCcWcfvs}\label{thm:cc-wcfvs}
  Let $w_{\max}\le n^{O(1)}$. There is a randomized algorithm that, given a graph $G$ with $n$
  vertices and weights $w\colon V(G)\to\{1,\dots,w_{\max}\}$, reports correctly that $G$ has no connected
  feedback vertex set or outputs one, in time $\Ostar(1.8^n)$ and polynomial space. The output has
  minimum weight with probability at least $1-1/e$. In particular, the decision version (``is there a
  connected feedback vertex set of weight at most $\theta$?'') is solved with one-sided error.
\end{restatable}

\noindent The proof is in \cref{sec:app-weighted}.
{}
\section{Further applications}\label{sec:further}

The deterministic algorithm of \cref{sec:det} and Algorithm RS-$\theta$ of \cref{sec:rstheta} use the problem through two ingredients: an extension algorithm whose
running time is $\Ostar(2^{g(X)})$ for a computable running-time exponent $g$, written with the shorthand
$g(X)=g(I,X,k)$ declared after \cref{def:fw-cost}; and a pointwise affine bound $g(X)\le\alpha-w(X)$
for the samples $X$ of a solution $S$, with a number $\alpha$ and nonnegative integer weights
$w\colon S\to\mathbb Z_{\ge0}$ whose total $w(S)$ is known (or, for RS-$\theta$, the tree potential of
\cref{lem:rst-E}). This section transfers them to four variants of
\pname{Connected Dominating Set} and then treats \pname{Directed Steiner Tree} and \pname{Paired
Domination} (\cref{sec:fa-dst,sec:fa-pd}). The four variants are distance-$r$ domination
(\cref{sec:rdom}), vertex weights (\cref{sec:cv-weighted}), weak connectivity (\cref{sec:cv-wcds}) and
directed graphs (\cref{sec:cv-dir}). All algorithms use polynomial space. All four extension algorithms rest on the
vertex-weighted directed Steiner lemma of \cref{sec:cv-steiner} (\cref{lem:cv-W}): the first three through the
directed Steiner encoding of \cref{lem:cds-dir-steiner}, applied to states of red--blue graphs
(\cref{lem:cv-redblue}), and the directed variant through the extension algorithm for \pname{Directed
Steiner Tree} (\cref{lem:fa-dst-ext}). For each variant we state the
problem and the results, and say what changes relative to \pname{Connected Dominating Set} and why the
analysis transfers; the lemmas and proofs are in \cref{sec:app-variants}.

\pname{Directed Steiner Tree} and \pname{Paired Domination} share the feature of the earlier applications: the running time of the finishing step
is exponential in a count that a sample can reduce, while a global constraint on the solution does not
enter the exponent. In \pname{Directed Steiner Tree} the count is the number of source components left
by a sample, and the finishing step is inclusion--exclusion over them. In \pname{Paired Domination} the
count is the number of undominated vertices, and the global constraint, a perfect matching inside the
solution, is handled by a Tutte matrix at no increase of the exponent. For both problems Algorithm~\MLS{}
runs in polynomial space, with one-sided error, and below the best previous bound that we know
(\cref{tab:intro-applications}); the analysis is tight for \pname{Paired Domination} (\cref{prop:fa-pd-tight}) and appears
close to tight, numerically, for \pname{Directed Steiner Tree}.
{}

\subsection{Connected \texorpdfstring{$r$}{r}-domination}\label{sec:rdom}

The two ingredients of \cref{sec:further} transfer to \emph{domination relations}, which include
distance-$r$ domination.

\begin{definition}[Domination relations]\label{def:rdom-relation}
Let $G$ be a graph. A \emph{domination relation} on $G$ assigns to every vertex $v$ a set
$\cN[v]\subseteq V$, the vertices that \emph{dominate}~$v$, such that $N[v]\subseteq\cN[v]$
for every $v$, and such that $\cN[v]$ can be computed from $G$ (and the description of $\cN$) in
time polynomial in their total length. A set $S\subseteq V$ is a \emph{connected $\cN$-dominating set} if it
is nonempty, $G[S]$ is connected, and $\cN[v]\cap S\neq\emptyset$ for every $v\in V$; for a
connected graph we write $\gc^\cN(G)$ for the minimum size of one. For $X\subseteq V$ let
$\operatorname{und}_\cN(X)=\{v\in V:\cN[v]\cap X=\emptyset\}$ be the vertices not dominated by~$X$.
\end{definition}

With $\cN[v]=N[v]$ this is connected domination. With $\cN[v]=\{u:\mathrm{dist}_G(u,v)\le r\}$
for an integer $r\ge1$, which may be part of the input, it is \emph{connected $r$-domination}, and
\pname{Connected $r$-Dominating Set} asks for $\gc^\cN(G)$. Every superset of an $\cN$-dominating
set is $\cN$-dominating, so a connected $\cN$-dominating set exists if $G$ is connected ($V$
itself is one), and for distance-$r$ domination one exists only if $G$ is connected. As in
\cref{sec:cds}, $G$ is connected and has $n\ge2$ vertices.

\paragraph{What changes.} Both ingredients change, and each keeps its constants. First, the number of
terminals $|\Tm(J_X)|$ of~\eqref{eq:cds-TX} is replaced by $g_\cN(X)=|\operatorname{und}_\cN(X)|+\comp(G[X])$,
and the Steiner extension of \cref{cor:cds-steiner} transfers with this running-time exponent
(\cref{lem:rdom-steiner}): the completions of $X$ are the solutions of a state (\cref{def:cds-state}) of a
red--blue graph in which the red copy of $v$ is joined to the blue copies of the vertices of $\cN[v]$,
while connectivity is still measured by adjacency in $G$ (\cref{lem:cv-redblue}). Second, every vertex
outside the solution is assigned to a solution vertex that dominates
it, which need not be a neighbor, and this assignment is treated as a pendant edge of a virtual tree,
the \emph{anchored tree}, in which the solution is a connected dominating set. The running-time exponent of a
sample is at most its number of terminals in the anchored tree (\cref{lem:rdom-tree}), so the rate of
\cref{prop:rst-rate} and the certificate~\eqref{eq:det-cert}, applied to the anchored tree, bound the
exponential moments and the integer certificate for $g_\cN$ with the same constants as for connected
domination, and the analyses of \cref{thm:rstheta,thm:det-cds} transfer pointwise.

\begin{restatable}[Connected $\cN$-domination]{theorem}{RestThmRdom}\label{thm:rdom}
Let $\cN$ be a domination relation. For connected $n$-vertex graphs $G$, the value $\gc^\cN(G)$,
in particular the minimum size of a connected $r$-dominating set for every $r\ge1$ (with $r$ part of
the input), can be computed
\begin{enumerate}[label=(\alph*)]
\item by a randomized algorithm with one-sided error in time $O(1.7204^n)$ and polynomial space;
\item deterministically in time $O\bigl((\sqrt5-\frac12)^{n}\,2^{O(n(\log\log n)^{2}/\log n)}\bigr)
=O(1.7361^n)$ and polynomial space.
\end{enumerate}
\end{restatable}

The bases are those of \cref{thm:rstheta,thm:det-cds}. The proof is in \cref{sec:app-rdom};
\cref{rem:cv-limits} lists what does not transfer.

\paragraph{Related work.} Fomin, Grandoni and Kratsch~\cite{FominGK08} and Abu-Khzam, Mouawad and
Liedloff~\cite{Abu-KhzamML11} solve \pname{Connected Dominating Set} by branching on closed
neighborhoods, and their measures do not apply directly to $r\ge2$. The red--blue graph $H_\cN$ of
\cref{lem:rdom-steiner}, on $2n$ vertices, reduces the problem to \pname{Connected Red-Blue Dominating Set},
so the algorithm of Abu-Khzam, Mouawad and Liedloff~\cite{Abu-KhzamML11} computes $\gc^\cN(G)$
deterministically in time $\Ostar(1.3645^{2n})=\Ostar(1.8619^{n})$ and polynomial space. For $r\ge2$ we know
of no published exact algorithm below $2^n$ for \pname{Connected $r$-Dominating Set}, also known as
connected distance-$r$ or $d$-hop connected domination; the reduction above is the only one known to us.
{}

\subsection{Vertex-weighted connected domination and weighted MaxLeaf}\label{sec:cv-weighted}

In this subsection $G$ is a connected graph with $n\ge2$ vertices and $\varphi\colon V\to\{0,1,\dots,\varphi_{\max}\}$
is a weight function; we write $\varphi(X)=\sum_{x\in X}\varphi(x)$. To keep the notation of the paper, the
weights are part of the instance in unary, so that the input length $N$ is at least $n+\varphi_{\max}$ and $\Ostar$
hides factors polynomial in $n+\varphi_{\max}$. \pname{Vertex-Weighted Connected Dominating Set} asks for
$\gc^\varphi(G)=\min\{\varphi(S):S\text{ a connected dominating set of }G\}$, and \pname{Weighted Max Leaf Spanning
Tree} asks for the maximum of $\varphi(\operatorname{leaves}(F))$ over the spanning trees $F$ of $G$, where $\operatorname{leaves}(F)$ is the set of leaves of $F$.

As in the unweighted case, for $n\ge3$ the maximum leaf weight of a spanning tree equals
$\varphi(V)-\gc^\varphi(G)$ (\cref{lem:cv-wmaxleaf}).

\paragraph{What changes.} The extension now has to find a completion of least weight. This
is the minimum weight of a solution of the state $J_X$ of~\eqref{eq:cds-JX}, and the directed Steiner
encoding of \cref{lem:cds-dir-steiner}, which already allows weights on the blue vertices, computes it in time
$2^{|\Tm(J_X)|}(n+\varphi_{\max})^{O(1)}$ and polynomial space. So the running-time exponent is
still $|\Tm(J_X)|$, and it does not involve the weights. A completion of least weight may be large, so this
extension does not control the number of added vertices; the moment and derandomized theorems need
only two properties of the extension, which it has (\cref{lem:cv-relaxed}). The algorithms run the analyses
of \cref{thm:rstheta,thm:det-cds} for every weight budget and every solution size (\cref{lem:cv-budget-loop}).
The tree potential and the certificate~\eqref{eq:det-cert} concern the solution and not its weight, so they
hold unchanged; a minimum-weight solution itself has some size $s^*$, at which the run for the optimal
budget succeeds.

\begin{restatable}[Vertex-weighted \pname{Connected Dominating Set} and weighted MaxLeaf]{theorem}{RestThmCvWeighted}\label{thm:cv-weighted}
For connected $n$-vertex graphs with vertex weights in $\{0,\dots,\varphi_{\max}\}$, the minimum weight of a
connected dominating set, and the maximum leaf weight of a spanning tree (by \cref{lem:cv-wmaxleaf} for $n\ge3$;
for $n\le2$ it is $0$ if $n=1$ and $\varphi(V)$ if $n=2$), can be computed
\begin{enumerate}[label=(\alph*)]
\item by a randomized algorithm with one-sided error in time $O(1.7204^n)(n+\varphi_{\max})^{O(1)}$ and polynomial space;
\item deterministically in time $\bigl(\sqrt5-\frac12\bigr)^{n}\,2^{O(n(\log\log n)^{2}/\log n)}(n+\varphi_{\max})^{O(1)}
=O(1.7361^n)(n+\varphi_{\max})^{O(1)}$ and polynomial space.
\end{enumerate}
\end{restatable}

The bases are those of \cref{thm:rstheta,thm:det-cds}. The proof is in \cref{sec:app-cv-weighted};
\cref{rem:cv-limits} lists what does not transfer.

\paragraph{Related work.} The measure-and-conquer algorithms for \pname{Connected Dominating
Set}~\cite{FominGK08,Abu-KhzamML11} are stated for the cardinality version. With nonnegative weights a
minimum-weight connected dominating set can be chosen inclusion-wise minimal, so enumerating all minimal
connected dominating sets solves the vertex-weighted problem, and by \cref{lem:cv-wmaxleaf} weighted MaxLeaf
as well. Lokshtanov, Pilipczuk and Saurabh~\cite{LokshtanovPS18} enumerate them in time
$2^{(1-\varepsilon)n}n^{O(1)}$ for some $\varepsilon>10^{-50}$, and \cite{Abu-KhzamFGLM25} in time
$\Ostar(1.9896^n)$ and polynomial space. This gives $\Ostar(1.9896^n)$ for both weighted problems; we know of no faster
exact algorithm for them.

\subsection{Weakly connected domination}\label{sec:cv-wcds}

For a set $S\subseteq V$ of a graph $G$, the \emph{weakly induced subgraph} $\langle S\rangle_{\mathrm w}$
is the subgraph of $G$ with vertex set $N[S]$ whose edges are the edges of $G$ with at least one end in~$S$;
for a dominating set $S$ it is a spanning subgraph. A
\emph{weakly connected dominating set} is a dominating set $S$ for which $\langle S\rangle_{\mathrm w}$
is connected; see \cite{DunbarGHHM97}. A graph has one if
and only if it is connected ($V$ is one), and every connected dominating set is one. Throughout, $G$ is
connected with $n\ge2$ vertices, and $\gamma_{\mathrm w}(G)$ is the minimum size of a weakly connected
dominating set. For a set $X\subseteq V$ write $G^2[X]$ for the graph on $X$ in which two vertices are
adjacent when their distance in $G$ is at most $2$.

\paragraph{What changes.} Weak connectivity changes the running-time exponent. The number of terminals
$|\Tm(J_X)|=n-|N[X]|+\comp(G[X])$ is replaced by
\[
  g_{\mathrm w}(X)=n-|N[X]|+\comp(G^2[X]),
\]
because a set is weakly connected dominating exactly when it is dominating and $G^2[S]$ is connected
(\cref{lem:cv-wchar}); a red--blue graph whose blue copies are adjacent at distance at most $2$ turns the
completions of $X$ into the solutions of a state, and \cref{lem:cv-redblue} gives an extension in time
$2^{g_{\mathrm w}(X)}n^{O(1)}$ and polynomial space (\cref{lem:cv-wext-weak}). The pointwise certificate
charges each edge of a spanning tree of $\langle S\rangle_{\mathrm w}$ to an end in $S$ (\cref{lem:cv-wcert});
it has $\alpha=n$ and total weight $\Xi=n-1$ in place of $\alpha=n-1$ and $\Xi=n-2$, and the running time of
the integer-affine algorithm is the same for both pairs (\cref{lem:det-time}).

\begin{restatable}[\pname{Weakly Connected Dominating Set}]{theorem}{RestThmCvWcds}\label{thm:cv-wcds}
For connected $n$-vertex graphs, $\gamma_{\mathrm w}(G)$ can be computed deterministically in time
$O\bigl((\sqrt5-\frac12)^{n}\,2^{O(n(\log\log n)^{2}/\log n)}\bigr)=O(1.7361^n)$ and polynomial space.
\end{restatable}

The bound rests on \cref{lem:det-cert} and matches the deterministic bound for \pname{Connected Dominating
Set} (\cref{thm:det-cds}). The proof is in \cref{sec:app-cv-wcds}; \cref{rem:cv-limits} lists what does not
transfer.

\paragraph{Related work.} Weakly connected domination is studied in structural graph theory and, for
distributed and approximation algorithms, in wireless networks. We know of no published exact algorithm
for it; a reduction to \pname{Connected Red-Blue Dominating Set} and the algorithm of Abu-Khzam, Mouawad and
Liedloff~\cite{Abu-KhzamML11} give $\Ostar(1.8619^{n})$ (\cref{sec:app-cv-wcds}), the only bound below $2^n$
known to us.

\subsection{Directed Max Leaf Spanning Tree}\label{sec:cv-dir}

Let $\vec G$ be a digraph without loops or parallel arcs (opposite arcs are allowed) with $n\ge2$ vertices. An
\emph{out-branching} rooted at $r$ is a spanning subdigraph that is a tree in which every arc points away
from $r$; its \emph{leaves} are its vertices of out-degree $0$. \pname{Directed Max Leaf Spanning Tree}
asks for the maximum number of leaves of an out-branching. Binkele-Raible and
Fernau~\cite{Binkele-RaibleF12dir} solve it in time $\Ostar(1.9044^n)$ and polynomial space, and in time
$\Ostar(1.8139^n)$ with exponential space.
With the number
$k$ of leaves as the parameter, \cite{AlonFGKS09} decide in
time $2^{O(k\log^2k)}n^{O(1)}$ whether a strongly connected digraph has an out-branching with at least
$k$ leaves. For all digraphs, Kneis, Langer and Rossmanith~\cite{KneisLR11} decide this in time
$4^kn^{O(1)}$, and \cite{DaligaultGKY10} in time $3.72^kn^{O(1)}$, which also gives
$O(1.9973^n)$ for \pname{Directed Max Leaf Spanning Tree}. \cite{Binkele-RaibleFFLSV12}
give a kernel with $O(k^3)$ vertices when the root is part of the input, and show that without the root
there is no polynomial kernel unless $\mathrm{coNP}\subseteq\mathrm{NP/poly}$.

For $r\in V(\vec G)$, a set $S\subseteq V(\vec G)$ is \emph{$r$-internal} if $r\in S$, every vertex of $S$ is
reachable from $r$ in $\vec G[S]$, and every $v\notin S$ has an in-neighbor in $S$.

By \cref{lem:cv-dir-identity}, the maximum number of leaves of an out-branching is $n-\min|S|$, the minimum
over all roots $r$ and all $r$-internal sets $S$.

\paragraph{What changes.} For a fixed root $r$, the role of the number of terminals is taken by
$g_r(X)=|\operatorname{und}_r(X)|+|\mathrm{Src}_r(X)|$, which counts the vertices outside $X\cup\{r\}$ that have no
in-neighbor in $X$, and the source strong components of $\vec G[X]$ that do not contain $r$. The extension
problem is an instance of \pname{Directed Steiner Tree} on a digraph with a blue and a red copy of every
vertex, and the extension algorithm of \cref{lem:fa-dst-ext} solves it in time $2^{g_r(X)}n^{O(1)}$ and
polynomial space (\cref{lem:cv-dir-ext}). The certificate
is the directed form of the tree inequality: in an out-branching the weight $d_F(u)-1$ of
\cref{lem:cds-pointwise} becomes the number of children, minus $1$ at the root, and with these weights the
certificate~\eqref{eq:det-cert} holds with the same constants $\alpha=n-1$ and $\Xi=n-2$
(\cref{lem:cv-dir-cert}). So the algorithm of \cref{thm:det-cds} applies for each root.

\begin{restatable}[\pname{Directed Max Leaf Spanning Tree}]{theorem}{RestThmCvDir}\label{thm:cv-dir}
For digraphs with $n$ vertices, the maximum number of leaves of an out-branching (or the fact that there is
none) can be computed deterministically in time
$O\bigl((\sqrt5-\frac12)^{n}\,2^{O(n(\log\log n)^{2}/\log n)}\bigr)=O(1.7361^n)$ and polynomial space.
\end{restatable}

The bound rests on \cref{lem:det-cert}. The proof is in \cref{sec:app-cv-dir}.

\begin{remark}[What does not transfer]\label{rem:cv-limits}
The closure sampling RSC of \cref{sec:rsc} is claimed for none of the four variants: its closure rule and
its penalty for false candidates use the neighborhoods $N(u)$ of vertices outside the sample, which an
anchor at distance $r\ge2$, a weight or an arc direction do not respect; we leave open whether a variant of
the closure rule gives $1.6991^n$ for \pname{Connected $r$-Dominating Set}. The exponential-moment analysis
of RS-$\theta$ (\cref{sec:rstheta}) is claimed for $\cN$-domination and for weights (\cref{thm:rdom}(a) and
\cref{thm:cv-weighted}(a)). For weak connectivity we do not expect it to beat the affine certificate of
\cref{lem:cv-wcert}: on spiders that certificate is an equality, and there the moment bound appears to
decrease toward $\sqrt5-\frac12$ (\cref{prop:cv-spider} and the discussion after it). For out-branchings the tree potential of \cref{lem:rst-E}
does not apply, since it uses that a sampled vertex dominates its parent in the spanning tree of the
solution, while in an out-branching a vertex dominates only its children. The equivalence with
\pname{Max Leaf Spanning Tree} (\cref{lem:cds-maxleaf}) is special to $r=1$.
\end{remark}
{}

\subsection{Directed Steiner Tree}\label{sec:fa-dst}

An instance of \pname{Directed Steiner Tree} is a digraph $\vec G$ with $n$ vertices, a root $r$ and a set
$K\subseteq V(\vec G)\setminus\{r\}$ of terminals. Let $U=V(\vec G)\setminus(\{r\}\cup K)$ be the set of
non-terminals and $n_U=|U|$. A set $S\subseteq U$ is a \emph{Steiner set} if every terminal is reachable
from $r$ in $\vec G[\{r\}\cup K\cup S]$; the task is to find a Steiner set of least size. Two algorithms are
immediate: inclusion--exclusion over the terminals (\cref{lem:cv-W}, with unit weights; see
also~\cite{MisraPRSS12} and~\cite[Lemma~3.5]{JonesLRSS17}) takes time $2^{|K|}n^{O(1)}$, and enumeration takes
time $2^{n_U}n^{O(1)}$. Since $|K|+n_U=n-1$, the better of the two runs in time $2^{(n-1)/2}n^{O(1)}$, and we
know of no better bound in terms of $n$ for digraphs. Feldmann and Marx~\cite{FeldmannM23} classify the
directed Steiner network problems that are fixed-parameter tractable in the number of terminals. For undirected graphs and unit weights, \cite[Theorem~5]{FominGKLS13} solve \pname{Steiner Tree} in time
$O(1.3533^n)$ and polynomial space, which is below the bound of \cref{thm:fa-dst} on this special case; their
branching uses a contraction lemma for adjacent terminals that is proved for undirected graphs.

The sampling algorithm works on rooted Steiner sets. Call $S\subseteq U$ \emph{rooted} if every vertex
of $K\cup S$ is reachable from $r$ in $\vec G[\{r\}\cup K\cup S]$, and let $\Phi_{\mathrm{DST}}$ be the
implicit set system with universe $U$ whose solutions are the rooted Steiner sets. Every rooted set is
a Steiner set, and the vertices of a Steiner set $S$ that are reachable from $r$ in
$\vec G[\{r\}\cup K\cup S]$ form a rooted Steiner set; so the least size of a solution of
$\Phi_{\mathrm{DST}}$ is the optimum. In particular, every inclusion-minimal Steiner set is rooted, so
the restriction to rooted sets loses no minimum solution; it makes the running-time exponent of a sample,
defined next, a count that can be read off the digraph.

\paragraph{The running-time exponent of a sample.} For $X\subseteq U$ let $\bar K_X=\{r\}\cup K\cup X$ and let $\bar K^{r}_X$
be the set of vertices reachable from $r$ in $\vec G[\bar K_X]$. A \emph{source component} of
$\vec G[\bar K_X\setminus \bar K^{r}_X]$ is a strong component that no arc of $\vec G[\bar K_X\setminus \bar K^{r}_X]$ enters from outside.
For the instance $I=(\vec G,r,K)$, a set $X\subseteq U$ and $k\ge0$, let $g(I,X,k)$ be the number of source
components of $\vec G[\bar K_X\setminus \bar K^{r}_X]$, and $0$ if $\bar K_X=\bar K^{r}_X$. It does not depend on $k$, and for fixed $I$
we write $g(X)$, in agreement with the shorthand of \cref{def:fw-cost}. Every vertex of
$\bar K_X\setminus \bar K^{r}_X$ is reachable within
$\vec G[\bar K_X\setminus \bar K^{r}_X]$ from some source component, since the condensation of a digraph is acyclic.

\begin{lemma}[Extension]\label{lem:fa-dst-ext}
There is a deterministic algorithm that, given $(\vec G,r,K)$, a set $X\subseteq U$ and weights
$w\colon U\to\{0,1,\dots,w_{\max}\}$, computes the least weight $w(Y)$ over the sets $Y\subseteq U\setminus X$
for which $X\cup Y$ is a rooted Steiner set ($\infty$ if there is none), in time
$2^{g(X)}(n+w_{\max})^{O(1)}$ and polynomial space. With unit weights it decides, for every integer
$k\ge0$, whether some $Y\subseteq U\setminus X$ with $|Y|\le k-|X|$ makes $X\cup Y$ a rooted Steiner set,
in time $2^{g(X)}n^{O(1)}$. Moreover, $g(X)$ is computable in polynomial time, so $g$ is a running-time
exponent for this algorithm for $\Phi_{\mathrm{DST}}$-Extension.
\end{lemma}
\begin{proof}
The set $\bar K^{r}_X$, the strong components of $\vec G[\bar K_X\setminus \bar K^{r}_X]$ and the arcs between them can be computed
in polynomial time, and hence so can $g(X)$. Choose one vertex in each source component of
$\vec G[\bar K_X\setminus \bar K^{r}_X]$, and let $Q$ be the set of chosen
vertices, so $|Q|=g(X)$. Give every vertex of $\bar K_X$ the weight $0$ and every vertex $v\in U\setminus X$
the weight $w(v)$, and compute $\OPT$, the least weight of a set $D\subseteq V(\vec G)\setminus(\{r\}\cup Q)$
such that every vertex of $Q$ is reachable from $r$ in $\vec G[\{r\}\cup Q\cup D]$, by \cref{lem:cv-W}
(root $r$, terminals $Q$, weights in $\{0,\dots,w_{\max}\}$). This takes time $2^{g(X)}(n+w_{\max})^{O(1)}$
and polynomial space. We show that $\OPT$ is the least $w(Y)$ over the sets $Y$ in question; with unit
weights it is the least $|Y|$, and one accepts if and only if $\OPT\le k-|X|$.

Let $X\cup Y$ be rooted, with $Y\subseteq U\setminus X$. Then every vertex of $\bar K_X$ is reachable from
$r$ in $\vec G[\bar K_X\cup Y]$, and $D=(\bar K_X\cup Y)\setminus(\{r\}\cup Q)$ is feasible, of weight $w(Y)$.
Conversely, let $D$ be feasible, and put $Y=D\cap(U\setminus X)$, so that $w(Y)$ is the weight of $D$
and $\{r\}\cup Q\cup D\subseteq \bar K_X\cup Y$. In $\vec G_Y=\vec G[\bar K_X\cup Y]$ every vertex of $Q$ is reachable from
$r$. Each source component contains a vertex of $Q$ and is strongly connected within $\vec G[\bar K_X]$, so it
is reachable in $\vec G_Y$; every other vertex of $\bar K_X\setminus \bar K^{r}_X$ is reachable from a source component
within $\vec G[\bar K_X\setminus \bar K^{r}_X]$; and $\bar K^{r}_X$ is reachable by definition. So all of $\bar K_X$ is reachable from
$r$ in $\vec G_Y$. Let $Y'\subseteq Y$ be the set of vertices of $Y$ reachable from $r$ in $\vec G_Y$. A path from $r$
in $\vec G_Y$ visits only reachable vertices, so every vertex of $\bar K_X\cup Y'$ is reachable from $r$ in
$\vec G[\bar K_X\cup Y']$, that is, $X\cup Y'$ is rooted, and $w(Y')\le w(Y)$, as weights are nonnegative.
\end{proof}

\paragraph{The analysis.} The proof of the following theorem is in \cref{sec:app-fa-dst}. It fixes an
inclusion-minimal rooted Steiner set $S$ and an out-arborescence $A$ of $\vec G[\{r\}\cup K\cup S]$ rooted
at $r$, whose leaves other than $r$ are terminals. The running-time exponent $g(X)$ of a sample
$X\subseteq S$ is at most the number of trees of the forest $A[\bar K_X]$ that do not contain $r$
(\cref{lem:fa-dst-point}). Whether a vertex is counted depends only on whether it and its parent lie in
$\bar K_X$, so the resulting lower bound on the generating function is computed by a two-state recursion
along $A$ (\cref{lem:fa-dst-rec}). As for RS-$\theta$ (\cref{lem:rst-abstract}), a region of the plane
that the recursion maps into itself, up to known scale factors, bounds the generating function for all
arborescences at once (\cref{lem:fa-dst-cert}); three such regions, combined with rational weights, are
checked by computer (\cref{prop:fa-dst-certs}). The finishing step uses \cref{lem:cv-W}.

\begin{restatable}[Directed Steiner Tree]{theorem}{RestThmFaDst}\label{thm:fa-dst}
\pname{Directed Steiner Tree} on digraphs with $n$ vertices can be solved by a randomized algorithm
with one-sided error in time $2^{0.457n}n^{O(1)}=O(1.3727^n)$ and polynomial space.
\end{restatable}

Numerically, the analysis appears close to tight for this algorithm: on a family of digraphs built from a
directed path with pendant terminals, the bound that the generating function gives is
$\Omega(1.3716^n)$, against the proved $2^{0.457n}<1.3727^n$ (\cref{sec:anc-dst}).

\paragraph{Vertex weights.} With polynomially bounded integer weights on the non-terminals, the same
algorithm computes the least weight of a Steiner set within the same time bound, up to a factor polynomial in $w_{\max}$ (\cref{cor:fa-dst-weighted},
proved in \cref{sec:app-weighted-fa}).

\subsection{Paired Domination}\label{sec:fa-pd}

Let $G$ be a graph with $n$ vertices and no isolated vertex. A \emph{paired dominating set} is a
dominating set $S$ such that $G[S]$ has a perfect matching, and $\gamma_{\mathrm{pr}}(G)$ is the least
size of one; it exists, as the vertex set of a maximal matching is one, since $G$ has no isolated
vertex. \cite{AndreevBKSTV24} give an exact algorithm in time $O(1.7159^n)$ and polynomial space;
$1.7159^n$ is also the bound of \cite{FominGPS08} on the number
of minimal dominating sets. We use the implicit set system $\Phi_{\mathrm{PD}}$ with universe $V$ whose
solutions are the paired dominating sets, and for $X\subseteq V$ we write
$\operatorname{und}(X)=V\setminus N[X]$ for the set of vertices that $X$ leaves undominated.

\paragraph{The finishing step.} The undominated vertices are handled by inclusion--exclusion and the
perfect matching by the Tutte matrix~\cite{Tutte47}. Fix an orientation of the edges of $G$, and let
$\mathbf T$ be the skew-symmetric $V\times V$ matrix with $\mathbf T_{uv}=x_e=-\mathbf T_{vu}$ for every
edge $e=uv$ oriented from $u$ to $v$, where the $x_e$ are distinct indeterminates, and with all other
entries $0$. For $A\subseteq V$ let $\mathbf T_A$ be its principal submatrix on $A$. Then
$\det\mathbf T_A=\mathrm{Pf}(\mathbf T_A)^2$, and the Pfaffian is
$\mathrm{Pf}(\mathbf T_A)=\sum_M\pm x_M$, summed over the perfect matchings $M$ of $G[A]$, where
$x_M=\prod_{e\in M}x_e$ and $\mathrm{Pf}(\mathbf T_\emptyset)=1$. Distinct matchings give distinct
monomials.

\begin{restatable}[Tutte-matrix finishing]{lemma}{RestLemFaPdExt}\label{lem:fa-pd-ext}
There is a randomized algorithm that, given $G$, weights $w\colon V\to\{0,\dots,w_{\max}\}$, a budget
$\varpi\ge0$, $X\subseteq V$ and $k\ge0$, never accepts if no $Y\subseteq V\setminus X$ with $|Y|\le k-|X|$
makes $X\cup Y$ a paired dominating set of weight at most $\varpi$, accepts with probability at least
$\frac12$ otherwise, and runs in time $2^{|\operatorname{und}(X)|}(n+w_{\max})^{O(1)}$ and polynomial space for
every outcome of its coins. With $w\equiv0$ and $\varpi=0$ this is an algorithm for
$\Phi_{\mathrm{PD}}$-Extension with the same error bounds, in time $2^{|\operatorname{und}(X)|}n^{O(1)}$.
\end{restatable}

So $g(I,X,k)=|\operatorname{und}(X)|=n-|N[X]|$, for the instance $I=G$ (with or without weights), is a running-time
exponent, with integer values in
$\{0,\dots,n\}$, for a one-sided-error extension algorithm as in \cref{cor:fw-onesided}; it does not
depend on $k$, and we write $g(X)$ as in \cref{def:fw-cost}. It is the number of undominated vertices,
the exponent of the inclusion--exclusion finishing step for \pname{Dominating Set}: the perfect-matching
requirement does not increase it.

\paragraph{The analysis.} The proofs of \cref{lem:fa-pd-ext} and of the next two statements are in
\cref{sec:app-fa-pd}. They fix a paired dominating set $S$, a perfect matching $M$ of $G[S]$ and, for every
vertex outside $S$, an anchor in $S$ among its neighbors. A sample $X\subseteq S$ dominates a vertex outside $S$
once it contains its anchor, and a vertex of $S$ once it contains that vertex or its partner in $M$; so the
lower bound on the generating function that this gives factorizes over the pairs of $M$
(\cref{lem:fa-pd-point}). A pair responsible for $\bar n$ vertices, the pair itself and the vertices
anchored at it, contributes at most $\bar n\,h_{\bar n}(y)$ to $\log\bigl((1+y)^n/W_S(y)\bigr)$, where for an
integer $\bar n\ge2$
\[
  h_{\bar n}(y)=\log(1+y)-\frac1{\bar n}\log\bigl(y^2+4\cdot2^{-\bar n/2}y+2^{-\bar n}\bigr).
\]
Let $y^*=(3\sqrt3-1)/16=0.2622595\ldots$ and $\psi_{\mathrm{pr}}=h_6(y^*)=0.7050097\ldots$. The worst pairs
are responsible for six vertices (\cref{lem:fa-pd-rate}), as is a pair $uv$ with two vertices anchored at each
of $u$ and $v$, as in a double star, and $y^*$ is the minimizer of $h_6$ (\cref{prop:fa-pd-tight}).

\begin{restatable}[Paired Domination]{theorem}{RestThmFaPd}\label{thm:fa-pd}
For graphs $G$ with $n$ vertices and no isolated vertex, $\gamma_{\mathrm{pr}}(G)$ can be computed by a randomized algorithm with one-sided error in time
$O(1.6302^n)$ and polynomial space.
\end{restatable}

The only randomness in the extension algorithm of \cref{lem:fa-pd-ext} is the identity test by the
Schwartz--Zippel lemma. A deterministic version cannot rely on generic identity testing alone:
Kabanets and Impagliazzo~\cite{KabanetsI04} show that a deterministic polynomial-time identity test for
polynomials given by arithmetic circuits would imply circuit lower bounds that are open.

We show next that this analysis of Algorithm~\MLS{} with the running-time exponent $|\operatorname{und}(X)|$ is tight.

\begin{restatable}[Tightness]{proposition}{RestPropFaPdTight}\label{prop:fa-pd-tight}
The only critical point of $h_6$ in $(0,\infty)$ is $y^*$, and $\min_{y\ge0}h_6(y)=h_6(y^*)$.
Let $G_L$ be the disjoint union of $L$ double stars, each an edge $uv$ with two pendant vertices at $u$
and two at $v$. Then the centers form the only paired dominating set $S^*$ of $G_L$ of size at most $2L$,
and if Algorithm~\MLS{} of \cref{thm:fa-pd} is run with a rate $\psi'<\psi_{\mathrm{pr}}$, it accepts
$k=2L$ with probability tending to $0$ as $L\to\infty$.
\end{restatable}

So the base $2^{\psi_{\mathrm{pr}}}=1.63015\ldots$ is exact for Algorithm~\MLS{} with the running-time
exponent $|\operatorname{und}(X)|$; a better bound needs a different running-time exponent or extension algorithm.

\paragraph{Vertex weights.} With vertex weights in $\{0,\dots,w_{\max}\}$, written in unary, the least weight of a
paired dominating set can be computed within the same time bound, up to a factor polynomial in
$w_{\max}$ (\cref{thm:fa-pd-weighted}, proved in \cref{sec:app-weighted-fa}).

{}
\section{Conclusion and open problems}\label{sec:conclusion}

Monotone Local Search needs only large exponential moments $\mu_t(S)$ of the running-time exponent over the
subsets of one solution (\cref{def:fw-moments,thm:fw-main}), rather than a worst-case running-time
bound for every sample of a given size; the criterion is necessary for Algorithm~\MLS{}
(\cref{prop:fw-necessary}), the algorithm is optimal within the class of \cref{thm:fw-optimal}, and
for product-form running-time exponents it certifies the multivariate bound and, up to polynomial factors, nothing better
(\cref{cor:co-multivariate,prop:co-product}). The gains come from extension algorithms whose running time depends on
the structure of the sample, such as the number of Steiner terminals or of components of the sample.
Several questions remain open.

\begin{enumerate}[leftmargin=*]
\item \emph{Derandomization.} The black-box barrier of \cref{sec:derandomisation} rules out a general
  derandomization, and balanced set-inclusion families derandomize only the bounds that come from integer
  affine certificates (\cref{sec:det}). Can RS-$\theta$ or RSC be derandomized with a factor $2^{o(n)}$? Their
  analyses rest on tests along an unknown spanning tree of the solution; a counting argument shows that
  against every family of $2^{O(n)}$ partitions into blocks of size $n^{o(1)}$ some such tree has almost all
  of its edges between blocks, so these tests are not separable over the blocks of a product construction (\cref{prop:dr-tree-cut}).
\item \emph{Better trials for Connected Dominating Set.} Algorithm~\MLS{} is optimal within the class
  of \cref{thm:fw-optimal}, so with the same extension algorithm and running-time exponent further gains
  must come from trials outside this class: a non-uniform sampling distribution
  (\cref{sec:fw-levers}), a rule that adds vertices to the sample, as RSC does, or an accounting that
  charges a trial only for the extension call that it actually makes. For \pname{Connected Dominating Set}
  we do not know whether any of these improves on \cref{thm:rsc}.
\item \emph{Component-count running times.} For $c=3$ the base $1.8$ is exact for Algorithm~\MLS{}
  (\cref{prop:cc-tight,prop:cc-cfvs-tight}), so further progress for \pname{Connected Feedback Vertex
  Set} needs a different extension algorithm or a different trial. For $c=2$ the proved base $1.6444$
  is within $2\cdot10^{-4}$ of the value $2^{\psi_2}\approx1.64429$ that \cref{thm:fw-main} certifies for a
  solution whose spanning tree is a path (\cref{sec:cc-cost}), but we have no family of instances of
  \pname{Graph Motif} or \pname{Connected Vertex Cover} on which this value is tight. No deterministic
  algorithm for \pname{Connected Feedback Vertex Set} in time $O((2-\varepsilon)^n)$ is known to us; it would
  need to derandomize both the sampling and the Cut~\&~Count step. Nor do we know of an analysis that bounds a
  natural branching extension algorithm (branching on whether a vertex on a cycle joins the solution,
  followed by a Steiner phase) below $2^n$.
\item \emph{Closure rules elsewhere.} Does a closure rule, as in RSC, lower the running time for
  \pname{Directed Steiner Tree}, or give $O(1.6991^n)$ for \pname{Connected $r$-Dominating Set} with
  $r\ge2$ (\cref{rem:cv-limits})?
\item \emph{Other problems and enumeration.} Our method applies whenever the running time of the
  extension algorithm is exponential in a polynomial-time computable structural parameter of the sample
  whose exponential moments over the subsets of a solution are large. It would be interesting to
  characterize the problems for which \cref{thm:fw-main} is exponentially better than every bound from a
  product-form running-time exponent. \cite{FominGLS19} also turn product-form Monotone
  Local Search into enumeration algorithms; we have not studied enumeration with sample-dependent running
  times.
\end{enumerate}
{}
\section*{Acknowledgments}
This research was supported by the Australian Government through the Australian Research Council's
Discovery Projects funding scheme (project DP210103849). This research includes computations using the
computational cluster Katana supported by Research Technology Services at UNSW Sydney~\cite{Katana}, which was used to
build and check the certificate for \cref{thm:rsc} (\cref{sec:ancillary}). We thank the UNSW AI Institute for supporting
the co-algorithmist framework, a framework for human--AI research collaboration in algorithms and theoretical
computer science that is under development.
{}

\appendix
\crefalias{section}{appendix}
\crefalias{subsection}{subappendix}
\crefalias{subsubsection}{subsubappendix}
\section{Other levers}\label{sec:app-levers}\label{sec:fw-levers}

This appendix extends \cref{thm:fw-main} to every samplable distribution and examines two changes of the
class of \cref{thm:fw-optimal}, whose proof is in \cref{sec:app-optimal}. Sampling from a
non-uniform distribution can lower the running time by an exponential factor
(\cref{prop:fw-twoclass}), whereas sampling in several adaptive stages cannot
(\cref{prop:fw-multistage}). Neither has yet given a proven improvement of the running-time bound
for \pname{Connected Dominating Set}, and no result of the main text depends on this appendix. We keep
the notation of \cref{sec:aprime}, in particular the extendable sets $\cE$, the counts $q_\theta(t)$ and
the class optimum $\pi^*$ of \cref{def:fw-extendable}.

\subsection{Samplable distributions}\label{sec:app-samplable}

Uniform $t$-sets are one choice of sampling distribution. The theorem extends to every distribution
that can be sampled in polynomial time; the success criterion is then the exponential moment of
the distribution over the extendable sets of \cref{def:fw-extendable}. For a distribution $\mathcal X$ on
subsets of $U=U_I$ define
\[
  M_{\mathcal X}=\sum_{X\in\cE}\mathcal X(X)\,2^{-g(X)} .
\]

\begin{algorithm}[\MLS{} for a distribution $\mathcal X$]\label{alg:fw-samplable}
  \Input an instance $I$ of $\Phi$ and an integer $k\le n$.
  \Parameters $\psi$, $\Gamma$, $\kappa$
  and a distribution $\mathcal X=\mathcal X_{I,k}$ on subsets of $U_I$.
  \Body
  Take a cap $\Theta$ and trial counts $m_\theta$ as in~\eqref{eq:fw-params} (with a single index in
  place of $t$). For every integer $\theta\in\{0,\dots,\Theta\}$, repeat $m_\theta$ times: draw $X\sim \mathcal X$; if
  $|X|\le k$ and $g(X)\le\theta$, run $\cA$ on $(I,X,k)$, and answer \YES{} if it accepts. Answer
  \NO{} after all trials.
\end{algorithm}

\begin{theorem}[Exponential-moment Monotone Local Search for samplable distributions]\label{thm:fw-samplable}
  Let $\cA$ solve $\Phi$-Extension with running-time exponent $g$, and let $\mathcal X=\mathcal X_{I,k}$ be a distribution on
  subsets of $U_I$ that can be sampled exactly in time $N^{O(1)}$. Let $\psi,\Gamma,\kappa$ be as
  in \cref{thm:fw-main}. Then \cref{alg:fw-samplable} is a randomized algorithm that runs in time
  $2^{\psi n}N^{O(1)}$ and in space $N^{O(1)}$ plus the space of $\cA$ and of the sampler, answers
  \NO{} if $\cF_I$ contains no set of size at most $k$, and answers \YES{} with probability at
  least $1-e^{-\kappa}$ if
  \[
    M_{\mathcal X}\ \ge\ \frac{2^{-\psi n}}\Gamma ,
  \]
  in particular if $\sum_{X\subseteq S}\mathcal X(X)2^{-g(X)}\ge2^{-\psi n}/\Gamma$ for some $S\in\cF_I$
  with $|S|=k$.
\end{theorem}

\begin{proof}
  This is \cref{thm:fw-budgeted} with $\cT$ a single index, the trial of \cref{alg:fw-samplable} as
  $\mathrm{Tr}_\theta$, and $\xi=g(X)$ if $X\in\cE$ and $\xi=\infty$ otherwise, for $X\sim\mathcal X$:
  a trial takes time $N^{O(1)}$ plus at most $2^\theta N^{O(1)}$ (\ref{it:fw-T1}), an acceptance of $\cA$
  certifies a solution of size at most $k$ (\ref{it:fw-T2}), and on the event $\xi\le\theta$ we have
  $X\in\cE$, hence $|X|\le k$ and acceptance by $\cA$ (\ref{it:fw-T3}). Finally
  $\E[2^{-\xi}]=M_{\mathcal X}$, and the sufficient condition holds because every $X\subseteq S$ lies in
  $\cE$.
\end{proof}

The proof uses only that $\cA$ accepts on $\cE$. The same statement therefore holds in \emph{event
form}, with $\cE$ replaced by any set of samples on which a possibly different extension algorithm
is guaranteed to accept, provided that every acceptance certifies the desired output. If $\mathcal X$ is uniform on $\binom Ut$, then $M_{\mathcal X}=\mu_t(\cE)/\binom nt\ge\mu_t(S)/\binom nt$,
which gives item~(iii) of \cref{thm:fw-main} for the algorithm that runs \cref{alg:fw-samplable} once for
each $t\le k$: the run for a value of $t$ that attains the maximum in~\eqref{eq:fw-star} alone succeeds
with probability at least $1-e^{-\kappa}$, and the $k+1$ runs multiply the running time by $k+1\le n+1$. If every element is included independently with
probability $y/(1+y)$, for a rational $y\ge0$, then $\mathcal X(X)=y^{|X|}/(1+y)^n$ and
$M_{\mathcal X}\ge W_S(y)/(1+y)^n$, which gives the generating-function condition of \cref{thm:fw-main}
directly. The Bernoulli closure of Algorithm RSC (\cref{sec:rsc}) is an instance whose distribution
is not a product distribution.

\subsection{A bound on the success probability}

The lower bounds below compare algorithms by their total \emph{charge}, a running-time bound that is
fixed for each trial. The following bound is the counterpart of \cref{thm:fw-optimal}(a) for
arbitrary trials.

\begin{lemma}[Charge bound]\label{lem:fw-charge}
  Let $\pi\ge0$. Consider an algorithm that performs trials one after another and answers
  \YES{} at the first success, where trial $i$ is any randomized procedure, chosen as a possibly
  randomized function of the past, that succeeds or fails and is charged an amount
  $\mathrm{chg}_i\ge1$ determined by its outcome. Suppose that, conditioned on the history before trial
  $i$ (including the choice of the trial), trial $i$ succeeds with probability at most
  $\pi\,\E[\mathrm{chg}_i\mid\text{history}]$, and that the total charge never exceeds $\mathsf B'$. Then the
  algorithm answers \YES{} with probability at most $\pi\mathsf B'$.
\end{lemma}

\begin{proof}
  Let $\mathcal G_{i-1}$ be the history before trial $i$ together with the choice of trial $i$,
  let $s_i=\Pr[\text{trial $i$ succeeds}\mid\mathcal G_{i-1}]$ and $c_i=\E[\mathrm{chg}_i\mid\mathcal G_{i-1}]$,
  and let $i_{\max}$ be the number of trials. Every charge is at least $1$, so $i_{\max}\le \mathsf B'$, and
  $\{i\le i_{\max}\}$ is determined by $\mathcal G_{i-1}$. By hypothesis $s_i\le\pi c_i$, so the union bound
  gives
  \[
    \Pr[\YES]\le\E\Bigl[\sum_{i\le i_{\max}}s_i\Bigr]\le\pi\,\E\Bigl[\sum_{i\le i_{\max}}c_i\Bigr]
    =\pi\,\E\Bigl[\sum_{i\le i_{\max}}\mathrm{chg}_i\Bigr]\le\pi\mathsf B' ,
  \]
  where the equality holds term by term, since
  $\E[\mathbf 1[i\le i_{\max}]\mathrm{chg}_i]=\E[\mathbf 1[i\le i_{\max}]c_i]$ by the tower property.
\end{proof}

\subsection{Non-uniform sampling}

Sampling from a non-uniform distribution can lower the running time when the
instance exposes classes of elements on which including an element in the sample has different
effects on the probability that the sample lies in a solution and on the running-time exponent. A synthetic family below shows that this can change the worst-case exponent;
its analysis uses product distributions.

For a product distribution $\pi_p(X)=\prod_{u\in X}p_u\prod_{u\notin X}(1-p_u)$ with rational $p_u$
computed from the instance, the condition of \cref{thm:fw-samplable} becomes
$W^p_S:=\sum_{X\subseteq S}\pi_p(X)2^{-g(X)}\ge2^{-\psi n}/\Gamma$ for some $S\in\cF_I$ with
$|S|=k$. For running-time exponents that are sums of per-element terms on subsets of $S$, namely
$g(X)=\gamma+\sum_{u\in S\setminus X}a_u+\sum_{u\in X}b_u$ for $X\subseteq S$, this sum factorizes.
Writing $\pi_p(X)=\prod_{u\notin S}(1-p_u)\cdot\prod_{u\in S}\pi_u(X)$ with $\pi_u(X)=p_u$ if
$u\in X$ and $1-p_u$ otherwise, and expanding the product over $u\in S$ according to whether
$u\in X$, gives
\begin{equation}\label{eq:fw-product-measure}
  W^p_S=2^{-\gamma}\prod_{u\notin S}(1-p_u)\prod_{u\in S}\bigl((1-p_u)2^{-a_u}+p_u2^{-b_u}\bigr).
\end{equation}
In the following family the classes are given with the instance.

\begin{definition}[Two-class family]\label{def:fw-twoclass}
  Let $n$ be even. The \emph{two-class family} consists of instances that expose a partition
  $U=U_A\sqcup U_B$ with $|U_A|=|U_B|=n/2$ and have a unique solution $S$, so that the extendable sets
  are exactly the subsets of $S$; put $k_A=|S\cap U_A|$ and $k_B=|S\cap U_B|$. The extension
  algorithm takes two size bounds $(\ell_A,\ell_B)$, one for each class, and, for this pair, has running-time exponent
  \[
    g(X)=(\ell_A-|X\cap U_A|)\log c_A+(\ell_B-|X\cap U_B|)\log c_B
  \]
  on samples $X$ with $|X\cap U_A|\le\ell_A$ and $|X\cap U_B|\le\ell_B$, with $(c_A,c_B)=(2,4)$.
  Algorithms for the family try all of the polynomially many pairs of size bounds.
\end{definition}

\begin{proposition}[A two-class family]\label{prop:fw-twoclass}
  Let $n$ be even, and consider the two-class family of \cref{def:fw-twoclass}. Then:
  \begin{enumerate}[label=\textup{(\alph*)}]
    \item the product distribution with $p_u=1/3$ on $U_A$ and $p_u=3/7$ on $U_B$ gives, by
      \cref{thm:fw-samplable}, time $\Ostar\bigl((3/2)^{n/2}(7/4)^{n/2}\bigr)=\Ostar\bigl(\sqrt{21/8}^{\,n}\bigr)$,
      with $\sqrt{21/8}=1.6201\ldots$, on every instance of the family;
    \item on the instance with $S=U_B$, every algorithm that, trial by trial, chooses $t$, $\theta$ and a
      pair of size bounds, draws $X$ uniformly from $\binom Ut$, runs the extension algorithm when its running-time
      exponent is at most $\theta$, and is charged $2^\theta$ per trial, needs total charge at least
      $\sqrt3^{\,n}/(2(n+1))$ for success probability $1/2$; here $\sqrt3=1.7320\ldots$.
  \end{enumerate}
\end{proposition}

\begin{proof}
  (a) Fix the correct pair of size bounds $(k_A,k_B)$, and write $p_j$ for the
  probability on class $U_j$, $j\in\{A,B\}$. On subsets of $S$ the running-time exponent is a sum of
  per-element terms, with $a_u=\log c_j$ for $u\in S\cap U_j$, $b_u=0$ and $\gamma=0$. The factorization~\eqref{eq:fw-product-measure}
  gives
  \[
    W^p_S=\prod_{j\in\{A,B\}}(1-p_j)^{n/2-k_j}\Bigl(\frac{1-p_j}{c_j}+p_j\Bigr)^{k_j}.
  \]
  With $p_j=y_j/(1+y_j)$ this equals $\prod_j(1+y_j)^{-n/2}c_j^{-k_j}(1+y_jc_j)^{k_j}$. At
  $y_j=(c_j-1)/c_j$, that is $p_A=1/3$ and $p_B=3/7$, we have $1+y_jc_j=c_j$, so
  $W^p_S=\prod_j(1+y_j)^{-n/2}=\prod_j(2-1/c_j)^{-n/2}=(21/8)^{-n/2}$ for every $(k_A,k_B)$. Since
  the extendable sets are the subsets of $S$, $M_{\mathcal X}=W^p_S$, and \cref{thm:fw-samplable} with the
  correct pair of size bounds, $\Gamma=1$ and $2^{\psi n}=(21/8)^{n/2}$ gives the claim.

  (b) Here $k_A=0$ and $k_B=n/2$. A sample is accepted under a pair of size bounds $(\ell_A,\ell_B)$ only if it
  is a subset of $U_B$ and $\ell_B\ge n/2$; bounds above the class sizes are not used, so $\ell_B=n/2$,
  and the running-time exponent is then at least that of the pair $(0,n/2)$. Let $q_\theta(t)$ and $\pi^*$
  be computed for the pair $(0,n/2)$. A trial with $t$, $\theta$ and a pair of size bounds chosen before it
  succeeds only if its uniform sample $X\in\binom Ut$ is a subset of $S$ and passes the threshold
  $\theta$ under the pair $(0,n/2)$, so, conditioned on the history, it succeeds with probability at
  most $q_\theta(t)/\binom nt\le\pi^*2^\theta$, while its charge is $2^\theta\ge1$. By
  \cref{lem:fw-charge} with $\pi=\pi^*$, success probability $1/2$ needs total charge
  at least $1/(2\pi^*)$. By \cref{thm:fw-optimal}(c), $\pi^*\le\max_t\mu_t(\cE)/\binom nt$, and here
  $\cE_t=\binom St$, so $\mu_t(\cE)=\mu_t(S)=\binom{n/2}t4^{-(n/2-t)}$ for $0\le t\le n/2$. With
  $\tau=t/n\in[0,\frac12]$, the entropy estimates~\eqref{eq:fw-binom} give
  \[
    \frac{\mu_t(S)}{\binom nt}\le(n+1)\,2^{\frac n2H(2\tau)-nH(\tau)-n+2\tau n}=(n+1)\,2^{-nf_{1/2}(\tau)},
    \qquad f_{1/2}(\tau)=H(\tau)-\tfrac12H(2\tau)+1-2\tau,
  \]
  where $f_{1/2}$ is the function $f_\sigma$ of~\eqref{eq:co-fsigma} with $(\beta,\rho,\sigma)=(1,2,\frac12)$,
  the same function as in the proof of \cref{cor:dr-bb-rs}. By \cref{rem:co-fsigma}, since
  $p=2^{-\rho}=\frac14<\sigma$, its minimum over $[0,\frac12]$ is
  $F^*(\frac12)=1+\frac12\log\frac12+\frac12\log\frac32=\frac12\log3$ by~\eqref{eq:co-Fstar}. Hence
  $\max_t\mu_t(S)/\binom nt\le(n+1)\sqrt3^{\,-n}$, and the charge is at least $\sqrt3^{\,n}/(2(n+1))$.
\end{proof}

The family is synthetic: a sensible algorithm would solve the two parts separately, in time
bounded by the larger of the two running times rather than by their product. Its purpose is to
show that, for a fixed extension algorithm, the sampling distribution can change the worst-case exponent of Monotone Local Search. For
\pname{Connected Dominating Set}, where degrees are visible, it is open whether importance sampling helps.

\subsection{Multi-stage sampling}

A \emph{uniform multi-stage process} starts with $D=\emptyset$ and
repeatedly either stops, outputting $D$, or chooses a size $\ell\ge1$ and adds a uniform $\ell$-subset of
$U\setminus D$. Its choices may depend on everything seen so far, including earlier draws, running-time
exponents of intermediate sets, its own coins and even an oracle for $\cE$; it may also abort
without output. The next proposition shows that no such process outputs a given set more often
than a uniform draw of the same size does, so multi-stage trials are neutral.

\begin{proposition}[Multi-stage sampling]\label{prop:fw-multistage}
  For every uniform multi-stage process and every $Y\subseteq U$,
  $\Pr[\text{output}=Y]\le1/\binom n{|Y|}$. Consequently, suppose each multi-stage trial is charged
  $2^\theta$ for a threshold $\theta$ fixed before its first stage, and runs $\cA$ on its output $Y$
  when $|Y|\le k$ and $g(Y)\le\theta$. Then the trial succeeds with probability at most
  $\sum_{t\le k}q_\theta(t)/\binom nt\le(k+1)2^\theta\pi^*$, and \cref{thm:fw-optimal}(a) holds for
  algorithms using such trials with $\mathsf B\pi^*$ replaced by $(k+1)\mathsf B\pi^*$.
\end{proposition}

\begin{proof}
  We prove by induction on $|U\setminus D|$ that, from any point of the process at which the
  current set is $D\subseteq Y$, the probability of eventually outputting $Y$ is at most
  $1/\binom{n-|D|}{|Y\setminus D|}$, whatever the history and the later choices. If the process
  stops, it outputs $Y$ only if $D=Y$, with probability at most $1=1/\binom{n-|D|}0$. Suppose it adds
  a uniform $\ell$-subset $E$ of $U\setminus D$, and put $n'=n-|D|$ and $j=|Y\setminus D|$. It can output
  $Y$ only if $E\subseteq Y\setminus D$, which is impossible if $j<\ell$. Otherwise, by the induction
  hypothesis applied at $D\cup E$, the probability is at most
  \[
    \sum_{E\in\binom{Y\setminus D}\ell}\frac1{\binom{n'}\ell}\cdot\frac1{\binom{n'-\ell}{j-\ell}}
    =\frac{\binom j\ell}{\binom{n'}\ell\binom{n'-\ell}{j-\ell}}=\frac1{\binom{n'}j},
  \]
  using $\binom{n'}\ell\binom{n'-\ell}{j-\ell}=\binom{n'}j\binom j\ell$. A randomized choice of the action mixes
  these bounds, which are all at most $1/\binom{n'}j$, and aborting only lowers the probability. The
  process terminates, because $|D|$ grows at every step that does not stop. Taking $D=\emptyset$
  gives the first claim. For the consequence, sum the pointwise bound over the extendable passing
  sets $Y$, layer by layer, and use $q_\theta(t)/\binom nt\le2^\theta\pi^*$ and $t\le k$. The
  pointwise bound holds whatever the history, so conditioned on the history before a multi-stage trial
  with threshold $\theta$ the trial succeeds with probability at most $(k+1)\pi^*2^\theta$, while its
  charge is $2^\theta\ge1$; \cref{lem:fw-charge} with $\pi=(k+1)\pi^*$ gives the last claim.
\end{proof}

Neutrality needs the charge of a trial to be fixed before the trial starts. If the threshold may be
chosen after an early stage, the early stages act as a filter that is not charged $2^\theta$, which
we do not analyze further.
{}
\section{Related mechanisms}\label{sec:fw-mechanisms}

This appendix compares the probabilistic core of \cref{thm:fw-main} with related mechanisms
outside Monotone Local Search, and its generating function with earlier uses in Monotone Local Search
(\cref{sec:fw-related}).

\paragraph{Restarts with a cutoff.}
Luby, Sinclair and Zuckerman~\cite{LubySZ93} analyzed restart strategies for Las Vegas algorithms,
in which runs are cut off after a prescribed number of steps and restarted; \cite{AltGMKW96} truncate runs to obtain small tail probabilities. Skipping
samples whose running-time exponent exceeds a threshold is this mechanism.

\paragraph{The Markov cutoff.}
Zamir~\cite[Lemma~2]{Zamir23jensen} shows that a Las Vegas algorithm whose running time,
conditioned on a random variable $Y$, is $2^Y$ in expectation can be run in expected time
$O(2^{\E[Y]}\cdot\E[Y])$ when $\E[Y]$ is known, by repeating runs truncated at the threshold
$\E[Y]+1$ and applying Markov's inequality. The Markov-cutoff algorithm of \cref{rem:co-markov}, which
gives the mean-exponent theorem (\cref{thm:co-mean}) directly, is this mechanism transplanted into Monotone
Local Search, with the same polynomial factor in the running time. Zamir's main
theorem~\cite[Theorem~1]{Zamir23jensen} removes the polynomial factor and the knowledge of the
distribution, giving $O(2^{\E[Y]})$; Zamir states it with base $e$, and it gives this form when applied to $X=Y\ln2$, as $2^Y=e^{X}$. \Cref{thm:fw-main} differs in what it is compared with: up
to polynomial factors its running-time bound is $1/\E[2^{-g}]$, which is at most $2^{\E[g]}$ by Jensen's
inequality and can be exponentially smaller.

\paragraph{Exponential moments for restart strategies.}
Scaman~\cite[Lemma~4]{Scaman23} shows that for a Las Vegas algorithm with running time
$\mathrm{rt}\in[1,\infty]$ the best expected running time of a restart strategy lies between
$1/\E[1/\mathrm{rt}]$ and $e^2\bigl(1+\ln(1/\E[1/\mathrm{rt}])\bigr)/\E[1/\mathrm{rt}]$. For one sample size $t$, put
$\mathrm{rt}=2^{g(I,X,k)}$ if a uniform $t$-subset $X$ of the universe lies in the solution $S$ and $\mathrm{rt}=\infty$ otherwise; then
$\E[1/\mathrm{rt}]=\mu_t(S)/\binom nt$. So Scaman's Lemma~4 is the restart-strategy analogue of the
probabilistic core of \cref{thm:fw-main}: its lower bound corresponds to
\cref{thm:fw-optimal}(a), and its upper bound to \cref{thm:fw-main} for one fixed $t$, with each run cut off after the
running time that its threshold allows in place of a computed running-time exponent. Scaman bounds the expected running time, whereas we bound
the success probability within a fixed budget; the two forms convert into each other up to
constant factors (by Markov's inequality, expected time $T$ gives success probability at least $1/2$ within
budget $2T$; conversely, success probability $p$ within budget $B$ gives expected time at most $B/p$ by
independent restarts). His upper bound has the factor $e^2(1+\ln(1/\E[1/\mathrm{rt}]))$ where
our threshold lemma (\cref{lem:fw-threshold-prob}) has a factor linear in the largest threshold; both are polynomial. Scaman's impossibility result for
universal strategies, which use one cutoff sequence for every distribution, concerns a different
question, since our algorithm is given its target running time $2^{\psi n}$.

\paragraph{Exponential moments in exact algorithms.}
As Zamir~\cite{Zamir23jensen} discusses, the success probability of the PPSZ algorithm of \cite{PaturiPSZ05} is an exponential moment $\E[2^{-G}]$ of the number $G$
of guessed variables, which Jensen's inequality bounds from below by $2^{-\E[G]}$. The normalized
moments $\mu_t(S)/\binom nt=\E\left[\mathbf 1[X\subseteq S]\,2^{-g(I,X,k)}\right]$, for a uniform
$t$-subset $X$ of the universe and with $\mathbf 1[X\subseteq S]$ equal to $1$ if $X\subseteq S$ and $0$
otherwise, have the same form.

\paragraph{Generating functions.}
\cite{FominGLS19} and Gaspers and Lee~\cite{GaspersL17} already used generating
functions in Monotone Local Search, for product-form running times, in particular a negative-binomial series in $k-t$ followed by the binomial
theorem. Our $W_S(y)$ is a different series, the
binomial series over the subsets of one solution, and it is defined for arbitrary running-time exponents;
the comparison $W_S(y)\le\max_t(\mu_t(S)/\binom nt)(1+y)^n$ in the proof of \cref{thm:fw-main} is
what makes it usable.

\paragraph{Summary.} \Cref{tab:fw-mechanisms} summarizes which features of our setting the mechanisms above
have. The features are the five of the paragraph ``What is new compared with restarts and cutoffs'' in
\cref{sec:intro}: (i) the running time of a call is computed before the call, so a call above the threshold is
skipped rather than cut off; (ii) a call can succeed only if the sample lies in an unknown solution; (iii) the
running time depends on the structure of the sample, not only on its size; (iv) one generating function covers
all sample sizes; and (v) that generating function is bounded by combinatorial recursions over the sampled
structure.

\begin{table}[!htb]
  \centering
  \renewcommand{\arraystretch}{1.12}
  \begin{tabular}{@{}lccccc@{}}
    \toprule
    Mechanism & (i) & (ii) & (iii) & (iv) & (v) \\
    \midrule
    Restarts with a cutoff~\cite{LubySZ93,AltGMKW96} & no & no & no & -- & no \\
    Markov cutoff~\cite{Zamir23jensen} & no & no & no & -- & no \\
    Restart strategies via $\E[1/\mathrm{rt}]$~\cite{Scaman23} & no & no & no & -- & no \\
    Monotone Local Search, product form~\cite{FominGLS19,GaspersL17} & yes & yes & no & yes & no \\
    \Cref{thm:fw-main} with \cref{lem:fw-gf} & yes & yes & yes & yes & yes \\
    \bottomrule
  \end{tabular}
  \caption{Features (i)--(v) of the paragraph above. A dash means that the mechanism has no sample size.}
  \label{tab:fw-mechanisms}
\end{table}
{}
\section{Linear savings, product-form running times and Max Leaf Spanning Tree}\label{sec:app-co-hardness}

This appendix explains why \pname{Max Leaf Spanning Tree} needs sample-dependent running times: its natural
extension problem is W[1]-hard (\cref{sec:co-w1}), and product-form running-time exponents certify nothing below
a single unsampled call when $k/n\le d/(bc)$, nor, when $c=1$, for the worst solution size. The product-form analyses we tried had
their numerical optimum at $c=1$ (\cref{sec:co-limits}). The proofs of \cref{cor:co-linear} and of the converse part of \cref{prop:co-product} are in
\cref{sec:app-co-linear,sec:app-co-product}. We keep the notation of \cref{sec:corollaries}.

\subsection{MaxLeaf-Extension is W[1]-hard}\label{sec:co-w1}

The applications of the paper concern \pname{Max Leaf Spanning Tree} (MaxLeaf), here in its
decision version: given a connected graph $G$ and an integer $\ell$, decide whether $G$ has a spanning
tree with at least $\ell$ leaves. A \emph{leaf} of a tree is a vertex of degree~$1$, and $\operatorname{leaves}(T)$ denotes the set of
leaves of a tree $T$. The most direct way to apply Monotone Local Search to MaxLeaf is to sample
vertices that are to become leaves and to complete the sample with an algorithm for the following
extension problem.

\begin{definition}[MaxLeaf-Extension]\label{def:co-mle}
\pname{MaxLeaf-Extension}: given a graph $G$, a set $X\subseteq V(G)$ and an integer $\ell'\ge0$, decide
whether $G$ has a spanning tree $T$ with $X\subseteq \operatorname{leaves}(T)$ and $|\operatorname{leaves}(T)\setminus X|\ge\ell'$. The parameter
is $\ell'$.
\end{definition}

The words ``at least'' matter: if exactly $\ell'$ further leaves were required, the forward direction of
the reduction below could fail, because vertices that receive no pendant subdivision vertex become
additional leaves. MaxLeaf parameterized by $\ell$ is fixed-parameter tractable, with an
$\Ostar(3.188^{\ell})$-time algorithm by Zehavi~\cite{Zehavi18}. We show that its extension problem
parameterized by $\ell'$ is not, unless $\mathrm{FPT}=\mathrm{W}[1]$; in particular it has no algorithm with
running time $c^{\ell'}N^{O(1)}$, which is what the original Monotone Local Search~\cite{FominGLS19} would
need.

In the framework of \cref{def:fw-iss}, MaxLeaf is the implicit set system $\Phi_{\mathrm{ML}}$ whose
instances are pairs $(G,\ell)$ of a graph $G$ and an integer $\ell\ge0$, with universe $V(G)$ and with
solutions the sets $L\subseteq V(G)$ with $|L|=\ell$ and $L\subseteq\operatorname{leaves}(T)$ for some
spanning tree $T$ of $G$; on connected graphs, $\Phi_{\mathrm{ML}}$-Subset is the decision version above. Membership is decided in
polynomial time: graphs with at most two vertices are checked directly (a one-vertex tree has no leaf, and both
vertices of a two-vertex tree are leaves), and if $|V(G)|\ge3$, such a tree
$T$ exists exactly when $G-L$ is connected and every vertex of $L$ has a neighbor outside $L$. On an
input $((G,\ell),X,k)$, $\Phi_{\mathrm{ML}}$-Extension asks for $Y\subseteq V(G)\setminus X$ with
$|Y|\le k-|X|$ and $|X\cup Y|=\ell$ such that $X\cup Y\subseteq\operatorname{leaves}(T)$ for some spanning tree
$T$. So the input is a \NO{}-instance if $\ell<|X|$ or $\ell>k$, and otherwise it is equivalent to
the instance $(G,X,\ell-|X|)$ of \pname{MaxLeaf-Extension}, since a spanning tree with at least $\ell-|X|$
leaves outside $X$ has a set of exactly $\ell-|X|$ of them, and every such set is a valid $Y$.

\begin{theorem}[MaxLeaf-Extension is W{[1]}-hard]\label{thm:co-w1}
\textup{\pname{MaxLeaf-Extension}} parameterized by $\ell'$ is W[1]-hard. Consequently it has no algorithm with
running time $f(\ell')N^{O(1)}$, for any computable function $f$, unless $\mathrm{FPT}=\mathrm{W}[1]$.
\end{theorem}

\begin{proof}
We give a parameterized reduction from \pname{Independent Set} parameterized by the solution size,
which is W[1]-hard~\cite{CyganFKLMPPS15}. The idea is to subdivide every edge and to prescribe every
subdivision vertex as a leaf: a leaf's neighbor in a tree with at least three vertices is not a
leaf, so the further leaves lie in $V(G)$ and are pairwise non-adjacent in $G$. Let $(G,k)$ be an instance, asking whether $G$ has an
independent set of size $k$. If $k\le0$ the answer is \YES{}; if $k\ge1$ and $V(G)=\emptyset$ it is \NO{}. In
these cases we output a single edge with $X=\emptyset$ and $\ell'=0$ (a yes-instance) or $\ell'=3$ (a
no-instance). So let
$k\ge1$ and $|V(G)|\ge1$; the case $k>|V(G)|$ needs no separate treatment.

\emph{Construction.} Let $G'$ be obtained from $G$ by subdividing every edge $e=uv\in E(G)$ with a new
vertex $x_e$ (adjacent to $u$ and $v$ only), adding a vertex $r$ adjacent to every vertex of $V(G)$,
and adding a vertex $z$ adjacent only to $r$. Let $X=\{z\}\cup\{x_e:e\in E(G)\}$, and output
$(G',X,\ell')$ with $\ell'=k$. This takes polynomial time, $G'$ is connected, and the parameter is unchanged. Note that
$V(G')\setminus X=V(G)\cup\{r\}$.

\emph{Forward direction.} Let $S$ be an independent set of $G$ with $|S|=k$. Let $T$ consist of the
edge $rz$, the edges $rv$ for all $v\in V(G)$, and, for each $e=uv\in E(G)$, one edge $x_ey_e$ with
$y_e\in\{u,v\}\setminus S$; such a $y_e$ exists because $S$ is independent, so $u$ and $v$ are not both
in $S$. Then $T$ is connected, since every vertex is joined to $r$ directly or through a vertex of
$V(G)$, and it has $1+|V(G)|+|E(G)|=|V(G')|-1$ edges; hence $T$ is a spanning tree of $G'$. In $T$
the vertex $z$ and every $x_e$ have degree~$1$, and every $v\in S$ has degree~$1$, because its only
edge in $T$ is $rv$ (no $x_e$ is attached to a vertex of $S$). So $X\cup S\subseteq \operatorname{leaves}(T)$, and
$|\operatorname{leaves}(T)\setminus X|\ge|S|=\ell'$ since $S\cap X=\emptyset$.

\emph{Backward direction.} Let $T$ be a spanning tree of $G'$ with $X\subseteq \operatorname{leaves}(T)$ and
$|\operatorname{leaves}(T)\setminus X|\ge\ell'$, and put $S'=\operatorname{leaves}(T)\setminus X$. Since $|V(G)|\ge1$, the tree $T$ has at least
three vertices, so no two leaves of $T$ are adjacent in $T$ (two adjacent leaves would form a
component with two vertices). The only neighbor of $z$ in $G'$ is $r$, so $rz\in E(T)$, and $r$ is not
a leaf because its neighbor $z\in X$ is a leaf. Hence $S'\subseteq V(G')\setminus(X\cup\{r\})=V(G)$. Let $e=uv\in E(G)$. The vertex $x_e$ is a
leaf of $T$, and its neighbor in $T$ is $u$ or $v$; this neighbor is not a leaf, so $u$ and $v$ are
not both in $S'$. Thus $S'$ is an independent set of $G$ with $|S'|\ge\ell'=k$, and any $k$ of its vertices
form an independent set of size $k$.

Since parameterized reductions preserve membership in FPT, an algorithm with running time
$f(\ell')N^{O(1)}$ for \pname{MaxLeaf-Extension} would give one with running time $f(k)N^{O(1)}$ for \pname{Independent Set}, and hence
$\mathrm{FPT}=\mathrm{W}[1]$.
\end{proof}

The map $(G',X,\ell')\mapsto((G',|X|+\ell'),X,|X|+\ell')$ takes polynomial time, keeps the parameter
(the number $k-|X|$ of further elements equals $\ell'$), and by the description of $\Phi_{\mathrm{ML}}$-Extension after \cref{def:co-mle} maps
yes-instances of \pname{MaxLeaf-Extension} to yes-instances of $\Phi_{\mathrm{ML}}$-Extension and
no-instances to no-instances. So $\Phi_{\mathrm{ML}}$-Extension on inputs $(I,X,k)$, parameterized by $k-|X|$, is W[1]-hard as
well, which rules out the case $b=d=1$ of \cref{cor:co-multivariate} for $\Phi_{\mathrm{ML}}$,
under $\mathrm{FPT}\ne\mathrm{W}[1]$. It does not rule out extension algorithms with $b>1$ or $d>1$,
which the multivariate version allows, and an earlier unpublished attempt pursued exactly these. The
next subsection records what is known about this route.

\subsection{Limits of product-form running times}\label{sec:co-limits}

The product form is the case of \cref{cor:co-linear} with the following parameters, which the discussion after
\cref{prop:co-limits} uses.

\begin{remark}[Product form through linear savings]\label{rem:co-product-linear}
Let $1\le d\le bc$ and $0<k<n$, and put $\sigma=k/n$. Let $S\in\cF_I$ with $|S|=k$ be such that $g$ has
the product form \eqref{eq:co-product-g} on all inputs $(I,X,k)$ with $X\subseteq S$. Then every
$X\in\binom St$ has $g(I,X,k)=n\log b+k\log c-t\log(bc/d)$, so \eqref{eq:co-linear-hyp} holds with
equality for every $t$, with
\[
  \beta=\log b+\sigma\log c,\qquad\rho=\log\frac{bc}{d}\ge0,\qquad C=0,\qquad p=2^{-\rho}=\frac{d}{bc}.
\]
Since $g$ is constant on each $\binom St$, Jensen's inequality in the proof of \cref{cor:co-linear}(1)
is an equality, so the upper bound computed there is exact: $G_k=2^{nF^*(\sigma)}$. The criterion
$\rho>\log(1/\sigma)$ of \cref{cor:co-linear}(2) reads $\sigma>d/(bc)$.
\end{remark}

Two exact statements limit what product-form running-time exponents can certify.

\begin{proposition}[Two limits of the product form]\label{prop:co-limits}
Let $b,c,d\ge1$, $0\le k\le n$, and let $M_k$ be as in \cref{prop:co-product}. Then:
\begin{enumerate}[label=(\alph*)]
\item $\mathrm{base}(b,c,d)\ge b$, with equality if and only if $c=1$. If $c=1$, then $\max_{0\le k\le n}M_k=M_0=b^n$.
\item Let $1\le k\le n$. If $k/n\le d/(bc)$, then $M_k=b^nc^k$; if $k/n>d/(bc)$, then $M_k<b^nc^k$.
\end{enumerate}
\end{proposition}

By~(a), if $c=1$, the main theorem and multivariate Monotone Local Search certify no running-time bound
below the bound $b^nN^{O(1)}$ of a single call of the extension algorithm with $X=\emptyset$. Part~(b)
is a sampling threshold: $b^nc^k$ is the value of $2^{g(I,\emptyset,k)}$ for the unsampled call.

\begin{proof}
(a) Since $\min(d,bc)\ge1>0$, $\mathrm{base}(b,c,d)-b=\min(d,bc)(1-1/c)$ vanishes exactly when $c=1$. If $c=1$,
then $M_0=b^n$ by \eqref{eq:co-product-mu}, and $M_k\le \mathrm{base}(b,1,d)^n=b^n$ for all $k$ by
\cref{prop:co-product}(b).

(b) Write $a_t=\binom nt\binom kt^{-1}(d/(bc))^t$, so that $M_k=b^nc^k\min_{0\le t\le k}a_t$ and
$a_0=1$. For $0\le t<k$,
\[
  \frac{a_{t+1}}{a_t}=\frac{n-t}{k-t}\cdot\frac{d}{bc},
\]
because $\binom n{t+1}/\binom nt=(n-t)/(t+1)$ and $\binom k{t+1}/\binom kt=(k-t)/(t+1)$. As $k\le n$,
the factor $(n-t)/(k-t)$ is at least $n/k$ and is nondecreasing in $t$. If $k/n\le d/(bc)$, every ratio is at
least $(n/k)(d/(bc))\ge1$, so the sequence $(a_t)$ is nondecreasing and $\min_ta_t=a_0=1$. If
$k/n>d/(bc)$, then $a_1=(n/k)(d/(bc))<1=a_0$, so $M_k\le b^nc^ka_1<b^nc^k$.
\end{proof}

Part (b) is the exact, finite-$n$ form of the following asymptotic observation: the limit of
$\frac1n\log$ of the term $\binom nt\binom kt^{-1}b^{\,n-t}c^{\,k-t}d^{\,t}$ of the classical expression,
for fixed $k/n$ and $t/n$, is a convex function of the sampling fraction $t/n$ and is minimized at
$t=0$ exactly when $k/n\le d/(bc)$. Both are the product-form case of \cref{cor:co-linear}(2): with
$\rho=\log(bc/d)$ and $\sigma=k/n$ (\cref{rem:co-product-linear}), the condition $\rho>\log(1/\sigma)$
reads $k/n>d/(bc)$.

We complement \cref{prop:co-limits} with numerical evidence from a local numerical optimization
(\cref{sec:ancillary}): with the measure-and-conquer branching algorithm of Binkele-Raible and
Fernau~\cite{Binkele-RaibleF14} as a \pname{MaxLeaf-Extension} algorithm, the prescribed leaves $X$ as
initial floating leaves, and a general linear measure optimized for the base $b+d-d/c$, every variant
we tried had its optimum at $c=1$, where by \cref{prop:co-limits}(a) multivariate Monotone Local
Search certifies nothing below a single call of the extension algorithm with $X=\emptyset$. A
first-order sensitivity analysis indicates the reason: each branching that determines the base has a
branch that creates no leaves and does not decrease $\ell'$. This is evidence about the analyses we
tried, not a theorem, and it is not used elsewhere; the resulting bound, $\Ostar(1.8957^n)$ with a
numerically computed base, is also weaker than $\Ostar(1.8619^n)$~\cite{Abu-KhzamML11}.
{}
\section{Deferred proofs}\label{sec:app-proofs}
\statementswithin{subsection}

This appendix proves the statements of the main text whose proofs were deferred. Each statement is
restated, with its number, before its proof, together with the lemmas that only these proofs use. The
subsections follow the order of the main text: optimality in its class (\cref{sec:app-optimal}), linear
savings and product-form running times (\cref{sec:app-co-proofs}), the Steiner computations
(\cref{sec:app-steiner}), Algorithm RS-$\theta$ (\cref{sec:app-rst}), the closure bound of Algorithm RSC
(\cref{sec:app-rsc}), the black-box barrier (\cref{sec:app-barrier}), balanced set-inclusion families
(\cref{sec:app-balanced}), the deterministic algorithms (\cref{sec:app-det}), \pname{Connected Feedback
Vertex Set} (\cref{sec:app-fvs}), the weighted finishing steps (\cref{sec:app-weighted}) and the further
applications (\cref{sec:app-variants}).

\subsection{Optimality in its class}\label{sec:app-optimal}

\RestThmFwOptimal*

\begin{proof}
  (a) Let $\mathcal G_{i-1}$ be everything the algorithm has observed before trial $i$, including
  the coins it uses to choose $(t_i,\theta_i)$, and let $i_{\max}$ be the number of trials. Every trial is
  charged at least $2^0=1$, so $i_{\max}\le \mathsf B$. The event $\{i\le i_{\max}\}$ is decided before trial $i$, so it is
  determined by $\mathcal G_{i-1}$. Because $\cA$ is correct, trial $i$ succeeds exactly when
  $X_i\in\cE_{t_i}$ and $g(X_i)\le\theta_i$. As $X_i$ is uniform and independent of
  $\mathcal G_{i-1}$,
  \[
    \Pr\bigl[\text{trial $i$ succeeds}\bigm|\mathcal G_{i-1}\bigr]
    =\frac{q_{\theta_i}(t_i)}{\binom n{t_i}}\le\pi^*2^{\theta_i}.
  \]
  By the union bound and the tower property,
  \[
    \Pr[\YES]\le\sum_{i\ge1}\E\Bigl[\mathbf 1[i\le i_{\max}]\,
    \Pr\bigl[\text{trial $i$ succeeds}\bigm|\mathcal G_{i-1}\bigr]\Bigr]
    \le\pi^*\,\E\Bigl[\sum_{i\le i_{\max}}2^{\theta_i}\Bigr]\le\pi^*\mathsf B .
  \]

  (b) Let $(t,\theta)$ attain $\pi^*$ and put $p=q_\theta(t)/\binom nt=\pi^*2^\theta\ge
  2^{\theta-\psi n}/\Gamma$. Since $2^{\theta-\psi n}/\Gamma\le p\le1$, we have $2^\theta\le\Gamma2^{\psi n}$, so
  $\theta\le\psi n+\log\Gamma$, which is below the cap $\Theta\ge\psi n+\log\Gamma+1$.
  So the algorithm runs trials of the pair $(t,\theta)$, and their number is
  $m_{t,\theta}\ge M_{t,\theta}\ge\kappa\Gamma2^{\psi n-\theta}\ge\kappa/p$. Each such trial succeeds with
  probability $p$, because $\cA$ accepts every $X\in\cE_t$. So all of them fail with probability at
  most $(1-p)^{m_{t,\theta}}\le e^{-pm_{t,\theta}}\le e^{-\kappa}$.

  (c) For every $\theta$,
  \[
    q_\theta(t)2^{-\theta}=\sum_{X\in\cE_t,\ g(X)\le\theta}2^{-\theta}
    \le\sum_{X\in\cE_t,\ g(X)\le\theta}2^{-g(X)}\le\mu_t(\cE).
  \]
  In the integer case, with $q_{-1}(t)=0$, exactly $q_\theta(t)-q_{\theta-1}(t)$ sets in $\cE_t$
  have running-time exponent $\theta$, so
  $\mu_t(\cE)=\sum_{\theta=0}^{g_{\max}}(q_\theta(t)-q_{\theta-1}(t))2^{-\theta}
  \le\sum_{\theta=0}^{g_{\max}}q_\theta(t)2^{-\theta}$, and the largest of these $g_{\max}+1$ terms is at
  least $\mu_t(\cE)/(g_{\max}+1)$.
\end{proof}

\subsection{Linear savings and product-form running times}\label{sec:app-co-proofs}

This subsection proves \cref{cor:co-linear} and the converse part of \cref{prop:co-product}; the rest of
what is known about product-form running times is in \cref{sec:app-co-hardness}.

\subsubsection{Proof of the linear-savings corollary}\label{sec:app-co-linear}

We prove \cref{cor:co-linear} here, with the notation of \cref{sec:co-linear}.

\RestCorCoLinear*

\begin{proof}
(1) Let $S\in\cF_I$ with $|S|=k$, and put $u=2^{\rho}=1/p$. As in the proof of
\cref{thm:co-mean}, Jensen's inequality and \eqref{eq:co-linear-hyp} give
$\mu_t(S)\ge\binom kt2^{-(\beta n-\rho t+C)}$ for every $t\in\{0,\dots,k\}$. Hence, by the binomial
theorem, for every $y\ge0$
\[
  W_S(y)\ge2^{-\beta n-C}\sum_{t=0}^{k}\binom kt(uy)^t=2^{-\beta n-C}(1+uy)^k,
  \qquad\text{so}\qquad
  \frac{(1+y)^n}{W_S(y)}\le2^{\beta n+C}\,2^{E(y)},
\]
where $E(y)=n\log(1+y)-k\log(1+uy)$. The derivative
$E'(y)=\frac{1}{\ln2}\left(\frac{n}{1+y}-\frac{ku}{1+uy}\right)$ has the sign of
\[
  n(1+uy)-ku(1+y)=n(1-\sigma u)+uy(n-k),
\]
which is affine in $y$ with nonnegative slope $u(n-k)$.

If $\sigma\le p$, then $\sigma u\le1$, so $E'\ge0$ on $[0,\infty)$ and
$\min_{y\ge0}E(y)=E(0)=0$. At $y=0$ the upper bound is therefore
$2^{\beta n+C}=2^{C+nF^*(\sigma)}$.

If $\sigma>p$, then $\sigma u>1$, so $E$ decreases on $[0,y_0]$ and increases on $[y_0,\infty)$,
where $y_0=(\sigma u-1)/(u(1-\sigma))>0$ is the root of the affine function above. From
$1+y_0=(u-1)/(u(1-\sigma))$ and $1+uy_0=\sigma(u-1)/(1-\sigma)$,
\[
  \begin{aligned}
  \frac{E(y_0)}{n}&=\log(1+y_0)-\sigma\log(1+uy_0)=H(\sigma)+(1-\sigma)\log(u-1)-\log u\\
  &=H(\sigma)+(1-\sigma)\log(1-p)-\sigma\rho ,
  \end{aligned}
\]
where the last step uses $\log(u-1)=\log(1-p)+\rho$ and $\log u=\rho$. By \eqref{eq:co-Fstar} this
is $F^*(\sigma)-\beta$, so $(1+y_0)^n/W_S(y_0)\le2^{C+nF^*(\sigma)}$.

The algorithm of~(1) computes a rational number $\psi$ with
$C+nF^*(\sigma)\le\psi n\le C+nF^*(\sigma)+1$, in time $N^{O(1)}$ from approximations of $\beta$,
$\rho$, $C$ and the logarithms in $F^*(\sigma)$, and runs Algorithm \MLS{} with rate
$\psi^+=\max(0,\psi)$, $\Gamma=1$ and confidence $\kappa$. In both cases above, the
generating-function condition holds at $y=0$ if $\sigma\le p$ and at $y=y_0$ if $\sigma>p$, with
$\Gamma=1$ and $2^{C+nF^*(\sigma)}$ in place of $2^{\psi n}$, hence also with the larger value
$2^{\psi^+n}$, so \eqref{eq:fw-star} holds with rate $\psi^+$ by the first inequality of
\cref{lem:fw-gf}. The rate $\psi^+$ is rational, computable and nonnegative, and
$\psi^+\le\max(0,(C+1)/n+\beta)\le N^{O(1)}$ because $F^*(\sigma)\le\beta$ and $n\ge1$. So
\cref{thm:fw-main} applies, and its time bound
$2^{\psi^+n}N^{O(1)}\le2^{\max(0,\,C+nF^*(\sigma)+1)}N^{O(1)}\le2\cdot2^{\max(0,\,C+nF^*(\sigma))}N^{O(1)}$
gives the claims.

(2) By the remark after \eqref{eq:co-Fstar}, $F^*(\sigma)<\beta$ holds exactly when $\sigma>p=2^{-\rho}$,
which is equivalent to $\rho>\log(1/\sigma)$.
\end{proof}

\subsubsection{Proof of the converse for product-form running times}\label{sec:app-co-product}

We prove here part~(c) of \cref{prop:co-product} and the final inequalities stated after
it; parts~(a) and~(b) are proved in \cref{sec:co-product}. We keep the notation of that proposition.

\RestPropCoProduct*

\begin{proof}[Proof of \cref{prop:co-product}(c) and of the final inequalities]
(c) Put $u=bc/d$ and, for real $x\in[0,n]$ and $y\ge0$,
$E_x(y)=n\log(1+y)-x\log(1+uy)$, so that by (a)
$(1+y)^n/W(y)=b^nc^k2^{E_k(y)}$ for integers $k$. As in the proof of \cref{cor:co-linear}, the
derivative $E_x'(y)$ has the sign of $n(1-(x/n)u)+uy(n-x)$.

\emph{The case $c=1$.} For $k=0$ only $X=\emptyset$ occurs: $W(y)=b^{-n}$, so
$G_0=\inf_{y\ge0}b^n(1+y)^n=b^n$, attained at $y=0$, and $M_0=b^n$. Also $\mathrm{base}(b,1,d)=b$.

\emph{The case $c>1$ and $d<bc$.} Here $\sigma^*>0$, and $\sigma^*<1$ because $cd<bc+cd-d$ is
equivalent to $d<bc$. Moreover $\sigma^*u=bc^2/(bc+cd-d)>1$, because this is equivalent to
$bc^2-bc>cd-d$, that is $bc(c-1)>d(c-1)$, which holds as $c>1$ and $d<bc$. Take $x=\sigma^*n$.
Then $E_x'(0)<0$, and since $x<n$ the affine function that carries the sign of
$E_x'$ has positive slope, so $E_x$ has a unique minimizer on $[0,\infty)$,
the root of
$n(1+uy)=xu(1+y)$. Since $1+uy^*=c$ and
$\sigma^*u(1+y^*)=\frac{bc^2}{bc+cd-d}\cdot\frac{bc+cd-d}{bc}=c$, the root is $y^*$. Therefore
\begin{equation}\label{eq:co-psi-star}
  \inf_{y\ge0}\Bigl[n\log\bigl(b(1+y)\bigr)+x\log\frac{c}{1+uy}\Bigr]
  =n\log\bigl(b(1+y^*)\bigr)+x\log\frac{c}{c}=n\log\Bigl(b+d-\frac dc\Bigr),
\end{equation}
and if $\sigma^*n$ is an integer $k$, the left-hand side is $\log G_k$, so $G_k=(b+d-d/c)^n$.

For general $n\ge1$, define for real $x\in[0,n]$
\[
  \mathsf f(x)=\inf_{y\ge0}\Bigl[n\log\bigl(b(1+y)\bigr)+x\log\frac{c}{1+uy}\Bigr],
\]
so that $\mathsf f(k)=\log G_k$ for integers $0\le k\le n$, by (a). The function $\mathsf f$ is finite: since
$u>1$ we have $1+uy\le u(1+y)$, so the bracket is at least
$n\log b+(n-x)\log(1+y)+x\log(c/u)\ge n\log b-n|\log(c/u)|$. As an infimum of functions
that are affine in $x$, $\mathsf f$ is concave on $[0,n]$. We have $\mathsf f(0)=n\log b$ (the bracket is
$n\log(b(1+y))$, minimized at $y=0$), and $\mathsf f(\sigma^*n)=n\log(b+d-d/c)$ by
\eqref{eq:co-psi-star}; note that $b+d-d/c\ge b$, because $c\ge1$. Let $k=\lfloor\sigma^*n\rfloor$, so
$\sigma^*n-1<k\le\sigma^*n$. By concavity on $[0,\sigma^*n]$,
\[
  \begin{aligned}
  \log G_k=\mathsf f(k)&\ge\Bigl(1-\frac{k}{\sigma^*n}\Bigr)\mathsf f(0)+\frac{k}{\sigma^*n}\,\mathsf f(\sigma^*n)
  =n\log b+\frac{k}{\sigma^*}\log\frac{b+d-d/c}{b}\\
  &\ge n\log\Bigl(b+d-\frac dc\Bigr)-\frac{1}{\sigma^*}\log\frac{b+d-d/c}{b},
  \end{aligned}
\]
which is the claimed bound.

\emph{The case $c>1$ and $d\ge bc$.} Take $k=n$. Now $u\le1$, so $1+uy\le1+y$ and
\[
  \frac{(1+y)^n}{W(y)}=(bc)^n\Bigl(\frac{1+y}{1+uy}\Bigr)^n\ge(bc)^n,
\]
with equality at $y=0$; so $G_n=(bc)^n$. Also
$M_n=\min_{0\le t\le n}(bc)^{n-t}d^t=(bc)^n$, attained at $t=0$, since $d\ge bc$.

\emph{The final inequalities.} The upper bound $\max_kM_k\le \mathrm{base}(b,c,d)^n$ follows from (b). For the lower bound,
\cref{lem:fw-gf} gives $M_k\ge G_k/(n+1)$ for every $k$, and the three cases give some $k$ with
$G_k\ge \mathrm{base}(b,c,d)^n/\mathsf K$: in the second case $\mathrm{base}(b,c,d)=b+d-d/c$ because $d<bc$.
\end{proof}
{}
\subsection{The Steiner computations}\label{sec:app-steiner}

This subsection proves the statements of \cref{sec:cv-steiner,sec:cds-states}, with the notation of those
subsections.

\RestLemCvW*

\begin{proof}
If $K=\emptyset$, then $D=\emptyset$ is feasible and the answer is $0$; let $K\neq\emptyset$. Let
$\varphi'$ agree with $\varphi$ outside $\{r\}\cup K$ and vanish on $\{r\}\cup K$. Give every arc $uv$ the
length $\mathrm{len}(uv)=n\,\varphi'(v)+1$, a positive integer.

\emph{Branching walks.} A \emph{branching walk from $v$} is a finite sequence
$\bigl((u_1,\mathcal W_1),\dots,(u_q,\mathcal W_q)\bigr)$, $q\ge0$, in which $vu_i\in A(\vec G)$ and $\mathcal W_i$ is a
branching walk from $u_i$, for every $i$; this is an inductive definition, and the empty sequence is
the branching walk with $q=0$. Its \emph{image} is $\{v\}$ together with the images of the $\mathcal W_i$,
its \emph{arcs} are the arcs $vu_i$ together with the arcs of the $\mathcal W_i$ (counted with
multiplicity), and its \emph{length} is the sum of the lengths of its arcs. Equivalently, a branching
walk is a rooted ordered tree mapped into $\vec G$ so that the root goes to $v$ and every tree edge, directed
away from the root, goes to an arc of $\vec G$. By induction, every vertex of the image other than $v$ is the
head of one of its arcs, and every vertex of the image is reachable from $v$ in the subdigraph induced
by the image.

For $Y\subseteq V(\vec G)$, $v\in V(\vec G)$ and an integer $\mathsf L\ge0$ let $b_Y(v,\mathsf L)$ be the number of branching walks
from $v$ of length $\mathsf L$ whose image is disjoint from $Y$. Then $b_Y(v,\mathsf L)=0$ if $v\in Y$, and for
$v\notin Y$
\[
  b_Y(v,0)=1,\qquad
  b_Y(v,\mathsf L)=\sum_{\substack{vu\in A(\vec G)\\u\notin Y}}\ \sum_{\mathsf L_1=0}^{\mathsf L-\mathrm{len}(vu)}b_Y(u,\mathsf L_1)\,b_Y\bigl(v,\mathsf L-\mathrm{len}(vu)-\mathsf L_1\bigr)
  \quad(\mathsf L\ge1).
\]
Indeed, the only branching walk of length $0$ is the empty one, since every arc has positive length.
A nonempty branching walk from $v$ splits uniquely into its first pair $(u_1,\mathcal W_1)$ and the
branching walk formed by the remaining pairs, which again starts at $v$; the length is
$\mathrm{len}(vu_1)$ plus the two lengths, and the image avoids $Y$ exactly when $u_1\notin Y$ and both pieces
avoid $Y$ (the image of the remainder contains $v$). The right-hand side refers only to lengths below
$\mathsf L$, so the table can be filled in order of increasing $\mathsf L$.

\emph{Inclusion--exclusion.} For $\mathsf L\ge0$ put $b(\mathsf L)=\sum_{Y\subseteq K}(-1)^{|Y|}b_Y(r,\mathsf L)$. A
branching walk from $r$ with image $\mathcal I$ is counted in $b_Y(r,\mathsf L)$ exactly for the sets
$Y\subseteq K\setminus \mathcal I$, and $\sum_{Y\subseteq K\setminus \mathcal I}(-1)^{|Y|}$ is $1$ if $K\subseteq \mathcal I$ and
$0$ otherwise. So $b(\mathsf L)$ is the number of branching walks from $r$ of length $\mathsf L$ whose image contains
$K$.

\emph{Least length.} We show that $b(\mathsf L)>0$ for some $\mathsf L$ if and only if $\OPT=\OPT_\varphi(\vec G,r,K)<\infty$,
and that the least such $\mathsf L$, say $\mathsf L^*$, satisfies $\lfloor \mathsf L^*/n\rfloor=\OPT$.
First let a branching walk from $r$ have image $\mathcal I\supseteq K$ and length $\mathsf L$. Every vertex of
$\mathcal I\setminus\{r\}$ is the head of one of its arcs, and the arcs have lengths at least $n\,\varphi'$ of their
heads, so, as $\varphi'\ge0$ and $\varphi'(r)=0$,
\[
  \mathsf L\ \ge\ n\sum_{v\in \mathcal I}\varphi'(v)\ =\ n\,\varphi(D_{\mathcal I}),\qquad D_{\mathcal I}=\mathcal I\setminus(\{r\}\cup K).
\]
Every terminal is reachable from $r$ in $\vec G[\mathcal I]=\vec G[\{r\}\cup K\cup D_{\mathcal I}]$, so $D_{\mathcal I}$ is feasible and
$\lfloor \mathsf L/n\rfloor\ge \varphi(D_{\mathcal I})\ge\OPT$. Conversely, let $D$ be feasible with $\varphi(D)=\OPT$, and let $T$ be an
out-tree of $\vec G[\{r\}\cup K\cup D]$ rooted at $r$ that spans the vertices reachable from $r$ there,
for instance a breadth-first search tree. Ordering the children of every vertex arbitrarily turns $T$ into
a branching walk from $r$ whose image contains $K$. It has $|A(T)|\le n-1$ arcs, each non-root vertex of $T$
is the head of exactly one of them, and the non-root vertices of $T$ outside $K$ lie in $D$; so its
length is at most $n\,\varphi(D)+|A(T)|\le n\,\OPT+n-1$. Hence $\mathsf L^*$ exists and $\lfloor \mathsf L^*/n\rfloor\le\OPT$, and
with the first part $\lfloor \mathsf L^*/n\rfloor=\OPT$. As $\OPT\le \varphi(V(\vec G))\le n\,\varphi_{\max}$, it suffices to compute
$b(\mathsf L)$ for $0\le \mathsf L\le \mathsf L_{\max}:=n^2\varphi_{\max}+n-1$; if all these values are $0$, then $\OPT=\infty$.

\emph{Running time and space.} A branching walk of length at most $\mathsf L$ has at most $\mathsf L$ arcs; it is determined by a rooted
ordered tree with $e\le \mathsf L$ edges, of which there are at most $4^e$, and by the images of the $e$ non-root
vertices. So $b_Y(v,\mathsf L)\le(\mathsf L+1)(4n)^{\mathsf L}$, and all numbers involved, including the partial sums of $b(\mathsf L)$, have
$O(\mathsf L_{\max}\log n+n)$ bits, which is polynomial in $n+\varphi_{\max}$. For one set $Y$ the table
$\bigl(b_Y(v,\mathsf L)\bigr)_{v,\,\mathsf L\le \mathsf L_{\max}}$ has $n(\mathsf L_{\max}+1)$ entries, each computed with
$O(n\mathsf L_{\max})$ arithmetic operations. The algorithm runs through the $2^{|K|}$ sets $Y\subseteq K$ one at
a time, keeps one table and the $\mathsf L_{\max}+1$ partial sums of $b(\mathsf L)$, and finally returns
$\lfloor \mathsf L^*/n\rfloor$ or $\infty$. This takes time $2^{|K|}(n+\varphi_{\max})^{O(1)}$ and space $(n+\varphi_{\max})^{O(1)}$.
\end{proof}

\RestLemCdsDirSteiner*

\begin{proof}
Let $C_1\in\cD$ and put $\rho=\hat C_1$ and $K=K_1$. The non-root, non-terminal vertices of $\vec G_J$
are the vertices $\hat b$ with $b\in B$, so a feasible set for $(\vec G_J,\rho,K)$ in the sense of
\cref{sec:cv-steiner} has the form $\hat Z=\{\hat b:b\in Z\}$ with $Z\subseteq B$, and its weight is
$\varphi(Z)$. We show that
\begin{equation}\label{eq:cds-dir-match}
  \OPT_\varphi(\vec G_J,\rho,K)=\min\bigl\{\varphi(Y):Y\text{ a solution of }J\bigr\},
\end{equation}
which is the claim.

\emph{From solutions to feasible sets.} Let $Y$ be a solution of $J$. Let $H$ be the graph obtained from
the connected graph $G[V(\cD)\cup Y]$ by contracting each member of $\cD$ into one vertex; it is
connected, and its vertices correspond to the vertices $\hat C$ ($C\in\cD$) and $\hat y$ ($y\in Y$) of
$\vec G_J$, among them $\rho$. Every edge of $H$ joins two vertices $y,y'\in Y\subseteq B$ or a vertex
$C\in\cD$ to a vertex $y\in N(C)\cap Y$, since distinct members of $\cD$ are non-adjacent; in both cases
$\vec G_J$ has the two opposite arcs between the corresponding vertices. So every $\hat C$ ($C\in\cD$) and
every $\hat y$ ($y\in Y$) is reachable from $\rho$ in $\vec G_J[\{\rho\}\cup K\cup\hat Y]$. Every $r\in R$
lies in $N[Y]$, so some $y\in Y\cap N[r]\subseteq N[r]\cap B$ exists, $\hat y$ is reachable from $\rho$,
and $\check r$ is reached through the arc $\hat y\check r$. Hence $\hat Y$ is feasible, of weight
$\varphi(Y)$, and the left-hand side of~\eqref{eq:cds-dir-match} is at most the right-hand side.

\emph{From feasible sets to solutions.} Let $\hat Z$ be feasible, let $Q$ be the set of vertices reachable
from $\rho$ in $\vec G_J[\{\rho\}\cup K\cup\hat Z]$, which contains $\rho$ and $K$, and let
$Y'=\{b\in Z:\hat b\in Q\}$. Then $Y'\subseteq B$ and $\varphi(Y')\le\varphi(Z)$. The
vertices of $\check R$ have no outgoing arcs, so every vertex of $Q\setminus\check R$ is reached from $\rho$
by a path all of whose vertices lie in $Q\setminus\check R$, that is, are of the form $\hat C$ ($C\in\cD$) or
$\hat b$ ($b\in Y'$). Each arc of such a path joins two vertices whose underlying sets ($C$ or $\{b\}$) are
joined by an edge of $G$, and every $C\in\cD$ induces a connected subgraph of $G$; as $Q\setminus\check R$
consists of all $\hat C$ ($C\in\cD$) and all $\hat b$ ($b\in Y'$), the graph $G[V(\cD)\cup Y']$ is connected.
It is nonempty, because it contains $C_1$. Every $\check r$ with $r\in R$ lies in $Q$, and its
predecessor on a path from $\rho$ is some $\hat b\in Q$ with $b\in N[r]\cap B$, so $b\in Y'$; hence
$R\subseteq N[Y']$. So $Y'$ is a solution of $J$, and the right-hand side
of~\eqref{eq:cds-dir-match} is at most $\varphi(Y')\le\varphi(Z)$. This
proves~\eqref{eq:cds-dir-match}.

For the running time, let first $\cD\neq\emptyset$. The digraph $\vec G_J$ is built in polynomial time, has at
most $2n$ vertices and weights at most $\varphi_{\max}$, and \cref{lem:cv-W} is applied once, with
$|K_1|=|\cD|-1+|R|=|\Tm(J)|-1$ terminals; this takes time $2^{|\Tm(J)|}(n+\varphi_{\max})^{O(1)}$ and space
$(n+\varphi_{\max})^{O(1)}$, deterministically. If $\cD=\emptyset$, every solution $Y$ of $J$ is nonempty,
because $V(\cD)\cup Y\neq\emptyset$. For $b\in B$ put $J^{(b)}=(\{\{b\}\},B\setminus\{b\},R\setminus N[b])$.
This is a state ($\{b\}$ is connected, $B\setminus\{b\}$ avoids it and $R\setminus N[b]$ avoids $N[b]$), and a
set $Y\subseteq B$ with $b\in Y$ is a solution of $J$ if and only if $Y\setminus\{b\}$ is a solution of $J^{(b)}$: $G[Y]=G[\{b\}\cup(Y\setminus\{b\})]$, and $R\subseteq N[Y]=N[b]\cup N[Y\setminus\{b\}]$
holds if and only if $R\setminus N[b]\subseteq N[Y\setminus\{b\}]$. Hence
$\OPT_\varphi(J)=\min_{b\in B}\bigl(\varphi(b)+\OPT_\varphi(J^{(b)})\bigr)$, with $\min\emptyset=\infty$, and
$|\Tm(J^{(b)})|=|R\setminus N[b]|+1\le|\Tm(J)|+1$. Applying the first part to the at most $n$ states $J^{(b)}$
gives the same bounds.
\end{proof}

\RestCorCdsSteiner*

\begin{proof}
The first statement is \cref{lem:cds-dir-steiner} with $\varphi\equiv1$, so that $\varphi_{\max}=1$ and the
bounds become $2^{|\Tm(J)|}n^{O(1)}$ and $n^{O(1)}$.

For the second statement, $\Phi_{\mathrm{CDS}}$ is an implicit set system (\cref{def:fw-iss}): $V(G)$
is listed and connected domination is tested in polynomial time, and strings that do not encode a
graph get an empty universe. By the description of the solutions of $J_X$ after~\eqref{eq:cds-JX},
$\cA_{\mathrm{St}}$ accepts $(G,X,k)$ exactly when some $Y\subseteq V\setminus X$ with $|Y|\le k-|X|$
makes $X\cup Y$ a connected dominating set. Its running time on every input, whether or not $X$ lies
in a solution, is $2^{|\Tm(J_X)|}n^{O(1)}$ by the first statement, and $|\Tm(J_X)|$ is computed in
polynomial time by~\eqref{eq:cds-TX}, which also gives the range.
\end{proof}
{}
\subsection{Algorithm RS-\texorpdfstring{$\theta$}{theta}}\label{sec:app-rst}

This subsection proves the statements of \cref{sec:rst-tree,sec:rst-recursion,sec:rst-antinorm}, with the
notation of \cref{sec:rstheta}, and the lower bound of \cref{sec:rst-caterpillar} on uniform lobsters
(\cref{sec:app-lobster}).

\subsubsection{The tree reduction and the three-state recursion}\label{sec:app-rst-tree}

\RestLemRstE*

\begin{proof}
By~\eqref{eq:cds-TX}, $|\Tm(J_X)|=|(V\setminus S)\setminus N[X]|+|S\setminus N[X]|+\comp(G[X])$, and
we bound its three parts by the three terms of $\Pi_{F,\nu}(X)$ in turn.
(1) If $u\in V\setminus S$ is not in $N[X]$, then its neighbor $o(u)$ is not in $X$; so
$|(V\setminus S)\setminus N[X]|\le\sum_{v\in S\setminus X}|o^{-1}(v)|=\sum_{v\in S\setminus X}\nu_v$.
(2) If $v\in S$ is not in $N[X]$, then $N_G[v]\cap X=\emptyset$, so $N_F[v]\cap X=\emptyset$ because
$N_F[v]\subseteq N_G[v]$.
(3) $F[X]$ is a spanning subgraph of $G[X]$, as $X\subseteq S=V(F)$; so $\comp(G[X])\le\comp(F[X])$.

For the equality case, suppose that $o(u)$ is the only neighbor of $u$ in $S$ for every $u\notin S$.
Then, as $X\subseteq S$, $u\notin N[X]$ if and only if $o(u)\notin X$, so (1) is an equality. If
moreover $G[S]=F$, then $N_G[v]\cap S=N_F[v]$ for $v\in S$ and $G[X]=F[X]$, so (2) and (3) are
equalities. The last claim holds termwise, as $2^{-|\Tm(J_X)|}\ge2^{-\Pi_{F,\nu}(X)}$.
\end{proof}

\RestLemRstTreerec*

\begin{proof}
The proof is a double induction whose core is a decomposition of the potential of a partial subtree
into the potentials of its two pieces, with one correction for the edge joining $v$ to the new child
and one for the domination of that child; the three states carry the information that these
corrections need.

Let $F'$ be the subtree consisting of a vertex $v$ and the subtrees $F_{w_1},\dots,F_{w_i}$ of its first
$i$ children ($0\le i\le d$). For $X\subseteq V(F')$ let
\[
  \Pi^{\mathrm{in}}_{F'}(X)=\sum_{u\in V(F')\setminus X}\nu_u+\bigl|\{u\in V(F')\setminus\{v\}:N_F[u]\cap X=\emptyset\}\bigr|
  +\comp(F'[X]);
\]
for $u\neq v$ in $F'$ the set $N_F[u]$ lies inside $V(F')$, since the parent of $u$ is $v$ or lies in
the same subtree $F_{w_k}$ for some $k\le i$. Let $P'=(P'_1,P'_2,P'_3)$ be the sums of $y^{|X|}2^{-\Pi^{\mathrm{in}}_{F'}(X)}$ over the
sets $X\subseteq V(F')$ in states $1$, $2$ and $3$ of $v$, where ``child'' means one of
$w_1,\dots,w_i$. We show, by induction on the height of $v$ and then on $i$, that
$P'=\bigl(\cdots(\mathrm{Init}_y(\nu_v)\circ\mathrm{msg}_{w_1})\circ\cdots\bigr)\circ\mathrm{msg}_{w_i}$; for $i=d$
this identifies $\mathrm{msg}_v$ as the triple of $F_v$, whose entries are nonnegative. Write $\Pi^{\mathrm{in}}_{F_v}$ for
$\Pi^{\mathrm{in}}_{F'}$ when $F'=F_v$.

For $i=0$, $F'$ consists of $v$ alone. The set $X=\{v\}$ has $\Pi^{\mathrm{in}}_{F'}(X)=1$ (one component, no owner
term) and $X=\emptyset$ has $\Pi^{\mathrm{in}}_{F'}(X)=\nu_v$, and state~$2$ is empty because $v$ has no children yet;
so the triple is $(y/2,0,2^{-\nu_v})=\mathrm{Init}_y(\nu_v)$. For the step from $i-1$ to $i$, let $F''$ be the subtree for $i-1$, with triple $P''$ and potential
$\Pi^{\mathrm{in}}_{F''}$, let $w=w_i$, and let $Q=\mathrm{msg}_w$ be the triple of $F_w$, by the induction hypothesis. Every
$X\subseteq V(F')$ splits uniquely as $X=X''\cup X_w$ with $X''\subseteq V(F'')$ and
$X_w\subseteq V(F_w)$, and
\[
\begin{split}
  \Pi^{\mathrm{in}}_{F'}(X)&=\Pi^{\mathrm{in}}_{F''}(X'')+\Pi^{\mathrm{in}}_{F_w}(X_w)-[v\in X\text{ and }w\in X]\\
  &\quad+[v\notin X,\ w\notin X\text{ and no child of $w$ is in }X].
\end{split}
\]
Indeed, the owner terms are additive over the split. A vertex $u\notin\{v,w\}$ has its closed
neighborhood inside $F''$ or inside $F_w$, so it is counted in $\Pi^{\mathrm{in}}_{F'}$ exactly when it is counted in
$\Pi^{\mathrm{in}}_{F''}$ or $\Pi^{\mathrm{in}}_{F_w}$. The
vertex $w$ is not counted in $\Pi^{\mathrm{in}}_{F_w}$, and it is counted in $\Pi^{\mathrm{in}}_{F'}$ exactly when
$N_F[w]\cap X=\emptyset$, that is, when $v$, $w$ and the children of $w$ are not in $X$; the vertex $v$
is counted in neither. Finally, $F'[X]$ is the disjoint union of $F''[X'']$ and $F_w[X_w]$, together
with the edge $vw$ if $v,w\in X$, which joins two components into one.

Now we sum the weights over the states of $v$ and $w$. If $v\in X$, then $w$ contributes the factor
$2Q_1$ if $w\in X$ (one component fewer) and $Q_2+Q_3$ otherwise ($w$ is dominated by $v$); so
$P'_1=P''_1(2Q_1+Q_2+Q_3)$. If
$v$ is in state $2$ for $F''$, then $v$ stays in state $2$, and $w$ contributes $Q_1$, $Q_2$ or
$Q_3/2$ according to its state (in state $3$, $w$ is not dominated, which adds $1$ to the potential
and so multiplies the weight by $\frac12$).
If $v$ is in state $3$ for $F''$, then $w\in X$ moves $v$ to state $2$ (term $P''_3Q_1$), and
otherwise $v$ stays in state $3$, with the factor $Q_2+Q_3/2$. So $P'=P''\circ Q$.

At the root, $N_F[r]$ consists of $r$ and its children, so $\Pi_{F,\nu}(X)=\Pi^{\mathrm{in}}_{F_r}(X)+[r\text{ is in
state }3]$, and $W_{F,\nu}(y)=\mathrm{msg}_{r,1}+\mathrm{msg}_{r,2}+\mathrm{msg}_{r,3}/2=\mathrm{fin}(\mathrm{msg}_r)$.
\end{proof}

\subsubsection{Antinorm certificates in the cumulative cone}\label{sec:app-rst-antinorm}

We collect the facts about $\mathcal K_{\mathrm{cum}}$ that the certificate argument uses, and then the
abstract mechanism, which \cref{sec:rsc} reuses with another cone.

\begin{lemma}[Closure facts]\label{lem:rst-closure}
\begin{enumerate}[label=(\alph*)]
\item If $Q\in\R^3_{\ge0}$ and $\Delta\in\mathcal K_{\mathrm{cum}}$, then $\Delta\circ Q\in\mathcal K_{\mathrm{cum}}$ and $Q\circ \Delta\in\mathcal K_{\mathrm{cum}}$.
\item $\mathrm{fin}(\Delta)\ge0$ for every $\Delta\in\mathcal K_{\mathrm{cum}}$.
\end{enumerate}
\end{lemma}
\begin{proof}
(a) The operation $\circ$ is bilinear and $\mathcal K_{\mathrm{cum}}$ is a convex cone, so it suffices to check the three
generators of $\mathcal K_{\mathrm{cum}}$. For $Q\ge0$ we list each product and its cumulative coordinates:
$(1,-1,0)\circ Q=\bigl(2Q_1+Q_2+Q_3,\,-Q_1-Q_2-\tfrac{Q_3}2,\,0\bigr)$, with cumulative coordinates
$\bigl(2Q_1+Q_2+Q_3,\,Q_1+\tfrac{Q_3}2,\,Q_1+\tfrac{Q_3}2\bigr)$;
$(0,1,-1)\circ Q=\bigl(0,\,Q_2+\tfrac{Q_3}2,\,-Q_2-\tfrac{Q_3}2\bigr)$, with
$\bigl(0,\,Q_2+\tfrac{Q_3}2,\,0\bigr)$; and $(0,0,1)\circ Q=\bigl(0,\,Q_1,\,Q_2+\tfrac{Q_3}2\bigr)\ge0$.
In the other order,
$Q\circ(1,-1,0)=(Q_1,\,Q_3,\,-Q_3)$, with cumulative coordinates $(Q_1,\,Q_1+Q_3,\,Q_1)$;
$Q\circ(0,1,-1)=\bigl(0,\,\tfrac{Q_2}2,\,\tfrac{Q_3}2\bigr)\ge0$; and
$Q\circ(0,0,1)=\bigl(Q_1,\,\tfrac{Q_2}2,\,\tfrac{Q_3}2\bigr)\ge0$. All cumulative coordinates are
nonnegative.
(b) $\mathrm{fin}(\Delta)=\tfrac12(\Delta_1+\Delta_2)+\tfrac12(\Delta_1+\Delta_2+\Delta_3)$, and both brackets are nonnegative for
$\Delta\in\mathcal K_{\mathrm{cum}}$.
\end{proof}

The nonnegativity of the other factor in (a) is needed: $(1,-1,0)\circ(0,1,-1)=(0,-\tfrac12,0)\notin\mathcal K_{\mathrm{cum}}$.

The next lemma isolates the mechanism: a convex region that the merge operation maps into itself
up to a known scale, for any bilinear operation and any cone with the closure property of
\cref{lem:rst-closure}(a).

\begin{lemma}[Antinorm certificates for a bilinear merge]\label{lem:rst-abstract}
Let $\circ$ be a bilinear operation on $\R^m$, let $\mathcal K\subseteq\R^m$ be a convex cone
containing $\R^m_{\ge0}$ such that $\Delta\circ Q\in\mathcal K$ and $Q\circ\Delta\in\mathcal K$ for every
$\Delta\in\mathcal K$ and every $Q\in\R^m_{\ge0}$, and for a finite nonempty set $Z\subseteq\R^m_{\ge0}$
put $\mathcal P_Z=\mathrm{conv}(Z)+\mathcal K$. Then:
\begin{enumerate}[label=(\alph*)]
\item $\mathcal P_Z+\mathcal K=\mathcal P_Z$, and $\eta\mathcal P_Z\subseteq\eta'\mathcal P_Z$ whenever
$0<\eta'\le\eta$;
\item let $Y,Y',Z\subseteq\R^m_{\ge0}$ be finite and nonempty with $y\circ z\in\mathcal P_{Y'}$ for all
$y\in Y$ and $z\in Z$; if $P\in\eta\mathcal P_Y$ and $Q\in\eta'\mathcal P_Z$ with $\eta,\eta'>0$ and
$Q\ge0$, then $P\circ Q\in\eta\eta'\mathcal P_{Y'}$.
\end{enumerate}
\end{lemma}
\begin{proof}
(a) The first claim holds because $\mathcal K+\mathcal K=\mathcal K$. For the second, let
$P=\eta(z+\Delta)$ with $z\in\mathrm{conv}(Z)$ and $\Delta\in\mathcal K$. Then $P=\eta'(z+\Delta'')$ with
$\Delta''=\frac{\eta-\eta'}{\eta'}z+\frac\eta{\eta'}\Delta$, and $\Delta''\in\mathcal K$ because $z\ge0$
and $\mathcal K\supseteq\R^m_{\ge0}$ is a convex cone.

(b) Write $P=\eta(y+\Delta)$ and $Q=\eta'\hat Q$ with $\hat Q=z+\Delta'\ge0$, where
$y\in\mathrm{conv}(Y)$, $z\in\mathrm{conv}(Z)$ and $\Delta,\Delta'\in\mathcal K$. By bilinearity,
\[
  \frac{P\circ Q}{\eta\eta'}=y\circ z+y\circ \Delta'+\Delta\circ\hat Q .
\]
With convex combinations $y=\sum_i\alpha_iy_i$ and $z=\sum_j\beta_jz_j$ of points of $Y$ and of $Z$
we get $y\circ z=\sum_{i,j}\alpha_i\beta_j\,y_i\circ z_j\in\mathcal P_{Y'}$, as $\mathcal P_{Y'}$ is
convex. Since $y\ge0$ and $\hat Q\ge0$, the closure property gives $y\circ \Delta'\in\mathcal K$ and
$\Delta\circ\hat Q\in\mathcal K$, and (a) gives $P\circ Q\in\eta\eta'\mathcal P_{Y'}$.
\end{proof}

\RestLemRstF*

\begin{proof}
Root $F$ anywhere and let $\mathrm{msg}_v$ be the messages of \cref{lem:rst-treerec}, which are
nonnegative. By induction from the leaves,
$\mathrm{msg}_v\in e_1^{|V(F_v)|}e_2^{\,\nu(F_v)}\mathcal P_Z$, where $\nu(F_v)=\sum_{u\in V(F_v)}\nu_u$: the
start $\mathrm{Init}_y(\nu_v)$ lies in $e_1e_2^{\,\nu_v}\mathcal P_Z$ by~(b), and each merge with a child
$w$ multiplies the scale by $e_1^{|V(F_w)|}e_2^{\,\nu(F_w)}$, by \cref{lem:rst-abstract}(b) with
$Y=Y'=Z$, applied with $Q=\mathrm{msg}_w\ge0$, whose closure hypothesis is \cref{lem:rst-closure}(a) and
whose hypothesis on $Z$ is~(a). At the
root, $\mathrm{msg}_r=\eta(z+\Delta)$ with $\eta=e_1^se_2^{\nu(F)}$, $z\in\mathrm{conv}(Z)$ and $\Delta\in\mathcal K_{\mathrm{cum}}$. Since
$\mathrm{fin}$ is linear, \cref{lem:rst-treerec} and \cref{lem:rst-closure}(b) give
\[
  W_{F,\nu}(y)=\mathrm{fin}(\mathrm{msg}_r)=\eta\bigl(\mathrm{fin}(z)+\mathrm{fin}(\Delta)\bigr)\ge\eta\,\mathrm{fin}(z)\ge\eta\,c_Z .
\]
For the last claim let $j>j_0$. Then $\mathrm{Init}_y(j)-(y/2,0,0)=(0,0,2^{-j})\in\mathcal K_{\mathrm{cum}}$, so by
\cref{lem:rst-abstract}(a)
$\mathrm{Init}_y(j)\in e_1e_2^{\,j_0+1}\mathcal P_Z+\mathcal K_{\mathrm{cum}}=e_1e_2^{\,j_0+1}\mathcal P_Z\subseteq
e_1e_2^{\,j}\mathcal P_Z$, because $e_1e_2^{\,j}\le e_1e_2^{\,j_0+1}$.
\end{proof}

\RestCorRstF*

\begin{proof}
Since $G[S]$ is connected, it has a spanning tree $F$; since $S$ is dominating, an owner map exists,
and its counts satisfy $\sum_v\nu_v=n-s$. \Cref{lem:rst-E,lem:rst-F} give the first inequality. The second
is the first inequality of \cref{lem:fw-gf}.
\end{proof}

\begin{remark}[Why the cumulative cone]\label{rem:rst-cone}
By \cref{lem:rst-closure}(a) and~(b), moving weight from state $2$ to state $1$, or from state $3$
to state $2$, never decreases the final value $\mathrm{fin}(\mathrm{msg}_r)$. The nonnegative orthant is also
preserved by $\circ$, so \cref{lem:rst-F} holds verbatim with $\R^3_{\ge0}$ in place of
$\mathcal K_{\mathrm{cum}}$. The cumulative cone is larger: in its order a deficit in state $3$ can
be compensated by weight in states $1$ or $2$ (an attempt with the orthant is reported in
\cref{sec:anc-rstheta}).
\end{remark}
{}
\subsubsection{A lower bound on uniform lobsters}\label{sec:app-lobster}

We prove here the following proposition of \cref{sec:rst-caterpillar}, with the notation of that section.

\RestPropRstCaterpillar*

\begin{proof}
(a) Let $S$ be a connected dominating set of $\Lambda_L$, where $L\ge1$, so that $n\ge7$. If a
leg $u$ is not in $S$, then its two claws, whose only neighbor is $u$, must lie in $S$ to be dominated; they are not adjacent and every
path between them passes through $u$, contradicting the connectivity of $G[S]$. So all legs lie in
$S$. The neighbors of a leg of $x_i$ are $x_i$ and its two claws, which have no other
neighbors, and $\Lambda_L$ has at least two legs; so a path in $G[S]$ from a leg of $x_i$ to
another leg passes through $x_i$, and $x_i\in S$. Hence $S\supseteq S^*$. Conversely, $S^*$ induces
a tree (a caterpillar: $\Lambda_L$ without its claws) and dominates every claw through its leg,
so it is a connected dominating set.

(b) By (a), the only connected dominating set of size at most $s=\gc(\Lambda_L)$ is $S^*$. A
trial accepts only if $\cA_{\mathrm{St}}$ is run on a sample $X$ for which some $Y$ with $|Y|\le s-t$
makes $X\cup Y$ a connected dominating set, so only if $X\cup Y=S^*$ and in particular $X\subseteq
S^*$. Hence the probability that $s$ is accepted is at most the expected number of trials that run
$\cA_{\mathrm{St}}$ on a subset of $S^*$, which \cref{prop:fw-necessary} with $\Gamma=2$ bounds by the displayed
quantity.

(c) $G[S^*]$ is a tree and every vertex outside $S^*$ is a claw, so \cref{lem:rst-E} holds with
equality, with $\nu=2$ at the legs and $\nu=0$ on the spine, and $W_{S^*}=W_{F,\nu}$ for
$F=\Lambda_L[S^*]$. Root $F$ at $x_L$, with $x_{i-1}$ a child of $x_i$. By
\cref{lem:rst-treerec}, the message of $x_1$ is
\[
  \mathbf u_y=\mathrm{Init}_y(0)\circ\mathrm{Init}_y(2)\circ\mathrm{Init}_y(2)
  =\Bigl(\tfrac y2\bigl(y+\tfrac14\bigr)^2,\ \tfrac{y^2}4+\tfrac y8,\ \tfrac1{64}\Bigr),
\]
and the message of $x_i$ is $\mathbf u_y\circ\mathrm{msg}_{x_{i-1}}$ for $i\ge2$, which is linear in $\mathrm{msg}_{x_{i-1}}$
with matrix
\[
  \mathbf M_y=\begin{pmatrix}2u_1&u_1&u_1\\u_2+u_3&u_2&u_2/2\\0&u_3&u_3/2\end{pmatrix},\qquad (u_1,u_2,u_3)=\mathbf u_y .
\]
So $W_{S^*}(y)=\mathrm{fin}(\mathbf M_y^{L-1}\mathbf u_y)$. The entries of $\mathbf u_y$ and $\mathbf M_y$
are polynomials in $y$ with nonnegative coefficients, so they are nonnegative and nondecreasing in $y\ge0$.
Write $f=(1,1,\frac12)$, so that $\mathrm{fin}(P)=f^{\mathsf T}P$.

For the first bound, $\mu_t(S^*)=0$ for $t>3L$. For $0\le t\le3L$ put $y_t=t/(n-t)\in[0,\frac34]$. As in
the proof of \cref{lem:fw-gf}, $\mu_t(S^*)\,y_t^{\,t}\le W_{S^*}(y_t)$ and
$\binom nt y_t^{\,t}\ge(1+y_t)^n/(n+1)$, so (also for $t=0$, where $y_t=0$)
\[
  \mu_t(S^*)\Big/\binom nt\le(n+1)\,W_{S^*}(y_t)/(1+y_t)^n .
\]
The certificate of \cref{sec:anc-lobster} covers $[0,\frac34]$ by finitely many intervals
$[a,b]$ and gives for each a rational vector $w>0$ and a rational $\lambda>0$ with
$w^{\mathsf T}\mathbf M_b\le\lambda w^{\mathsf T}$ componentwise, $c\,w^{\mathsf T}\mathbf u_b\le\lambda$ for
$c=\max_i f_i/w_i$, and $1.71927^7\lambda\le(1+a)^7$. For $y\in[a,b]$ this gives
\[
\begin{split}
  W_{S^*}(y)&\le f^{\mathsf T}\mathbf M_b^{L-1}\mathbf u_b\le c\,w^{\mathsf T}\mathbf M_b^{L-1}\mathbf u_b
  \le c\,\lambda^{L-1}w^{\mathsf T}\mathbf u_b\le\lambda^L\\
  &\le(1+a)^{7L}\,1.71927^{-n}\le(1+y)^n\,1.71927^{-n},
\end{split}
\]
and with $y=y_t$ the first bound of (c) follows. For the converse, at $y^*=0.3453$ the same certificate gives a
rational vector $v>0$ and a rational $\lambda'>0$ with $\mathbf M_{y^*}v\ge\lambda'v$ componentwise and
$(1+y^*)^7\le1.71928^7\lambda'$. With $c'=\min_iu_i/v_i>0$ for $\mathbf u_{y^*}=(u_1,u_2,u_3)$, we get
$W_{S^*}(y^*)\ge c'f^{\mathsf T}\mathbf M_{y^*}^{L-1}v\ge c'\lambda'^{L-1}f^{\mathsf T}v$, and by \cref{lem:fw-gf}
$\min_t\binom nt/\mu_t(S^*)\le(1+y^*)^n/W_{S^*}(y^*)\le C\cdot1.71928^n$ with $C=\lambda'/(c'f^{\mathsf T}v)$.
The numerical value in~(c) and values for finite $L$ are reported in \cref{sec:anc-lobster}; they
are not used in the proof.

For the consequence, suppose $2^\psi<1.71927$. By (c), $\max_t\mu_t(S^*)/\binom nt\le
(n+1)\,1.71927^{-n}$ for all $L$, so by (b) the probability that $\gc$ is accepted is at most
$n^{O(1)}(2^\psi/1.71927)^n$, which tends to $0$. As RS-$\theta$ never accepts a value below $\gc$, it
then outputs a value above $\gc$ with probability tending to $1$.
\end{proof}

The bound does not rely on $\Lambda_L$ being a tree, on which \pname{Max Leaf Spanning Tree} is easy:
\cref{prop:rst-caterpillar} holds for every graph $G$, with $L\ge2$, whose unique
maximum-leaf spanning tree is $\Lambda_L$: part~(a) by the argument below, which replaces its proof, and
parts~(b) and~(c) with their proofs verbatim. Indeed, a minimum connected dominating set $D$ of $G$ yields a
spanning tree with at least $n-|D|$ leaves (\cref{lem:cds-maxleaf}), which must be $\Lambda_L$, so
$D\supseteq S^*$ and $D=S^*$. Next, every vertex of $S^*$ has degree at least $3$ in $\Lambda_L$ (a leg has
its spine vertex and two claws; a spine vertex has two legs and, as $L\ge2$, a spine neighbor), and the
leaves of $\Lambda_L$ are exactly the claws. Let $e$ be an edge of $G$ that is not in $\Lambda_L$. If both ends
of $e$ lie in $S^*$, the cycle of $\Lambda_L+e$ runs through the subtree $\Lambda_L[S^*]$; delete an edge
$f\ne e$ of it. Each end of $f$ lies in $S^*$ and loses at most one unit of degree, so keeps degree at
least $2$, and no other degree decreases, so the new spanning tree has exactly the claws as leaves. If
$e=cv$ joins a claw $c$ with leg $\ell$ to a vertex $v\in S^*\setminus\{\ell\}$, delete the edge $c\ell$, which
lies on the cycle of $\Lambda_L+e$: then $c$ is still a leaf, $\ell$ keeps degree $2$, and again the leaves
are exactly the claws. In both cases we obtain a spanning tree other than $\Lambda_L$ with as many leaves,
contrary to uniqueness. The remaining edges of $G$ outside $\Lambda_L$ join two claws; they change neither
$G[S^*]$ nor the neighbors in $S^*$ of any claw. So $G[S^*]=\Lambda_L[S^*]$ and each claw has exactly one
neighbor in $S^*$, which is what parts~(b) and~(c) use.

The extremal structure is neither the star of stars of \cref{prop:cds-tight}, which is tight for
\cref{lem:cds-C} but has a smaller base than $\Lambda_L$ under exponential moments, nor a path. With
at most $8$ pendant vertices per solution vertex, the base given by the generating function, computed
as in \cref{prop:rst-caterpillar}(c), is at most $1.5878$ for stars of stars and at most $1.6972$ for
paths (numerical; further families in \cref{sec:anc-lobster}).

{}
\subsection{The closure bound of Algorithm RSC}\label{sec:app-rsc}

This subsection proves \cref{prop:rsc-bound}, the lower bound on the closure generating function
$W^{\mathrm{cl}}_S$ of a minimum connected dominating set $S$, in the three steps described in
\cref{sec:rsc}: the tree model (\cref{sec:rsc-tree}), the ten-state recursion and the light owners
(\cref{sec:rsc-rec,sec:rsc-light}), and the antinorm certificate (\cref{sec:rsc-cert}). The notation
is that of \cref{sec:rsc}.

\RestPropRscBound*

\subsubsection{From the graph to a tree model}\label{sec:rsc-tree}

From now on $G$ is a connected graph, $S$ a connected dominating set of $G$ with $|S|=s$, $F$ a
spanning tree of $G[S]$, and $O=V\setminus S$ the set of \emph{outside} vertices. For $u\in O$ let
$N_S(u)=N(u)\cap S$, which is nonempty because $S$ is dominating. An outside vertex $u$ is
\emph{private} if $|N_S(u)|=1$ and \emph{shared} otherwise; a shared vertex has \emph{type}
$\tau(u)=2$ if $|N_S(u)|=2$ and $\tau(u)=3$ if $|N_S(u)|\ge3$. For $X'\subseteq S$ and $v\in S$ let
$k_F(v)=|N_F(v)\cap X'|$ be the number of sampled $F$-neighbors of $v$ (the dependence on $X'$ is
suppressed), and let
\[
  C_F(X')=\bigl\{v\in S\setminus X':\ k_F(v)\ge2\bigr\}
\]
be the candidates that the tree $F$ certifies. For $\Upsilon\subseteq C_F(X')$ let
\[
  \pi_{X'}(\Upsilon)=\prod_{v\in\Upsilon}p_{k_F(v)}\prod_{v\in C_F(X')\setminus\Upsilon}q_{k_F(v)},
\]
the probability that a random subset of $C_F(X')$ which contains each $v$ independently with
probability $p_{k_F(v)}$ equals $\Upsilon$. For $X\subseteq S$ put
\[
  \Psi_F(X)=\bigl|\{v\in S:\ N_F[v]\cap X=\emptyset\}\bigr|+\comp(F[X]),
\]
and for $u\in O$ and $X'\subseteq X\subseteq S$ put, with $k(u)=|N(u)\cap X'|$,
\[
  h_u(X',X)=2^{-\mathbf 1[N(u)\cap X=\emptyset]}\,q_{k(u)}^{\mathbf 1[k(u)\ge2]} .
\]
The factor $h_u$ combines the factor $\frac12$ that an outside vertex not dominated by the closed
sample $X$ contributes to $2^{-|\Tm(J_X)|}$ with the probability $q_{k(u)}$ that a candidate outside
the solution with $k(u)$ sampled neighbors is not put into $\Upsilon$. Note that $k(u)\le|N_S(u)|$, so
$k(u)\le2$ for a vertex of type $2$.

We first record that more sampled vertices never mean more terminals. This is why adding candidates
inside the solution never decreases $2^{-|\Tm(J_X)|}$.

\begin{lemma}[Monotonicity of the number of terminals]\label{lem:rsc-mono}
  For $X\subseteq Y\subseteq V$ we have $|\Tm(J_Y)|\le|\Tm(J_X)|$.
\end{lemma}

\begin{proof}
  It suffices to add one vertex $v\notin X$, and we use~\eqref{eq:cds-TX}. If $v\notin N[X]$, then
  $|N[X\cup\{v\}]|\ge|N[X]|+1$, because $v$ is new in the closed neighborhood, and
  $\comp(G[X\cup\{v\}])\le\comp(G[X])+1$. If $v\in N(X)$, then $v$ is adjacent to at least one
  component of $G[X]$, so $\comp(G[X\cup\{v\}])\le\comp(G[X])$, and $N[X\cup\{v\}]\supseteq N[X]$. In
  both cases $n-|N[\cdot]|+\comp(G[\cdot])$ does not increase.
\end{proof}

We shall use the following consequence of monotonicity several times. If $f$ is a nonnegative
function on the subsets of a finite set $C$ that is nondecreasing under inclusion, and $\mu_{\mathbf p}$
is the product distribution on the subsets of $C$ that contains each $v\in C$ independently with
probability $\mathbf p_v$, then $\sum_\Upsilon\mu_{\mathbf p}(\Upsilon)f(\Upsilon)$ is nondecreasing in each
$\mathbf p_v$: writing the sum as $\mathbf p_v\,\E[f(\Upsilon'\cup\{v\})]+(1-\mathbf p_v)\E[f(\Upsilon')]$
with $\Upsilon'$ the random subset of $C\setminus\{v\}$, its derivative in $\mathbf p_v$ is
$\E[f(\Upsilon'\cup\{v\})-f(\Upsilon')]\ge0$. We call this the \emph{monotone coupling}.

The next lemma replaces the graph by the tree $F$. It keeps the candidates that $F$ certifies, with
the (smaller) closure probabilities that $F$ certifies, drops the other candidates inside $S$, and
turns every candidate outside $S$ into a factor $q_{k(u)}$.

\begin{lemma}[Closure tree model]\label{lem:rsc-model}
  For $0\le p_2\le p_3\le1$ and $y\ge0$,
  \[
    W^{\mathrm{cl}}_S(y)\ \ge\ \sum_{X'\subseteq S}\ \sum_{\Upsilon\subseteq C_F(X')}
    y^{|X'|}\,\pi_{X'}(\Upsilon)\,2^{-\Psi_F(X'\cup\Upsilon)}\prod_{u\in O}h_u(X',X'\cup\Upsilon).
  \]
\end{lemma}

\begin{proof}
  Fix $X'\subseteq S$ and let $\Upsilon$ be the random closure of \cref{def:rsc-gf}. Split the candidate set
  $C(X')$ into three disjoint parts: $C^{\mathrm{out}}=C(X')\cap O$, the set $C_F(X')$, and
  $C^{\mathrm{rest}}=(C(X')\cap S)\setminus C_F(X')$. These parts cover $C(X')$ because $C_F(X')\subseteq C(X')\cap S$: a
  vertex $v\in S\setminus X'$ with two $F$-neighbors in $X'$ has two $G$-neighbors in $X'$, as
  $F\subseteq G$. The sets
  $\Upsilon\cap C^{\mathrm{out}}$, $\Upsilon'=\Upsilon\cap C_F(X')$ and $\Upsilon''=\Upsilon\cap C^{\mathrm{rest}}$ are independent, and the
  event $\Upsilon\subseteq S$ is the event $\Upsilon\cap C^{\mathrm{out}}=\emptyset$, which has probability
  $\prod_{u\in C^{\mathrm{out}}}q_{|N(u)\cap X'|}$. On this event $X'\cup\Upsilon\supseteq X'\cup\Upsilon'$, so
  $2^{-|\Tm(J_{X'\cup\Upsilon})|}\ge2^{-|\Tm(J_{X'\cup\Upsilon'})|}$ by \cref{lem:rsc-mono}. The set $\Upsilon'$ contains
  each $v\in C_F(X')$ independently with probability $p_{|N(v)\cap X'|}\ge p_{k_F(v)}$, because
  $|N(v)\cap X'|\ge k_F(v)$ and $p_2\le p_3$, and $\Upsilon'\mapsto2^{-|\Tm(J_{X'\cup\Upsilon'})|}$ is nondecreasing by
  \cref{lem:rsc-mono}; so the monotone coupling gives
  \[
    \E_\Upsilon\Bigl[\mathbf 1[\Upsilon\subseteq S]\,2^{-|\Tm(J_{X'\cup\Upsilon})|}\Bigr]
    \ \ge\ \prod_{u\in C^{\mathrm{out}}}q_{|N(u)\cap X'|}\ \sum_{\Upsilon'\subseteq C_F(X')}\pi_{X'}(\Upsilon')\,2^{-|\Tm(J_{X'\cup\Upsilon'})|} .
  \]
  A vertex $u\in O$ is not in $X'\subseteq S$, so $u\in C^{\mathrm{out}}$ if and only if $k(u)\ge2$, and
  the product over $C^{\mathrm{out}}$ equals $\prod_{u\in O}q_{k(u)}^{\mathbf 1[k(u)\ge2]}$.

  Now fix $X\subseteq S$ (we shall take $X=X'\cup\Upsilon'$). By~\eqref{eq:cds-TX},
  $|\Tm(J_X)|=|S\setminus N[X]|+|O\setminus N[X]|+\comp(G[X])$. First, if $v\in S$ is not in $N[X]$, then
  $N_G[v]\cap X=\emptyset$ and hence $N_F[v]\cap X=\emptyset$, since $N_F[v]\subseteq N_G[v]$. Second, a
  vertex $u\in O$ is not in $X$, so $u\notin N[X]$ if and only if $N(u)\cap X=\emptyset$. Third,
  $F[X]$ is a spanning subgraph of $G[X]$, because $X\subseteq S=V(F)$, so $\comp(G[X])\le\comp(F[X])$;
  for $X=\emptyset$ both sides are $0$. Together,
  \[
    |\Tm(J_X)|\ \le\ \Psi_F(X)+\sum_{u\in O}\mathbf 1[N(u)\cap X=\emptyset] .
  \]
  Inserting this with $X=X'\cup\Upsilon'$ into the previous display, multiplying by $y^{|X'|}$ and summing
  over $X'\subseteq S$ gives the claim.
\end{proof}

The factors $h_u$ couple the outside vertices with all of their neighbors in $S$. To obtain a
recursion along $F$ we assign every outside vertex to one of its neighbors in $S$, its owner.

\begin{definition}[Owners]\label{def:rsc-owner}
  An \emph{owner map} is a map $o:O\to S$ with $o(u)\in N_S(u)$ for all $u\in O$. For $v\in S$ let
  $\nu_v=|o^{-1}(v)|$ be the number of vertices that $v$ owns, so $\sum_{v\in S}\nu_v=n-s$. For $u\in O$
  and $X\subseteq S$ let $\bar h_u(X)=2^{-\mathbf 1[o(u)\notin X]}$.
\end{definition}

The following pointwise inequality compares $h_u$ with the owner factor $\bar h_u$; it is the only
place where the values of $h_u$ are examined. Put
\[
  \varepsilon_2=-\ln q_2,\qquad \varepsilon_3=-\ln q_3,\qquad \delta_2=\ln2,\qquad \delta_3=\ln(2q_3),
\]
so $0\le\varepsilon_2\le\varepsilon_3$ and $0\le\delta_3\le\delta_2$ (recall $q_2\ge q_3>\frac12$).
For $u\in O$ let $N'_u=N_S(u)\setminus\{o(u)\}$, and for $X'\subseteq S$ let $\mathbf 1_u=\mathbf 1[N'_u\cap X'\ne\emptyset]$
indicate that some other neighbor of $u$ in $S$ is sampled.

\begin{lemma}[Pointwise bound]\label{lem:rsc-pointwise}
  Let $0<p_2\le p_3<\frac12$, let $o$ be an owner map, let $u\in O$ have the owner $o=o(u)$, and let
  $X'\subseteq X\subseteq S$. Then $h_u(X',X)\ge\bar h_u(X)$ whenever $o\notin X$, and:
  \begin{enumerate}[label=(\roman*)]
    \item if $u$ is private, then $h_u(X',X)\ge\bar h_u(X)$;
    \item if $u$ is shared of type $\tau$, then $h_u(X',X)\ge\bar h_u(X)\,q_\tau^{\mathbf 1[o\in X]}$;
    \item if $u$ is shared of type $\tau$, $o\notin X$ and $N'_u\cap X\ne\emptyset$, then
      $h_u(X',X)\ge e^{\delta_\tau}\,\bar h_u(X)$;
    \item if $u$ is shared of type $3$, then
      $\ln\bigl(h_u(X',X)/\bar h_u(X)\bigr)\ge\mathbf 1_u\bigl(\delta_3\,\mathbf 1[o\notin X]-\varepsilon_3\,\mathbf 1[o\in X]\bigr)$;
    \item if $u$ is shared of type $2$, then
      $\ln\bigl(h_u(X',X)/\bar h_u(X)\bigr)\ge\mathbf 1_u\bigl(\delta_2\,\mathbf 1[o\notin X]-\varepsilon_2\,\mathbf 1[o\in X']\bigr)$.
  \end{enumerate}
\end{lemma}

\begin{proof}
  Suppose first that $o\notin X$, so $\bar h_u(X)=\frac12$ and $o\notin X'$. If $N'_u\cap X\ne\emptyset$,
  then $u$ is dominated and $h_u(X',X)=q_{k(u)}^{\mathbf 1[k(u)\ge2]}\ge q_3$; so $h_u/\bar h_u\ge2q_3=e^{\delta_3}$,
  and if moreover $\tau(u)=2$, then $k(u)\le|N'_u|=1$ gives $h_u=1$ and $h_u/\bar h_u=2=e^{\delta_2}$.
  This proves (iii), and it covers the case $\mathbf 1_u=1$ of (iv) and (v), since $X'\subseteq X$. If
  $N'_u\cap X=\emptyset$, then $N(u)\cap X'=\emptyset$, so
  $h_u(X',X)=2^{-\mathbf 1[N(u)\cap X=\emptyset]}\ge\frac12=\bar h_u(X)$. In all cases $h_u\ge\bar h_u$ when $o\notin X$.

  Suppose now that $o\in X$, so $\bar h_u(X)=1$ and $u$ is dominated; then
  $h_u(X',X)=q_{k(u)}^{\mathbf 1[k(u)\ge2]}$. If $\mathbf 1_u=0$, then $N(u)\cap X'\subseteq\{o\}$, so
  $h_u=1$. If $\mathbf 1_u=1$, then $h_u\ge q_3$, which gives (iv); if moreover $\tau(u)=2$, then
  $k(u)\le2$ and $k(u)=2$ only if $o\in X'$, so $h_u\ge q_2^{\mathbf 1[o\in X']}$, which gives (v); and
  in both cases $h_u\ge q_{\tau(u)}$, which gives (ii), as $q_\tau\le1$. For (i), a private $u$ has
  $N'_u=\emptyset$, so $k(u)\le1$ and $h_u=1$ when $o\in X$.
\end{proof}

Replacing $h_u$ by $\bar h_u q_\tau^{\mathbf 1[o(u)\in X]}$ for every shared vertex would give a model of product
form along the tree. We call the factor $q_\tau^{\mathbf 1[o(u)\in X]}$ an \emph{exposure} of the owner of
$u$, of type $\tau=\tau(u)$. Exposing every shared vertex loses too much (for $p_2=p_3=1/20$, about $2^{0.012n}$ in the
running time on the uniform $(2,2)$-lobster $\Lambda_L$ of \cref{prop:rst-caterpillar}, numerically), because the exposure applies the factor $q_\tau$ to a shared vertex whenever its
owner is in $X$, although $h_u<1$ requires at least two sampled neighbors. For some shared vertices
we therefore keep the exact factor $h_u$ and bound its effect by a correction term, using Jensen's
inequality. For the others, (iii) recovers part of the loss: when the owner is outside the closed
sample, an exposed vertex is usually dominated by one of its other neighbors, and then $h_u$ is
$e^{\delta_\tau}$ times the owner factor $\frac12$. The tree model records this as a \emph{credit}
$\varpi_\tau\ge1$ on the factor of an owner outside the closed sample.

\begin{definition}[Tree model]\label{def:rsc-tm}
  Let $o$ be an owner map, let the shared vertices be partitioned into a set $O^{\mathrm{co}}$ of
  \emph{corrected} vertices and a set $O^{\mathrm{ex}}$ of \emph{exposed} vertices, and let
  $\varpi_2,\varpi_3\ge1$ be real numbers, the \emph{credits}. For $v\in S$ let $\nu_v$ be as in
  \cref{def:rsc-owner}, let $c_{v,\tau}=|\{u\in o^{-1}(v)\cap O^{\mathrm{ex}}:\tau(u)=\tau\}|$ for $\tau\in\{2,3\}$
  be the numbers of exposed vertices of each type that $v$ owns, and $c_v=c_{v,2}+c_{v,3}\le\nu_v$. For $X\subseteq S$ let
  \[
    \hat h_v(X)=
    \begin{cases}q_2^{c_{v,2}}q_3^{c_{v,3}}&\text{if }v\in X,\\[2pt]
      2^{-\nu_v}\varpi_2^{c_{v,2}}\varpi_3^{c_{v,3}}&\text{if }v\notin X,
    \end{cases}
  \]
  and for $X'\subseteq S$ and $\Upsilon\subseteq C_F(X')$ let
  \[
    \mathrm{wt}(X',\Upsilon)=y^{|X'|}\,\pi_{X'}(\Upsilon)\,2^{-\Psi_F(X'\cup\Upsilon)}\prod_{v\in S}\hat h_v(X'\cup\Upsilon),\qquad
    W^{\mathrm{tm}}=\sum_{X'\subseteq S}\ \sum_{\Upsilon\subseteq C_F(X')}\mathrm{wt}(X',\Upsilon).
  \]
  Let $\Pr_{\mathrm{tm}}$ be the probability distribution on the pairs $(X',\Upsilon)$ with $\Pr_{\mathrm{tm}}(X',\Upsilon)=\mathrm{wt}(X',\Upsilon)/W^{\mathrm{tm}}$, and
  write $X=X'\cup\Upsilon$ for the closed sample under $\Pr_{\mathrm{tm}}$; we write $\E_{\mathrm{tm}}$ for the expectation under $\Pr_{\mathrm{tm}}$.
\end{definition}

The value $W^{\mathrm{tm}}$ depends only on $F$, $y$, $p_2$, $p_3$, $\varpi_2$, $\varpi_3$ and the numbers $\nu_v$, $c_{v,2}$ and $c_{v,3}$; we write
$W^{\mathrm{tm}}(F,\nu,c)$ when this matters. It is positive: the pair $(\emptyset,\emptyset)$ has
weight $2^{-s}\prod_v2^{-\nu_v}\varpi_2^{c_{v,2}}\varpi_3^{c_{v,3}}>0$, since $C_F(\emptyset)=\emptyset$. For $u\in O$ put
$\tilde h_u=h_u(X',X)/\bar h_u(X)$, a random variable under $\Pr_{\mathrm{tm}}$; it is positive when $p_3<1$. For
an exposed $u$ of type $\tau$ with owner $o$ put also
\[
  \tilde h^{\mathrm{ex}}_u=\frac{h_u(X',X)}{\bar h_u(X)\,q_\tau^{\mathbf 1[o\in X]}\,\varpi_\tau^{\mathbf 1[o\notin X]}} .
\]

\begin{lemma}[Exposures, credits and the Jensen correction]\label{lem:rsc-jensen}
  Let $0<p_2\le p_3<\frac12$, $y\ge0$ and $\varpi_2,\varpi_3\ge1$. In the setting of \cref{def:rsc-tm},
  \[
    W^{\mathrm{cl}}_S(y)\ \ge\ W^{\mathrm{tm}}\exp\Bigl(\sum_{u\in O^{\mathrm{co}}}\E_{\mathrm{tm}}[\ln\tilde h_u]
    +\sum_{u\in O^{\mathrm{ex}}}\E_{\mathrm{tm}}[\ln\tilde h^{\mathrm{ex}}_u]\Bigr).
  \]
  In particular, if all these expectations are nonnegative, then $W^{\mathrm{cl}}_S(y)\ge W^{\mathrm{tm}}$.
\end{lemma}

\begin{proof}
  Let $X'\subseteq S$, $\Upsilon\subseteq C_F(X')$ and $X=X'\cup\Upsilon$. Grouping the outside vertices by their
  owners,
  \[
    \prod_{u\in O}\bar h_u(X)\prod_{u\in O^{\mathrm{ex}}}q_{\tau(u)}^{\mathbf 1[o(u)\in X]}\varpi_{\tau(u)}^{\mathbf 1[o(u)\notin X]}=\prod_{v\in S}\hat h_v(X),
  \]
  because the vertices owned by $v\in X$ contribute $1$ each, except the exposed ones, which
  contribute $q_{\tau(u)}$, and the $\nu_v$ vertices owned by $v\notin X$ contribute $\frac12$ each, the exposed ones
  with the extra factor $\varpi_{\tau(u)}$. Hence the
  summand for $(X',\Upsilon)$ on the right-hand side of \cref{lem:rsc-model} equals
  \[
    \mathrm{wt}(X',\Upsilon)\prod_{u\in O^{\mathrm{ex}}}\tilde h^{\mathrm{ex}}_u\ \prod_{u\in O^{\mathrm{co}}}\tilde h_u
    \prod_{u\text{ private}}\frac{h_u(X',X)}{\bar h_u(X)}
    \ \ge\ \mathrm{wt}(X',\Upsilon)\prod_{u\in O^{\mathrm{ex}}}\tilde h^{\mathrm{ex}}_u\prod_{u\in O^{\mathrm{co}}}\tilde h_u ,
  \]
  because the private factors are at least $1$ by \cref{lem:rsc-pointwise}(i). Summing over $(X',\Upsilon)$ and using \cref{lem:rsc-model},
  \[
    W^{\mathrm{cl}}_S(y)\ \ge\ W^{\mathrm{tm}}\,\E_{\mathrm{tm}}\Bigl[\exp\Bigl(\sum_{u\in O^{\mathrm{co}}}\ln\tilde h_u+\sum_{u\in O^{\mathrm{ex}}}\ln\tilde h^{\mathrm{ex}}_u\Bigr)\Bigr]
    \ \ge\ W^{\mathrm{tm}}\exp\Bigl(\E_{\mathrm{tm}}\Bigl[\sum_{u\in O^{\mathrm{co}}}\ln\tilde h_u+\sum_{u\in O^{\mathrm{ex}}}\ln\tilde h^{\mathrm{ex}}_u\Bigr]\Bigr),
  \]
  where the last step is Jensen's inequality for the convex function $\exp$; all the random variables are positive
  because $q_3>0$.
\end{proof}

Taking expectations in \cref{lem:rsc-pointwise}(iv) and (v), for every corrected $u$,
\begin{align}
  \E_{\mathrm{tm}}[\ln\tilde h_u]&\ \ge\ \delta_3\,\Pr_{\mathrm{tm}}\bigl(\mathbf 1_u=1,\ o(u)\notin X\bigr)-\varepsilon_3\,\Pr_{\mathrm{tm}}\bigl(\mathbf 1_u=1,\ o(u)\in X\bigr)
  &&\text{if }\tau(u)=3,\label{eq:rsc-EQ3}\\
  \E_{\mathrm{tm}}[\ln\tilde h_u]&\ \ge\ \delta_2\,\Pr_{\mathrm{tm}}\bigl(\mathbf 1_u=1,\ o(u)\notin X\bigr)-\varepsilon_2\,\Pr_{\mathrm{tm}}\bigl(\mathbf 1_u=1,\ o(u)\in X'\bigr)
  &&\text{if }\tau(u)=2.\label{eq:rsc-EQ2}
\end{align}
For an exposed $u$ of type $\tau$ with owner $o$, \cref{lem:rsc-pointwise}(ii) gives $\ln\tilde h^{\mathrm{ex}}_u\ge0$ on the
event $o\in X$, and on the event $o\notin X$ we have $\ln\tilde h^{\mathrm{ex}}_u=\ln\tilde h_u-\ln\varpi_\tau$ with
$\ln\tilde h_u\ge\delta_\tau\mathbf 1[N'_u\cap X\ne\emptyset]$ by (iii) and the first assertion of the lemma; hence
\begin{equation}\label{eq:rsc-EQex}
  \E_{\mathrm{tm}}[\ln\tilde h^{\mathrm{ex}}_u]\ \ge\ \Pr_{\mathrm{tm}}(o\notin X)\Bigl(\delta_\tau\Pr_{\mathrm{tm}}\bigl(N'_u\cap X\ne\emptyset\bigm|o\notin X\bigr)-\ln\varpi_\tau\Bigr)
\end{equation}
Every event of the form $U\cap X=\emptyset$ has positive probability under $\Pr_{\mathrm{tm}}$, because the pair $(\emptyset,\emptyset)$ has positive weight; in particular $\Pr_{\mathrm{tm}}(o\notin X)>0$, so the conditional probability is defined.
\Cref{sec:rsc-light} shows that the right-hand sides of \eqref{eq:rsc-EQ3} and \eqref{eq:rsc-EQ2} are nonnegative when the owner $o(u)$ is
\emph{light} for the type of $u$ (\cref{def:rsc-light}): it owns few vertices and has small degree in $F$. It also
bounds the conditional probability in \eqref{eq:rsc-EQex} from below, which yields the admissible credits. To compute the
probabilities we need a recursion for the tree model.

\subsubsection{The ten-state recursion}\label{sec:rsc-rec}

As in \cref{lem:rst-treerec}, the tree model is a sum over configurations of a product of local
factors, which can be evaluated from the leaves of $F$ upwards. The weight $\mathrm{wt}(X',\Upsilon)$ factorizes over
the vertices and edges of $F$. Since $F[X]$ is a forest, $\comp(F[X])=|X|-|E(F[X])|$, so
\[
  2^{-\Psi_F(X)}=\prod_{v\in X}\tfrac12\ \prod_{vw\in E(F[X])}2\ \prod_{v\in S:\,N_F[v]\cap X=\emptyset}\tfrac12 ;
\]
and $\pi_{X'}(\Upsilon)$ is the product, over the vertices $v\in S\setminus X'$, of $p_{k_F(v)}$ if $v\in\Upsilon$, of $q_{k_F(v)}$
if $v\in C_F(X')\setminus\Upsilon$, and of $1$ otherwise. The factor of a vertex depends on its own status
and on how many of its $F$-neighbors lie in $X'$ and in $X$; since $p_k$ and $q_k$ depend on $k$ only
through $\min(k,3)$, a vertex has to count its sampled neighbors up to three. Write
\[
  \lambda=p_3/p_2\ \ge1 .
\]
A message records the status of a vertex together with the information about its children that is
still needed once the status of its parent is known.

\paragraph{States.} Root $F$ at a vertex $r$. Fix a vertex $v$ and a configuration on the subtree
formed by $v$ and some of its children's subtrees. The \emph{state} of $v$ is one of the following
ten; ``child'' refers to the children included so far, and ``sampled'' means ``in $X'$.''
\begin{itemize}
  \item $\mathsf A$: $v\in X'$.
  \item $\mathsf C_0$, $\mathsf C_1$, $\mathsf C_2$, $\mathsf C_3$: $v\in\Upsilon$, and $v$ has $0$, $1$, $2$, or at least $3$
    sampled children. When the configuration is completed, $v\in\Upsilon$ is allowed only if $v\in C_F(X')$,
    that is, if the number of sampled children plus $\mathbf 1[\text{parent}\in X']$ is at least $2$.
  \item $\mathsf O_{00}$: $v\notin X$, and no child of $v$ is in $X$.
  \item $\mathsf O_{01}$: $v\notin X$, no child of $v$ is sampled, and some child is in $\Upsilon$.
  \item $\mathsf O_1$, $\mathsf O_2$, $\mathsf O_3$: $v\notin X$, and exactly one, exactly two, or at least three children of $v$ are sampled.
\end{itemize}
A \emph{message} is a vector $P\in\R^{10}$ indexed by these states, in this order; we write its
coordinates as $P=(A,C_0,C_1,C_2,C_3,O_{00},O_{01},O_1,O_2,O_3)$ for a parent and
$Q=(a,c_0,c_1,c_2,c_3,o_{00},o_{01},o_1,o_2,o_3)$ for a child. (These letters denote numbers; the
states $\mathsf C_j$ are those of the vertices in the closure~$\Upsilon$.) The weight of a vertex in
$\Upsilon$ is carried as $\frac{p_2}2$, and the factor $\lambda$ that turns $p_2$ into $p_3$ is applied when
the vertex turns out to have at least three sampled neighbors.

\paragraph{Merge.} For a message $Q$ of a child let
\begin{gather*}
  \chi_{\mathsf S}(Q)=2a+2c_1+2\lambda(c_2+c_3)+o_{00}+o_{01}+q_2o_1+q_3(o_2+o_3),\\
  \chi^{\mathrm s}_{\mathsf C}(Q)=2a,\qquad \chi^{\mathrm u}_{\mathsf C}(Q)=2c_2+2\lambda c_3+o_{00}+o_{01}+o_1+q_2o_2+q_3o_3,\\
  \chi^{\mathrm s}_{\mathsf O}(Q)=a,\qquad \chi^{\mathrm x}_{\mathsf O}(Q)=c_2+\lambda c_3,\qquad \chi^{\mathrm u}_{\mathsf O}(Q)=\tfrac12o_{00}+o_{01}+o_1+q_2o_2+q_3o_3,
\end{gather*}
and $\chi_{\mathsf O}(Q)=\chi^{\mathrm s}_{\mathsf O}(Q)+\chi^{\mathrm x}_{\mathsf O}(Q)+\chi^{\mathrm u}_{\mathsf O}(Q)=a+c_2+\lambda c_3+\tfrac12o_{00}+o_{01}+o_1+q_2o_2+q_3o_3$.
These are the total weights that the child contributes when the parent is sampled ($\chi_{\mathsf S}$);
when the parent is in $\Upsilon$, split by whether the child is sampled ($\chi^{\mathrm s}_{\mathsf C}$) or not
($\chi^{\mathrm u}_{\mathsf C}$); and when the parent is not in $X$, split by whether the child is sampled
($\chi^{\mathrm s}_{\mathsf O}$), is in $\Upsilon$ ($\chi^{\mathrm x}_{\mathsf O}$), or neither ($\chi^{\mathrm u}_{\mathsf O}$). Writing $\chi^{\mathrm s}_{\mathsf C}$
for $\chi^{\mathrm s}_{\mathsf C}(Q)$ and so on, define
\[
  P\circ Q=
  \begin{pmatrix}
    A\,\chi_{\mathsf S}(Q)\\
    C_0\chi^{\mathrm u}_{\mathsf C}\\
    C_1\chi^{\mathrm u}_{\mathsf C}+C_0\chi^{\mathrm s}_{\mathsf C}\\
    C_2\chi^{\mathrm u}_{\mathsf C}+C_1\chi^{\mathrm s}_{\mathsf C}\\
    C_3(\chi^{\mathrm u}_{\mathsf C}+\chi^{\mathrm s}_{\mathsf C})+C_2\chi^{\mathrm s}_{\mathsf C}\\
    O_{00}\chi^{\mathrm u}_{\mathsf O}\\
    O_{01}(\chi^{\mathrm u}_{\mathsf O}+\chi^{\mathrm x}_{\mathsf O})+O_{00}\chi^{\mathrm x}_{\mathsf O}\\
    O_1(\chi^{\mathrm u}_{\mathsf O}+\chi^{\mathrm x}_{\mathsf O})+(O_{00}+O_{01})\chi^{\mathrm s}_{\mathsf O}\\
    O_2(\chi^{\mathrm u}_{\mathsf O}+\chi^{\mathrm x}_{\mathsf O})+O_1\chi^{\mathrm s}_{\mathsf O}\\
    O_3(\chi^{\mathrm u}_{\mathsf O}+\chi^{\mathrm x}_{\mathsf O}+\chi^{\mathrm s}_{\mathsf O})+O_2\chi^{\mathrm s}_{\mathsf O}
  \end{pmatrix}^{\!\top},
\]
and
\[
  \mathrm{fin}(P)=A+C_2+\lambda C_3+\tfrac12O_{00}+O_{01}+O_1+q_2O_2+q_3O_3 .
\]
The operation $\circ$ is bilinear and maps nonnegative messages to nonnegative messages.

\paragraph{Initial messages and forcing.} For integers $\nu\ge0$ and $c_2,c_3\ge0$ let
\[
  \mathrm{Init}(\nu;c_2,c_3)=\bigl(q_2^{c_2}q_3^{c_3}\tfrac y2,\ q_2^{c_2}q_3^{c_3}\tfrac{p_2}2,\ 0,\ 0,\ 0,\ 2^{-\nu}\varpi_2^{c_2}\varpi_3^{c_3},\ 0,\ 0,\ 0,\ 0\bigr).
\]
A \emph{forcing} is a map $f$ from a subset of $S$ to $\{\mathsf{in},\mathsf{out},\mathsf{excl}\}$; it restricts the
configurations to those with $v\in X'$ if $f(v)=\mathsf{in}$, with $v\notin X'$ if $f(v)=\mathsf{out}$, and with
$v\notin X$ if $f(v)=\mathsf{excl}$ (the vertex is \emph{excluded}; it may be a candidate, but it is not put
into $\Upsilon$). For a forced vertex we use the \emph{forced initial message}: $\mathrm{Init}(\nu;c_2,c_3)$
with all coordinates except the first set to $0$ if $f(v)=\mathsf{in}$, with the first coordinate set to
$0$ if $f(v)=\mathsf{out}$, and with the first two coordinates set to $0$ if $f(v)=\mathsf{excl}$. Note that
$f(v)=\mathsf{out}$ still allows $v\in\Upsilon$.

\begin{definition}[Messages]\label{def:rsc-messages}
  Let $F$ be rooted at $r$, let $\nu,c_2,c_3:S\to\mathbb Z_{\ge0}$, let $0<p_2\le p_3<1$, $y\ge0$, $\varpi_2,\varpi_3\ge1$, and let $f$ be a
  forcing. For every $v\in S$ fix an order $w_1,\dots,w_d$ of its children. The \emph{messages} $M_v$,
  $v\in S$, are defined from the leaves upwards by
  \[
    M_v=\Bigl(\cdots\bigl(\mathrm{Init}_f(\nu_v;c_{v,2},c_{v,3})\circ M_{w_1}\bigr)\circ\cdots\Bigr)\circ M_{w_d},
  \]
  where $\mathrm{Init}_f$ is the forced initial message if $v$ is forced and $\mathrm{Init}$ otherwise.
\end{definition}

\begin{lemma}[Ten-state recursion]\label{lem:rsc-recursion}
  Let $M_v$, $v\in S$, be the messages of \cref{def:rsc-messages}, for any orders of the children.
  Then every $M_v$ is nonnegative, and with the weights $\mathrm{wt}$ of \cref{def:rsc-tm}
  \begin{align*}
    A_r&=\sum_{(X',\Upsilon)\models f,\ r\in X'}\mathrm{wt}(X',\Upsilon),\\
    (C_2)_r+\lambda(C_3)_r&=\sum_{(X',\Upsilon)\models f,\ r\in\Upsilon}\mathrm{wt}(X',\Upsilon),\\
    \tfrac12(O_{00})_r+(O_{01})_r+(O_1)_r+q_2(O_2)_r+q_3(O_3)_r&=\sum_{(X',\Upsilon)\models f,\ r\notin X}\mathrm{wt}(X',\Upsilon),
  \end{align*}
  where $M_r=(A_r,\dots,(O_3)_r)$ and $(X',\Upsilon)\models f$ ranges over the pairs with $X'\subseteq S$ and
  $\Upsilon\subseteq C_F(X')$ that respect the forcing. In particular $\mathrm{fin}(M_r)$ is the sum of $\mathrm{wt}$ over
  these pairs, and $\mathrm{fin}(M_r)=W^{\mathrm{tm}}$ for the empty forcing.
\end{lemma}

\begin{proof}
  The proof assigns to every partial configuration on a subtree a weight that collects those factors of
  $\mathrm{wt}(X',\Upsilon)$ which are already determined inside the subtree, and it shows by induction along the merges
  that each message sums these weights state by state. At the root nothing remains undetermined, and
  the displayed sums follow.

  For a vertex $v$ and $0\le i\le d$ let $F_v^i$ be the subtree formed by $v$ and the subtrees of
  $w_1,\dots,w_i$; write $F_v$ for $F_v^d$. A \emph{partial configuration} on a vertex set $U$ is a pair
  $(X'_U,\Upsilon_U)$ of disjoint subsets of $U$ respecting the forcing. For a partial configuration on
  $V(F_v^i)$, define its \emph{partial weight} as the product of the following factors, where $k$ denotes
  the number of sampled $F$-neighbors of the vertex in question:
  \begin{itemize}
    \item for every vertex $x\ne v$ of $F_v^i$ (whose $F$-neighbors all lie in $F_v^i$, since the
      parent of $x$ is in $F_v^i$): the \emph{complete vertex factor}, namely $\frac y2q_2^{c_{x,2}}q_3^{c_{x,3}}$ if
      $x\in X'$; $\frac{p_k}2q_2^{c_{x,2}}q_3^{c_{x,3}}$ if $x\in\Upsilon$ and $k\ge2$, and $0$ if
      $x\in\Upsilon$ and $k\le1$; if $x\notin X'\cup\Upsilon$, the product of $2^{-\nu_x}\varpi_2^{c_{x,2}}\varpi_3^{c_{x,3}}$, of $q_k$ if $k\ge2$,
      and of $\frac12$ if no $F$-neighbor of $x$ is in $X'\cup\Upsilon$;
    \item for $v$ itself, the \emph{partial vertex factor}: $\frac y2q_2^{c_{v,2}}q_3^{c_{v,3}}$ if $v\in X'$, $\frac{p_2}2q_2^{c_{v,2}}q_3^{c_{v,3}}$
      if $v\in\Upsilon$, and $2^{-\nu_v}\varpi_2^{c_{v,2}}\varpi_3^{c_{v,3}}$ if $v\notin X'\cup\Upsilon$;
    \item a factor $2$ for every edge of $F_v^i$ with both ends in $X'\cup\Upsilon$.
  \end{itemize}
  By the factorization displayed before the definition of the states, the product of the complete
  vertex factors of all vertices of $S$ and of the edge factors is $\mathrm{wt}(X',\Upsilon)$ for $\Upsilon\subseteq C_F(X')$,
  and $0$ if $\Upsilon$ contains a vertex with fewer than two sampled $F$-neighbors. (For the vertex factor of
    $x\in\Upsilon$, the requirement $x\in C_F(X')$ is exactly that $x$ has at least two sampled $F$-neighbors; the
    factor $\frac12$ per vertex of $X$, the factor $p_k$ or $q_k$ of $\pi_{X'}$, and the factor $\hat h_x$ are
  all included.) Let $M_v^i\in\R^{10}$ be the vector whose coordinate for a state is the sum of the partial
  weights of the partial configurations on $V(F_v^i)$ in which $v$ is in that state. We show by
  induction that $M_v^i$ is the message obtained after merging $w_1,\dots,w_i$; for $i=d$ this is
  $M_v$, which is then nonnegative.

  For $i=0$, $F_v^0=\{v\}$; the configurations are $v\in X'$ (state $\mathsf A$, weight $\frac y2q_2^{c_{v,2}}q_3^{c_{v,3}}$),
  $v\in\Upsilon$ (state $\mathsf C_0$, weight $\frac{p_2}2q_2^{c_{v,2}}q_3^{c_{v,3}}$) and $v\notin X'\cup\Upsilon$ (state $\mathsf O_{00}$,
  weight $2^{-\nu_v}\varpi_2^{c_{v,2}}\varpi_3^{c_{v,3}}$), restricted by the forcing. This is $\mathrm{Init}_f(\nu_v;c_{v,2},c_{v,3})$.

  For the step, let $w=w_i$ and $Q=M_w$, which by induction collects the partial configurations of
  $F_w$ with $w$ as the special vertex. A partial configuration on $V(F_v^i)$ is a pair of partial
  configurations on $V(F_v^{i-1})$ and $V(F_w)$. Its partial weight is the product of their partial
  weights, of the factor that completes the vertex factor of $w$ (now that the status of its parent $v$
  is known), and of the factor $2$ if $v,w\in X$; the partial vertex factor of $v$ does not change. We
  list, for each status of $v$ and each state of $w$, the completing factor, the edge factor, and the
  new state of $v$. If $w$ is in state $\mathsf C_j$ or $\mathsf O_j$, it has $j$ sampled children
  (counted up to $3$, with $j=0$ for $\mathsf O_{00}$ and $\mathsf O_{01}$).
  \begin{itemize}
    \item $v\in X'$. If $w$ is in state $\mathsf A$, the edge gives $2$. If $w$ is in state $\mathsf C_j$,
      it has $k=j+1$ sampled neighbors, so it is a valid candidate if and
      only if $j\ge1$, with the completing factor $p_k/p_2$, which is $1$ for $j=1$ and $\lambda$ for $j\ge2$; the edge gives $2$. If $w$ is in an $\mathsf O$-state, it is dominated by $v$, and it
      has $k=j+1$ sampled neighbors, so it is a candidate
      not put into $\Upsilon$, with factor $q_k$, if and only if $j\ge1$: $q_2$ in $\mathsf O_1$ and $q_3$ in $\mathsf O_2$ and $\mathsf O_3$. The state
      of $v$ remains $\mathsf A$. The total is $A\,\chi_{\mathsf S}(Q)$.
    \item $v\in\Upsilon$. A sampled child contributes $2a=\chi^{\mathrm s}_{\mathsf C}$ (edge factor $2$) and raises the
      number of sampled children of $v$ by one. A child in $\Upsilon$ has an unsampled parent, so it has $k=j$ sampled
      neighbors and is valid only
      in the states $\mathsf C_2$ (factor $1$) and $\mathsf C_3$ (factor $\lambda$), with edge factor $2$; a child not in $X$ is dominated by $v$, has $k=j$ sampled
      neighbors, and so gets the factor $q_2$ in state $\mathsf O_2$ and $q_3$ in state $\mathsf O_3$.
      These cases contribute $\chi^{\mathrm u}_{\mathsf C}$ and leave the count of $v$ unchanged. This gives the second to
      fifth coordinates of $P\circ Q$, with the count capped at $3$.
    \item $v\notin X$. A sampled child contributes $a=\chi^{\mathrm s}_{\mathsf O}$ and moves $v$ from $\mathsf O_{00}$
      or $\mathsf O_{01}$ to $\mathsf O_1$, from $\mathsf O_1$ to $\mathsf O_2$, and from $\mathsf O_2$ or $\mathsf O_3$ to $\mathsf O_3$. A child in $\Upsilon$ is valid
      only in the states $\mathsf C_2$ and $\mathsf C_3$, contributes $c_2+\lambda c_3=\chi^{\mathrm x}_{\mathsf O}$, dominates $v$, and moves $v$ from
      $\mathsf O_{00}$ to $\mathsf O_{01}$. A child not in $X$ has no dominating parent: it gets $\frac12$ in
      state $\mathsf O_{00}$, $q_2$ in state $\mathsf O_2$ and $q_3$ in state $\mathsf O_3$; this contributes $\chi^{\mathrm u}_{\mathsf O}$ and leaves the
      state of $v$ unchanged. This gives the last five coordinates of $P\circ Q$.
  \end{itemize}
  In all cases no other vertex factor changes, because only $v$ and $w$ gain a neighbor with known
  status. So $M_v^i=M_v^{i-1}\circ M_w$.

  At the root, the configurations on $V(F_r)=S$ are complete once the partial vertex factor of $r$ is
  completed, and $r$ has no parent. In state $\mathsf A$ nothing remains. In state $\mathsf C_j$, $r$ is
  valid if and only if $j\ge2$, with the completing factor $1$ for $j=2$ and $\lambda$ for $j=3$. In state $\mathsf O_{00}$, $r$ is undominated (factor $\frac12$); in
  $\mathsf O_{01}$ and $\mathsf O_1$ it is dominated and has at most one sampled neighbor; in
  $\mathsf O_2$ and $\mathsf O_3$ it is a candidate not put into $\Upsilon$ (factor $q_2$, respectively $q_3$). This gives the displayed sums, whose
  total is $\mathrm{fin}(M_r)$.
\end{proof}

For $p_2=p_3$ the factor $\lambda$ is $1$ and the pairs of states $\mathsf C_2,\mathsf C_3$ and
$\mathsf O_2,\mathsf O_3$ can be merged, which gives an eight-state recursion; in the limit $p_2=p_3\to0$ the $\mathsf C$-coordinates vanish, the states $\mathsf O_{01}$, $\mathsf O_1$,
$\mathsf O_2$ and $\mathsf O_3$ play the role of one state, and the recursion reduces to the three-state recursion of
\cref{lem:rst-treerec}.

\subsubsection{Light owners and credits}\label{sec:rsc-light}

The probabilities in~\eqref{eq:rsc-EQ3}--\eqref{eq:rsc-EQex} are ratios of messages at the root, once $F$ is rooted at the
owner or at a neighbor of the exposed vertex. The next lemma compares coordinates of messages; it is the core of the argument. Put
\[
  \mathsf b=2+\frac{2\max(p_2,\,p_3-p_2)}y .
\]
Throughout this subsection $0<p_2\le p_3<\frac12$, $y>0$, $1\le\varpi_\tau\le2q_\tau$ for $\tau\in\{2,3\}$, and
$\nu,c_2,c_3:S\to\mathbb Z_{\ge0}$ with $c_{v,2}+c_{v,3}\le\nu_v$ for all $v$; these are the standing
assumptions on the tree model. Note that $\varpi_\tau\le2q_\tau$ and $c_v\le\nu_v$ give
\begin{equation}\label{eq:rsc-credit-small}
  2^{-\nu_v}\varpi_2^{c_{v,2}}\varpi_3^{c_{v,3}}\ \le\ 2^{-\nu_v}(2q_2)^{c_{v,2}}(2q_3)^{c_{v,3}}\ \le\ q_2^{c_{v,2}}q_3^{c_{v,3}} ,
\end{equation}
that is, the factor $\hat h_v$ of \cref{def:rsc-tm} is not larger for $v\notin X$ than for $v\in X$.

\begin{lemma}[Message inequalities]\label{lem:rsc-messages}
  Let $F$ be rooted anywhere and let $f$ be a forcing. Then the messages $M_v=(A,C_0,\dots,O_3)$
  of \cref{def:rsc-messages} satisfy:
  \begin{enumerate}[label=(\alph*)]
    \item $C_0\ge\frac{p_2}2(O_{00}+O_{01})$ and $C_j\ge\frac{p_2}2O_j$ for $j=1,2,3$, for every $v$ with $f(v)\ne\mathsf{excl}$;
    \item $C_0+C_1+C_2+C_3\le\frac{p_2}y\,A$, for every $v$ with $f(v)\ne\mathsf{out}$ none of whose children is excluded;
    \item $\chi_{\mathsf S}(M_v)\le\mathsf b\,\chi_{\mathsf O}(M_v)$, for every $v$ with $f(v)\ne\mathsf{out}$ none of whose children is excluded;
    \item $\chi_{\mathsf S}(M_v)\ge\chi_{\mathsf O}(M_v)$ for every $v$ with $f(v)\ne\mathsf{excl}$, and
      $\chi_{\mathsf S}(M_v)\ge q_3\,\chi_{\mathsf O}(M_v)$ for every $v$ with $f(v)=\mathsf{excl}$.
  \end{enumerate}
\end{lemma}

\begin{proof}
  All messages are nonnegative by \cref{lem:rsc-recursion}. We prove (a) and (b) by induction along the
  merges that produce $M_v$, and then derive (c) and (d).

  (a) For an initial message with $f(v)=\mathsf{in}$ all coordinates involved are $0$. Otherwise
  $C_1=C_2=C_3=O_1=O_2=O_3=O_{01}=0$, $C_0=\frac{p_2}2q_2^{c_{v,2}}q_3^{c_{v,3}}$ and
  $O_{00}=2^{-\nu_v}\varpi_2^{c_{v,2}}\varpi_3^{c_{v,3}}$, so $C_0\ge\frac{p_2}2O_{00}$ by~\eqref{eq:rsc-credit-small}. For a merge with a
  child message $Q\ge0$, the merge formulas give
  \begin{gather*}
    C_0'=C_0\,\chi^{\mathrm u}_{\mathsf C},\qquad O_{00}'+O_{01}'=(O_{00}+O_{01})(\chi^{\mathrm u}_{\mathsf O}+\chi^{\mathrm x}_{\mathsf O}),\\
    C_1'=C_1\chi^{\mathrm u}_{\mathsf C}+C_0\chi^{\mathrm s}_{\mathsf C},\qquad O_1'=O_1(\chi^{\mathrm u}_{\mathsf O}+\chi^{\mathrm x}_{\mathsf O})+(O_{00}+O_{01})\chi^{\mathrm s}_{\mathsf O},\\
    C_2'=C_2\chi^{\mathrm u}_{\mathsf C}+C_1\chi^{\mathrm s}_{\mathsf C},\qquad O_2'=O_2(\chi^{\mathrm u}_{\mathsf O}+\chi^{\mathrm x}_{\mathsf O})+O_1\chi^{\mathrm s}_{\mathsf O},\\
    C_3'=C_3(\chi^{\mathrm u}_{\mathsf C}+\chi^{\mathrm s}_{\mathsf C})+C_2\chi^{\mathrm s}_{\mathsf C},\qquad O_3'=O_3(\chi^{\mathrm u}_{\mathsf O}+\chi^{\mathrm x}_{\mathsf O}+\chi^{\mathrm s}_{\mathsf O})+O_2\chi^{\mathrm s}_{\mathsf O},
  \end{gather*}
  and $\chi^{\mathrm u}_{\mathsf C}-(\chi^{\mathrm u}_{\mathsf O}+\chi^{\mathrm x}_{\mathsf O})=c_2+\lambda c_3+\frac12o_{00}\ge0$, $\chi^{\mathrm s}_{\mathsf C}=2\chi^{\mathrm s}_{\mathsf O}\ge\chi^{\mathrm s}_{\mathsf O}$.
  Hence each primed $\mathsf C$-coordinate is at least $\frac{p_2}2$ times the corresponding primed
  $\mathsf O$-coordinate, term by term; for instance
  $C_1'\ge\frac{p_2}2O_1(\chi^{\mathrm u}_{\mathsf O}+\chi^{\mathrm x}_{\mathsf O})+\frac{p_2}2(O_{00}+O_{01})\cdot2\chi^{\mathrm s}_{\mathsf O}\ge\frac{p_2}2O_1'$.

  (b) If $f(v)=\mathsf{in}$, the $\mathsf C$-coordinates of the initial message are $0$, and they stay
  $0$ under merging; if $f(v)=\mathsf{excl}$, both sides are $0$. Otherwise the initial message has $C_0=\frac{p_2}2q_2^{c_{v,2}}q_3^{c_{v,3}}=\frac{p_2}y A$. For a merge with a child message $Q$, summing the second to fifth coordinates of $P\circ Q$
  gives
  \[
    C_0'+C_1'+C_2'+C_3'=(C_0+C_1+C_2+C_3)(\chi^{\mathrm u}_{\mathsf C}+\chi^{\mathrm s}_{\mathsf C}),\qquad A'=A\,\chi_{\mathsf S}(Q),
  \]
  and
  \[
    \chi_{\mathsf S}(Q)-(\chi^{\mathrm u}_{\mathsf C}+\chi^{\mathrm s}_{\mathsf C})=(2c_1-p_2\,o_1)+\bigl(2(\lambda-1)c_2-(p_3-p_2)\,o_2\bigr)\ \ge\ 0
  \]
  by (a) for the child, which is not excluded: $2c_1\ge p_2o_1$, and $2(\lambda-1)c_2\ge(\lambda-1)p_2o_2=(p_3-p_2)o_2$. So the inequality is preserved.

  (c) For $M_v=(a,c_0,c_1,c_2,c_3,o_{00},o_{01},o_1,o_2,o_3)$,
  \[
    \chi_{\mathsf S}(M_v)-2\chi_{\mathsf O}(M_v)=2c_1+2(\lambda-1)c_2-o_{01}-(2-q_2)o_1-(2q_2-q_3)o_2-q_3o_3\ \le\ 2\max(1,\lambda-1)(c_1+c_2),
  \]
  since $2q_2\ge q_3$. By (b), $c_1+c_2\le\frac{p_2}ya\le\frac{p_2}y\chi_{\mathsf O}(M_v)$, since all terms of $\chi_{\mathsf O}$ are nonnegative, and
  $2\max(1,\lambda-1)p_2=2\max(p_2,p_3-p_2)$. Hence $\chi_{\mathsf S}(M_v)\le\mathsf b\,\chi_{\mathsf O}(M_v)$.

  (d) In the same notation,
  \[
    \chi_{\mathsf S}(M_v)-\chi_{\mathsf O}(M_v)=a+\tfrac12o_{00}+\lambda c_3+(2c_1-p_2o_1)+\bigl((2\lambda-1)c_2-(p_3-p_2)o_2\bigr),
  \]
  which is nonnegative by (a) if $f(v)\ne\mathsf{excl}$, because $(2\lambda-1)c_2\ge(2\lambda-1)\frac{p_2}2o_2=(p_3-\frac{p_2}2)o_2\ge(p_3-p_2)o_2$.
  If $f(v)=\mathsf{excl}$, then $a=c_0=\dots=c_3=0$ (the initial message has these coordinates $0$, and they stay $0$), so
  \[
    \chi_{\mathsf S}(M_v)-q_3\chi_{\mathsf O}(M_v)=\bigl(1-\tfrac{q_3}2\bigr)o_{00}+(1-q_3)o_{01}+(q_2-q_3)o_1+q_3(1-q_2)o_2+q_3(1-q_3)o_3\ \ge\ 0 .\qedhere
  \]
\end{proof}

\begin{definition}[Light vertices]\label{def:rsc-light}
  A pair $(\nu,d)$ of nonnegative integers is \emph{$3$-light} if
  \[
    2^{-\nu}\mathsf b^{-d}\ \ge\ \Bigl(1+\frac{p_3}y\Bigr)\frac{\varepsilon_3\,y}{\delta_3} ,
  \]
  and \emph{$2$-light} if
  \[
    2^{-\nu}\mathsf b^{-d}\ \ge\ \frac{\varepsilon_2\,y}{\delta_2} ;
  \]
  otherwise it is \emph{$3$-failing}, respectively \emph{$2$-failing}. Given an owner map, a vertex $v\in S$ is $\tau$-light if $(\nu_v,d_F(v))$ is
  $\tau$-light, and $\tau$-failing otherwise.
\end{definition}

Since $\varepsilon_3\ge\varepsilon_2$ and $\delta_3\le\delta_2$, a $3$-light pair is $2$-light. The right-hand
sides are positive because $q_3>\frac12$; both are small when $p_3$ is small, because $\varepsilon_\tau\approx p_\tau$.

The following lemma is the place where the shape of the tree enters. Let $T_u$ be the smallest
subtree of $F$ that contains $N_S(u)$. If $u$ is shared, $T_u$ has at least two vertices, and its
leaves lie in $N_S(u)$; otherwise a leaf outside $N_S(u)$ could be removed.

\begin{lemma}[Nonnegative correction at light owners]\label{lem:rsc-jlemma}
  Under the standing assumptions of \cref{sec:rsc-light} (in particular $y>0$), in the setting of \cref{def:rsc-tm}, let $u$ be a shared vertex of type $\tau$ whose
  owner $o=o(u)$ is $\tau$-light and is a leaf of $T_u$. Then $\E_{\mathrm{tm}}[\ln\tilde h_u]\ge0$.
\end{lemma}

\begin{proof}
  The proof decomposes the event $\mathbf 1_u=1$ into disjoint events, each given by a forcing that uses
  only the values $\mathsf{in}$ and $\mathsf{out}$ and forces
  neither $o$ nor any $F$-neighbor of $o$ to be $\mathsf{out}$. For each event it bounds the two
  probabilities in~\eqref{eq:rsc-EQ3}, respectively~\eqref{eq:rsc-EQ2}, by the same product of child messages at $o$, using
  \cref{lem:rsc-messages}; the lightness of $o$ then gives the sign.

  \emph{The other neighbors lie in one branch.} Since $o$ is a leaf of $T_u$, it has exactly one
  neighbor $w$ in $T_u$, and $T_u-o$ is a tree that contains $w$ and all of $N'_u=N_S(u)\setminus\{o\}$.
  So $N'_u$ lies in the component of $F-o$ that contains $w$, and $w$ is the only $F$-neighbor of $o$
  in that component. Hence $N'_u\cap N_F(o)\subseteq\{w\}$.

  \emph{Decomposition of the event $\mathbf 1_u=1$.} Let $N^{\circ}_u=N'_u\setminus\{w\}=\{v_1,\dots,v_h\}$ in any
  order, so no $v_i$ is an $F$-neighbor of $o$. For $1\le i\le h$ let $f_i$ be the forcing with
  $f_i(v_i)=\mathsf{in}$ and $f_i(v_j)=\mathsf{out}$ for $j<i$; if $w\in N'_u$, let $f_{h+1}$ be the forcing
  with $f_{h+1}(w)=\mathsf{in}$ and $f_{h+1}(v_j)=\mathsf{out}$ for all $j\le h$. The events ``$(X',\Upsilon)$
  respects $f_i$'' are pairwise disjoint, and their union is the event $N'_u\cap X'\ne\emptyset$, that
  is, $\mathbf 1_u=1$: a sampled vertex of $N^{\circ}_u$ is matched by the first such $v_i$, and otherwise $w$ is the
  sampled vertex of $N'_u$. No forcing $f_i$ forces an $F$-neighbor of $o$ to be $\mathsf{out}$, none
  forces $o$, and none excludes a vertex. If $\tau=2$, then $N'_u$ consists of one vertex and there is a single forcing $f_1$.

  \emph{One forced event.} Fix one of the forcings $f=f_i$, and root $F$ at $o$; its children are the
  $d=d_F(o)$ neighbors of $o$. Let $M_o=(A,C_0,\dots,O_3)$ be the message of $o$ under
  $f$. By \cref{lem:rsc-recursion},
  \begin{gather*}
    W^{\mathrm{tm}}\Pr_{\mathrm{tm}}\bigl(f,\ o\in X'\bigr)=A,\qquad
    W^{\mathrm{tm}}\Pr_{\mathrm{tm}}\bigl(f,\ o\in X\bigr)=A+C_2+\lambda C_3,\\
    W^{\mathrm{tm}}\Pr_{\mathrm{tm}}\bigl(f,\ o\notin X\bigr)=\tfrac12O_{00}+O_{01}+O_1+q_2O_2+q_3O_3 ,
  \end{gather*}
  where ``$f$'' denotes the event that $(X',\Upsilon)$ respects $f$. We bound all three by the product
  $\mathrm{pr}_o=\prod_{w'}\chi_{\mathsf O}(M_{w'})$ over the children $w'$ of $o$. Since $o$ is not forced,
  the merge formulas give
  \begin{gather*}
    A=q_2^{c_{o,2}}q_3^{c_{o,3}}\,\frac y2\prod_{w'}\chi_{\mathsf S}(M_{w'})\ \le\ \frac y2\prod_{w'}\chi_{\mathsf S}(M_{w'}),\\
    O_{00}+O_{01}+O_1+O_2+O_3=2^{-\nu_o}\varpi_2^{c_{o,2}}\varpi_3^{c_{o,3}}\,\mathrm{pr}_o\ \ge\ 2^{-\nu_o}\,\mathrm{pr}_o ;
  \end{gather*}
  the first inequality holds because $q_\tau\le1$, the second because $\varpi_\tau\ge1$, and the identity holds because the sum of the last five coordinates of $P\circ Q$ is
  $(O_{00}+O_{01}+O_1+O_2+O_3)\,\chi_{\mathsf O}(Q)$. No child $w'$ of $o$ is forced $\mathsf{out}$ and no vertex is excluded, so
  \cref{lem:rsc-messages}(c) gives $\chi_{\mathsf S}(M_{w'})\le\mathsf b\,\chi_{\mathsf O}(M_{w'})$, and
  \cref{lem:rsc-messages}(b) at $o$ gives $C_2+\lambda C_3\le\lambda(C_0+C_1+C_2+C_3)\le\frac{\lambda p_2}yA=\frac{p_3}yA$. Since $\frac12,1,q_2,q_3\ge\frac12$,
  \[
    A\le\frac y2\,\mathsf b^{d}\,\mathrm{pr}_o,\qquad
    A+C_2+\lambda C_3\le\Bigl(1+\frac{p_3}y\Bigr)\frac y2\,\mathsf b^{d}\,\mathrm{pr}_o,\qquad
    \tfrac12O_{00}+O_{01}+O_1+q_2O_2+q_3O_3\ge\frac12\,2^{-\nu_o}\,\mathrm{pr}_o .
  \]
  If $\tau=3$, then, as $(\nu_o,d)$ is $3$-light,
  \[
    \delta_3\bigl(\tfrac12O_{00}+O_{01}+O_1+q_2O_2+q_3O_3\bigr)-\varepsilon_3(A+C_2+\lambda C_3)
    \ \ge\ \frac{\mathrm{pr}_o\,\mathsf b^d}2\Bigl(\delta_3\,2^{-\nu_o}\mathsf b^{-d}-\varepsilon_3\Bigl(1+\frac{p_3}y\Bigr)y\Bigr)\ \ge\ 0 ;
  \]
  if $\tau=2$, then, as $(\nu_o,d)$ is $2$-light,
  \[
    \delta_2\bigl(\tfrac12O_{00}+O_{01}+O_1+q_2O_2+q_3O_3\bigr)-\varepsilon_2A
    \ \ge\ \frac{\mathrm{pr}_o\,\mathsf b^d}2\Bigl(\delta_2\,2^{-\nu_o}\mathsf b^{-d}-\varepsilon_2\,y\Bigr)\ \ge\ 0 .
  \]

  \emph{Conclusion.} Dividing by $W^{\mathrm{tm}}$ and summing over the disjoint events $f_i$ gives
  $\delta_3\,\Pr_{\mathrm{tm}}(\mathbf 1_u=1,\ o\notin X)\ge\varepsilon_3\,\Pr_{\mathrm{tm}}(\mathbf 1_u=1,\ o\in X)$ if $\tau=3$, and
  $\delta_2\,\Pr_{\mathrm{tm}}(\mathbf 1_u=1,\ o\notin X)\ge\varepsilon_2\,\Pr_{\mathrm{tm}}(\mathbf 1_u=1,\ o\in X')$ if $\tau=2$; then~\eqref{eq:rsc-EQ3}, respectively~\eqref{eq:rsc-EQ2}, gives
  $\E_{\mathrm{tm}}[\ln\tilde h_u]\ge0$.
\end{proof}

The credits rest on a lower bound for the odds that a vertex of $S$ is sampled, given that some
other vertices are outside the closed sample.

\begin{lemma}[Odds of being sampled]\label{lem:rsc-odds}
  In the setting of \cref{def:rsc-tm}, let $U\subseteq S$ and $w\in S\setminus U$ with
  $|U\cap N_F(w)|\le1$, and let $E$ be the event that $U\cap X=\emptyset$. Then
  \[
    \Pr_{\mathrm{tm}}\bigl(E,\ w\in X'\bigr)\ \ge\ \frac y2\,q_3^{|U\cap N_F(w)|}\,\Pr_{\mathrm{tm}}\bigl(E,\ w\notin X\bigr).
  \]
\end{lemma}

\begin{proof}
  The event $E$ is the event that $(X',\Upsilon)$ respects the forcing $f$ with $f(v)=\mathsf{excl}$ for $v\in U$.
  Root $F$ at $w$ and let $M_w=(A,C_0,\dots,O_3)$ be its message under $f$; $w$ itself is not forced. By
  \cref{lem:rsc-recursion} and the identities used in the proof of \cref{lem:rsc-jlemma},
  \begin{gather*}
    W^{\mathrm{tm}}\Pr_{\mathrm{tm}}(E,\ w\in X')=A=\frac y2\,q_2^{c_{w,2}}q_3^{c_{w,3}}\prod_{x}\chi_{\mathsf S}(M_x),\\
    W^{\mathrm{tm}}\Pr_{\mathrm{tm}}(E,\ w\notin X)\le O_{00}+\dots+O_3=2^{-\nu_w}\varpi_2^{c_{w,2}}\varpi_3^{c_{w,3}}\prod_{x}\chi_{\mathsf O}(M_x),
  \end{gather*}
  the products being over the children $x$ of $w$, that is, over $N_F(w)$. By
  \cref{lem:rsc-messages}(d), $\chi_{\mathsf S}(M_x)\ge\chi_{\mathsf O}(M_x)$ for $x\notin U$ and
  $\chi_{\mathsf S}(M_x)\ge q_3\chi_{\mathsf O}(M_x)$ for $x\in U$, and by~\eqref{eq:rsc-credit-small}
  $q_2^{c_{w,2}}q_3^{c_{w,3}}\ge2^{-\nu_w}\varpi_2^{c_{w,2}}\varpi_3^{c_{w,3}}$. The claim follows.
\end{proof}

\begin{lemma}[Credits]\label{lem:rsc-credit}
  Put $\vartheta=\frac y2q_3$ and $\rho_1=(1+\vartheta)^{-1}$. Under the standing assumptions of \cref{sec:rsc-light}, in the setting of \cref{def:rsc-tm}, let $u$ be an exposed vertex of type $\tau$ with owner $o$.
  Then
  \[
    \Pr_{\mathrm{tm}}\bigl(N'_u\cap X\ne\emptyset\bigm|o\notin X\bigr)\ \ge\
    \begin{cases}1-\rho_1&\text{if }\tau=2,\\ 1-\rho_1^2&\text{if }\tau=3.
    \end{cases}
  \]
  Consequently, if
  \begin{equation}\label{eq:rsc-credits}
    \ln\varpi_2\le(1-\rho_1)\,\delta_2\qquad\text{and}\qquad \ln\varpi_3\le(1-\rho_1^2)\,\delta_3 ,
  \end{equation}
  then $\E_{\mathrm{tm}}[\ln\tilde h^{\mathrm{ex}}_u]\ge0$ for every exposed vertex $u$.
\end{lemma}

\begin{proof}
  Let $E_0$ be the event $o\notin X$. For $w\in N'_u$, \cref{lem:rsc-odds} with $U=\{o\}$ gives
  $\Pr(E_0,w\in X')\ge\vartheta\Pr(E_0,w\notin X)$, where we write $\Pr$ for $\Pr_{\mathrm{tm}}$. Since
  $\Pr(E_0,w\in X')\le\Pr(E_0)-\Pr(E_0,w\notin X)$, this gives $(1+\vartheta)\Pr(E_0,w\notin X)\le\Pr(E_0)$, that is,
  \begin{equation}\label{eq:rsc-odds1}
    \Pr\bigl(w\notin X\bigm|E_0\bigr)\le\rho_1 .
  \end{equation}
  If $\tau=2$, then $N'_u=\{w\}$ and the claim is~\eqref{eq:rsc-odds1}. If $\tau=3$, choose $w_1,w_2\in N'_u$ with
  $\mathrm{dist}_F(o,w_1)\le\mathrm{dist}_F(o,w_2)$, and let $E_1$ be the event $o,w_1\notin X$, which has positive probability. We apply \cref{lem:rsc-odds} to $w=w_2$ and $U=\{o,w_1\}$: both $o$ and $w_1$
  cannot be $F$-neighbors of $w_2$, for then the path $o,w_2,w_1$ would give $\mathrm{dist}_F(o,w_1)=2>1=\mathrm{dist}_F(o,w_2)$.
  So $|U\cap N_F(w_2)|\le1$, and as above $\Pr(w_2\notin X\mid E_1)\le\rho_1$. Hence
  \[
    \Pr\bigl(w_1\notin X,\ w_2\notin X\bigm|E_0\bigr)=\Pr\bigl(w_1\notin X\bigm|E_0\bigr)\Pr\bigl(w_2\notin X\bigm|E_1\bigr)\le\rho_1^2 ,
  \]
  and $\Pr(N'_u\cap X\ne\emptyset\mid E_0)\ge1-\rho_1^2$. The last assertion follows from~\eqref{eq:rsc-EQex}:
  its right-hand side is at least $\Pr(E_0)\bigl(\delta_\tau(1-\rho_1^{\tau-1})-\ln\varpi_\tau\bigr)\ge0$ under~\eqref{eq:rsc-credits}.
\end{proof}

The certificate needs one restriction at the leaves of $F$, which follows from the minimality of $S$.

\begin{lemma}[Private neighbors of leaves]\label{lem:rsc-leaf}
  If $S$ is a minimum connected dominating set with $|S|\ge2$, then every leaf $v$ of $F$ has a private
  neighbor, that is, some $u\in O$ with $N_S(u)=\{v\}$.
\end{lemma}

\begin{proof}
  Since $F-v$ is a spanning tree of $S\setminus\{v\}\ne\emptyset$, the set $S\setminus\{v\}$ induces a
  connected subgraph, and it dominates $v$, which is adjacent to its $F$-neighbor. If every outside
  neighbor of $v$ had a second neighbor in $S$, then $S\setminus\{v\}$ would also dominate $O$, so it
  would be a connected dominating set of size $|S|-1$, contradicting minimality.
\end{proof}

The certificate of \cref{sec:rsc-cert} bounds $W^{\mathrm{tm}}$ for the following class of quadruples.

\begin{definition}[Admissible quadruples]\label{def:rsc-admissible}
  A quadruple $(F,\nu,c_2,c_3)$ of a tree $F$ and maps
  $\nu,c_2,c_3:V(F)\to\mathbb Z_{\ge0}$ is \emph{admissible} if, for every $v\in V(F)$ with $d=d_F(v)$ and $c_v=c_{v,2}+c_{v,3}$:
  $c_v\le\nu_v$; $c_{v,\tau}=0$ whenever $(\nu_v,d)$ is $\tau$-light or $d=0$, for $\tau\in\{2,3\}$; and $c_v\le\nu_v-1$ whenever
  $d=1$ and $c_v\ge1$.
\end{definition}

The analysis assigns owners and classifies them as follows. A leaf of $T_u$ always exists in
$N_S(u)$, so the rule is well defined.

\begin{definition}[Owner rule]\label{def:rsc-rule}
  Choose an owner map $o$ such that $o(u)$ is a leaf of $T_u$ for every
  shared $u$ (for a private $u$ the owner is its unique neighbor in $S$). Let $\nu_v$ be the resulting
  numbers of owned vertices, let $O^{\mathrm{co}}$ be the set of shared vertices $u$ whose owner is $\tau(u)$-light and $O^{\mathrm{ex}}$ the set of
  shared vertices $u$ whose owner is $\tau(u)$-failing. So $c_{v,\tau}=0$ if $v$ is $\tau$-light, and $c_{v,\tau}$ is the number of shared
  vertices of type $\tau$ owned by $v$ if $v$ is $\tau$-failing.
\end{definition}

\begin{corollary}[Reduction to the tree model]\label{cor:rsc-reduction}
  Assume the standing assumptions of \cref{sec:rsc-light} (in particular $y>0$). Let $\varpi_2,\varpi_3$ satisfy $1\le\varpi_\tau\le2q_\tau$ and~\eqref{eq:rsc-credits}, let $S$ be a minimum connected dominating set with $|S|=s$, and let $o$,
  $O^{\mathrm{co}}$, $O^{\mathrm{ex}}$, $\nu$, $c_2$, $c_3$ be given by \cref{def:rsc-rule}. Then
  $(F,\nu,c_2,c_3)$ is admissible, $s+\sum_{v\in S}\nu_v=n$, and
  $W^{\mathrm{cl}}_S(y)\ge W^{\mathrm{tm}}(F,\nu,c)$.
\end{corollary}

\begin{proof}
  For $u\in O^{\mathrm{co}}$, the owner $o(u)$ is $\tau(u)$-light and is a leaf of $T_u$, so \cref{lem:rsc-jlemma} gives
  $\E_{\mathrm{tm}}[\ln\tilde h_u]\ge0$; for $u\in O^{\mathrm{ex}}$, \cref{lem:rsc-credit} gives $\E_{\mathrm{tm}}[\ln\tilde h^{\mathrm{ex}}_u]\ge0$.
  So \cref{lem:rsc-jensen} gives the inequality. The identity
  $\sum_{v\in S}\nu_v=n-s$ is from \cref{def:rsc-owner}. For admissibility, $c_v\le\nu_v$ because
  the exposed vertices owned by $v$ are among the vertices it owns, and $c_{v,\tau}=0$ for $\tau$-light $v$ by
  \cref{def:rsc-rule}. Now let $c_v\ge1$. Then $v$ owns a shared vertex $u$, so $|N_S(u)|\ge2$,
  hence $s\ge2$ and $d_F(v)\ge1$, as $F$ is a tree with at least two vertices. If $d_F(v)=1$, then $v$ is a
  leaf of $F$, and \cref{lem:rsc-leaf} gives a private neighbor $u'$ of $v$; by \cref{def:rsc-rule} $v$
  owns $u'$, and $u'$ is not shared, so at most $\nu_v-1$ of the vertices owned by $v$ are exposed.
\end{proof}

\subsubsection{The antinorm certificate}\label{sec:rsc-cert}

The last step bounds $W^{\mathrm{tm}}(F,\nu,c)$ from below for every admissible
quadruple (\cref{def:rsc-admissible}), by the method of \cref{lem:rst-F}: a convex region that the merge maps into itself up to a
known scale. The certificate works with eight states rather than ten: it forgets whether a vertex
has more than two sampled children, except where a sampled parent makes the count matter. This is a
relaxation of the tree model, which never overestimates it.

\paragraph{The relaxed model.} Root $F$ at $r$; the relaxed model depends on this choice, and everything below holds for every root. For $X'\subseteq S$ and a candidate $v\in C_F(X')$ put
$\tilde p(v)=p_3$ if $v\ne r$, the parent of $v$ is in $X'$ and $v$ has at least two sampled children,
and $\tilde p(v)=p_2$ otherwise; so $\tilde p(v)\le p_{k_F(v)}$, since $\tilde p(v)=p_3$ only if $k_F(v)\ge3$. Let
$\tilde\pi_{X'}(\Upsilon)=\prod_{v\in\Upsilon}\tilde p(v)\prod_{v\in C_F(X')\setminus\Upsilon}(1-\tilde p(v))$, let $\widetilde{\mathrm{wt}}(X',\Upsilon)$ be
$\mathrm{wt}(X',\Upsilon)$ of \cref{def:rsc-tm} with $\tilde\pi_{X'}$ in place of $\pi_{X'}$, and let
$\widetilde W(F,\nu,c)=\sum_{X'}\sum_{\Upsilon\subseteq C_F(X')}\widetilde{\mathrm{wt}}(X',\Upsilon)$. For a message
$P\in\R^8$ indexed by $\mathsf A,\mathsf C_0,\mathsf C_1,\mathsf C_2,\mathsf O_{00},\mathsf O_{01},\mathsf O_1,\mathsf O_2$,
where now $\mathsf C_2$ and $\mathsf O_2$ mean ``at least two sampled children,'' define the
\emph{eight-state merge} $P\mathbin{\tilde\circ}Q$ by the formula for $P\circ Q$ of \cref{sec:rsc-rec} with the pairs of
coordinates $(C_2,C_3)$ and $(O_2,O_3)$ of a parent merged into $C_2$ and $O_2$, that is, by the merge formula of
\cref{sec:rsc-rec} with the functionals
\begin{gather*}
  \tilde\chi_{\mathsf S}(Q)=2a+2c_1+2\lambda c_2+o_{00}+o_{01}+q_2o_1+q_3o_2,\\
  \tilde\chi^{\mathrm s}_{\mathsf C}(Q)=2a,\qquad \tilde\chi^{\mathrm u}_{\mathsf C}(Q)=2c_2+o_{00}+o_{01}+o_1+q_2o_2,\\
  \tilde\chi^{\mathrm s}_{\mathsf O}(Q)=a,\qquad \tilde\chi^{\mathrm x}_{\mathsf O}(Q)=c_2,\qquad \tilde\chi^{\mathrm u}_{\mathsf O}(Q)=\tfrac12o_{00}+o_{01}+o_1+q_2o_2,
\end{gather*}
the fourth coordinate $C_2(\tilde\chi^{\mathrm u}_{\mathsf C}+\tilde\chi^{\mathrm s}_{\mathsf C})+C_1\tilde\chi^{\mathrm s}_{\mathsf C}$, the eighth
coordinate $O_2(\tilde\chi^{\mathrm u}_{\mathsf O}+\tilde\chi^{\mathrm x}_{\mathsf O}+\tilde\chi^{\mathrm s}_{\mathsf O})+O_1\tilde\chi^{\mathrm s}_{\mathsf O}$, and
\begin{gather*}
  \widetilde{\mathrm{fin}}(P)=A+C_2+\tfrac12O_{00}+O_{01}+O_1+q_2O_2,\\
  \widetilde{\mathrm{Init}}(\nu;c_2,c_3)=\bigl(q_2^{c_2}q_3^{c_3}\tfrac y2,\ q_2^{c_2}q_3^{c_3}\tfrac{p_2}2,\ 0,\ 0,\ 2^{-\nu}\varpi_2^{c_2}\varpi_3^{c_3},\ 0,\ 0,\ 0\bigr).
\end{gather*}
The \emph{relaxed messages} are $\widetilde M_v=\widetilde{\mathrm{Init}}(\nu_v;c_{v,2},c_{v,3})\mathbin{\tilde\circ}\widetilde M_{w_1}\mathbin{\tilde\circ}\cdots\mathbin{\tilde\circ}\widetilde M_{w_d}$, for any orders of the children, merged from left to right as in \cref{def:rsc-messages}; no forcing is used.
In the merge, the closure probability $p_3$ enters only through the terms $2\lambda c_2$, with $\lambda=p_3/p_2$, and $q_3o_2$ of
$\tilde\chi_{\mathsf S}$: a child with at least two sampled children and a sampled parent is closed with probability $p_3$,
and everywhere else the probability is $p_2$.

\begin{lemma}[The relaxed recursion]\label{lem:rsc-relaxed}
  Under the standing assumptions of \cref{sec:rsc-light}, for every tree $F$ and every choice of the root $r$,
  $\widetilde W(F,\nu,c)\le W^{\mathrm{tm}}(F,\nu,c)$ and $\widetilde W(F,\nu,c)=\widetilde{\mathrm{fin}}(\widetilde M_r)$.
\end{lemma}

\begin{proof}
  For the inequality, fix $X'\subseteq S$ and put $g(\Upsilon)=y^{|X'|}2^{-\Psi_F(X'\cup\Upsilon)}\prod_{v\in S}\hat h_v(X'\cup\Upsilon)$ for
  $\Upsilon\subseteq C_F(X')$, so that $\sum_\Upsilon\pi_{X'}(\Upsilon)g(\Upsilon)$ and $\sum_\Upsilon\tilde\pi_{X'}(\Upsilon)g(\Upsilon)$ are the
  contributions of $X'$ to $W^{\mathrm{tm}}$ and to $\widetilde W$. The function $g$ is nondecreasing under inclusion: let
  $v\in C_F(X')\setminus\Upsilon$ and $X=X'\cup\Upsilon$. Adding $v$ to $X$ does not increase the number of vertices of $S$
  with $N_F[\cdot]\cap X=\emptyset$, and it decreases $\comp(F[X])$, because the at least two $F$-neighbors of $v$ in
  $X'\subseteq X$ lie in distinct components of $F[X]$ (a path between two of them in $F[X]$ together with $v$ would be a
  cycle of $F$); so $2^{-\Psi_F}$ does not decrease. The factor $\hat h_v$ changes from $2^{-\nu_v}\varpi_2^{c_{v,2}}\varpi_3^{c_{v,3}}$
  to $q_2^{c_{v,2}}q_3^{c_{v,3}}$, which is not smaller by~\eqref{eq:rsc-credit-small}, and no other factor changes. Since
  $\pi_{X'}$ and $\tilde\pi_{X'}$ are product distributions on the subsets of $C_F(X')$ with $\tilde p(v)\le p_{k_F(v)}$ for every
  $v$, the monotone coupling of \cref{sec:rsc-tree} gives $\sum_\Upsilon\tilde\pi_{X'}(\Upsilon)g(\Upsilon)\le\sum_\Upsilon\pi_{X'}(\Upsilon)g(\Upsilon)$;
  summing over $X'$ gives $\widetilde W\le W^{\mathrm{tm}}$.

  For the identity, the proof of \cref{lem:rsc-recursion} applies with $\tilde p(v)$ in place of $p_{k_F(v)}$ and with
  the empty forcing: the factor of a vertex $w$ in $\Upsilon$ or in $C_F(X')\setminus\Upsilon$ is now $\tilde p(w)$ or
  $1-\tilde p(w)$, which is $p_3$, respectively $q_3$, exactly when $w$ has a sampled parent and at least two sampled
  children, and $p_2$, respectively $q_2$, otherwise; so a vertex has to count its sampled children only up to two, the
  states $\mathsf C_2,\mathsf C_3$ and $\mathsf O_2,\mathsf O_3$ can be merged, and in the case analysis of the merge the
  completing factors become those of $\tilde\chi_{\mathsf S}$, $\tilde\chi^{\mathrm u}_{\mathsf C}$ and $\tilde\chi^{\mathrm u}_{\mathsf O}$,
  $\tilde\chi^{\mathrm x}_{\mathsf O}$ above (a child in $\mathsf C_2$ or $\mathsf O_2$ of a sampled parent gets $\lambda$, respectively $q_3$;
  of an unsampled parent it gets $1$, respectively $q_2$). The root has no parent, so $\tilde p(r)=p_2$, which gives $\widetilde{\mathrm{fin}}$.
\end{proof}

In the rest of this subsection all messages are eight-dimensional, and we write $\circ$, $\mathrm{fin}$ and
$\mathrm{Init}$ for $\tilde\circ$, $\widetilde{\mathrm{fin}}$ and $\widetilde{\mathrm{Init}}$.
The region of the certificate must be closed upwards in an order of the states that the merge respects. The
order is the reflexive and transitive closure of
\[
  \mathsf C_2\succ\mathsf C_1\succ\mathsf C_0,\qquad \mathsf C_2\succ\mathsf O_{01}\succ\mathsf O_{00},
  \qquad \mathsf O_{01}\succ\mathsf O_1\succ\mathsf O_2 ,
\]
with $\mathsf A$ incomparable to the other states. Informally, a higher state is never more
restrictive: for instance, a vertex in $\mathsf C_1$ is valid in every situation in which one in
$\mathsf C_0$ is, and a vertex in
$\mathsf O_{01}$ is dominated and has no sampled child, so it imposes nothing on its parent that a
vertex in $\mathsf O_{00}$, $\mathsf O_1$ or $\mathsf O_2$ does not. A set $U$ of states is an
\emph{up-set} if $\mathsf x\in U$ and $\mathsf x'\succ\mathsf x$ imply $\mathsf x'\in U$. Let
\[
  \mathcal K_{\mathrm{pos}}=\Bigl\{P\in\R^8:\ \sum_{\mathsf x\in U}P_{\mathsf x}\ge0\ \text{for every up-set }U\Bigr\},
\]
and write $P\succeq P'$ if $P-P'\in\mathcal K_{\mathrm{pos}}$. Let $e_{\mathsf x}$ denote the unit vector of the
state $\mathsf x$. The next lemma collects the properties of $\mathcal K_{\mathrm{pos}}$ that
the certificate uses: a merge with a nonnegative message, on either side, maps $\mathcal K_{\mathrm{pos}}$ into
itself, and $\mathrm{fin}$ is nonnegative on $\mathcal K_{\mathrm{pos}}$. As for the cumulative cone of
\cref{lem:rst-closure}, both properties are checked on a finite set of generators.

\begin{lemma}[The poset cone]\label{lem:rsc-cone}
  Let $0<p_2\le p_3<1$. The cone $\mathcal K_{\mathrm{pos}}$ defined above, together with the eight-state merge $\circ$ and the functional
  $\mathrm{fin}$, has the following properties.
  \begin{enumerate}[label=(\alph*)]
    \item $\mathcal K_{\mathrm{pos}}$ is a closed convex cone containing $\R^8_{\ge0}$, and it is the convex cone
      generated by the eight unit vectors $e_{\mathsf x}$ and the six vectors $e_{\mathsf x'}-e_{\mathsf x}$
      for the covering pairs $\mathsf x'\succ\mathsf x$ displayed above.
    \item If $Q\in\R^8_{\ge0}$ and $\Delta\in\mathcal K_{\mathrm{pos}}$, then $\Delta\circ Q\in\mathcal K_{\mathrm{pos}}$ and $Q\circ\Delta\in\mathcal K_{\mathrm{pos}}$.
    \item $\mathrm{fin}(\Delta)\ge0$ for every $\Delta\in\mathcal K_{\mathrm{pos}}$.
  \end{enumerate}
\end{lemma}

\begin{proof}
  (a) $\mathcal K_{\mathrm{pos}}$ is an intersection of closed half-spaces through $0$, hence a closed convex
  cone, and it contains the orthant. Each of the fourteen vectors lies in $\mathcal K_{\mathrm{pos}}$: the up-set
  sums of $e_{\mathsf x}$ are $0$ or $1$, and so are those of $e_{\mathsf x'}-e_{\mathsf x}$ with
  $\mathsf x'\succ\mathsf x$, because an up-set containing $\mathsf x$ contains $\mathsf x'$. So the cone they
  generate is contained in $\mathcal K_{\mathrm{pos}}$.

  For the converse, note that the seven states other than $\mathsf A$ form a rooted tree with root
  $\mathsf C_2$, in which $\mathsf C_1$ and $\mathsf O_{01}$ are the children of $\mathsf C_2$, $\mathsf C_0$ is
  the child of $\mathsf C_1$, $\mathsf O_{00}$ and $\mathsf O_1$ are the children of $\mathsf O_{01}$, and
  $\mathsf O_2$ is the child of $\mathsf O_1$; and $\mathsf x\prec\mathsf x'$ holds exactly when $\mathsf x'$ is
  a proper ancestor of $\mathsf x$. The up-sets are the sets closed under taking parents, with $\mathsf A$
  added or not. Let $P\in\mathcal K_{\mathrm{pos}}$. The up-set $\{\mathsf A\}$ gives $P_{\mathsf A}\ge0$, and
  $P-P_{\mathsf A}e_{\mathsf A}$ again lies in $\mathcal K_{\mathrm{pos}}$, so we may assume $P_{\mathsf A}=0$. We
  show by induction on $|T|$ that a vector $P$ supported on a subtree $T$ containing the root, whose
  sums over the subsets of $T$ closed under taking parents are nonnegative, is a nonnegative
  combination of the vectors $e_{\mathsf x}$ with $\mathsf x\in T$ and $e_{\mathsf y}-e_{\mathsf x}$ with
  $\mathsf x\in T$ and parent $\mathsf y$. If $T=\{\mathsf C_2\}$, then $P=P_{\mathsf C_2}e_{\mathsf C_2}$ with
  $P_{\mathsf C_2}\ge0$. Otherwise let $\mathsf x$ be a leaf of $T$ with parent $\mathsf y$, and put
  $P'=P-P_{\mathsf x}e_{\mathsf x}$ if $P_{\mathsf x}\ge0$ and $P'=P-(-P_{\mathsf x})(e_{\mathsf y}-e_{\mathsf x})$
  otherwise. In both cases $P'$ is supported on $T-\mathsf x$, and $P-P'$ is a nonnegative multiple of a
  generator. Let $U'\subseteq T-\mathsf x$ be closed under taking parents; it is also closed under taking
  parents in $T$, as $\mathsf x$ is a leaf. The sum of $P'$ over $U'$ equals the sum of $P$ over $U'$,
  except in the second case with $\mathsf y\in U'$, where it equals the sum of $P$ over
  $U'\cup\{\mathsf x\}$, which is again closed under taking parents. So all these sums are nonnegative,
  and the induction hypothesis applies to $P'$ and $T-\mathsf x$.

  (b) The operation $\circ$ is bilinear and $\mathcal K_{\mathrm{pos}}$ is a convex cone, so by (a) it suffices
  to check the fourteen generators $\Delta$. For a unit vector $\Delta=e_{\mathsf x}$ both products are
  nonnegative, since $\circ$ maps nonnegative messages to nonnegative messages. For the six generators
  $e_{\mathsf x'}-e_{\mathsf x}$ we read both products off the merge formula, with
  $Q=(a,c_0,c_1,c_2,o_{00},o_{01},o_1,o_2)$ and with the functionals
  $\tilde\chi_{\mathsf S},\tilde\chi^{\mathrm s}_{\mathsf C},\dots$ evaluated at $Q$, where they are nonnegative (we drop the tildes):
  \begin{align*}
    (e_{\mathsf C_2}-e_{\mathsf C_1})\circ Q&=\chi^{\mathrm u}_{\mathsf C}\,(e_{\mathsf C_2}-e_{\mathsf C_1}),\\
    (e_{\mathsf C_1}-e_{\mathsf C_0})\circ Q&=\chi^{\mathrm u}_{\mathsf C}\,(e_{\mathsf C_1}-e_{\mathsf C_0})+\chi^{\mathrm s}_{\mathsf C}\,(e_{\mathsf C_2}-e_{\mathsf C_1}),\\
    (e_{\mathsf C_2}-e_{\mathsf O_{01}})\circ Q&=(\chi^{\mathrm u}_{\mathsf O}+\chi^{\mathrm x}_{\mathsf O})(e_{\mathsf C_2}-e_{\mathsf O_{01}})
    +\chi^{\mathrm s}_{\mathsf O}\bigl[(e_{\mathsf C_2}-e_{\mathsf O_{01}})+(e_{\mathsf O_{01}}-e_{\mathsf O_1})\bigr]\\
    &\qquad+\bigl(a+c_2+\tfrac12o_{00}\bigr)e_{\mathsf C_2},\\
    (e_{\mathsf O_{01}}-e_{\mathsf O_{00}})\circ Q&=\chi^{\mathrm u}_{\mathsf O}\,(e_{\mathsf O_{01}}-e_{\mathsf O_{00}}),\\
    (e_{\mathsf O_{01}}-e_{\mathsf O_1})\circ Q&=(\chi^{\mathrm u}_{\mathsf O}+\chi^{\mathrm x}_{\mathsf O})(e_{\mathsf O_{01}}-e_{\mathsf O_1})+\chi^{\mathrm s}_{\mathsf O}\,(e_{\mathsf O_1}-e_{\mathsf O_2}),\\
    (e_{\mathsf O_1}-e_{\mathsf O_2})\circ Q&=(\chi^{\mathrm u}_{\mathsf O}+\chi^{\mathrm x}_{\mathsf O})(e_{\mathsf O_1}-e_{\mathsf O_2}),
  \end{align*}
  and
  \begin{align*}
    Q\circ(e_{\mathsf C_2}-e_{\mathsf C_1})&=(2(\lambda-1)a,\,2c_0,\,2c_1,\,2c_2,\,0,\,o_{00}+o_{01},\,o_1,\,o_2),\\
    Q\circ(e_{\mathsf C_1}-e_{\mathsf C_0})&=2a\,e_{\mathsf A},\\
    Q\circ(e_{\mathsf C_2}-e_{\mathsf O_{01}})&=((2\lambda-1)a,\,c_0,\,c_1,\,c_2,\,0,\,0,\,0,\,0)+o_{00}\,(e_{\mathsf O_{01}}-e_{\mathsf O_{00}}),\\
    Q\circ(e_{\mathsf O_{01}}-e_{\mathsf O_{00}})&=\tfrac12\,(0,\,0,\,0,\,0,\,o_{00},\,o_{01},\,o_1,\,o_2),\\
    Q\circ(e_{\mathsf O_{01}}-e_{\mathsf O_1})&=p_2\,a\,e_{\mathsf A},\\
    Q\circ(e_{\mathsf O_1}-e_{\mathsf O_2})&=\bigl((p_3-p_2)a,\,p_2c_0,\,p_2c_1,\,p_2c_2,\,p_2o_{00},\,p_2o_{01},\,p_2o_1,\,p_2o_2\bigr).
  \end{align*}
  In the third line of the first list, the coefficient of $e_{\mathsf C_2}$ in
  $(e_{\mathsf C_2}-e_{\mathsf O_{01}})\circ Q$ is $\chi^{\mathrm u}_{\mathsf C}+\chi^{\mathrm s}_{\mathsf C}$, and
  $\chi^{\mathrm u}_{\mathsf C}+\chi^{\mathrm s}_{\mathsf C}-\chi_{\mathsf O}=a+c_2+\frac12o_{00}$. In the second list, the entries involving
  $\lambda$ and $p_3$ come from the terms $2\lambda c_2$ and $q_3o_2$ of $\tilde\chi_{\mathsf S}$; they are nonnegative because $\lambda\ge1$ and $p_3\ge p_2$. So every
  product is a nonnegative combination of generators, and it lies in $\mathcal K_{\mathrm{pos}}$.

  (c) $\mathrm{fin}$ is linear with the coefficients $(1,0,0,1,\frac12,1,1,q_2)$, so by (a) it suffices that
  it is nonnegative on the generators: its values on the unit vectors are these coefficients, and its
  values on the six vectors $e_{\mathsf x'}-e_{\mathsf x}$ are $1$, $0$, $0$, $\frac12$, $0$ and $p_2$, in
  the order of the covering pairs displayed above.
\end{proof}

For a finite nonempty set $Z\subseteq\R^8_{\ge0}$ let $\mathcal P_Z=\mathrm{conv}(Z)+\mathcal K_{\mathrm{pos}}$ and
$c_Z=\min_{z\in Z}\mathrm{fin}(z)$. For
$\mathcal Y\subseteq\R^8$ and $P\in\R^8$ we write $P\in\mathcal Y+\mathcal K_{\mathrm{pos}}$ also as ``$P$ dominates a point of $\mathcal Y$.''
Write $\mathrm{Init}(\nu)=\mathrm{Init}(\nu;0,0)$ and $\mathrm{Init}(\infty)=(\frac y2,\frac{p_2}2,0,0,0,0,0,0)$, the limit of $\mathrm{Init}(\nu)$ as
$\nu\to\infty$, and let
\[
  L(\nu)=\bigl\{(c_2,c_3)\in\mathbb Z_{\ge0}^2:\ 1\le c_2+c_3\le\nu\bigr\}
\]
be the \emph{exposure lattice}. The initial message is linear in the pair
$(q_2^{c_2}q_3^{c_3},\ \varpi_2^{c_2}\varpi_3^{c_3})$:
\begin{equation}\label{eq:rsc-init-linear}
  \mathrm{Init}(\nu;c_2,c_3)=q_2^{c_2}q_3^{c_3}\,\mathrm{Init}(\infty)+\varpi_2^{c_2}\varpi_3^{c_3}\,2^{-\nu}e_{\mathsf O_{00}} .
\end{equation}

The next lemma states the conditions that the certificate verifies. The first two conditions are the
analogues of the two conditions of the antinorm certificate for RS-$\theta$ (\cref{lem:rst-F}), with
one scale $\hat e$ for solution vertices and owned vertices in place of $e_1,e_2$, so that the bound
$c_Z\hat e^n$ does not depend on the solution fraction $s/n$ and no case distinction on $s/n$ is
needed; the remaining four control the initial messages of vertices with exposures, which the proof places in a scaled copy of
$\mathcal P_Z$ either directly or after their first few merges.

\begin{lemma}[Certificate conditions]\label{lem:rsc-cert}
  Let $0<p_2\le p_3<\frac12$, $y>0$, $1\le\varpi_\tau\le2q_\tau$ and $0<\hat e<q_3$. Let $Z\subseteq\R^8_{\ge0}$ be finite and nonempty,
  let $\nu_0\ge0$ and $\nu_E\ge1$ be integers, and let $d^{(3)}_{\min},d^{(2)}_{\min}:\{1,\dots,\nu_E\}\to\mathbb Z_{\ge1}$ with
  $d^{(3)}_{\min}\le d^{(2)}_{\min}$; put $j_\tau(\nu)=\max(1,d^{(\tau)}_{\min}(\nu)-1)$.
  Suppose that:
  \begin{enumerate}[label=(C\arabic*)]
    \item\label{it:rsc-C1} $z\circ z'\in\mathcal P_Z$ for all $z,z'\in Z$;
    \item\label{it:rsc-C2} $\mathrm{Init}(\nu)\in \hat e^{1+\nu}\mathcal P_Z$ for $0\le\nu\le\nu_0$, and
      $\mathrm{Init}(\infty)\in \hat e^{\nu_0+2}\mathcal P_Z$;
    \item\label{it:rsc-C3} for $1\le\nu\le\nu_E$ and $\tau\in\{2,3\}$, every pair $(\nu,d)$ with $1\le d<d^{(\tau)}_{\min}(\nu)$ is $\tau$-light;
    \item\label{it:rsc-C4} for $2\le\nu\le\nu_E$: if $d^{(3)}_{\min}(\nu)\le1$, then
      $\mathrm{Init}(\nu;0,c_3)\in \hat e^{1+\nu}\mathcal P_Z$ for $1\le c_3\le\nu-1$, and if moreover $d^{(2)}_{\min}(\nu)\le1$, then
      $\mathrm{Init}(\nu;c_2,c_3)\in \hat e^{1+\nu}\mathcal P_Z$ for all $(c_2,c_3)\in L(\nu-1)$ with $c_2\ge1$;
    \item\label{it:rsc-C5} for $1\le\nu\le\nu_E$ and $(c_2,c_3)\in L(\nu)$, with $j=j_3(\nu)$ if $c_2=0$ and $j=j_2(\nu)$ if $c_2\ge1$:
      $\hat e^{-1-\nu}\mathrm{Init}(\nu;c_2,c_3)\circ Q_1\circ\cdots\circ Q_j\in\mathcal P_Z$ (merged from left to right) for all $Q_1,\dots,Q_j\in\mathcal P_Z\cap\R^8_{\ge0}$;
    \item\label{it:rsc-C6} $(q_3/\hat e)^{\nu_E+1}\ge \hat e^{-(\nu_0+1)}$.
  \end{enumerate}
  Then every admissible quadruple $(F,\nu,c_2,c_3)$ in which $F$ has $s$ vertices satisfies, with
  $\nu(F)=\sum_{v\in V(F)}\nu_v$,
  \[
    W^{\mathrm{tm}}(F,\nu,c)\ \ge\ \widetilde W(F,\nu,c)\ \ge\ c_Z\,\hat e^{\,s+\nu(F)} .
  \]
\end{lemma}

\begin{proof}
  The first inequality is \cref{lem:rsc-relaxed}, for any root. Step 1 shows that membership in scaled copies of $\mathcal P_Z$ is supermultiplicative under the
  merge. Step 2 places the initial message $\mathrm{Init}(\nu_v;c_{v,2},c_{v,3})$ of most vertices $v$ in
  $\hat e^{1+\nu_v}\mathcal P_Z$, and Step 3 treats the remaining vertices, whose message reaches the
  right scaled copy only after its first few merges. Step 4 propagates this up the tree and reads off
  the bound at the root.

  \emph{Step 1: supermultiplicativity.} The merge $\circ$, the cone $\mathcal K_{\mathrm{pos}}$ and the set
  $Z$ satisfy the hypotheses of \cref{lem:rst-abstract}: $\circ$ is bilinear; by \cref{lem:rsc-cone}(a),
  $\mathcal K_{\mathrm{pos}}$ is a convex cone containing $\R^8_{\ge0}$; by \cref{lem:rsc-cone}(b),
  $\Delta\circ Q$ and $Q\circ\Delta$ lie in $\mathcal K_{\mathrm{pos}}$ for $\Delta\in\mathcal K_{\mathrm{pos}}$ and
  $Q\in\R^8_{\ge0}$; and $z\circ z'\in\mathcal P_Z$ for $z,z'\in Z$ by \ref{it:rsc-C1}. Hence
  \cref{lem:rst-abstract} gives $\mathcal P_Z+\mathcal K_{\mathrm{pos}}=\mathcal P_Z$,
  $\eta\mathcal P_Z\subseteq\eta'\mathcal P_Z$ for $0<\eta'\le\eta$, and (its part (b) with $Y=Y'=Z$)
  $P\circ Q\in\eta\eta'\mathcal P_Z$ whenever
  $P\in\eta\mathcal P_Z$, $Q\in\eta'\mathcal P_Z$ and $Q\ge0$. In particular, if $P\ge0$ and $P\in\eta\mathcal P_Z$,
  then $\eta''P\in\eta''\eta\,\mathcal P_Z\subseteq\eta\mathcal P_Z$ for every $\eta''\ge1$.

  \emph{Step 2: initial messages.} Let $(F,\nu,c_2,c_3)$ be admissible and $v\in V(F)$, with $\nu=\nu_v$,
  $(c_2,c_3)=(c_{v,2},c_{v,3})$, $c=c_2+c_3$ and $d=d_F(v)$. We claim that $\mathrm{Init}(\nu;c_2,c_3)\in \hat e^{1+\nu}\mathcal P_Z$, unless
  $c\ge1$, $\nu\le\nu_E$ and $d\ge2$; that remaining case is handled in Step 3.
  \begin{itemize}
    \item \emph{Case $c=0$ and $\nu\le\nu_0$}: this is \ref{it:rsc-C2}.
    \item \emph{Case $c=0$ and $\nu>\nu_0$}: $\mathrm{Init}(\nu)-\mathrm{Init}(\infty)=2^{-\nu}e_{\mathsf O_{00}}\in\mathcal K_{\mathrm{pos}}$,
      so $\mathrm{Init}(\nu)\in \hat e^{\nu_0+2}\mathcal P_Z+\mathcal K_{\mathrm{pos}}=\hat e^{\nu_0+2}\mathcal P_Z\subseteq
      \hat e^{1+\nu}\mathcal P_Z$ by Step 1, as $\hat e^{1+\nu}\le \hat e^{\nu_0+2}$ (since $\hat e<1$).
    \item \emph{Case $c\ge1$ and $\nu>\nu_E$}: by~\eqref{eq:rsc-init-linear},
      $\mathrm{Init}(\nu;c_2,c_3)-q_3^\nu\mathrm{Init}(\infty)=(q_2^{c_2}q_3^{c_3}-q_3^\nu)\mathrm{Init}(\infty)+\varpi_2^{c_2}\varpi_3^{c_3}2^{-\nu}e_{\mathsf O_{00}}\ge0$,
      because $q_2^{c_2}q_3^{c_3}\ge q_3^{c}\ge q_3^\nu$; so $\mathrm{Init}(\nu;c_2,c_3)\succeq q_3^\nu\mathrm{Init}(\infty)$. Moreover
      $\hat e^{-1-\nu}q_3^\nu=\hat e^{-1}(q_3/\hat e)^\nu\ge \hat e^{-1}(q_3/\hat e)^{\nu_E+1}\ge \hat e^{-\nu_0-2}$ by \ref{it:rsc-C6}, as
      $q_3/\hat e>1$. So $\hat e^{-1-\nu}\mathrm{Init}(\nu;c_2,c_3)$ dominates a multiple $\eta''\,\hat e^{-\nu_0-2}\mathrm{Init}(\infty)$ with
      $\eta''\ge1$, which lies in $\mathcal P_Z$ by \ref{it:rsc-C2} and Step 1.
    \item \emph{Case $c\ge1$, $\nu\le\nu_E$ and $d\le1$}: admissibility excludes $d=0$ and gives $c\le\nu-1$
      for $d=1$, so $\nu\ge2$; it also says that $(\nu,1)$ is $\tau$-failing for every $\tau$ with $c_\tau\ge1$, so by \ref{it:rsc-C3}
      $d^{(\tau)}_{\min}(\nu)\le1$ for these $\tau$. If $c_2=0$, then $c_3\ge1$, $d^{(3)}_{\min}(\nu)\le1$ and the first part of \ref{it:rsc-C4} applies.
      If $c_2\ge1$, then $d^{(2)}_{\min}(\nu)\le1$, hence also $d^{(3)}_{\min}(\nu)\le1$, and $(c_2,c_3)\in L(\nu-1)$, so the second part of \ref{it:rsc-C4} applies.
  \end{itemize}

  \emph{Step 3: failing internal vertices with few owned vertices.} Let $c\ge1$, $\nu\le\nu_E$ and $d\ge2$. If $c_2\ge1$,
  then $(\nu,d)$ is $2$-failing by admissibility, so $d\ge d^{(2)}_{\min}(\nu)$ by \ref{it:rsc-C3} and
  $d-1\ge\max(1,d^{(2)}_{\min}(\nu)-1)=j_2(\nu)$; if $c_2=0$, then $(\nu,d)$ is $3$-failing, and likewise $d-1\ge j_3(\nu)$. Let $j$ be
  $j_2(\nu)$ or $j_3(\nu)$ accordingly. With $F$
  rooted as in Step 4, $v$ has at least $d-1\ge j$ children $w_1,\dots,w_j,\dots$, in the
  order of merging. Suppose that $\widetilde M_{w_i}\in\eta_i\mathcal P_Z$ with $\eta_i>0$ for $i\le j$ (this
  will be the induction hypothesis of Step 4). The messages $\widetilde M_{w_i}/\eta_i$ are nonnegative and lie in $\mathcal P_Z$, so
  \ref{it:rsc-C5} gives, after $j$ merges and by bilinearity,
  \[
    \mathrm{Init}(\nu;c_2,c_3)\circ\widetilde M_{w_1}\circ\cdots\circ\widetilde M_{w_j}\ \in\ \hat e^{1+\nu}\eta_1\cdots\eta_j\,\mathcal P_Z .
  \]

  \emph{Step 4: induction over the tree.} Root $F$ at any vertex $r$, and let $\eta(F_v)=\hat e^{|V(F_v)|+\nu(F_v)}$
  with $\nu(F_v)=\sum_{x\in V(F_v)}\nu_x$. We show $\widetilde M_v\in\eta(F_v)\mathcal P_Z$ by induction from the
  leaves. If Step 2 applies to $v$, start from $\mathrm{Init}(\nu_v;c_{v,2},c_{v,3})\in \hat e^{1+\nu_v}\mathcal P_Z$ and
  merge the children one by one; each merge with a child $w$ multiplies the scale by $\eta(F_w)$, by
  Step 1, as $\widetilde M_w\ge0$. Otherwise Step 3 handles the first $j$ children and Step 1 the remaining ones.
  At the root, $\widetilde M_r=\hat e^{s+\nu(F)}(z+\Delta)$ with $z\in\mathrm{conv}(Z)$ and $\Delta\in\mathcal K_{\mathrm{pos}}$, so by
  \cref{lem:rsc-relaxed} and \cref{lem:rsc-cone}(c)
  \[
    \widetilde W(F,\nu,c)=\mathrm{fin}(\widetilde M_r)=\hat e^{s+\nu(F)}\bigl(\mathrm{fin}(z)+\mathrm{fin}(\Delta)\bigr)\ \ge\ \hat e^{s+\nu(F)}c_Z ,
  \]
  since $\mathrm{fin}$ is linear and $\mathrm{fin}(z)\ge c_Z$ for $z\in\mathrm{conv}(Z)$.
\end{proof}

Condition \ref{it:rsc-C5} is established by the certificate in two steps. First, for a point $P\ge0$ and finite
sets $Y_1,\dots,Y_{j-1}\subseteq\R^8_{\ge0}$ (the \emph{stages}), if every product $x\circ z$ with $x\in Y_{i-1}$ and $z\in Z$ lies in
$\mathrm{conv}(Y_i)+\mathcal K_{\mathrm{pos}}$ for $i=1,\dots,j$, where $Y_0=\{P\}$ and $Y_j=Z$, then
$P\circ Q_1\circ\cdots\circ Q_j\in\mathcal P_Z$ for all $Q_i\in\mathcal P_Z\cap\R^8_{\ge0}$: by induction on $i$,
\cref{lem:rst-abstract}(b) applied to the sets $(Y_{i-1},Z,Y_i)$ in place of $(Y,Z,Y')$, with $\eta=\eta'=1$, gives
$P\circ Q_1\circ\cdots\circ Q_i\in\mathrm{conv}(Y_i)+\mathcal K_{\mathrm{pos}}$ (the closure hypothesis of that lemma was checked in Step 1 above).
Second, the property in \ref{it:rsc-C5} is preserved under convex combinations of $P$, because $P\mapsto P\circ Q_1\circ\cdots\circ Q_j$
is linear and $\mathcal P_Z$ is convex; it is preserved when $j$ is increased, by Step 1; and by~\eqref{eq:rsc-init-linear},
$\mathrm{Init}(\nu;c_2,c_3)$ for $(c_2,c_3)\in L(\nu)$ is a convex combination of the points $\mathrm{Init}(\nu;c_2',c_3')$ over the
vertices $(c'_2,c'_3)$ of the convex hull of $\bigl\{(q_2^{c_2}q_3^{c_3},\varpi_2^{c_2}\varpi_3^{c_3}):0\le c_2+c_3\le\nu\bigr\}$, with the same
coefficients that express the pair of $(c_2,c_3)$ through the hull vertices; the vertex $(0,0)$, if used, contributes $\mathrm{Init}(\nu)$,
which lies in $\hat e^{1+\nu}\mathcal P_Z$ by the first two cases of Step 2. Since $j_3(\nu)\le j_2(\nu)$, it therefore suffices to
provide stages for $\hat e^{-1-\nu}\mathrm{Init}(\nu;0,c_3)$ with $j=j_3(\nu)$ for $1\le c_3\le\nu$ and for
$\hat e^{-1-\nu}\mathrm{Init}(\nu;c_2,c_3)$ with $j=j_2(\nu)$ for the hull vertices with $c_2\ge1$. The same convexity argument reduces
the second part of \ref{it:rsc-C4} to the hull vertices of the lattice for $\nu-1$.

The certificate verifies the conditions of \cref{lem:rsc-cert} and the hypotheses of \cref{cor:rsc-reduction} for the parameters of RSC. Condition
\ref{it:rsc-C3} does not involve the certificate: it follows from \cref{lem:rsc-lightpairs}, which can be checked by hand
and is also checked in Lean (\cref{sec:anc-lean}).

\begin{lemma}[Light pairs for RSC]\label{lem:rsc-lightpairs}
  For $p_2=0.07$, $p_3=0.077$ and $y=31/100$ the pairs $(1,1)$, $(1,2)$, $(2,1)$ and $(3,1)$ are $3$-light, and the
  pairs $(1,3)$ and $(2,2)$ are $2$-light.
\end{lemma}

\begin{proof}
  Here $q_2=\frac{93}{100}$, $q_3=\frac{923}{1000}$, $\max(p_2,p_3-p_2)=p_2$ and $\mathsf b=2+\frac{14}{31}=\frac{76}{31}$. Since
  $-\ln(1-p)=\sum_{k\ge1}p^k/k\le p+\frac{p^2}2+\frac{p^3}{3(1-p)}$ for $0<p<1$, we have
  $\varepsilon_3\le\frac{443756467}{5538000000}<0.08013$ and $\varepsilon_2\le\frac{404957}{5580000}<0.07258$. Since
  $\ln\frac{1+z}{1-z}=2\sum_{k\ge0}\frac{z^{2k+1}}{2k+1}\ge2z+\frac23z^3$ for $0\le z<1$, with $z=\frac{423}{1423}$ we have
  $\delta_3=\ln\frac{1846}{1000}>0.61202$, and with $z=\frac13$ and four terms, $\delta_2=\ln2>0.69313$. Hence, with $1+\frac{p_3}y=\frac{387}{310}$,
  \[
    \Bigl(1+\frac{p_3}y\Bigr)\frac{\varepsilon_3\,y}{\delta_3}<\frac{387}{310}\cdot\frac{0.08013\cdot0.31}{0.61202}<0.0507,\qquad
    \frac{\varepsilon_2\,y}{\delta_2}<\frac{0.07258\cdot0.31}{0.69313}<0.0325 .
  \]
  The left-hand side $2^{-\nu}\mathsf b^{-d}=2^{-\nu}(\frac{31}{76})^d$ is decreasing in $\nu$ and
  $d$, and its values are $\frac{961}{11552}>0.0831$ at $(1,2)$, $\frac{31}{304}>0.1019$ at $(2,1)$, $\frac{31}{608}>0.0509$ at
  $(3,1)$, which gives the $3$-light pairs, and $\frac{29791}{877952}>0.0339$ at $(1,3)$ and $\frac{961}{23104}>0.0415$ at $(2,2)$, which gives the $2$-light pairs.
\end{proof}

\begin{lemma}[Credits for RSC]\label{lem:rsc-rsccredits}
  For $p_2=0.07$, $p_3=0.077$, $y=31/100$, $\varpi_2=\frac{272657}{250000}=1.090628$ and $\varpi_3=\frac{288677}{250000}=1.154708$, we have $1\le\varpi_\tau\le2q_\tau$ and~\eqref{eq:rsc-credits} holds.
\end{lemma}

\begin{proof}
  $2q_2=1.86$ and $2q_3=1.846$. Here $\vartheta=\frac{31}{200}\cdot\frac{923}{1000}$ and $\rho_1=\frac{200000}{228613}$, so
  $1-\rho_1=\frac{28613}{228613}$ and $1-\rho_1^2=\frac{12263903769}{52263903769}$. The two inequalities of~\eqref{eq:rsc-credits} read
  $\ln\varpi_2\le\frac{28613}{228613}\ln2$ and $\ln\varpi_3\le\frac{12263903769}{52263903769}(\ln2+\ln q_3)$; numerically the two sides are
  $0.086753678$ versus $0.086753685$ and $0.143847499$ versus $0.143847506$ (left-hand sides rounded up,
  right-hand sides rounded down), so the credits were chosen as large as the
  inequalities allow, to six decimals. They follow from the truncated-series bounds
  $\ln\varpi_\tau\le\sum_{k=1}^{15}(-1)^{k+1}(\varpi_\tau-1)^k/k$, an alternating series with decreasing terms cut after a positive term;
  $\ln2\ge2\sum_{k=0}^{6}3^{-2k-1}/(2k+1)$; and $-\ln q_3\le\sum_{k=1}^{6}p_3^k/k+p_3^7/(7q_3)$: with these bounds,
  evaluated exactly in rational arithmetic (\cref{sec:anc-rsc}), the right-hand sides
  exceed the left-hand sides by more than $6\cdot10^{-9}$ and $5\cdot10^{-9}$.
\end{proof}

\paragraph{What the certificate proves.} The certificate is a finite set $Z$ of $1024$ points of $\R^8$, and
$\mathcal P_Z=\mathrm{conv}(Z)+\mathcal K_{\mathrm{pos}}$ is the region that they span together with the poset cone. Each
of \ref{it:rsc-C1}, \ref{it:rsc-C2}, \ref{it:rsc-C4} and \ref{it:rsc-C5} is a finite list of statements of the
form ``this rational vector lies in this region''. Together they say two things: the message of a single vertex
lies in a scaled copy of $\mathcal P_Z$ (\ref{it:rsc-C2}, \ref{it:rsc-C4} and \ref{it:rsc-C5}, the last after a few
merges), and the merge of two messages in scaled copies of $\mathcal P_Z$ lies in the product scaled copy
(\ref{it:rsc-C1}, extended from the points of $Z$ to the whole region by bilinearity and Step~1 of the proof of
\cref{lem:rsc-cert}). By induction from the leaves, the message of every tree, of any size and shape, then lies in
$\hat e^{s+\nu(F)}\mathcal P_Z$, and since $\mathrm{fin}\ge c_Z$ on $\mathcal P_Z$, the generating function is at least
$c_Z\hat e^{s+\nu(F)}$. Finitely many exact checks thus prove a bound for arbitrarily large trees, in the same way
as an invariant region of a map bounds all of its iterates.

The next proposition records the parameters of the
certificate for the remaining conditions and the sizes of its point sets.

\begin{proposition}[The certificate]\label{prop:rsc-cert}
  Let $p_2=0.07$, $p_3=0.077$, $y=31/100$, $\varpi_2=272657/250000$, $\varpi_3=288677/250000$, $\hat e=145729958/189014575$, $\nu_0=24$, $\nu_E=36$,
  $d^{(3)}_{\min}(1)=3$, $d^{(3)}_{\min}(2)=d^{(3)}_{\min}(3)=2$, $d^{(2)}_{\min}(1)=4$, $d^{(2)}_{\min}(2)=3$, $d^{(2)}_{\min}(3)=2$ and
  $d^{(3)}_{\min}(\nu)=d^{(2)}_{\min}(\nu)=1$ for $4\le\nu\le36$. There is a set $Z$ of $1024$
  nonnegative rational points with $c_Z>0.6060$, and there are nonnegative rational stages for \ref{it:rsc-C5}, namely for $\nu=1$ a set $Y_1$ of $62$ points for $(c_2,c_3)=(0,1)$ and sets $Y_1$, $Y_2$ of $62$ and $304$ points for $(1,0)$,
  and for $\nu=2$ sets $Y_1$ of $61$, $62$ and $62$ points for $(1,0)$, $(1,1)$ and $(2,0)$, such that
  conditions \ref{it:rsc-C1}--\ref{it:rsc-C6} of \cref{lem:rsc-cert} hold.
\end{proposition}

For all other pairs $(\nu,(c_2,c_3))$ we have $j=1$ in \ref{it:rsc-C5}, and no intermediate stages are needed.

\begin{proof}
Condition \ref{it:rsc-C3} is \cref{lem:rsc-lightpairs}, together with the fact, noted after \cref{def:rsc-light}, that a $3$-light pair is $2$-light, and the
hypotheses on the credits are \cref{lem:rsc-rsccredits}. Condition \ref{it:rsc-C6} is an inequality between explicit rationals,
$(q_3/\hat e)^{37}\ge \hat e^{-25}$; numerically the two sides are about $779$ and $666$. The values of
$d^{(3)}_{\min}$ and $d^{(2)}_{\min}$ satisfy $d^{(3)}_{\min}\le d^{(2)}_{\min}$. The certificate file of \cref{sec:anc-rsc} stores the values stated above for $1\le\nu\le4$ and the value $0$ in place of $1$
for $5\le\nu\le36$, and its check uses these stored values; they give the same conditions: for $d^{(\tau)}_{\min}(\nu)\in\{0,1\}$ no integer $d$ satisfies
$1\le d<d^{(\tau)}_{\min}(\nu)$, so \ref{it:rsc-C3} requires nothing at $\nu$; the hypotheses $d^{(\tau)}_{\min}(\nu)\le1$ in
\ref{it:rsc-C4} hold; and $j_\tau(\nu)=\max(1,d^{(\tau)}_{\min}(\nu)-1)=1$ in \ref{it:rsc-C5}; these are the only places where the
hypotheses of \cref{lem:rsc-cert} use the values besides $d^{(3)}_{\min}\le d^{(2)}_{\min}$.

Conditions \ref{it:rsc-C1} and \ref{it:rsc-C2} are finite lists of relations of the form $P\in\mathrm{conv}(\mathcal Y)+\mathcal K_{\mathrm{pos}}$,
with $\mathcal Y=Z$. By the convexity argument after \cref{lem:rsc-cert}, \ref{it:rsc-C4} follows from \ref{it:rsc-C1}, \ref{it:rsc-C2},
the first part of (C4) and the second part at the hull vertices with $c_2\ge1$ of the lattice for $\nu-1$, which are again such
relations. By the argument on stages and the convexity argument after \cref{lem:rsc-cert}, \ref{it:rsc-C5} follows from
\ref{it:rsc-C1}, \ref{it:rsc-C2} and the relations $x\circ z\in\mathrm{conv}(Y_i)+\mathcal K_{\mathrm{pos}}$ for $x\in Y_{i-1}$,
$z\in Z$ and $1\le i\le j$, for the chains $Y_0=\{P\},Y_1,\dots,Y_j=Z$ that start at the scaled initial messages
$P$ listed there and use the stages of the proposition when $j\ge2$. So \ref{it:rsc-C1}, \ref{it:rsc-C2},
\ref{it:rsc-C4} and \ref{it:rsc-C5} follow from finitely many relations, each asserting that a given rational vector $P$ (a product $z\circ z'$, a scaled initial message,
or a staged product $x\circ z$) lies in $\mathrm{conv}(\mathcal Y)+\mathcal K_{\mathrm{pos}}$ for a given finite set $\mathcal Y\subseteq\R^8_{\ge0}$.
For each such $P$ the certificate gives nonnegative
rational coefficients $\alpha_i$ with $\sum_i\alpha_i\ge1$ and points $z_i\in\mathcal Y$ such that
$P\succeq\sum_i\alpha_iz_i$; this implies
$P\in\mathrm{conv}(\mathcal Y)+\mathcal K_{\mathrm{pos}}$, because $\sum_i\alpha_iz_i-\sum_i\frac{\alpha_i}{\sum_j\alpha_j}z_i$ is a nonnegative
vector. All these relations, the nonnegativity of the points of $Z$ and of the stages, and the value
$c_Z=0.60605\ldots$ have been verified in exact rational arithmetic (\cref{sec:anc-rsc}).
\end{proof}

\begin{proof}[Proof of \cref{prop:rsc-bound}]
  Let $S$ be a minimum connected dominating set and $s=|S|$. By \cref{lem:rsc-rsccredits} and \cref{cor:rsc-reduction}, the owner
  rule produces an admissible quadruple $(F,\nu,c_2,c_3)$ with $s+\sum_v\nu_v=n$ and
  $W^{\mathrm{cl}}_S(y)\ge W^{\mathrm{tm}}(F,\nu,c)$. By \cref{prop:rsc-cert} the hypotheses of
  \cref{lem:rsc-cert} hold (note $0<\hat e<q_3$, as $\hat e<0.771<0.923$), so
  $W^{\mathrm{cl}}_S(y)\ge c_Z\,\hat e^n\ge0.6060\,\hat e^n$, since $c_Z>0.6060$.
\end{proof}

{}
\subsection{The black-box barrier}\label{sec:app-barrier}

We work in the black-box model of \cref{def:dr-blackbox}, with the notation of \cref{sec:dr-barrier}, and
assume $n\ge1$. \Cref{sec:app-bb-mean} states the lower bounds for mean exponents, which
\cref{cor:dr-bb-rs} evaluates for \pname{Connected Dominating Set}, and \cref{sec:app-bb-proofs} proves them
together with \cref{cor:dr-bb-rs,thm:dr-bb-moments}.

\subsubsection{Lower bounds for mean exponents}\label{sec:app-bb-mean}

\paragraph{One sample size.} The first lower bound concerns the hypothesis~\eqref{eq:co-mean-hyp} of
\cref{thm:co-mean} at a single sample size $t$, that is, the class $\cC_{\mathrm{mean}}(t,\lambda)$. For
$h>0$ put $\mathsf N(h)=\lfloor\lambda\binom kt/h\rfloor+1$ and
\[
  \Sigma(h)=\frac{\binom nt}{\binom kt}\cdot
  \begin{cases}2^{h}&\text{if }0<h\le\lambda,\\ \min\bigl(2^{h},\mathsf N(h)\bigr)&\text{if }h>\lambda.
  \end{cases}
\]
The idea is that the adversary gives touched $t$-sets the running-time exponent $h$ and plants the running-time
exponent $0$ on the untouched ones: as long as fewer than $\mathsf N(h)$ subsets of the planted solution are
touched, the planted instance lies in $\cC_{\mathrm{mean}}(t,\lambda)$. Since $\binom kt$ is a
positive integer, $\mathsf N(h)>\binom kt$ holds exactly when $h\le\lambda$; so in the first case the
planted instance lies in the class whatever the algorithm touches, and only the $\mathrm{EXT}$ charges
count. In particular $\Sigma(\lambda)=\mathsf R_t$ when $\lambda>0$.

\begin{theorem}[Black-box lower bound, one sample size]\label{thm:dr-bb-one}
  Let $0\le t\le k\le n$ and $\lambda\ge0$, and let $\mathcal H$ be a deterministic algorithm in the black-box
  model (\cref{def:dr-blackbox}) that is correct for $\cC_{\mathrm{mean}}(t,\lambda)$. For every
  $h>0$, the cost of $\mathcal H$ on some no-instance is at least
  $\frac13\min\{\binom nk,\Sigma(h)\}$. In particular, with
  $h^{*}=\max\left(1,\lceil\log((\lambda+1)\binom kt)\rceil\right)$,
  \[
    \mathrm{cost}(\mathcal H)\ \ge\ \frac13\min\Bigl\{\binom nk,\ \max\Bigl(\mathsf R_t,\ \binom nt\frac{\lambda}{h^{*}}\Bigr)\Bigr\}.
  \]
\end{theorem}

So, up to the factor $\lambda/h^{*}$ and the cap $\binom nk$, the cost of a deterministic black-box
algorithm is at least $\binom nt$, whereas the randomized algorithm of the mean-exponent theorem has cost
$\mathsf R_t$ up to a polynomial factor; the ratio is exponential in $n$ whenever
$\log\binom kt$ exceeds $\lambda$ by a linear amount (\cref{cor:dr-bb-gap}). \Cref{thm:dr-bb-one} follows from \cref{thm:dr-bb-all} below applied to the subclass in which the
mean bound at every level other than $t$ is $n$, and the proof below proceeds in this way.

The adversary can use integer running-time exponents in $\{0,\dots,n\}$, as for the number of Steiner
terminals in \cref{sec:rs}: whenever $(\lambda+1)\binom kt\le2^{n}$, the second claim of
\cref{thm:dr-bb-one} holds with $\mathsf R_t$ replaced by $\mathsf R_t/2$, and by $\mathsf R_t$ itself
for integer $\lambda$, against algorithms that are correct for the instances of
$\cC_{\mathrm{mean}}(t,\lambda)$ with integer running-time exponents (see the paragraph on integer running-time exponents in \cref{sec:app-bb-proofs}). We record the
gap in quantitative form.

\begin{corollary}[Gap]\label{cor:dr-bb-gap}
  Let $\lambda\ge1$ and $\varepsilon>0$ with $\log\binom kt\ge\lambda+\varepsilon n$. Every deterministic
  black-box algorithm that is correct for $\cC_{\mathrm{mean}}(t,\lambda)$ has cost at least
  $\min\{\binom nk,2^{\varepsilon n}\mathsf R_t\}/(3(n+2))$ on some no-instance, while the Markov-cutoff
  algorithm of \cref{rem:co-markov}, with the exact threshold $\lambda+1$, decides the class with one-sided error at cost at most
  $6(\lambda+1)\mathsf R_t$.
\end{corollary}

In applications the mean bound usually holds for every sample size, and the lower bound persists
under this stronger hypothesis.

\paragraph{All sample sizes.} For $0\le j\le k$ and $h>0$ let $\mathsf N_j(h)$, $\Sigma_j(h)$ and
$h^{*}_j$ be $\mathsf N(h)$, $\Sigma(h)$ and $h^{*}$ with $(t,\lambda)$ replaced by $(j,\lambda_j)$, and put
$\mathsf R_j=\binom nj\binom kj^{-1}2^{\lambda_j}$. The planting argument applies at all levels at once,
because the condition of $\cC_{\mathrm{mean}}^{\mathrm{all}}$ at level $j$ involves only the $j$-subsets
of the solution.

\begin{theorem}[Black-box lower bound, all sample sizes]\label{thm:dr-bb-all}
  Let $\lambda_0,\dots,\lambda_k\ge0$, and let $\mathcal H$ be a deterministic black-box algorithm that is
  correct for $\cC_{\mathrm{mean}}^{\mathrm{all}}(\lambda_0,\dots,\lambda_k)$. For all
  $h_0,\dots,h_k>0$, the cost of $\mathcal H$ on some no-instance is at least
  $\frac13\min\{\binom nk,\min_{0\le j\le k}\Sigma_j(h_j)\}$. In particular,
  \[
    \mathrm{cost}(\mathcal H)\ \ge\ \frac13\min\Bigl\{\binom nk,\ \min_{0\le j\le k}\max\Bigl(\mathsf R_j,\ \binom nj\frac{\lambda_j}{h^{*}_j}\Bigr)\Bigr\},
  \]
  and $\lambda_j/h^{*}_j\ge1/(n+2)$ for every $j$ with $\lambda_j\ge1$.
\end{theorem}

\subsubsection{Proofs}\label{sec:app-bb-proofs}

\paragraph{The planting step.} All lower bounds use the same adversary. Fix a deterministic
algorithm $\mathcal H$ and run it on the \emph{no-instance} $\cF=\emptyset$, $g(X,\ell)=h_{|X|}$ for every $\ell$, where the
reals $h_0,\dots,h_n\ge0$ are chosen later. Since $\mathcal H$ is deterministic, its run is a fixed sequence
of calls. Call a set $X$
\emph{touched} if $\mathrm{COST}$ or $\mathrm{EXT}$ was called on $(X,\ell)$ for some $\ell$. Let $\mathsf T_j$
be the number of touched $j$-sets, $\gamma_j$ and $e_j$ the numbers of $\mathrm{COST}$ and $\mathrm{EXT}$
calls on $j$-sets, and $z$ the number of $\mathrm{MEM}$ calls. Then $\mathsf T_j\le \gamma_j+e_j$, and the cost of
the run is
\begin{equation}\label{eq:dr-cost}
  \mathrm{cost}=\sum_j\gamma_j+\sum_je_j2^{h_j}+z .
\end{equation}
For a $k$-set $S$ let $q_j(S)$ be the number of touched $j$-subsets of $S$, and call $S$
\emph{clean} if $\mathrm{MEM}(S)$ was not called and no $\mathrm{EXT}$ call was made on a subset of
$S$.

\begin{lemma}[Planting]\label{lem:dr-planting}
Let $S$ be a clean $k$-set and let $g'\colon2^U\times\N\to\R_{\ge0}$ agree with $g$ on $(X,\ell)$ for
every touched $X$ and every $\ell$. Then $\mathcal H$ answers \NO{} on the instance $(\{S\},g')$. Moreover, for a
uniformly random $S\in\binom Uk$,
\begin{gather*}
  \Pr[\mathrm{MEM}(S)\text{ called}]\le\frac{z}{\binom nk},\qquad
  \Pr[\text{some EXT call on a subset of }S]\le\sum_je_j\frac{\binom kj}{\binom nj},\\
  \text{and}\quad\E\,q_j(S)=\mathsf T_j\frac{\binom kj}{\binom nj}\quad\text{for every }j.
\end{gather*}
\end{lemma}

\begin{proof}
We show, by induction along the run, that the instance $(\{S\},g')$ answers every call exactly as the
no-instance does. Since $\mathcal H$ is deterministic, it then makes the same calls on both instances and
returns the same answer, and on the no-instance that answer is \NO{} by correctness. Consider a call of
the run. A $\mathrm{COST}$ call concerns a touched set, on which $g'$ and $g$ agree. A call
$\mathrm{MEM}(X)$ answers \YES{} on $(\{S\},g')$ only for $X=S$, which was not queried, so it answers
\NO{} on both instances. A call $\mathrm{EXT}(X,\ell)$ answers \NO{} on $(\{S\},g')$ if
$X\not\subseteq S$, which cleanness ensures, so it too answers \NO{} on both instances. For the
probabilities, the $z$ calls of $\mathrm{MEM}$
query at most $z$ of the $\binom nk$ equally likely sets; a fixed $j$-set lies in $S$ with probability
$\binom{n-j}{k-j}/\binom nk=\binom kj/\binom nj$, which gives the second bound by the union bound and
the expectation by linearity.
\end{proof}

The three lower bounds differ only in which planted instances lie in the class. The next lemma turns
this into one counting inequality.

\begin{lemma}[Counting]\label{lem:dr-counting}
Let $\mathcal H$ be correct for a class $\cC$, and let $M_0,\dots,M_k\ge1$ be integers such that for
every clean $k$-set $S$ with $q_j(S)<M_j$ for all $j\in\{0,\dots,k\}$ there is a $g'$ as in
\cref{lem:dr-planting} with $(\{S\},g')\in\cC$. Then
\begin{equation}\label{eq:dr-counting}
  1\le\frac{z}{\binom nk}+\sum_{j=0}^{k}\Bigl(e_j\frac{\binom kj}{\binom nj}
  +\mathbf 1\bigl[M_j\le\tbinom kj\bigr]\,\frac{\mathsf T_j\binom kj}{M_j\binom nj}\Bigr).
\end{equation}
\end{lemma}

\begin{proof}
Let $S$ be a clean $k$-set with $q_j(S)<M_j$ for all $j\le k$. The instance $(\{S\},g')$ lies in
$\cC$ and its family $\{S\}$ contains a set of size exactly $k$, so correctness forces $\mathcal H$ to
answer \YES{} on it, which contradicts \cref{lem:dr-planting}. Hence, for uniformly random
$S\in\binom Uk$, the event that $S$ is clean and $q_j(S)<M_j$ for all $j\le k$ has probability zero,
and the union bound gives
\[
  1\le\Pr[\mathrm{MEM}(S)\text{ called}]+\Pr[\text{some EXT call on a subset of }S]
  +\sum_{j=0}^{k}\Pr[q_j(S)\ge M_j].
\]
\Cref{lem:dr-planting} bounds the first two terms. For the third, $q_j(S)\le\binom kj$, so
$\Pr[q_j(S)\ge M_j]=0$ if $M_j>\binom kj$, and otherwise Markov's inequality and
\cref{lem:dr-planting} give $\Pr[q_j(S)\ge M_j]\le\E\,q_j(S)/M_j=\mathsf T_j\binom kj/(M_j\binom nj)$.
\end{proof}

We use the following facts about $\Sigma$.

\begin{lemma}[The function $\Sigma$]\label{lem:dr-sigma}
Let $0\le t\le k\le n$, $\lambda\ge0$, and let $\mathsf N$, $\Sigma$, $\mathsf R_t$ and $h^{*}$ be as in
\cref{sec:dr-barrier,sec:app-bb-mean}.
\begin{enumerate}[label=\textup{(\alph*)}]
\item $\Sigma(h)\le\binom nt2^{h}/\binom kt$ for every $h>0$, and
$\Sigma(h)\le\binom nt\mathsf N(h)/\binom kt$ for every $h>\lambda$.
\item If $\lambda>0$ then $\Sigma(\lambda)=\mathsf R_t$; if $\lambda=0$ then $\Sigma(h)=\mathsf R_t$ for
every $h>0$. In both cases some $h>0$ has $\Sigma(h)=\mathsf R_t$.
\item $\Sigma(h^{*})\ge\binom nt\lambda/h^{*}$.
\item If $\lambda\ge1$ then $\lambda/h^{*}\ge1/(n+2)$.
\end{enumerate}
\end{lemma}

\begin{proof}
(a) Both cases of the definition of $\Sigma$ are at most $\binom nt2^{h}/\binom kt$, and the second
case is at most $\binom nt\mathsf N(h)/\binom kt$.

(b) If $\lambda>0$, the first case applies at $h=\lambda$ and gives
$\Sigma(\lambda)=\binom nt2^{\lambda}/\binom kt=\mathsf R_t$. If $\lambda=0$, every $h>0$ falls in the
second case with $\mathsf N(h)=1$, so $\Sigma(h)=\binom nt\binom kt^{-1}\min(2^{h},1)=\binom nt/\binom kt=\mathsf R_t$.

(c) By definition, $h^{*}\ge1$ and $2^{h^{*}}\ge(\lambda+1)\binom kt$. If $h^{*}\le\lambda$, then
$\Sigma(h^{*})=\binom nt2^{h^{*}}/\binom kt\ge\binom nt(\lambda+1)\ge\binom nt\lambda/h^{*}$. If
$h^{*}>\lambda$, then $\mathsf N(h^{*})\le\lambda\binom kt/h^{*}+1\le(\lambda+1)\binom kt\le2^{h^{*}}$, so
$\Sigma(h^{*})=\binom nt\mathsf N(h^{*})/\binom kt$, and $\mathsf N(h^{*})>\lambda\binom kt/h^{*}$ gives the claim.

(d) Since $\lceil x\rceil\le x+1$ and $\log\binom kt\le n$,
$h^{*}\le\max(1,\log(\lambda+1)+n+1)=\log(\lambda+1)+n+1$. For $\lambda\ge1$ we have
$2^{\lambda}\ge1+\lambda$ (equality at $\lambda=1$, and the left side grows faster), that is
$\log(\lambda+1)\le\lambda$, so $h^{*}\le\lambda+n+1\le(n+2)\lambda$.
\end{proof}

\begin{proof}[Proof of \cref{thm:dr-bb-all}]
Run the planting step with the given values $h_0,\dots,h_k$ at levels $0,\dots,k$ and, say,
$h_j=0$ at levels $j>k$, which play no role, as a set of size $j>k$ is not a subset of a $k$-set. We
apply \cref{lem:dr-counting} with $M_j=\mathsf N_j(h_j)$. Let $S$ be a clean $k$-set with
$q_j(S)<\mathsf N_j(h_j)$ for all $j\le k$, and define $g'$ to be $0$ on the untouched subsets of $S$ of
every size, with every second argument, and $g$ elsewhere. The hypothesis of
$\cC_{\mathrm{mean}}^{\mathrm{all}}$ at level $j$ involves only $g'(X,k)$ for $X\in\binom Sj$; every
touched $j$-subset of $S$ has $g'=h_j$ and every untouched one has $g'=0$, so the mean of
$g'(\cdot,k)$ over $\binom Sj$ is $q_j(S)h_j/\binom kj\le(\mathsf N_j(h_j)-1)h_j/\binom kj\le\lambda_j$. Hence
$(\{S\},g')\in\cC_{\mathrm{mean}}^{\mathrm{all}}(\lambda_0,\dots,\lambda_k)$, and \eqref{eq:dr-counting} holds.

We bound the term of \eqref{eq:dr-counting} at level $j$. By \cref{lem:dr-sigma}(a),
$e_j\binom kj/\binom nj\le e_j2^{h_j}/\Sigma_j(h_j)$. The indicator is $1$ exactly when $h_j>\lambda_j$,
since $\mathsf N_j(h_j)>\binom kj$ holds exactly when $h_j\le\lambda_j$; then
$\Sigma_j(h_j)\le\binom nj\mathsf N_j(h_j)/\binom kj$ by \cref{lem:dr-sigma}(a), so the Markov term is at
most $\mathsf T_j/\Sigma_j(h_j)$. As $h_j>0$,
$\mathsf T_j+e_j2^{h_j}\le\gamma_j+e_j+e_j2^{h_j}\le2(\gamma_j+e_j2^{h_j})$. So the level-$j$ term is at most
$2\,\mathrm{cost}_j/\Sigma_j(h_j)$ with $\mathrm{cost}_j=\gamma_j+e_j2^{h_j}$, and by \eqref{eq:dr-cost} the
right-hand side of \eqref{eq:dr-counting} is at most
$\mathrm{cost}/\binom nk+2\,\mathrm{cost}/\min_j\Sigma_j(h_j)\le3\,\mathrm{cost}/\min\{\binom nk,\min_j\Sigma_j(h_j)\}$,
where $\mathrm{cost}$ is the cost of the run on the no-instance. This proves the first claim.

For the second claim, choose at each level $j$ the better of two values of $h_j$: a value with
$\Sigma_j(h_j)=\mathsf R_j$, which exists by \cref{lem:dr-sigma}(b), and $h_j=h^{*}_j$, for which
$\Sigma_j(h^{*}_j)\ge\binom nj\lambda_j/h^{*}_j$ by \cref{lem:dr-sigma}(c). The first claim with these
values gives the displayed bound, and \cref{lem:dr-sigma}(d) gives $\lambda_j/h^{*}_j\ge1/(n+2)$ for
$\lambda_j\ge1$.
\end{proof}

\begin{proof}[Proof of \cref{thm:dr-bb-one}]
Let $\mathcal H$ be correct for $\cC_{\mathrm{mean}}(t,\lambda)$, and let
$\lambda_t=\lambda$ and $\lambda_j=n$ for $j\ne t$. Every instance of
$\cC_{\mathrm{mean}}^{\mathrm{all}}(\lambda_0,\dots,\lambda_k)$ satisfies the level-$t$ condition, so
it lies in $\cC_{\mathrm{mean}}(t,\lambda)$, and $\mathcal H$ is correct for the subclass. Apply
\cref{thm:dr-bb-all} with $h_t=h$ and $h_j=n$ for $j\ne t$. At a level $j\ne t$ we have
$h_j=n\le\lambda_j$, so $\Sigma_j(n)=\binom nj2^{n}/\binom kj\ge2^{n}\ge\binom nk$, and likewise
$\mathsf R_j=\binom nj2^{n}/\binom kj\ge\binom nk$. Hence in both bounds of \cref{thm:dr-bb-all} the
levels $j\ne t$ do not lower the minimum below $\binom nk$, and the bounds read
$\frac13\min\{\binom nk,\Sigma(h)\}$ and $\frac13\min\{\binom nk,\max(\mathsf R_t,\binom nt\lambda/h^{*})\}$.
\end{proof}

\paragraph{Integer running-time exponents.} In the proof of \cref{thm:dr-bb-one} the adversary uses the running-time
exponents $n$ at the levels $j\ne t$ and $h$ at level $t$. The second claim uses $h=h^{*}$, an integer,
which is at most $n$ exactly when $(\lambda+1)\binom kt\le2^{n}$, and a value $h$ with $\Sigma(h)=\mathsf R_t$,
which is $h=\lambda$ when $\lambda>0$ and need not be an integer. An integer replacement in
$\{1,\dots,n\}$ loses at most a factor $2$: if $\lambda\ge n$, take $h=n\le\lambda$, for which
$\Sigma(n)=\binom nt2^{n}/\binom kt\ge2^{n}\ge\binom nk$, so the capped bound $\frac13\binom nk$ is
attained in full; if $1\le\lambda<n$, take $h=\lfloor\lambda\rfloor\le\lambda$, for which
$\Sigma(h)=\binom nt2^{\lfloor\lambda\rfloor}/\binom kt\ge\mathsf R_t/2$; and if $0\le\lambda<1$, take
$h=1>\lambda$, for which $\mathsf N(1)\ge1$ and
$\Sigma(1)=\binom nt\binom kt^{-1}\min(2,\mathsf N(1))\ge\binom nt/\binom kt\ge\mathsf R_t/2$, as
$2^{\lambda}<2$. So, whenever $(\lambda+1)\binom kt\le2^{n}$, the second claim of \cref{thm:dr-bb-one}
holds with $\mathsf R_t$ replaced by $\mathsf R_t/2$ against algorithms that are correct for the instances of
$\cC_{\mathrm{mean}}(t,\lambda)$ with integer running-time exponents in $\{0,\dots,n\}$; the planted running-time
exponents are $0$, so the planted instances have integer running-time exponents as well. If $\lambda$ is an
integer, the three choices give $\Sigma(h)\ge\mathsf R_t$ or the full cap, so $\mathsf R_t$ itself remains.

\begin{proof}[Proof of \cref{cor:dr-bb-gap}]
First, $\binom nt=\mathsf R_t\binom kt2^{-\lambda}\ge2^{\varepsilon n}\mathsf R_t$, and
$\lambda/h^{*}\ge1/(n+2)$ by \cref{lem:dr-sigma}(d), so
$\max(\mathsf R_t,\binom nt\lambda/h^{*})\ge2^{\varepsilon n}\mathsf R_t/(n+2)$, and \cref{thm:dr-bb-one}
gives the lower bound. For the upper bound, computation is free in the model, so the algorithm of
\cref{rem:co-markov} can use the exact threshold $\lambda+1$. Each trial then makes one $\mathrm{COST}$
call and at most one $\mathrm{EXT}$ call on a set with $g\le\lambda+1$, so it costs at most
$1+2^{\lambda+1}\le3\cdot2^{\lambda}$. The number of trials is
$\lceil(\lambda+1)\binom nt\binom kt^{-1}\rceil\le2(\lambda+1)\binom nt\binom kt^{-1}$, since the argument
of the ceiling is at least $1$. The total cost is therefore at most $6(\lambda+1)\mathsf R_t$, and
\cref{rem:co-markov} gives the one-sided error.
\end{proof}

\RestCorDrBbRs*

For its proof we use the function $f_\sigma$ of~\eqref{eq:co-fsigma} with $\beta=1$,
$\rho=2$ and $\sigma=\frac12$, that is,
\[
  f_{1/2}(\tau)=H(\tau)-\tfrac12H(2\tau)+1-2\tau\qquad(0\le\tau\le\tfrac12).
\]

\begin{proof}
The lower bound comes from the second bound of \cref{thm:dr-bb-all}: we estimate $\mathsf R_j$ and $\binom nj$
by entropy and show that, at every level, the larger of the two exponents is at least $H(\frac14)$. The
upper bound is the linear-savings corollary.

For $1\le j\le k$, $\lambda_j$ is nonincreasing in $j$ and $\lambda_k=1$, so every $\lambda_j\ge1$ (also
$\lambda_0=n\ge1$), and \cref{thm:dr-bb-all} gives cost at least
$\frac13\min\{\binom nk,\min_j\max(\mathsf R_j,\binom nj/(n+2))\}$. We bound its terms. First,
$\binom n{n/2}\ge2^{n}/(n+1)$ by~\eqref{eq:fw-binom}, and $\mathsf R_0=2^{n}$. For $1\le j\le k$ put
$\tau=j/n\in(0,\frac12]$. Since $\lambda_j=n-1-2j+4j/n\ge n(1-2\tau)-1$,
\eqref{eq:fw-binom} gives
\[
  \mathsf R_j\ge\frac{2^{nH(\tau)}}{n+1}\cdot2^{-kH(j/k)}\cdot2^{n(1-2\tau)-1}=\frac{2^{nf_{1/2}(\tau)}}{2(n+1)},
  \qquad\frac{\binom nj}{n+2}\ge\frac{2^{nH(\tau)}}{(n+1)(n+2)} .
\]
So
\[
  \max\Bigl(\mathsf R_j,\frac{\binom nj}{n+2}\Bigr)\ge\frac{2^{n\max(f_{1/2}(\tau),H(\tau))}}{2(n+1)(n+2)} .
\]
We show that
$\max(f_{1/2}(\tau),H(\tau))\ge H(\frac14)$ for every $\tau\in(0,\frac12]$. For $\tau\ge\frac14$ this
holds because $H$ increases on $[0,\frac12]$. For $0<\tau\le\frac14$, \cref{rem:co-fsigma} with
$(\beta,\rho,\sigma)=(1,2,\frac12)$ and $p=\frac14$ shows that $f_{1/2}$ is convex with minimizer
$\tau^*=\frac13$, so it is nonincreasing on $[0,\frac13]$ and
$f_{1/2}(\tau)\ge f_{1/2}(\frac14)=H(\frac14)-\frac12H(\frac12)+1-\frac12=H(\frac14)$. (In terms of
binomial coefficients: the ratio of consecutive terms of $\binom nj\binom kj^{-1}2^{n-2j}$ is
$(n-j)/(4(k-j))$, which is at most $1$ for $j\le n/3$.)
Since $H(\frac14)<1$, all terms are at least $2^{nH(1/4)}/(2(n+1)(n+2))$, and \cref{thm:dr-bb-all}
gives the lower bound. Finally $H(\frac14)=\frac14\log4+\frac34\log\frac43=2-\frac34\log3$, so
$2^{H(1/4)}=4\cdot3^{-3/4}=1.7547\ldots$.

For the upper bound, the class satisfies~\eqref{eq:co-linear-hyp} with $(\beta,\rho,C)=(1,2,1)$
at every sample size: $\lambda_0=n\le n+1$, and for $1\le j\le k$,
$\lambda_j=n-1-2j+4j/n\le n-2j+1$ because $j\le n/2$. By \cref{cor:co-linear}(1), Algorithm
\MLS{} decides the class with one-sided error in time $2^{C+nF^{*}(1/2)}N^{O(1)}$, where
$F^{*}(\frac12)=1+\frac12\log\frac12+\frac12\log\frac32=\frac12\log3$
by~\eqref{eq:co-Fstar}, since $p=\frac14<\frac12$. In the model, a trial of $(t,\theta)$ in
Algorithm \MLS{} makes one $\mathrm{COST}$ call and at most one $\mathrm{EXT}$ call, on a set with
$g\le\theta$, so it costs at most $2^{\theta+1}$. The algorithm of \cref{cor:co-linear}(1) uses
$\Gamma=1$, confidence $\kappa=1$ and a rate $\psi$ with $\psi n\le C+nF^{*}(\frac12)+1\le n+2$.
By~\eqref{eq:fw-params}, $\Theta\le\psi n+3$ and $m_{t,\theta}2^{\theta+1}\le(2M_{t,\theta}+1)2^{\theta+1}
\le16\kappa(\Theta+1)2^{\psi n}+2^{\Theta+1}\le16(\kappa(\Theta+1)+1)2^{\psi n}$. Summing over the
$(k+1)(\Theta+1)$ pairs $(t,\theta)$, the cost is at most $16(k+1)(\Theta+1)(\kappa(\Theta+1)+1)2^{\psi n}$,
and $2^{\psi n}\le4\cdot\sqrt3^{\,n}$, so the cost is $\sqrt3^{\,n}n^{O(1)}$.
\end{proof}

\RestThmDrBbMoments*

\begin{proof}
Take $h_j=n$ for all $j$, so that $g\equiv n$. Fix a feasible $t$ and apply \cref{lem:dr-counting}
with $M_t=\binom kt-J_t+1$ and $M_j=\binom kj+1$ for $j\ne t$; these are integers, and
$M_t\ge1$ since $J_t\le\binom kt/2$. Let $S$ be a clean $k$-set with $q_j(S)<M_j$ for all $j\le k$;
the conditions at levels $j\ne t$ hold for every $S$, and the condition at level $t$ says that at
least $J_t$ sets of $\binom St$ are untouched. Let $g'$ be $0$ on $J_t$ of them, with every second
argument, and $g$ elsewhere. Then $\mu_t(S)\ge J_t\ge\binom nt2^{-\psi n}/\Gamma$ under $g'$, so
$(\{S\},g')\in\cC_{\mathrm{exp}}(\psi,\Gamma)$. In \eqref{eq:dr-counting} the indicator vanishes at the
levels $j\ne t$, and $M_t>\binom kt/2$ at level $t$, so, using $\binom kj/\binom nj\le1\le2^{n}/\binom nk$,
\[
  1\le\frac{z}{\binom nk}+\sum_je_j\frac{\binom kj}{\binom nj}+\frac{\mathsf T_t\binom kt}{M_t\binom nt}
  \le\frac{z+\sum_je_j2^{n}}{\binom nk}+\frac{2\mathsf T_t}{\binom nt}.
\]
By~\eqref{eq:dr-cost},
$z+\sum_je_j2^{n}\le\mathrm{cost}$ and $\mathsf T_t\le \gamma_t+e_t\le\mathrm{cost}$, so the right-hand side is at most
$3\,\mathrm{cost}/\min\{\binom nk,\binom nt\}$.
\end{proof}
{}
\subsection{Balanced set-inclusion families}\label{sec:app-balanced}

This subsection constructs the balanced families of \cref{def:dr-balanced} and proves the following theorem of
\cref{sec:dr-balanced}, whose notation we keep.

\RestThmDrBalanced*

The construction achieves exact balance. Its building block is a covering of \emph{ordered}
structures on small blocks by subsets that are good on every prefix: when a set is ordered by
non-increasing weight, a prefix-good subset captures its proportional share of the weight
(\cref{lem:dr-abel}), and at least a $1/p$ fraction of all ${\tilde t}$-subsets of a $p$-set is prefix-good
(\cref{lem:dr-rotation}). Formally, for a finite set $S$ with a linear order $u_1,\dots,u_p$ and an
integer $0\le {\tilde t}\le p$, call $Y\subseteq S$ with $|Y|=\tilde t$ \emph{prefix-good} if
$p\,|Y\cap\{u_1,\dots,u_i\}|\ge {\tilde t}\,i$ for $i=1,\dots,p$.

\begin{lemma}[Rotation]\label{lem:dr-rotation}
  Let $p\ge1$. For every $\xi\in\{0,1\}^{p}$ with $\sum_i\xi_i={\tilde t}$, some cyclic rotation of $\xi$ has
  $i$-th prefix sum at least $({\tilde t}/p)i$ for every $i$. Consequently, for a fixed order of $S$, a uniformly
  random ${\tilde t}$-subset of $S$ is prefix-good with probability at least $1/p$.
\end{lemma}

\begin{proof}
  Let $\pi_0=0$ and $\pi_i=\sum_{l\le i}(\xi_l-{\tilde t}/p)$, so $\pi_p=0$. Pick $m\in\{0,\dots,p-1\}$ minimizing
  $\pi_m$. The rotation that starts at $\xi_{m+1}$ has $i$-th prefix sum minus $({\tilde t}/p)i$ equal to
  $\pi_{m+i}-\pi_m\ge0$ if $m+i\le p$, and to $\pi_p+\pi_{m+i-p}-\pi_m=\pi_{m+i-p}-\pi_m\ge0$ if $m+i>p$. For the
  second claim, identify ${\tilde t}$-subsets of $S$ with their indicator vectors. The cyclic group of rotations
  acts on these vectors and the uniform distribution is invariant under it. Each orbit has size dividing
  $p$ and contains a prefix-good member, so at least a $1/p$ fraction of every orbit, and hence of all
  ${\tilde t}$-subsets, is prefix-good.
\end{proof}

The second ingredient is the weight guarantee of prefix-good sets.

\begin{lemma}[Abel summation]\label{lem:dr-abel}
  If $w\ge0$ is non-increasing along the order $u_1,\dots,u_p$ of $S$ and $Y\subseteq S$ is prefix-good
  with $|Y|=\tilde t$, then $w(Y)\ge({\tilde t}/p)w(S)$.
\end{lemma}

\begin{proof}
  Put $c_i=|Y\cap\{u_1,\dots,u_i\}|$, $c_0=0$ and $w(u_{p+1})=0$. Then
  $w(Y)=\sum_{i=1}^{p}w(u_i)(c_i-c_{i-1})=\sum_{i=1}^{p}c_i\bigl(w(u_i)-w(u_{i+1})\bigr)$. Each difference
  is nonnegative and $c_i\ge({\tilde t}/p)i$, so
  $w(Y)\ge({\tilde t}/p)\sum_{i=1}^{p}i\bigl(w(u_i)-w(u_{i+1})\bigr)=({\tilde t}/p)\sum_{i=1}^pw(u_i)=({\tilde t}/p)w(S)$.
\end{proof}

The block families are found by greedy \pname{Set Cover}, whose guarantee we use in the following
form: on an instance whose sets each cover at most $\Delta$ elements, greedy returns a cover of size at
most $1+\ln \Delta$ times the optimum~\cite{Johnson74,Lovasz75}.

\begin{lemma}[Block families]\label{lem:dr-block}
  Let $\iota\ge1$, let $q_1,\dots,q_{\mathsf r}\ge0$ be integers with $\sum_jq_j\le\iota$, and let $\tilde t_j$ be integers with
  $0\le \tilde t_j\le q_j$ and ${\tilde t}=\sum_j\tilde t_j$. A deterministic algorithm computes, in time polynomial
  in $\iota!\,2^{\iota}$, an $(\iota;q;{\tilde t})$-balanced family $\mathcal B$ over $[\iota]$ with
  \[
    |\mathcal B|\le N_{\mathrm{cov}}\,(1+\ln\iota!)^{2},\qquad N_{\mathrm{cov}}=\binom\iota {\tilde t}\prod_j\frac{\max(q_j,1)}{\binom{q_j}{\tilde t_j}} .
  \]
\end{lemma}

\begin{proof}
  An \emph{ordered structure} is a tuple of pairwise disjoint sets $S_j\subseteq[\iota]$ with
  $|S_j|=q_j$, each with a linear order; let $\mathcal O$ be the set of ordered structures, so that
  $|\mathcal O|=\iota!/(\iota-\sum_jq_j)!\le\iota!$. A
  candidate $X\in\binom{[\iota]}{\tilde t}$ \emph{covers} a structure if $|X\cap S_j|=\tilde t_j$ for every $j$ (so
  $X\subseteq\bigcup_jS_j$, as $\sum_j\tilde t_j={\tilde t}=|X|$) and $X\cap S_j$ is prefix-good in $S_j$ for every $j$
  with $q_j\ge1$.

  \emph{A random candidate covers a fixed structure with probability at least $1/N_{\mathrm{cov}}$.} A uniform
  $X\in\binom{[\iota]}{\tilde t}$ has the right class counts with probability $\prod_j\binom{q_j}{\tilde t_j}/\binom\iota {\tilde t}$,
  because the candidates with these counts are exactly the unions of a $\tilde t_j$-subset of each $S_j$.
  Conditioned on that event, the sets $X\cap S_j$ are independent and uniform among the $\tilde t_j$-subsets of
  $S_j$, so by \cref{lem:dr-rotation} each is prefix-good with probability at least $1/q_j$ (a class with
  $q_j=0$ contributes the factor $1$). In particular $N_{\mathrm{cov}}\ge\binom\iota {\tilde t}/\prod_j\binom{q_j}{\tilde t_j}\ge1$.

  \emph{Size.} Take $N_{\mathrm{cand}}=\lfloor N_{\mathrm{cov}}\ln|\mathcal O|\rfloor+1$ independent uniform candidates. A fixed structure
  is uncovered by all of them with probability at most $(1-1/N_{\mathrm{cov}})^{N_{\mathrm{cand}}}<e^{-N_{\mathrm{cand}}/N_{\mathrm{cov}}}<1/|\mathcal O|$, so by the
  union bound some choice covers all structures, and the minimum cover has at most
  $N_{\mathrm{cand}}\le N_{\mathrm{cov}}(1+\ln|\mathcal O|)$ members. A candidate covers at most $|\mathcal O|$ structures, so greedy
  \pname{Set Cover} returns a cover $\mathcal B$ with
  $|\mathcal B|\le(1+\ln|\mathcal O|)N_{\mathrm{cov}}(1+\ln|\mathcal O|)\le N_{\mathrm{cov}}(1+\ln\iota!)^2$.

  \emph{Balance.} Given disjoint $S_j\subseteq[\iota]$ with $|S_j|=q_j$ and $w\ge0$, order each $S_j$ by
  non-increasing $w$ (ties broken arbitrarily), take a member of $\mathcal B$ covering this ordered
  structure, and apply \cref{lem:dr-abel} to each class $j$ with $q_j\ge1$: it gives
  $w(X\cap S_j)\ge(\tilde t_j/q_j)w(S_j)$, simultaneously for all classes.

  \emph{Time.} The covering relation has $|\mathcal O|\le\iota!$ elements and at most $2^\iota$ candidate sets, and
  both it and the greedy computation take time polynomial in $\iota!\,2^\iota$.
\end{proof}

The block targets are distributed over the blocks by systematic rounding, which keeps every block
target within one of its proportional share and has mean equal to that share. Formally, let
$p_1,\dots,p_b\ge0$ be integers with $\sum_ip_i=k\ge1$, let $t\in\{0,\dots,k\}$ and $\delta=t/k$, and
put $p_{\le i}=p_1+\dots+p_i$ ($p_{\le0}=0$). For an offset $\upsilon\in[0,1)$, the \emph{systematic rounding}
of $(p_i)$ with target $t$ is
\[
  t_i(\upsilon)=\lfloor\delta p_{\le i}+\upsilon\rfloor-\lfloor\delta p_{\le i-1}+\upsilon\rfloor\qquad(1\le i\le b).
\]

\begin{lemma}[Systematic rounding]\label{lem:dr-rounding}
  In the setting above, for every $\upsilon\in[0,1)$:
  $t_i(\upsilon)\in\{\lfloor\delta p_i\rfloor,\lceil\delta p_i\rceil\}$, $0\le t_i(\upsilon)\le p_i$ and
  $\sum_it_i(\upsilon)=t$. For $\upsilon$ uniform in $[0,1)$, $\E\,t_i(\upsilon)=\delta p_i$.
\end{lemma}

\begin{proof}
  For real $x$ and $x'\ge0$, $\lfloor x+x'\rfloor-\lfloor x\rfloor\in\{\lfloor x'\rfloor,\lceil x'\rceil\}$; and
  $0\le\lfloor\delta p_i\rfloor\le\lceil\delta p_i\rceil\le p_i$ since $0\le\delta\le1$. The sum telescopes to
  $\lfloor t+\upsilon\rfloor-\lfloor\upsilon\rfloor=t$. Finally, writing $\{x\}$ for the fractional part,
  $\E\lfloor x+\upsilon\rfloor=\lfloor x\rfloor+\Pr[\upsilon\ge1-\{x\}]=\lfloor x\rfloor+\{x\}=x$, so
  $\E\,t_i(\upsilon)=\delta p_{\le i}-\delta p_{\le i-1}=\delta p_i$.
\end{proof}

Two entropy facts control the size of the product construction.

\begin{lemma}[Entropy facts]\label{lem:dr-entropy}
  The binary entropy function $H$ has the following properties.
  \begin{enumerate}[label=(E\arabic*)]
    \item\label{it:dr-E1} $|H(x)-H(x')|\le H(|x-x'|)$ for $x,x'\in[0,1]$.
    \item\label{it:dr-E2} For an integer $p\ge1$, a real $\delta\in[0,1]$ and an integer $q$ with
      $|q-\delta p|<1$ and $0\le q\le p$: $pH(q/p)\ge pH(\delta)-\log(ep)$.
  \end{enumerate}
\end{lemma}

\begin{proof}
  \ref{it:dr-E1} We may assume $x\le x'$ and put $h=x'-x$. Since $H$ is concave, $y\mapsto H(y+h)-H(y)$ is
  non-increasing on $[0,1-h]$, so $H(x')-H(x)\le H(h)-H(0)=H(h)$. Applying this to $1-x'\le1-x$ and using
  $H(y)=H(1-y)$ gives $H(x)-H(x')\le H(h)$.

  \ref{it:dr-E2} For $p=1$, $|H(q)-H(\delta)|\le1\le\log e$. Let $p\ge2$. By~\ref{it:dr-E1},
  $H(\delta)-H(q/p)\le H(|q/p-\delta|)$, and $|q/p-\delta|<1/p\le\frac12$, so by monotonicity of $H$ on
  $[0,\frac12]$ this is at most $H(1/p)$. Finally
  $pH(1/p)=\log p+(p-1)\log\frac p{p-1}\le\log p+(p-1)\frac{\log e}{p-1}=\log(ep)$, using $\ln(1+x)\le x$.
\end{proof}

\paragraph{The construction.} Let $\iota\ge2$ be a block length. Partition $U=[n]$ into
$b=\lceil n/\iota\rceil$ consecutive blocks $U_1,\dots,U_b$, all of size $n_i=\iota$ except possibly the
last, and put
$\delta_j=t_j/k_j$ ($\delta_j=0$ if $k_j=0$). The family $\mathcal B$ consists of all sets
$X=\bigcup_i\mathrm{ord}_i(Y_i)$ where:
\begin{enumerate}[label=(\arabic*)]
  \item\label{it:dr-c1} $(p_{ij})$ is a nonnegative integer $b\times \mathsf r$ matrix with $\sum_jp_{ij}\le n_i$ for every
    $i$ and $\sum_ip_{ij}=k_j$ for every $j$;
  \item\label{it:dr-c2} $(t_{ij})$ has $t_{ij}\in\{\lfloor\delta_jp_{ij}\rfloor,\lceil\delta_jp_{ij}\rceil\}$ and
    $\sum_it_{ij}=t_j$ for every $j$;
  \item\label{it:dr-c3} $Y_i$ ranges over the block family $\mathcal B_i=\mathcal B(n_i;(p_{ij})_j;(t_{ij})_j)$ of
    \cref{lem:dr-block} on $[n_i]$;
  \item\label{it:dr-c4} $\mathrm{ord}_i\colon[n_i]\to U_i$ is the order-preserving bijection.
\end{enumerate}
The matrix $(p_{ij})$ is a guess of how the unknown solution classes meet the blocks, and $(t_{ij})$
distributes the targets proportionally.

\begin{proof}[Proof of \cref{thm:dr-balanced}]
  Balance passes from the blocks to the whole universe through the rounding lemma; the size bound compares
  the product of the block bounds with the global binomial quotient, through an injection for the
  numerator and the entropy facts for the denominator; time and
  space follow from the order of enumeration.

  \emph{Balance.} Let $(S_j)$ and $w$ be given, and put $p_{ij}=|S_j\cap U_i|$; this matrix satisfies
  \ref{it:dr-c1}. Fix a class $j$ with $k_j\ge1$, and let $t_{ij}(\upsilon_j)$ be the systematic rounding
  of $(p_{ij})_i$ with target $t_j$ (\cref{lem:dr-rounding}). The quantity
  \[
    \sum_{i:\,p_{ij}\ge1}\frac{t_{ij}(\upsilon_j)}{p_{ij}}\,w(S_j\cap U_i)
  \]
  has expectation $\delta_jw(S_j)$ for uniform $\upsilon_j$ (blocks with $p_{ij}=0$ contain no element of
  $S_j$), so some $\upsilon_j$ makes it at least $\delta_jw(S_j)$. For a class with $k_j=0$ all $t_{ij}=0$.
  By \cref{lem:dr-rounding} the resulting matrix $(t_{ij})$ satisfies \ref{it:dr-c2}; the rounding
  parameters of different classes are chosen independently, and every combination occurs
  in~\ref{it:dr-c2}. By \cref{lem:dr-block}, applied in block $i$ to the sets
  $\mathrm{ord}_i^{-1}(S_j\cap U_i)$ and the weight $w\circ\mathrm{ord}_i$, the family $\mathcal B_i$ contains $Y_i$
  with $|Y_i\cap\mathrm{ord}_i^{-1}(S_j)|=t_{ij}$, $Y_i\subseteq\bigcup_j\mathrm{ord}_i^{-1}(S_j)$ and class-$j$ weight at
  least $(t_{ij}/p_{ij})w(S_j\cap U_i)$, simultaneously for all $j$ with $p_{ij}\ge1$. The union
  $X=\bigcup_i\mathrm{ord}_i(Y_i)$ lies in $\mathcal B$, satisfies $X\subseteq\bigcup_jS_j$ and
  $|X\cap S_j|=\sum_it_{ij}=t_j$, and $w(X\cap S_j)\ge\delta_jw(S_j)$ for every $j$ with $k_j\ge1$.

  \emph{Size.} Write $t_i=\sum_jt_{ij}$, so that $\sum_it_i=t$. The number of listed tuples is at most
  $\#(p_{ij})\cdot\#(t_{ij})\cdot\max\prod_i|\mathcal B_i|$, where $\#(p_{ij})\le(\iota+1)^{\mathsf rb}$ (each entry
  lies in $\{0,\dots,\iota\}$) and $\#(t_{ij})\le2^{\mathsf rb}$. By \cref{lem:dr-block},
  $|\mathcal B_i|\le\iota^{4}N_{\mathrm{cov},i}$, because $1+\ln n_i!\le1+\iota\ln\iota\le1+\iota(\iota-1)\le\iota^2$. Also
  $N_{\mathrm{cov},i}\le\iota^{\mathsf r}\binom{n_i}{t_i}/\prod_j\binom{p_{ij}}{t_{ij}}$. We bound the two products over $i$.
  \begin{itemize}
    \item \emph{Numerator.} $\prod_i\binom{n_i}{t_i}\le\binom nt$, because choosing $t_i$ elements in
      each block $U_i$ gives distinct $t$-subsets of $U$.
    \item \emph{Denominator.} For $p_{ij}\ge1$ we have $|t_{ij}-\delta_jp_{ij}|<1$, so~\eqref{eq:fw-binom}
      and~\ref{it:dr-E2} give
      \[
        \log\binom{p_{ij}}{t_{ij}}\ge p_{ij}H(\delta_j)-\log(ep_{ij})-\log(p_{ij}+1)\ge p_{ij}H(\delta_j)-\log\bigl(e\iota(\iota+1)\bigr),
      \]
      and the inequality between the outer terms also holds for $p_{ij}=0$. Summing over $i$ and using
      $\sum_ip_{ij}H(\delta_j)=k_jH(\delta_j)\ge\log\binom{k_j}{t_j}$ (by~\eqref{eq:fw-binom}) gives
      $\prod_{i,j}\binom{p_{ij}}{t_{ij}}\ge\prod_j\binom{k_j}{t_j}\cdot(e\iota(\iota+1))^{-\mathsf rb}$.
  \end{itemize}
  Altogether
  \[
    |\mathcal B|\le\frac{\binom nt}{\prod_j\binom{k_j}{t_j}}
    \bigl[2(\iota+1)\iota\cdot e\iota(\iota+1)\bigr]^{\mathsf rb}\iota^{4b}
    \le\frac{\binom nt}{\prod_j\binom{k_j}{t_j}}\,\iota^{12\mathsf rb},
  \]
  because $2e\iota^{2}(\iota+1)^{2}\le\iota^{8}$ for $\iota\ge2$ (at $\iota=2$ the left side is $72e<256$, and the
  ratio of the two sides decreases in $\iota$). Finally $b\le n/\iota+1$. For $n\ge16$ we have
  $\iota\ge2$, and $\iota\ge\log n/\log\log n-1\ge\frac12\log n/\log\log n$ once $\log n/\log\log n\ge2$, which
  holds for $n\ge16$; hence
  $\log\iota^{12\mathsf r(n/\iota+1)}\le12\mathsf r(n/\iota+1)\log\log n=O(\mathsf rn(\log\log n)^2/\log n)$.

  \emph{Time and space.} Only two block sizes occur, so there are at most $2(\iota+1)^{\mathsf r}2^{\mathsf r}$ distinct
  block parameters $(n_i;(p_{ij})_j;(t_{ij})_j)$, which is polynomial in $n$ for fixed $\mathsf r$. Each block
  family takes time polynomial in $\iota!\,2^{\iota}\le n^2$, because
  $\iota!\le\iota^{\iota}=2^{\iota\log\iota}\le2^{(\log n/\log\log n)\log\log n}=n$ and $2^\iota\le n$; all block families
  are computed once and stored in polynomial space. The matrices $(p_{ij})$ are enumerated by depth-first
  search over the blocks, extending a partial choice only if the remaining column sums fit into the
  remaining blocks, that is, $\sum_j(\text{remaining }k_j)\le\sum_{i'>i}n_{i'}$ after block $i$; this is the
  exact feasibility condition, because any distribution of the remaining demand that respects the block
  capacities is admissible. The matrices $(t_{ij})$ are enumerated by depth-first search, keeping each
  remaining column sum between the sums of the remaining floors and ceilings, which is again exact. The
  product over blocks is enumerated by an odometer; every block family is nonempty, because for its
  parameters at least one ordered structure exists. So every leaf of the search yields a listed set, the
  time is $n^{O(1)}$ per listed set plus $n^{O(1)}$, and only the search state is kept.
\end{proof}

\begin{remark}[Comparison with the construction of Fomin, Gaspers, Lokshtanov and Saurabh]\label{rem:dr-fgls-compare}
  Compared with~\cite{FominGLS19}, the construction of \cref{thm:dr-balanced} uses no hashing: a fixed
  partition into blocks suffices because the block targets follow the block counts: the numerator of the
  size bound is at most $\binom nt$ because distinct choices in the blocks give distinct $t$-sets, and the denominator loses only a factor
  $(e\iota(\iota+1))^{\mathsf r}=2^{O(\mathsf r\log\iota)}$ per block. Small blocks make the greedy step take polynomial
  time and space, and hence the whole construction runs in polynomial space. In exchange, the size
  bound has a larger subexponential factor, $2^{O(n(\log\log n)^{2}/\log n)}$.
\end{remark}

\paragraph{The subexponential factor.} With the explicit constants of \cref{thm:dr-balanced}, the factor
by which the balanced families multiply the running-time bound of \cref{thm:det-cds} is, for the two-class
families of \cref{alg:dr-integer} ($\mathsf r=2$), at most $\iota^{24(n/\iota+1)}$ with $\iota=\lfloor\log n/\log\log n\rfloor$, about $2^{12718}$ at $n=1000$
against $1.7361^{1000}\approx2^{796}$. The product of this factor and $1.7361^n$ is below the trivial $2^n$
if and only if $\iota\ge1203$, that is, if and only if $\log n\ge16895.338\ldots$. These are upper bounds, not measured family sizes;
for $n\le20$ and block lengths $4$ to $6$ the actual product families contain 80--90\% of all $t$-sets
(\cref{sec:ancillary}).

\subsubsection{Trees across block partitions}\label{sec:app-tree-cut}

The product construction of \cref{thm:dr-balanced} handles conditions that are separable over a
fixed partition of the universe into blocks. The analyses of RS-$\theta$ and RSC rest on tests
along an unknown spanning tree of the solution, and the following count shows that no family of
$2^{O(n)}$ fixed partitions into blocks of size $n^{o(1)}$ avoids cutting almost all edges of some
such tree. This is an obstacle for one construction strategy, not an impossibility result.

\begin{proposition}[Trees across block partitions]\label{prop:dr-tree-cut}
  Let $S$ be a set of $k$ elements, let $c\ge0$ and $B\ge1$, and let $\mathcal P$ be a family of at most
  $2^{ck}$ partitions of $S$ into blocks of size at most $B$. If $\varepsilon\in(0,1]$ satisfies
  $\varepsilon\log k>c+\log(B+1)+2+3(\log k)/k$, then some tree $T$ with vertex set $S$ has, for every
  partition in $\mathcal P$, at least $(1-\varepsilon)k$ edges whose ends lie in different blocks. In
  particular, if $B=k^{o(1)}$, then for every fixed $\varepsilon>0$ and all large $k$ some tree has
  $(1-\varepsilon)k$ such edges in every partition of the family.
\end{proposition}

\begin{proof}
  Fix a partition in $\mathcal P$. A tree $T$ on $S$ is determined by its restrictions to the blocks,
  which are forests, and by the set of its edges between blocks. A forest on $b$ labeled vertices
  becomes a tree on $b+1$ vertices when a new vertex is joined to one vertex of each component, and the
  forest is recovered by deleting the new vertex; so by Cayley's formula there are at most
  $(b+1)^{b-1}\le(B+1)^{b}$ forests on a block of size $b$, and at most $(B+1)^{k}$ choices of the
  restrictions in all. If $m$ edges lie between blocks, they form an $m$-subset of the fewer than
  $k^{2}$ pairs of elements in different blocks, which gives at most $\binom{k^2}{m}\le(ek^2/m)^{m}$
  choices for $m\ge1$, a bound that is nondecreasing in $m$ for $m\le k^{2}$. For $\varepsilon=1$
  there is nothing to prove, so let $\varepsilon<1$ and $\alpha=1-\varepsilon\in(0,1)$. Then at most
  $k(B+1)^{k}(ek/\alpha)^{\alpha k}$ trees
  have fewer than $\alpha k$ edges between blocks of this partition (there are at most $k$ values of
  $m<\alpha k$), and since
  $\alpha\log(e/\alpha)\le\log e+\max_{0<\alpha\le1}\alpha\log(1/\alpha)<2$, this number is at most
  $k(B+1)^{k}4^{k}k^{\alpha k}$. Over the family, at most
  $2^{ck}k(B+1)^{k}4^{k}k^{(1-\varepsilon)k}=k^{k-2}\cdot2^{k(c+\log(B+1)+2)+3\log k-\varepsilon k\log k}$
  trees have fewer than $(1-\varepsilon)k$ edges between blocks of some partition, and this is less
  than $k^{k-2}$ because the exponent of $2$ is negative by the hypothesis on $\varepsilon$. By
  Cayley's formula there are $k^{k-2}$ trees on $S$, so some tree has at least $(1-\varepsilon)k$
  edges between blocks of every partition in $\mathcal P$. The last sentence holds because $c+\log(B+1)+2+3(\log k)/k=o(\log k)$ when
  $B=k^{o(1)}$.
\end{proof}

For a solution $S$ of linear size $k=\Theta(n)$, a family of $2^{O(n)}$ partitions of the universe
into blocks of size $n^{o(1)}$ induces on $S$ a family as in \cref{prop:dr-tree-cut}, so some tree
$T$ on $S$ has $(1-o(1))k$ edges between blocks of every partition of the family; on an instance in
which $S$ induces exactly the tree $T$, these are edges of the only spanning tree of the solution.
{}
\subsection{The deterministic algorithms}\label{sec:app-det}

This subsection proves the statements of \cref{sec:dr-moments,sec:det}, with the notation of those sections.

\subsubsection{Exponential moments of integer affine certificates}\label{sec:app-det-moments}

\RestLemDrChord*

\begin{proof}
  The function $\mathrm{cv}(x)=\ln(1+y2^{x})$ is convex. Let $\mathrm{chord}$ be the affine function that agrees with
  $\mathrm{cv}$ at $a$ and $a+1$. A convex function lies on or above its chord outside the chord's interval, so
  $\mathrm{cv}(v_u)\ge \mathrm{chord}(v_u)$ for every $u$. Let $w^*$ have $k_a$ entries $a$ and $k_{a+1}$ entries $a+1$; its
  total is $ak+k_{a+1}=\Xi$. Since $\mathrm{cv}=\mathrm{chord}$ at $a$ and $a+1$, and $\mathrm{chord}$ is affine with
  $\sum_uv_u=\sum_uw^*_u$,
  \[
    \sum_u\mathrm{cv}(v_u)\ge\sum_u\mathrm{chord}(v_u)=\sum_u\mathrm{chord}(w^*_u)=\sum_u\mathrm{cv}(w^*_u)=\ln \mathsf P^*(y).
  \]
  The second claim is Jensen's inequality for $\mathrm{cv}$.
\end{proof}

\RestPropDrReal*

\begin{proof}
  For a profile $w$ let $W^w_S(y)=\sum_t\mu^w_t(S)y^t=2^{-\alpha}\prod_{u\in S}(1+y2^{w(u)})$ be the
  generating function of the surrogate exponent. \Cref{lem:fw-gf}, whose proof uses only
  $\mu_t(S)>0$ for $0\le t\le k$ and $W_S(y)=\sum_t\mu_t(S)y^t$, gives
  $\mathsf R(w)\le\inf_{y\ge0}(1+y)^n/W^w_S(y)\le(n+1)\mathsf R(w)$. By the second claim of
  \cref{lem:dr-chord}, $W^w_S(y)\ge W^{\bar w}_S(y)$ for every $y>0$, so
  $\mathsf R(w)\le\inf_{y\ge0}(1+y)^n/W^w_S(y)\le\inf_{y\ge0}(1+y)^n/W^{\bar w}_S(y)\le(n+1)\mathsf R(\bar w)$.
  If $w$ has integer values, it has no entry in $(a,a+1)$, and the first claim of \cref{lem:dr-chord}
  gives $\mathsf R(w)\le(n+1)\mathsf R(w^*)$ in the same way. Finally, for the uniform profile
  $\mu^{\bar w}_t(S)=\binom kt2^{-\alpha+t\Xi/k}$, which gives the formula for $\mathsf R(\bar w)$.
\end{proof}

\RestThmDrDetMoments*

\begin{proof}
  \emph{The early answer.} If $\alpha<\Xi$, no solution has an integer certificate: for $X=S$ it would
  give $0\le g(I,S,|S|)\le\alpha-\Xi<0$. So answering \NO{} is correct. From now on $\alpha\ge\Xi$.

  \emph{Number of guesses.} There are at most $k(\Xi+1)\le N^{O(1)}$ guesses, and for each the choice in
  step~(2) ranges over $O(k^2)$ pairs and takes polynomial time: the logarithm of each candidate is a
  rational plus logarithms of integers, so approximations to within $\frac14$ are computable in time
  $N^{O(1)}$, and a minimizer of the approximations is within a factor $2^{1/2}\le2$ of the minimum.

  \emph{Soundness.} The algorithm answers \YES{} only when $\cA$ accepts some $(I,X,k)$, which means
  that $X\cup Y\in\cF_I$ for some $Y$ with $|Y|\le k-t$, a solution of size at most $k$.

  \emph{Completeness.} Let $w$ be an integer certificate for $S$. Put $S^-=\{u\in S:w(u)\le a\}$ and
  $S^+=S\setminus S^-$; as the weights are integers, every $u\in S^+$ has $w(u)\ge a+1$. Then
  $S^-\ne\emptyset$, since otherwise $\Xi\ge(a+1)k>\Xi$. The pair $(k^-,\Xi^-)=(|S^-|,w(S^-))$
  satisfies $1\le k^-\le k$ and $\Xi^-\le ak^-$; moreover $\Xi^+=w(S^+)\ge(a+1)k^+$, and if $k^+=0$
  then $S^+=\emptyset$ and $\Xi^+=0$. So this guess is tried. For it, the balanced family of step~(3),
  applied to the classes $S^-$, $S^+$ and the weight $w$, contains some $X\subseteq S$ with
  $|X\cap S^-|=t^-$, $|X\cap S^+|=t^+$ and
  $w(X)\ge(t^-/k^-)\Xi^-+(t^+/k^+)\Xi^+=t^-\bar w^-+t^+\bar w^+$ (the second term is absent if $k^+=0$). As
  $t\ge1$, $X$ is nonempty, so $g(I,X,k)\le\alpha-w(X)\le\theta$, and $\cA$ is run on $(I,X,k)$ and
  accepts, with witness $S\setminus X$.

  \emph{The threshold is nonnegative.} For every guess, $\bar w^-,\bar w^+\ge0$, so
  $t^-\bar w^-+t^+\bar w^+\le k^-\bar w^-+k^+\bar w^+=\Xi^-+\Xi^+=\Xi\le\alpha$. (For $k^+=0$ this uses
  $\Xi^-=\Xi$, which is the condition $\Xi^+=0$.) Hence $\theta\ge0$, and each listed set takes time at
  most $N^{O(1)}+2^{\theta}N^{O(1)}\le2^{\theta}N^{O(1)}$.

  \emph{Time.} Fix a guess, and let $v\in\R^k$ have $k^-$ entries $\bar w^-$ and $k^+$ entries $\bar w^+$. Its total is
  $\Xi^-+\Xi^+=\Xi$, and it has no entry in $(a,a+1)$, because $\bar w^-\le a$ and $\bar w^+\ge a+1$. For
  $1\le t\le k$ put
  \[
    \mathsf P_t=\sum_{t^-+t^+=t}\binom{k^-}{t^-}\binom{k^+}{t^+}2^{t^-\bar w^-+t^+\bar w^+},
  \]
  a positive sum of at most $t+1$ terms. Its largest term is at least $\mathsf P_t/(t+1)$, so by the choice in
  step~(2), which is within a factor $2$ of the minimum, and by \cref{thm:dr-balanced}, the running time
  of the guess is at most
  \[
    2^{O(n(\log\log n)^{2}/\log n)}N^{O(1)}\min_{1\le t\le k}(t+1)\binom nt\frac{2^{\alpha}}{\mathsf P_t}.
  \]
  Expanding the product over the entries of $v$ by subsets and grouping the subsets by their numbers of
  light and heavy entries gives, for every $y>0$,
  \[
    \sum_{t=1}^{k}\mathsf P_ty^t=\prod_u\bigl(1+y2^{v_u}\bigr)-1\le\max_{1\le t\le k}\frac{\mathsf P_t}{\binom nt}\sum_{t=1}^{k}\binom nty^t
    \le\max_{1\le t\le k}\frac{\mathsf P_t}{\binom nt}\bigl((1+y)^{n}-1\bigr),
  \]
  so $\min_{1\le t\le k}\binom nt/\mathsf P_t\le((1+y)^{n}-1)/(\prod_u(1+y2^{v_u})-1)$. \Cref{lem:dr-chord} bounds the
  product from below by $\mathsf P^*(y)$, and $\mathsf P^*(y)>1$ for $y>0$. As this holds for every $y>0$, and
  $t+1\le n+1$, the running time of the guess is at most
  $2^{O(n(\log\log n)^{2}/\log n)}N^{O(1)}\inf_{y>0}((1+y)^{n}-1)2^{\alpha}/(\mathsf P^*(y)-1)$. Multiplying by the
  polynomial number of guesses, and adding the polynomial time of the early answer and of the choices
  in steps (1) and (2), which the maximum with $1$ absorbs, gives the time bound.

  \emph{Space.} The guesses are run one after another, each in the space of \cref{thm:dr-balanced} plus that
  of $\cA$.
\end{proof}

\subsubsection{Deterministic Connected Dominating Set}\label{sec:app-det-cds}

\RestLemDetCert*

\begin{proof}
For fixed $y$ the function $\chi_y$ is affine in $u$ on each interval
$[\frac1{a+1},\frac1a]$, so it suffices to check endpoints. The proof has an algebraic part, on one
interval, where a single $y$ gives equality at both endpoints, and a rational part, on five
intervals, where the endpoint inequalities are inequalities between rational numbers.

\emph{Reduction to endpoints.} On each interval $u\in[\frac1{a+1},\frac1a]$ we use the formula with this
value of $a$. For $u$ in the interior or at the right endpoint this is $\lfloor1/u\rfloor$; at the left
endpoint $u=\frac1{a+1}$ we have $\lfloor1/u\rfloor=a+1$, but both formulas give the same value there,
namely $\log(1+y)+1-\frac1{a+1}\log(1+y2^{a+1})$, since the coefficients are $(0,\frac1{a+1})$ for $a$ and
$(\frac1{a+1},0)$ for $a+1$. For fixed $y$ and $a$ the two coefficients $(a+1)u-1$ and $1-au$ are affine in
$u$, so $\chi_y$ is affine on the interval and attains its maximum at an endpoint. The intervals that meet
$[0.15,0.85]$ are those with $a=1,\dots,6$.

\emph{The interval $a=2$, $u\in[\frac13,\frac12]$ (algebraic).} Take $y_0=(1+\sqrt5)/8$. Then
$1+4y_0=\frac{3+\sqrt5}2=\left(\frac{1+\sqrt5}2\right)^2$, $1+8y_0=2+\sqrt5=\left(\frac{1+\sqrt5}2\right)^3$ and $2(1+y_0)=(9+\sqrt5)/4$. At $u=\frac12$ the
coefficients are $\frac12$ and $0$, so $2^{\chi_{y_0}(1/2)}=2(1+y_0)/(1+4y_0)^{1/2}$. At $u=\frac13$ they are $0$
and $\frac13$, so $2^{\chi_{y_0}(1/3)}=2(1+y_0)/(1+8y_0)^{1/3}$. Both equal
\[
  \frac{9+\sqrt5}{4}\cdot\frac{2}{1+\sqrt5}=\frac{9+\sqrt5}{2(1+\sqrt5)}=\frac{(9+\sqrt5)(\sqrt5-1)}{8}=\frac{8\sqrt5-4}8=\sqrt5-\frac12 .
\]
Since $\chi_{y_0}$ is affine on the interval, $\chi_{y_0}(u)=\log(\sqrt5-\frac12)$ for every $u\in[\frac13,\frac12]$.

\emph{The other intervals (rational).} Use $y=\frac12$ for $a=1$ on $[\frac12,\frac{17}{20}]$,
$y=\frac5{16}$ for $a=3$ on $[\frac14,\frac13]$, $y=\frac14$ for $a=4$ on $[\frac15,\frac14]$,
$y=\frac{27}{128}$ for $a=5$ on $[\frac16,\frac15]$, and $y=\frac{29}{160}$ for $a=6$ on
$[\frac3{20},\frac16]$. At each of the ten endpoints $u$, both coefficients are nonnegative rationals; with
$h$ their least common denominator, the inequality $\chi_y(u)\le\log\frac{1736}{1000}$ is
equivalent to
\[
  \bigl(2(1+y)\bigr)^{h}\ \le\ \Bigl(\frac{1736}{1000}\Bigr)^{h}\,\bigl(1+y2^{a}\bigr)^{h((a+1)u-1)}\bigl(1+y2^{a+1}\bigr)^{h(1-au)},
\]
an inequality between rational numbers with nonnegative integer exponents. It holds at all ten endpoints,
by an exact evaluation (\cref{sec:anc-det}). The largest endpoint value of $2^{\chi_y(u)}$ on these
intervals is $\sqrt3=1.7320\ldots$, at $u=\frac12$ for $a=1$. Since $\frac{1736}{1000}<\sqrt5-\frac12$, the lemma holds
on these intervals. As the five intervals and $[\frac13,\frac12]$ cover $[0.15,0.85]$,
the lemma follows.
\end{proof}

\RestLemDetTime*

\begin{proof}
Write $u=s/\Xi$ and $a=\lfloor1/u\rfloor=\lfloor\Xi/k\rfloor$. In the notation of
\cref{thm:dr-det-moments}, $k_{a+1}=\Xi-as=\Xi(1-au)$ and $k_a=s-k_{a+1}=\Xi((a+1)u-1)$, so
\[
  \log \mathsf P^*(y)=\Xi\,\mathsf p_y,\qquad \mathsf p_y=\bigl((a+1)u-1\bigr)\log\bigl(1+y2^{a}\bigr)+(1-au)\log\bigl(1+y2^{a+1}\bigr).
\]
The two coefficients of $\mathsf p_y$ are nonnegative and sum to $u\le1$, so $0\le \mathsf p_y\le\log(1+y2^{a+1})$. If
$\mathsf P^*(y)\ge2$, then $\mathsf P^*(y)-1\ge \mathsf P^*(y)/2$, and, as $(1+y)^{n}-1\le(1+y)^{n}$ and $\alpha+1\le n+1$,
\[
  \frac1n\log\frac{\bigl((1+y)^{n}-1\bigr)\,2^{\alpha}}{\mathsf P^*(y)-1}
  \le\frac1n\bigl(n\log(1+y)+\alpha+1-\Xi\mathsf p_y\bigr)\le\chi_y(u)+\frac{1+2\mathsf p_y}n ,
\]
using $\chi_y(u)=\log(1+y)+1-\mathsf p_y$ in the notation of \cref{lem:det-cert}, $\alpha+1=\Xi+2\le n+1$
and $\Xi\mathsf p_y\ge(n-2)\mathsf p_y$.

We check the range of $u$. Since $\Xi\le n-1<n$ and $n\ge34$, $u>0.15n/\Xi>0.15$, and
$u<0.8n/\Xi\le0.8n/(n-2)\le0.85$, where the last inequality is equivalent to $n\ge34$. So
$u\in[0.15,0.85]$, and $1\le a\le6$ because $1/u\in[1/0.85,1/0.15]$. \Cref{lem:det-cert} gives, for
this $u$, some $y$ from its list with $\chi_y(u)\le\log(\sqrt5-\frac12)$. Every listed $y$ is at least
$\frac{29}{160}$, so
\[
  \mathsf P^*(y)\ge1+(k_a+k_{a+1})\,y2^{a}=1+sy2^{a}\ge1+0.15\cdot34\cdot2\cdot\tfrac{29}{160}>2 .
\]
Moreover $y\le\frac12$ and $a\le6$, so $\mathsf p_y\le\log(1+2^{6})$ is bounded by a constant. Hence
\cref{thm:dr-det-moments}, evaluated at this $y$, bounds the time of the run by
\[
  N^{O(1)}\,2^{O(n(\log\log n)^{2}/\log n)}\max\bigl(1,2^{n\chi_y(u)+1+2\mathsf p_y}\bigr)
  \le N^{O(1)}\,2^{O(n(\log\log n)^{2}/\log n)}\,\bigl(\sqrt5-\tfrac12\bigr)^{n},
\]
where the last step uses $\chi_y(u)\le\log(\sqrt5-\frac12)$, $\mathsf p_y\ge0$ and $\sqrt5-\frac12>1$.
\end{proof}

\RestLemDetDecide*

\begin{proof}
The enumeration is deterministic, uses polynomial space, since membership in $\cF_I$ is tested in
polynomial time (\cref{def:fw-iss}), and answers \YES{} exactly when $\cF_I$ has a solution of size
$s$. It takes time $\binom nsN^{O(1)}$, which is $N^{O(1)}$ for $n<34$ and otherwise at most
$2^{H(0.15)n}N^{O(1)}=O(1.5261^n)N^{O(1)}$ for $s\le0.15n$ and $2^{H(0.2)n}N^{O(1)}=O(1.6494^n)N^{O(1)}$
for $s\ge0.8n$, by~\eqref{eq:fw-binom}, since the binary entropy function $H$ is increasing on
$[0,\frac12]$ and symmetric about $\frac12$; both bases are below $\sqrt5-\frac12$. In the other case,
$\alpha$ is a rational and $\Xi$ a nonnegative integer at most $n$, both computable in polynomial time,
and $\alpha\ge\Xi$, so \cref{alg:dr-integer} does not stop early. \Cref{thm:dr-det-moments} gives that
the run is deterministic, uses space $N^{O(1)}$ plus the space of $\cA$, answers \NO{} if $\cF_I$
has no solution of size at most $s$, and answers \YES{} if a solution of size $s$ has an integer
certificate; \cref{lem:det-time} gives the time bound.
\end{proof}

\RestThmDetCds*

\begin{proof}
For $s=1,\dots,n$ run the procedure $\mathrm{Det}(G,s)$ of \cref{lem:det-decide} for the implicit set
system $\Phi_{\mathrm{CDS}}$, the extension algorithm $\cA_{\mathrm{St}}$ with its running-time exponent
$g(G,X,k)=|\Tm(J_X)|$, which is deterministic, uses polynomial space and has values in
$\{0,\dots,n\}$ by \cref{cor:cds-steiner}, and $(\alpha,\Xi)=(n-1,n-2)$; output the least $s$ at which
it answers \YES{}. By \cref{lem:det-decide}, $\mathrm{Det}(G,s)$ answers \NO{} if $\gc(G)>s$.
At $s=\gc(G)$ a minimum connected dominating set $S$ has size exactly $s$, and, for $n\ge2$, the
weights $w=d_F-1$ of~\eqref{eq:det-cert} are an integer certificate for $S$: they are nonnegative
integers with $w(S)=n-2=\Xi$ and $g(G,X,s)=|\Tm(J_X)|\le\alpha-w(X)$ for every nonempty
$X\subseteq S$. So $\mathrm{Det}(G,\gc(G))$ answers \YES{}, and the deterministic form of
\cref{lem:cds-first-accepted} shows that the output is $\gc(G)$, and that the same algorithm solves
\pname{Max Leaf Spanning Tree}. The time is $n$ times the bound of \cref{lem:det-decide}, which is
$O(1.7361^n)$ because $2^{O(n(\log\log n)^2/\log n)}=2^{o(n)}$ and $\sqrt5-\frac12<1.7361$, and the
space is polynomial.
\end{proof}
{}
\subsection{\pname{Connected Feedback Vertex Set}}\label{sec:app-fvs}

This subsection proves the two hardness facts of \cref{sec:fvs-problem} and constructs the extension
algorithm of \cref{lem:fvs-ext}. We keep the notation of \cref{sec:cfvs}.

\subsubsection{Hardness of the extension problem}

\begin{lemma}[The extension problem is W{[2]}-hard]\label{lem:fvs-w2}
  \pname{Connected Feedback Vertex Set Extension}, parameterized by the number $k-|X|$ of further
  vertices, is W[2]-hard.
\end{lemma}

\begin{proof}
  We reduce from \pname{Set Cover} parameterized by the number of sets, which is
  W[2]-hard~\cite{CyganFKLMPPS15}. Let $(\mathcal U,\mathcal S,k)$ be an instance with ground set
  $\mathcal U$, a family $\mathcal S$ of subsets of $\mathcal U$ and an integer $k$. We may assume that
  $\mathcal U\ne\emptyset$, that $\mathcal S\neq\emptyset$ and that every element lies in some set, since
  otherwise the instance is trivial. Build a graph $G$ with a vertex for every element, a vertex for
  every set and a root $r$; join $r$ to every set vertex, and join each set vertex to the elements
  it contains. Then $G$ is connected. Put $X=\{r\}\cup\mathcal U$ and take the size bound
  $|X|+k$, so that $k$ further vertices may be added.

  The vertices outside $X$ are the set vertices. Let $Y$ be a set of set vertices. The graph
  $G-(X\cup Y)$ consists of set vertices only, which are pairwise non-adjacent, so it is a forest.
  In $G[X\cup Y]$ the vertices of $X$ are pairwise non-adjacent and every vertex of $Y$ is adjacent
  to $r$. Hence $G[X\cup Y]$ is connected if and only if every element vertex has a neighbor in
  $Y$, that is, if and only if $Y$ covers $\mathcal U$. So $(G,X,|X|+k)$ is a \YES{}-instance if and only if
  $\mathcal U$ has a cover by at most $k$ sets. The reduction is polynomial and keeps the parameter.
\end{proof}

The next statement uses the Set Cover Conjecture of \cite{CyganDLMNOPSW16}: for every
$\varepsilon>0$ there is an integer $h$ such that \pname{Set Cover} with sets of size at most
$h$, on a universe $\mathcal U$ of $n_{\mathcal U}$ elements and with $|\mathcal S|$ sets, cannot be solved in time
$(2-\varepsilon)^{n_{\mathcal U}}(n_{\mathcal U}+|\mathcal S|)^{O(1)}$.

\begin{proposition}[Product-form running times cannot help]\label{prop:fvs-sc}
  Assume the Set Cover Conjecture. Let $c,d\ge1$ be constants such that \pname{Connected Feedback
  Vertex Set Extension} can be solved in time $c^{k-|X|}d^{|X|}N^{O(1)}$ on inputs $(G,X,k)$. Then $c\ge2$ and $d\ge2$, and the running-time bound
  of \cref{cor:co-multivariate} with $(b,c,d)=(1,c,d)$ has a base at least $2$.
\end{proposition}

\begin{proof}
  Suppose that $d<2$. Put $\varepsilon=(2-d)/2>0$ and fix an integer $j\ge1$ with
  $c^{1/j}d\le2-\varepsilon$; it exists because $c^{1/j}\to1$ and $d<2-\varepsilon$. Neither
  $\varepsilon$ nor $j$ depends on $h$. Let $h$ be arbitrary and let
  $(\mathcal U,\mathcal S,k)$ be an instance of \pname{Set Cover} with sets of size at most $h$ and
  $n_{\mathcal U}=|\mathcal U|$; we may assume $k\le n_{\mathcal U}$, since otherwise the answer is \YES{} if and only if every
  element lies in some set.
  \begin{enumerate}
    \item \emph{Padding.} Let $\hat k=\lceil k/j\rceil$ and $p=j\hat k-k<j$. Add $p$ new elements, each
      with its own singleton set. Every cover of the new instance contains the $p$ singletons, because
      they are the only sets containing the new elements. Hence the new instance has a cover with at most
      $j\hat k$ sets if and only if the old one has a cover with at most $k$ sets.
    \item \emph{Unions.} Replace the sets by all unions of at most $j$ of them. There are at most
      $(|\mathcal S|+j+1)^{j}$ such unions, a polynomial for fixed $j$. The padded instance has
      a cover with at most $j\hat k$ sets if and only if the union instance has a cover with at most $\hat k$
      sets: group the sets of a cover into $\hat k$ groups of at most $j$, and conversely split each union.
    \item \emph{Reduction.} Apply the reduction of \cref{lem:fvs-w2} to the union instance. It produces
      a forced set with $|X|\le n_{\mathcal U}+j+1$ and $\hat k\le n_{\mathcal U}/j+1$ further vertices, so the assumed algorithm
      decides it in time
      \[
        c^{\hat k}d^{\,n_{\mathcal U}+j+1}N^{O(1)}\le c\,d^{j+1}\bigl(c^{1/j}d\bigr)^{n_{\mathcal U}}N^{O(1)}
        \le(2-\varepsilon)^{n_{\mathcal U}}(n_{\mathcal U}+|\mathcal S|)^{O(1)},
      \]
      since $N$ is polynomial in $n_{\mathcal U}+|\mathcal S|$ for fixed $j$.
  \end{enumerate}
  This holds for every $h$ with the same $\varepsilon$, contradicting the conjecture. Hence
  $d\ge2$.

  Now suppose that $c<2$. Put $\varepsilon=(2-c)/2>0$ and fix an integer $j\ge1$ with
  $c^{1+1/j}\le2-\varepsilon$; it exists because $c^{1/j}\to1$ and $c<2-\varepsilon$. Let $h$ be
  arbitrary, let $(\mathcal U,\mathcal S,k)$ be an instance of \pname{Set Cover} with sets of size at most
  $h$, $n_{\mathcal U}=|\mathcal U|\ge1$ and $k\le n_{\mathcal U}$, in which every element lies in some set (otherwise the answer
  is \NO{}), and apply the padding and union steps above. They yield an
  instance $(\mathcal U',\mathcal S',\hat k)$ with $n_{\mathcal U}\le|\mathcal U'|<n_{\mathcal U}+j$ and $\hat k\le n_{\mathcal U}/j+1$, in which every
  element lies in some set, and which has a cover with at most $\hat k$ sets if and only if the original
  instance has a cover with at most $k$ sets. Build the graph of the proof of \cref{lem:fvs-w2} for
  $(\mathcal U',\mathcal S')$, with a root $r$, one vertex for every set of $\mathcal S'$ and one vertex for
  every element of $\mathcal U'$, and attach to $r$ and to every element vertex $v$ a \emph{private
  triangle}: two new vertices $v',v''$, adjacent to $v$ and to each other. Call $r$ and the element
  vertices \emph{anchors}; as $\mathcal U'\ne\emptyset$, there are at least two of them. Let $G''$ be the
  resulting connected graph, put $X=\emptyset$ and $s=|\mathcal U'|+1+\hat k$. We claim that $G''$ has a
  connected feedback vertex set of size at most $s$ if and only if $\mathcal U'$ has a cover by at most $\hat k$
  sets of $\mathcal S'$.

  If $Y$ is such a cover, then $A=\{r\}\cup\mathcal U'\cup Y$, viewed as a vertex set, is a connected feedback
  vertex set of size at most $s$: $G''[A]$ is connected, because every set vertex is adjacent to $r$ and
  every element vertex is adjacent to a set vertex in $Y$; and $G''-A$ consists of isolated set
  vertices and the edges $v'v''$, so it is a forest. Conversely, let $A$ be a connected feedback vertex
  set of $G''$ with $|A|\le s$. Every anchor lies in $A$: if an anchor $v$ is not in $A$, then $v'$ or
  $v''$ is, because $A$ meets the triangle at $v$; the neighbors of $v'$ and of $v''$ are $v$ and each
  other, so the connectivity of $G''[A]$ forces $A\subseteq\{v',v''\}$, and then $G''-A$ contains the
  triangle of another anchor, a contradiction. Removing all triangle vertices from $A$ keeps it a
  connected feedback vertex set of no larger size: a triangle vertex is adjacent only to its anchor,
  which lies in $A$, and to its twin, so it is an interior vertex of no path of $G''[A]$ between other
  vertices, and returning $v'$ and $v''$ to $G''-A$ adds at most the isolated edge $v'v''$. So we may
  assume that $A=\{r\}\cup\mathcal U'\cup Y$ for a set $Y$ of at most $s-|\mathcal U'|-1=\hat k$ set vertices, and
  the connectivity of $G''[A]$ means that every element vertex has a neighbor in $Y$, that is, that $Y$
  covers $\mathcal U'$. This proves the claim.

  The assumed algorithm decides $(G'',\emptyset,s)$ in time
  \[
    c^{s}N^{O(1)}\le c^{\,n_{\mathcal U}+j+n_{\mathcal U}/j+2}N^{O(1)}=c^{j+2}\bigl(c^{1+1/j}\bigr)^{n_{\mathcal U}}N^{O(1)}
    \le(2-\varepsilon)^{n_{\mathcal U}}(n_{\mathcal U}+|\mathcal S|)^{O(1)},
  \]
  again for every $h$ with the same $\varepsilon$, contradicting the conjecture. Hence $c\ge2$.

  With $b=1$ the base of \cref{cor:co-multivariate} is $1+\min(d,c)(1-1/c)$. If $d\ge c$, it equals
  $c\ge2$. If $d<c$, it equals $1+d-d/c>1+d-1=d\ge2$.
\end{proof}

\Cref{prop:fvs-sc} concerns extension algorithms whose running-time bound depends only on $k-|X|$
and $|X|$ (the case $b=1$); it says nothing about bounds with a further factor $b^{n-|X|}$, $b>1$.

\subsubsection{Contractions}\label{sec:fvs-graphs}

The extension algorithm contracts the components of the sample, and its iterative compression
runs along a sequence of edge contractions. This subsection collects the facts about contractions
that the proofs use. For a graph $\tilde G$ and an edge $uv\in E(\tilde G)$, the graph $\tilde G/uv$ is obtained by
replacing $u$ and $v$ by a new vertex $w$ adjacent to $N_{\tilde G}(u)\cup N_{\tilde G}(v)\setminus\{u,v\}$; it is
simple. For a set $C\subseteq V(\tilde G)$ with $\tilde G[C]$ connected, contracting $C$ means replacing $C$ by
one vertex adjacent to $N_{\tilde G}(C)$; it is the result of $|C|-1$ edge contractions inside $C$. We use
the standard count: a graph with $p$ vertices, $e$ edges and $q$ components satisfies $e\ge p-q$,
with equality if and only if it is a forest, since a connected graph on $p_i$ vertices has at least
$p_i-1$ edges, with equality if and only if it is a tree. The lemma below records that an edge
contraction preserves the number of components and the property of being a forest, and that
contracting the sample does not change the extension problem.

Let $G$ be connected, $X\subseteq V(G)$ nonempty, and let $X_1,\dots,X_q$ be the vertex sets of
the components of $G[X]$, so $q=\comp(G[X])$. The \emph{contracted graph} $G_X$ is obtained
from $G$ by contracting each $X_j$ to a new vertex $x_j$. Its vertex set is
$(V(G)\setminus X)\cup \hat X$ with $\hat X=\{x_1,\dots,x_q\}$; we call the vertices of $\hat X$ \emph{forced}.
Distinct components of $G[X]$ are not adjacent, so $\hat X$ is an independent set of $G_X$.

\begin{lemma}[Contractions]\label{lem:fvs-contract}
  \begin{enumerate}[label=(\arabic*)]
    \item Let $\tilde G$ be a graph and $uv\in E(\tilde G)$. Then $\comp(\tilde G/uv)=\comp(\tilde G)$;
      in particular, $\tilde G$ is connected if and only if $\tilde G/uv$ is. If $\tilde G$ is a forest,
      then $\tilde G/uv$ is a forest.
    \item With $G$, $X$, $G_X$ and $\hat X$ as above, $G_X$ is connected, and for every
      $Y\subseteq V(G)\setminus X$, the set $X\cup Y$ is a connected feedback vertex set of $G$ if
      and only if $\hat X\cup Y$ is a connected feedback vertex set of $G_X$.
  \end{enumerate}
\end{lemma}

\begin{proof}
  (1) The map $h$ that sends $u$ and $v$ to $w$ and fixes the other vertices sends every edge of
  $\tilde G$ other than $uv$ to an edge of $\tilde G/uv$, and every edge of $\tilde G/uv$ is the image of an
  edge of $\tilde G$; since $h^{-1}(w)=\{u,v\}$ is connected in $\tilde G$, $h$ induces a bijection between
  the components. If $\tilde G$ is a forest, then $\tilde G/uv$ has one vertex fewer, at least one edge
  fewer and the same number of components; as $|E(\tilde G)|=|V(\tilde G)|-\comp(\tilde G)$, the count
  $e\ge p-q$ holds with equality for $\tilde G/uv$, so it is a forest.
  (2) $G_X$ arises from the connected graph $G$ by edge contractions, so it is connected by (1). The
  graphs $G-(X\cup Y)$ and $G_X-(\hat X\cup Y)$ are identical: both have vertex set
  $V(G)\setminus(X\cup Y)$, and contracting the sets $X_j$ does not change the edges between vertices
  outside $X$. The graph $G_X[\hat X\cup Y]$ is obtained from $G[X\cup Y]$ by contracting the connected sets
  $X_1,\dots,X_q$, so by (1) one is connected if and only if the other is.
\end{proof}

For the iterative compression we need contraction closure in the presence of forced vertices,
which adapts the observation in the proof of Theorem~B.3 of~\cite{CyganNPPRW11arxiv}. Let $\tilde G$ be
a graph with a forced set $\hat X$, let $uv\in E(\tilde G)$, let $\tilde G_{uv}=\tilde G/uv$ with new vertex $w$, and let
the forced set $\hat X_{uv}$ of $\tilde G_{uv}$ be the image of $\hat X$, so $w\in \hat X_{uv}$ if and only if $u\in \hat X$ or
$v\in \hat X$. The \emph{excess} of a set $A$ in a graph with forced set $\hat X$ is $|A\setminus \hat X|$.

\begin{lemma}[Contraction closure with forced vertices]\label{lem:fvs-closure}
  Let $\tilde G$, $\hat X$, $uv$, $\tilde G_{uv}$, $w$ and $\hat X_{uv}$ be as above, and let $A$ be a connected feedback vertex
  set of $\tilde G$ with $\hat X\subseteq A$. Put
  $A'=A$ if $u,v\notin A$, and $A'=(A\setminus\{u,v\})\cup\{w\}$ otherwise. Then $A'$ is a
  connected feedback vertex set of $\tilde G_{uv}$, $\hat X_{uv}\subseteq A'$, and $|A'\setminus \hat X_{uv}|\le|A\setminus \hat X|$.
\end{lemma}

\begin{proof}
  Suppose first that $u,v\notin A$. Then $A'=A$, $\tilde G_{uv}[A]=\tilde G[A]$, which is connected, and
  $\tilde G_{uv}-A=(\tilde G-A)/uv$, which is a forest by \cref{lem:fvs-contract}(1). Since $w\notin A'$ and
  $\hat X\subseteq A$ contains neither $u$ nor $v$, we have $\hat X_{uv}=\hat X$, so $\hat X_{uv}\subseteq A'$ and
  $A'\setminus \hat X_{uv}=A\setminus \hat X$.

  Suppose now that $u\in A$ or $v\in A$. Then $\tilde G_{uv}-A'=\tilde G-(A\cup\{u,v\})$ is an induced subgraph of
  the forest $\tilde G-A$. The set $A\cup\{u,v\}$ induces a connected subgraph of $\tilde G$, because $\tilde G[A]$
  is connected and $uv$ is an edge with an end in $A$; and $\tilde G_{uv}[A']=\tilde G[A\cup\{u,v\}]/uv$ is connected
  by \cref{lem:fvs-contract}(1). Every forced vertex of $\tilde G_{uv}$ other than $w$ lies in $\hat X\setminus
  \{u,v\}\subseteq A\setminus\{u,v\}\subseteq A'$, and $w\in A'$; so $\hat X_{uv}\subseteq A'$. Finally,
  $A'\setminus \hat X_{uv}\subseteq(A\setminus \hat X\setminus\{u,v\})\cup(\{w\}\setminus \hat X_{uv})$. If $w\notin \hat X_{uv}$,
  then $u,v\notin \hat X$, and at least one of them lies in $A\setminus \hat X$, so the excess does not
  increase. If $w\in \hat X_{uv}$, the excess is at most $|A\setminus \hat X|$ directly.
\end{proof}

\subsubsection{Cut \& Count with forced vertices}\label{sec:fvs-cc}

The counting step of the extension algorithm is the Cut~\&~Count algorithm of~\cite{CyganNPPRW11arxiv,CyganNPPRW22} for the constrained problem, run on a tree
decomposition of the form ``forest plus a fixed modulator,'' as in their proof of the bound
$3^kn^{O(1)}$ for \pname{Connected Feedback Vertex Set}. The journal version~\cite{CyganNPPRW22}
states these bounds but leaves their proofs to the extended arXiv version~\cite{CyganNPPRW11arxiv} (see~\cite[Section~1.5]{CyganNPPRW22}),
so the numbers below that concern \pname{Connected Feedback Vertex Set} refer
to~\cite{CyganNPPRW11arxiv}; the corresponding results for \pname{Feedback Vertex Set}
are~\cite[Theorems~4.6 and~5.1]{CyganNPPRW22}. \pname{Constrained Connected
Feedback Vertex Set} asks, given a graph $\tilde G$ with $\tilde n$ vertices, a nonempty set
$Y_{\mathrm{in}}\subseteq V(\tilde G)$ and an integer $\eta$, whether $\tilde G$ has a connected
feedback vertex set $Y\supseteq Y_{\mathrm{in}}$ with $|Y|=\eta$. We use the following facts
from these papers.
\begin{enumerate}[label=(CC\arabic*)]
  \item\label{it:fvs-CC1} (\cite[Theorem~A.13]{CyganNPPRW11arxiv}.) Given a tree decomposition of $\tilde G$ of width $w$, a Monte
    Carlo algorithm decides \pname{Constrained Connected Feedback Vertex Set} in time
    $4^w\tilde n^{O(1)}$. It never accepts a \NO{}-instance, and it rejects a \YES{}-instance with
    probability at most $1/2$.
  \item\label{it:fvs-CC2} (Proof of \cite[Theorem~A.13]{CyganNPPRW11arxiv}.) The algorithm fixes a vertex $r\in Y_{\mathrm{in}}$ and
    computes, modulo $2$, the number of candidate solutions with a marked candidate forest
    (markers on side~$1$ of its cut), together with a consistent cut of the
    candidate forest and a consistent cut of the candidate solution that puts $r$ on side~$1$. Its
    dynamic program gives every vertex of a bag one of four states, which we write
    $\mathsf y_1,\mathsf y_2$ (in the solution, on side $1$ or $2$ of its cut) and
    $\mathsf z_1,\mathsf z_2$ (in the forest, on side $1$ or $2$ of its cut; these are the
    states $0_1,0_2,1_1,1_2$ of~\cite{CyganNPPRW11arxiv}). A counted object gives a state
    $\mathsf y_1$ or $\mathsf y_2$ to every vertex of $Y_{\mathrm{in}}$ and the state $\mathsf y_1$
    to~$r$, and no edge has the pair of states $\{\mathsf y_1,\mathsf y_2\}$ or $\{\mathsf
    z_1,\mathsf z_2\}$, because both cuts are consistent.
  \item\label{it:fvs-CC3} (Proof of~\cite[Theorem~5.1]{CyganNPPRW22} for \pname{Feedback Vertex Set}, which the proof of
    Theorem~B.3 of~\cite{CyganNPPRW11arxiv} repeats for the connected problem.) Let $\mathrm{Mod}\subseteq V(\tilde G)$ be such
    that $\tilde G-\mathrm{Mod}$ is a forest, and use the tree decomposition obtained from a
    decomposition of width~$1$ of $\tilde G-\mathrm{Mod}$ by adding $\mathrm{Mod}$ to every bag. Below the
    bags that forget the vertices of $\mathrm{Mod}$, each table entry
    for a coloring $\varsigma$ of $\mathrm{Mod}$ refers only to entries with the same coloring of $\mathrm{Mod}$. So
    the count is the sum over the colorings $\varsigma\colon\mathrm{Mod}\to\{\mathsf y_1,\mathsf y_2,\mathsf
    z_1,\mathsf z_2\}$ of the counts with $\mathrm{Mod}$ colored by $\varsigma$ (the ``core evaluations''),
    and each of these is computed in time and space polynomial in~$\tilde n$.
\end{enumerate}
By~\ref{it:fvs-CC2}, only the following colorings of $\mathrm{Mod}$ can have a nonzero count. A coloring
$\varsigma\colon \mathrm{Mod}\to\{\mathsf y_1,\mathsf y_2,\mathsf z_1,\mathsf z_2\}$ is \emph{reasonable} if
$\varsigma(v)\in\{\mathsf y_1,\mathsf y_2\}$ for $v\in Y_{\mathrm{in}}\cap \mathrm{Mod}$, $\varsigma(r)=\mathsf y_1$ when
$r\in \mathrm{Mod}$, and no edge $uv$ of $\tilde G[\mathrm{Mod}]$ has $\{\varsigma(u),\varsigma(v)\}=\{\mathsf y_1,\mathsf
y_2\}$ or $\{\varsigma(u),\varsigma(v)\}=\{\mathsf z_1,\mathsf z_2\}$. Summing the core evaluations over the
reasonable colorings therefore gives the same parity as the algorithm of~\ref{it:fvs-CC1}. The
following lemma bounds the number of reasonable colorings for the modulators produced by iterative
compression. It is the forced-vertex version of the count $4^33^{k-1}$ in the proof of Theorem~B.3
of~\cite{CyganNPPRW11arxiv}, in which only the vertex $r$ and the at most two vertices outside the
connected part of the modulator have four possible states.

\begin{lemma}[Number of reasonable colorings]\label{lem:fvs-cc-evals}
  Let $\tilde G$, $Y_{\mathrm{in}}$ and $r$ be as above. Let $\mathrm{Mod}=C\cup C'$ with $C\cap C'=\emptyset$, $\tilde G[C]$ connected and
  $|C'|\le2$, and let $\hat X\subseteq Y_{\mathrm{in}}\cap C$ with $r\in \hat X$. Then $\mathrm{Mod}$ has at most
  $4^{|C'|}3^{|C\setminus \hat X|}2^{|\hat X|-1}$ reasonable colorings, and they can be listed in time
  $4^{|C'|}3^{|C\setminus \hat X|}2^{|\hat X|-1}\tilde n^{O(1)}$ and polynomial space. In particular, if
  $|\mathrm{Mod}\setminus \hat X|\le k'+2$, there are at most $8\cdot3^{k'}2^{|\hat X|}$ of them.
\end{lemma}

\begin{proof}
  Fix a spanning tree of $\tilde G[C]$ rooted at $r$, and order $C$ so that every vertex comes after its
  parent. Let $\varsigma$ be reasonable. Then $\varsigma(r)=\mathsf y_1$. Let $v\in C\setminus\{r\}$ have
  parent $p$. If $\varsigma(p)=\mathsf y_j$, then $\varsigma(v)\in\{\mathsf y_j,\mathsf z_1,\mathsf z_2\}$,
  because $\{\varsigma(p),\varsigma(v)\}\neq\{\mathsf y_1,\mathsf y_2\}$; if $\varsigma(p)=\mathsf z_j$, then
  $\varsigma(v)\in\{\mathsf z_j,\mathsf y_1,\mathsf y_2\}$. If moreover $v\in \hat X\subseteq Y_{\mathrm{in}}$, then
  $\varsigma(v)\in\{\mathsf y_1,\mathsf y_2\}$, which leaves one option when $\varsigma(p)$ is a
  $\mathsf y$-state and two when it is a $\mathsf z$-state. So, given the values on the earlier
  vertices, $\varsigma(v)$ lies in an explicit set of at most $3$ values, and at most $2$ if $v\in \hat X$.
  The vertices of $C'$ take at most $4$ values. Enumerating all choices in this order lists a superset
  of the reasonable colorings, of size at most $4^{|C'|}3^{|C\setminus \hat X|}2^{|\hat X|-1}$; testing each
  candidate takes polynomial time, and only the current candidate is stored.

  For the last claim, put $n_1=|C\setminus \hat X|$ and $n_2=|C'|$. Then $n_1+n_2=|\mathrm{Mod}\setminus \hat X|\le k'+2$ and
  $n_2\le2$, so $4^{n_2}3^{n_1}=(4/3)^{n_2}3^{n_1+n_2}\le(16/9)3^{k'+2}=16\cdot3^{k'}$, and
  $16\cdot3^{k'}2^{|\hat X|-1}=8\cdot3^{k'}2^{|\hat X|}$.
\end{proof}

A \emph{structured modulator} for $(\tilde G,\hat X,k')$, where $\hat X\subseteq V(\tilde G)$ is nonempty
and $k'\ge0$, is a set $\mathrm{Mod}=C\cup C'$ such that $\tilde G-\mathrm{Mod}$ is a forest, $\tilde G[C]$ is
connected, $\hat X\subseteq C$, $|C'|\le2$, $C\cap C'=\emptyset$ and $|\mathrm{Mod}\setminus \hat X|\le k'+2$.

\begin{lemma}[The test]\label{lem:fvs-test}
  Let $\tilde G$ be a graph with $\tilde n$ vertices, let $\emptyset\ne\hat X\subseteq Y_{\mathrm{in}}\subseteq
  V(\tilde G)$, let $k'\ge0$, and let $\mathrm{Mod}$ be a structured modulator for $(\tilde G,\hat X,k')$. One can
  decide in time $3^{k'}2^{|\hat X|}\tilde n^{O(1)}$ and polynomial space, with one-sided error,
  whether $\tilde G$ has a connected feedback vertex set $Y\supseteq Y_{\mathrm{in}}$ with $|Y|\le|\hat X|+k'$:
  the algorithm never accepts if there is no such set, and it accepts with probability at least
  $1/2$ if there is one.
\end{lemma}

\begin{proof}
  Fix $r\in\hat X$. For every $\eta$ with $|Y_{\mathrm{in}}|\le\eta\le\min(|\hat X|+k',\tilde n)$, run the
  algorithm of~\ref{it:fvs-CC1} on $(\tilde G,Y_{\mathrm{in}},\eta)$ with the root $r$ and the tree
  decomposition of~\ref{it:fvs-CC3}, computing the count as the sum of the core evaluations over the
  reasonable colorings of $\mathrm{Mod}$, listed by \cref{lem:fvs-cc-evals}; accept if one of these runs
  accepts. By the remark before \cref{lem:fvs-cc-evals}, each run decides as the algorithm
  of~\ref{it:fvs-CC1} does. Hence no run accepts if there is no set $Y$ as in the lemma, and if there
  is one, the run with $\eta=|Y|$, which lies in the range tried, accepts with probability at least
  $1/2$. There are at most $\tilde n+1$ runs, each sums at most $8\cdot3^{k'}2^{|\hat X|}$ core
  evaluations by \cref{lem:fvs-cc-evals}, and each core evaluation takes polynomial time and space
  by~\ref{it:fvs-CC3}; the sum modulo $2$ is accumulated coloring by coloring.
\end{proof}

\subsubsection{The extension algorithm}\label{sec:app-fvs-ext}

\RestLemFvsExt*

The extension algorithm contracts the sample and then follows the iterative compression of the
proof of Theorem~B.3 of~\cite{CyganNPPRW11arxiv} along a contraction sequence, keeping track of the
images of the forced vertices. \Cref{lem:fvs-closure} guarantees that every intermediate graph has a
solution of no larger excess, and the previous solution, expanded, is a structured modulator for
\cref{lem:fvs-test}. Running times are stated for every outcome of the coins, as
\cref{cor:fw-onesided} requires. The extension algorithm solves a \emph{forced problem}, in which the
sample has been contracted: given a connected graph $\tilde G$ with $\tilde n$ vertices, a nonempty
set $\hat X\subseteq V(\tilde G)$, weights $\tilde w\colon V(\tilde G)\to\{0,\dots,w_{\max}\}$ with
$\tilde w=0$ on $\hat X$, and integers $k'\ge0$ (the number of further vertices) and $\theta'$, find a connected feedback vertex set
$A\supseteq\hat X$ of $\tilde G$ with $|A\setminus\hat X|\le k'$ and $\tilde w(A)\le\theta'$; we call
such an $A$ a \emph{solution} of the forced instance $(\tilde G,\hat X,\tilde w,k',\theta')$. The
weights are carried so that \cref{sec:app-weighted} can reuse the following wrapper, which turns a
procedure for the forced problem into an extension algorithm. For $X=\emptyset$ it guesses one
solution vertex, as in Remark~A.9 of~\cite{CyganNPPRW11arxiv}.

\begin{lemma}[From the forced problem to the extension problem]\label{lem:fvs-wrapper}
  Let $G$ be a connected graph with $n$ vertices, $w\colon V(G)\to\{0,\dots,w_{\max}\}$, $\theta$ an
  integer, $X\subseteq V(G)$ and $k\ge|X|$, and put $k'=k-|X|$. Let $p\in(0,1]$, and let $\mathrm{Forced}$ be a randomized
  procedure that, given a forced instance $(\tilde G,\hat X,\tilde w,k'',\theta')$ with
  $\tilde n\le n$, runs in polynomial space and, for every outcome of its coins, in time
  $3^{k''}2^{|\hat X|}(\tilde n+w_{\max})^{O(1)}$, outputs only solutions of the instance, and outputs
  one with probability at least $p$ whenever one exists. Consider the following algorithm.
  \begin{enumerate}
    \item If $X\neq\emptyset$: compute the contracted graph $G_X$ with forced set $\hat X$, put
      $\tilde w=w$ on $V(G)\setminus X$ and $\tilde w=0$ on $\hat X$, and run $\mathrm{Forced}$ on
      $(G_X,\hat X,\tilde w,k',\theta-w(X))$. If it outputs $A$, accept with $Y=A\setminus\hat X$;
      otherwise reject.
    \item If $X=\emptyset$: if $G$ is a forest, accept with $Y=\emptyset$ when $\theta\ge0$ and reject
      otherwise. If $G$ is not a forest and $k'=0$, reject. Otherwise, for every $v\in V(G)$, let
      $\tilde w_v$ agree with $w$ except that $\tilde w_v(v)=0$, and run $\mathrm{Forced}$ on
      $(G,\{v\},\tilde w_v,k'-1,\theta-w(v))$; if one run outputs $A$, accept with $Y=A$. Otherwise
      reject.
  \end{enumerate}
  The algorithm runs in polynomial space and, for every outcome of its coins, in time
  $3^{k'}2^{\comp(G[X])}(n+w_{\max})^{O(1)}$. If it accepts, then $Y\subseteq V(G)\setminus X$,
  $|Y|\le k'$, and $X\cup Y$ is a connected feedback vertex set of $G$ with $w(X\cup Y)\le\theta$; if
  such a set $Y$ exists, it accepts with probability at least $p$.
\end{lemma}

\begin{proof}
  \emph{Case $X\neq\emptyset$.} The graph $G_X$ is connected with at most $n$ vertices, and
  $\tilde w=0$ on $\hat X$. For $Y\subseteq V(G)\setminus X$, by \cref{lem:fvs-contract}(2) the set
  $X\cup Y$ is a connected feedback vertex set of $G$ if and only if $\hat X\cup Y$ is one of $G_X$, and
  $w(X\cup Y)=w(X)+\tilde w(\hat X\cup Y)$. So the sets $Y$ in question are exactly the sets
  $A\setminus\hat X$ for the sets $A$ that $\mathrm{Forced}$ may output on
  $(G_X,\hat X,\tilde w,k',\theta-w(X))$, and the claims follow from the properties of
  $\mathrm{Forced}$, with $|\hat X|=\comp(G[X])$.

  \emph{Case $X=\emptyset$.} If $G$ is a forest, then $\emptyset$ is a connected feedback vertex set,
  and no set has smaller weight than $w(\emptyset)=0$, so the answer is correct. Otherwise every
  connected feedback vertex set is nonempty. If a run for $v$ outputs $A$, then $A\ni v$ is a
  connected feedback vertex set with $|A|\le1+(k'-1)$ and $w(A)=\tilde w_v(A)+w(v)\le\theta$.
  Conversely, let $Y^*$ be a connected feedback vertex set with $|Y^*|\le k'$ and $w(Y^*)\le\theta$.
  Then $Y^*\neq\emptyset$, so $k'\ge1$; for $v\in Y^*$, the set $Y^*$ contains $\{v\}$, has
  $|Y^*\setminus\{v\}|\le k'-1$ and $\tilde w_v(Y^*)=w(Y^*)-w(v)\le\theta-w(v)$, so the run for $v$
  outputs a set with probability at least $p$. The time is
  $n\cdot3^{k'-1}2^1(n+w_{\max})^{O(1)}\le3^{k'}(n+w_{\max})^{O(1)}$, with $\comp(G[\emptyset])=0$.
\end{proof}

\begin{proof}[Proof of \cref{lem:fvs-ext}]
  \emph{The forced problem.} We first solve the forced problem without weights: given a connected
  graph $\tilde G$ with $\tilde n\le n$ vertices, a nonempty set $\hat X\subseteq V(\tilde G)$ and
  $k'\ge0$, find a connected feedback vertex set $A\supseteq\hat X$ of $\tilde G$ with
  $|A\setminus\hat X|\le k'$. Compute
  graphs $\tilde G_{\tilde n}=\tilde G,\tilde G_{\tilde n-1},\dots,\tilde G_1$ with $\tilde G_i=\tilde
  G_{i+1}/u_iv_i$ for an edge $u_iv_i$ of $\tilde G_{i+1}$ and new vertex $w_i$; all of them are
  connected by \cref{lem:fvs-contract}(1). Let $\hat X_i$ be the image of $\hat X$ in $\tilde G_i$; then
  $|\hat X_i|\le|\hat X|$, $\tilde G_1$ has one vertex, $\hat X_1=V(\tilde G_1)$, and we set $A_1=V(\tilde
  G_1)$. For $i=1,\dots,\tilde n-1$, given a connected feedback vertex set $A_i\supseteq\hat X_i$ of $\tilde
  G_i$ with $|A_i\setminus\hat X_i|\le k'$, let $\mathrm{Mod}=(A_i\setminus\{w_i\})\cup\{u_i,v_i\}$ in $\tilde
  G_{i+1}$, and let $C=\mathrm{Mod}$, $C'=\emptyset$ if $w_i\in A_i$, and $C=A_i$, $C'=\{u_i,v_i\}$ otherwise. Start
  with $Y_{\mathrm{in}}=\hat X_{i+1}$ and go through the vertices $v\notin\hat X_{i+1}$ in a fixed order:
  run the test of \cref{lem:fvs-test} on $(\tilde G_{i+1},Y_{\mathrm{in}}\cup\{v\},\hat X_{i+1},k',\mathrm{Mod})$
  independently $\lceil\log(4\tilde n^2)\rceil$ times, and if one run accepts, add $v$ to
  $Y_{\mathrm{in}}$. Afterward, if $Y_{\mathrm{in}}$ is a connected feedback vertex set of $\tilde G_{i+1}$ with
  $|Y_{\mathrm{in}}\setminus\hat X_{i+1}|\le k'$, set $A_{i+1}=Y_{\mathrm{in}}$; otherwise stop without a solution.
  Output $A_{\tilde n}$. This is the self-reduction of the proof of~\cite[Theorem~5.1]{CyganNPPRW22} and of
  Theorem~B.3 of~\cite{CyganNPPRW11arxiv}, with the forced vertices kept in $Y_{\mathrm{in}}$ throughout.

  \emph{The modulator.} Write $u,v,w$ for $u_i,v_i,w_i$. The graphs $\tilde G_{i+1}-\{u,v\}$ and
  $\tilde G_i-w$ are identical, so $\tilde G_{i+1}-\mathrm{Mod}=\tilde G_i-(A_i\cup\{w\})$ is an induced subgraph of
  the forest $\tilde G_i-A_i$. Every forced vertex of $\tilde G_{i+1}$ other than $u,v$ is a forced vertex
  of $\tilde G_i$ other than $w$, hence lies in $A_i\setminus\{w\}$. If $w\in A_i$, then $\tilde G_{i+1}[\mathrm{Mod}]$ is
  connected by \cref{lem:fvs-contract}(1), because $\tilde G_{i+1}[\mathrm{Mod}]/uv=\tilde G_i[A_i]$, and $\hat
  X_{i+1}\subseteq\mathrm{Mod}=C$. If $w\notin A_i$, then $w\notin\hat X_i$, so $u,v\notin\hat X_{i+1}$, $\hat
  X_{i+1}\subseteq A_i=C$, and $\tilde G_{i+1}[A_i]=\tilde G_i[A_i]$ is connected. Finally $|\mathrm{Mod}\setminus\hat
  X_{i+1}|\le|A_i\setminus\hat X_i|+2\le k'+2$. So $\mathrm{Mod}$ is a structured modulator for $(\tilde
  G_{i+1},\hat X_{i+1},k')$, and $\hat X_{i+1}\subseteq Y_{\mathrm{in}}\cup\{v\}$ in every test.

  \emph{Correctness.} Every set $A_i$ is checked before it is kept, so an output is a connected
  feedback vertex set $A\supseteq\hat X$ of $\tilde G$ with $|A\setminus\hat X|\le k'$. Suppose such a set
  exists. By \cref{lem:fvs-closure}, applied along the contraction sequence, so does every $\tilde G_i$
  for $\hat X_i$. Step $i$ is the self-reduction of \cref{lem:wt-self-reduction} with
  $\Omega=V(\tilde G_{i+1})$, $Z_0=\hat X_{i+1}$, the family $\mathcal P$ of the connected feedback vertex
  sets $Y\supseteq\hat X_{i+1}$ of $\tilde G_{i+1}$ with $|Y\setminus\hat X_{i+1}|\le k'$, which is nonempty,
  the test $\mathsf T(Y_{\mathrm{in}})$ of \cref{lem:fvs-test} on
  $(\tilde G_{i+1},Y_{\mathrm{in}},\hat X_{i+1},k',\mathrm{Mod})$, $\delta=1/(4\tilde n)$ and
  $\rho=\lceil\log(4\tilde n^2)\rceil\ge\log(|\Omega|/\delta)$, since $|\Omega|\le\tilde n$. By
  \cref{lem:fvs-test}, the test rejects $Y_{\mathrm{in}}$ if no member of $\mathcal P$ contains it
  (for $Y\supseteq\hat X_{i+1}$ the conditions $|Y|\le|\hat X_{i+1}|+k'$ and
  $|Y\setminus\hat X_{i+1}|\le k'$ agree), and with probability at least $1/2$ if one does. So step $i$
  fails to produce a member of $\mathcal P$ with probability at most $1/(4\tilde n)$; there are fewer
  than $\tilde n$ steps, so by the union bound a solution is found with probability at least $3/4$.
  Each step runs at most $\tilde n\lceil\log(4\tilde n^2)\rceil$ tests, each in time
  $3^{k'}2^{|\hat X_{i+1}|}\tilde n^{O(1)}\le3^{k'}2^{|\hat X|}\tilde n^{O(1)}$ and polynomial space by
  \cref{lem:fvs-test}; only the current $A_i$ is stored between steps. So the forced problem is solved
  in time $3^{k'}2^{|\hat X|}\tilde n^{O(1)}$ and polynomial space.

  \emph{The algorithm $\cA$.} Let $\cA$ be the algorithm of \cref{lem:fvs-wrapper} with $w\equiv0$,
  $\theta=0$ and, as $\mathrm{Forced}$, the procedure above, which ignores $\tilde w$ and $\theta'$;
  every call then has $\tilde w\equiv0$ and $\theta'=0$, so every output satisfies
  $\tilde w(A)\le\theta'$, and the hypotheses of \cref{lem:fvs-wrapper} hold with $w_{\max}=0$ and
  $p=3/4$. The lemma gives the time, space and correctness claims. The running-time
  exponent~\eqref{eq:fvs-g} is nonnegative and computable in polynomial time; comparisons of
  $g(G,X,k)$ with numbers of the form $i\log3+j$ with integers $i,j$, which are the only ones the
  algorithms below make, reduce to integer comparisons.
\end{proof}

\subsubsection{Tightness of the base \texorpdfstring{$1.8$}{1.8}}\label{sec:app-fvs-tight}

This subsection proves \cref{prop:cc-tight,prop:cc-cfvs-tight}. The proof of \cref{prop:cc-tight} uses the
following evaluation of the infimum over $y$ of the function $\phi_{3,\sigma}$ of~\eqref{eq:cc-phi}.

\begin{lemma}[The critical fraction for $c=3$]\label{lem:cc-inf}
  For $\sigma^*=26/45$, $\inf_{y\ge0}\phi_{3,\sigma^*}(y)=\log1.8$, attained at $y=4/5$.
\end{lemma}

\begin{proof}
  The map $y\mapsto u=\lambda_{\max}(M_{3y/2})$ is an increasing bijection from $[0,\infty)$ onto
  $[1,\infty)$. Its inverse is $\zeta=3y/2=(u^2-u)/(2u-1)$, because $u$ is a root of
  $x^2-(1+2\zeta)x+\zeta$ (and the larger one, as the other root $\zeta/u=(u-1)/(2u-1)$ is less
  than~$1$). Substituting, $1+y=(2u^2+4u-3)/(3(2u-1))$, and
  $\phi_{3,\sigma^*}(y)\ln2=\hat\phi(u)$ with
  \[
    \hat\phi(u)=\ln(2u^2+4u-3)-\ln(2u-1)-\ln3-\sigma^*\ln\frac u3\qquad(u\ge1).
  \]
  Differentiating and bringing the three terms to a common denominator gives
  \[
    u\,\hat\phi'(u)=\frac{4u^2+4u}{2u^2+4u-3}-\frac{2u}{2u-1}-\frac{26}{45}
    =\frac{(u-3)(76u^2-108u+26)}{45\,(2u^2+4u-3)(2u-1)} .
  \]
  The denominator is positive for $u\ge1$. The quadratic $76u^2-108u+26$ is negative at
  $u=1$ and has exactly one root $u_1=(54+\sqrt{940})/76\in(1,3)$ in $[1,\infty)$. So $\hat\phi$
  increases on $[1,u_1]$, decreases on $[u_1,3]$ and increases on $[3,\infty)$, and its
  minimum on $[1,\infty)$ is $\min(\hat\phi(1),\hat\phi(3))$. Now $\hat\phi(3)=\ln(27/15)=\ln1.8$, which is
  the value at $y=4/5$, and $\hat\phi(1)=\sigma^*\ln3$. Finally $\sigma^*\ln3>\ln1.8$ is equivalent to
  $5^{45}>3^{64}$, which holds: $5^{45}=125^{15}>(\tfrac32\cdot81)^{15}=(\tfrac32)^{15}3^{60}$, and
  $(\tfrac32)^{15}>(\tfrac32)^{12}=(\tfrac{81}{16})^3>5^3>3^4$.
\end{proof}

\RestPropCcTight*

\begin{proof}
  (a) By the assumed lower bound on $g$, $W_S(y)\le2^{\gamma'}3^{-|S|}\mathcal Z_F(3y/2)$ for $y\ge0$, by the
  computation in the proof of \cref{cor:cc-T}(a) with the inequality reversed. By \cref{lem:cc-path} with $\zeta=3y/2$,
  $3^{-|S|}\mathcal Z_F(\zeta)\le2(\lambda_{\max}(M_\zeta)/3)^{|S|}$, so, with $\sigma^*=26/45$,
  $W_S(y)/(1+y)^n\le2^{\gamma'+1}2^{-n\phi_{3,\sigma^*}(y)}\le2^{\gamma'+1}(5/9)^n$ by \cref{lem:cc-inf}.
  By \cref{lem:fw-gf}, $\max_t\mu_t(S)/\binom nt\le(n+1)\sup_{y\ge0}W_S(y)/(1+y)^n$, which proves~(a).

  (b) Let $|S|\le k\le n$. By \cref{prop:fw-necessary} for $S$ and $k$, the expected number of trials of the run on $(I,k)$
  that run $\cA$ on a subset of $S$ is at most $P_{\mathrm{nec}}\cdot2^{\psi n}\max_{t\le|S|}\mu^{(k)}_t(S)/\binom
  nt$. As $g$ is nondecreasing in its budget, \eqref{eq:fw-moment-monotone} gives
  $\mu^{(k)}_t(S)\le\mu_t(S)$, and (a) bounds the expectation by
  $P_{\mathrm{nec}}\cdot2^{\gamma'+1}(n+1)(2^\psi\cdot5/9)^n$. By Markov's inequality, this also bounds the
  probability that at least one such trial occurs.

  (c) A run answers \YES{}, or returns a set, only after a trial with some sample $X$ of size $t$ in
  which $\cA$ accepts $(I,X,k)$. Since $\cA$ never accepts a \NO{}-instance, some
  $Y\subseteq U_I\setminus X$ with $|Y|\le k-t$ satisfies $X\cup Y\in\cF_I$, and a returned set is
  such a set $X\cup Y$. If $k\le|S|$, then $|X\cup Y|\le k\le|S|$; if the run returns a set of size at
  most $|S|$, then this set is an $X\cup Y$ of size at most $|S|$. In both cases $X\cup Y$ is a set in
  $\cF_I$ of size at most $|S|$, so $X\cup Y=S$ and $X\subseteq S$.

  (d) A run with $k<|S|$ never answers \YES{}, by item~(ii) of \cref{thm:fw-main}, since $S$ is the
  only set in $\cF_I$ of size at most $|S|$ and $|S|>k$. By (c), a \YES{} at $k=|S|$ or a returned set
  of size at most $|S|$ requires a run that runs $\cA$ on a subset of $S$, and this run has
  $k\ge|S|$, because a returned set $X\cup Y$ has size at most $|X|+(k-|X|)=k$ and equals $S$ by the
  proof of (c). By (b) and the union bound over the at most $n$ runs with $k\ge|S|$, the probability
  is at most $nP_{\mathrm{nec}}2^{\gamma'+1}(n+1)(2^\psi\cdot5/9)^n$, and
  $2^\psi\cdot5/9\le2^{-\varepsilon}\cdot1.8\cdot5/9=2^{-\varepsilon}$.
\end{proof}

\RestPropCcCfvsTight*

\begin{proof}
  Let $L\ge2$ and let $T\subseteq\{0,\dots,L-1\}$ contain $0$ and $L-1$. Let $G_{L,T}$ consist of a
  path $v_0v_1\dots v_L$ and, for each $i\in T$, a vertex $w_i$ adjacent to $v_i$ and $v_{i+1}$, and
  put $S=\{v_1,\dots,v_{L-1}\}$. We show that $S$ is the unique minimum connected feedback vertex set
  of $G_{L,T}$, that $G_{L,T}[S]$ is a path, and that
  $g(G_{L,T},X,|S|)=(|S|-|X|)\log3+|X|-e(G_{L,T}[X])$ for every $X\subseteq S$. The family
  consists of the graphs with $L=26j+1$ and $|T|=19j-2$, $j\ge1$.

  Write $G=G_{L,T}$. The set $S$ is a connected feedback vertex set: $G[S]$ is the path $v_1\dots v_{L-1}$, and $G-S$
  consists of the edges $v_0w_0$ and $v_Lw_{L-1}$ and isolated vertices. Let $A$ be any connected
  feedback vertex set. It meets the triangle $v_0v_1w_0$ in a vertex $x$ and the triangle
  $v_{L-1}v_Lw_{L-1}$ in a vertex $z$. For $1\le h\le L-1$, deleting $v_h$ separates
  $\{v_0,\dots,v_{h-1}\}\cup\{w_i:i<h\}$ from $\{v_{h+1},\dots,v_L\}\cup\{w_i:i\ge h\}$. Either
  $v_h\in\{x,z\}$, or $x$ lies in the first part and $z$ in the second; in the latter case
  $v_h\in A$ because $G[A]$ is connected. Hence $S\subseteq A$, so $S$ is the unique minimum
  solution. For $X\subseteq S$, $G[X]$ is a subgraph of a path, so $\comp(G[X])=|X|-e(G[X])$, and
  the formula for $g$ is~\eqref{eq:fvs-g}. With $L=26j+1$ and $|T|=19j-2$ (possible, since
  $2\le19j-2\le L$) we get $|S|=26j$ and $n=L+1+|T|=45j$, so $|S|/n=26/45$.

  We check the hypotheses of \cref{prop:cc-tight} for $I=G$, the path $F=G[S]$ and $\gamma'=0$. The
  running-time exponent $g(G,X,k)=(k-|X|)\log3+\comp(G[X])$ is nondecreasing in $k$, and the formula above
  gives the lower bound on $g(G,X,|S|)$ with equality. By \cref{lem:fvs-ext}, the extension
  algorithm never accepts a \NO{}-instance and returns sets $Y$ such that $X\cup Y$ is a connected
  feedback vertex set. Every connected feedback vertex set contains $S$, so $S$ is the only one of
  size at most $|S|$. The algorithm of \cref{thm:cc-cfvs} outputs $S$ only if some run returns a set
  of size at most $|S|$ (its default output $V(G)$ has size $n>|S|$), so by \cref{prop:cc-tight}(d)
  with $\gamma'=0$ this has probability at most $2nP_{\mathrm{nec}}(n+1)2^{-\varepsilon n}$, which tends
  to~$0$ as $j\to\infty$, because the input length $N$ is polynomial in $n$ for the graphs $G_{L,T}$.
\end{proof}
{}
\subsection{Weighted finishing steps}\label{sec:app-weighted}

This subsection proves \cref{thm:cc-mwcs,thm:cc-wcfvs}. In both cases the sampling analysis is that of
\cref{sec:cc-apps}, and only the extension algorithm is new. Both extension algorithms turn a randomized
decision procedure into one that outputs a witness, by the following self-reduction, which is also the
one used in the proof of \cref{lem:fvs-ext}.

\begin{lemma}[Witness by self-reduction]\label{lem:wt-self-reduction}
  Let $\Omega$ be a set with $m\ge1$ elements, let $\mathcal P$ be a nonempty family of subsets of
  $\Omega$, let $Z_0$ be a subset of some member of $\mathcal P$, let $0<\delta<1$ and let $\rho\ge\log(m/\delta)$
  be an integer. Let $\mathsf T$ be a randomized test that, on input $Z\subseteq\Omega$, never accepts if no
  member of $\mathcal P$ contains $Z$, and accepts with probability at least $1/2$ otherwise, with
  independent coins in every run. Starting from $Z=Z_0$, go through the elements $v\in\Omega\setminus Z_0$ in
  a fixed order and add $v$ to $Z$ if one of $\rho$ independent runs of $\mathsf T$ on $Z\cup\{v\}$ accepts.
  Then $\mathsf T$ is run at most $m\rho$ times, and the final set $Z$ is a member of $\mathcal P$ with
  probability at least $1-\delta$.
\end{lemma}

\begin{proof}
  Throughout, $Z$ is contained in a member of $\mathcal P$: this holds for $Z_0$, and $v$ is added only
  after an accepting run on $Z\cup\{v\}$, which by the soundness of $\mathsf T$ is contained in a member.
  For $v\in\Omega\setminus Z_0$ let $E_v$ be the event that, when $v$ is processed with the current set $Z$,
  the set $Z\cup\{v\}$ is contained in a member of $\mathcal P$ and all $\rho$ runs reject. Conditioned on
  the history before $v$ is processed, each run then rejects with probability at most $1/2$,
  independently, so $\Pr[E_v]\le2^{-\rho}\le\delta/m$, and by the union bound some $E_v$ occurs with
  probability at most $\delta$. Suppose that no $E_v$ occurs, and let $P\in\mathcal P$ contain the final set
  $Z$. If some $v\in P\setminus Z$ existed, then $v\notin Z_0$, and when $v$ was processed the current set
  was contained in $Z$, so the tested set was contained in $P$ and all runs rejected it, that is, $E_v$
  occurred. Hence $Z=P$. There are at most $m$ elements to process, each with $\rho$ runs.
\end{proof}

\subsubsection{Maximum-Weight Connected \texorpdfstring{$k$}{k}-Subgraph}\label{sec:app-weighted-mwcs}

\RestThmCcMwcs*

The finishing step is the algebraic sieve of Björklund, Kaski and Kowalik~\cite{BjorklundKK15motif}
for \pname{Graph Motif}, with one more indeterminate that records the weight, as they do for the
edit costs of \pname{Closest Graph Motif}. Their cost-constrained lemma~\cite[Lemma~3]{BjorklundKK15motif} lets the coloring vary, whereas we need the
coloring fixed and a cost attached to each vertex; so we derive the variant we need from the proofs
of their Lemmas~1 and~2. As in the proof of \cref{lem:mot-ext}, the components of the sample are
contracted and given fresh colors, which the motif demands exactly once.

\paragraph{Notation from~\cite{BjorklundKK15motif}.}
Let $H$ be a graph with vertex set $\{1,\dots,n_H\}$, let $D\subseteq V(H)$, let
$w'\colon V(H)\to\{-w_{\max},\dots,w_{\max}\}$ and let $p$ be an integer with $\max(1,|D|)\le p\le
n_H$. A \emph{branching walk} in $H$ is a pair $(T,\varphi)$ of an ordered rooted tree $T$, whose
vertices are numbered in preorder, and a homomorphism $\varphi\colon V(T)\to V(H)$; it is \emph{simple}
if $\varphi$ is injective, \emph{properly ordered} if siblings $a<b$ satisfy $\varphi(a)<\varphi(b)$,
it has \emph{size} $|V(T)|$, and it \emph{spans} $\varphi(V(T))$. With a variable $x_u$ for every
vertex $u$ and variables $y_{(u,u')},y_{(u',u)}$ for every edge $uu'$, the \emph{fingerprint} of a
walk is $\prod_{ab\in E(T),\,a<b}y_{(\varphi(a),\varphi(b))}x_{\varphi(b)}$, and
\[
  P_p(\vec x,\vec y)=\sum_{s\in V(H)}x_s\sum_{(T,\varphi)}\prod_{ab\in E(T),\,a<b}y_{(\varphi(a),\varphi(b))}x_{\varphi(b)},
\]
the inner sum over the properly ordered branching walks of size $p$ that start at $s$
(\cite[Section~3.2]{BjorklundKK15motif}). Every monomial of $P_p$ has degree $p$ in $\vec x$ and
degree $p-1$ in $\vec y$. By~\cite[Lemma~7]{BjorklundKK15motif}, a monomial of $P_p$ is multilinear
in $\vec x$ if and only if it comes from a simple walk, and a simple walk can be reconstructed from
its monomial; so the monomials of $P_p$ that are multilinear in $\vec x$ are in bijection with the
simple properly ordered branching walks of size $p$, each has coefficient~$1$, and its
$\vec x$-part is $\prod_{u\in K}x_u$ for the set $K$ that the walk spans, which induces a connected
subgraph. Conversely, every set $K$ with $|K|=p$ and $H[K]$ connected is spanned by such a walk:
take a spanning tree of $H[K]$, root it, order siblings by their vertex numbers and number the vertices
in preorder.

Color the vertices of $H$ as follows: each $u\in D$ gets its own color $c_u$ with multiplicity
$m(c_u)=1$, and every other vertex gets the color $\star$ with multiplicity $m(\star)=p-|D|$. Give
each color $c$ a set $\Sigma_c$ of $m(c)$ \emph{shades}, these sets pairwise disjoint, put
$\Sigma=\bigcup_c\Sigma_c$, so $|\Sigma|=p$, and let $\Lambda$ be a set of $p$ \emph{labels}. A set
$K\subseteq V(H)$ is \emph{properly colored} if every color $c$ occurs on at most $m(c)$ vertices
of $K$. A set $K$ with $|K|=p$ is properly colored if and only if $D\subseteq K$: if $D\subseteq K$,
each $c_u$ occurs once and $\star$ occurs $p-|D|$ times; conversely, at most $p-|D|$ vertices of $K$
have color $\star$, so at least $|D|$ vertices of $K$ lie in $D$, that is, $D\subseteq K$. The
\emph{cost} of a vertex is $\operatorname{cost}(u)=w_{\max}-w'(u)\in\{0,\dots,2w_{\max}\}$, and
$\operatorname{cost}(K)=\sum_{u\in K}\operatorname{cost}(u)=p\,w_{\max}-w'(K)$ for $|K|=p$.

Work over the field $\mathbb F=\mathbb F_{2^b}$. Introduce variables $v_{u,d}$ for $u\in V(H)$ and
$d\in\Sigma_{\mathrm{col}(u)}$, variables $\zeta_{d,j}$ for $d\in\Sigma$ and $j\in\Lambda$, and one more
variable $\lambda$. Put $\upsilon_{u,j}=\sum_{d\in\Sigma_{\mathrm{col}(u)}}v_{u,d}\zeta_{d,j}$,
$\upsilon^A_u=\sum_{j\in A}\upsilon_{u,j}$ for $A\subseteq\Lambda$, and
\begin{equation}\label{eq:mwcs-Q}
  Q(\vec v,\vec\zeta,\vec y,\lambda)=\sum_{A\subseteq\Lambda}P_p\bigl((\lambda^{\operatorname{cost}(u)}\upsilon^A_u)_{u\in V(H)},\vec y\bigr).
\end{equation}
This is the sieve~(2) of~\cite{BjorklundKK15motif} applied to $P_p(\lambda^{\operatorname{cost}(1)}x_1,\dots,
\lambda^{\operatorname{cost}(n_H)}x_{n_H},\vec y)$. Write $Q=\sum_\sigma Q_\sigma(\vec v,\vec\zeta,\vec y)\lambda^\sigma$.

\begin{lemma}[Weighted sieve identity]\label{lem:mwcs-sieve}
  Let $H$, $D$, $w'$, $p$ and the coloring be as above, and let $\sigma\ge0$ be an integer. The
  polynomial $Q_\sigma$ is nonzero if and only if some set $K\subseteq V(H)$ with $|K|=p$, $D\subseteq
  K$ and $H[K]$ connected has $\operatorname{cost}(K)=\sigma$, that is, $w'(K)=pw_{\max}-\sigma$. Moreover $Q_\sigma$
  has total degree at most $3p-1$, and $Q_\sigma=0$ for $\sigma>2pw_{\max}$.
\end{lemma}

\begin{proof}
  The idea is that the sieve of~\cite{BjorklundKK15motif} acts on each monomial of $P_p$ separately,
  and the substitution $x_u\mapsto\lambda^{\operatorname{cost}(u)}x_u$ only multiplies each monomial by a power of
  $\lambda$ determined by its $\vec x$-part. Write $P_p=\sum_m a_m\vec x^{\,\delta_m}\vec y^{\,\epsilon_m}$ with
  distinct exponent pairs $(\delta_m,\epsilon_m)$ and nonzero coefficients $a_m$. Then
  \[
    Q=\sum_m a_m\,\lambda^{\sum_u\operatorname{cost}(u)\delta_{m,u}}\,\vec y^{\,\epsilon_m}\sum_{A\subseteq\Lambda}\prod_{u}\bigl(\upsilon^A_u\bigr)^{\delta_{m,u}} .
  \]
  Fix $m$ and write $\delta=\delta_m$; then $\sum_u\delta_u=p=|\Lambda|$. The proofs of Lemmas~1 and~2
  of~\cite{BjorklundKK15motif} (equations (5)--(8)) show, in characteristic~$2$, that
  $\sum_A\prod_u(\upsilon^A_u)^{\delta_u}=0$ unless $\delta\in\{0,1\}^{V(H)}$; that argument uses only
  $\upsilon^A_u=\sum_{j\in A}\upsilon_{u,j}$ and $\sum_u\delta_u=|\Lambda|$. If $\delta$ is the indicator
  vector of a set $K$, then by~(8) and~(9) of~\cite{BjorklundKK15motif} and the paragraph after~(9),
  \[
    \sum_{A\subseteq\Lambda}\prod_{u\in K}\upsilon^A_u=R_K:=\sum_{h}\ \sum_{g}\ \prod_{u\in K}v_{u,h(u)}\zeta_{h(u),g(u)},
  \]
  where $h$ ranges over the injective maps $K\to\Sigma$ with $h(u)\in\Sigma_{\mathrm{col}(u)}$ and $g$
  over the bijections $K\to\Lambda$; the triple $(K,h,g)$ can be read off from the monomial
  $\prod_{u\in K}v_{u,h(u)}\zeta_{h(u),g(u)}$, so these monomials are distinct and $R_K\neq0$ if and
  only if such an $h$ exists, that is, if and only if $K$ is properly colored.

  By the facts quoted from~\cite[Lemma~7]{BjorklundKK15motif}, the monomials of $P_p$ that are
  multilinear in $\vec x$ are $\prod_{u\in K_W}x_u\,\vec y^{\,\epsilon_W}$ with coefficient $1$, one for
  each simple properly ordered branching walk $W$ of size $p$, where $K_W$ is the set that $W$ spans
  and $\vec y^{\,\epsilon_W}$ the $\vec y$-part of its fingerprint. Hence
  \[
    Q_\sigma=\sum_{W\,:\,\operatorname{cost}(K_W)=\sigma,\ K_W\text{ properly colored}}\vec y^{\,\epsilon_W}\,R_{K_W}.
  \]
  The monomials $\vec y^{\,\epsilon_W}\prod_{u\in K_W}v_{u,h(u)}\zeta_{h(u),g(u)}$ over all such $W$ and
  all $(h,g)$ are pairwise distinct: the $\vec v$-variables determine $K_W$ and $h$, the
  $\vec\zeta$-variables then determine $g$, and $K_W$ together with $\vec y^{\,\epsilon_W}$ is the
  monomial of $W$ in $P_p$, which determines $W$. So no cancellation occurs, and $Q_\sigma\neq0$ if and
  only if some simple walk of size $p$ spans a properly colored set $K$ with $\operatorname{cost}(K)=\sigma$.
  By the facts quoted above and the characterization of properly colored sets, this holds if and
  only if some $K$ with $|K|=p$, $D\subseteq K$ and $H[K]$ connected has $\operatorname{cost}(K)=\sigma$. Each
  monomial of $Q_\sigma$ has degree $p-1$ in $\vec y$ and $2p$ in $(\vec v,\vec\zeta)$, and
  $\operatorname{cost}(K)\le2pw_{\max}$.
\end{proof}

\begin{algorithm}[Weighted sieve]\label{alg:mwcs-sieve}
  \Input $H$, $D$, $w'$, $p$ as above and an integer $\theta'$.
  \Body
  If $p>n_H$ or
  $\theta'>pw_{\max}$, reject. Let $b=\lceil\log(6pw_{\max})\rceil$, and find an irreducible polynomial of
  degree $b$ over $\mathbb F_2$ by trying all $2^b$ monic polynomials of degree $b$ and testing each by
  division by all polynomials of degree between $1$ and $\lfloor b/2\rfloor$; it defines $\mathbb F=\mathbb F_{2^b}$.
  Draw values for all variables $\vec v,\vec\zeta,\vec y$ independently and uniformly from $\mathbb F$.
  For each of $D_\lambda+1$ distinct nonzero elements $\lambda_0\in\mathbb F$, where $D_\lambda=2pw_{\max}$, compute
  $Q(\vec v,\vec\zeta,\vec y,\lambda_0)$ by evaluating, for every $A\subseteq\Lambda$, the polynomial
  $P_p$ at the point $x_u=\lambda_0^{\operatorname{cost}(u)}\upsilon^A_u$ with the recurrence~(11)--(13)
  of~\cite{BjorklundKK15motif}, and summing over $A$. Recover the coefficients $Q_\sigma(\vec v,\vec\zeta,\vec
  y)$, $0\le\sigma\le D_\lambda$, of the univariate polynomial $\lambda\mapsto Q(\vec v,\vec\zeta,\vec y,\lambda)$ by
  Lagrange interpolation. Accept if $Q_\sigma(\vec v,\vec\zeta,\vec y)\neq0$ for some
  $\sigma\le pw_{\max}-\theta'$; otherwise reject.
\end{algorithm}

\begin{lemma}[Weighted sieve, guarantees]\label{lem:mwcs-sieve-alg}
  Let $H$ have $n_H$ vertices. \Cref{alg:mwcs-sieve} runs in polynomial space and, for every outcome of
  its coins, in time $2^p(n_Hw_{\max})^{O(1)}$. It rejects if no $K\subseteq V(H)$ with $|K|=p$,
  $D\subseteq K$ and $H[K]$ connected has $w'(K)\ge\theta'$, and if such a $K$ exists, it accepts with
  probability at least $1/2$.
\end{lemma}

\begin{proof}
  Since $w_{\max}\ge1$, we have $2^b\ge6pw_{\max}\ge\max(6p,D_\lambda+2)$, so the field has $D_\lambda+1$ nonzero
  distinct elements. By \cref{lem:mwcs-sieve}, $\lambda\mapsto Q(\vec v,\vec\zeta,\vec y,\lambda)$ is a
  polynomial of degree at most $D_\lambda$ whose coefficients are the values $Q_\sigma(\vec v,\vec\zeta,\vec
  y)$, so the interpolation recovers them. If the algorithm accepts, some $Q_\sigma$ with
  $\sigma\le pw_{\max}-\theta'$ is a nonzero polynomial, and \cref{lem:mwcs-sieve} gives a set $K$ as in
  the lemma with $w'(K)=pw_{\max}-\sigma\ge\theta'$. Conversely, let $K^*$ be such a set and
  $\sigma^*=\operatorname{cost}(K^*)=pw_{\max}-w'(K^*)\le pw_{\max}-\theta'$. By \cref{lem:mwcs-sieve}, $Q_{\sigma^*}$ is a nonzero
  polynomial of total degree at most $3p-1$, so by the lemma of DeMillo, Lipton, Schwartz and Zippel
  in the form of~\cite[Lemma~8]{BjorklundKK15motif} its value at a uniformly random point is zero with
  probability at most $(3p-1)/2^b<1/2$.

  The search for the irreducible polynomial takes time $2^{3b/2}b^{O(1)}=(pw_{\max})^{O(1)}$, as
  $2^b<12pw_{\max}$, and every field operation takes time $b^{O(1)}$. For fixed $\lambda_0$ and $A$, the
  point $(\lambda_0^{\operatorname{cost}(u)}\upsilon^A_u)_u$ is computed with $O(p^2n_H+n_Hw_{\max})$ field operations from the
  values $\upsilon_{u,j}$, which are computed once, and $P_p$ is evaluated with $O(p^2|E(H)|)$ field
  operations by the recurrence~(11)--(13) (\cite[Section~3.5]{BjorklundKK15motif}). There are
  $(D_\lambda+1)2^p$ such pairs, and the interpolation takes $D_\lambda^{O(1)}$ field operations. The
  algorithm stores the $D_\lambda+1$ sums and the tables of one evaluation of the recurrence, which is
  polynomial space.
\end{proof}

The sieve only decides. A set is found by deleting vertices, as follows.

\begin{algorithm}[Weighted finishing with a witness]\label{alg:mwcs-fin}
  \Input $H$, $D$, $w'$, $p$, $\theta'$ as in \cref{alg:mwcs-sieve}.
  \Body
  Put
  $\rho=\lceil\log(4n_H)\rceil$ and $H_{\mathrm{cur}}=H$. For every $u\in V(H)\setminus D$ in a fixed order,
  run \cref{alg:mwcs-sieve} on $(H_{\mathrm{cur}}-u,D,w',p,\theta')$ independently $\rho$ times, and if one
  run accepts, replace $H_{\mathrm{cur}}$ by $H_{\mathrm{cur}}-u$. Afterward, if $K=V(H_{\mathrm{cur}})$ satisfies
  $|K|=p$, $H[K]$ connected and $w'(K)\ge\theta'$, output $K$; otherwise reject.
\end{algorithm}

\begin{lemma}[Weighted finishing step]\label{lem:mwcs-fin}
  \Cref{alg:mwcs-fin} runs in polynomial space and, for every outcome of its coins, in time
  $2^p(n_Hw_{\max})^{O(1)}$. Every output $K$ satisfies $|K|=p$, $D\subseteq K$, $H[K]$ connected and
  $w'(K)\ge\theta'$. If such a set exists, the algorithm outputs one with probability at least $3/4$.
\end{lemma}

\begin{proof}
  The output is checked explicitly. Suppose that a set as in the lemma exists. If $V(H)=D$, then this set
  is $V(H)$, the loop is empty, and the final check outputs it. Otherwise, with $Z$ the set of
  deleted vertices, so that $H_{\mathrm{cur}}=H-Z$, the algorithm is the self-reduction of
  \cref{lem:wt-self-reduction} with $\Omega=V(H)\setminus D$, $Z_0=\emptyset$, the family $\mathcal P$ of the
  sets $V(H)\setminus K$ over the sets $K$ with $|K|=p$, $D\subseteq K$, $H[K]$ connected and
  $w'(K)\ge\theta'$, the test $\mathsf T(Z)$ given by \cref{alg:mwcs-sieve} on $(H-Z,D,w',p,\theta')$, $\delta=1/4$
  and $\rho=\lceil\log(4n_H)\rceil\ge\log(|\Omega|/\delta)$. The family is nonempty by assumption, and
  $Z_0$ lies in every member. By \cref{lem:mwcs-sieve-alg}, the test rejects $Z$ if $H-Z$ contains
  no such set $K$, that is, if $Z$ lies in no member of $\mathcal P$, and it accepts
  with probability at least $1/2$ if this holds, since then $K\subseteq V(H-Z)$ and $(H-Z)[K]=H[K]$. So,
  with probability at least $3/4$, the final $Z$ is $V(H)\setminus K$ for such a $K$, and $K$ is output.
  There are at most $n_H\rho$ runs of \cref{alg:mwcs-sieve}, each within the bounds of
  \cref{lem:mwcs-sieve-alg}.
\end{proof}

\paragraph{The extension problem.}
For an instance $I=(G,w,k)$ and an integer $\theta$, let $\Phi_\theta$ be the implicit set system with
universe $V(G)$ whose solutions are the connected $k$-sets $S$ with $w(S)\ge\theta$. Membership is
decided in polynomial time, and every solution has size~$k$. For a nonempty $X\subseteq V(G)$, let
$X_1,\dots,X_q$ be the vertex sets of the components of $G[X]$, and let $G_X$ be obtained from $G$ by
contracting each $X_i$ to a new vertex $d_i$, as in \cref{sec:fvs-graphs}; here $G$ need not be
connected, and $\hat X=\{d_1,\dots,d_q\}$ is an independent set of $G_X$. For $X=\emptyset$ put $G_X=G$
and $\hat X=\emptyset$. Define $w_X\colon V(G_X)\to\{-w_{\max},\dots,w_{\max}\}$ by $w_X=w$ on $V(G)\setminus X$
and $w_X(d_i)=0$.

For $Y\subseteq V(G)\setminus X$, the graph $G_X[\hat X\cup Y]$ is obtained from $G[X\cup Y]$ by
contracting the connected sets $X_1,\dots,X_q$: the vertex $d_i$ is adjacent to $y\in Y$ exactly when
$y$ has a neighbor in $X_i$, edges inside $Y$ are unchanged, and there are no edges between
different $X_i$. So, by \cref{lem:fvs-contract}(1), $G[X\cup Y]$ is connected if and only if
$G_X[\hat X\cup Y]$ is. Moreover $|\hat X\cup Y|=|Y|+q$ and $w_X(\hat X\cup Y)=w(X\cup Y)-w(X)$. Every set
$K\subseteq V(G_X)$ with $\hat X\subseteq K$ is $\hat X\cup Y$ for $Y=K\setminus\hat X\subseteq V(G)\setminus X$.
Consequently, for every integer $\ell\ge0$, some $Y\subseteq V(G)\setminus X$ with $|Y|\le\ell-|X|$ makes $X\cup
Y$ a solution of $\Phi_\theta$ if and only if $|X|\le k\le\ell$ and $G_X$ has a set $K\supseteq\hat X$ with
$|K|=k-|X|+q$, $G_X[K]$ connected and $w_X(K)\ge\theta-w(X)$, and then $Y=K\setminus\hat X$ is such a
completion.

The extension algorithm $\cA_{\mathrm{mw}}$, given $(I,X,\ell)$ for $\Phi_\theta$, where the size bound $\ell$ of
the extension problem is distinct from the size $k$ in the instance, rejects if $|X|>k$ or
$\ell<k$; otherwise it computes $G_X$, $\hat X$ and $w_X$, runs \cref{alg:mwcs-fin} on
$(G_X,\hat X,w_X,k-|X|+q,\theta-w(X))$, where $q=\comp(G[X])$, and if this outputs $K$, accepts with
$Y=K\setminus\hat X$. (The input of \cref{alg:mwcs-fin} satisfies $p=k-|X|+q\ge\max(1,q)$ because $k\ge1$.)
Its running-time exponent is
\begin{equation}\label{eq:mwcs-g}
  g_{\mathrm{mw}}(I,X,\ell)=
  \begin{cases}(k-|X|)+\comp(G[X])&\text{if }|X|\le k\le\ell,\\0&\text{otherwise.}
  \end{cases}
\end{equation}

\begin{lemma}[Extension for \pname{Maximum-Weight Connected $k$-Subgraph}]\label{lem:mwcs-ext}
  Let $1\le w_{\max}\le n^{O(1)}$. The algorithm $\cA_{\mathrm{mw}}$ solves $\Phi_\theta$-Extension with
  one-sided error: if it accepts, $X\cup Y$ is a connected $k$-set of weight at least $\theta$ with
  $Y\subseteq V(G)\setminus X$ and $|Y|\le\ell-|X|$, and on a \YES{}-instance it accepts with probability at
  least $3/4$. It runs in polynomial space and, for every outcome of its coins, in time
  $2^{g_{\mathrm{mw}}(I,X,\ell)}n^{O(1)}$.
\end{lemma}

\begin{proof}
  Correctness follows from the equivalence before~\eqref{eq:mwcs-g} and \cref{lem:mwcs-fin}. If the
  algorithm runs \cref{alg:mwcs-fin}, then $p=k-|X|+\comp(G[X])=g_{\mathrm{mw}}(I,X,\ell)$ and $G_X$ has at
  most $n$ vertices, so \cref{lem:mwcs-fin} gives time $2^{g_{\mathrm{mw}}}(nw_{\max})^{O(1)}=2^{g_{\mathrm{mw}}}n^{O(1)}$;
  otherwise it takes polynomial time. The running-time exponent is an integer computable in polynomial time.
\end{proof}

\begin{proof}[Proof of \cref{thm:cc-mwcs}]
  A connected $k$-set exists if and only if some component of $G$ has at least $k$ vertices; then the
  first $k$ vertices of a breadth-first search order of that component form one. If none exists,
  report this. Otherwise, for every $\theta\in\{-kw_{\max},\dots,kw_{\max}\}$, run Algorithm~\MLS{} for
  $\Phi_\theta$ and the size $k$ as in \cref{cor:cc-rate}, with the extension algorithm $\cA_{\mathrm{mw}}$,
  $c=2$, $\gamma=0$, the rate $\psi=0.71755$ of \cref{lem:cc-cert}, $\Gamma_0=16$ and $\kappa=1$. Output the heaviest set returned
  by these runs, and, if no run returns a set, the connected $k$-set found by breadth-first search.

  By \cref{lem:mwcs-ext}, $\cA_{\mathrm{mw}}$ satisfies the hypotheses of \cref{cor:fw-onesided}, with running-time
  exponent $g_{\mathrm{mw}}$. It has a component-count bound with constants $(2,0)$: for a solution $S$
  with $|S|=k$, let $F$ be a spanning tree of $G[S]$; for $X\subseteq S$ and the size bound $\ell=k$ we have $|X|\le k\le\ell$,
  so $g_{\mathrm{mw}}(I,X,k)=(k-|X|)+\comp(G[X])\le(k-|X|)\log2+|X|-e(F[X])$, because $F[X]$ is a
  spanning forest of a subgraph of $G[X]$. By \cref{lem:cc-cert}, every $\sigma\in(0,1]$ satisfies one
  of~\ref{it:cc-R1}--\ref{it:cc-R3} with these constants. So, by \cref{cor:cc-rate}, every run takes
  time $2^{0.71755n}n^{O(1)}$ and polynomial space, every returned set is a solution of $\Phi_\theta$, hence a
  connected $k$-set, and the run for $\theta$ equal to the maximum weight returns a set with
  probability at least $1-1/e$; that set has maximum weight. There are $2kw_{\max}+1\le n^{O(1)}$
  runs, so the total time is $2^{0.71755n}n^{O(1)}$, which is $O(1.6444^n)$ because $0.71755<\log1.6444$
  (\cref{lem:cc-cert}).
\end{proof}

\subsubsection{Vertex-weighted Connected Feedback Vertex Set}\label{sec:app-weighted-wcfvs}

\RestThmCcWcfvs*

The weighted extension algorithm combines two ingredients. The unweighted extension algorithm
(\cref{lem:fvs-ext}) provides a connected feedback vertex set of small excess, which serves as a
structured modulator; it is only needed as a modulator, so the intermediate graphs of its iterative
compression never need weights. The weighted test then runs the counting of~\ref{it:fvs-CC1}
and~\ref{it:fvs-CC3} with a weight function that combines the isolating weights with the vertex
weights.

\paragraph{The counted objects.}
We use the families of the proof of Theorem~A.13 of~\cite{CyganNPPRW11arxiv}, with a general weight function,
and we prove the two properties of them that the weighted test needs, so that this subsection does not
depend on their dynamic program. Let $\tilde G$ be a graph with $\tilde n$ vertices, $Y_{\mathrm{in}}\subseteq V(\tilde G)$
nonempty and $r\in Y_{\mathrm{in}}$, and let $\tilde U=V(\tilde G)\times\{\mathrm F,\mathrm M\}$. For a weight function
$\mathrm{wt}\colon\tilde U\to\{1,\dots,N\}$ and integers $a,e,c,W$, let $\mathcal R^{a,e,c}_W$ be the family of pairs
$(Z,M)$ with $Z\subseteq V(\tilde G)\setminus Y_{\mathrm{in}}$, $|Z|=a$, $|E(\tilde G[Z])|=e$, $M\subseteq Z$, $|M|=c$ and
$\mathrm{wt}(Z\times\{\mathrm F\})+\mathrm{wt}(M\times\{\mathrm M\})=W$. Let $\mathcal S^{a,e,c}_W$ consist of the pairs in
$\mathcal R^{a,e,c}_W$ such that $\tilde G[Z]$ is a forest with a vertex of $M$ in each of its components
and $\tilde G-Z$ is connected. Let $\mathcal C^{a,e,c}_W$ consist of the triples of a pair $(Z,M)\in
\mathcal R^{a,e,c}_W$, a consistent cut $(Z_1,Z_2)$ of $\tilde G[Z]$ with $M\subseteq Z_1$, and a consistent cut
$(Y_1,Y_2)$ of $\tilde G-Z$ with $r\in Y_1$; a cut of a graph is \emph{consistent} if no edge joins its two sides.
(We write $Z$ for the candidate forest, which~\cite{CyganNPPRW11arxiv} call $X$, and $Y_{\mathrm{in}}$ for their $S$.)

\begin{lemma}[Cut parity]\label{lem:wfvs-parity}
  For every weight function $\mathrm{wt}$ and all integers $a,e,c,W$ with $c\le a-e$,
  $|\mathcal S^{a,e,c}_W|\equiv|\mathcal C^{a,e,c}_W|\pmod2$.
\end{lemma}

\begin{proof}
  Fix $(Z,M)\in\mathcal R^{a,e,c}_W$. A cut of $\tilde G[Z]$ is consistent exactly when it puts each component of
  $\tilde G[Z]$ on one side. The components that contain a vertex of $M$ must lie on side~$1$, and each of the
  $\ell$ other components can lie on either side, so there are $2^\ell$ cuts $(Z_1,Z_2)$. Likewise there are
  $2^{\comp(\tilde G-Z)-1}$ cuts $(Y_1,Y_2)$, since the component of $r\in Y_{\mathrm{in}}\subseteq V(\tilde G)\setminus Z$ must lie on
  side~$1$. So $(Z,M)$ lies in $2^{\comp(\tilde G-Z)-1+\ell}$ triples, and this number is odd if and only if
  $\tilde G-Z$ is connected and every component of $\tilde G[Z]$ contains a vertex of $M$. In that case
  $\comp(\tilde G[Z])\le|M|=c\le a-e=|Z|-|E(\tilde G[Z])|$. Every graph $H$ satisfies $\comp(H)\ge|V(H)|-|E(H)|$, with
  equality if and only if $H$ is a forest, so $\tilde G[Z]$ is a forest and $(Z,M)\in\mathcal S^{a,e,c}_W$. Conversely,
  every pair in $\mathcal S^{a,e,c}_W$ lies in exactly one triple. Summing over the pairs gives the claim.
\end{proof}

A triple in $\mathcal C^{a,e,c}_W$ is determined by its \emph{state map} $\sigma\colon V(\tilde G)\to\Sigma$,
$\Sigma=\{\mathsf y_1,\mathsf y_2,\mathsf z_1,\mathsf z_2\}$, which sends $v$ to $\mathsf y_j$ if $v\in Y_j$ and to $\mathsf z_j$ if
$v\in Z_j$, together with the marker set $M$. A map $\sigma$ and a set $M$ come from a triple, in the family with
the indices computed from them, if and only if
\begin{enumerate}[label=(S\arabic*)]
  \item\label{it:wfvs-S1} $\sigma(v)\in\{\mathsf y_1,\mathsf y_2\}$ for $v\in Y_{\mathrm{in}}$, and $\sigma(r)=\mathsf y_1$;
  \item\label{it:wfvs-S2} no edge $uv$ has $\{\sigma(u),\sigma(v)\}=\{\mathsf y_1,\mathsf y_2\}$ or
    $\{\sigma(u),\sigma(v)\}=\{\mathsf z_1,\mathsf z_2\}$;
  \item\label{it:wfvs-S3} $M\subseteq\sigma^{-1}(\mathsf z_1)$.
\end{enumerate}
The indices are $a=|\sigma^{-1}(\{\mathsf z_1,\mathsf z_2\})|$, the number $e$ of edges with both ends in
$\sigma^{-1}(\{\mathsf z_1,\mathsf z_2\})$, $c=|M|$ and $W=\sum_{v\in\sigma^{-1}(\{\mathsf z_1,\mathsf z_2\})}\mathrm{wt}(v,\mathrm F)+\sum_{v\in M}
\mathrm{wt}(v,\mathrm M)$. These are the four states $0_1,0_2,1_1,1_2$ of the dynamic program
of~\cite{CyganNPPRW11arxiv} (see~\ref{it:fvs-CC2}). For $\mathrm{Mod}\subseteq V(\tilde G)$ and a coloring
$\varsigma\colon\mathrm{Mod}\to\Sigma$, the \emph{core evaluation} is the family of numbers $|\mathcal C^{a,e,c}_W(\varsigma)|\bmod 2$, where
$\mathcal C^{a,e,c}_W(\varsigma)$ consists of the triples in $\mathcal C^{a,e,c}_W$ whose state map agrees with $\varsigma$ on
$\mathrm{Mod}$. A coloring that violates~\ref{it:wfvs-S1} or~\ref{it:wfvs-S2} on $\mathrm{Mod}$ has empty families, so
$|\mathcal C^{a,e,c}_W|$ is the sum of $|\mathcal C^{a,e,c}_W(\varsigma)|$ over the reasonable colorings $\varsigma$ of
$\mathrm{Mod}$ (\cref{sec:fvs-cc}).

\begin{lemma}[Weighted core evaluation]\label{lem:wfvs-core}
  Let $\mathrm{wt}\colon\tilde U\to\{1,\dots,N\}$, let $\mathrm{Mod}\subseteq V(\tilde G)$ be such that $\tilde G-\mathrm{Mod}$ is a forest, and
  let $\varsigma\colon\mathrm{Mod}\to\Sigma$. Then the numbers $|\mathcal C^{a,e,c}_W(\varsigma)|\bmod2$ for all $0\le a,c\le\tilde n$,
  $0\le e\le\tilde n^2$ and $0\le W\le2N\tilde n$ (all other values are $0$) can be computed in time and space
  polynomial in $\tilde n$ and $N$. The dynamic program has four states for each vertex of the forest, whatever
  the weight function; $\mathrm{wt}$ affects only the range of the index $W$ and the values added to it.
\end{lemma}

\begin{proof}
  We compute the counts exactly and reduce them modulo $2$. A \emph{table} is an array $T(a,e,c,W)$ of integers
  indexed as in the lemma, and $T*T'$ denotes the convolution
  $(T*T')(a,e,c,W)=\sum T(a',e',c',W')\,T'(a-a',e-e',c-c',W-W')$, the sum over all index vectors with both arguments
  in range. For an index vector $\iota$, $[\iota]$ is the table with entry $1$ at $\iota$ and $0$ elsewhere, and
  $T^{+e}$ is $T$ shifted by $1$ in the index $e$.

  Write $Z(\sigma)=\sigma^{-1}(\{\mathsf z_1,\mathsf z_2\})$. We assign each vertex and each edge of $\tilde G$ to one
  part of the computation. Let $F=\tilde G-\mathrm{Mod}$, root each tree of $F$ at a vertex, and let $T_v$ be the
  subtree of $F$ rooted at $v$. Then every edge of $\tilde G$ lies inside $\mathrm{Mod}$, joins a forest vertex to
  $\mathrm{Mod}$, or joins a forest vertex $u$ to its parent $v$ in $F$. We call the last kind the \emph{parent
  edge} of $u$.

  \emph{The modulator.} If $\varsigma$ violates~\ref{it:wfvs-S1} or~\ref{it:wfvs-S2} on $\mathrm{Mod}$, all counts are $0$.
  Otherwise let $e_0$ be the number of edges of $\tilde G[\mathrm{Mod}]$ with both ends in $\varsigma^{-1}(\{\mathsf z_1,\mathsf z_2\})$, and put
  \[
    T_{\mathrm{Mod}}=[(0,e_0,0,0)]*\mathop{\ast}_{m\in\mathrm{Mod}}L_m,
  \]
  where $L_m=[(0,0,0,0)]$ if $\varsigma(m)\in\{\mathsf y_1,\mathsf y_2\}$, $L_m=[(1,0,0,\mathrm{wt}(m,\mathrm F))]$ if
  $\varsigma(m)=\mathsf z_2$, and $L_m=[(1,0,0,\mathrm{wt}(m,\mathrm F))]+[(1,0,1,\mathrm{wt}(m,\mathrm F)+\mathrm{wt}(m,\mathrm M))]$ if
  $\varsigma(m)=\mathsf z_1$. The two terms for $\mathsf z_1$ are the choices $m\notin M$ and $m\in M$.

  \emph{A forest vertex alone.} For a vertex $v$ of $F$ and $s\in\Sigma$, let $L_v[s]=0$ if $s$ violates~\ref{it:wfvs-S1} at
  $v$ (that is, $v\in Y_{\mathrm{in}}$ and $s\notin\{\mathsf y_1,\mathsf y_2\}$, or $v=r$ and $s\neq\mathsf y_1$), or if some edge $vm$ with $m\in\mathrm{Mod}$ has $\{s,\varsigma(m)\}=\{\mathsf y_1,\mathsf y_2\}$ or
  $\{\mathsf z_1,\mathsf z_2\}$. Otherwise let $d_v$ be the number of neighbors $m\in\mathrm{Mod}$ of $v$ with
  $\varsigma(m)\in\{\mathsf z_1,\mathsf z_2\}$, and let $L_v[s]$ be $[(0,0,0,0)]$ for $s\in\{\mathsf y_1,\mathsf y_2\}$,
  $[(1,d_v,0,\mathrm{wt}(v,\mathrm F))]$ for $s=\mathsf z_2$, and
  $[(1,d_v,0,\mathrm{wt}(v,\mathrm F))]+[(1,d_v,1,\mathrm{wt}(v,\mathrm F)+\mathrm{wt}(v,\mathrm M))]$ for $s=\mathsf z_1$.

  \emph{Subtrees.} Call $s,s'\in\Sigma$ \emph{compatible} if $\{s,s'\}$ is neither $\{\mathsf y_1,\mathsf y_2\}$ nor
  $\{\mathsf z_1,\mathsf z_2\}$. Going up from the leaves of $F$, for each vertex $v$ with children $u_1,\dots,u_d$ and each
  $s\in\Sigma$ put
  \[
    A_v[s]=L_v[s]*\mathop{\ast}_{i=1}^{d}\ \sum_{s'\text{ compatible with }s}A_{u_i}[s']^{(s,s')},
  \]
  where $A^{(s,s')}=A^{+e}$ if $s,s'\in\{\mathsf z_1,\mathsf z_2\}$ and $A^{(s,s')}=A$ otherwise. The shift counts the
  parent edge of $u_i$ when both of its ends lie in $Z(\sigma)$. Finally, with $\rho_1,\dots,\rho_t$ the roots of the
  trees of $F$, output
  \[
    T_\varsigma=T_{\mathrm{Mod}}*\mathop{\ast}_{j=1}^{t}\ \sum_{s\in\Sigma}A_{\rho_j}[s]
  \]
  reduced modulo~$2$.

  \emph{Correctness.} By induction from the leaves, $A_v[s](a,e,c,W)$ is the number of pairs $(\sigma_v,M_v)$ with
  the following properties. Here $\sigma_v\colon V(T_v)\to\Sigma$ satisfies $\sigma_v(v)=s$,
  satisfies~\ref{it:wfvs-S1} on $V(T_v)$, and satisfies~\ref{it:wfvs-S2} on the edges inside $T_v$ and on the edges
  from $T_v$ to $\mathrm{Mod}$, with $\varsigma$ on $\mathrm{Mod}$. The set $M_v$ is a subset of $\sigma_v^{-1}(\mathsf z_1)$. The
  index $a$ is $|Z(\sigma_v)|$, and $e$ is the number of those edges with both ends in
  $Z(\sigma_v)\cup\varsigma^{-1}(\{\mathsf z_1,\mathsf z_2\})$. The index $c$ is $|M_v|$, and $W$ is the
  $\mathrm{wt}$-weight of $Z(\sigma_v)\times\{\mathrm F\}$ and $M_v\times\{\mathrm M\}$. The induction step holds because
  such a pair for $T_v$ is exactly a choice at $v$, counted by $L_v[s]$, together with one such pair for each
  $T_{u_i}$ whose root state is compatible with $s$, so that~\ref{it:wfvs-S2} holds on the parent edge of $u_i$.
  All four indices add over these parts, and the parent edge of $u_i$ is added to $e$ exactly when both of its
  ends lie in $Z(\sigma_v)$. In the same way, $T_\varsigma$ before reduction counts the pairs $(\sigma,M)$ with
  $\sigma|_{\mathrm{Mod}}=\varsigma$ that satisfy~\ref{it:wfvs-S1}--\ref{it:wfvs-S3}, by their indices. Each vertex
  of $\tilde G$ is counted in exactly one of the parts, and so is each edge: an edge inside $\mathrm{Mod}$ in $e_0$, an edge
  from a forest vertex $v$ to $\mathrm{Mod}$ in $L_v$, and a parent edge in the shift. By the correspondence between
  triples and pairs $(\sigma,M)$, this number is $|\mathcal C^{a,e,c}_W(\varsigma)|$. All indices stay in range,
  since $a,c\le\tilde n$, $e\le\tilde n^2$ and $W\le2N\tilde n$, so the truncation in the convolution loses
  nothing.

  \emph{Time and space.} A table has $(\tilde n+1)^2(\tilde n^2+1)(2N\tilde n+1)$ entries. Each entry is a
  nonnegative integer at most $5^{\tilde n}$, a bound on the number of pairs $(\sigma,M)$ (each vertex has state
  $\mathsf y_1$, $\mathsf y_2$ or $\mathsf z_2$, or state $\mathsf z_1$ with or without a marker), so it has $O(\tilde n)$ bits. One
  convolution therefore takes time polynomial in $\tilde n$ and $N$. The computation keeps four tables $A_v[s]$
  for each vertex whose parent has not been processed, at most $4\tilde n$ tables in all, and performs
  $O(\tilde n+|\mathrm{Mod}|)$ convolutions and sums for each of the at most $\tilde n$ vertices. The state set
  $\Sigma$, the marker choice and the recursion do not depend on $\mathrm{wt}$. The weight function enters only through
  the values $\mathrm{wt}(v,\mathrm F)$ and $\mathrm{wt}(v,\mathrm M)$ placed in the index $W$ by the tables $L_m$ and
  $L_v$, and through the range $0\le W\le2N\tilde n$ of that index.
\end{proof}

For $N=4\tilde n$ and the weights of~\ref{it:fvs-CC1}, \cref{lem:wfvs-core} is a direct proof of the core
evaluation of~\ref{it:fvs-CC3}. In the weighted test we use it with a weight function that also encodes the
vertex weights.

\begin{algorithm}[Weighted test]\label{alg:wfvs-test}
  \Input a graph $\tilde G$ with $\tilde n$ vertices, sets $\emptyset\neq\hat X\subseteq Y_{\mathrm{in}}\subseteq V(\tilde G)$,
  weights $\tilde w\colon V(\tilde G)\to\{0,\dots,w_{\max}\}$, integers $k'\ge0$ and $\theta'$, and a
  structured modulator $\mathrm{Mod}$ for $(\tilde G,\hat X,k')$.
  \Body
  Fix $r\in\hat X$ and put $\beta=8\tilde n^2+1$ and
  $N'=\beta w_{\max}+4\tilde n$. For every $\eta$ with $|Y_{\mathrm{in}}|\le\eta\le\min(|\hat X|+k',\tilde n)$:
  \begin{enumerate}
    \item draw $\mathrm{iso}\colon\tilde U\to\{1,\dots,4\tilde n\}$ uniformly at random, and put
      $\mathrm{wt}'(v,\mathrm F)=\beta\tilde w(v)+\mathrm{iso}(v,\mathrm F)$ and $\mathrm{wt}'(v,\mathrm M)=\mathrm{iso}(v,\mathrm M)$, so that
      $\mathrm{wt}'$ takes values in $\{1,\dots,N'\}$;
    \item compute $|\mathcal C^{a,e,a-e}_{W}|\bmod2$ for $a=\tilde n-\eta$, all $e$ and all $W$, for the weight
      function $\mathrm{wt}'$, as the sum over the reasonable colorings of $\mathrm{Mod}$, listed by
      \cref{lem:fvs-cc-evals}, of the core evaluations of \cref{lem:wfvs-core};
    \item if $|\mathcal C^{a,e,a-e}_{W}|$ is odd for some $e$ and some $W$ with
      $\tilde w(V(\tilde G))-\lfloor W/\beta\rfloor\le\theta'$, accept.
  \end{enumerate}
  If no $\eta$ leads to acceptance, reject.
\end{algorithm}

\begin{lemma}[Weighted test]\label{lem:wfvs-test}
  \Cref{alg:wfvs-test} runs in polynomial space and, for every outcome of its coins, in time
  $3^{k'}2^{|\hat X|}(\tilde nw_{\max})^{O(1)}$. If it accepts, then $\tilde G$ has a connected feedback vertex
  set $Y\supseteq Y_{\mathrm{in}}$ with $|Y|\le|\hat X|+k'$ and $\tilde w(Y)\le\theta'$. If $\tilde G$ has such a set, the
  algorithm accepts with probability at least $1/2$.
\end{lemma}

\begin{proof}
  For a pair $(Z,M)$ write $\mathrm{iso}(Z,M)=\mathrm{iso}(Z\times\{\mathrm F\})+\mathrm{iso}(M\times\{\mathrm M\})$, and
  similarly $\mathrm{wt}'(Z,M)$. Then $\mathrm{wt}'(Z,M)=\beta\tilde w(Z)+\mathrm{iso}(Z,M)$ with $0\le\mathrm{iso}(Z,M)\le2\tilde
  n\cdot4\tilde n<\beta$, so $\lfloor\mathrm{wt}'(Z,M)/\beta\rfloor=\tilde w(Z)$ and $\mathrm{wt}'(Z,M)\bmod\beta=\mathrm{iso}(Z,M)$.
  Fix $\eta$ and put $a=\tilde n-\eta$. By \cref{lem:wfvs-core} the algorithm computes the parities
  $|\mathcal C^{a,e,a-e}_W|\bmod2$ for $\mathrm{wt}'$, which equal $|\mathcal S^{a,e,a-e}_W|\bmod2$
  by \cref{lem:wfvs-parity}.

  \emph{Soundness.} Suppose the algorithm accepts at $\eta$, $e$ and $W$. Then $\mathcal S^{a,e,a-e}_W$
  contains a pair $(Z,M)$. The set $Y=V(\tilde G)\setminus Z$ contains $Y_{\mathrm{in}}$, has size $\eta\le|\hat X|+k'$,
  $\tilde G-Y=\tilde G[Z]$ is a forest and $\tilde G[Y]=\tilde G-Z$ is connected; so $Y$ is a connected feedback vertex
  set. Its weight is $\tilde w(Y)=\tilde w(V(\tilde G))-\tilde w(Z)=\tilde w(V(\tilde G))-\lfloor W/\beta\rfloor\le\theta'$.

  \emph{Completeness.} Let $Y^*$ be a set as in the lemma and $\eta=|Y^*|$; then $|Y_{\mathrm{in}}|\le\eta\le
  \min(|\hat X|+k',\tilde n)$, so this $\eta$ is tried. Let $\mathcal P$ be the family of pairs $(Z,M)$ such
  that $Y=V(\tilde G)\setminus Z$ is a connected feedback vertex set with $Y\supseteq Y_{\mathrm{in}}$, $|Y|=\eta$ and
  $\tilde w(Y)\le\theta'$, and $M$ contains exactly one vertex of each component of the forest $\tilde G[Z]$.
  Identify $(Z,M)$ with $(Z\times\{\mathrm F\})\cup(M\times\{\mathrm M\})\subseteq\tilde U$; this is injective
  and maps $\mathrm{iso}(Z,M)$ to the weight of the set. The family $\mathcal P$ is nonempty, since $Y^*$ with one
  marker per component of $\tilde G-Y^*$ belongs to it. As $|\tilde U|=2\tilde n$, the Isolation
  Lemma~\cite[Lemma~2.5]{CyganNPPRW22} with weights drawn from $\{1,\dots,4\tilde n\}$ shows that with
  probability at least $1/2$ a unique pair $(Z^\circ,M^\circ)\in\mathcal P$ minimizes $\mathrm{iso}$ over $\mathcal P$.
  Assume this, and put $e^\circ=|E(\tilde G[Z^\circ])|$ and $W^\circ=\mathrm{wt}'(Z^\circ,M^\circ)$. We claim
  $\mathcal S^{a,e^\circ,a-e^\circ}_{W^\circ}=\{(Z^\circ,M^\circ)\}$. The pair $(Z^\circ,M^\circ)$ belongs to it,
  since $|M^\circ|=\comp(\tilde G[Z^\circ])=|Z^\circ|-e^\circ$ for a forest. Let $(Z,M)$ belong to it. Then $\tilde G[Z]$ is a
  forest with $|Z|-e^\circ$ components, each containing a vertex of $M$, and $|M|=|Z|-e^\circ$, so each
  component contains exactly one vertex of $M$; the set $Y=V(\tilde G)\setminus Z$ is a connected feedback
  vertex set containing $Y_{\mathrm{in}}$ with $|Y|=\eta$; and $\tilde w(Z)=\lfloor W^\circ/\beta\rfloor=\tilde w(Z^\circ)$, so
  $\tilde w(Y)=\tilde w(V(\tilde G)\setminus Z^\circ)\le\theta'$. Hence $(Z,M)\in\mathcal P$ with $\mathrm{iso}(Z,M)=W^\circ\bmod\beta=
  \mathrm{iso}(Z^\circ,M^\circ)$, and uniqueness gives $(Z,M)=(Z^\circ,M^\circ)$. So $|\mathcal C^{a,e^\circ,a-e^\circ}_{W^\circ}|$
  is odd, and $\tilde w(V(\tilde G))-\lfloor W^\circ/\beta\rfloor=\tilde w(V(\tilde G)\setminus Z^\circ)\le\theta'$, so the
  algorithm accepts.

  \emph{Time and space.} There are at most $\tilde n+1$ values of $\eta$. By \cref{lem:fvs-cc-evals}, $\mathrm{Mod}$
  has at most $8\cdot3^{k'}2^{|\hat X|}$ reasonable colorings, listed in that time and polynomial space;
  its hypotheses hold because $\mathrm{Mod}=C\cup C'$ is a structured modulator and $r\in\hat X\subseteq Y_{\mathrm{in}}\cap C$. By
  \cref{lem:wfvs-core} each core evaluation takes time and space polynomial in $\tilde n$ and
  $N'\le(8\tilde n^2+1)w_{\max}+4\tilde n$, and the parities are accumulated coloring by coloring.
\end{proof}

We now solve the forced problem of \cref{sec:app-fvs-ext} with weights.

\begin{algorithm}[Forced weighted extension]\label{alg:wfvs-forced}
  \Input a forced instance $(\tilde G,\hat X,\tilde w,k',\theta')$, with $\tilde G$ on $\tilde n$
  vertices.
  \Body
  If $\theta'<0$, reject.
  \begin{enumerate}
    \item Run the algorithm $\cA$ of \cref{lem:fvs-ext} on $(\tilde G,\hat X,|\hat X|+k')$. If it rejects, reject.
      Otherwise it outputs $Y_0$; put $A_0=\hat X\cup Y_0$ and $\mathrm{Mod}=C\cup C'$ with $C=A_0$ and $C'=\emptyset$.
    \item Start with $Y_{\mathrm{in}}=\hat X$ and go through the vertices $v\in V(\tilde G)\setminus\hat X$ in a fixed
      order: run \cref{alg:wfvs-test} on $(\tilde G,Y_{\mathrm{in}}\cup\{v\},\hat X,\tilde w,k',\theta',\mathrm{Mod})$ independently
      $\lceil\log(8\tilde n)\rceil$ times, and if one run accepts, add $v$ to $Y_{\mathrm{in}}$.
    \item If $Y_{\mathrm{in}}$ is a connected feedback vertex set of $\tilde G$ with $|Y_{\mathrm{in}}\setminus\hat X|\le k'$ and
      $\tilde w(Y_{\mathrm{in}})\le\theta'$, output $A=Y_{\mathrm{in}}$; otherwise reject.
  \end{enumerate}
\end{algorithm}

\begin{lemma}[Forced weighted extension]\label{lem:wfvs-forced}
  \Cref{alg:wfvs-forced} runs in polynomial space and, for every outcome of its coins, in time
  $3^{k'}2^{|\hat X|}(\tilde nw_{\max})^{O(1)}$. Every output is a solution of the input instance, and if
  one exists, the algorithm outputs one with probability at least $5/8$.
\end{lemma}

\begin{proof}
  The output is checked in step~3.

  \emph{The modulator.} If step~1 succeeds, then by \cref{lem:fvs-ext} (with $G=\tilde G$ and $X=\hat X$)
  the set $A_0$ is a connected feedback vertex set of $\tilde G$ with $\hat X\subseteq A_0$ and $|A_0\setminus\hat
  X|\le k'$. So $\tilde G-\mathrm{Mod}$ is a forest, $\tilde G[C]$ is connected, $\hat X\subseteq C$, $C'=\emptyset$ and $|\mathrm{Mod}\setminus\hat
  X|\le k'\le k'+2$: $\mathrm{Mod}$ is a structured modulator for $(\tilde G,\hat X,k')$, and every set $Y_{\mathrm{in}}\cup\{v\}$
  tested in step~2 contains $\hat X$, as \cref{alg:wfvs-test} requires.

  \emph{Success.} Let $A^*$ be a solution of the instance. Then $Y_0=A^*\setminus\hat X$ shows that $(\tilde G,\hat
  X,|\hat X|+k')$ is a \YES{}-instance of \pname{Connected Feedback Vertex Set Extension}, so step~1 succeeds with
  probability at least $3/4$ by \cref{lem:fvs-ext}. Step~2 is the self-reduction of
  \cref{lem:wt-self-reduction} with $\Omega=V(\tilde G)$, $Z_0=\hat X$, the family $\mathcal P$ of the connected
  feedback vertex sets $Y\supseteq\hat X$ of $\tilde G$ with $|Y\setminus\hat X|\le k'$ and $\tilde w(Y)\le\theta'$,
  which contains $A^*$, the test $\mathsf T(Y_{\mathrm{in}})$ given by \cref{alg:wfvs-test} on
  $(\tilde G,Y_{\mathrm{in}},\hat X,\tilde w,k',\theta',\mathrm{Mod})$, $\delta=1/8$ and
  $\rho=\lceil\log(8\tilde n)\rceil\ge\log(\tilde n/\delta)$. By \cref{lem:wfvs-test}, the test rejects
  $Y_{\mathrm{in}}$ if no member of $\mathcal P$ contains $Y_{\mathrm{in}}$ (for $Y\supseteq\hat X$ the conditions
  $|Y|\le|\hat X|+k'$ and $|Y\setminus\hat X|\le k'$ agree), and with probability at least $1/2$ if one
  does. So, with probability at least $7/8$, the final $Y_{\mathrm{in}}$ is a member of $\mathcal P$, and step~3
  outputs it. Hence the algorithm fails with probability at most $1/4+1/8$.

  \emph{Time and space.} Step~1 takes time $3^{k'}2^{\comp(\tilde G[\hat X])}\tilde n^{O(1)}\le3^{k'}2^{|\hat
  X|}\tilde n^{O(1)}$ and polynomial space by \cref{lem:fvs-ext}. Step~2 runs \cref{alg:wfvs-test} at most
  $\tilde n\lceil\log(8\tilde n)\rceil$ times, each within the bounds of \cref{lem:wfvs-test}.
\end{proof}

For an instance $I=(G,w)$ with $G$ connected and not a forest, and an integer $\theta$, let
$\Phi^{\mathrm{w}}_\theta$ be the implicit set system with universe $V(G)$ whose solutions are the
connected feedback vertex sets $S$ of $G$ with $w(S)\le\theta$. Membership is decided in polynomial
time, and every solution is nonempty.

\begin{proof}[Proof of \cref{thm:cc-wcfvs}]
  If $G$ is a forest, output $\emptyset$. If $G$ is disconnected, then, as in \cref{sec:fvs-problem}, a
  connected feedback vertex set exists only if exactly one component $G_1$ of $G$ contains a cycle (as
  $G$ is not a forest, at least one does), and
  then the connected feedback vertex sets of $G$ are those of $G_1$; report that none exists or replace
  $G$ by $G_1$. So let $G$ be connected and not a forest.

  Let $\cA_{\mathrm w}$ be the algorithm of \cref{lem:fvs-wrapper} with \cref{alg:wfvs-forced} as
  $\mathrm{Forced}$; by \cref{lem:wfvs-forced} the hypotheses of \cref{lem:fvs-wrapper} hold with $p=5/8$,
  as $(\tilde nw_{\max})^{O(1)}\le(\tilde n+w_{\max})^{O(1)}$. So $\cA_{\mathrm w}$ solves
  $\Phi^{\mathrm{w}}_\theta$-Extension with one-sided error, accepts a \YES{}-instance with probability at
  least $5/8$, and runs in polynomial space and in time $3^{k-|X|}2^{\comp(G[X])}n^{O(1)}$ for every outcome
  of its coins, because $w_{\max}\le n^{O(1)}$. For every $\theta\in\{1,\dots,nw_{\max}\}$ and every
  $k\in\{1,\dots,n\}$, run Algorithm~\MLS{} for $\Phi^{\mathrm{w}}_\theta$ and $k$ as in \cref{cor:cc-rate},
  with the extension algorithm $\cA_{\mathrm w}$, $c=3$, $\gamma=0$, $\psi=\log1.8$ and $\kappa=1$.
  Output the lightest set returned by these runs, and $V(G)$ if no run returns a set; $V(G)$ is a
  connected feedback vertex set because $G$ is connected.

  The algorithm $\cA_{\mathrm w}$ satisfies the hypotheses of \cref{cor:fw-onesided} with the running-time
  exponent $g(G,X,k)=(k-|X|)\log3+\comp(G[X])$ of~\eqref{eq:fvs-g}, whose comparisons with the thresholds of
  Algorithm~\MLS{} reduce to integer comparisons (\cref{lem:fvs-ext}). Every solution $S$ induces a
  connected subgraph, so a spanning tree of $G[S]$ gives the component-count bound with constants
  $(3,0)$, as in the proof of \cref{thm:cc-cfvs}. By the last claim of \cref{cor:cc-rate}, every run
  takes time $1.8^nn^{O(1)}$ and polynomial space, and every returned set is a connected feedback vertex
  set of weight at most~$\theta$. Let $S^*$ be a connected feedback vertex set of minimum weight
  $\theta^*\le nw_{\max}$; the run for $\theta=\theta^*$ and $k=|S^*|$ returns a set with probability at least
  $1-1/e$, and then the output has weight $\theta^*$. There are $n^2w_{\max}\le n^{O(1)}$ runs.
\end{proof}
{}
\subsection{The further applications}\label{sec:app-variants}

This subsection contains the lemmas and proofs for \cref{sec:rdom,sec:cv-weighted,sec:cv-wcds,sec:cv-dir}, the
analyses of \pname{Directed Steiner Tree} and \pname{Paired Domination} (\cref{sec:app-fa-dst,sec:app-fa-pd}), and the
proofs of their weighted versions.

\subsubsection{Red--blue states}\label{sec:app-redblue}

For a finite set $V$ let $V^{\mathrm b}=\{v^{\mathrm b}:v\in V\}$ and $V^{\mathrm r}=\{v^{\mathrm r}:v\in V\}$ be
two disjoint copies of $V$, the \emph{blue} and the \emph{red} copy, and for $X\subseteq V$ write
$X^{\mathrm b}=\{x^{\mathrm b}:x\in X\}$. A \emph{red--blue graph} on $V$ is a graph $H$ with vertex set
$V^{\mathrm b}\cup V^{\mathrm r}$ in which $v^{\mathrm b}v^{\mathrm r}$ is an edge for every $v\in V$. A set
$S\subseteq V$ is \emph{connected red--blue dominating} in $H$ if $S\neq\emptyset$, $H[S^{\mathrm b}]$ is
connected and every red vertex has a neighbor in $S^{\mathrm b}$. This is the problem that Abu-Khzam,
Mouawad and Liedloff~\cite{Abu-KhzamML11} solve in time $\Ostar(1.3645^{|V^{\mathrm b}|+|V^{\mathrm r}|})$
and polynomial space; here it serves to formulate the extension problems of the variants as states of
\cref{def:cds-state}.

\begin{lemma}[Red--blue states]\label{lem:cv-redblue}
Let $H$ be a red--blue graph on a set $V$ with $n\ge1$ elements, and for $X\subseteq V$ put
\[
  g_H(X)=\bigl|\{v\in V:v^{\mathrm r}\text{ has no neighbor in }X^{\mathrm b}\}\bigr|+\comp\bigl(H[X^{\mathrm b}]\bigr).
\]
Then $g_H$ is computed in polynomial time and takes integer values in $\{0,\dots,n\}$, with
$g_H(\emptyset)=n$ and $g_H(S)=1$ for every connected red--blue dominating set $S$. For every
$X\subseteq V$ and every $\varphi\colon V\setminus X\to\{0,1,\dots,\varphi_{\max}\}$, the least weight
$\varphi(Y)$ over the sets $Y\subseteq V\setminus X$ for which $X\cup Y$ is connected red--blue dominating
in $H$ ($\infty$ if there is none) is computed deterministically in time
$2^{g_H(X)}(n+\varphi_{\max})^{O(1)}$ and polynomial space; with $\varphi\equiv1$, this decides for every
integer $k\ge0$, in time $2^{g_H(X)}n^{O(1)}$, whether some such $Y$ has $|Y|\le k-|X|$. Consequently, if
$\Phi$ is an implicit set system whose instance $I$ determines in polynomial time a red--blue graph
$H_I$ on $U_I$ whose connected red--blue dominating sets are the solutions, then this decision
algorithm solves $\Phi$-Extension deterministically in polynomial space, with the running-time exponent
$(I,X,k)\mapsto g_{H_I}(X)$ (\cref{def:fw-cost}).
\end{lemma}
\begin{proof}
Let $\cD$ be the family of vertex sets of the components of $H[X^{\mathrm b}]$, let
$B=(V\setminus X)^{\mathrm b}$, and let $R$ be the set of red vertices without a neighbor in $X^{\mathrm b}$.
Then $J=(\cD,B,R)$ is a state of $H$ in the sense of \cref{def:cds-state}: the members of $\cD$ are
disjoint, nonempty and connected, no edge of $H$ joins two distinct components of an induced subgraph,
$V(\cD)=X^{\mathrm b}$, $B$ avoids $X^{\mathrm b}$, and $R$ avoids $N_H[X^{\mathrm b}]$ because a red vertex
is not in $X^{\mathrm b}$ and the vertices of $R$ are not adjacent to $X^{\mathrm b}$. A subset of $B$ has
the form $Y^{\mathrm b}$ with $Y\subseteq V\setminus X$, and as no red vertex lies in $Y^{\mathrm b}$, the
condition $R\subseteq N_H[Y^{\mathrm b}]$ says that every red vertex without a neighbor in $X^{\mathrm b}$
has one in $Y^{\mathrm b}$, that is, that every red vertex has a neighbor in $(X\cup Y)^{\mathrm b}$. So the
solutions of $J$ are exactly the sets $Y^{\mathrm b}$ with $Y\subseteq V\setminus X$ for which $X\cup Y$ is
connected red--blue dominating, and $|\Tm(J)|=|R|+|\cD|=g_H(X)$. Give the blue vertex $y^{\mathrm b}$ of
$B$ the weight $\varphi(y)$. By \cref{lem:cds-dir-steiner}, applied to the state $J$ of the graph $H$ with
$2n$ vertices, the value $\OPT_\varphi(J)$, which is the least weight in question, is computed
deterministically in time $2^{g_H(X)}(2n+\varphi_{\max})^{O(1)}$ and polynomial space. With $\varphi\equiv1$
this is $\OPT(J)$, the least $|Y|$, computed in time $2^{g_H(X)}n^{O(1)}$ by the first statement of
\cref{cor:cds-steiner}, and one accepts if and only if $\OPT(J)\le k-|X|$.

For the range, every $v\in X$ has the neighbor $v^{\mathrm b}\in X^{\mathrm b}$ of $v^{\mathrm r}$, so the
first term of $g_H(X)$ is at most $n-|X|$, and $\comp(H[X^{\mathrm b}])\le|X|$. For $X=\emptyset$ the first
term is $n$ and the second is $0$. For a connected red--blue dominating set $S$ the first term is $0$ and
the second is $1$. Both terms are computed in polynomial time from $H$ and $X$.

For the last claim, membership in $\cF_I$ is tested in polynomial time through $H_I$, so $\Phi$ is as in
\cref{def:fw-iss}, and $g_{H_I}$ is computable in polynomial time and bounds the running time of the
algorithm as \cref{def:fw-cost} requires.
\end{proof}

\subsubsection{Connected \texorpdfstring{$r$}{r}-domination}\label{sec:app-rdom}

\RestThmRdom*

Throughout, $\cN$ is a domination relation (\cref{def:rdom-relation}) on a connected graph $G$ with
$n\ge2$ vertices.

\paragraph{The Steiner extension.}

For $X\subseteq V$ put
\[
  g_\cN(X)=|\operatorname{und}_\cN(X)|+\comp(G[X]).
\]
For $\cN[v]=N[v]$ this is $|\Tm(J_X)|$ by~\eqref{eq:cds-TX}. The instance $I$ consists of $G$ and the
description of $\cN$, which are fixed, and $g_\cN(X)$ does not depend on $k$; in the notation of
\cref{def:fw-cost}, the running-time exponent is $g(I,X,k)=g_\cN(X)$. The next lemma shows that the Steiner
extension transfers with this running-time exponent.

\begin{lemma}[Steiner extension for domination relations]\label{lem:rdom-steiner}
Let $H_\cN$ be the red--blue graph on $V$ with the edges $u^{\mathrm b}v^{\mathrm b}$ for the edges $uv$
of $G$ and the edges $u^{\mathrm b}v^{\mathrm r}$ for all $u,v$ with $u\in\cN[v]$. The connected red--blue
dominating sets of $H_\cN$ are the connected $\cN$-dominating sets of $G$, and $g_{H_\cN}=g_\cN$. So
\cref{lem:cv-redblue} gives a deterministic extension algorithm for the implicit set system of
connected $\cN$-dominating sets, in time $2^{g_\cN(X)}n^{O(1)}$ and polynomial space, with the running-time
exponent $(I,X,k)\mapsto g_\cN(X)$, which takes values in $\{0,\dots,n\}$.
\end{lemma}
\begin{proof}
The graph $H_\cN$ is built in polynomial time, because every $\cN[v]$ is, and $v^{\mathrm b}v^{\mathrm r}$
is an edge since $v\in N[v]\subseteq\cN[v]$. For $S\subseteq V$ the graph $H_\cN[S^{\mathrm b}]$ is a copy
of $G[S]$, and $v^{\mathrm r}$ has a neighbor in $S^{\mathrm b}$ if and only if $\cN[v]\cap S\neq\emptyset$.
So $S$ is connected red--blue dominating in $H_\cN$ if and only if $S$ is a connected $\cN$-dominating
set, and $g_{H_\cN}(X)=g_\cN(X)$ for every $X\subseteq V$. Now \cref{lem:cv-redblue} applies.
\end{proof}

\paragraph{Anchored trees.}

The analyses of \cref{sec:cds-sampling,sec:rst-tree} assign every vertex outside the solution to a
neighbor inside it. For a domination relation we assign it to a solution vertex that dominates it,
which need not be a neighbor, and we treat the assignment as a pendant edge of a virtual tree. In
this tree the solution is a connected dominating set, and the running-time exponent of a sample is at most its
number of terminals in the tree. For a graph $T'$ on the vertex set $V$ and $X\subseteq V$ we write
\[
  g_{T'}(X)=n-|N_{T'}[X]|+\comp(T'[X])
\]
for the number of terminals $|\Tm(J_X)|$ of the state $J_X$ of~\eqref{eq:cds-JX} formed in the graph
$T'$, by~\eqref{eq:cds-TX}.

\begin{lemma}[Anchored tree]\label{lem:rdom-tree}
Let $S$ be a connected $\cN$-dominating set, let $F$ be a spanning tree of $G[S]$, and let
$o:V\setminus S\to S$ be an \emph{anchor map}, that is, $o(v)\in\cN[v]\cap S$ for every
$v\notin S$. Let $T'$ be the graph on $V$ formed by $F$ and the edges $v\,o(v)$ for $v\notin S$. Then
$T'$ is a tree, $S$ is a connected dominating set of $T'$, every vertex outside $S$ is a leaf of $T'$,
and
\[
  g_\cN(X)\ \le\ g_{T'}(X)\qquad\text{for every }X\subseteq S .
\]
Consequently, in the notation of \cref{def:fw-moments}, the exponential moments $\mu_t(S)$ for the running-time
exponent $g_\cN$ and $\mu^{T'}_t(S)$ for the running-time exponent $g_{T'}$ of $\Phi_{\mathrm{CDS}}$ on the
graph $T'$ satisfy $\mu_t(S)\ge\mu^{T'}_t(S)$ termwise, and every integer certificate
(\cref{thm:dr-det-moments}) for $S$ with respect to $g_{T'}$ is one with respect to $g_\cN$.
\end{lemma}
\begin{proof}
The graph $T'$ is the tree $F$ with the $n-|S|$ vertices outside $S$ attached as pendant vertices, so it
is a tree, $T'[S]=F$ is connected, every $v\notin S$ has exactly one neighbor, namely $o(v)\in S$, and
an anchor map exists because $S$ is $\cN$-dominating. Let $X\subseteq S$. We show
$\operatorname{und}_\cN(X)\subseteq V\setminus N_{T'}[X]$. Let $v\in N_{T'}[X]$. If $v\in X$, then
$v\in N[v]\subseteq\cN[v]$ meets $X$. Otherwise $v$ is joined in $T'$ to some $x\in X\subseteq S$; if the
edge $vx$ belongs to $F$, then $x\in N[v]\subseteq\cN[v]$, and if it is a pendant edge, then $v\notin S$
and $x=o(v)\in\cN[v]$, because the only pendant edge at $v$ is $v\,o(v)$. In both cases $v$ is dominated
by $X$. Finally $\comp(G[X])\le\comp(F[X])=\comp(T'[X])$, because $F[X]$ is a spanning subgraph of
$G[X]$ and $T'[X]=F[X]$ for $X\subseteq S$. Adding the two bounds gives
$g_\cN(X)\le|V\setminus N_{T'}[X]|+\comp(T'[X])=g_{T'}(X)$. The last claims follow from
$2^{-g_\cN(X)}\ge2^{-g_{T'}(X)}$ for every $X\subseteq S$, and from $g_\cN(X)\le g_{T'}(X)\le\alpha-w(X)$ for
every nonempty $X\subseteq S$ when $w$ is an integer certificate for $S$ with respect to $g_{T'}$.
\end{proof}

\begin{proof}[Proof of \cref{thm:rdom}]
Let $\Phi^\cN$ be the implicit set system of connected $\cN$-dominating sets, with the extension
algorithm and the running-time exponent $g_\cN$ of \cref{lem:rdom-steiner}; $G$ has a solution, and
$\gc^\cN(G)$ is the least solution size. Let $S$ be a minimum connected $\cN$-dominating set,
$s^*=|S|$, and let $T'$ be an anchored tree of \cref{lem:rdom-tree} for $S$.

(a) Let $\mathrm{Dec}(G,s)$ be Algorithm \MLS{} for $\Phi^\cN$, with $k=s$, $\psi=\psi^*$,
$\Gamma=2$ and confidence $\kappa$, that is, Algorithm RS-$\theta$ (\cref{alg:rst-algorithm}) with the
extension algorithm of \cref{lem:rdom-steiner} in place of $\cA_{\mathrm{St}}$ and $g_\cN$ in place of
$|\Tm(J_X)|$. By \cref{thm:fw-main}, it takes time $2^{\psi^*n}n^{O(1)}$ and polynomial space, and it
rejects if $\gc^\cN(G)>s$. As $T'$ is a connected graph and $S$ is a connected dominating set
of $T'$, \cref{prop:rst-rate} applied to $T'$ gives $\max_t\mu^{T'}_t(S)/\binom nt\ge2^{-\psi^*n}/2$,
and by \cref{lem:rdom-tree} the set $S$ satisfies condition~\eqref{eq:fw-star} for $g_\cN$; so
$\mathrm{Dec}(G,s^*)$ accepts with probability at least $1-e^{-\kappa}$, by \cref{thm:fw-main}. Now
\cref{lem:cds-first-accepted} gives the claim; the bound $2^{\psi^*}<1.7204$ is as in
\cref{thm:rstheta}.

(b) Run the procedure $\mathrm{Det}$ of \cref{lem:det-decide} for $\Phi^\cN$, the extension algorithm
and running-time exponent of \cref{lem:rdom-steiner} and $(\alpha,\Xi)=(n-1,n-2)$, for $s=1,\dots,n$, and output
the least $s$ at which it answers \YES{}. By \cref{lem:det-decide}, it answers \NO{} if
$\gc^\cN(G)>s$. At $s=s^*$ the set $S$ has size exactly $s$, and it has an integer certificate for
$(n-1,n-2)$: the tree $T'$ is a spanning tree of the graph $T'$ in which every vertex outside $S$ is a
leaf, so the weights $w(u)=d_{T'}(u)-1$ on $S$ form the certificate~\eqref{eq:det-cert} for $S$ in the
graph $T'$, that is, they are nonnegative integers with $w(S)=n-2$ and $g_{T'}(X)\le n-1-w(X)$ for every
nonempty $X\subseteq S$, by \cref{lem:cds-pointwise} and~\eqref{eq:cds-TX} applied to $T'$, and by
\cref{lem:rdom-tree} they form an integer certificate for $S$ with respect to $g_\cN$. So
$\mathrm{Det}$ answers \YES{} at $s=s^*$, and \cref{lem:cds-first-accepted}, in its deterministic form
and with the time and space bounds of \cref{lem:det-decide}, gives the theorem.
\end{proof}

\subsubsection{Vertex-weighted connected domination}\label{sec:app-cv-weighted}

\RestThmCvWeighted*

The notation is that of \cref{sec:cv-weighted}.

\begin{lemma}[Weighted max leaf and connected domination]\label{lem:cv-wmaxleaf}
If $n\ge3$, the maximum leaf weight of a spanning tree of $G$ equals $\varphi(V)-\gc^\varphi(G)$.
\end{lemma}
\begin{proof}
If $F$ is a spanning tree, then $V\setminus\operatorname{leaves}(F)$ is a connected dominating set, by the
first half of the proof of \cref{lem:cds-maxleaf}, so $\varphi(\operatorname{leaves}(F))=\varphi(V)-\varphi(V\setminus\operatorname{leaves}(F))\le \varphi(V)-\gc^\varphi(G)$.
Conversely, for a connected dominating set $S$ the second half of that proof gives a spanning tree in
which every vertex of $V\setminus S$ is a leaf; as $\varphi\ge0$, its leaf weight is at least
$\varphi(V\setminus S)=\varphi(V)-\varphi(S)$.
\end{proof}

\paragraph{The weighted Steiner extension.} For $X\subseteq V$ let
\[
  \OPT_\varphi(X)=\min\{\varphi(Y):Y\subseteq V\setminus X,\ X\cup Y\text{ is a connected dominating set}\}.
\]
Since the solutions of the state $J_X$ of~\eqref{eq:cds-JX} are the sets $Y\subseteq V\setminus X$ for which
$X\cup Y$ is a connected dominating set (also for $X=\emptyset$), this is $\OPT_\varphi(J_X)$ in the notation of
\cref{def:cds-dir-encoding}, with $\varphi$ restricted to the blue vertices $V\setminus X$, and
\cref{lem:cds-dir-steiner} computes it deterministically in time $2^{|\Tm(J_X)|}(n+\varphi_{\max})^{O(1)}$ and
polynomial space.

\paragraph{Relaxed extension.} The weighted extension does not control the size of the completion:
\cref{lem:cv-W} minimizes weight, and a completion of least weight may be large. We show that this does
not affect \cref{thm:fw-main,thm:dr-det-moments}: their proofs use the correctness of the
extension algorithm in only two places. The following lemma extends the two theorems accordingly.

\begin{lemma}[Relaxed extension algorithms]\label{lem:cv-relaxed}
Let $\Phi$ be an implicit set system, and let $\cA$ be a deterministic algorithm on inputs $(I,X,k)$ such
that
\begin{enumerate}[label=\textup{(R\arabic*)}]
\item\label{it:cv-R1} $\cA$ accepts $(I,X,k)$ whenever some $Y\subseteq U_I\setminus X$ with $|Y|\le k-|X|$
has $X\cup Y\in\cF_I$, and
\item\label{it:cv-R2} $\cA$ rejects $(I,X,k)$ if $\cF_I=\emptyset$.
\end{enumerate}
Let $g$ satisfy \cref{def:fw-cost} for $\cA$. Then \cref{thm:fw-main}, with its ranges for the cap and
the trial counts, and \cref{thm:dr-det-moments} hold for $\cA$
and $g$, with the soundness statement ``answers \NO{} if $\cF_I$ contains no set of size at most
$k$'' replaced by ``answers \NO{} if $\cF_I=\emptyset$.''
\end{lemma}
\begin{proof}
An algorithm for $\Phi$-Extension satisfies \ref{it:cv-R1} and~\ref{it:cv-R2}, and the proofs of the two
theorems use the correctness of $\cA$ only through these two properties, in the following places. The
soundness parts (item~(ii) of \cref{thm:fw-main}, and the paragraph called soundness in the proof of
\cref{thm:dr-det-moments}) conclude from an accepting run of $\cA$ that a solution
exists; by~\ref{it:cv-R2}, $\cF_I\neq\emptyset$. The completeness parts (item~(iii) of \cref{thm:fw-main},
and the paragraph called completeness) run $\cA$ on $(I,X,k)$ with $X\subseteq S$, $S\in\cF_I$ and
$|S|=k$, and conclude that it accepts because $Y=S\setminus X$ is a witness; this is~\ref{it:cv-R1}. The
statements about time and space use only that $\cA$ runs in time $2^{g(I,X,k)}N^{O(1)}$ on every input
$(I,X,k)$, and the
early answer of \cref{alg:dr-integer}, the rounding arguments, \cref{lem:fw-threshold-prob},
\cref{thm:dr-balanced} and \cref{lem:dr-chord} do not involve $\cA$.
\end{proof}

\paragraph{Budget loop.} The weighted problems of this paper are solved by running a decision algorithm
for every weight budget and every solution size. The following lemma states this wrapper once.

\begin{lemma}[Budget loop]\label{lem:cv-budget-loop}
Let $\Phi$ be an implicit set system whose instances $I$ include a weight function
$w\colon U_I\to\{0,1,\dots,w_{\max}\}$, written in unary, and write $w(S)=\sum_{u\in S}w(u)$. For an
integer $\varpi\ge0$ let $\Phi_\varpi$ be the implicit set system with the same instances and universes
whose solutions are the solutions $S$ of $\Phi$ with $w(S)\le\varpi$. Let $p\in(0,1]$, and suppose that for
every $\varpi$ an algorithm $\mathcal{B}_\varpi$, given $I$ and an integer $k\in\{0,\dots,n\}$, answers \YES{} or
\NO{} within time $T$ and polynomial space, such that
\begin{enumerate}[label=\textup{(B\arabic*)}]
\item\label{it:cv-B1} $\mathcal{B}_\varpi(I,k)$ answers \NO{} if $\Phi_\varpi$ has no solution on $I$, and
\item\label{it:cv-B2} if $\Phi$ has a solution on $I$, then some solution $S^*$ of minimum weight has
the property that $\mathcal{B}_{w(S^*)}(I,|S^*|)$ answers \YES{} with probability at least $p$.
\end{enumerate}
Let the algorithm $\mathcal{B}$ run $\mathcal{B}_\varpi(I,k)$ for every $\varpi\in\{0,\dots,w(U_I)\}$ and every
$k\in\{0,\dots,n\}$, output the least $\varpi$ for which some run answers \YES{}, and report that
$\Phi$ has no solution on $I$ if no run answers \YES{}. Then $\mathcal{B}$ runs in time
$(w(U_I)+1)(n+1)T+N^{O(1)}$ and polynomial space, it never outputs a value below the minimum weight of a
solution, and with probability at least $p$ it outputs the minimum weight of a solution (respectively
reports correctly that there is none).
\end{lemma}
\begin{proof}
If $\Phi$ has no solution on $I$, then no $\Phi_\varpi$ has one, so by~\ref{it:cv-B1} no run answers \YES{}
and the report is correct. Otherwise let $S^*$ be as in~\ref{it:cv-B2} and $\varpi^*=w(S^*)$. Every
solution has weight at least $\varpi^*$, so for $\varpi<\varpi^*$ the system $\Phi_\varpi$ has no solution,
and by~\ref{it:cv-B1} no run for such a $\varpi$ answers \YES{}; hence the output, if any, is at least
$\varpi^*$. The run $\mathcal{B}_{\varpi^*}(I,|S^*|)$ is among the runs, because $\varpi^*\le w(U_I)$ and $|S^*|\le n$,
and it answers \YES{} with probability at least $p$ by~\ref{it:cv-B2}; then the output is $\varpi^*$. There
are $(w(U_I)+1)(n+1)$ runs, and $w(U_I)\le N$ because $w$ is written in unary.
\end{proof}

In the applications, $\mathcal{B}_\varpi$ is Algorithm RS-$\theta$, \cref{alg:dr-integer} or Algorithm~\MLS{} for
$\Phi_\varpi$, with a relaxed extension algorithm (\cref{lem:cv-relaxed}) or an extension algorithm with
one-sided error (\cref{cor:fw-onesided}); property~\ref{it:cv-B1} is the soundness statement of these
theorems, and property~\ref{it:cv-B2} is their completeness at $k=|S^*|$, from condition~\eqref{eq:fw-star}
with $p=1-e^{-\kappa}$ or from an integer certificate with $p=1$; the completeness is used only at the
size of $S^*$.

\paragraph{The algorithms.} For a budget $\varpi\in\{0,\dots,\varphi(V)\}$ let $\Phi_{\mathrm{wCDS}}$ be the
implicit set system whose instances are triples $(G,\varphi,\varpi)$, with universe $V$ and with the connected
dominating sets $S$ with $\varphi(S)\le\varpi$ as solutions. Let $\cA_\varphi$ be the algorithm that, on
$((G,\varphi,\varpi),X,k)$, computes $\OPT_\varphi(X)$ by \cref{lem:cds-dir-steiner} and accepts if and only if
$\varphi(X)+\OPT_\varphi(X)\le\varpi$; it ignores~$k$. It satisfies~\ref{it:cv-R1} and~\ref{it:cv-R2}: a witness $Y$ in
\ref{it:cv-R1} gives $\varphi(X)+\OPT_\varphi(X)\le \varphi(X\cup Y)\le\varpi$, and an acceptance yields a connected
dominating set $X\cup Y$ of weight at most $\varpi$. By \cref{lem:cds-dir-steiner}, the function $g$ with
$g((G,\varphi,\varpi),X,k)=|\Tm(J_X)|$ is a running-time exponent for $\cA_\varphi$ in the sense of \cref{def:fw-cost}, with
integer values in $\{0,\dots,n\}$; it depends neither on $k$ nor on $\varphi$ and $\varpi$, and in the shorthand
declared after \cref{def:fw-cost} it is $g(X)=|\Tm(J_X)|$.

\begin{proof}[Proof of \cref{thm:cv-weighted}]
For $n\le2$ the maximum leaf weight is output directly, as stated in the theorem, since $G$ is its own only
spanning tree. For $n\ge3$, by \cref{lem:cv-wmaxleaf} it suffices to compute $\gc^\varphi(G)$, which we do by \cref{lem:cv-budget-loop}
for the implicit set system of connected dominating sets with the weights $\varphi$, so that $\Phi_\varpi$ is
$\Phi_{\mathrm{wCDS}}$ on $(G,\varphi,\varpi)$ and $w(U_I)=\varphi(V)\le n\varphi_{\max}$. In~(a),
$\mathcal{B}_\varpi(G,\varphi,s)$ is the run for the candidate size $s$ of \cref{thm:rstheta} (Algorithm
RS-$\theta$, \cref{alg:rst-algorithm}) for $\Phi_{\mathrm{wCDS}}$ on $(G,\varphi,\varpi)$, with the relaxed
extension algorithm $\cA_\varphi$ in place of $\cA_{\mathrm{St}}$ and the running-time exponent $|\Tm(J_X)|$. In~(b),
$\mathcal{B}_\varpi(G,\varphi,s)$ is the procedure $\mathrm{Det}$ of \cref{lem:det-decide} for
$\Phi_{\mathrm{wCDS}}$ on $(G,\varphi,\varpi)$ with $\cA_\varphi$, its running-time exponent $|\Tm(J_X)|$ and
$(\alpha,\Xi)=(n-1,n-2)$: by \cref{lem:cv-relaxed}, \cref{thm:dr-det-moments} holds for $\cA_\varphi$ with
the soundness statement ``only if $\Phi_{\mathrm{wCDS}}$ has a solution,'' so \cref{lem:det-decide}, whose
proof uses $\cA_\varphi$ only through \cref{thm:dr-det-moments,lem:det-time}, holds for $\cA_\varphi$ with
the same change; its enumeration tests membership in $\Phi_{\mathrm{wCDS}}$, that is, connected domination
and weight at most $\varpi$, in polynomial time per set.

\emph{Property~\ref{it:cv-B1}.} By \cref{lem:cv-relaxed} with \cref{thm:fw-main} in~(a), and by
\cref{lem:det-decide} as just stated in~(b), $\mathcal{B}_\varpi(G,\varphi,s)$ answers \NO{} if
$\Phi_{\mathrm{wCDS}}$ has no solution on $(G,\varphi,\varpi)$, that is, if $G$ has no connected
dominating set of weight at most $\varpi$.

\emph{Property~\ref{it:cv-B2}.} Let $S^*$ be a connected dominating set with $\varphi(S^*)=\gc^\varphi(G)$, and
put $s^*=|S^*|$ and $\varpi^*=\varphi(S^*)$; $S^*$ is a solution of $\Phi_{\mathrm{wCDS}}$ on $(G,\varphi,\varpi^*)$ of
size exactly $s^*$. In~(a), \cref{prop:rst-rate} gives condition~\eqref{eq:fw-star} for $S^*$ with the running-time
exponent $|\Tm(J_X)|$, which does not involve the weights, so by \cref{lem:cv-relaxed} with
\cref{thm:fw-main}(iii) the run $\mathcal{B}_{\varpi^*}(G,\varphi,s^*)$ answers \YES{} with probability at least
$1-e^{-\kappa}$. In~(b), the certificate~\eqref{eq:det-cert} for $S^*$ is an integer certificate for
$(n-1,n-2)$, so $\mathcal{B}_{\varpi^*}(G,\varphi,s^*)$ answers \YES{} by \cref{lem:det-decide}.

\emph{Time and space.} By \cref{lem:cv-relaxed}, each run in~(a) has the time and space bounds of the
corresponding run in \cref{thm:rstheta}, and each run in~(b) those of \cref{lem:det-decide}, with
$N^{O(1)}=(n+\varphi_{\max})^{O(1)}$. So \cref{lem:cv-budget-loop} gives the claims, with $p=1-e^{-\kappa}$
in~(a) and $p=1$ in~(b), and a factor $(n\varphi_{\max}+1)(n+1)$ in the time.
\end{proof}

\subsubsection{Weakly connected domination}\label{sec:app-cv-wcds}

\RestThmCvWcds*

The notation is that of \cref{sec:cv-wcds}. For $X\subseteq V$ the running-time exponent is
\begin{equation}\label{eq:cv-gw}
  g_{\mathrm w}(X)=n-|N[X]|+\comp(G^2[X]),
\end{equation}
the analogue of~\eqref{eq:cds-TX} with $G^2[X]$ in place of $G[X]$.

\begin{lemma}[Characterization]\label{lem:cv-wchar}
A set $S\subseteq V$ is a weakly connected dominating set of $G$ if and only if $S$ is dominating and
$G^2[S]$ is connected.
\end{lemma}
\begin{proof}
Let $S$ be a weakly connected dominating set, and let $x,x'\in S$. The graph $\langle S\rangle_{\mathrm w}$
is connected, so it contains a path from $x$ to $x'$. A vertex $v\notin S$ on this path is an end of its
two path edges, each of which has an end in $S$; so both path neighbors of $v$ lie in $S$. Hence any two
consecutive vertices of $S$ on the path are adjacent in $G$ or have a common neighbor, that is, they are
adjacent in $G^2[S]$, and $x$ and $x'$ lie in one component of $G^2[S]$.

Conversely, let $S$ be dominating with $G^2[S]$ connected. Then $\langle S\rangle_{\mathrm w}$ has vertex
set $N[S]=V$. Every $v\notin S$ has a neighbor $u\in S$, and $uv$ is an edge of $\langle S\rangle_{\mathrm w}$.
If $x,x'\in S$ are adjacent in $G^2[S]$, then $xx'$ is an edge of $\langle S\rangle_{\mathrm w}$, or $x$ and
$x'$ have a common neighbor $u$ and $xu$, $ux'$ are edges of $\langle S\rangle_{\mathrm w}$. So all of $S$
lies in one component of $\langle S\rangle_{\mathrm w}$, and every other vertex is adjacent to $S$ in it.
\end{proof}

By \cref{lem:cv-wchar} and the graph $H_{\mathrm w}$ of \cref{lem:cv-wext-weak} below, the weakly
connected dominating sets of $G$ are the connected red--blue dominating sets of $H_{\mathrm w}$, so the
algorithm of Abu-Khzam, Mouawad and Liedloff~\cite{Abu-KhzamML11} for \pname{Connected Red-Blue Dominating
Set} solves \pname{Weakly Connected Dominating Set} in time $\Ostar(1.3645^{2n})=\Ostar(1.8619^{n})$ and polynomial space.

\begin{lemma}[Extension for weak connectivity]\label{lem:cv-wext-weak}
Let $H_{\mathrm w}$ be the red--blue graph on $V$ with the edges $u^{\mathrm b}v^{\mathrm b}$ for all $u\neq v$
at distance at most $2$ in $G$ and the edges $u^{\mathrm b}v^{\mathrm r}$ for all $u,v$ with $u\in N[v]$. The
connected red--blue dominating sets of $H_{\mathrm w}$ are the weakly connected dominating sets of $G$, and
$g_{H_{\mathrm w}}=g_{\mathrm w}$. So \cref{lem:cv-redblue} gives a deterministic extension algorithm for
the implicit set system of weakly connected dominating sets, in time $2^{g_{\mathrm w}(X)}n^{O(1)}$ and
polynomial space, with the running-time exponent $(G,X,k)\mapsto g_{\mathrm w}(X)$, which takes values in
$\{0,\dots,n\}$.
\end{lemma}
\begin{proof}
The graph $H_{\mathrm w}$ is built in polynomial time. For $S\subseteq V$ the graph $H_{\mathrm w}[S^{\mathrm b}]$
is a copy of $G^2[S]$, and $v^{\mathrm r}$ has a neighbor in $S^{\mathrm b}$ if and only if
$N[v]\cap S\neq\emptyset$; so by \cref{lem:cv-wchar} the set $S$ is connected red--blue dominating in
$H_{\mathrm w}$ if and only if it is weakly connected dominating, and
$g_{H_{\mathrm w}}(X)=n-|N[X]|+\comp(G^2[X])=g_{\mathrm w}(X)$ by~\eqref{eq:cv-gw}. Now
\cref{lem:cv-redblue} applies, for the implicit set system of weakly connected dominating sets.
\end{proof}

The next lemma is the analogue of \cref{lem:cds-pointwise}; unlike~\eqref{eq:det-cert}, it also holds
for the empty sample.

\begin{lemma}[Pointwise certificate for weak connectivity]\label{lem:cv-wcert}
Let $S$ be a weakly connected dominating set, let $T$ be a spanning tree of $\langle S\rangle_{\mathrm w}$,
rooted anywhere, and for $u\in S$ let $w(u)$ be $d_T(u)$ minus the number of children of $u$ in
$T$ that lie in $S$. Then $w$ takes nonnegative integer values, $w(S)=n-1$, and
\[
  g_{\mathrm w}(X)\ \le\ n-w(X)\qquad\text{for every }X\subseteq S .
\]
\end{lemma}
\begin{proof}
Every edge of $\langle S\rangle_{\mathrm w}$ has an end in $S$, so $S$ is a vertex cover of $T$. Charge every
edge of $T$, joining a vertex $v$ to its parent $p$, to $v$ if $v\in S$ and to $p$ otherwise; in the second
case $p\in S$, because $S$ covers the edge. A vertex $u\in S$ is charged with the edge to its parent (if
$u$ is not the root) and with the edges to its children outside $S$, that is, with exactly $w(u)\ge0$ edges.
So $w(S)=|E(T)|=n-1$. Let $E_T(X)$ be the set of edges of $T$ with an end in $X$. An edge charged to a
vertex of $X$ has that vertex as an end, so $w(X)\le|E_T(X)|$.

It remains to show $g_{\mathrm w}(X)\le n-|E_T(X)|$. Let $T_X$ be the graph with vertex set $N_T[X]$ and
edge set $E_T(X)$; every edge of $E_T(X)$ has both ends in $N_T[X]$, and $T_X$ is a forest, being a
subgraph of $T$, so it has $|N_T[X]|-|E_T(X)|$ components. Let $x,x'\in X$ lie in one component of $T_X$,
and take a path of $T_X$ from $x$ to $x'$. A vertex $v\notin X$ on this path is an end of its two path
edges, each of which has an end in $X$; so both path neighbors of $v$ lie in $X$. Hence consecutive
vertices of $X$ on the path are at distance at most $2$ in $T$, and so in $G$, that is, they are
adjacent in $G^2[X]$; thus $x$ and $x'$ lie in one component of $G^2[X]$. Therefore
$\comp(G^2[X])\le\comp(T_X)=|N_T[X]|-|E_T(X)|$, and with $N_T[X]\subseteq N[X]$, as $T\subseteq G$,
\begin{align*}
  g_{\mathrm w}(X)&=n-|N[X]|+\comp(G^2[X])\le n-|N_T[X]|+|N_T[X]|-|E_T(X)|\\
  &=n-|E_T(X)|\le n-w(X).\qedhere
\end{align*}
\end{proof}

\begin{proof}[Proof of \cref{thm:cv-wcds}]
Run the procedure $\mathrm{Det}$ of \cref{lem:det-decide} for the implicit set system of weakly connected
dominating sets, the extension algorithm and running-time exponent $g_{\mathrm w}$ of \cref{lem:cv-wext-weak} and
$(\alpha,\Xi)=(n,n-1)$, for $s=1,\dots,n$, and output the least $s$ at which it answers \YES{}; $V$ is a
solution, and $\gamma_{\mathrm w}(G)$ is the least solution size. By \cref{lem:det-decide}, $\mathrm{Det}$
answers \NO{} if $\gamma_{\mathrm w}(G)>s$. At $s=\gamma_{\mathrm w}(G)$ a minimum weakly connected
dominating set $S$ has size exactly $s$, and the weights of \cref{lem:cv-wcert} are an integer certificate
for $S$ and $(n,n-1)$: they are nonnegative integers with $w(S)=n-1=\Xi$ and $g_{\mathrm w}(X)\le\alpha-w(X)$
for every $X\subseteq S$. So $\mathrm{Det}$ answers \YES{} at $s=\gamma_{\mathrm w}(G)$, and
\cref{lem:cds-first-accepted}, in its deterministic form and with the bounds of \cref{lem:det-decide},
gives the theorem.
\end{proof}

Unlike for \pname{Connected Dominating Set}, we do not expect the exponential moments of the running-time exponent $g_{\mathrm w}$ to give a
smaller running time than its affine certificate: on the following instances the inequality of
\cref{lem:cv-wcert} is an equality. For integers $s\ge2$ and $d_1,\dots,d_s\ge2$, the \emph{spider} $T$ is
the tree with a center $z$ adjacent to vertices $v_1,\dots,v_s$, where $v_i$ carries $d_i-1$ pendant
vertices; so $T$ has $n=1+\sum_id_i$ vertices.

\begin{proposition}[Spiders]\label{prop:cv-spider}
Let $s\ge2$ and $d_1,\dots,d_s\ge2$ be integers, let $T$ be the spider with these parameters, and let
$S=\{v_1,\dots,v_s\}$. Then:
\begin{enumerate}[label=(\alph*)]
\item $S$ is the unique minimum weakly connected dominating set of $T$.
\item $g_{\mathrm w}(X)=n-\sum_{v_i\in X}d_i$ for every $X\subseteq S$, so the generating function of
\cref{def:fw-moments} for the running-time exponent $g_{\mathrm w}$ is
$W_S(y)=2^{-n}\prod_{i=1}^s\bigl(1+y2^{d_i}\bigr)$.
\end{enumerate}
\end{proposition}
\begin{proof}
(a) The sets $\{v_i\}\cup \operatorname{pend}(v_i)$, where $\operatorname{pend}(v_i)$ is the set of pendant vertices of $v_i$, are disjoint, and each
meets every weakly connected dominating set, since a pendant vertex must be dominated. So a set of size at
most $s$ has exactly one vertex in each of them and does not contain $z$. If such a set $S'$ contains a
pendant vertex of $v_i$ instead of $v_i$, then no edge at $v_i$ other than those to $\operatorname{pend}(v_i)$ has an end in
$S'$, so $\{v_i\}\cup \operatorname{pend}(v_i)$ is a union of components of $\langle S'\rangle_{\mathrm w}$ that misses $z$, and
$\langle S'\rangle_{\mathrm w}$ is disconnected. Hence $S'=S$. And $S$ dominates $T$ and every edge of $T$
has an end in $S$, so $\langle S\rangle_{\mathrm w}=T$ is connected.

(b) For $X=\emptyset$ both sides equal $n$. For nonempty $X\subseteq S$, $N[X]$ consists of $z$ and the
sets $\{v_i\}\cup \operatorname{pend}(v_i)$ with $v_i\in X$, so $|N[X]|=1+\sum_{v_i\in X}d_i$, and $G^2[X]$ is connected, as any
two vertices of $X$ have the common neighbor $z$. Now~\eqref{eq:cv-gw} gives the formula, and the product
form of $W_S$ follows by expanding the product over the subsets of $S$.
\end{proof}

By \cref{prop:cv-spider}(b), on spiders $g_{\mathrm w}$ equals the integer affine surrogate exponent
$n-w(X)$ of \cref{lem:cv-wcert} for $T$ rooted at $z$, whose weights $d_i\in\{2,3\}$ form the worst
integer profile of \cref{lem:dr-chord} when $s/\Xi\in[\frac13,\frac12]$. Let
$\beta_s$ be the largest value of $\bigl(\inf_{y>0}(1+y)^n/W_S(y)\bigr)^{1/n}$ over
$d_1,\dots,d_s\in\{2,3\}$. Numerical computation (\cref{sec:anc-det}) gives values of $\beta_s$ that appear to decrease
toward $\sqrt5-\frac12$; this is not proved. By
\cref{prop:cv-spider}(a), \cref{lem:fw-gf} and \cref{prop:fw-necessary}, if $\beta_s$ tends to
$\sqrt5-\frac12$, then Algorithm~\MLS{} with the running-time exponent $g_{\mathrm w}$ needs time
$(\sqrt5-\frac12)^{n-o(n)}$ on spiders and cannot improve on \cref{thm:cv-wcds} by more than a
subexponential factor.

\subsubsection{Directed Max Leaf Spanning Tree}\label{sec:app-cv-dir}

\RestThmCvDir*

The notation is that of \cref{sec:cv-dir}. Write $N^-(v)$ for the set of
in-neighbors of $v$, and $\mathrm{ch}_F(x)$ for the set of children of $x$ in an out-branching~$F$.

\begin{lemma}[Internal sets]\label{lem:cv-dir-identity}
Let $r\in V(\vec G)$. Then:
\begin{enumerate}[label=(\alph*)]
\item The set of non-leaves of an out-branching rooted at $r$ is $r$-internal.
\item From an $r$-internal set $S$ one computes in polynomial time an out-branching rooted at $r$ in which
every vertex outside $S$ is a leaf.
\item Every superset of an $r$-internal set is $r$-internal.
\end{enumerate}
Hence the maximum number of leaves of an out-branching is $n-\min|S|$, the minimum over all $r$ and all
$r$-internal sets $S$, and an out-branching rooted at $r$ exists if and only if $V(\vec G)$ is $r$-internal.
\end{lemma}
\begin{proof}
(a) Let $F$ be an out-branching rooted at $r$ and $S$ its set of non-leaves. As $n\ge2$, $r$ has a child, so
$r\in S$. For $v\in S$, every vertex of the path of $F$ from $r$ to $v$ has a child or is $v$, so the path
lies in $\vec G[S]$. A leaf $v$ is not $r$, and its parent is a non-leaf and an in-neighbor of~$v$.
(b) Take a breadth-first search out-tree of $\vec G[S]$ from $r$, which spans $S$, and attach every $v\notin S$
by an arc from an in-neighbor in $S$.
(c) If $S\subseteq S'$, then a vertex of $S'\setminus S$ has an in-neighbor in $S$, which is reachable from
$r$ in $\vec G[S]\subseteq \vec G[S']$; the other conditions are inherited.
The final statements follow from (a) and (b), and from (c) applied to $S'=V(\vec G)$.
\end{proof}

For a fixed root $r$, let $\Phi_r$ be the implicit set system whose instances are pairs $(\vec G,r)$, with
universe $V(\vec G)$ and the $r$-internal sets as solutions. For $X\subseteq V(\vec G)$ let
\[
  \operatorname{und}_r(X)=\{v\notin X\cup\{r\}:N^-(v)\cap X=\emptyset\},
\]
let $\mathrm{Src}_r(X)$ be the set of strong components $C$ of $\vec G[X]$ that are \emph{sources} (no arc of $\vec G[X]$
enters $C$ from $X\setminus C$) and do not contain $r$, and put
\[
  g_r(X)=|\operatorname{und}_r(X)|+|\mathrm{Src}_r(X)| .
\]
The members of $\mathrm{Src}_r(X)$ are disjoint, and each contains a vertex of $X\setminus\{r\}$, so $g_r(X)\le(n-|X\cup\{r\}|)+|X\setminus\{r\}|=n-1$; also
$g_r(\emptyset)=n-1$, and $g_r(S)=0$ for every $r$-internal set $S$, because $\operatorname{und}_r(S)=\emptyset$ and the
strong component of $r$ is the only source of $\vec G[S]$.

The extension problem for $\Phi_r$ is an instance of \pname{Directed Steiner Tree}, on a digraph with a
blue and a red copy of every vertex, and \cref{lem:fa-dst-ext} solves it. Let $\vec G^{\,r}$ be the digraph
with a vertex $v^{\mathrm b}$ for every $v\in V(\vec G)$ and a vertex $v^{\mathrm r}$ for every $v\neq r$, with
the arcs $u^{\mathrm b}v^{\mathrm b}$ for every arc $uv$ of $\vec G$, the arcs $u^{\mathrm b}v^{\mathrm r}$ for every
arc $uv$ of $\vec G$ with $v\neq r$, and the arcs $v^{\mathrm b}v^{\mathrm r}$ for $v\neq r$. We consider the
instance $(\vec G^{\,r},r^{\mathrm b},K)$ of \pname{Directed Steiner Tree} with root $r^{\mathrm b}$ and
terminals $K=\{v^{\mathrm r}:v\neq r\}$; its non-terminals are $U=\{v^{\mathrm b}:v\neq r\}$, and we write
$g_{\mathrm{DST}}$ for the running-time exponent $g$ of \cref{sec:fa-dst} on this instance. Red vertices have no
outgoing arcs. So for $S\subseteq V(\vec G)$ with $r\in S$, in the digraph
$\vec G^{\,r}[\{r^{\mathrm b}\}\cup K\cup(S\setminus\{r\})^{\mathrm b}]$ the blue vertices reachable from $r^{\mathrm b}$
are the copies of the vertices reachable from $r$ in $\vec G[S]$, and a red vertex $v^{\mathrm r}$ is
reachable if and only if one of its in-neighbors in this digraph, which are $v^{\mathrm b}$ if $v\in S$
and $u^{\mathrm b}$ for $u\in S\cap N^-(v)$, is reachable. Hence $(S\setminus\{r\})^{\mathrm b}$ is a rooted
Steiner set of $(\vec G^{\,r},r^{\mathrm b},K)$ if and only if $S$ is $r$-internal: the blue vertices of
$S\setminus\{r\}$ are reachable if and only if every vertex of $S$ is reachable from $r$ in $\vec G[S]$, and
then a red vertex $v^{\mathrm r}$ is reachable if and only if $v\in S$ or $v$ has an in-neighbor in $S$.

\begin{lemma}[Directed extension]\label{lem:cv-dir-ext}
For every $X\subseteq V(\vec G)$ and integer $k\ge0$ one can decide in time $2^{g_r(X)}n^{O(1)}$ and polynomial
space, deterministically, whether some $Y\subseteq V(\vec G)\setminus X$ with $|Y|\le k-|X|$ makes $X\cup Y$ an
$r$-internal set. So $((\vec G,r),X,k)\mapsto g_r(X)$ is a running-time exponent (\cref{def:fw-cost}) for $\Phi_r$, with
values in $\{0,\dots,n-1\}$.
\end{lemma}
\begin{proof}
Put $X'=(X\setminus\{r\})^{\mathrm b}\subseteq U$ and $k'=k-|X|-[r\notin X]$; the root $r$ belongs to every
$r$-internal set, so if $r\notin X$ it must be added, using one of the $k-|X|$ further elements. If $k'<0$,
reject. Otherwise run the algorithm of \cref{lem:fa-dst-ext} on the instance $(\vec G^{\,r},r^{\mathrm b},K)$
with the set $X'$ and the size bound $|X'|+k'$, and return its answer. This is correct: the sets $Y\subseteq
V(\vec G)\setminus X$ with $|Y|\le k-|X|$ for which $X\cup Y$ is $r$-internal correspond, by
$Y\mapsto Y'=(Y\setminus\{r\})^{\mathrm b}$, to the sets $Y'\subseteq U\setminus X'$ with $|Y'|\le k'$ for which
$X'\cup Y'$ is a rooted Steiner set. Indeed, if $X\cup Y$ is $r$-internal, then $r\in X\cup Y$, so $r\in Y$
exactly when $r\notin X$, $|Y'|=|Y|-[r\notin X]$, and $(X\cup Y\setminus\{r\})^{\mathrm b}=X'\cup Y'$ is rooted
by the equivalence above; conversely, for $Y'\subseteq U\setminus X'$ put $Y=\{v:v^{\mathrm b}\in Y'\}\cup(\{r\}\setminus X)$,
so that $|Y|=|Y'|+[r\notin X]$, $r\in X\cup Y$ and $(X\cup Y\setminus\{r\})^{\mathrm b}=X'\cup Y'$, and $X\cup Y$
is $r$-internal when $X'\cup Y'$ is rooted.

The algorithm of \cref{lem:fa-dst-ext} runs in time $2^{g_{\mathrm{DST}}(X')}n^{O(1)}$ and polynomial space,
as $\vec G^{\,r}$ has $2n-1$ vertices. We show $g_{\mathrm{DST}}(X')\le g_r(X)$. Let $W=\{r^{\mathrm b}\}\cup K\cup X'$,
whose blue vertices are the copies of $X\cup\{r\}$, and let $W^{r}$ be the set of vertices reachable from
$r^{\mathrm b}$ in $\vec G^{\,r}[W]$. Let $A$ be the set of vertices reachable from $r$ in $\vec G[X\cup\{r\}]$;
since red vertices have no outgoing arcs, the blue vertices of $W^{r}$ are exactly $A^{\mathrm b}$, and the
blue vertices of $W\setminus W^{r}$ are $(X\setminus A)^{\mathrm b}$, which induce a copy of $\vec G[X\setminus A]$.
Let $C$ be a source component of $\vec G^{\,r}[W\setminus W^{r}]$; we assign to it an element of
$\operatorname{und}_r(X)$ or of $\mathrm{Src}_r(X)$, injectively.

If $C$ contains a red vertex $v^{\mathrm r}$, then $C=\{v^{\mathrm r}\}$, because every vertex of a strong
component with at least two vertices has an outgoing arc inside it. Let $u^{\mathrm b}$ be an in-neighbor
of $v^{\mathrm r}$ in $\vec G^{\,r}$ with $u\in X\cup\{r\}$, so that $u^{\mathrm b}\in W$. If $u^{\mathrm b}\in W^{r}$, then
$v^{\mathrm r}\in W^{r}$, and otherwise the arc $u^{\mathrm b}v^{\mathrm r}$ enters $C$ from $(W\setminus W^{r})\setminus C$;
both contradict the choice of $C$. So no in-neighbor of $v^{\mathrm r}$ has its underlying vertex in
$X\cup\{r\}$: $v\notin X$ (as $v^{\mathrm b}$ is an in-neighbor), $v\neq r$, and $N^-(v)\cap X=\emptyset$. Hence
$v\in\operatorname{und}_r(X)$.

Otherwise $C$ consists of blue vertices, $C=D^{\mathrm b}$ with $D\subseteq X\setminus A$, and $D$ is strongly
connected in $\vec G[X\setminus A]$. No arc of $\vec G[X]$ enters $D$ from $X\setminus D$: an arc from
$X\setminus(A\cup D)$ would enter $C$ from $(W\setminus W^{r})\setminus C$ in $\vec G^{\,r}[W\setminus W^{r}]$, and an
arc from a vertex of $A$ to a vertex $d\in D$ would make $d$ reachable from $r$ in $\vec G[X\cup\{r\}]$, that
is, $d\in A$. So $D$ is a strong component of $\vec G[X]$ (a vertex of $X\setminus D$ in the same strong
component would have a path to $D$ in $\vec G[X]$, whose last arc enters $D$), it is a source, and $r\notin D$
because $r\in A$. Hence $D\in\mathrm{Src}_r(X)$.

Distinct source components give distinct elements, so $g_{\mathrm{DST}}(X')\le|\operatorname{und}_r(X)|+|\mathrm{Src}_r(X)|=g_r(X)$.
The value $g_r(X)$, and membership in $\Phi_r$, are computed in polynomial time, and the range was shown
before the lemma.
\end{proof}

The proof of the certificate is the directed analogue of \cref{lem:fa-dst-point}: the source components of $\vec G[X]$ are bounded by the vertices of $X$
whose parent is not in $X$.

\begin{lemma}[Pointwise certificate for out-branchings]\label{lem:cv-dir-cert}
Let $S$ be an $r$-internal set, and let $F$ be an out-branching rooted at $r$ in which every vertex outside
$S$ is a leaf (\cref{lem:cv-dir-identity}(b)). Then for every $X\subseteq S$,
\[
  g_r(X)\ \le\ n-1-\sum_{x\in X}|\mathrm{ch}_F(x)| .
\]
Consequently the weights $w(x)=|\mathrm{ch}_F(x)|-[x=r]$ on $S$ are nonnegative integers with $w(S)=n-2$,
and $g_r(X)\le n-1-w(X)$ for every $X\subseteq S$; this is the certificate~\eqref{eq:det-cert}.
\end{lemma}
\begin{proof}
Let $Q$ be the set of vertices $v\neq r$ whose parent in $F$ is not in $X$. Every vertex other than $r$
has exactly one parent, and the children sets $\mathrm{ch}_F(x)$ are disjoint, so
$|Q|=n-1-\sum_{x\in X}|\mathrm{ch}_F(x)|$. We show that $\operatorname{und}_r(X)\subseteq Q\setminus X$ and that
$\mathrm{Src}_r(X)$ maps injectively into $Q\cap X$; adding the two bounds gives $g_r(X)\le|Q|$.

A vertex $v\in\operatorname{und}_r(X)$ is not $r$ and not in $X$, and its parent is an in-neighbor of $v$, so
the parent is not in $X$; hence $v\in Q\setminus X$. Let $C\in\mathrm{Src}_r(X)$, let $x\in C$, and let $u$ be
the ancestor of $x$ in $F$ closest to $r$ such that the path of $F$ from $u$ to $x$ lies in $X$. This path
is a path of $\vec G[X]$ ending in $C$, and no arc of $\vec G[X]$ enters the source component $C$ from outside,
so $u\in C$. As $r\notin C$, $u\neq r$, and by the choice of $u$ its parent is not in $X$; so $u\in Q\cap
X\cap C$. As the members of $\mathrm{Src}_r(X)$ are disjoint, choosing such a $u$ for each of them is injective.

Every vertex other than $r$ has a parent, which is a non-leaf and so lies in $S$; hence
$\sum_{x\in S}|\mathrm{ch}_F(x)|=n-1$ and $w(S)=n-2$. As $n\ge2$, $r$ has a child, so $w\ge0$, and
$w(X)\le\sum_{x\in X}|\mathrm{ch}_F(x)|$ gives $g_r(X)\le n-1-w(X)$.
\end{proof}

\begin{proof}[Proof of \cref{thm:cv-dir}]
For $n=1$ the answer is $1$; let $n\ge2$. For every $r$ for which $V(\vec G)$ is $r$-internal (by
\cref{lem:cv-dir-identity}, the other roots carry no out-branching), compute the least size $s_r$ of an
$r$-internal set, and output $n-\min_rs_r$; if there is no such $r$, report that there is no out-branching.
This is correct by \cref{lem:cv-dir-identity}.

To compute $s_r$, run the procedure $\mathrm{Det}$ of \cref{lem:det-decide} for $\Phi_r$, the extension
algorithm and running-time exponent $g_r$ of \cref{lem:cv-dir-ext} and $(\alpha,\Xi)=(n-1,n-2)$, for $s=1,\dots,n$,
and let $s_r$ be the least $s$ at which it answers \YES{}; $V(\vec G)$ is a solution of $\Phi_r$, and $s_r$
is the least solution size. By \cref{lem:det-decide}, $\mathrm{Det}$ answers \NO{} if $s_r>s$. At
$s=s_r$ a minimum $r$-internal set $S$ has size exactly $s$, and \cref{lem:cv-dir-cert} gives its integer
certificate for $(n-1,n-2)$, so $\mathrm{Det}$ answers \YES{}. Hence \cref{lem:cds-first-accepted}, in its
deterministic form, computes $s_r$ within the time and space bounds of \cref{lem:det-decide}, and running
the computation for each of the at most $n$ roots multiplies the running time by at most~$n$.
\end{proof}

\subsubsection{Directed Steiner Tree}\label{sec:app-fa-dst}

This subsection proves \cref{thm:fa-dst}, with the notation of \cref{sec:fa-dst}. Fix an inclusion-minimal rooted Steiner set $S$, put $s=|S|$ and $p=|K|$, and let $A$ be an
out-arborescence of $\vec G[\{r\}\cup K\cup S]$ rooted at $r$; it exists because $S$ is rooted. By
minimality every leaf of $A$ other than $r$ is a terminal: if a vertex $v\in S$ were a leaf, $A-v$ would
show that $S\setminus\{v\}$ is rooted. For $u\neq r$ write $\mathrm{par}(u)$ for the parent of $u$ in
$A$. The trees of the forest $A[\bar K_X]$ that do not contain $r$ bound the running-time exponent $g(X)$ of a sample
$X\subseteq S$ from above.

\begin{lemma}[Pointwise bound]\label{lem:fa-dst-point}
For every $X\subseteq S$,
\[
  g(X)\ \le\ \Psi_A(X):=\bigl|\{u\in K\cup X:\ \mathrm{par}(u)\notin \bar K_X\}\bigr| .
\]
Equality holds if $\vec G[\bar K_X]=A[\bar K_X]$.
\end{lemma}
\begin{proof}
The forest $A[\bar K_X]$ consists of out-trees; the tree containing $r$ lies in $\bar K^{r}_X$, and every other tree
has a root $u\in K\cup X$ with $\mathrm{par}(u)\notin \bar K_X$. So at most $\Psi_A(X)$ trees miss $r$. A
tree whose root lies in $\bar K^{r}_X$ lies entirely in $\bar K^{r}_X$, as its arcs are arcs of $\vec G[\bar K_X]$. So every
$v\in \bar K_X\setminus \bar K^{r}_X$ lies in a tree with root $u\in \bar K_X\setminus \bar K^{r}_X$, and the tree path from $u$ to
$v$ avoids $\bar K^{r}_X$; hence $v$ is reachable from $u$ within $\vec G[\bar K_X\setminus \bar K^{r}_X]$. If $v$ lies in a source
component $C$, then so does $u$, since nothing outside $C$ reaches $C$. Assigning to each source
component the root of the tree of one of its vertices is therefore injective, and $g(X)\le\Psi_A(X)$.
If $\vec G[\bar K_X]=A[\bar K_X]$, then $\vec G[\bar K_X\setminus \bar K^{r}_X]$ is a forest of out-trees whose source components are
exactly the roots counted by $\Psi_A$.
\end{proof}

Consequently, with the quantities of \cref{def:fw-moments} for the universe $U$,
\begin{equation}\label{eq:fa-dst-W}
  W_S(y)=\sum_{X\subseteq S}y^{|X|}2^{-g(X)}\ \ge\ W_A(y):=\sum_{X\subseteq S}y^{|X|}2^{-\Psi_A(X)}
  \qquad(y\ge0).
\end{equation}
Whether a vertex $u$ is counted in $\Psi_A(X)$ depends only on whether $u$ and $\mathrm{par}(u)$ lie
in $\bar K_X$. This locality lets us compute $W_A$ by a recursion along $A$ (\cref{lem:fa-dst-rec}) with two
states: a vertex lies in $\bar K_X$ (a
terminal, the root, or a sampled vertex), or it is an unsampled vertex of $S$.
For $x,x'\in\R^2$ let
\[
  x\circ x'=\Bigl(x_1(x'_1+x'_2),\ x_2\bigl(\tfrac{x'_1}2+x'_2\bigr)\Bigr),
\]
a bilinear map with nonnegative coefficients. For a vertex $v$ of $A$ let $A_v$ be the subtree of $v$
and its descendants. For $y\ge0$ we define vectors $\mathrm{msg}_v\in\R^2$ for the vertices $v$ of $A$, from
the leaves up. Put $\mathrm{Init}_v=(1,0)$ if $v\in K\cup\{r\}$ and $\mathrm{Init}_v=(y,1)$ if $v\in S$, and for $v$ with
children $o_1,\dots,o_\ell$, in any order, let
$\mathrm{msg}_v=\bigl(\cdots(\mathrm{Init}_v\circ\mathrm{msg}_{o_1})\circ\cdots\bigr)\circ\mathrm{msg}_{o_\ell}$.

\begin{lemma}[Two-state recursion]\label{lem:fa-dst-rec}
Let $y\ge0$, and let the vectors $\mathrm{msg}_v$ be defined with this $y$ and any order of the children. Then
$W_A(y)=(\mathrm{msg}_r)_1$.
\end{lemma}
\begin{proof}
For a vertex $v$ and $X\subseteq S\cap V(A_v)$ put
$\Psi_v(X)=|\{u\in V(A_v)\setminus\{v\}:u\in K\cup X,\ \mathrm{par}(u)\notin \bar K_X\}|$; this depends only
on $X$, as $\mathrm{par}(u)\in V(A_v)$ for these $u$. Say that $X$ puts $v$ in state~$1$ if $v\in \bar K_X$ and in
state~$2$ otherwise. We show by induction that $(\mathrm{msg}_v)_i$ is the sum of $y^{|X|}2^{-\Psi_v(X)}$ over
the sets $X\subseteq S\cap V(A_v)$ that put $v$ in state $i$; more precisely, that the same holds after
the first $j$ children, for the subtree $A_v^j$ formed by $v$ and the subtrees of $o_1,\dots,o_j$, and
for $\Psi_v^j$ defined like $\Psi_v$ on $A_v^j$.

For $j=0$ the subtree is $\{v\}$ and $\Psi_v^0=0$. If $v\in K\cup\{r\}$, the only set is $\emptyset$, in
state~$1$, giving $(1,0)$; if $v\in S$, the set $\{v\}$ is in state~$1$ with weight $y$ and $\emptyset$ in
state~$2$ with weight $1$, giving $(y,1)$. For the step to $j$, write $o=o_j$ and split
$X=X'\cup X_o$ with $X'\subseteq V(A_v^{j-1})$ and $X_o\subseteq V(A_o)$. Then
$\Psi_v^j(X)=\Psi_v^{j-1}(X')+\Psi_o(X_o)+[o\in \bar K_X\text{ and }v\notin \bar K_X]$, because $o$ is the
only vertex counted in $\Psi_v^j$ whose parent lies in the other part. Summing the weights, a state-$1$
vertex $v$ receives the factor $(\mathrm{msg}_o)_1+(\mathrm{msg}_o)_2$ and a state-$2$ vertex $v$ the factor
$\frac12(\mathrm{msg}_o)_1+(\mathrm{msg}_o)_2$, which is the product $\circ$. Finally $r\in \bar K_X$ always and $r$ is not
counted, so $W_A(y)=\sum_Xy^{|X|}2^{-\Psi_r(X)}=(\mathrm{msg}_r)_1$.
\end{proof}

\paragraph{Antinorm certificates.} As for RS-$\theta$ (\cref{lem:rst-abstract}), a region of the plane that
$\circ$ maps into itself, up to known scale factors, bounds $W_A$ from below for all arborescences at
once. With two states the region lies in the plane, and the cone added to the convex hull can be the
nonnegative quadrant. For a finite set
$Z\subseteq\R^2_{\ge0}$ whose points have positive first coordinate, let
$\mathcal R_Z=\mathrm{conv}(Z)+\R^2_{\ge0}$ and $c_Z=\min_{z\in Z}z_1$. Minimality of $S$ enters
through the vertices of $S$: since none of them is a leaf of $A$, their initial vector $\mathrm{Init}_v=(y,1)$ is
never the vector $\mathrm{msg}_v$ of a complete subtree, and it is needed only as a left factor of $\circ$
(condition~\ref{it:fa-C3} below).

\begin{lemma}[Certificate]\label{lem:fa-dst-cert}
Let $y\ge0$, $e_1,e_2>0$, and let $Z\subseteq\R^2_{\ge0}$ be a finite set whose points have positive
first coordinate, with
\begin{enumerate}[label=\textup{(C\arabic*)}]
\item\label{it:fa-C1} $z\circ z'\in\mathcal R_Z$ for all $z,z'\in Z$;
\item\label{it:fa-C2} $(1,0)\in e_2\mathcal R_Z$;
\item\label{it:fa-C3} $(y,1)\circ z\in e_1\mathcal R_Z$ for all $z\in Z$.
\end{enumerate}
Then $W_A(y)\ge c_Z\,e_1^{\,s}e_2^{\,p+1}$ for every out-arborescence $A$ as above, that is, rooted at
$r$, with $s$ vertices in $S$, $p$ terminals, and no leaf in $S$.
\end{lemma}
\begin{proof}
First, if $x\in\eta\mathcal R_Z$ and $x'\in\eta'\mathcal R_Z$ with $\eta,\eta'>0$, then
$x\circ x'\in\eta\eta'\mathcal R_Z$, and $(y,1)\circ x'\in e_1\eta'\mathcal R_Z$. Indeed, write
$x=\eta(z+q)$ and $x'=\eta'(z'+q')$ with $z,z'\in\mathrm{conv}(Z)$ and $q,q'\in\R^2_{\ge0}$. By
bilinearity, $x\circ x'=\eta\eta'\bigl(z\circ z'+z\circ q'+q\circ(z'+q')\bigr)$. Writing $z$ and $z'$
as convex combinations of points of $Z$ shows that $z\circ z'$ is a convex combination of products of
points of $Z$, so it lies in the convex set $\mathcal R_Z$ by~\ref{it:fa-C1}; the other two terms are
nonnegative, because $\circ$ maps pairs of nonnegative vectors to nonnegative vectors, and
$\mathcal R_Z+\R^2_{\ge0}=\mathcal R_Z$. Likewise
$(y,1)\circ x'=\eta'\bigl((y,1)\circ z'+(y,1)\circ q'\bigr)$ with $(y,1)\circ z'\in e_1\mathcal R_Z$
by~\ref{it:fa-C3} and convexity, and $(y,1)\circ q'\ge0$.

Let $s_v$ and $p_v$ be the numbers of vertices of $A_v$ in $S$ and in $K\cup\{r\}$. We show
$\mathrm{msg}_v\in e_1^{\,s_v}e_2^{\,p_v}\mathcal R_Z$ by induction from the leaves. If $v\in K\cup\{r\}$, then
$\mathrm{Init}_v=(1,0)\in e_2\mathcal R_Z$ by~\ref{it:fa-C2}, and each child $o$ multiplies the scale by that of
$\mathrm{msg}_o$, by the first claim. If $v\in S$, then $v$ is not a leaf; with its first child $o_1$, the second
claim gives $\mathrm{Init}_v\circ\mathrm{msg}_{o_1}\in e_1e_1^{\,s_{o_1}}e_2^{\,p_{o_1}}\mathcal R_Z$, and the remaining
children are handled by the first claim. At the root, $s_r=s$ and $p_r=p+1$. Every point of
$\mathcal R_Z$ has first coordinate at least $c_Z$, so \cref{lem:fa-dst-rec} gives
$W_A(y)=(\mathrm{msg}_r)_1\ge c_Ze_1^{\,s}e_2^{\,p+1}$.
\end{proof}

\begin{proposition}[Certificates for Directed Steiner Tree]\label{prop:fa-dst-certs}
The certificates of \cref{sec:anc-dst} are three rational certificates
$(y_j,e_{1,j},e_{2,j},Z_j)$, $j=0,1,2$, each of which satisfies~\ref{it:fa-C1}--\ref{it:fa-C3} with
$y=y_j$, $e_1=e_{1,j}$, $e_2=e_{2,j}$ and $Z=Z_j$, and rational weights $\lambda_j\ge0$ with
$\sum_j\lambda_j=1$. With the exponent vectors
$\boldsymbol\alpha_j=(\alpha_{j,1},\alpha_{j,2},\alpha_{j,3})=\bigl(\log(1+y_j)-\log e_{1,j},\ -\log e_{2,j},\ \log(1+y_j)\bigr)$, the data are, approximately,
\begin{center}
\setlength{\tabcolsep}{4pt}
\begin{tabular}{@{}llllllll@{}}
\toprule
$j$ & $\lambda_j\approx$ & $y_j\approx$ & $e_{1,j}\approx$ & $e_{2,j}\approx$ & $|Z_j|$ & $c_{Z_j}e_{2,j}\ge$ & $(\alpha_{j,1},\alpha_{j,2},\alpha_{j,3})\approx$ \\
\midrule
$0$ & $0.7458$ & $0.45494$ & $1.003883$ & $0.78$ & $27$ & $0.40$ & $(0.535367,\,0.358454,\,0.540958)$ \\
$1$ & $0.0858$ & $0.48223$ & $1.041122$ & $0.78$ & $10$ & $0.41$ & $(0.509635,\,0.358454,\,0.567774)$ \\
$2$ & $0.1684$ & $0.02$ & $0.963015$ & $0.52$ & $28$ & $0.011$ & $(0.082938,\,0.943416,\,0.028569)$ \\
\bottomrule
\end{tabular}
\end{center}
and
\[
  \max_{i\in\{1,2,3\}}\ \sum_{j=0}^2\lambda_j\alpha_{j,i}\ =\ 0.4569669\ldots\ <\ 0.457 .
\]
\end{proposition}

\begin{proof}
The hypotheses~\ref{it:fa-C1}--\ref{it:fa-C3}, the displayed maximum, which lies more than
$3.3\cdot10^{-5}$ below $0.457$, and the inequality $0.457<\log1.3727$ are checked in \cref{sec:anc-dst}.
\end{proof}

\RestThmFaDst*

\begin{proof}
Let $\psi=0.457$; we may assume $n_U\ge1$. Run Algorithm~\MLS{}
for $\Phi_{\mathrm{DST}}$, the extension algorithm of \cref{lem:fa-dst-ext} and the running-time exponent $g$,
with universe size $n_U$, rate $\psi n/n_U$, $\Gamma=100$ and confidence $\kappa$, for $k=0,1,\dots,n_U$, and
output the least accepted $k$, or report that no Steiner set exists if no $k$ is accepted. By \cref{thm:fw-main} and \cref{rem:fw-decision}, this takes time
$2^{\psi n}n^{O(1)}$ and polynomial space, and no accepted $k$ is below the optimum; in particular, no $k$ is
accepted if there is no Steiner set. Let $S$ be a
minimum Steiner set; it is rooted and inclusion-minimal, so for $k=|S|$ it remains to verify
condition~\eqref{eq:fw-star}. Let $r_0=n_U-s$, so that $s+p+r_0=n-1$. For each $j$,
\cref{eq:fa-dst-W,lem:fa-dst-cert} and the first inequality of \cref{lem:fw-gf} give
\[
  \max_{0\le t\le s}\frac{\mu_t(S)}{\binom{n_U}t}\ \ge\ \frac{W_S(y_j)}{(1+y_j)^{n_U}}
  \ \ge\ c_{Z_j}e_{2,j}\,\frac{e_{1,j}^{\,s}e_{2,j}^{\,p}}{(1+y_j)^{s+r_0}}
  \ =\ c_{Z_j}e_{2,j}\,2^{-(\alpha_{j,1}s+\alpha_{j,2}p+\alpha_{j,3}r_0)} .
\]
Choose $j$ with the least exponent. By \cref{prop:fa-dst-certs},
$\alpha_{j,1}s+\alpha_{j,2}p+\alpha_{j,3}r_0\le\sum_i\lambda_i(\alpha_{i,1}s+\alpha_{i,2}p+\alpha_{i,3}r_0)\le\psi(n-1)$, and
$c_{Z_j}e_{2,j}\ge1/100$. So $\max_t\mu_t(S)/\binom{n_U}t\ge2^{-(\psi n/n_U)n_U}/\Gamma$, which
is~\eqref{eq:fw-star}. Finally $2^{0.457}=1.37268\ldots$.
\end{proof}

\subsubsection{Paired Domination}\label{sec:app-fa-pd}

This subsection proves \cref{lem:fa-pd-ext,thm:fa-pd,prop:fa-pd-tight}, with the notation of \cref{sec:fa-pd}.

\RestLemFaPdExt*

\begin{proof}
Fix a target size $s\in\{|X|,\dots,\min(n,k)\}$ and a target weight
\[
  b\in\bigl\{0,\dots,\lfloor\min(\varpi,w(V))\rfloor\bigr\};
\]
there are at most $(n+1)(nw_{\max}+1)$ such pairs. A set
$A\supseteq X$ dominates $G$ if and only if it dominates every vertex of $\operatorname{und}(X)$, that is,
$A\cap N[u]\neq\emptyset$ for all $u\in \operatorname{und}(X)$. For $F\subseteq \operatorname{und}(X)$ let
$V_F=V\setminus N[F]$, and consider the polynomial
\[
  \mathsf F_{s,b}=\sum_{F\subseteq \operatorname{und}(X)}(-1)^{|F|}\sum_{\substack{X\subseteq A\subseteq V_F\\|A|=s,\ w(A)=b}}\det\mathbf T_A .
\]
Exchanging the sums, the coefficient of $\det\mathbf T_A$, for $A\supseteq X$ with $|A|=s$ and $w(A)=b$, is
\[
  \sum_{F\subseteq \operatorname{und}(X)}(-1)^{|F|}\prod_{u\in F}[A\cap N[u]=\emptyset]
  =\prod_{u\in \operatorname{und}(X)}\bigl(1-[A\cap N[u]=\emptyset]\bigr),
\]
which is $1$ if $A$ dominates $\operatorname{und}(X)$ and $0$ otherwise. So
\[
  \mathsf F_{s,b}=\sum_{A}\mathrm{Pf}(\mathbf T_A)^2,
\]
summed over the dominating sets $A\supseteq X$ of size $s$ and weight $b$.

\emph{No cancellation.} Let $M$ be a perfect matching of $G[A]$ for such an $A$. In
$\mathrm{Pf}(\mathbf T_{A'})^2=\sum_{M_1,M_2}\pm x_{M_1}x_{M_2}$, a product $x_{M_1}x_{M_2}$ equals
$x_M^2$ only if every edge of $M$ lies in both $M_1$ and $M_2$ and no other edge occurs, that is,
$M_1=M_2=M$ and $A'=V(M)=A$; this term has coefficient $(\pm1)^2=1$. So the coefficient of $x_M^2$ in
$\mathsf F_{s,b}$ is exactly $1$, and $\mathsf F_{s,b}\neq0$ if and only if some dominating $A\supseteq X$ of size
$s$ and weight $b$ has a perfect matching in $G[A]$, that is, if $A\setminus X$ is a completion of size
$s-|X|$ with $w(A)=b$.

\emph{Evaluation.} For $F\subseteq \operatorname{und}(X)$ we have $X\subseteq V_F$, because no vertex of $F$ lies in
$N[X]$. Let $\mathbf E$ be the diagonal matrix on $V_F$ with $\mathbf E_{vv}=z^{w(v)}$ for $v\notin X$ and
$\mathbf E_{vv}=0$ for $v\in X$, with a further indeterminate $z$. Expanding the determinant along the
diagonal, with one more indeterminate $x_\star$,
\[
  \det\bigl(\mathbf T_{V_F}+x_\star\mathbf E\bigr)=\sum_{Q\subseteq V_F\setminus X}x_\star^{|Q|}z^{w(Q)}
  \det\mathbf T_{V_F\setminus Q},
\]
so the inner sum of $\mathsf F_{s,b}$ is the coefficient of $x_\star^{|V_F|-s}z^{w(V_F)-b}$. Choose a prime
$q$ with $4n+nw_{\max}\le q\le8n+2nw_{\max}$, by trial division, and substitute independent uniform
elements of $\mathbb F_q$ for the $x_e$. The determinant is then a polynomial in $x_\star$ and $z$ of degree
at most $n$ in $x_\star$ and at most $nw_{\max}<q$ in $z$, so evaluating it at $(n+1)(nw_{\max}+1)$ points
and interpolating gives the coefficient over $\mathbb F_q$ in polynomial time, and summing over $F$ gives
$\mathsf F_{s,b}$ at the random point, in time $2^{|\operatorname{und}(X)|}(n+w_{\max})^{O(1)}$ and polynomial space.
Accept if the value is nonzero for some $s$ and $b$.

If no completion of weight at most $\varpi$ exists, every $\mathsf F_{s,b}$ with $b\le\varpi$ is the zero
polynomial and the algorithm never accepts. If a completion of size $s-|X|$ and weight $b\le\varpi$
exists, then $\mathsf F_{s,b}$ is a nonzero polynomial of degree $s\le n$ in the $x_e$ with a coefficient
$1$, so it is nonzero over $\mathbb F_q$, and by the Schwartz--Zippel lemma~\cite{Schwartz80,Zippel79} it
vanishes at the random point with probability at most $n/q\le\frac14$. With $w\equiv0$ and $\varpi=0$ only
$b=0$ occurs, the determinant does not involve $z$, and the bounds are polynomial in $n$.
\end{proof}

Fix a paired dominating set $S$ and a perfect matching $M$ of $G[S]$, and
choose an \emph{anchor} $\mathrm{anc}(v)\in S\cap N(v)$ for every $v\notin S$. For $u\in S$ let $d_u=|\mathrm{anc}^{-1}(u)|$ be the number of vertices anchored at $u$,
and for a pair $uv\in M$ let $n_{uv}=2+d_u+d_v$, the number of vertices the pair is responsible for; so
$\sum_{uv\in M}n_{uv}=n$.

\begin{lemma}[Pointwise certificate]\label{lem:fa-pd-point}
For every $X\subseteq S$,
\[
  |\operatorname{und}(X)|\ \le\ \bigl|\{v\notin S:\mathrm{anc}(v)\notin X\}\bigr|+2\,\bigl|\{uv\in M:u,v\notin X\}\bigr| ,
\]
and consequently, for every $y\ge0$,
\[
  W_S(y)=\sum_{X\subseteq S}y^{|X|}2^{-|\operatorname{und}(X)|}\ \ge\ \prod_{uv\in M}\Omega_{uv}(y),\qquad
  \Omega_{uv}(y)=y^2+y\bigl(2^{-d_u}+2^{-d_v}\bigr)+2^{-n_{uv}} .
\]
\end{lemma}
\begin{proof}
A vertex $v\notin S$ is dominated once $\mathrm{anc}(v)\in X$. A vertex $u\in S$ is dominated once $u$ or its partner
in $M$ lies in $X$. This gives the first inequality. The right-hand side is a sum over the pairs of $M$
of a term that depends only on $X\cap\{u,v\}$: it is $0$ if $u,v\in X$, it is $d_v$ if only $u\in X$, it is
$d_u$ if only $v\in X$, and it is $n_{uv}$ if neither is in $X$. So
$\sum_{X\subseteq S}y^{|X|}2^{-(\text{right-hand side})}$ factorizes over the pairs into the
$\Omega_{uv}(y)$.
\end{proof}

By the inequality of arithmetic and geometric means,
$2^{-d_u}+2^{-d_v}\ge2\cdot2^{-(n_{uv}-2)/2}$, so $\Omega_{uv}(y)\ge y^2+4\cdot2^{-n_{uv}/2}y+2^{-n_{uv}}$ for $y\ge0$. The
functions $h_{\bar n}$, the point $y^*$ and the rate $\psi_{\mathrm{pr}}$ are those of \cref{sec:fa-pd}.

\begin{lemma}[Rate]\label{lem:fa-pd-rate}
The functions $h_{\bar n}$ and the point $y^*$ defined above satisfy
\[
  \max_{\bar n\in\{2,3,\dots\}}h_{\bar n}(y^*)=h_6(y^*)=\psi_{\mathrm{pr}}=0.7050097\ldots,
\]
and for every graph
$G$ without isolated vertices and every paired dominating set $S$ of $G$,
\[
  W_S(y^*)\ \ge\ (1+y^*)^n\,2^{-\psi_{\mathrm{pr}}n} .
\]
\end{lemma}
\begin{proof}
For $\bar n\ge41$ we have $y^2+4\cdot2^{-\bar n/2}y+2^{-\bar n}\ge y^2$ and $0<y^*<1$, so
$h_{\bar n}(y^*)\le\log(1+y^*)+\frac2{41}\log\frac1{y^*}<0.431$. For $2\le \bar n\le40$, $\bar n\neq6$, direct evaluation (\cref{sec:anc-pd})
gives $h_6(y^*)-h_{\bar n}(y^*)>0.00297$, the smallest gap occurring at $\bar n=7$; it also gives
$\log(1+y^*)+\frac2{41}\log\frac1{y^*}<0.431<h_6(y^*)$, which completes the tail bound. This
proves the first claim. By \cref{lem:fa-pd-point} and the bound
$\Omega_{uv}(y)\ge y^2+4\cdot2^{-n_{uv}/2}y+2^{-n_{uv}}$,
\[
\begin{split}
  \log\frac{(1+y^*)^n}{W_S(y^*)}\ &\le\ \sum_{uv\in M}\Bigl(n_{uv}\log(1+y^*)-\log\Omega_{uv}(y^*)\Bigr)\\
  &\le\ \sum_{uv\in M}n_{uv}\,h_{n_{uv}}(y^*)\ \le\ \psi_{\mathrm{pr}}\sum_{uv\in M}n_{uv}
  =\psi_{\mathrm{pr}}\,n .
\end{split}\qedhere
\]
\end{proof}

\RestThmFaPd*

\begin{proof}
Let $\psi=0.70502$, a rational with $\psi_{\mathrm{pr}}<\psi$ and $2^\psi<1.6302$. For
$k=2,4,\dots$ up to $n$, run Algorithm~\MLS{} for
$\Phi_{\mathrm{PD}}$, the extension algorithm of \cref{lem:fa-pd-ext} (with $w\equiv0$ and $\varpi=0$), the
running-time exponent $|\operatorname{und}(X)|$, rate
$\psi$, $\Gamma=1$ and confidence $\kappa$; as this extension algorithm has one-sided error, the cap and the trial
counts are those of \cref{cor:fw-onesided}, that is, those of \cref{thm:fw-main} for $2\Gamma=2$. Output the least
accepted $k$, or $n$ if no $k$ is accepted (as in \cref{lem:cds-first-accepted}). By \cref{thm:fw-main} with \cref{cor:fw-onesided}, each run takes time $2^{\psi n}n^{O(1)}$ and polynomial space and
rejects if no paired dominating set of size at most $k$ exists. For $k=\gamma_{\mathrm{pr}}(G)$, a
minimum paired dominating set satisfies the generating-function condition of \cref{thm:fw-main} at
$y=y^*$ with $\Gamma=1$, by \cref{lem:fa-pd-rate}, so by \cref{cor:fw-onesided} this $k$ is accepted with probability at least
$1-e^{-\kappa}$.
\end{proof}

\RestPropFaPdTight*

\begin{proof}
The derivative of $h_6$ vanishes exactly where
$\frac1{1+y}=\frac{2y+1/2}{6(y^2+y/2+1/64)}$, that is, where $4y^2+\frac y2-\frac{13}{32}=0$; the only
positive root is $y^*$. As $h_6(0)=1$ and $h_6(y)\to\infty$ as $y\to\infty$, while
$h_6(y^*)<1$, the minimum is attained at $y^*$.

In $G_L$, a pendant vertex at $u$ is dominated only by itself or by $u$; if it lies in a paired
dominating set, its partner is its only neighbor $u$. So every paired dominating set contains all
centers, and the centers form one. Hence $S^*$ is the only one of size at most $2L$, and a trial for
$k=2L$ accepts only if its sample lies in $S^*$. With the anchors at the centers,
\cref{lem:fa-pd-point} holds with equality, and $W_{S^*}(y)=(y^2+\frac y2+\frac1{64})^L$, so
$(1+y)^n/W_{S^*}(y)=2^{n\,h_6(y)}\ge2^{n\,\psi_{\mathrm{pr}}}$ for all $y\ge0$, with $n=6L$. By
the second inequality of \cref{lem:fw-gf}, $\max_t\mu_t(S^*)/\binom nt\le(n+1)2^{-\psi_{\mathrm{pr}}n}$, and by
\cref{prop:fw-necessary} the expected number of trials that run the extension on a subset of $S^*$, and so the
probability of acceptance, is at most $n^{O(1)}2^{(\psi'-\psi_{\mathrm{pr}})n}\to0$.
\end{proof}

\subsubsection{Vertex weights for Directed Steiner Tree and Paired Domination}\label{sec:app-weighted-fa}

The two weighted results stated in \cref{sec:fa-dst,sec:fa-pd} follow from the unweighted proofs through
the budget loop of \cref{lem:cv-budget-loop}, with the notation of those subsections.

The proof of \cref{thm:fa-dst} uses only two properties of $S$, that $S$ is
rooted and that $A$ has no leaf in $S$, and not that $|S|$ is least.

\begin{corollary}[Vertex-weighted Directed Steiner Tree]\label{cor:fa-dst-weighted}
For non-terminal weights $w\colon U\to\{0,1,\dots,w_{\max}\}$, written in unary, the least weight of a Steiner
set can be computed by a randomized algorithm with one-sided error in time $2^{0.457n}(n+w_{\max})^{O(1)}$ and
polynomial space.
\end{corollary}
\begin{proof}
We apply \cref{lem:cv-budget-loop} to $\Phi_{\mathrm{DST}}$ with the weights $w$ on its universe $U$; so
$\Phi_\varpi$ has as solutions the rooted Steiner sets of weight at most $\varpi$, and its least solution
weight is the least weight of a Steiner set, because the vertices of a Steiner set that are reachable
from $r$ form a rooted Steiner set of no larger weight. The extension algorithm $\cA_w$ computes the
least weight $\OPT$ of a completion of $X$ by \cref{lem:fa-dst-ext} and accepts if and only if
$w(X)+\OPT\le \varpi$; it ignores $k$. A witness $Y$ in~\ref{it:cv-R1} gives $w(X)+\OPT\le w(X\cup Y)\le\varpi$,
and an acceptance yields a rooted Steiner set of weight at most $\varpi$; so $\cA_w$
satisfies~\ref{it:cv-R1} and~\ref{it:cv-R2} of \cref{lem:cv-relaxed}, and $g$ is a running-time exponent for it,
with $N^{O(1)}=(n+w_{\max})^{O(1)}$. Let $\mathcal{B}_\varpi(I,k)$ be
Algorithm~\MLS{} for $\Phi_\varpi$ with $\cA_w$ and $g$, with the parameters of the proof of
\cref{thm:fa-dst}. By \cref{lem:cv-relaxed} with
\cref{thm:fw-main}, it runs in time $2^{0.457n}(n+w_{\max})^{O(1)}$ and polynomial space, and it answers
\NO{} if $\Phi_\varpi$ has no solution, which is property~\ref{it:cv-B1}. For
property~\ref{it:cv-B2}, let $S$ be a Steiner set of least weight $\varpi^*$ that is inclusion-minimal
among those; it is rooted, and $A$ has no leaf in $S$, since deleting such a leaf would keep a Steiner set
of no larger weight. The proof of \cref{thm:fa-dst} uses only these two properties of $S$ to verify
condition~\eqref{eq:fw-star} at $k=|S|$, so $\mathcal{B}_{\varpi^*}(I,|S|)$ answers \YES{} with probability at least
$1-e^{-\kappa}$. Now \cref{lem:cv-budget-loop} gives the claim, with $w(U)+1\le n_Uw_{\max}+1$ budgets.
\end{proof}

For unit weights this is \cref{thm:fa-dst}. We
know of no algorithm for vertex-weighted \pname{Directed Steiner Tree}, or for its undirected special case,
with running time $O(c^n)$ for a constant $c<\sqrt2$; for undirected \pname{Vertex-Weighted Steiner Tree},
Buchanan, Wang and Butenko~\cite{BuchananWB18} give $O(1.5296^n)$.

Let $w\colon V\to\{0,\dots,w_{\max}\}$ be vertex weights, and let
$\gamma^w_{\mathrm{pr}}(G)$ be the least weight of a paired dominating set.

\begin{theorem}[Weighted Paired Domination]\label{thm:fa-pd-weighted}
For graphs $G$ with $n$ vertices and no isolated vertex and weights $w\colon V\to\{0,\dots,w_{\max}\}$ in
unary, $\gamma^w_{\mathrm{pr}}(G)$ can be computed by a randomized algorithm with one-sided error in time
$O(1.6302^n)(n+w_{\max})^{O(1)}$ and polynomial space.
\end{theorem}
\begin{proof}
We apply \cref{lem:cv-budget-loop} to $\Phi_{\mathrm{PD}}$ with the weights $w$, so that the solutions of
$\Phi_\varpi$ are the paired dominating sets of weight at most $\varpi$; \cref{lem:fa-pd-ext} with the weights $w$
and the budget $\varpi$ is a one-sided-error extension algorithm for $\Phi_\varpi$ with running-time exponent
$|\operatorname{und}(X)|$, as in \cref{cor:fw-onesided}.
Let $\mathcal{B}_\varpi(G,k)$ be Algorithm~\MLS{} for $\Phi_\varpi$ with this extension algorithm and the parameters of
the proof of \cref{thm:fa-pd}, for even $k$ (and the answer \NO{} for odd $k$, as a paired dominating set
has even size). By \cref{thm:fw-main} with \cref{cor:fw-onesided}, it runs in time
$2^{\psi n}(n+w_{\max})^{O(1)}=O(1.6302^n)(n+w_{\max})^{O(1)}$, with $\psi=0.70502$ as in that proof, and in
polynomial space. It answers \NO{} if $\Phi_\varpi$
has no solution, which is property~\ref{it:cv-B1}. For property~\ref{it:cv-B2}, let $S$ be a paired dominating
set of least weight; \cref{lem:fa-pd-rate} holds for every paired dominating set, so $S$ satisfies
condition~\eqref{eq:fw-star} at $k=|S|$ and $\mathcal{B}_{w(S)}(G,|S|)$ answers \YES{} with probability at least
$1-e^{-\kappa}$. Now \cref{lem:cv-budget-loop} gives the claim, with $w(V)+1\le nw_{\max}+1$ budgets.
\end{proof}
{}
\statementswithin{section}%
\section{Audit of the applications}\label{sec:app-audit}

This appendix collects, for each algorithmic theorem of the paper, what its proof imports, where its
randomness comes from, how it treats weights, and what it outputs. Every algorithm in the tables runs in
polynomial space, and every randomized one has one-sided error: it never outputs a value below the optimum,
or never accepts a \NO-instance, and it succeeds with probability at least $1-e^{-\kappa}$ for the confidence
$\kappa$ of the theorem, or at least $1-1/e$ where the theorem says so.

\paragraph{Polynomial factors.} Every extension algorithm runs in time $2^{g(I,X,k)}N^{O(1)}$ on every input, with
one polynomial for all inputs (\cref{def:fw-cost}); for randomized extension algorithms the bound holds for every
outcome of the coins. The polynomial factor of \cref{thm:fw-main,thm:fw-budgeted} depends only on $N$, since the
rate, the loss factor and the confidence are at most $N^{O(1)}$, and the trials of RSC run in time
$2^\theta r_0(n)$ for one polynomial $r_0$ that does not depend on $G$, $s$, $t$ or $\theta$
(\cref{lem:rsc-trials}). The deterministic algorithms use \cref{lem:det-time}, which holds uniformly for every
implicit set system, extension algorithm and running-time exponent with values in $\{0,\dots,n\}$, and they
carry the further factor $2^{O(n(\log\log n)^2/\log n)}$ of the balanced families. Weighted versions carry the
factor $(n+w_{\max})^{O(1)}$ stated in their theorems, which is $n^{O(1)}$ when $w_{\max}\le n^{O(1)}$.

\paragraph{From decisions to optimal values.} The algorithms for \pname{Connected Dominating Set} and its
variants run the decision procedure for $s=1,2,\dots$ and output the first accepted $s$
(\cref{lem:cds-first-accepted}); vertex-weighted versions run it for every budget $\varpi$ and output the least
accepted budget (\cref{lem:cv-budget-loop}). The randomized ones output a value that is never below the optimum and
equals it with the success probability of their theorem; the deterministic ones output the optimal value. The
algorithms for
\pname{Connected Feedback Vertex Set}, its weighted version, \pname{Connected Vertex Cover} and
\pname{Maximum-Weight Connected $k$-Subgraph} also output a solution, which is optimal with probability at least
$1-1/e$ (at least $1-e^{-\kappa}$ for \pname{Connected Vertex Cover}), through the witness of the
extension algorithm and, where the black box gives only a decision, a self-reduction
(\cref{lem:wt-self-reduction}). \pname{Graph Motif} is a decision problem.

\begin{table}[htbp]
  \centering
  \renewcommand{\arraystretch}{1.12}
  \begin{tabularx}{\textwidth}{@{}>{\raggedright\arraybackslash}p{2.5cm}>{\raggedright\arraybackslash}X>{\raggedright\arraybackslash}p{3.0cm}>{\raggedright\arraybackslash}p{3.3cm}@{}}
    \toprule
    Theorem & Extension step and imported results & Randomness & Weights \\
    \midrule
    \Cref{thm:rs,thm:rstheta} & Steiner extension \cref{cor:cds-steiner}, from \cref{lem:cv-W}, proved in \cref{sec:app-steiner}; parameter: the number of terminals & uniform samples & none \\
    \Cref{thm:rsc} & the same, on the closed sample & uniform samples and the closure & none \\
    \Cref{thm:det-cds} & the same & deterministic (balanced families, \cref{thm:dr-balanced}) & none \\
    \Cref{thm:rdom} & \cref{lem:cv-W} on the red--blue graph with $2n$ vertices (\cref{lem:cv-redblue}) & (a) uniform samples; (b) deterministic & none \\
    \Cref{thm:cv-weighted} & \cref{lem:cv-W} with vertex weights, encoded as arc lengths (\cref{sec:app-steiner}) & (a) uniform samples; (b) deterministic & $\{0,\dots,\varphi_{\max}\}$, unary \\
    \Cref{thm:cv-wcds} & \cref{lem:cv-W} via \cref{lem:cv-redblue} & deterministic & none \\
    \Cref{thm:cv-dir} & \cref{lem:fa-dst-ext}, from \cref{lem:cv-W}, once for every root & deterministic & none \\
    \bottomrule
  \end{tabularx}
  \caption{Algorithms for \pname{Connected Dominating Set}, \pname{Max Leaf Spanning Tree} and their variants.
    Each outputs the optimal value, with probability at least $1-e^{-\kappa}$ for the randomized ones
    (\cref{lem:cds-first-accepted,lem:cv-budget-loop}). No external algorithm is
    used as a black box: the vertex-weighted directed Steiner lemma (\cref{lem:cv-W}) is proved in this paper
    (\cref{sec:app-steiner}), following inclusion--exclusion over branching walks~\cite{Nederlof13}.}
  \label{tab:audit-cds}
\end{table}

\begin{table}[htbp]
  \centering
  \renewcommand{\arraystretch}{1.12}
  \begin{tabularx}{\textwidth}{@{}>{\raggedright\arraybackslash}p{2.5cm}>{\raggedright\arraybackslash}X>{\raggedright\arraybackslash}p{2.6cm}>{\raggedright\arraybackslash}p{3.6cm}@{}}
    \toprule
    Theorem & Extension step and imported results & Randomness & Weights \\
    \midrule
    \Cref{thm:cc-cfvs} & Cut \& Count with forced vertices (\cref{lem:fvs-ext}), from the facts (CC1)--(CC3): Theorem~A.13 of~\cite{CyganNPPRW11arxiv}, its proof, and the proof of Theorem~5.1 of~\cite{CyganNPPRW22} (\cref{sec:fvs-cc}) & uniform samples; Cut \& Count & none \\
    \Cref{thm:cc-wcfvs} & \cref{lem:fvs-ext} as a modulator, the cut parity and weighted core evaluation (\cref{lem:wfvs-parity,lem:wfvs-core}), proved in \cref{sec:app-weighted}, and the isolation lemma~\cite[Lemma~2.5]{CyganNPPRW22} & uniform samples; Cut \& Count; isolation & $1$ to $w_{\max}\le n^{O(1)}$; forced vertices weigh $0$ \\
    \Cref{thm:cc-motif} & \cite[Theorem~4 and Section~3.7]{BjorklundKK15motif} on the instance in which each component of the sample is contracted to a vertex with a fresh color (\cref{lem:mot-ext}) & uniform samples; the black box & none \\
    \Cref{thm:cc-cvc} & \cite[Theorem~3]{Cygan12}, deterministic, on the contracted graph with pendant vertices (\cref{lem:cc-cvc-ext}) & uniform samples only & $1$ to $w_{\max}\le n^{O(1)}$; contracted and pendant vertices weigh $1$ \\
    \Cref{thm:cc-mwcs} & a weighted sieve adapted from~\cite{BjorklundKK15motif}, proved in \cref{sec:app-weighted} & uniform samples; polynomial identity test & $-w_{\max}$ to $w_{\max}\le n^{O(1)}$; contracted vertices weigh $0$ \\
    \bottomrule
  \end{tabularx}
  \caption{Algorithms with component-count running times. The algorithms for \pname{Connected Feedback Vertex
    Set} and \pname{Maximum-Weight Connected $k$-Subgraph} output a solution that is optimal with probability
    at least $1-1/e$, the one for \pname{Connected Vertex Cover} a weight and a solution that are optimal with
    probability at least $1-e^{-\kappa}$, and the one for \pname{Graph Motif} a decision with one-sided error. Each extension step runs in time exponential in the number $k-|X|$ of elements still to be added
    and in the number $\comp(G[X])$ of components of the sample (\cref{sec:cc}).}
  \label{tab:audit-cc}
\end{table}

\begin{table}[htbp]
  \centering
  \renewcommand{\arraystretch}{1.12}
  \begin{tabularx}{\textwidth}{@{}>{\raggedright\arraybackslash}p{2.95cm}>{\raggedright\arraybackslash}X>{\raggedright\arraybackslash}p{3.0cm}>{\raggedright\arraybackslash}p{3.3cm}@{}}
    \toprule
    Theorem & Extension step and imported results & Randomness & Weights \\
    \midrule
    \Cref{thm:fa-dst} & \cref{lem:fa-dst-ext}, from \cref{lem:cv-W} & uniform samples & none \\
    \Cref{cor:fa-dst-weighted} & the same with weights, and \cref{lem:cv-budget-loop} & uniform samples & $\{0,\dots,w_{\max}\}$, unary \\
    \Cref{thm:fa-pd} & \cref{lem:fa-pd-ext}: Tutte matrix~\cite{Tutte47} and the identity test of Schwartz and Zippel~\cite{Schwartz80,Zippel79} & uniform samples; identity test & none \\
    \Cref{thm:fa-pd-weighted} & the same with an indeterminate for the weights, and \cref{lem:cv-budget-loop} & uniform samples; identity test & $\{0,\dots,w_{\max}\}$, unary \\
    \bottomrule
  \end{tabularx}
  \caption{\pname{Directed Steiner Tree} and \pname{Paired Domination}. Each algorithm outputs a value that is never
    below the optimum and equals it with probability at least $1-e^{-\kappa}$; the one for \pname{Directed Steiner Tree}
    may instead report, correctly, that no Steiner set exists.}
  \label{tab:audit-further}
\end{table}

\paragraph{Imported results.} The results that the proofs use as black boxes are: the Cut \& Count facts
(CC1)--(CC3) for \pname{Connected Feedback Vertex Set}, which its weighted version uses only through the
unweighted extension algorithm that supplies the modulator, together with the isolation lemma
in the form of~\cite[Lemma~2.5]{CyganNPPRW22}; the algorithm of Björklund, Kaski and Kowalik for \pname{Graph Motif}; Cygan's algorithm
for weighted \pname{Connected Vertex Cover}; and the Tutte matrix with the identity test of Schwartz and
Zippel. The Steiner computations for connected domination and \pname{Directed Steiner Tree}, the balanced
families, the weighted counting for \pname{Connected Feedback Vertex Set} (\cref{lem:wfvs-parity,lem:wfvs-core}),
and the weighted sieve for \pname{Maximum-Weight Connected $k$-Subgraph} are proved in this paper.
{}
\section{Computer certificates}\label{sec:ancillary}

Several constants in the bases of our running times rest on finite computations. The first table
lists every ancillary file that a proof in the paper depends on, together with what it certifies and
the arithmetic it uses; \cref{sec:anc-rs,sec:anc-rstheta,sec:anc-lobster,sec:anc-rsc,sec:anc-det,sec:anc-cc,sec:anc-dst,sec:anc-pd}
describe, result by result, what is checked and how. Every check is either in exact rational arithmetic or, for logarithms, exponentials
and square roots, in ball arithmetic with outward rounding (Arb~\cite{Johansson17arb}, through
python-flint, at $256$ bits): each operation returns a ball that provably contains the exact real
result, and an inequality is accepted only if it holds for the whole ball. Inputs such as
decimal endpoints and parameters are read as exact rationals. So each certificate is a proof, up to
the correctness of the software.

\begin{center}
\renewcommand{\arraystretch}{1.0}\setlength{\tabcolsep}{3pt}%
\begin{tabular}{@{}>{\raggedright\arraybackslash}p{5.5cm}>{\raggedright\arraybackslash}p{6.58cm}>{\raggedright\arraybackslash}p{4.0cm}@{}}
\toprule
\multicolumn{3}{@{}l}{\emph{Files that a proof depends on}}\\
\midrule
File & Certifies & Arithmetic \\
\midrule
\ancfile{rs\_closed\_form\_interval.py} & the constant $1.744835$ of RS and the two bounds on
$F^*$ used by RS-$\theta$ (\cref{prop:rs-rate}(c)), by adaptive bisection into $9608$, $60$ and $73$
intervals & exact rationals; Arb \\
\ancfile{cert\_final\_0.json}, \ancfile{cert\_final\_1.json}, \ancfile{verify\_cert.py},
\ancfile{verify\_cert\_indep.py}, \ancfile{v\_cert\_primal.py}, \ancfile{cz\_bounds.py} & the two antinorm certificates of
RS-$\theta$ (\cref{prop:rst-certs}), checked three independent ways, and the bounds on $c_{Z_i}$ & exact rationals \\
\ancfile{rst\_rate\_interval.py} & the rates of the RS-$\theta$ certificates at the endpoints of their
ranges and $2^{\psi^*}<1.7204$ (\cref{prop:rst-certs,prop:rst-rate}) & exact
rationals; Arb \\
\ancfile{lobster\_rate\_certificate.py} & the bounds $1.71927$ and $1.71928$ for the uniform lobsters
(\cref{prop:rst-caterpillar}(c)), on $4764$ intervals & exact rationals \\
\ancfile{rsc/}\allowbreak\ancfile{rsc-certificate.json.gz}, \ancfile{rsc/rsc-verify.py} (with \ancfile{rsc\_verify\_core.py} and \ancfile{rsc\_modelg.py} in \ancfile{rsc/}),
\ancfile{rsc/}\allowbreak\ancfile{rsc-credits.py}, \ancfile{rsc/}\allowbreak\ancfile{rsc-verify-independent.py} & the antinorm certificate of RSC (\cref{prop:rsc-cert}), checked by two independent programs, and the credit inequalities of
\cref{lem:rsc-rsccredits} & exact rationals \\
\ancfile{rates.py} (part 4), \ancfile{v\_rates.py} (part V1) & the inequalities at the ten endpoints
of the five intervals of the deterministic rate certificate (\cref{lem:det-cert}) & exact rationals \\
\ancfile{component\_count\_rate\_}\allowbreak\ancfile{interval.py} & the constant $1.6444$ for
\pname{Graph Motif}, \pname{Connected Vertex Cover} and its variants (\cref{lem:cc-cert}, with the rate
$0.71755<\log1.6444$)
& exact rationals, including Collatz--Wielandt bounds on spectral radii; Arb \\
\ancfile{dst\_certificates.json}, \ancfile{dst\_certificate\_}\allowbreak\ancfile{verify.py},
\ancfile{cz\_bounds.py} & the three certificates, the bounds on $c_{Z_j}e_{2,j}$ and the
constant $1.3727$ for \pname{Directed Steiner Tree} (\cref{prop:fa-dst-certs}) & exact rationals; Arb
for logarithms \\
\ancfile{paired\_rate\_certificate.py} & the constant $1.6302$ for \pname{Paired Domination}
(\cref{lem:fa-pd-rate}) & Arb \\
\bottomrule
\end{tabular}
\end{center}

The programs that produce the data files search in floating-point or decimal arithmetic; only the
checks in the table are used (of these producers, only
\ancfile{dst\_certificate\_}\allowbreak\ancfile{build.py} is included). The second table lists the programs behind the
numerical statements of the paper. Each of these statements is marked as numerical where it is made,
and no proof uses them.

\begin{center}
\renewcommand{\arraystretch}{1.12}
\begin{tabular}{@{}>{\raggedright\arraybackslash}p{5.4cm}>{\raggedright\arraybackslash}p{10.2cm}@{}}
\toprule
\multicolumn{2}{@{}l}{\emph{Numerical evidence, not used in any proof}}\\
\midrule
File & Computes \\
\midrule
\ancfile{rs\_closed\_form\_interval.py} (numerical part) & the maximum of the closed form $F^*$ for RS, about $0.8030893$, so that the constant
$1.744835$ of \cref{prop:rs-rate}(c) exceeds the base $2^{\max F^*}$ by less than $2\cdot10^{-6}$ \\
\ancfile{orthant/orthant\_try.py} & the orthant antinorm search for RS-$\theta$ that did not close
(\cref{rem:rst-cone}) \\
\ancfile{lower\_bound\_caterpillar.py} & the finite values for uniform lobsters
(\cref{prop:rst-caterpillar}) \\
\ancfile{families.py}, \ancfile{v\_extremal.py} & the bases of stars of stars and paths, and the search
over $58\,800$ periodic families (\cref{sec:rst-caterpillar}) \\
\ancfile{v\_rates.py} (part V5: $1.9440$ at $\sigma=0.4172629$, the value used in the text, and $1.9443$ at $\sigma=0.4176$, as in \ancfile{rates.py} part 3) & the black-box lower bound of about $1.9440^n$
for RS-$\theta$ (\cref{thm:dr-bb-moments}) \\
\ancfile{rates.py} (part 4, grid), \ancfile{v\_rates.py} (part V2; floating point, tolerance $10^{-11}$) & numerical evidence that other choices of $y$ would not lower the bound of
\cref{lem:det-cert}: $\max_{u}\inf_{y>0}\chi_y(u)\approx0.795823$, with grid maximizer $u\approx0.436$ \\
\ancfile{size\_demo.py} & the sizes of the balanced product families for $n\le20$ and block lengths $4$ to $6$: they contain 80--90\% of all $t$-sets (\cref{sec:app-balanced}) \\
\ancfile{motif\_path\_extremal.py} & $\mathcal Z_F\ge\mathcal Z_{\text{path}}$ on trees with at most $8$
vertices (after \cref{thm:cc-motif}) \\
\ancfile{wcds\_spider.py} & the spider values after \cref{prop:cv-spider} (\cref{sec:anc-det}) \\
\ancfile{dst\_obstruction.py} & the bound $\Omega(1.3716^n)$ for a family of instances of \pname{Directed Steiner Tree}
(\cref{sec:fa-dst}) \\
\ancfile{maxleaf-measure/}\allowbreak\ancfile{optimise.py}, \ancfile{maxleaf-measure/}\allowbreak\ancfile{sensitivity.py} & the measure
optimization for the algorithm of Binkele-Raible and Fernau (after \cref{prop:co-limits}) \\
\bottomrule
\end{tabular}
\end{center}

The files of the measure optimization, of the orthant search for RS-$\theta$ and of RSC are in
the subfolders \ancfile{maxleaf-measure/}, \ancfile{orthant/} and \ancfile{rsc/} of the
ancillary folder; the tables give the paths of these files. All other files are at
its top level. Each program is run as \texttt{python3 <program>} from the folder that contains it,
with the data files of its table row as arguments where it reads them. \ancfile{verify\_cert.py}, \ancfile{verify\_cert\_indep.py} and \ancfile{v\_cert\_primal.py}
take \ancfile{cert\_final\_0.json} and \ancfile{cert\_final\_1.json}; \ancfile{rsc-verify.py} and
\ancfile{rsc-verify-independent.py} take \ancfile{rsc-certificate.json.gz} and an optional number of
processes; and \ancfile{dst\_certificate\_build.py} takes \ancfile{dst\_certificate\_params.json}. The other programs find their data themselves. The output of each program is recorded in
the file of the same name with extension \texttt{.out}. For RSC, the outputs are
\ancfile{rsc-verify.out}, \ancfile{rsc-verify-independent.out} (with
\ancfile{rsc-verify-independent-negative.out} for the perturbed copies) and \ancfile{rsc-credits.out}. The programs need Python~3 with python-flint (tested with version $0.9.0$); some also need NumPy and SciPy. The
file \ancfile{SHA256SUMS} lists the SHA-256 checksums of all ancillary files; the certificate of RSC,
\ancfile{rsc/rsc-certificate.json.gz}, has SHA-256 checksum
\texttt{9465bd0f7f74dc3a\allowbreak 6376724645bfbde9\allowbreak 7b417035583c89ba\allowbreak 787e347813177651}.
This file ($45$\,MB) is not among the ancillary files on arXiv, which would store it unpacked; it is in the
archived copy of this release~\cite{Gaspers26companion} and on the companion site, at the same path.
The script \ancfile{verify\_all.sh} checks the checksums and runs every program of the first table
except the two that check the certificate of RSC with one command; with the option \texttt{--full} it also runs these two,
which took about $46$ and $16$ minutes with four processes in our runs, while all other checks
together took about a minute and a half on one core. The file \ancfile{README.md} lists which file checks which statement.
The ancillary files are released under the MIT license (file \ancfile{LICENSE}).

\subsection{The closed form of RS}\label{sec:anc-rs}

For \cref{prop:rs-rate}(c), the program \ancfile{rs\_closed\_form\_interval.py} covers $[0.2,0.8]$ by
$9608$ closed intervals with exact rational endpoints, obtained by adaptive bisection, and encloses $F^*$
on each of them in ball arithmetic; every enclosure lies below the enclosure of $\log1.744835$. The
largest upper bound obtained is $0.80309061427$ (rounded up), while $\log1.744835=0.80309061469\ldots$, a margin above
$4.2\cdot10^{-10}$. The same program shows $F^*\le0.7807$ on $[0.15,0.31]$ and $F^*\le0.7802$ on
$[0.54,0.8]$ in the same way, by two further coverings of these intervals, with $60$ and $73$ intervals,
separate from the first one because the targets are lower; \cref{prop:rs-rate}(c) states all three bounds,
and RS-$\theta$ uses the last two (\cref{tab:rst-coverage}).

\subsection{The certificates of RS-\texorpdfstring{$\theta$}{theta}}\label{sec:anc-rstheta}

For $i\in\{0,1\}$, the file \ancfile{cert\_final\_$i$.json} stores the data $(y_i,e_{1,i},e_{2,i},Z_i)$
of certificate~$i$ of \cref{prop:rst-certs}. Three independent programs check the hypotheses of
\cref{lem:rst-F} for both certificates in exact rational arithmetic: \ancfile{verify\_cert.py},
\ancfile{verify\_cert\_indep.py}, which uses linear programming duality instead of the stored convex
combinations, and \ancfile{v\_cert\_primal.py}. The program \ancfile{cz\_bounds.py} computes
$c_{Z_i}=\min_{z\in Z_i}\mathrm{fin}(z)$ exactly, using the code of \ancfile{verify\_cert.py}, and checks
the bounds $c_{Z_0}\ge0.6212$ and $c_{Z_1}\ge0.6293$ of \cref{prop:rst-certs} (the values are
$0.62120\ldots$ and $0.62935\ldots$), hence $c_{Z_i}\ge0.62$, the bound used in \cref{prop:rst-rate}. The program \ancfile{rst\_rate\_interval.py} encloses the
coefficients of $\psi_0$ and $\psi_1$ and their values at the four endpoints $0.31$, $0.4172629$
(twice) and $0.54$ in ball arithmetic, starting from the exact rationals $y_i,e_{1,i},e_{2,i}$; every
enclosure lies below $0.78270152<\psi^*$, with a margin $\psi^*-\max>8\cdot10^{-8}$. It also checks
$2^{\psi^*}<1.7204$.

\subsection{Uniform lobsters}\label{sec:anc-lobster}

For \cref{prop:rst-caterpillar}(c), the program \ancfile{lobster\_rate\_certificate.py} produces the
$4764$ intervals $[a,b]$ covering $[0,\frac34]$, the rational vectors $w>0$ and the rationals $\lambda>0$,
and the rational data $v>0$ and $\lambda'>0$ at $y^*=0.3453$, and checks all the inequalities stated in
the proof of \cref{prop:rst-caterpillar} for these data in exact rational arithmetic.

The following computations are in floating point and are not used in any proof. The minimum over
$y>0$ of $\log(1+y)-\frac17\log\lambda_{\max}(\mathbf M_y)$, for the nonnegative matrix $\mathbf M_y$ of the proof of
\cref{prop:rst-caterpillar} and its largest real eigenvalue $\lambda_{\max}(\mathbf M_y)$, which is its spectral
radius by the Perron--Frobenius theorem, is $0.7817982553$ at $y^*=0.34531$, which gives the numerical value
$1.7192725\ldots$ of \cref{prop:rst-caterpillar}(c). The programs \ancfile{families.py} and
\ancfile{v\_extremal.py} compute the bases of stars of stars and of paths stated after the proof, and
search $58\,800$ periodic families; none has a larger base than $\Lambda_L$.

\subsection{The certificate of RSC}\label{sec:anc-rsc}

The data of \cref{prop:rsc-cert} are stored in \ancfile{rsc/rsc-certificate.json.gz}. The points of $Z$ and of
the stages are stored as exact rationals. For each vector $P$ whose membership in
$\mathrm{conv}(\mathcal Y)+\mathcal K_{\mathrm{pos}}$ is asserted, the file lists the points $z_i$ and stores the
weights $\alpha_i$ as floating-point numbers, which the checking program converts to rationals with
denominators at most $10^{12}$ (the closest such rationals, by \texttt{Fraction.limit\_denominator}), dropping
nonpositive weights; these rationals are the coefficients $\alpha_i$ of the proof of \cref{prop:rsc-cert}, and
the program checks exactly that $\sum_i\alpha_i\ge1$ and that $P\succeq\sum_i\alpha_iz_i$, the latter by evaluating all up-set sums; the poset
cone has $43$ nonempty up-sets. There are $1024^2=1{,}048{,}576$ pair conditions for \ref{it:rsc-C1}; $26$ conditions
for \ref{it:rsc-C2}; $1848$ conditions for \ref{it:rsc-C4} ($4\le\nu\le36$, for the points $(0,c_3)$ and
the hull vertices with $c_2\ge1$ of the lattice for $\nu-1$, which the program recomputes exactly); and
$1962$ staged chains for \ref{it:rsc-C5} (the points $(0,c_3)$, $1\le c_3\le\nu$, and the hull vertices
with $c_2\ge1$ of the lattice for $\nu$, $1\le\nu\le36$), with $2{,}636{,}800$ conditions $x\circ z$ in these chains. The file stores the values of $d^{(3)}_{\min}(\nu)$ and $d^{(2)}_{\min}(\nu)$ stated in \cref{prop:rsc-cert} for $1\le\nu\le4$, and the value $0$ for $5\le\nu\le36$, where \cref{prop:rsc-cert} states the value $1$; the proof of \cref{prop:rsc-cert} shows that both values give the same conditions.
The check also confirms that all points of $Z$ are nonnegative and that
$d^{(3)}_{\min}\le d^{(2)}_{\min}$, and it computes $c_Z=0.60605\ldots$ exactly. Both programs described below also
check exactly the inequality $(q_3/\hat e)^{37}\ge\hat e^{-25}$ of~\ref{it:rsc-C6}, which covers the
tail $\nu>36$.

The checking program (\ancfile{rsc/rsc-verify.py}, with \ancfile{rsc\_verify\_core.py} and
\ancfile{rsc\_modelg.py} in the same folder) works in rational arithmetic throughout; it also evaluates the inequalities of
\cref{def:rsc-light} and~\eqref{eq:rsc-credits} in $60$-digit decimal arithmetic; this evaluation is
not used, since \cref{lem:rsc-lightpairs,lem:rsc-rsccredits} prove these inequalities. The truncated-series
bounds in the proof of \cref{lem:rsc-rsccredits} are evaluated exactly by \ancfile{rsc/rsc-credits.py},
which also rechecks the light pairs of \cref{lem:rsc-lightpairs}. A second program
(\ancfile{rsc/rsc-verify-independent.py}), written from the definitions of \cref{sec:rsc} without reusing any
code of the first, re-derives the eight-state recursion, checks it exactly against a brute-force
enumeration of the relaxed model and checks $\widetilde W\le W^{\mathrm{tm}}$ on random trees, recomputes
the lattice hulls, confirms that all $613$ stage points are nonnegative, and verifies the same
certificate in rational arithmetic, with rigorous rational bounds for the logarithms in
\cref{def:rsc-light} and~\eqref{eq:rsc-credits}; perturbed copies of the certificate are rejected by it
at the expected condition. The first program checks every relation with the rationals above as the
coefficients $\alpha_i$ of the proof. The second accepts a relation if it holds with these rationals or with
the exact binary values of the stored weights; either set of nonnegative coefficients proves the relation, so
both programs confirm the same relations, possibly with different coefficients.

The points were found by an iterative floating-point search and then rounded to rationals; only the
exact check is used.

\subsection{The deterministic rate certificate}\label{sec:anc-det}

For \cref{lem:det-cert}, the inequalities between rational numbers at the ten endpoints of the five
rational intervals, against $\log\frac{1736}{1000}$, are checked in exact rational arithmetic by \ancfile{rates.py} (part~4) and,
independently, by \ancfile{v\_rates.py} (part~V1).

Separately, for the spiders of \cref{prop:cv-spider}, the program \ancfile{wcds\_spider.py} computes $\beta_s$, the largest
value of $\bigl(\inf_{y>0}(1+y)^n/W_S(y)\bigr)^{1/n}$ over $d_1,\dots,d_s\in\{2,3\}$, by one-dimensional
minimization over $y$ in floating point. It gives $\beta_{50}=1.743507$, $\beta_{200}=1.737898$ and
$\beta_{1000}=1.736432$, decreasing toward $\sqrt5-\frac12=1.736068\ldots$. These values are numerical
evidence and are not used in any proof.

\subsection{The component-count certificate}\label{sec:anc-cc}

For \cref{lem:cc-cert}, the program \ancfile{component\_count\_rate\_interval.py} checks the inequalities
(o)--(iii) of its proof with exact rational parameters and ball arithmetic for the logarithms. In~(iii),
$\lambda_{\max}(M_{y_j})$ is bracketed between rationals by the Collatz--Wielandt inequalities
$\min_i(M_{y_j}v)_i/v_i\le\lambda_{\max}(M_{y_j})\le\max_i(M_{y_j}v)_i/v_i$ for positive
vectors~$v$, evaluated exactly; the lower bracket enters $\phi_{2,\sigma}(y_j)$, which is decreasing in
$\lambda_{\max}(M_{y_j})$, and the upper one the prefactor. Part~B of the same program rechecks the
identities and the inequality behind the value $1.8$ (\cref{cor:cc-T,lem:cc-inf}), including
$5^{45}>3^{64}$ in exact integer arithmetic; the proofs of these statements do not use it.
The largest value of $\max\bigl(\phi_{2,\sigma^-_j}(y_j),\phi_{2,\sigma^+_j}(y_j)\bigr)$ in~(iii) is
$0.7174350\ldots$, against $\psi=0.71755$ and $\log1.6444=0.7175613\ldots$; the largest prefactor
$\lambda_{\max}(M_{y_j})/\min(1,y_j)$ is at most $5.933$. The program \ancfile{motif\_path\_extremal.py} checks
the inequality $\mathcal Z_F\ge\mathcal Z_{\text{path}}$ raised after \cref{thm:cc-motif} for all trees on at
most $8$ vertices at $\zeta\in\{\frac18,\frac14,\frac12,1,2,4\}$ in exact rational arithmetic; this is
evidence only.

\subsection{Directed Steiner Tree}\label{sec:anc-dst}

The file \ancfile{dst\_certificates.json} stores the three certificates of \cref{prop:fa-dst-certs}
and the weights $\lambda_j$. The program \ancfile{dst\_certificate\_verify.py} checks the
hypotheses~\ref{it:fa-C1}--\ref{it:fa-C3} in exact rational arithmetic and that the weights
$\lambda_j$ are nonnegative and sum to~$1$. The program \ancfile{cz\_bounds.py} computes $c_{Z_j}$
exactly ($0.52305\ldots$, $0.53544\ldots$ and $0.02153\ldots$) and checks
$c_{Z_j}e_{2,j}\ge0.40$, $0.41$ and $0.011$ (\cref{prop:fa-dst-certs}), so $c_{Z_j}e_{2,j}\ge1/100$
for every~$j$, as the proof of \cref{thm:fa-dst} uses. Membership of a point $x$ in
$\mathcal R_Z$ is certified by a witness, a point of $Z$ or a rational convex combination of two points
of $Z$ that lies coordinatewise below $x$; in the plane such a witness exists whenever
$x\in\mathcal R_Z$. The logarithms in $\boldsymbol\alpha_j$ and the weighted sums
$\sum_j\lambda_j\alpha_{j,i}$ are enclosed by the same program in ball arithmetic, starting from the
exact rationals $y_j,e_{1,j},e_{2,j},\lambda_j$; the enclosure of their maximum $0.4569669\ldots$ lies
more than $3.3\cdot10^{-5}$ below $0.457$, and $0.457<\log1.3727$ is checked in the same arithmetic.
The program \ancfile{dst\_obstruction.py} computes, in floating point, the value $1.37168\ldots$ behind the bound $\Omega(1.3716^n)$ stated after
\cref{thm:fa-dst}; this is evidence only. The family consists of a directed path
$r\to p_1\to\dots\to p_{2L}$ in which each odd $p_{2i-1}$ has one terminal out-neighbor of out-degree~$0$
and one further out-neighbor with two such terminals, each even $p_{2i}$ has two such terminals, and
$\lfloor\varkappa L\rfloor$ padding vertices have arcs to and from both terminals of $p_2$. The spine and
its further out-neighbors form the only Steiner set of size at most $3L$, and as $L\to\infty$ the bound
$\bigl(\min_t\binom{n_U}{t}/\mu_t(S^*)\bigr)^{1/n}$ for this set is largest at $\varkappa\approx2.9663$, where it
equals $2^{5/(8+\varkappa)}=1.37168\ldots$.

\subsection{Paired Domination}\label{sec:anc-pd}

For the first claim of \cref{lem:fa-pd-rate}, the program \ancfile{paired\_rate\_certificate.py} encloses $y^*$ and
$h_{\bar n}(y^*)$ for $2\le\bar n\le40$ in ball arithmetic and checks $h_6(y^*)-h_{\bar n}(y^*)>0.00297$
for $\bar n\neq6$, and it checks the tail bound
$\log(1+y^*)+\frac2{41}\log\frac1{y^*}<0.431<h_6(y^*)$ in the same arithmetic, starting from
$y^*=(3\sqrt3-1)/16$. It also checks $h_6(y^*)<0.70502$ and $2^{0.70502}<1.6302$, the two inequalities
used in the proof of \cref{thm:fa-pd}.

\subsection{Formal verification}\label{sec:anc-lean}

Part of the analysis is also checked in Lean~4 with Mathlib. Except for the data of the RSC certificate
(see the end of this subsection), it is checked by the Lean kernel, using only the standard
axioms (propositional extensionality, choice and quotient soundness) and without compiled evaluation
(\texttt{native\_decide}). The statements below refer to the Lean files in the folder \texttt{formal/} of the ancillary files, a copy of the development as of 1 November 2026; the files
\texttt{lean-toolchain} and \texttt{lake-manifest.json} there pin the versions of Lean and Mathlib. They cover the following statements.
\begin{itemize}
  \item Soundness and success probability, items~(ii) and~(iii), of \cref{thm:fw-main}, for the cap and trial
    counts stated after \cref{alg:fw-mls}; \cref{lem:fw-threshold-prob}; and soundness and success
    probability of \cref{thm:fw-budgeted} for abstract trials. Also both inequalities of \cref{lem:fw-gf}: the first in the
    form of the last sentence of \cref{thm:fw-main}, the second in the pointwise form of its proof; and \cref{prop:fw-necessary} for the same cap and trial counts, with the probability
    that some trial runs $\cA$ on a subset of $S$ in place of the expected number of such trials.
  \item The Jensen step of the proof of \cref{thm:co-mean} and the success probability it gives; the
    minimization in \cref{rem:co-fsigma} ($f_\sigma$ is strictly convex on $[0,\sigma]$ and attains its minimum
    exactly at $\tau^*$); the inequality of part~(1) of \cref{cor:co-linear}, and its part~(2).
  \item \Cref{lem:cds-maxleaf}, with the spanning tree shown to exist rather than computed;
    \cref{lem:cds-pointwise,lem:cds-C}; the inequality of \cref{lem:cds-tree}, without its
    equality case; the claims of \cref{lem:cds-first-accepted} about the output; and \cref{cor:cds-budgeted} except~(T1), running time and space, for abstract trials, with its claim for
    \pname{Max Leaf Spanning Tree} when $n\ge3$.
  \item For RS-$\theta$: \cref{lem:rst-E,lem:rst-treerec,lem:rst-closure,lem:rst-F}; both certificates of
    \cref{prop:rst-certs}, checked in exact rational arithmetic, with the numerical claims of
    \cref{prop:rst-certs} and $2^{\psi^*}<1.7204$; the constants of \cref{thm:rstheta}; and \cref{prop:rst-rate}.
  \item For RS: \cref{prop:rs-rate}, with part~(c) by interval covers checked in exact rational arithmetic;
    \cref{cor:rs-moments}; and the moment bound in the proof of \cref{thm:rs}, with
    $\log1.744835<\psi_{\mathrm{RS}}$.
  \item For RSC: \cref{prop:rsc-bound} from conditions \ref{it:rsc-C1}, \ref{it:rsc-C2}, \ref{it:rsc-C4}
    and~\ref{it:rsc-C5} of a certificate with $c_Z>0.6060$, through all statements of \cref{sec:app-rsc} except
    \cref{prop:rsc-cert}; the claim of \cref{lem:rsc-trials}(a) that a trial rejects if $\gc(G)>s$, and
    \cref{lem:rsc-trials}(b); and, from the same conditions, that the output of \cref{thm:rsc} is never below $\gc(G)$, its success probability, and its claim for
    \pname{Max Leaf Spanning Tree} when $n\ge3$. The bound $\varrho<1.6991$ is checked as
    an exact rational comparison. The two Lean definitions of the closure generating function of
    \cref{def:rsc-gf} that these results use are proved equal.
  \item For the Steiner extension: the correctness of the computed value in
    \cref{lem:cv-W,lem:cds-dir-steiner,cor:cds-steiner}, that is, the inclusion--exclusion over branching walks
    on the directed encoding returns $\OPT(J)$ for every state $J$, and $\cA_{\mathrm{St}}$ solves
    $\Phi_{\mathrm{CDS}}$-Extension. So the acceptance test of step~(c) of \cref{alg:rsc-algorithm}, as the
    algorithm computes it, accepts exactly the samples that its specification accepts, and the two RSC results
    above are also proved for that test. No Lean function returns a connected dominating set or a spanning tree.
  \item \Cref{lem:det-cert}; \cref{lem:cc-cert}, with $\lambda_{\max}(M_\zeta)$ given by its closed form in
    \cref{def:cc-Z}; \cref{lem:cc-gibbs} for every finite tree and \cref{lem:cc-path} for
    paths, both with the same closed form, which is proved to be the largest eigenvalue of $M_\zeta$; part~(a) of \cref{cor:cc-T}; the final inequality $5^{45}>3^{64}$ in the proof of \cref{lem:cc-inf}; \cref{prop:fa-dst-certs}, \cref{lem:fa-dst-point,lem:fa-dst-rec,lem:fa-dst-cert} and
    \cref{eq:fa-dst-W}, which give the lower bound on $W_S(y_j)$ in the proof of \cref{thm:fa-dst}, and the constants of that proof; \cref{lem:fa-pd-point,lem:fa-pd-rate},
    the constants of \cref{thm:fa-pd} and condition~\eqref{eq:fw-star} for every paired dominating set, and \cref{prop:fa-pd-tight}, whose last claim is checked for the cap and
    trial counts stated after \cref{alg:fw-mls} and an extension algorithm that accepts exactly the \YES{}-instances
    (the step to the algorithm of \cref{lem:fa-pd-ext} is not in Lean: that algorithm accepts only \YES{}-instances,
    so a run with it accepts only on samples on which the run with the exact algorithm accepts); and the two bounds of part~(c) of \cref{prop:rst-caterpillar}, for $\Lambda_L$ and for every graph that has the
    same edges as $\Lambda_L$ at the vertices of $S^*$.
\end{itemize}
The data of the RSC certificate are checked in Lean as well, but not by the kernel alone. The program
\texttt{formal/scripts/export\_rsc\_cert.py} translates \ancfile{rsc/rsc-certificate.json.gz} into Lean data
files. Checking functions written in Lean, whose correctness is proved in the kernel with the standard axioms,
are evaluated on these data by compiled code (\texttt{native\_decide}, in $472$ generated theorems). From
this Lean proves \cref{prop:rsc-cert} as stated: the $1024$ points of $Z$ satisfy
\ref{it:rsc-C1}--\ref{it:rsc-C6}, and for the same $Z$ there are nonnegative rational stages of the stated
sizes that satisfy the relations defining stages. The same theorem contains the conditions on the
parameters in \cref{lem:rsc-cert}, which the proof of \cref{prop:rsc-cert} checks. Lean also proves the
exact value of $c_Z$, a fraction equal to $0.60605\ldots$, and proves \ref{it:rsc-C5} directly for every
sequence of the required length, the form that \cref{lem:rsc-cert} uses. With the RSC item above, this gives the success probability in
\cref{thm:rsc} with no assumption on the certificate. This check trusts the Lean compiler in addition to
the kernel; it is a third check of the data, beside the two programs of \cref{sec:anc-rsc}. The large
generated data files (about $170$\,MB) are not included: the program regenerates them, and
\texttt{formal/MaxLeafFormal/RSCData/MANIFEST.sha256} lists the SHA-256 checksums of the files that Lean checked.
Files regenerated with the exporter shipped here differ from those only in one header comment, the checksum of
the exporter; the \texttt{README.md} of the ancillary files shows how to restore it, after which all checksums match.

Not checked in Lean are running times and space bounds, the algorithms other than the Steiner tree
computation, parts~(b) and~(c) of \cref{cor:cc-T}, parts~(a) and~(b) of \cref{prop:rst-caterpillar}, the last
sentence of its part~(c) (the bracketing up to factors and the numerical limit) and the consequence stated
after it, and the graph-theoretic parts of the other applications.
{}
\section*{AI methodology}
The author used Anthropic's Claude (Claude Opus 5.5 as the main research collaborator, and Claude Fable 5.1 for
part of the analysis of Algorithm RSC), OpenAI Codex and GitHub Copilot. These tools materially affected every
section of the paper.
\begin{itemize}
  \item Claude proposed and developed ideas in discussion with the author, carried out mathematical
    derivations and proofs, wrote the code for the experiments, the computer certificates and their checking
    programs (\cref{sec:ancillary}) and the Lean development (\cref{sec:anc-lean}), and drafted and revised
    the text.
  \item OpenAI Codex was used for independent reviews of the manuscript and the code, and for developing the
    research-workflow tooling of the project.
  \item GitHub Copilot was used for code completion, reviews of pull requests and commit messages. Pull
    requests were also reviewed by Claude Haiku 5.5.
\end{itemize}
The author directed the research, contributed ideas, among them graded closure (\cref{sec:rsc}), checked the
proofs and the Lean statements, edited the manuscript and extended the literature search. The project started
from work of the author from 2020: the W[1]-hardness of \pname{MaxLeaf-Extension}
(\cref{thm:co-w1}) and the product form of multivariate Monotone Local Search, which were the basis of an
Honours thesis~\cite{Lu20}. The author verified the correctness and originality of all content, including
references.

{}

\clearpage
\bibliographystyle{alphaurl}%
\phantomsection\addcontentsline{toc}{section}{References}
\bibliography{references}
\end{document}